\documentclass[a4paper,12pt,titlepage,twoside,openright]{report}
\usepackage[english]{babel}
\usepackage[dvips]{graphicx}
\usepackage{verbatim}
\usepackage{makeidx}
\usepackage{color}
\global\oddsidemargin=6mm
\global\evensidemargin=-0.5mm

\newcommand{\sqrtsNN}{\sqrt{s_{\scriptscriptstyle{{\rm NN}}}}}
\newcommand{\av}[1]{\left\langle #1 \right\rangle}

\newcommand{\mev}{\mathrm{MeV}}
\newcommand{\gev}{\mathrm{GeV}}
\newcommand{\tev}{\mathrm{TeV}}
\newcommand{\fm}{\mathrm{fm}}
\newcommand{\mm}{\mathrm{mm}}
\newcommand{\cm}{\mathrm{cm}}
\newcommand{\m}{\mathrm{m}}
\newcommand{\mum}{\mathrm{\mu m}}
\newcommand{\s}{\mathrm{s}}

\newcommand{\mb}{\mathrm{mb}}

\newcommand{\PbPb}{\mbox{Pb--Pb}}
\newcommand{\pPb}{\mbox{p--Pb}}

\newcommand{\NN}{\mbox{nucleon--nucleon}}
\newcommand{\pp}{\mbox{proton--proton}}
\newcommand{\pA}{\mbox{proton--nucleus}}
\renewcommand{\AA}{\mbox{nucleus--nucleus}}
\newcommand{\RAA}{R_{\rm AA}}
\newcommand{\RDh}{R_{{\rm D}/h}}
\newcommand{\pt}{p_{\rm t}}
\renewcommand{\d}{{\rm d}}
\newcommand{\dEdx}{{\rm d}E/{\rm d}x}
\newcommand{\dNdy}{{\rm d}N_{\rm ch}/{\rm d}y}
\newcommand{\dNdeta}{{\rm d}N_{\rm ch}/{\rm d}\eta}

\newcommand{\QQbar}{\mbox{$\mathrm {Q\overline{Q}}$}}
\newcommand{\ppbar}{\mbox{$\mathrm {p\overline{p}}$}}
\newcommand{\ccbar}{\mbox{$\mathrm {c\overline{c}}$}}
\newcommand{\bbbar}{\mbox{$\mathrm {b\overline{b}}$}}

\newcommand{\sbbbar}{\mbox{$\scriptstyle\mathrm {b\overline{b}}$}}
\newcommand{\Dz}{\mbox{$\mathrm {D^0}$}}
\newcommand{\DtoKpi}{\mbox{${\rm D^0\to K^-\pi^+}$}}

\newcommand{\K}{\mbox{$\mathrm {K}$}}

\newcommand{\Jpsi} {\mbox{J\kern-0.05em /\kern-0.05em$\psi$}\xspace}

\newcommand{\mychapter}[1]
	{
	\chapter{#1}	
	\markboth{\thechapter.~~{#1}}{\thechapter.~~{#1}}
	}
\newcommand{\mysection}[1]
	{
	\section{#1}	
	\markright{\thesection.~~{#1}}
	}

\newcommand{\myappendix}
	{
	\renewcommand\chaptername{Appendix}
	\renewcommand{\thechapter}{\Alph{chapter}}
	}

\author{Andrea Dainese\\~\\Universit\`a degli Studi di Padova}
\title{\mbox{Charm production and in-medium QCD energy loss} 
       \mbox{in nucleus--nucleus collisions with ALICE.}\\ 
       \mbox{\Large A performance study.}}
\date{October 31$^{\rm st}$, 2003}
\begin{document}
\pagestyle{empty}
\begin{figure}[!t]
  \begin{center}
    \includegraphics[width=2.5cm]{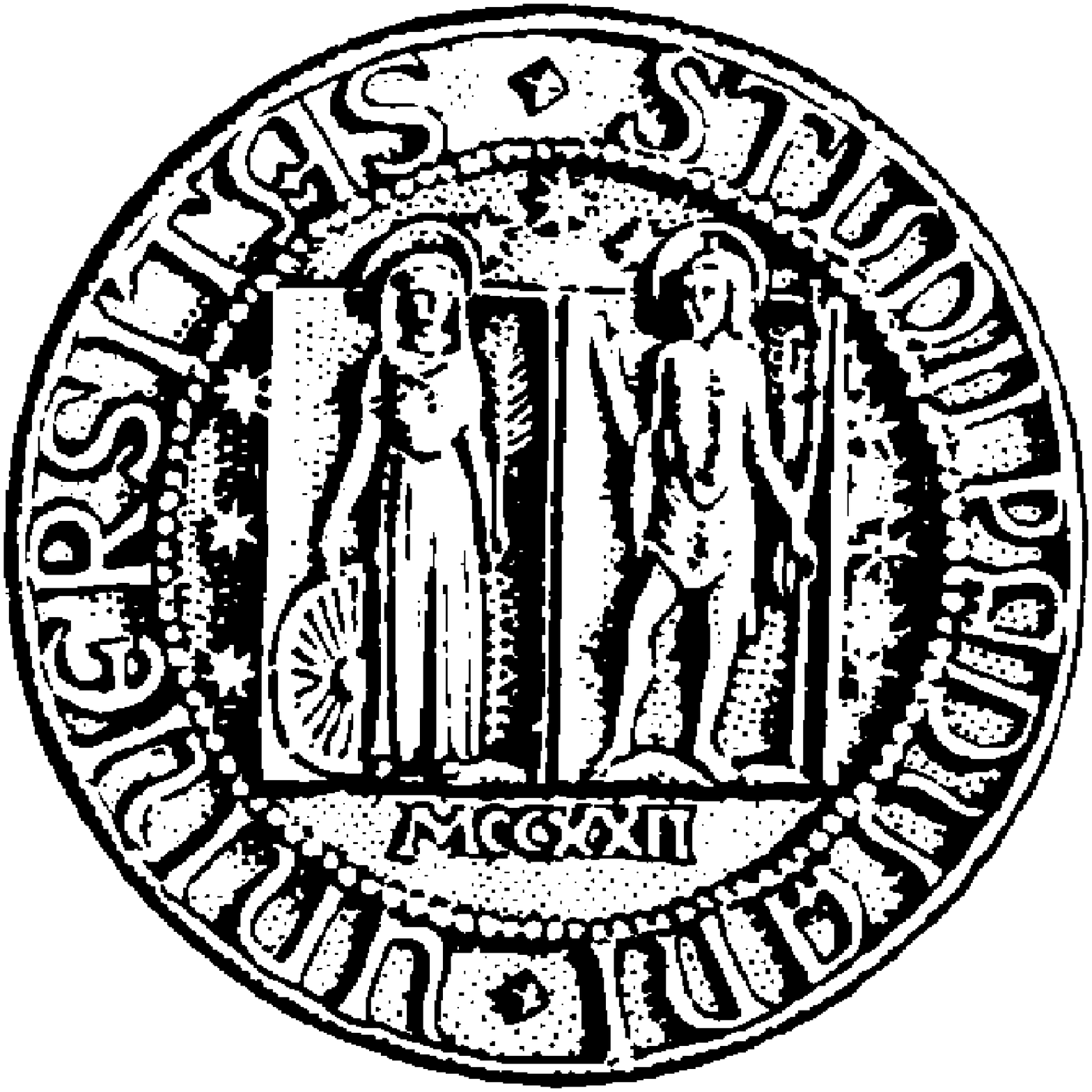}
  \end{center}
\end{figure}

\begin{center}
{\LARGE UNIVERSIT\`A DEGLI STUDI DI PADOVA}

\vglue0.2cm

{\large Sede Amministrativa: Universit\`a degli Studi di Padova}

\vglue0.2cm

{\large Dipartimento di Fisica ``G. Galilei''}

\vglue1cm

{\large DOTTORATO DI RICERCA IN FISICA}

{\large CICLO XVI}
\end{center}

\vglue1cm

\begin{center}
       \mbox{\LARGE Charm production and in-medium QCD energy loss}\\
       \vglue0.2cm 
       \mbox{\LARGE in nucleus--nucleus collisions with ALICE.}\\ 
       \vglue0.2cm 
       \mbox{\Large A performance study.}
\end{center}

\vglue3cm
\begin{tabular}{ll}
{\large {\bf Coordinatore:}} & {\large Ch.mo Prof. Attilio Stella} \\
{\large {\bf Supervisore:}}  & {\large Ch.mo Prof. Maurizio Morando}
\end{tabular}

\vglue2cm

\rightline{\large {\bf Dottorando:} Andrea Dainese}

\vglue2cm

\begin{center}
{\large 31 Ottobre 2003} 
\end{center}

\cleardoublepage
\maketitle
\cleardoublepage
\pagestyle{plain}
\pagenumbering{roman}
\setcounter{page}{1}
\tableofcontents
\cleardoublepage
\pagenumbering{arabic}
\setcounter{page}{1}
\chapter*{\centering Introduction}
\addcontentsline{toc}{chapter}{Introduction}

The search for quark--gluon plasma ---the state of deconfined strongly 
interacting matter which is thought to have constituted the 1-$\mu$s-old 
Universe--- received 
a big boost in the 1990s with the acceleration 
of heavy ions in the Super Proton Synchrotron at CERN. There, several 
fixed-target experiments gave results, on different physical observables,
indicating that a new state of matter with unusual properties is 
formed in the early stage of the collisions.
Heavy ion physics has now entered the collider era. 
Results from experiments at the Relativistic 
Heavy Ion Collider (RHIC) have provided further evidence for the long-sought 
quark--gluon plasma and encourage the study of its properties at the 
Large Hadron Collider (LHC), where energy densities of
$\sim100$-$600$ times the density of atomic nuclei will be reached
in the collisions of lead nuclei at 5.5~TeV per nucleon--nucleon pair.

The recent results from RHIC suggest that it is possible to probe
the dense medium formed in nucleus--nucleus collisions 
through the reduction in the 
production of high-momentum particles. This effect may be, indeed,
due to an energy loss, or quenching, of the partons as 
they propagate through the medium. If this is the case, the new deconfined 
phase can 
be probed and investigated by means of a `tomography' with beams of energetic 
partons.

At the LHC the probes being used at RHIC, light quarks and gluons,
will extend their energy range by one order of magnitude and a new 
type of probe will become available with fairly high cross sections:
heavy quarks. 

The large masses of the charm and beauty quarks make them qualitatively 
different probes, since, on well-established quantum chromodynamics grounds,
in-medium energy loss off massive partons is expected to be significantly
smaller than off massless partons. Therefore, a comparative study 
of the attenuation of massless (gluons and light quarks) and massive
probes is a promising tool to test the coherence of the interpretation
of quenching effects as energy loss in a deconfined medium and to further 
investigate the properties of such medium.

In this work we focus on charm physics with 
ALICE\footnote{A Large Ion Collider Experiment}, the heavy ion dedicated
experiment at the LHC. The aim is to study the ALICE capability to measure
charm production with good precision (small statistical errors) and accuracy
(small systematic errors) even in the high track-multiplicity environment 
of central lead--lead collisions and to carry out the above-mentioned 
comparative quenching studies.

The physics framework is outlined in the first part of the thesis
(Chapters~\ref{CHAP1} and~\ref{CHAP2}), where
we present the status of the experimental study of deconfinement in
heavy ion collisions and the qualitative improvement expected in this field
at the LHC collider and we detail how charm particles can serve as probes of
deconfined matter.

The experimental framework, ALICE, is described in Chapter~\ref{CHAP4}, 
in terms of layout, main sub-systems and their expected performance.

The activity carried out for this thesis can be summarized in the following 
four parts.   
\begin{itemize}
\item {\sl Definition of a baseline for heavy quarks production cross 
      sections and kinematical distributions.} 
      The HVQMNR computer program 
      for perturbative quantum chromodynamics calculations was  
      deployed to obtain and compare results at different energies  
      and for different colliding systems, taking into account known 
      nuclear collective effects. The Monte Carlo event generator PYTHIA
      was tuned in order to reproduce such results. This item is covered in 
      Chapter~\ref{CHAP3}.
\item {\sl Study of the experimental issues related to the identification 
      of the displaced decay vertices of charm mesons.} 
      Since charm particles
      have decay lengths of few tenths of a millimeter, 
      a precise reconstruction 
      of the event topology in the interaction region is mandatory for a 
      high-quality charm physics programme. The ALICE Inner Tracking System
      was designed to provide the required precision. Using the latest 
      detector geometry/response parameters and track reconstruction 
      algorithms, we carried out a systematic study of the track 
      impact parameter resolution for different particle species and 
      in different multiplicity environments, from central lead--lead to 
      proton--proton collisions. For the latter case, we developed 
      and tested a dedicated algorithm for the reconstruction of the 
      interaction vertex position in three dimensions. These items are 
      discussed in Chapter~\ref{CHAP5}.   
\item {\sl Definition of a strategy for the exclusive reconstruction of 
      charm mesons with ALICE and evaluation of the performance in terms of 
      momentum range, precision and accuracy of the measurement.}
      A preliminary study of the reconstruction of $\DtoKpi$ decays 
      had been carried out for the ALICE Proposal and Technical Design 
      Reports, using a schematic description of the detector geometry/response 
      and of the backgrounds, and only a momentum-integrated signal 
      significance had been estimated. We improved the strategy outlined 
      in those documents
      and carried out a complete and realistic simulation, from the heavy quark
      generation to the momentum-dependent estimate of statistical and 
      systematic uncertainties. Chapters~\ref{CHAP6} and~\ref{CHAP7} cover
      these topics.
\item {\sl Study and simulation of the predicted energy loss effect. 
      Assessment of a strategy to carry out comparative quenching 
      measurements and evaluation of the attainable sensitivity.}
      We considered one of the most advanced phenomenological models of 
      parton energy loss and we calculated, for different quark--gluon 
      medium densities, the effects on charm mesons and on
      hadrons originating from massless partons. We included a detailed 
      description of the \AA~collision geometry and an algorithm to take  
      into account the predicted reduced loss for heavy quarks. 
      The results of the study on the $\DtoKpi$ detection were then 
      used to assess the ALICE potential to investigate the medium with 
      massive probes.
      This part of the work, which was carried out in close collaboration with 
      the heavy ion group of the CERN Theory Division, is presented in 
      Chapter~\ref{CHAP8}.
\end{itemize}

\chapter*{\centering Introduzione}

La ricerca sperimentale del quark--gluon plasma ---lo stato 
deconfinato della materia nucleare che si ipotizza aver costituito 
l'Universo 1 $\mu$s circa 
dopo il Big Bang--- ha ricevuto un notevole impulso negli anni Novanta con 
l'accelerazione di ioni pesanti nel Super Proton Sinchrotron del CERN. L\'\i, 
numerosi esperimenti a bersaglio fisso hanno dato risultati, su diverse
osservabili fisiche, che indicano la formazione di un nuovo stato della
materia con propriet\`a insolite. Ora, la fisica degli ioni pesanti \`e 
entrata nell'era dei collisori. I risultati dagli esperimenti al Relativistic
Heavy Ion Collider (RHIC) hanno fornito ulteriore evidenza per il 
lungamente cercato quark--gluon plasma e incoraggiano lo studio delle sue 
propriet\`a al Large Hadron Collider (LHC), dove densit\`a di energia
pari a 100-600 volte quella dei nuclei atomici saranno raggiunte nelle
collisioni di nuclei di piombo a 5.5~TeV nel centro di massa 
per coppia nucleone--nucleone.

I recenti risultati del RHIC suggeriscono che \`e possibile studiare il 
mezzo denso formato in collisioni nucleo--nucleo per mezzo della 
riduzione nella produzione di particelle ad alto momento. Questo effetto
potrebbe, infatti, essere dovuto a una perdita di energia, o attenuazione, 
dei partoni mentre attraversano il mezzo. Se questo \`e il caso, la nuova
fase deconfinata pu\`o essere investigata per mezzo di una 
`tomografia' con fasci di partoni molto energetici.

A LHC le sonde usate al RHIC, quark leggeri e gluoni, estenderanno il loro 
intervallo in energia di un ordine di grandezza e un nuovo tipo di sonda
diverr\`a disponibile con sezioni d'urto elevate: i quark pesanti.

Le masse dei quark charm e beauty li rendono sonde qualitativamente 
diverse, dato che, su ben consolidate basi di cromodinamica quantistica,
ci si aspetta per i partoni pesanti una perdita di energia nel mezzo
significativamente minore che per partoni di massa trascurabile. 
Di conseguenza, uno studio comparativo con sonde leggere 
e pesanti \`e un promettente strumento per testare la coerenza 
dell'interpretazione degli effetti di attenuazione come perdita di energia
in un mezzo deconfinato e per investigare ulteriormente le propriet\`a 
del mezzo stesso.

Questo lavoro \`e incentrato sulla fisica del charm in ALICE\footnote{A
Large Ion Collider Experiment}, l'esperimento dedicato agli ioni 
pesanti a LHC. Lo scopo \`e quello di studiare la capacit\`a di ALICE di 
misurare la produzione di charm con buona precisione (basso errore
statistico) e accuratezza (basso errore sistematico) anche nell'ambiente
ad alta molteplicit\`a di tracce di una collisione piombo--piombo centrale
e di portare a termine i menzionati studi comparativi di attenuazione.

Lo scenario di fisica \`e delineato nella prima parte della tesi 
(Capitoli~\ref{CHAP1} e~\ref{CHAP2}), dove presentiamo lo stato dello 
studio sperimentale del deconfinamento in collisioni di ioni pesanti e
il miglioramento qualitativo che ci si aspetta in questo settore 
al collisore LHC e spieghiamo come le particelle con charm possano servire
da sonde della materia deconfinata.

Lo scenario sperimentale, ALICE, \`e descritto nel Capitolo~\ref{CHAP4},
in termini di apparato, sue principali componenti e le loro attese 
prestazioni. 

L'attivit\`a svolta per questa tesi pu\`o essere riassunta nelle seguenti 
quattro parti.

\begin{itemize}
\item {\sl Definizione delle sezioni d'urto di produzione 
      di quark pesanti e delle loro distribuzioni cinematiche.} Il programma
      HVQMNR per calcoli perturbativi di cromodinamica quantistica \`e stato 
      impiegato per ottenere e confrontare risultati a diverse energie 
      e per diversi sistemi ione--ione, includendo gli effetti nucleari
      noti. Il generatore di eventi Monte Carlo PYTHIA \`e stato tunato
      in modo da riprodurre questi risultati. Questo argomento \`e trattato
      nel Capitolo~\ref{CHAP3}.
\item {\sl Studio degli aspetti sperimentali legati all'identificazione
      dei vertici di decadimento di mesoni con charm.} Dato che le particelle
      con charm hanno lunghezze di decadimento di pochi decimi di millimetro,
      una precisa ricostruzione della topologia dell'evento nella regione 
      di interazione \`e necessaria per un programma di alta qualit\`a di
      fisica del charm. Il Sistema di Tracciamento Interno di ALICE \`e stato
      progettato in modo da fornire la precisione richiesta. Usando i pi\`u 
      recenti parametri sulla geometria e sulla risposta del rivelatore e 
      gli algoritmi di ricostruzione delle tracce, si \`e portato a termine
      uno studio sistematico della risoluzione sul parametro d'impatto al 
      vertice delle tracce, per diversi tipi di particelle e diversi scenari
      di molteplicit\`a, da collisioni piombo--piombo centrali a collisioni 
      protone--protone. Per quest'ultimo caso, si \`e sviluppato e testato 
      un algoritmo specifico per la ricostruzione in tre dimensioni della
      posizione del vertice di interazione. Questi argomenti sono discussi 
      nel Capitolo~\ref{CHAP5}.
\item {\sl Definizione di una strategia per la ricostruzione esclusiva di 
      mesoni con charm in ALICE e valutazione della resa in termini di 
      intervallo di momento, precisione e accuratezza della misura.} Uno 
      studio preliminare della ricostruzione di decadimenti $\DtoKpi$ era
      stato condotto per il Technical Proposal ed i Technical Design Reports
      di ALICE, usando una descrizione schematica del rivelatore e delle 
      sorgenti di fondo, e solo una significativit\`a globale (integrata 
      in momento) del segnale era stata stimata. Abbiamo migliorato la 
      strategia delineata in quei documenti e svolto una simulazione completa
      e realistica, dalla generazione dei quark pesanti fino a una 
      stima in funzione del momento di incertezze statistiche e sistematiche.
      I Capitoli~\ref{CHAP6} e~\ref{CHAP7} coprono questi soggetti.
\item {\sl Studio e simulazione del predetto effetto di perdita di energia.
      Elaborazione di una strategia per portare a termine misure comparative 
      di attenuazione e valutazione del livello di sensitivit\`a 
      raggiungibile.} Abbiamo considerato uno dei pi\`u avanzati modelli 
      fenomenologici di perdita di energia partonica e calcolato, per 
      diverse densit\`a della materia di quark e gluoni, gli effetti su mesoni
      con charm e su adroni prodotti da partoni senza massa. Abbiamo incluso 
      una descrizione dettagliata della geometria delle collisioni 
      nucleo--nucleo e un algoritmo per tenere conto della minore 
      perdita di energia predetta per i quark pesanti. I risultati dello
      studio sulla rivelazione del decadimento $\DtoKpi$ sono stati poi
      utilizzati per valutare il potenziale di ALICE per investigare
      il mezzo con sonde pesanti. Questa parte del lavoro, svolta in 
      stretta collaborazione con il gruppo che si occupa di ioni pesanti 
      della Divisione Teorica del CERN, \`e presentata nel 
      Capitolo~\ref{CHAP8}.
\end{itemize}

\setcounter{chapter}{0}
\mychapter{Heavy ion physics at the LHC: \mbox{study of deconfined QCD matter}}
\label{CHAP1}

\pagestyle{myheadings}

The aim of ultra-relativistic heavy ion physics is to study strongly 
interacting matter in 
conditions of high density and temperature; high with respect to the 
conditions characterizing the ordinary nuclear matter that constitutes
the known Universe. 

The fundamental questions in this field are: 
{\sl What is the limit of ordinary hadronic matter? What are the conditions 
beyond which separate hadrons do not retain their identity? }
In a more modern and specific language, where we talk about 
coloured quarks and their confinement into colourless hadrons, 
the questions read: 
{\sl What are the limits of confinement? 
Can the quarks be liberated from their hadronic `prison'? }

Once these questions have found answers, the next question is: 
{\sl What are the properties of de-confined matter? }
We know from cosmology that the Universe was in a deconfined state,
a soup, or plasma, of quarks and gluons, a few microseconds after its
formation. The above question is, therefore, a very fundamental one, not only 
on the nature of matter but also on the evolution of the Universe.
\\~

Even before quantum chromodynamics (QCD) had been established as the 
fundamental
theory of strong interactions, it had been argued that the basic properties
of strongly interacting hadrons must lead to some form of 
critical behaviour at high temperature and/or density.
Since a hadron has a finite size of \mbox{$\sim1~\fm^3$} (for pions), there 
is a limit to the density (and, thus, to the temperature) of a hadronic 
system beyond which hadrons start to `superimpose'~\cite{pomeranchuk}. 
Moreover, the exponentially-growing number of observed hadronic resonances as 
the energy (temperature) of the system increases indicates the existence of 
a limit temperature for 
hadronic matter~\cite{hagedorn}. 
The subsequent formulation of QCD led to the suggestion that this 
should be the limit between confined matter and a new phase of 
strongly interacting matter, the Quark--Gluon Plasma 
(QGP)~\cite{cabibboparisi}.  
More recently, lattice QCD calculations~\cite{karsch} 
have predicted that at a critical 
temperature of order $170~\mev$, corresponding to an energy density 
$\varepsilon_c\simeq 1~\gev/\fm^3$, nuclear matter undergoes a 
phase transition to a deconfined state of quarks and gluons, the QGP.
In addition, chiral symmetry is approximately restored and consequently 
quark masses are reduced from their large effective values in hadronic 
matter to small bare ones.

How to produce such phase transition in the laboratory?\\This could happen 
in ultra-relativistic heavy ion collisions, where one expects to attain energy 
densities which reach and exceed the critical value $\varepsilon_c$, thus 
making possible the transition to the deconfined state in laboratory 
experiments. The main objective of heavy ion physics is
to study this phase transition of QCD and the properties 
of the new quark--gluon plasma state. 

Over the past fifteen years, the heavy ion programmes with 
fixed-target experiments, at the AGS (Brookhaven) and the SPS (CERN) 
accelerators, and, more recently with colliding-beams experiments, 
at the RHIC (Brookhaven), have allowed to establish experimental evidence
of the phase transition. 
The Large Hadron Collider, with Pb beams collided at a 
centre-of-mass energy per nucleon pair $\sqrtsNN=5.5~\tev$
(more than a factor 20 larger than the RHIC energy), will be the next 
generation facility for the physics of deconfined QCD matter and should allow
a significant qualitative improvement with respect to the previous programmes.
After the SPS and RHIC experiments have provided answers to the first set of 
questions,
showing that there is a limit to confined matter, the task of the LHC 
heavy ion programme is to address the next question and investigate
the properties of deconfined quark--gluon matter. 

In this chapter, after a brief summary of the phenomenology of hot and 
dense nuclear matter (Section~\ref{CHAP1:qgp}) and of some of the most 
relevant results of the SPS and RHIC experiments 
(Sections~\ref{CHAP1:sps} and~\ref{CHAP1:rhic}), 
we address the specific and novel aspects 
of heavy ion physics at the LHC (Sections~\ref{CHAP1:lhc} 
and~\ref{CHAP1:novelaspects}) with particular 
focus on those which are related to the charm or, more generally, 
the hard probes sector.

\mysection{Phenomenology of hot and dense matter}
\label{CHAP1:qgp}

\subsection{The QCD phase diagram}
\label{CHAP1:phasediagram}

\begin{figure}[!b]
  \begin{center}
    \includegraphics[width=0.75\textwidth]{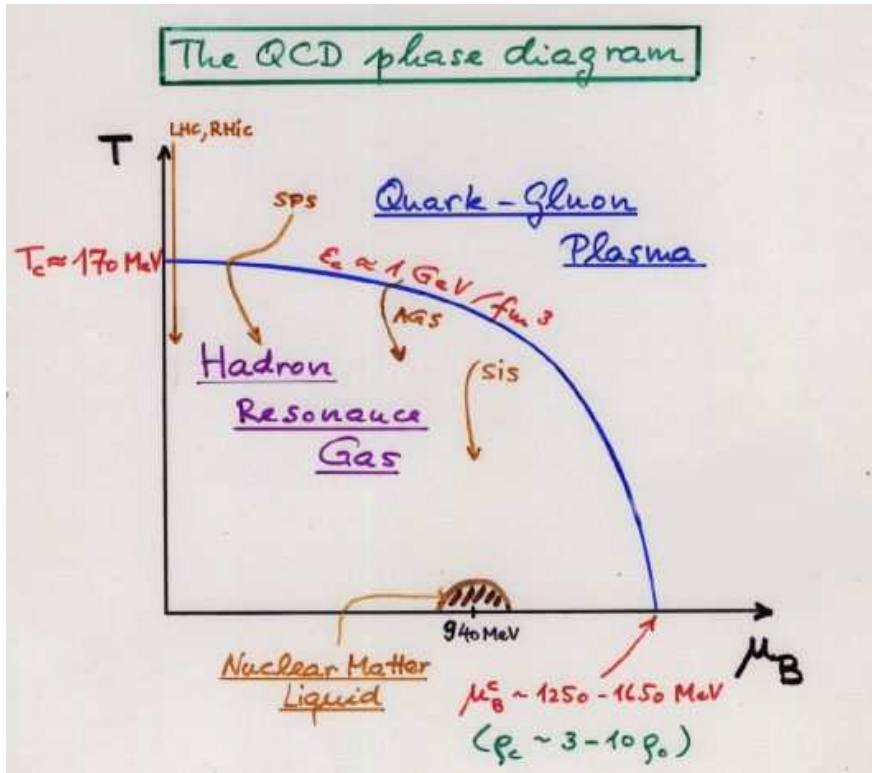}
  \end{center}
  \caption{Phase diagram of QCD matter~\cite{heinz}.}
  \label{fig:phasediagram}
\end{figure}

On the basis of thermodinamical considerations and of QCD calculations, 
strongly interacting matter is expected to exist in different states.
Its behaviour, as a function of the baryonic chemical 
potential\footnote{The baryonic chemical potential $\mu_{\rm B}$ of a system 
is defined as the change in the energy $E$ of the system when the total 
baryonic number $N_{\rm B}$ (baryons $-$ antibaryons) is increased by one 
unit: $\mu_{\rm B}=\partial E/\partial N_{\rm B}$.} 
$\mu_{\rm B}$ (a measure of the baryonic density) 
and of the temperature $T$, is displayed in the phase diagram 
reported in Fig.~\ref{fig:phasediagram}. At low temperatures 
and for $\mu_{\rm B}\simeq m_{\rm p}\simeq 940~\mev$, 
we have ordinary matter. Increasing the energy density
of the system, by `compression' (towards the right) or by `heating' 
(upward), a hadronic gas phase is reached in which nucleons interact and 
form pions, excited states of the proton and of the neutron 
($\Delta$ resonances) and other hadrons. If the energy density is
further increased, the transition to the deconfined QGP phase is predicted:
the density of partons (quarks and gluons) becomes so high that 
the confinement of quarks in hadrons vanishes. 

The phase transition can be reached along different `paths' on the 
$(\mu_{\rm B},\,T)$ plane. In the primordial Universe, 
the transition QGP-hadrons, from the deconfined to the confined phase, 
took place at $\mu_{\rm B}\approx 0$ (the global baryonic number was 
approximately zero) 
as a consequence of the expansion of the Universe and 
of the decrease of its temperature (path downward along the 
vertical axis)~\cite{cosmoastro}.
On the other hand, in the formation of neutron stars, the gravitational 
collapse causes an increase in the baryonic density at temperatures 
very close to zero (path towards the right along the 
horizontal axis)~\cite{cosmoastro}.
   
In heavy ion collisions, both temperature and density increase, possibly 
bringing the system to the phase transition.
In the diagram in Fig.~\ref{fig:phasediagram} the paths
estimated for the fixed-target (SIS, AGS, SPS) and collider (RHIC, LHC) 
experiments are shown. 

\subsection{Lattice QCD results}
\label{CHAP1:latticeqcd}

Exploring from a theoretical point of view the qualitative features 
of the QGP and making quantitative predictions about its properties is 
the central goal of the numerical studies of strongly interacting matter 
thermodynamics within the framework of lattice QCD~\cite{wilson,karsch}. 

Phase transitions are related to large-distance phenomena in a thermal medium.
Because of the increasing strength of QCD interactions with the 
distance, such phenomena cannot be treated using perturbative methods.
Lattice QCD provides a first-principle approach that allows to study 
large-distance aspects of QCD and to partially account for non-perturbative 
effects. 
However, at present, most calculations are limited by the
fact that they do not include a finite baryo-chemical potential $\mu_{\rm B}$
(i.e. they assume a baryonic density equal to zero). 
 
\begin{figure}[!t]
  \begin{center}
    \includegraphics[width=0.75\textwidth]{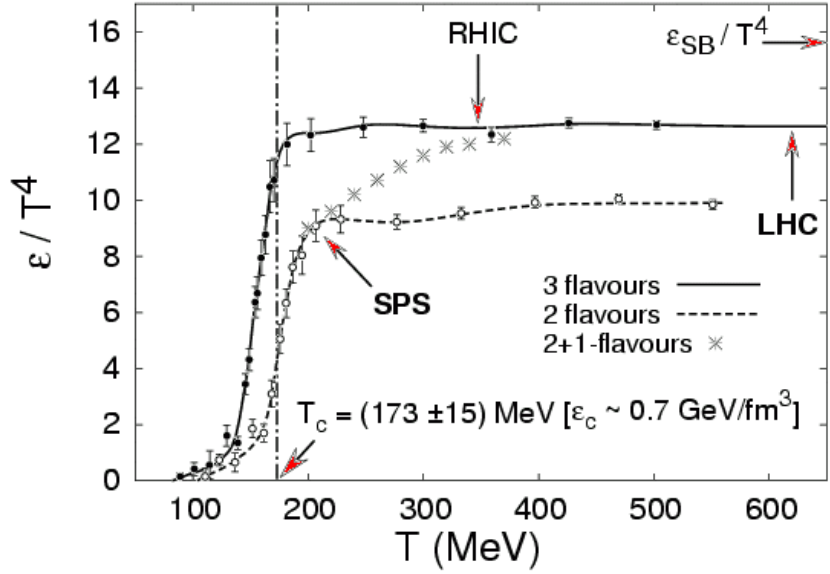}
  \end{center}
  \caption{The energy density in lattice QCD with 2 and 3 light quarks 
           and with 2 light plus 1 heavier (strange) quarks~\cite{karsch}.
           The calculation uses $\mu_{\rm B}=0$.}
  \label{fig:latticeqcd}
\end{figure}

Results of a recent calculation of $\varepsilon/T^4$ are shown in 
Fig.~\ref{fig:latticeqcd}, for 2- and 3-flavours QCD with light quarks
and for 2 light plus 1 heavier (strange) quark (indicated by the 
stars)~\cite{karsch}.
The latter case is likely to be the closest to the physically realized 
quark mass spectrum. 
The number of flavours and the masses of the quarks constitute the main 
uncertainties in the determination of the critical temperature 
and critical energy density. The critical temperature is estimated to be
$T_c=(175\pm 15)~\mev$ and the critical energy density 
$\varepsilon_c\simeq(6\pm 2)\,T_c^4\simeq(0.3$-$1.3)~\gev/\fm^3$.
Most of the uncertainty on $\varepsilon_c$ arises from the 10\% uncertainty 
on $T_c$.

Although the transition is not a first 
order one (which would be characterized by a discontinuity of 
$\varepsilon$ at $T=T_c$), a large `jump' of $\Delta\varepsilon/T_c^4\simeq 8$ 
in the energy density is observed in a temperature interval of 
only about $40~\mev$ (for the 2-flavours calculation). 
Considering that the energy density of an 
equilibrated ideal gas of particles with $n_{\rm dof}$ degrees of freedom is
\begin{equation}
\varepsilon = n_{\rm dof}\,\frac{\pi^2}{30}\,T^4,
\end{equation}
the dramatic increase of $\varepsilon/T^4$ can be interpreted as due to the 
change of $n_{\rm dof}$ from 3 in the pion gas phase to 37 (with 2 flavours)
in the deconfined phase, where the additional colour and quark flavour 
degrees of freedom are available\footnote{In a pion gas the degrees of freedom
are only the 3 values of the isospin for $\pi^+,\,\pi^0,\,\pi^-$. In a
QGP with 2 quark flavours the degrees of freedom are 
$n_g+7/8\,(n_q+n_{\bar{q}})=N_g(8)\,N_{\rm pol}(2)+7/8\,\times 2\times N_{\rm flav}(2)\,N_{\rm col}(3)\,N_{\rm spin}(2)=37$. The factor 7/8 accounts for the 
difference between Bose-Einstein (gluons) and Fermi-Dirac (quarks) 
statistics.}.

\mysection{Evidence for deconfinement in heavy ion collisions: the SPS 
programme}
\label{CHAP1:sps}

The desire to test this fascinating phase structure of 
strongly interacting matter first led to the fixed-target experiments at the 
AGS in Brookhaven (with $\sqrtsNN\simeq 5~\gev$) and at the CERN-SPS
(with $\sqrtsNN\simeq 17~\gev$). In 1986/87, the programme started with 
lighter ion beams (O, S, Si) on heavy ion targets (Au, Pb), and in 1994/95, 
heavy ion beams followed, with Au--Au collisions at the AGS and 
\PbPb~collisions at the SPS.

\begin{figure}[!t]
  \begin{center}
    \includegraphics[width=0.49\textwidth]{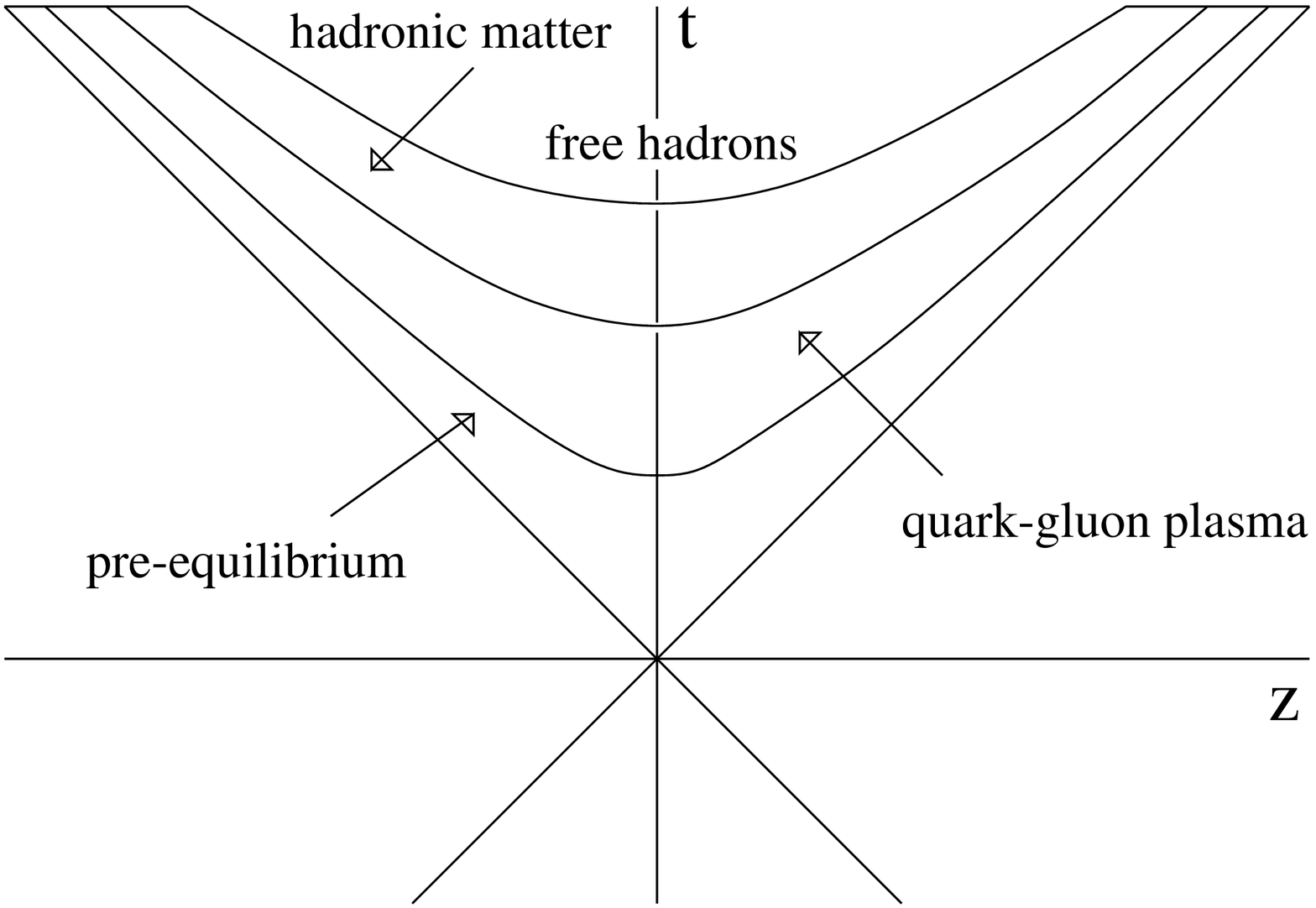}
    \includegraphics[width=0.49\textwidth]{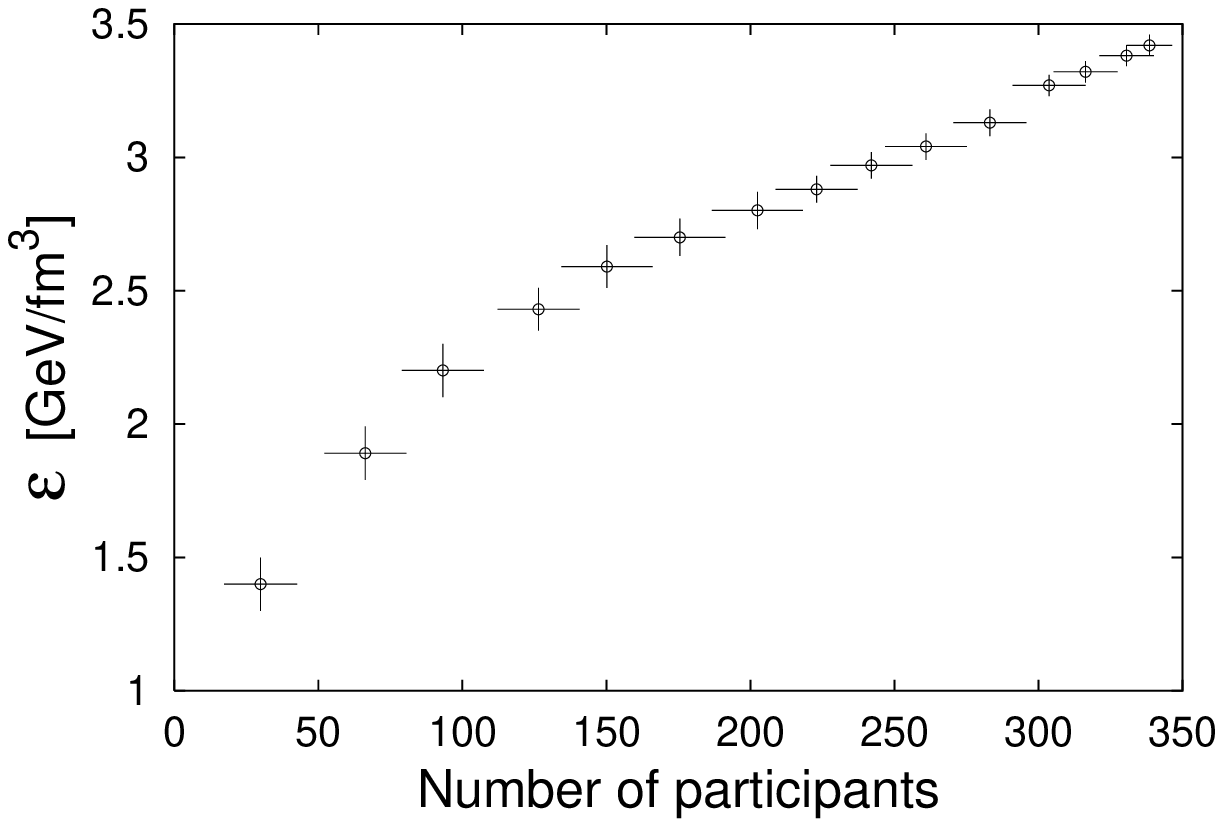}
  \end{center}
  \caption{Left: the expected evolution of a high-energy nuclear collision.
           Right: the energy density in \PbPb~collisions at the 
           SPS~\cite{epsilonNA50}.}
  \label{fig:collevolANDepsilon}
\end{figure}

The evolution of a high-energy \AA~collision is usually pictured
in the form shown in Fig.~\ref{fig:collevolANDepsilon} (left). 
After a rather short 
equilibration time $\tau_0\simeq 1~\fm/c$ (at the SPS), 
the presence of a thermalized medium is assumed, and for sufficiently-high 
energy densities, this medium would be in the quark--gluon plasma phase.
Afterwards, as the expansion reduces the energy density, the system goes 
through a hadron gas phase and finally reaches the freeze-out, when the 
final state hadrons do not interact with each other anymore. 
The choice of heavy nuclei allows to maximize the energy and the volume in 
which the energy density is very large. The energy density 
at the time of local thermal equilibration can be determined using the Bjorken 
estimate~\cite{bjorkenepsilon}:
\begin{equation}
\label{eq:bjorkenepsilon}
\varepsilon = \left(\frac{{\rm d}N_{\rm h}}{{\rm d}y}\right)_{y=0}\times \frac{w_{\rm h}}{\pi R_{\rm A}^2 \tau_0},
\end{equation}
where $({\rm d}N_{\rm h}/{\rm d}y)_{y=0}$ specifies the number of hadrons
emitted per unit of rapidity\footnote{The longitudinal rapidity 
of a particle with four-momentum $(E,\vec{p})$ is defined as 
\mbox{$y=\frac{1}{2}\,\ln\left(\frac{E+p_z}{E-p_z}\right)$}, 
being $z$ the direction of the beam(s).} at mid-rapidity and $w_{\rm h}$ their 
average energy in the direction transverse to the beam axis. The effective 
initial volume is determined in the transverse plane by the nuclear radius 
$R_{\rm A}$, and longitudinally by the formation time $\tau_0$ of the 
thermal medium.

The energy density was measured in \PbPb~collisions at $\sqrtsNN=17~\gev$
at the SPS by the NA50 experiment~\cite{epsilonNA50}. 
In Fig.~\ref{fig:collevolANDepsilon} (right) $\varepsilon$
is plotted as a function of the centrality of the collision, determined by the 
number of participant nucleons; it covers the range from 1 to 
$3.5~\gev/\fm^3$. Lattice calculations, as already mentioned, give for the 
energy density at deconfinement, $\varepsilon(T_c)$, values around 
or slightly below $1~\gev/\fm^3$.

We describe here the two clearest pieces of evidence for the production of a
deconfined medium in \PbPb~collisions at the SPS. Both of these 
effects were predicted in the eighties:
\begin{itemize}
\item enhancement of the production of strange and multi-strange 
      baryons (hyperons) with respect 
      to the rates extrapolated from pp data (predicted by 
      J.~Rafelski and B.~M\"uller in 1982~\cite{rafelskimuller});
\item suppression of the production of the J$/\psi$ meson (the lowest 
      $\ccbar$ bound state), always with respect
      to the rates extrapolated from pp (predicted by T.~Matsui and 
      H.~Satz in 1986~\cite{matsuisatz}). 
\end{itemize}

\begin{figure}[!t]
  \begin{center}
    \includegraphics[width=.9\textwidth]{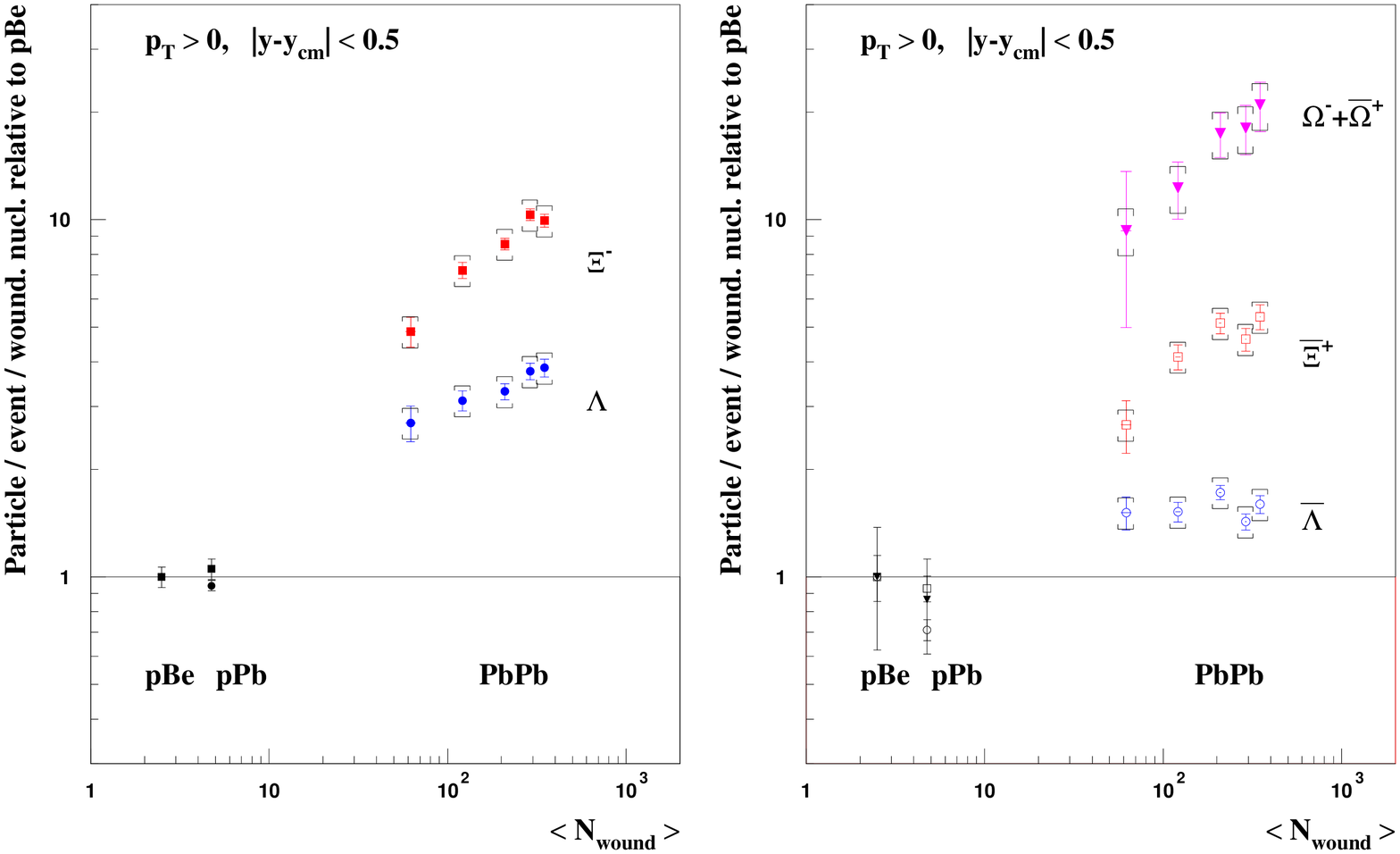}
  \caption{Strange baryon production in \PbPb~per participant nucleon, 
           normalized
           to the ratio from p--Be, as a function of the number of participant
           nucleons, as measured by NA57 at the SPS~\cite{strangenessNA57}.}
  \label{fig:strangeness}
   \vglue0.3cm
    \includegraphics[width=0.55\textwidth]{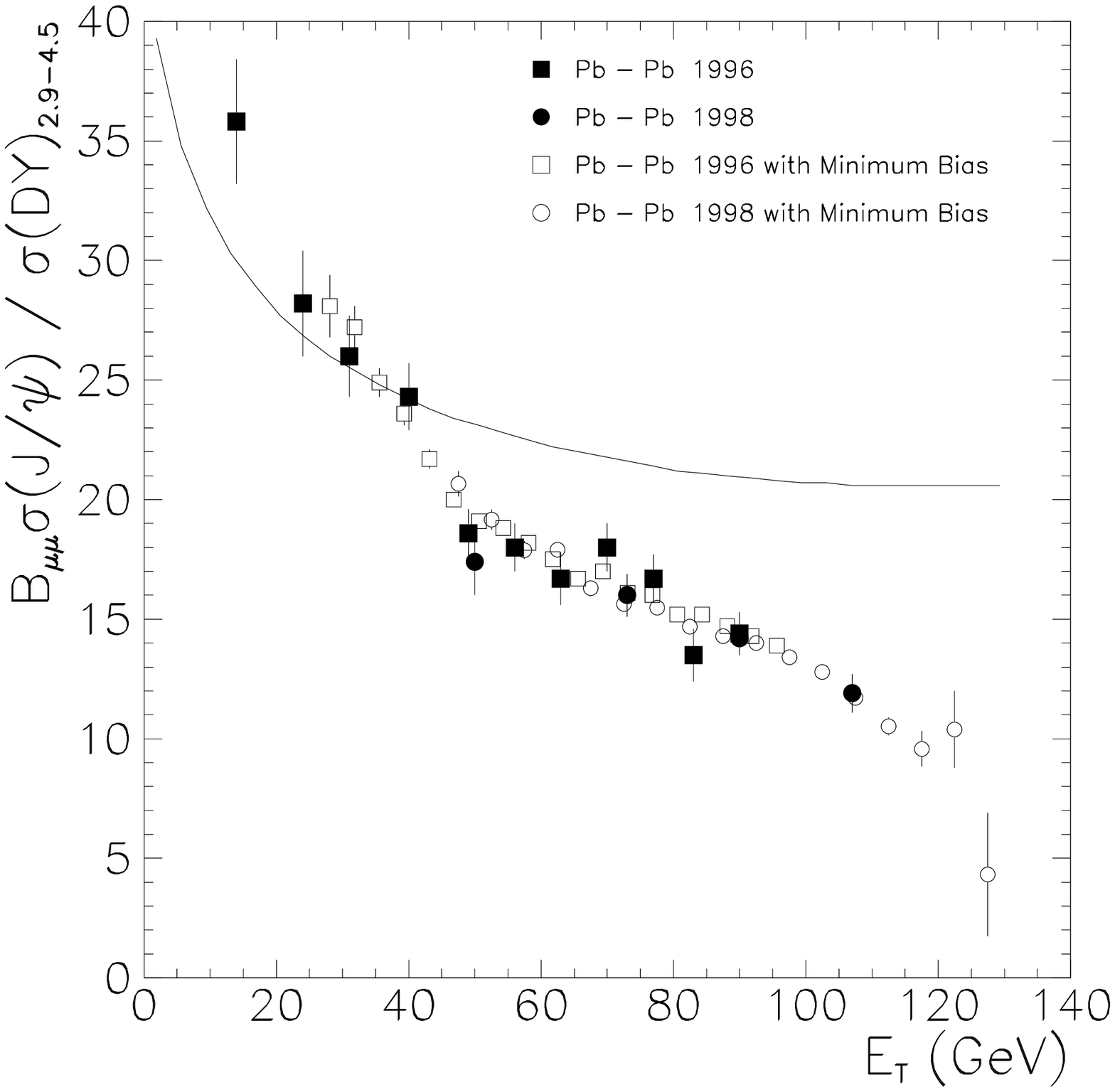}
  \caption{The ratio of J$/\psi$ to Drell-Yan production as a function of 
           the transverse energy, measured by NA50,
           in \PbPb~collisions at the SPS. The solid line indicates the 
           extrapolation of the normal nuclear absorption inferred from 
           pA collisions~\cite{epsilonNA50}.}
  \label{fig:jpsiNA50}
  \end{center}
\end{figure}

In the QGP, the chiral symmetry restoration decreases the threshold 
for the production of a $\rm s\overline{s}$ pair from twice the 
constituent mass of the s quark, $\approx 600~\mev$, to twice the {\sl bare}
mass of the s quark, $\approx 300~\mev$, which is less 
than half of the energy required to produce strange particles 
in hadronic interactions. 
In the QGP multi-strange baryons can be produced by statistical 
combination of strange (and non-strange) quarks, while in an hadronic gas 
they have to be produced through a chain of interactions that 
increase the strangeness content in steps of one unit. 
For this reason an hyperon enhancement growing with the strangeness content
was indicated as a signal for QGP formation.
This effect was, indeed, observed by the WA97/NA57
experiment: in Fig.~\ref{fig:strangeness} one can see that the production 
of strange and multi-strange baryons increases by 10 times and more
(up to 20 times for the $\Omega$) in central \PbPb~collisions in 
comparison to \mbox{p--Be}, where the QGP is not 
expected. As predicted, the enhancement $\mathcal{E}$ is increasing with 
the strangeness
content: $\mathcal{E}(\Lambda)<\mathcal{E}(\Xi)<\mathcal{E}(\Omega)$.

Also the other historic predicted signal of deconfinement was clearly 
observed, by the NA50 experiment: in Fig.~\ref{fig:jpsiNA50} the suppression 
of the J$/\psi$ particle with respect to the Drell-Yan process 
$q\overline{q}\to \ell^+\ell^-$, used as a reference, is shown as a
function of the centrality, measured by the energy $E_{\rm T}$ emitted in the 
transverse plane, in \PbPb~collisions. The line represents the 
expected trend of normal nuclear absorption extrapolated from 
\pA~measurements. The additional suppression, clearly visible for 
central collision ($E_{\rm T}>60$-$80~\gev$), is interpreted as due to 
the fact that, in the high colour-charge density environment of a QGP,   
the strong interaction between the two quarks of the $\ccbar$ pair is 
screened and the formation of their bound state is consequently prevented.

The results of the SPS programme at CERN, and in particular the enhancement 
of strangeness production and the J$/\psi$ suppression, allowed to 
conclude that in \PbPb~collisions at these energies a new state of matter 
is formed in which the effects of quark confinement appear to be 
removed~\cite{cernpressrelease,cernspsexperiments}.

\mysection{RHIC: focus on new observables}
\label{CHAP1:rhic}

The Relativistic Heavy Ion Collider (RHIC) in Brookhaven began 
operation during summer 2000. With a factor 10 increase in the centre-of-mass
energy with respect to the SPS, $\sqrtsNN$ up to $200~\gev$, 
the produced collisions are expected to be well above the phase transition 
threshold. Moreover, in this energy regime, the so-called 
`hard processes' ---production 
of energetic partons ($E>3$-$5~\gev$) out of the inelastic scattering of two 
partons from the colliding nuclei--- have a significantly large cross section 
and they become experimentally accessible. 

In this scenario, beyond the `traditional' observables
we have already introduced, the interesting phenomenon of in-medium
parton energy loss~\cite{bjorkenjets,gyulassywang,bdmps}, predicted 
for the first time by J.D.~Bjorken in 1982~\cite{bjorkenjets}, 
can be addressed. Since the study of the sensitivity for the 
measurement of charm quarks energy loss at the LHC is one of the physics goals
of this thesis work, a detailed description of the current theoretical 
view of this phenomenon will be given in Chapter~\ref{CHAP2}. For the moment
we will limit ourselves to a simplified description.

\begin{figure}[!t]
  \begin{center}
  \includegraphics[width=.49\textwidth]{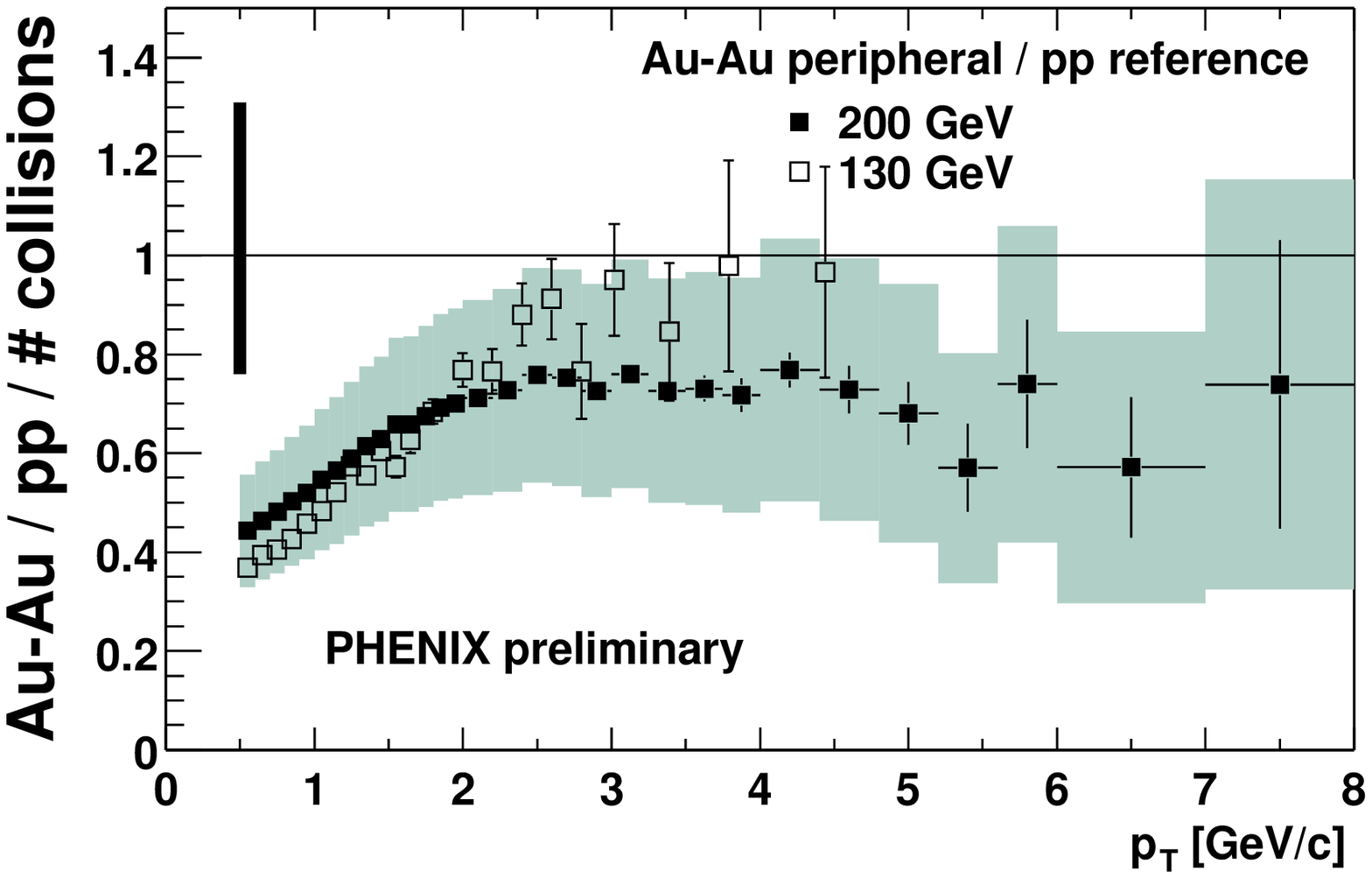}
  \includegraphics[width=.49\textwidth]{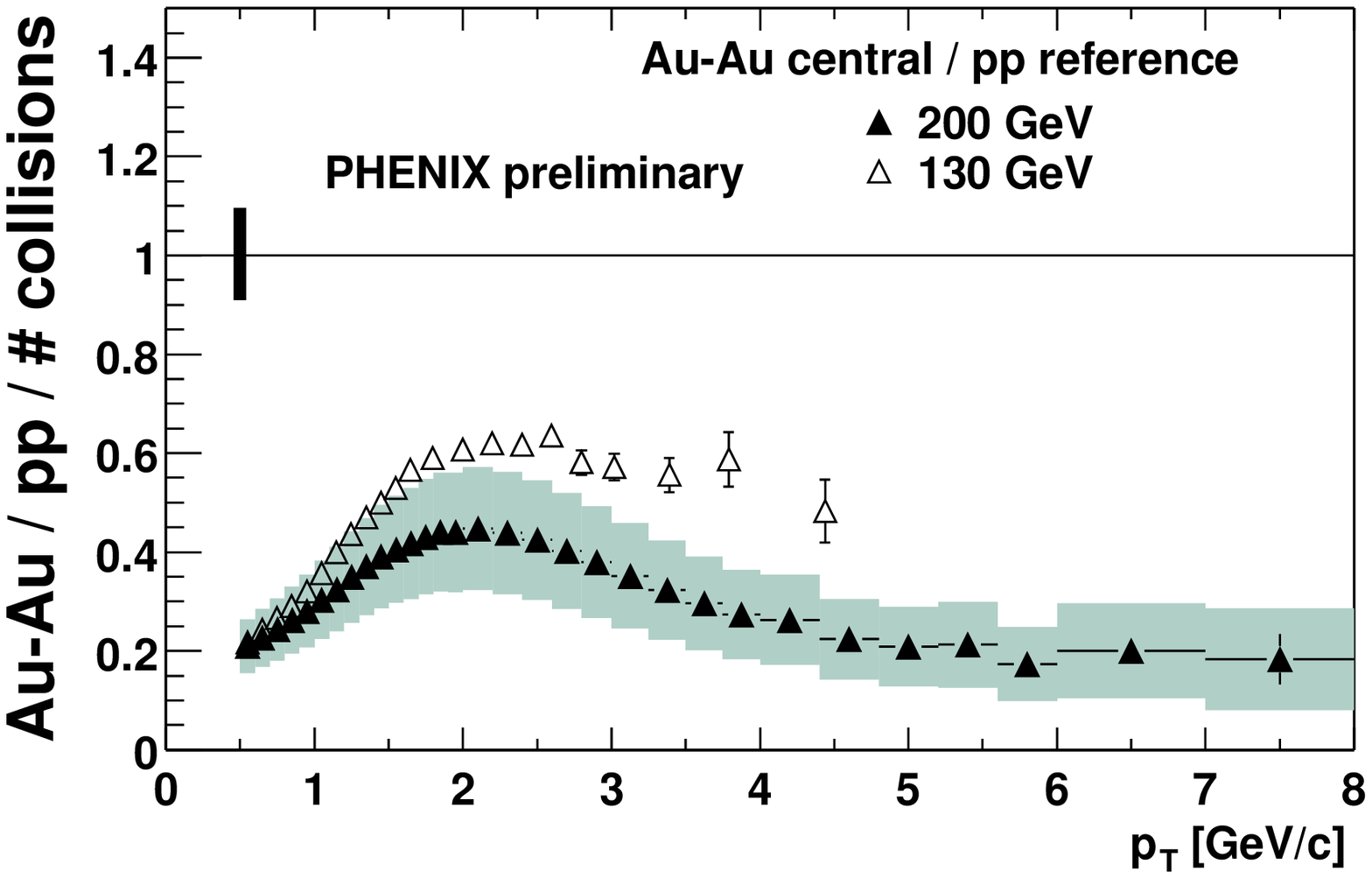}
  \caption{The ratio of transverse momentum distributions of 
           charged hadrons in Au--Au collisions and pp 
           collisions, scaled by the number of binary nucleon--nucleon 
           collisions, at $\sqrtsNN=130$ and $200~\gev$~\cite{phenixRAA}.}
  \label{fig:phenixRAA}
   \vglue0.3cm
  \includegraphics[width=\textwidth]{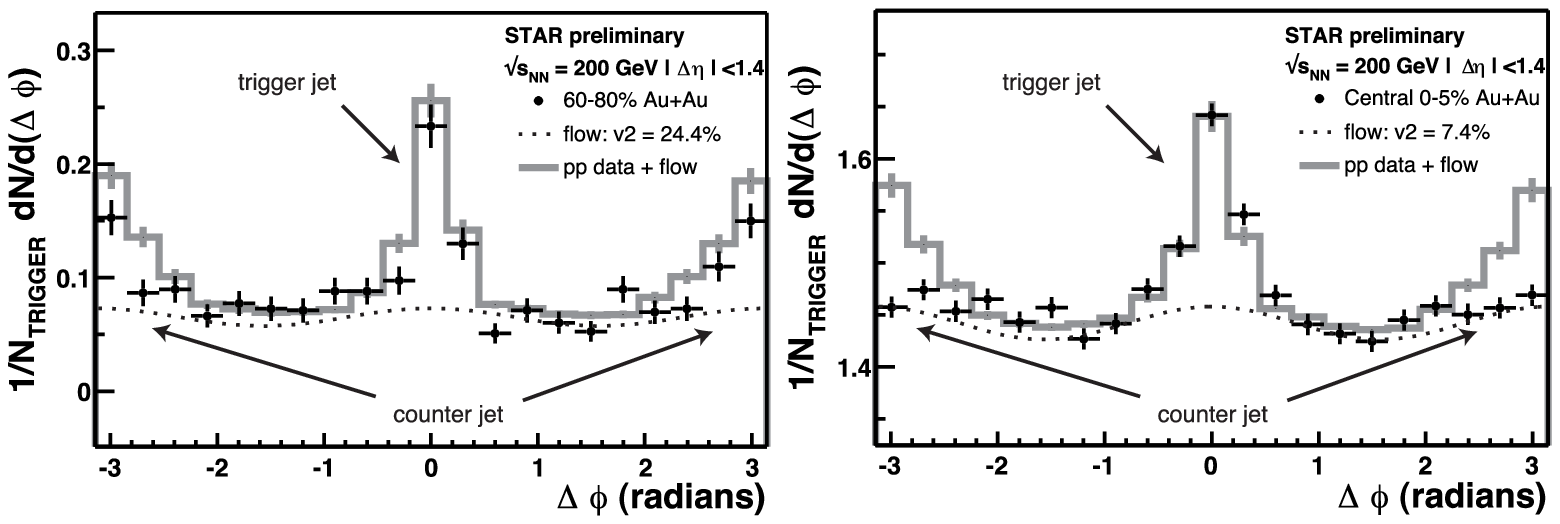}
  \caption{Azimuthal correlations of charged particles relative to a 
           high-$\pt$ trigger particle for peripheral (left) and 
           central (right) Au--Au collisions at 
           $\sqrtsNN=200~\gev$~\cite{starAzi}.}
  \label{fig:starAzi}
  \end{center}
\end{figure}

Hard partons are produced at the early stage of the collision and they 
propagate through the medium formed in the collision. During this 
propagation they undergo QCD interactions with the gluons present in the 
medium and they lose energy. Such energy loss is not peculiar of a 
deconfined medium, but, quantitatively, it is strongly dependent on the 
nature and on the properties of the medium, being predicted to be much 
larger in the case of deconfinement. This last point can be intuitively 
understood considering that, if the parton travels through a deconfined 
medium, it finds much harder gluons to interact with than it would in a
confined medium, where the gluons are constrained to carry only a very 
small fraction of the total hadron momentum, which is shared mainly among 
the valence quarks.

The measurement of high-$\pt$ (projection of the momentum on the plane 
transverse to the beam line) particle production is addressed at RHIC mainly 
by the PHENIX and STAR experiments. The results, although still 
preliminary, have aroused considerable interest.
Figure~\ref{fig:phenixRAA} reports the yield of charged hadrons measured 
by PHENIX~\cite{phenixRAA} in peripheral (left) and 
central (right) Au--Au collisions at 
$\sqrtsNN=130$ and $200~\gev$, divided by the yield in pp collisions
(scaled to the same energy) and by the estimated number of binary 
nucleon--nucleon collisions. This ratio should be 1 at high $\pt$ if no medium
effects are present. In central 
collisions the yield of high-$\pt$ ($8~\gev/c$)
hadrons is reduced of a factor 4 
with respect to what expected for incoherent
production in nucleon--nucleon collisions.

Another interesting result, obtained by STAR and PHENIX, is the gradual 
disappearing of the back-to-back azimuthal correlations of high-$\pt$ 
particles with increasing collision centrality~\cite{starAzi,phenixAzi}. 
In Fig.~\ref{fig:starAzi}
the azimuthal correlations of charged particles with respect to 
a high-$\pt$ trigger particle (black markers) 
are shown for peripheral (left) and central (right) collisions
and compared with reference data from pp collisions (grey histogram):
in central collisions the opposite-side ($\Delta\phi=\pm \pi$)
correlation is strongly suppressed with respect to the pp
and peripheral Au--Au cases. This effect suggests 
the absorption of one of the two jets (usually produced as 
back-to-back pairs) in the hot matter formed in central collisions.

The effects of leading particle and jet suppression shown in 
Figs.~\ref{fig:phenixRAA} and~\ref{fig:starAzi} are not observed 
in d--Au (deuteron--gold) collisions at 
$\sqrtsNN=200~\gev$~\cite{dGoldPhenix,dGoldStar}, 
where the formation of a dense medium is not expected. 

\mysection{LHC: study of `deeply deconfined' matter}
\label{CHAP1:lhc}

The Large Hadron Collider is scheduled to start operation in 2007. It 
will provide nuclear collisions at a centre-of-mass energy 30 times 
higher than at RHIC, opening a new era for the field, in which particle 
production will be dominated by hard processes, and the energy densities 
will possibly be high enough to treat the generated quark--gluon plasma
as an ideal gas. These qualitatively new features will allow to address
the task of the LHC heavy ion programme: a {\sl systematic study of the 
properties of the quark--gluon plasma state. }

\subsection{Systems, energies and expected multiplicity}
\label{CHAP1:sqrtsdNdy}

The ion beams will be accelerated in the LHC at a momentum of 
$7~\tev$ per unit of Z/A, where A and Z are the mass and atomic numbers
of the ions, respectively. Thus, a generic ion $({\rm A,Z})$ will have momentum
$p({\rm A,Z})={\rm (Z/A)}\, p^{\rm p}$, where \mbox{$p^{\rm p}=7~\tev$} 
is the momentum for a proton beam. The centre-of-mass (c.m.s.) energy
per nucleon--nucleon pair in the collision of two generic 
nuclei ${\rm (A_1,Z_1)}$ and ${\rm (A_2,Z_2)}$ is:
\begin{equation}
  \sqrtsNN = \sqrt{(E_1+E_2)^2-(\vec{p}_1+\vec{p}_2)^2}\simeq\sqrt{4\,p_1\,p_2} = \sqrt{\frac{\rm Z_1 Z_2}{\rm A_1 A_2}}\,14~\tev.
\end{equation}

The initial LHC running programme foresees~\cite{pprCh2}:
\begin{itemize}
\item Regular pp runs at $\sqrt{s}=14~\tev$
\item 1-2 years with \PbPb~runs at $\sqrtsNN=5.5~\tev$
\item 1 year with \pPb~runs at $\sqrtsNN=8.8~\tev$ (or \mbox{d--Pb} or
      \mbox{$\alpha$--Pb})
\item 1-2 years with \mbox{Ar--Ar} at $\sqrtsNN=6.3~\tev$
\end{itemize}

As we have seen for SPS and RHIC, the \pp~and \pA~runs are
mandatory for comparison of the results obtained with \PbPb~collisions;
we will detail this point during the discussion on the hard probes
in Section~\ref{CHAP1:hardprobes}. The runs with lighter ions (e.g. argon) 
will allow to vary the energy density and the volume of the produced system.
At least for what concerns the 
hard observables, the fact of having different c.m.s. energies for the 
different systems is not expected to introduce large uncertainties 
in the comparisons, because perturbative QCD (pQCD) calculations can be used
quite safely for the extrapolation to different energies 
(for example to scale the results measured in pp at $14~\tev$ to 
the energy of \PbPb, $5.5~\tev$). In Chapter~\ref{CHAP7}
a strategy for this extrapolation will be presented and discussed,
for charm production.\\~   

\begin{figure}[!t]
  \begin{center}
  \includegraphics[width=.7\textwidth]{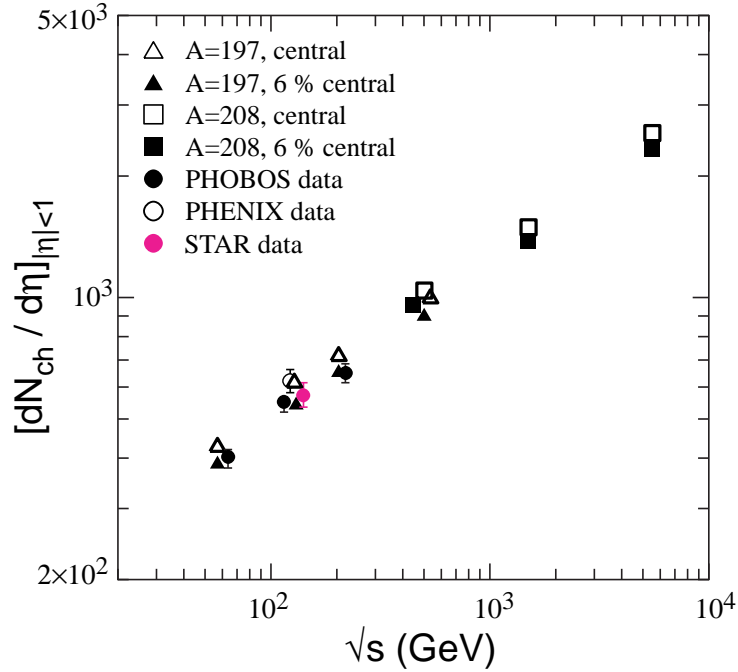}
  \caption{Charged multiplicity per unit of 
           pseudorapidity $\dNdeta$ 
           ($\simeq \dNdy$, at $y=0$) in AA collisions. 
           Square and triangle markers 
           are model predictions~\cite{eskola2}. 
           Circle markers are measurements at RHIC.}
  \label{fig:lhcdNchdy}
  \end{center}
\end{figure}

The most important global observable is the average charged particle 
multiplicity per rapidity 
unit ($\dNdy$) in central \PbPb~collisions. On the theoretical 
side, since it is related to the attained energy density (see Bjorken's formula
in equation~(\ref{eq:bjorkenepsilon})), it enters the calculation of most 
other observables. On the experimental side, the particle multiplicity
fixes the main unknown in the detector performance and the accuracy 
with which many observables can be measured.

There is no first principle calculation of $\dNdy$ starting from the QCD
Lagrangian, since particle production is dominated by soft non-perturbative
QCD. Therefore, the large variety of available models of heavy ion 
collisions gives a wide range of predicted multiplicities. 
Before RHIC, the predictions for the LHC reached up to more than 
8000 charged particles per unit of rapidity. The multiplicity measured 
at RHIC, $\dNdy\simeq 650$ at $\sqrtsNN=200~\gev$, is about a factor 2 
lower than what was predicted by most models. 
In the light of this result, the multiplicity 
at the LHC is not expected to be larger than $3000$-$4000$ charged particles
per unit of rapidity.
Figure~\ref{fig:lhcdNchdy} presents the result of a model~\cite{eskola2} 
that well reproduces the multiplicities measured at RHIC. It predicts 
\mbox{$\dNdy\simeq \dNdeta\simeq 2500$} for central \PbPb~at the LHC
\footnote{The pseudorapidity is defined as 
   $\eta=-\ln[\tan(\theta/2)]$, where $\theta$ is the polar angle with respect
   to the beam direction. For a particle with velocity 
   $v\to c$, $\eta\approx y$.}. 

Since this thesis work started before the first results from RHIC were 
available, the simulations were performed using $\dNdy=6000$. However, this 
(probably) over-estimated value provides a safety factor 
on the obtained results, which were, for completeness, 
extrapolated also to $\dNdy=3000$.   

\subsection{Why `deep deconfinement'?}
\label{CHAP1:deepdeconfinement}

Starting from the estimates of the charged multiplicity many
parameters of the medium produced in the collision 
can be inferred. Table~\ref{tab:spsrhiclhc} presents a comparison 
of the most relevant parameters for SPS, RHIC and LHC energies~\cite{eskola1}.

At the LHC, the high energy in the collision centre of mass is expected to 
determine a large energy density and an initial temperature at least a factor 
2 larger than at RHIC. This high initial temperature extends also the 
life-time and the volume of the deconfined medium, since it has to expand 
while cooling down to the freeze-out temperature, which is $\approx 170~\mev$
(it is independent of $\sqrt{s}$, above the SPS energy). In addition, 
the large expected number of gluons favours energy and momentum exchanges, 
thus considerably reducing the time needed for the thermal equilibration
of the medium. To summarize, the LHC will produce {\sl hotter, 
larger and longer-living} `drops' of QCD plasma than the present 
heavy ion facilities.

The key advantage in this new `deep deconfinement' scenario is that the 
quark--gluon plasma studied by the LHC experiments 
will be much more similar to the quark--gluon plasma that can be investigated 
from a theoretical point of view by means of lattice QCD. 

As mentioned, lattice calculations are mostly performed for a baryon-free  
system ($\mu_{\rm B}=0$). In general, $\mu_{\rm B}=0$ is not valid for heavy 
ion collisions, since the two colliding nuclei carry a total baryon number 
equal to twice their mass number. However, the baryon content of the system 
after the collision is expected to be concentrated rather near the rapidity of 
the two colliding nuclei. Therefore, the larger the rapidity of the 
beams, with respect to their center of mass, the lower the 
baryo-chemical potential in the central rapidity region. The rapidities 
of the beams at SPS, RHIC and LHC are 2.9, 5.3 and 8.6, respectively. 
Clearly, the LHC is expected to be much more baryon-free than RHIC and SPS 
and, thus, closer to the conditions simulated in lattice QCD.

\begin{table}[!t]
\caption{Comparison of the parameters characterizing central \AA~collisions
         at different energy regimes~\cite{eskola1}.}
\label{tab:spsrhiclhc}
\begin{center}
\begin{tabular}{ll|ccc}
\hline
\hline
Parameter & & SPS & RHIC & LHC \\
\hline
$\sqrtsNN$ & [GeV] & 17 & 200 & 5500 \\
d$N_{\rm gluons}/$d$y$ & & $\simeq 450$ & $\simeq 1200$ & $\simeq 5000$ \\
$\dNdy$ & & 400 & 650 & $\simeq 3000$ \\
Initial temperature & [MeV] & 200 & 350 & $>600$ \\
Energy density & [$\gev/\fm^3$] & 3 & 25 & 120 \\
Freeze-out volume & [$\fm^3$] & few $10^3$ & few $10^4$ & few $10^5$ \\
Life-time & [$\fm/c$] & $<2$ & $2$-$4$ & $>10$ \\ 
\hline
\hline
\end{tabular}
\end{center}
\end{table}

In addition to this effect, also the higher temperature predicted for the LHC
favours the comparison with theory.
This point can be better understood by going back to the lattice results for 
$\varepsilon/T^4$ (Fig.~\ref{fig:latticeqcd}). If we now concentrate 
on the result obtained with 2+1 flavours, 
2 light quarks plus a heavier one, we notice that $\varepsilon/T^4$ 
continues to rise for $T>T_c$, indicating that significant non-perturbative effects, not fully accounted 
for in the lattice formalism, are to be expected at least up to 
temperatures $T\simeq (2$-$3)\,T_c$. In Ref.~\cite{kajantieQM2002} the 
strong coupling constant in this range is estimated as
\begin{equation}
 \alpha_s(T) = \frac{4\pi}{18\,\ln(5\,T/\Lambda_{\rm QCD})}=\left\{
\begin{array}{ll}
0.43 & {\rm for}~T=T_c \\
0.3  & {\rm for}~T=2\,T_c \\
0.23 & {\rm for}~T=4\,T_c \\
\end{array}\right.
\end{equation}
using the fact that the QCD scaling constant $\Lambda_{\rm QCD}$
is of the same order of magnitude as $T_c$, $\approx 200~\mev$.
These values confirm that non-perturbative effects are larger in  
the range $T<2\,T_c$.

The conditions produced in heavy ion collisions at SPS and RHIC
are contained in this range ($T_{\rm SPS}\approx 1.2\times T_c$ and 
$T_{\rm RHIC}\approx 2\times T_c$), meaning that in these cases the comparison 
of experimentally determined quantities, such as temperature or 
energy density, to lattice QCD calculations is not fully reliable.
With an initial temperature of $\sim (4$-$5)\,T_c$ predicted for central 
\PbPb~collisions at $\sqrtsNN=5.5~\tev$, the LHC will 
provide closer-to-ideal conditions (i.e. with smaller non-perturbative 
effects), allowing a direct comparison to the theoretical calculations.
In this sense, the regime that will be realized at the LHC 
may be defined as `deep deconfinement'.

\mysection{Novel aspects of heavy ion physics at the LHC}
\label{CHAP1:novelaspects}

Heavy ion collisions at the LHC access not only a quantitatively different
regime of much higher energy density but also a qualitatively new regime, 
mainly because:
\begin{enumerate}
\item {\sl High-density parton distributions } are expected to dominate 
      particle production.
      The number of low-energy partons (mainly gluons) in the two colliding
      nuclei is, therefore, expected to be so large as to produce a 
      significant shadowing effect 
      (described later) that suppresses the inelastic scatterings 
      with low momentum transfer. 
\item {\sl Hard processes } should
      contribute significantly to the total AA cross section.
      The hard probes are at the LHC an ideal experimental tool for a 
      detailed characterization of the QGP medium.
\end{enumerate}

In the following we discuss these two aspects.

\subsection{Low-$x$ parton distribution functions}
\label{CHAP1:x}

In the inelastic collision of a proton (or, more generally, nucleon) with 
a particle, the Bjorken $x$ variable is defined 
as the fraction of the proton momentum carried by the parton that 
enters the hard scattering process. The distribution of $x$ for a given 
parton type (e.g. gluon, valence quark, sea quark) is called
Parton Distribution Function (PDF) and it gives the
probability to pick up a parton with momentum fraction $x$ from the proton.
The main experimental knowledge on the proton PDFs 
comes from Deep Inelastic Scattering (DIS) data, in particular from HERA data 
for the small-$x$ region. Several groups (MRST~\cite{mrst}, 
CTEQ~\cite{cteq}, GRV~\cite{grv}) have developed parameterizations 
of these data in the framework of DGLAP (Dokshitzer-Gribov-Lipatov, 
Altarelli-Parisi) QCD evolution~\cite{DGLAP}. An example of proton PDFs
will be shown at the end of the next paragraph.

\subsubsection{Accessible $x$ range}
\label{CHAP1:x1x2}

The LHC will allow to probe the parton distribution functions of the nucleon 
and, 
in the case of \pA~and \AA~collisions, also their modifications
in the nucleus, down to unprecedented low values of $x$. 
In this paragraph we compare the values of $x$ corresponding to the 
production of a $\ccbar$ pair at SPS, RHIC and LHC energies and we
estimate the $x$ range that can be 
accessed with ALICE for what concerns heavy flavour production. 
This information is particularly valuable because
the charm and beauty production cross sections 
at the LHC are significantly affected by parton dynamics in the 
small-$x$ region, as we will see in Chapter~\ref{CHAP3}. 
Therefore, the measurement of heavy flavour production may provide 
information on the nuclear parton densities. 

We can consider the simple case of the production of a heavy quark
pair, $Q\overline{Q}$, through the leading order\footnote{Leading order
(LO) is $\mathcal{O}(\alpha_s^2)$;
next-to-leading order (NLO) is $\mathcal{O}(\alpha_s^3)$. More 
details on QCD cross section calculations will be given in 
Section~\ref{CHAP2:pQCD}.}  
gluon--gluon fusion process
$gg\to Q\overline{Q}$ in the collision 
of two ions $({\rm A}_1,{\rm Z}_1)$ and $({\rm A}_2,{\rm Z}_2)$. 
The $x$ range actually probed depends on the value of the c.m.s. energy
per nucleon pair $\sqrtsNN$, 
on the invariant mass\footnote{For two particles with 
four-momenta $(E_1,\vec{p}_1)$ and $(E_2,\vec{p}_2)$,
the invariant mass is defined as the modulus of the total four-momentum:
$M=\sqrt{(E_1+E_2)^2-(\vec{p}_1+\vec{p}_2)^2}$.}
$M_{Q\overline{Q}}$ of the $Q\overline{Q}$
pair produced in the hard scattering and on the rapidity 
$y_{Q\overline{Q}}$ of the pair. If the parton intrinsic transverse momentum 
in the nucleon is neglected, the four-momenta of the two incoming gluons are 
$(x_1,0,0,x_1)\cdot ({\rm Z_1/A_1})\,\sqrt{s_{\rm pp}}/2$ and 
$(x_2,0,0,-x_2)\cdot ({\rm Z_2/A_2})\,\sqrt{s_{\rm pp}}/2$,
where $x_1$ and $x_2$ are the momentum fractions carried by the gluons, 
and $\sqrt{s_{\rm pp}}$ is the c.m.s. energy for pp 
collisions ($14~\tev$ at the LHC).
The square of the invariant mass of the $Q\overline{Q}$ pair is given by:
\begin{equation}
\label{eq:sx1x2M2}
  M^2_{Q\overline{Q}}=\hat{s}=x_1\,x_2\,s_{\scriptscriptstyle \rm NN}=x_1\,\frac{{\rm Z}_1}{{\rm A}_1}\,x_2\,\frac{{\rm Z}_2}{{\rm A}_2}\,s_{\rm pp};
\end{equation}
and its longitudinal rapidity in the laboratory is:
\begin{equation}
\label{eq:rapidityx1x2}
  y_{Q\overline{Q}} = \frac{1}{2}\ln \left[\frac{E+p_z}{E-p_z}\right] = \frac{1}{2}\ln\left[\frac{x_1}{x_2}\cdot\frac{{\rm Z}_1\,{\rm A}_2}{{\rm Z}_2\,{\rm A}_1}\right].
\end{equation}

From these two relations we can derive the dependence of $x_1$ and $x_2$ on 
colliding system, $M_{Q\overline{Q}}$ and $y_{Q\overline{Q}}$:
\begin{equation}
  x_1 = \frac{{\rm A}_1}{{\rm Z}_1}\cdot\frac{M_{Q\overline{Q}}}{\sqrt{s_{\rm pp}}}\exp\left({+y_{Q\overline{Q}}}\right)~~~~~~~~~~~~~~~~ 
  x_2 = \frac{{\rm A}_2}{{\rm Z}_2}\cdot\frac{M_{Q\overline{Q}}}{\sqrt{s_{\rm pp}}}\exp\left({-y_{Q\overline{Q}}}\right); 
\end{equation}
which simplifies to
\begin{equation}
\label{eq:yx1x2}
  x_1 = \frac{M_{Q\overline{Q}}}{\sqrtsNN}\exp\left({+y_{Q\overline{Q}}}\right)~~~~~~~~~~~~~~~~ 
  x_2 = \frac{M_{Q\overline{Q}}}{\sqrtsNN}\exp\left({-y_{Q\overline{Q}}}\right)
\end{equation}
for a symmetric colliding system ($\rm A_1=A_2$, $\rm Z_1=Z_2$).

At central rapidities we have $x_1\simeq x_2$ and their magnitude
is determined by the ratio of the pair invariant mass to the c.m.s. energy.
For production at the threshold 
(\mbox{$M_{\rm \scriptstyle {c\overline{c}}}=2\,m_{\rm c}\simeq 2.4$~GeV}, 
\mbox{$M_{\rm \scriptstyle {b\overline{b}}}=2\,m_{\rm b}\simeq 9$~GeV}) 
we obtain what reported in Table~\ref{tab:xtable}. The $x$ regime 
relevant to charm production at the LHC ($\sim 10^{-4}$) is about 2 orders 
of magnitude lower than at RHIC and 3 orders of magnitude lower than at 
the SPS.

\begin{table}[!t]
  \caption{Bjorken $x$ values corresponding to charm and beauty production 
           at threshold at central rapidity.}
  \label{tab:xtable}
  \begin{center}
  \begin{tabular}{c|cccc}
\hline
\hline 
Machine & SPS & RHIC & LHC & LHC \\
System & \PbPb & Au--Au & Pb--Pb & pp \\
$\sqrtsNN$ & 17 GeV & 200 GeV & 5.5 TeV & 14 TeV \\
\hline
$\ccbar$ & $x\simeq 10^{-1}$ & $x\simeq 10^{-2}$ & $x\simeq 4\cdot 10^{-4}$ &  $x\simeq 2\cdot 10^{-4}$ \\
$\bbbar$ & -- & -- & $x\simeq 2\cdot 10^{-3}$ & $x\simeq 6\cdot 10^{-4}$ \\
\hline
\hline
  \end{tabular}
  \end{center}
\end{table}

\begin{figure}[!t]
  \begin{center}
    \includegraphics[width=.7\textwidth]{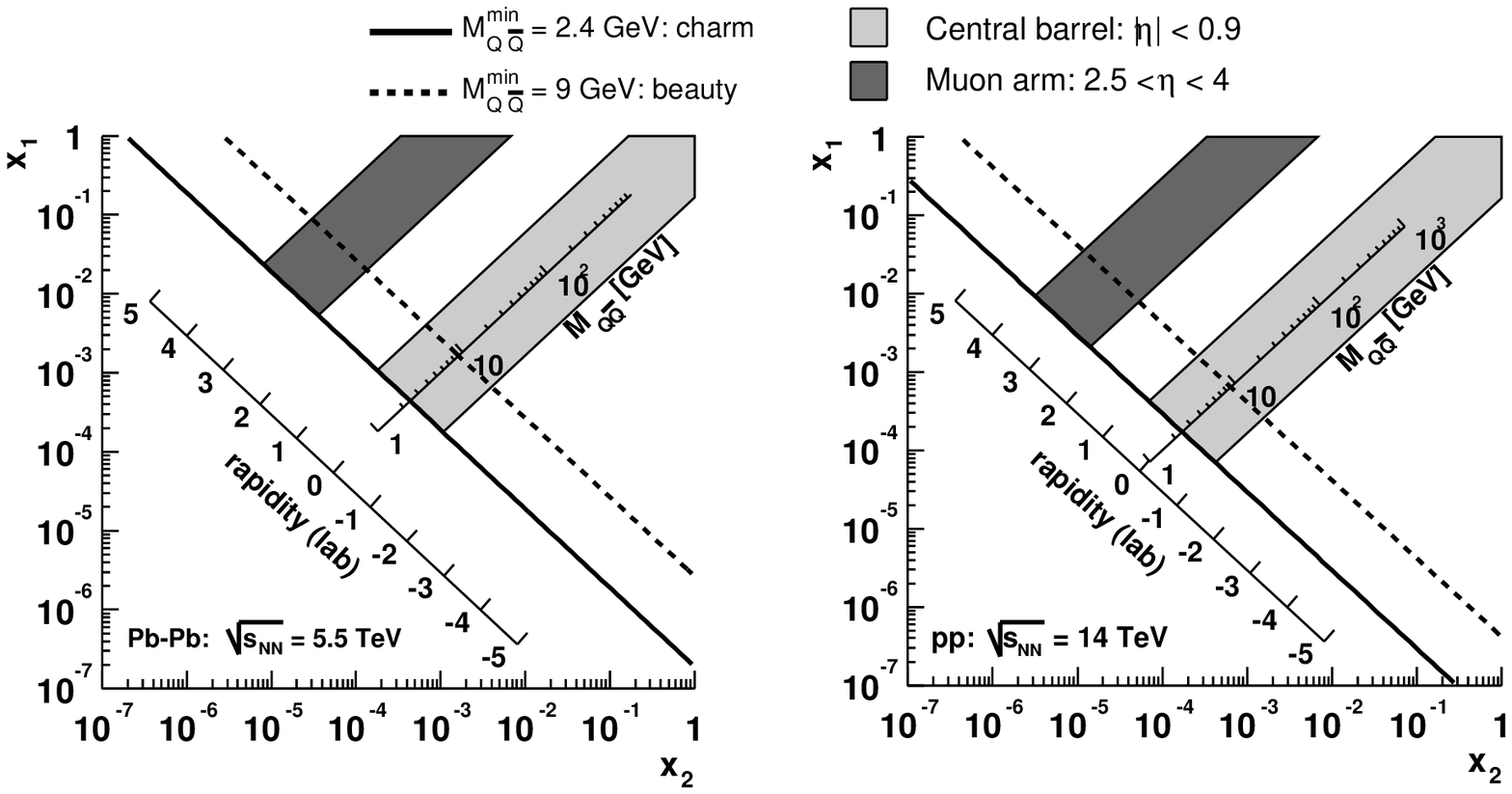}
  \caption{ALICE acceptance in the ($x_1$, $x_2$) plane for heavy flavours 
           in \mbox{Pb--Pb} (left) and in pp (right). The figure is explained 
           in detail in the text.}
  \label{fig:x1x2_AApp}
  \vglue0.2cm
    \includegraphics[width=.7\textwidth]{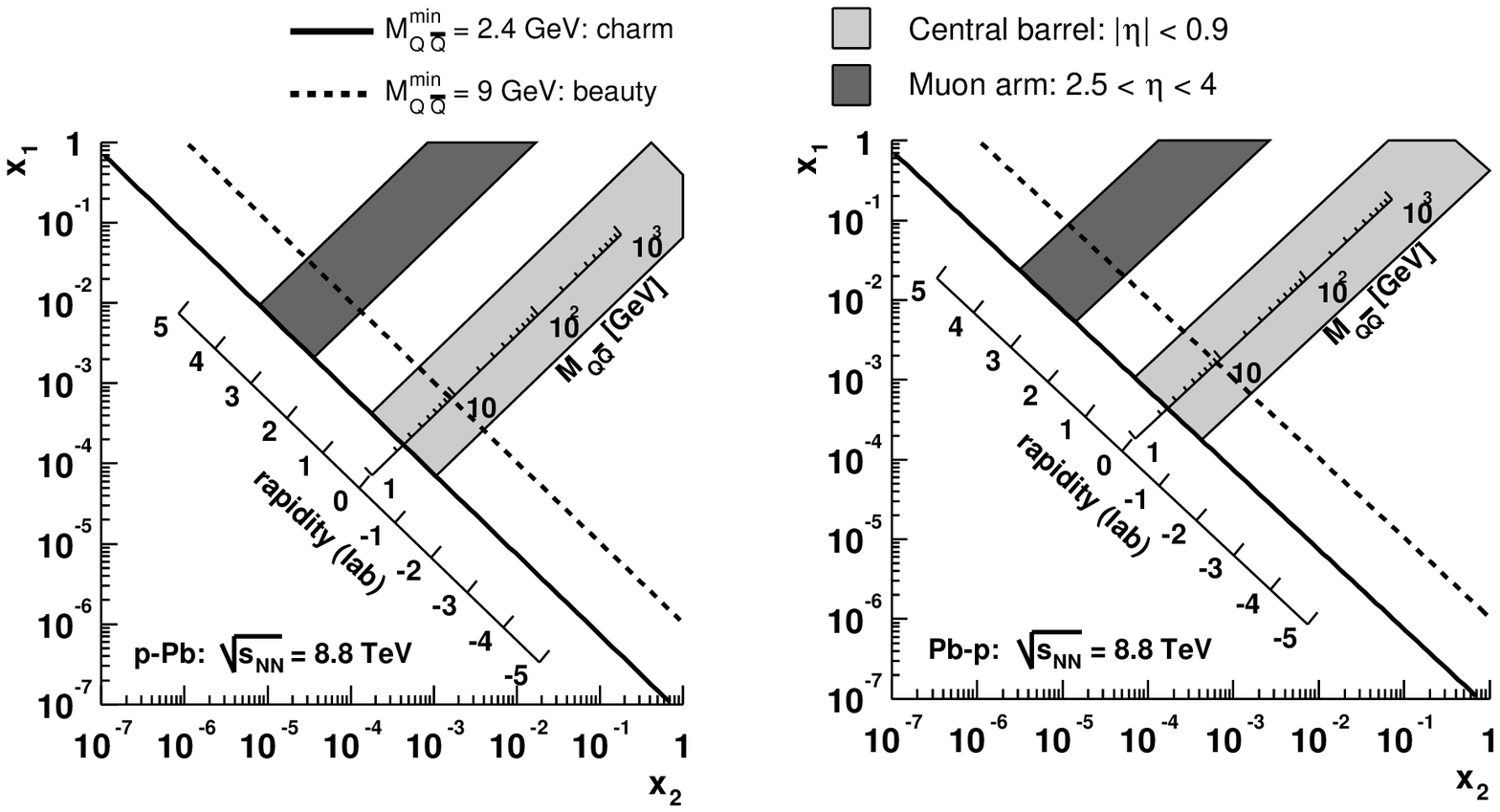}
  \caption{ALICE acceptance in the ($x_1$, $x_2$) plane for heavy flavours 
           in \mbox{p--Pb} (left) and in \mbox{Pb--p} (right).}
  \label{fig:x1x2_pAAp}
  \vglue0.2cm
    \includegraphics[width=.7\textwidth]{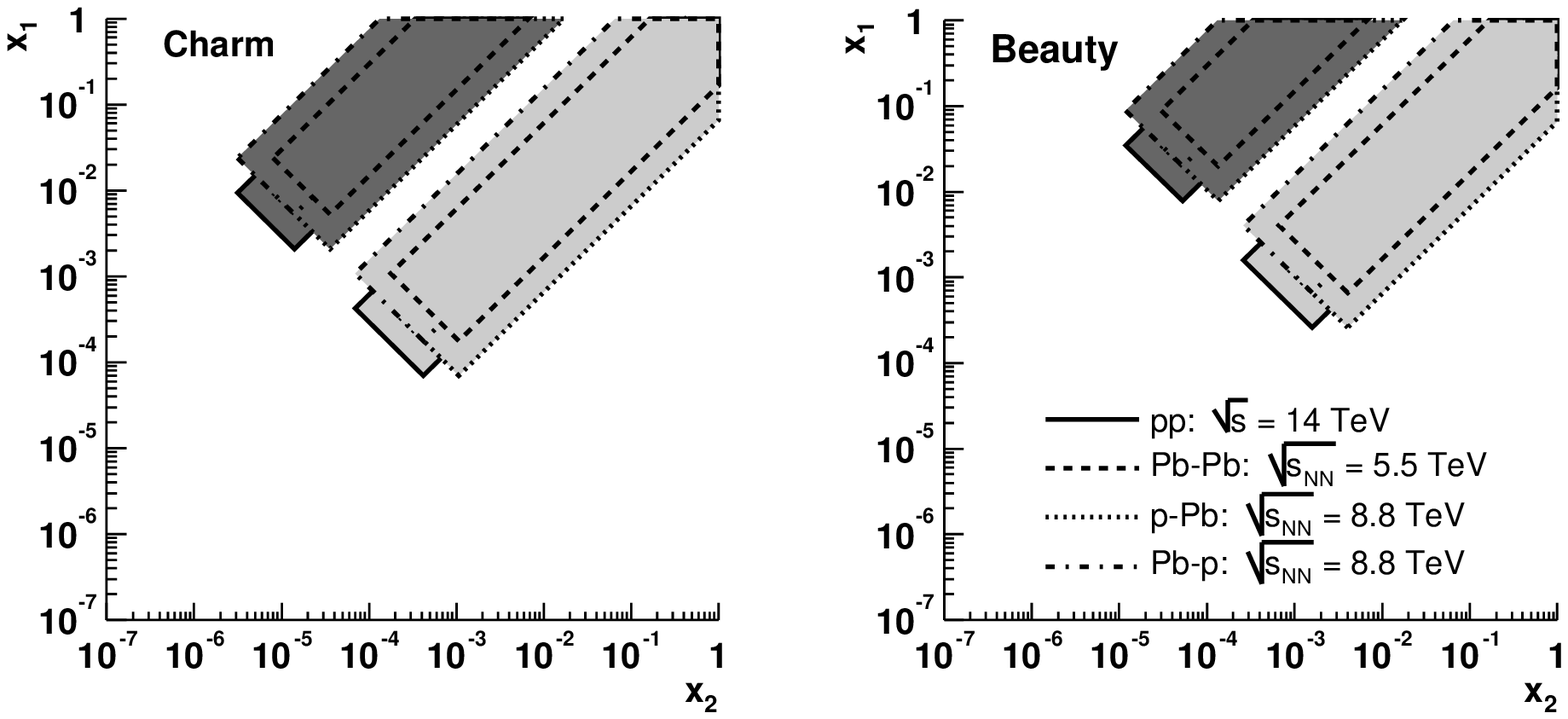}
  \caption{ALICE acceptance in the ($x_1$, $x_2$) plane for 
           charm (left) and beauty (right) in pp, \mbox{Pb--Pb}, 
           \mbox{p--Pb} and \mbox{Pb--p}.}
  \label{fig:x1x2_global}
  \end{center}
\end{figure}

Because of its lower mass, charm allows to probe lower $x$ values than beauty. 
The capability to measure charm and beauty particles in the forward 
rapidity region ($y\simeq 4$) would give access to $x$ regimes about 
2 orders of magnitude lower, down to $x\sim 10^{-6}$. 

In Fig.~\ref{fig:x1x2_AApp} we show the regions of the ($x_1$, $x_2$) plane 
covered for charm and beauty by the ALICE acceptance, 
in \mbox{Pb--Pb} at 5.5~TeV 
and in pp at 14~TeV. In this plane the points with constant invariant 
mass lie on hyperbolae 
($x_1=M^2_{Q\overline{Q}}/(x_2\,s_{\scriptscriptstyle \rm NN})$), 
straight lines in the log-log scale: 
we show those corresponding to the production of 
$\ccbar$ and $\bbbar$ pairs at the threshold; 
the points with constant rapidity lie
on straight lines ($x_1=x_2\exp(+2\,y_{Q\overline{Q}})$). The shadowed regions 
show the acceptance of the ALICE central barrel, covering the 
pseudorapidity
range $|\eta|<0.9$, and 
of the muon arm, $2.5<\eta<4$ (the ALICE experimental layout will be described 
in Chapter~\ref{CHAP4}).

In the case of asymmetric collisions, e.g. \mbox{p--Pb} and \mbox{Pb--p}, 
we have a rapidity shift: the centre of mass moves with a longitudinal 
rapidity
\begin{equation}
  y_{\rm c.m.} = \frac{1}{2}\ln\left(\frac{\rm Z_1 A_2}{\rm Z_2 A_1}\right),
\end{equation}
obtained from equation~(\ref{eq:rapidityx1x2}) for $x_1=x_2$. The rapidity 
window covered by the experiment is consequently shifted by
\begin{equation}
 \Delta y = y_{\rm lab.~system}-y_{\rm c.m.~system} = y_{\rm c.m.},
\end{equation}
corresponding to $+0.47$ ($-0.47$) for p--Pb (Pb--p) collisions.
Therefore, running with both \mbox{p--Pb} and \mbox{Pb--p} will allow 
to cover the largest interval in $x$. 
Figure~\ref{fig:x1x2_pAAp} shows the acceptances for \mbox{p--Pb} and 
\mbox{Pb--p}, while in Fig.~\ref{fig:x1x2_global} the coverages 
in pp, \mbox{Pb--Pb}, \mbox{p--Pb} and \mbox{Pb--p} are 
compared for charm (left) and 
beauty (right). 

\begin{figure}[!t]
  \begin{center}
  \includegraphics[width=.6\textwidth]{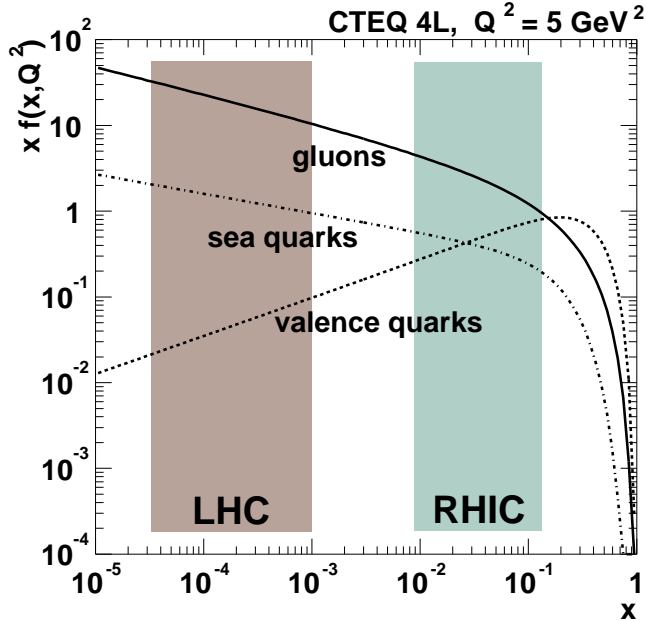}
  \caption{Parton distribution functions in the proton, in the CTEQ 4L 
           parameterization, for $Q^2=5~\gev^2$.} 
  \label{fig:pdf}
  \end{center}
\end{figure}
  
These figures are meant to give a first idea of the 
regimes accessible at ALICE; the simple relations for the leading order 
case were used, the ALICE rapidity acceptance cuts were applied to the 
rapidity of the $Q\overline{Q}$ pair, and not to that of the particles 
which are actually detected. In addition, no minimum $p_{\rm t}$ cuts 
were accounted for: such cuts will increase the minimum accessible value of 
$M_{Q\overline{Q}}$, thus increasing also the minimum accessible $x$. 
These approximations, however, are not too drastic, since 
there is a very strong correlation in rapidity between the 
initial $Q\overline{Q}$ pair and the heavy flavour particles it produces
and the minimum $p_{\rm t}$ 
cut will be quite low (lower than the mass of the hadron) for most of the 
channels studied at ALICE. This last point was demonstrated within this
thesis work for the specific case of open charm measurements at central 
rapidity (Chapter~\ref{CHAP6}).

The parton distribution functions $x\,f(x,Q^2)$ in the proton, in the CTEQ 
4L parameterization, are shown in Fig.~\ref{fig:pdf}. $Q^2$ is the virtuality, 
or QCD scale (in the case of the leading order heavy flavour 
production considered in this paragraph, 
$Q^2=M^2_{Q\overline{Q}}=s\,x_1\,x_2$). In the figure the value 
$Q^2=5~\gev^2$, corresponding to $\ccbar$ production at threshold, is used.
The regions in $x$ covered, at central rapidities, 
at RHIC and LHC are indicated by the shaded areas.

\subsubsection{Nuclear shadowing effect}
\label{CHAP1:shadowing}

The extension of the $x$ range down to $\sim 10^{-4}$ at the LHC means, in a
very simplified picture, that a large-$x$ parton in one of the two 
colliding Pb nuclei `sees' the other incoming nucleus as a superposition 
of $\sim {\rm A} \times 1/10^{-4} \sim 10^6$ gluons. These gluons are 
so many that the lower-momentum ones tend to merge together: 
two gluons with momentum fractions $x_1$ and $x_2$ merge in a gluon 
with momentum fraction $x_1+x_2$ ($g_{x_1}g_{x_2}\to g_{x_1+x_2}$).
As a consequence of this `migration towards larger $x$', that does not 
affect only gluons but all partons, the nuclear parton densities are 
depleted in the small-$x$ region (and slightly enhanced in the 
large-$x$ region) with respect to the proton parton densities.

This phenomenon is known as {\sl nuclear shadowing effect} and it has been 
experimentally studied in electron--nucleus DIS in the range 
$5\cdot 10^{-3}<x<1$~\cite{nDIS}. However, no data are available 
in the $x$ range covered by the LHC and the existing data provide only  
weak constraints for the gluon PDFs, which do not enter the measured 
structure functions at leading order.
Only two groups (EKS~\cite{EKS} and HKM~\cite{hkm}) 
have used the same approach as 
in the case of the proton to obtain a parameterization (and extrapolation 
to low $x$) of the nuclear-modified PDFs. Nuclear PDFs were also computed  
in several other 
models which tend to disagree where no experimental constraints 
are available. 

\begin{figure}[!t]
  \begin{center}
  \includegraphics[width=.65\textwidth]{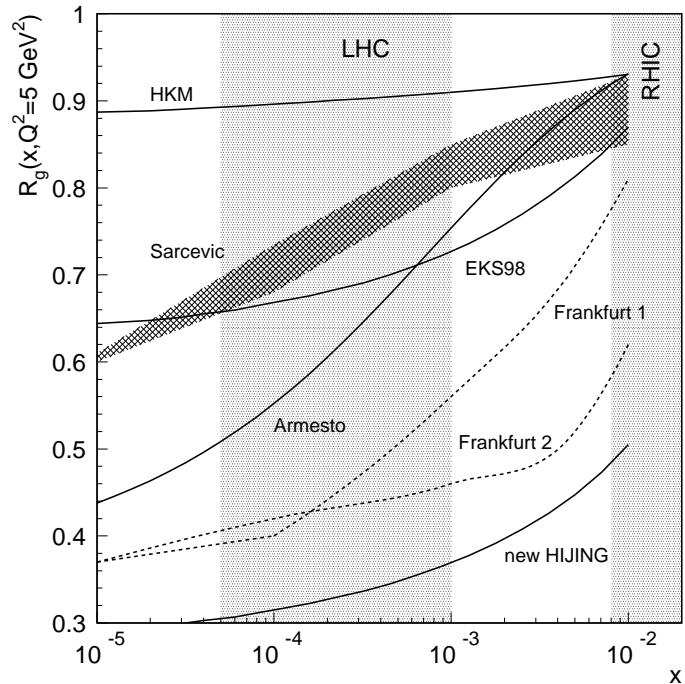}
  \caption{Ratio of the gluon distribution function for Pb to the one 
           for proton using different models at $Q^2=5~\gev^2$.}
  \label{fig:shadowingmodels}
  \end{center}
\end{figure}

The situation is summarized in Fig.~\ref{fig:shadowingmodels} that shows the 
results of the different models for the ratio of the gluon distribution in a 
Pb nucleus over the gluon distribution in a proton:
\begin{equation}
\label{eq:shadowing}
  R_g(x,Q^2)=\frac{g^{\rm Pb}(x,Q^2)}{g^{\rm p}(x,Q^2)}.
\end{equation}
In the figure the value $Q^2=5~\gev^2$, corresponding to $\ccbar$ production 
at threshold, is used.
The predictions for the gluon shadowing at the LHC ($R_g(x\sim 10^{-4})$) 
range from 30\% to 90\%. This large uncertainty will be reduced in the 
future by (a) more data in DIS with nuclei, (b) the pA data collected
at RHIC and, most important, (c) the measurements of charm and beauty 
production in p--Pb at the LHC.   

For the present work, we used the EKS98~\cite{EKS} parameterization 
since it is the one which includes most constraints from DIS data.
It gives $R_g(x\sim 10^{-4})\simeq 65\%$; in Chapter~\ref{CHAP3} we
will see that this determines a reduction of 35\% for the charm cross section 
per \NN~collision in Pb--Pb with respect to pp collisions 
at the same c.m.s. energy.

\subsection{Hard partons: probes of the QGP medium}
\label{CHAP1:hardprobes}

``Qualitatively, in minimum-bias \PbPb~(or Au--Au) 
collisions, SPS is 98\% soft and 2\% hard, RHIC is
50\% soft and 50\% hard and LHC is 2\% soft and 98\% 
hard'' (K.~Kajantie~\cite{kajantieQM2002}). 
This means that at the LHC practically in all minimum-bias events 
(no centrality selection applied) high-$\pt$ partons 
are expected to be produced in 
scattering processes involving a hard perturbative scale 
$Q\gg\Lambda_{\rm QCD}\simeq 200~\mev$. 

\begin{figure}[!t]
  \begin{center}
  \includegraphics[angle=-90,width=.6\textwidth]{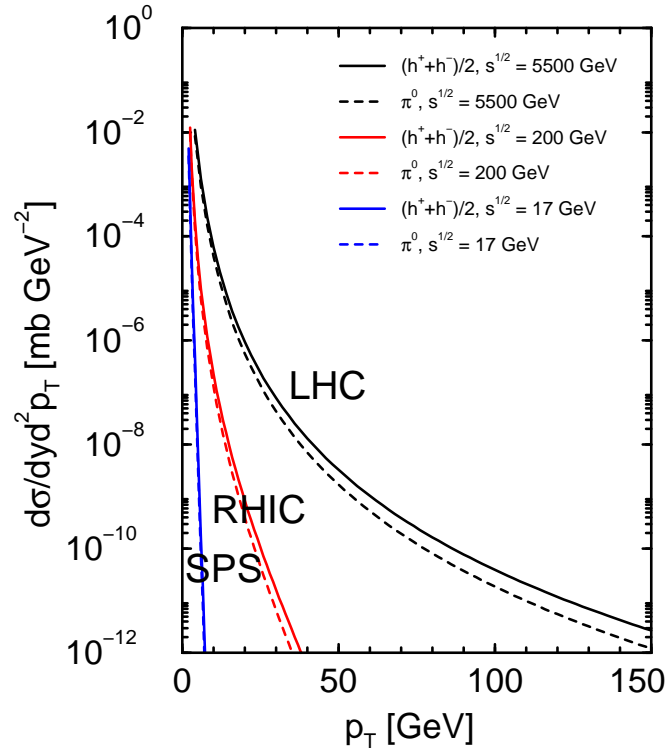}
  \caption{The predicted leading order (LO) differential cross section 
           for neutral pion and charged hadron 
           production is shown for pp collisions at 
           $\sqrt{s}=17,~200,~5500~\gev$.}
  \label{fig:hardptSPSRHICLHC}
  \end{center}
\end{figure}

We give two examples of this significant qualitative difference of the 
LHC with respect to SPS and RHIC.
The perturbative QCD (pQCD) results for the differential cross 
sections for charged hadrons and neutral pions at SPS, RHIC and
LHC energies are shown in Fig.~\ref{fig:hardptSPSRHICLHC}. 
The estimated yields for charm and beauty production are reported 
in Table~\ref{tab:hvqSPSRHICLHC}: the $\ccbar$ and $\bbbar$ yields are 
expected to be 10 and 100 times larger, respectively, at the LHC than at RHIC.

\begin{table}[!h]
  \caption{Charm and beauty production yields (per event) 
           estimated for central \PbPb~collisions at  
           $\sqrtsNN=17,~200,~5500~\gev$.}
  \label{tab:hvqSPSRHICLHC}
  \begin{center}
  \begin{tabular}{c|ccc}
\hline
\hline 
$\sqrt{s}$ & SPS, 17 GeV & RHIC, 200 GeV & LHC, 5.5 TeV \\
\hline
$N(\ccbar)$ & 0.2 & 10   & 120 \\ 
$N(\bbbar)$ & --  & 0.05 & 5   \\
\hline
\hline
  \end{tabular}
  \end{center}
\end{table}

\subsubsection{Initial state and final state effects}

In the absence of nuclear and medium effects, a \AA~collision 
can be considered as a superposition of independent \NN~collisions. 
Thus, the cross section for hard processes should scale from pp to AA 
proportionally to the number of inelastic \NN~collisions (binary scaling).

The effects that can modify this simple scaling are usually divided in two 
classes:
\begin{itemize}
\item {\sl initial state effects}, such as nuclear shadowing (described 
      in Section~\ref{CHAP1:x}), that 
      affect the hard cross section in a way which depends on the size and 
      energy of the colliding nuclei, but not on the medium formed 
      in the collision; 
\item {\sl final state effects}, induced by the medium, that can change 
      the yields and/or the kinematic distributions (e.g. $\pt$ and rapidity)
      of the produced hard partons; a typical example is the partonic 
      energy loss; these final state effects are not correlated to the 
      initial state effects, they depend strongly on the properties 
      (gluon density, temperature and volume) of the medium and they 
      can therefore provide information on such properties.
\end{itemize}

Initial state effects can be studied using pp and \pA~collisions and then 
reliably extrapolated to \AA. 
If a quark--gluon plasma is formed in AA collisions, the 
{\sl final state effects} will be significantly stronger than what is 
expected by an extrapolation from pA.  

\subsubsection{Why hard partons are good probes}

Primary hard quarks and gluons are very well suited to probe the medium
for three main reasons:
\begin{enumerate}
\item They are {\sl produced in the early stage of the collision} in 
      primary partonic scatterings, $gg\to gg$ or $gg\to q\overline{q}$, 
      with large virtuality $Q$ and, thus, on temporal and 
      spatial scales, $\Delta\tau\sim 1/Q$ and $\Delta r\sim 1/Q$, which are
      sufficiently small for the production to be 
      {\sl unaffected} by the properties of the 
      medium (i.e. {\sl by final state effects}).
\item Given the large virtuality, the production cross sections can be  
      reliably calculated with the perturbative approach of pQCD. 
      In fact, since
      \[\alpha_s(Q^2)\propto \frac{1}{\ln(Q^2/\Lambda^2_{\rm QCD})},\]
      in an expansion of the cross sections in powers of $\alpha_s$, 
      for large values of $Q^2$, the higher-order terms (in general 
      higher than next-to-leading order, $\mathcal{O}(\alpha_s^3)$) 
      are small and can be neglected.\\ 
      In this way, as already mentioned, one can safely use pQCD for 
      the energy interpolations needed to compare pp, pA and AA 
      and disentangle initial and final state effects. 
\item They are expected to be significantly attenuated, through the 
      QCD energy loss 
      mechanisms, when they propagate in the medium. The current
      theoretical understanding of these mechanisms and of the magnitude
      of the energy loss are extensively covered in the next chapter, 
      with particular focus on the predictions for charm quarks.         
\end{enumerate}

\begin{figure}[!t]
  \begin{center}
  \includegraphics[width=.75\textwidth]{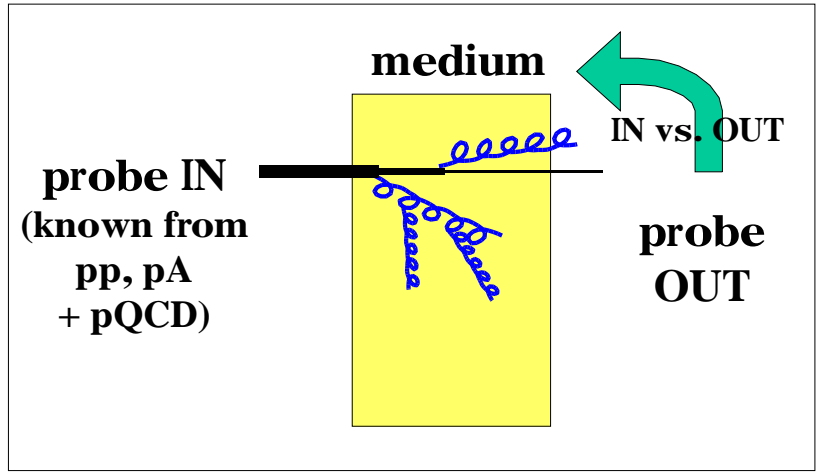}
  \caption{Schematic representation of how hard probes can be used 
           to investigate the properties of the medium.}
  \label{fig:probe}
  \end{center}
\end{figure}

Figure~\ref{fig:probe} shows a schematic view of how hard probes `work':
the input (yields and $\pt$ distributions) is known from the measurements
carried out in pp (and pA) interpolated to the AA energy by pQCD and
scaled according to the number of binary \NN~collisions.
The comparison of the output, measured `after the medium', to the input
allows to gain information on the medium itself.

\clearpage
\pagestyle{plain}

\setcounter{chapter}{1}
\mychapter{Charm in heavy ion collisions}
\label{CHAP2}

\pagestyle{myheadings}

Heavy quarks are sensitive probes of the medium produced in 
nucleus--nucleus collisions. In fact, they present all the features 
listed at the end of the previous chapter.
\begin{enumerate}
\item {\sl Initial production not affected by final state effects:} the minimum
      value of the virtuality $Q_{\rm min}=2\,m_Q$ in the production of 
      a $Q\overline{Q}$ pair implies very small space-time scales 
      of\footnote{Using $1~\gev^{-1}\approx 0.2~\fm$ in the natural units
      system with $c=\hbar=1$.} 
      $\sim 1/(2\,m_Q)\simeq 1/2.4~\gev^{-1}\simeq 0.1~\fm$ 
      (for charm), to be compared
      to the expected life-time of the QGP phase at the LHC, $>10~\fm$. 
      Thus, the initially-produced heavy quarks experience the full 
      collision history. 
\item {\sl Predictivity by pQCD:} this is another consequence of the large 
      mass (compared to $\Lambda_{\rm QCD}$). In Section~\ref{CHAP2:pQCD}
      we show the main production channels and the general lines followed 
      for the cross section calculations in $\pp$ 
      collisions. 
\item {\sl Strong nuclear effects:} both initial and final state effects 
      are expected to enter in the production and propagation of 
      heavy quarks, respectively; consequently, they are information-rich 
      probes.
      Such effects are summarized in Section~\ref{CHAP2:charminhic}.
      In particular, the study of the energy loss of heavy quarks at the LHC
      is very interesting, because of the prediction of 
      a significant mass-dependence of the effect 
      (Section~\ref{CHAP2:partonenergyloss}). 
\end{enumerate}

In the present work we concentrate on charm physics because (a) 
the production cross section is expected to be a factor $\simeq 20$  
larger for charm than for beauty and (b) charm mesons can be exclusively 
reconstructed even in \PbPb~collisions at the LHC via the decay 
channel $\DtoKpi$ (as we will show in Chapter~\ref{CHAP6}), 
while the feasibility of the exclusive reconstruction of beauty particles, 
e.g. via ${\rm B^0\to K_S^0\,J/\psi}$, is still unclear.
For these two reasons charm can be studied with better precision 
(smaller statistical uncertainties) and accuracy (smaller systematic 
uncertainties) than beauty.

Experimentally, charm production was extensively studied in pA collisions, 
for energies up to $\sqrtsNN\approx 40~\gev$. 
In AA collisions the present
knowledge (from SPS and RHIC) is quite poor, since up to now 
no dedicated experiments were performed. The situation is summarized in 
Section~\ref{CHAP2:data}.

At the end of the chapter we introduce the strategy aimed at the 
measurement and study of charm production in the LHC heavy ion programme
with the ALICE detector (Section~\ref{CHAP2:aliceD0}). The evaluation 
of the feasibility of this strategy and the attainable sensitivity for the
study of charm physics are the central subject of this thesis work.

\mysection{Heavy quark production in pQCD}
\label{CHAP2:pQCD}

\begin{figure}[!t]
  \begin{center}
    \includegraphics[width=.7\textwidth]{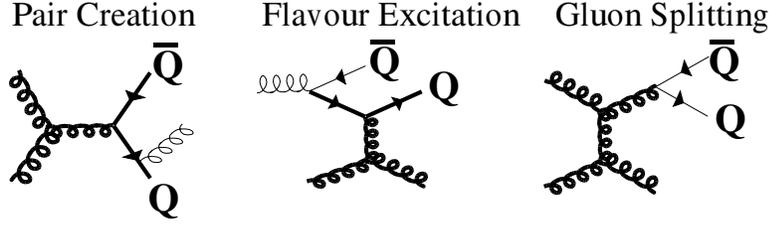}
    \caption{Some of the processes defined as pair creation, flavour
      excitation and gluon splitting. The thick lines correspond to the
      hard process.}
    \label{fig:processes}
  \end{center}
\end{figure}

At LHC energies, heavy quarks are produced, at leading order,
via {\sl pair creation} by gluon--gluon fusion ($gg\to Q\overline{Q}$), 
mostly, and $q\overline{q}$ annihilation 
($q\overline{q}\to Q\overline{Q}$)\footnote{$q$ indicates a light 
quark, $Q$ a charm or beauty quark.}.
At next-to-leading order more complicated topologies are included. 
Usually,
the processes are classified according to the number of heavy quarks in the 
final state of the hard
process, defined as the process with the highest
virtuality (i.e. highest invariant mass of the outgoing parton pair, 
as defined in equation (\ref{eq:sx1x2M2})). There are basically 
three classes of processes:
\begin{description}
\item[pair creation:] the hard process is one of the leading order
  graphs ($gg \rightarrow Q\overline{Q}$, $q\overline{q}
  \rightarrow Q\overline{Q}$); its final state contains two heavy
  quarks;
\item[flavour excitation:] an incoming heavy quark is put on mass shell
  by scattering on a parton of the other beam: $qQ\rightarrow qQ$ or
  $gQ\rightarrow gQ$; the incoming heavy quark must come from a $g
  \rightarrow Q\overline{Q}$ splitting in the PDF of the proton; this
  process is characterized by one heavy quark in the final state of
  the hard scattering;
\item[gluon splitting:] no heavy flavour is involved in the hard
  scattering, but a $Q\overline{Q}$ pair is produced in the
  final state from a $g \rightarrow Q\overline{Q}$
  branching.
\end{description}
Figure~\ref{fig:processes} shows some topologies belonging to the
processes specified above.

At any order, the partonic cross section can be expressed in terms of the 
dimensionless scaling functions $F^{(k,l)}_{ij}$ that depend only on the 
variable $\xi$~\cite{vogtXsec}
\begin{equation}
\hat{\sigma}_{ij}(\hat{s},m_Q^2,\mu_R,\mu_F)=
\frac{\alpha_s^2(\mu_F)}{m_Q^2}\sum_{k=0}^\infty(4\pi\alpha_s(\mu_F))^k
\sum_{l=0}^k F_{ij}^{(k,l)}(\xi)\ln^l\left(\frac{\mu_R^2}{m_Q^2}\right),
\end{equation}
where $\hat{s}$ is the partonic centre-of-mass energy squared for two 
partons $i$ and $j$ carrying momentum fractions $x_i$ and $x_j$
($\hat{s}=x_i\,x_j\,s$), 
$m_Q$ is the heavy quark mass, $\mu_F$ and $\mu_R$
are the factorization and renormalization scales, respectively, and
$\xi=\hat{s}/4m_Q^2-1$. The cross section is calculated as an 
expansion in powers of $\alpha_s$ with $k=0$ corresponding to the 
LO cross section. The first 
correction, $k=1$, corresponds to the NLO cross section. 
The complete calculation only 
exists up to NLO. However, as already mentioned, given the large value of
$m_Q$, the corrections above NLO are expected to be small.

The total hadronic cross section in pp collisions is obtained by 
convoluting the total partonic cross section with the parton 
distribution functions $f_i^{\rm p}|_{i=q,\overline{q},g}$ 
of the initial protons (factorization),
\begin{equation}
\label{eq:xsecpp}
\sigma^{\scriptstyle Q\overline{Q}}_{\rm pp}=
\sum_{i,j=q,\overline{q},g} \int_{4m_Q^2/s}^1\frac{{\rm d}\tau}{\tau}
\delta(x_i x_j-\tau)\,f_i^{\rm p}(x_i,\mu_F)\,f_j^{\rm p}(x_j,\mu_F)\,
\hat{\sigma}_{ij}(\tau s,m_Q^2,\mu_R,\mu_F),
\end{equation}
where $\tau=\hat{s}/s$ and the sum is over all massless partons.
In Ref.~\cite{vogtnew} it is shown how the differential cross sections
can be calculated in either one-particle inclusive kinematics or 
pair invariant mass kinematics.

In the next chapter, along with the results of the calculations,
we show that the main uncertainty on the cross sections comes from the 
values of the heavy quark masses and of the scales, rather than from the
choice of the PDF set.

\subsubsection{Which reference for ${\rm J}/\psi$ production at the LHC?}

Before going in the details of the physics motivations for charm 
measurements `per se', we point out that, at LHC energies,
these measurements are essential as a reference to study the effect of
the transition to a deconfined phase on charmonium production
(the states J$/\psi$ and $\psi'$ will be measured by ALICE).
At the SPS, where charm quarks are produced essentially via quark-antiquark
annihilation, the dilepton continuum 
produced in Drell-Yan processes ($q\overline{q}\to \ell^+\ell^-$) 
was used as a 
normalization for the ${\rm J}/\psi$ production (NA50 experiment). 
However, at LHC energies heavy quarks are mainly produced through gluon--gluon 
fusion processes and the Drell-Yan process does not provide an adequate
reference. A direct measurement of the ${\rm D}$ mesons yield 
would then give a natural normalization for charmonia production. 

\mysection{Physics of open charm in heavy ion collisions}
\label{CHAP2:charminhic}

The measurement of particles carrying open charm\footnote{Particles that 
contain c (or $\rm \overline{c}$) quarks and have Charm quantum 
number $\neq 0$ are called open charm particles. 
The $\ccbar$ bound states, that 
have Charm quantum number $=0$, are called hidden charm particles.}
(such as D mesons) allows to investigate the mechanisms that enter the 
charm quark production and in-medium propagation. 
We summarize here the most relevant issues.

\subsubsection{Parton intrinsic transverse momentum}
In pQCD calculations, in order to reproduce the pp data on charm 
production (in particular ${\rm D\overline{D}}$ azimuthal correlations), 
an intrinsic transverse 
momentum $k_{\rm t}$ has to be assigned to the two colliding partons. 
The value of $k_{\rm t}$ is usually sampled from a gaussian distribution with 
$\av{k_{\rm t}^2}=1~\gev^2$ (see Ref.~\cite{vogtnew} and references there in). 

In pA and AA interactions, the average intrinsic $k_{\rm t}^2$ is expected 
to increase; this effect is known as $k_{\rm t}$ broadening and it is 
observed in Drell-Yan, J$/\psi$ and $\Upsilon$ production. The 
broadening is interpreted as due to multiple scattering of the partons 
of one of the ions in the other ion. On average the effect is estimated to 
yield
$\av{k_{\rm t}^2}=1.35~\gev^2$ in pA and 
$\av{k_{\rm t}^2}=1.7~\gev^2$ in AA interactions~\cite{vogtnew}.

The $k_{\rm t}$ broadening should slightly change the shape of 
the c quark $\pt$ distribution at low $\pt$, given the moderate strength 
of the effect, while the total cross section should be unchanged.
For low-$\pt$ production, the $k_{\rm t}$ broadening is expected to 
reduce the back-to-back azimuthal correlation between the quark and the 
antiquark~\cite{vogtnew}. 

\subsubsection{Nuclear shadowing}
The suppression of the nuclear PDFs, with respect to the proton ones,
at low Bjorken $x$ determines in pA and AA collisions 
a reduction of the production cross section per \NN~collision 
in the low-$\pt$ region.
As a consequence of the factorization of the parton distributions
in the two colliding hadrons (seen in Eq.~(\ref{eq:xsecpp})), if the 
cross section per binary collision is reduced to a 
fraction $X$ in pA, it has to be reduced 
to a fraction $X^2$ in AA, at the same c.m.s. energy.

A simple estimate of the upper limit of the $\pt$-region affected by 
the shadowing in \PbPb~at the LHC is the 
following: for the back-to-back production of a $\ccbar$ pair at central 
rapidity, with transverse momenta 
$\pt^{\rm c}=\pt^{\rm \overline{c}}=5~\gev/c$, we have $Q\simeq 2\,\pt=10~\gev$
and $x\simeq Q/\sqrtsNN=10/5500\simeq 2\cdot 10^{-3}$; for these values of 
$x$ and $Q$, the EKS98~\cite{EKS} parameterization gives $R_g$ 
(defined in Eq.~(\ref{eq:shadowing})) $\simeq 90\%$.
This suppression is already quite small and it is partially compensated 
by the $k_{\rm t}$ broadening. Therefore, we can conclude that initial state
effects should modify the $\pt$ distribution of charm quarks 
only for $\pt<5$-$7~\gev/c$.

Final state effects are considered in the following. 

\subsubsection{Possible thermal charm production}
In addition to the hard primary production, secondary $\ccbar$ production 
in the quark--gluon plasma has been considered~\cite{mullerwang,levaivogt}. 
At high temperatures, thermal charm production 
might occur since the mass of the c quark, $\approx 1.2~\gev$, 
is not much larger than the highest predicted temperature at the LHC, 
$\simeq 0.6$-$0.8~\gev$. The thermal yield from a plasma of massless quarks and
gluons is probably not comparable with initial production. These thermal
charm pairs would have lower invariant masses than the initial $\ccbar$
pairs and would thus be accumulated in the low-$\pt$ region of the 
spectrum.
 
\subsubsection{Quenching}
Although predicted already twenty years ago by J.D.~Bjorken~\cite{bjorkenjets},
parton energy loss, which appears as a quenching (attenuation) 
of large-$\pt$ hadrons and jets, was revealed as one of the most interesting 
observables of heavy ion physics in the `collider era' only after the 
experimental evidences collected at RHIC in the last two years, which have
been reported in Chapter~\ref{CHAP1}.  


At the light of the RHIC results, combined measurements of 
relatively-large-$\pt$ ($5<\pt<20~\gev/c$) light flavour hadrons and 
heavy flavour hadrons are very promising tools for a detailed 
`tomography', or `colourimetry', of the deconfined medium 
that will be produced at the LHC.

The study of charm mesons quenching is particularly relevant because 
it is expected to be significantly lower than for hadrons containing 
only u and d quarks (and antiquarks). In fact, D mesons are originated by 
(c) quarks, while other hadrons come mainly from the fragmentation of 
gluons, which, due to their larger strong coupling, lose more 
energy than quarks. Moreover, partons with velocity $v<c$, like heavy quarks 
with moderate momentum, might lose less energy than very fast ($\approx$ 
massless) partons with $v\approx c$.

These topics are detailed in the next section, where one of the 
widely used energy loss models and its quantitative predictions 
are summarized. 

\mysection{Parton energy loss}
\label{CHAP2:partonenergyloss}

In the first formulation by J.D.~Bjorken~\cite{bjorkenjets} the arguments 
for the energy loss of partons in the quark--gluon plasma were based on 
elastic scattering of high-momentum partons from gluons in the QGP.
The resulting (`collisional') loss was estimated to be 
$\dEdx\simeq \alpha_s^2\sqrt{\varepsilon}$, with $\varepsilon$ the energy 
density of the QGP. This loss turns out to be quite low, of 
$\mathcal{O}(0.1~\gev/\fm)$~\cite{thoma}.

However, as in QED, bremsstrahlung (or, better, `gluon bremsstrahlung') 
is another important source of energy 
loss~\cite{gyulassy1}. 
Due to multiple (inelastic) scatterings and induced gluon 
radiation hard partons lose energy and become quenched. Such radiative 
loss, as we show in the following, is considerably larger than 
the collisional one. An intense theoretical activity has developed
around the subject~\cite{gyulassywang,gyulassy1,bdps,bdmps2,zakharov1,bdms,zakharov2,urs1,urs2,urs3,gyulassy2}. In the next section we 
present the general lines of the model proposed by R.~Baier, Yu.L.~Dokshitzer,
A.H.~Mueller, S.~Peign\'e and D.~Schiff~\cite{bdmps,bdmps2} (`BDMPS'). The 
quenching probabilities (or weights) for light quarks and gluons, 
as calculated by C.A.~Salgado and 
U.A.~Wiedemann~\cite{carlosurs} on the basis of the BDMPS model are presented
in Section~\ref{CHAP2:qw}.
Radiative energy loss off heavy quarks is considered in 
Section~\ref{CHAP2:deadcone}. 

\subsection{Medium-induced radiative energy loss}
\label{CHAP2:radenloss}

\begin{figure}[!t]
  \begin{center}
    \includegraphics[angle=-90,width=.7\textwidth]{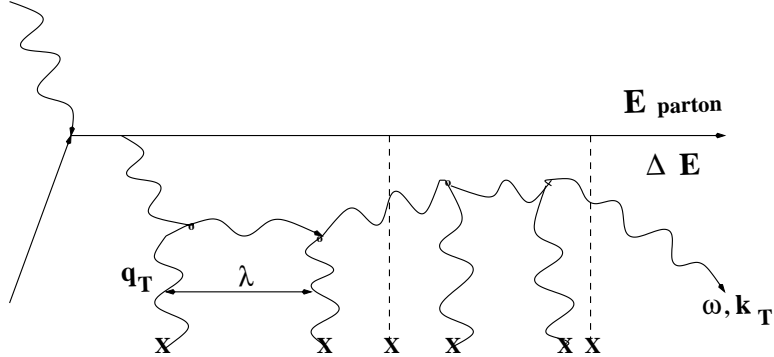}
    \caption{Typical gluon radiation diagram~\cite{baier}.}
    \label{fig:gluonsstrahlung}
  \end{center}
\end{figure}

After its production in a hard collision, an energetic parton radiates a
gluon with a probability which is proportional to its path length $L$
in the dense medium. Then (Fig.~\ref{fig:gluonsstrahlung})
the radiated gluon suffers multiple scatterings in the medium, in 
a Brownian-like motion with mean free path $\lambda$ which decreases 
as the density of the medium increases. The number of scatterings of the 
radiated gluon is also proportional to $L$. Therefore, the average energy loss
of the parton is proportional to $L^2$. This is the most distinctive 
feature of QCD energy loss (with respect to QED bremsstrahlung energy loss, 
$\propto L$)
and it is due to the fact that gluons interact with each other, 
while photons do not.

The scale of the energy loss is set by the `maximum' energy of the emitted
gluons, which depends on $L$ and on the properties of the 
medium~\cite{carlosurs}:
\begin{equation}
\omega_c = \frac{1}{2}\hat{q}L^2,
\end{equation} 
where $\hat{q}$ is the {\sl transport coefficient of the medium}, defined
as the average transverse momentum squared transferred to the projectile 
per unit path length
\begin{equation}
\hat{q} = \frac{\av{q_{\rm t}^2}_{\rm medium}}{\lambda}.
\end{equation}
In the case of a static medium, the distribution of the energy $\omega$ of the 
radiated gluons, for $\omega\ll \omega_c$, is of the form:
\begin{equation} 
\label{eq:wdIdw}
\omega\frac{{\rm d}I}{{\rm d}\omega}\simeq \frac{2\,\alpha_s\,C_R}{\pi}\sqrt{\frac{\omega_c}{2\omega}},
\end{equation}
where $C_R$ is the QCD coupling factor (Casimir factor), equal 
to 4/3 for quark--gluon coupling and to 3 for gluon--gluon coupling.
The integral of the energy distribution up to $\omega_c$ estimates the 
average energy loss of the initial parton:
\begin{equation}
\label{eq:avdE}
\av{\Delta E} = \int^{\omega_c} \omega \frac{{\rm d}I}{{\rm d}\omega}{\rm d}\omega
\propto \alpha_s\,C_R\,\omega_c \propto \alpha_s\,C_R\,\hat{q}\,L^2.
\end{equation}
The average energy loss is therefore:
\begin{itemize}
\item proportional to $\alpha_s\,C_R$ and, thus, larger by a factor 
      $9/4=2.25$ for gluons than for quarks;
\item proportional to the transport coefficient of the medium;
\item proportional to $L^2$; 
\item independent of the parton initial energy.  
\end{itemize}
The last point is peculiar to the BDMPS model. Other 
models~\cite{gyulassywang,gyulassy2} consider 
an explicit dependence of $\Delta E$ on the initial energy $E$. However, 
as we shall discuss in Chapter~\ref{CHAP8}, there is always an intrinsic 
dependence of the radiated energy on the initial energy, determined by the 
fact that the former cannot be larger than the latter, $\Delta E\leq E$.

The transport coefficient is proportional to the density $\rho$ of the 
scattering centres and to the typical momentum transfer in the 
gluon scattering off these centres. For cold nuclear matter, on the 
basis of the nuclear density $\rho_0=0.16~\fm^{-3}$ 
and of the gluon PDF in the nucleon, the 
value estimated in Ref.~\cite{bdmps} was:
\begin{equation}
\hat{q}_{\rm cold} \simeq 0.05~\gev^2/\fm\simeq 8\,\rho_0.
\end{equation}
This value is in agreement with the result of the analysis of gluon 
$k_{\rm t}$ broadening from experimental data on J$/\psi$ transverse momentum 
distributions~\cite{qcold}, which in the present notation yielded
\begin{equation}
\hat{q}=(9.4\pm0.7)\,\rho_0.
\end{equation}
An estimate~\cite{bdmps} for a hot medium based on perturbative treatment of 
gluon scattering in a QGP with $T\simeq 250~\mev$ resulted in the value of 
the transport coefficient of about a factor twenty larger than 
for cold matter:
\begin{equation}
\hat{q}_{\rm hot}\simeq 1~\gev^2/\fm\simeq 20\,\hat{q}_{\rm cold}.
\end{equation} 
Such large difference is due (a) to the higher density of colour charges, i.e.
shorter mean free path of the probe, in the 
QGP medium and (b), as already mentioned, to the fact that deconfined gluons
have harder momenta than confined gluons and, therefore, the typical 
momentum transfers are larger. 

\begin{figure}[!t]
  \begin{center}
    \includegraphics[width=.6\textwidth]{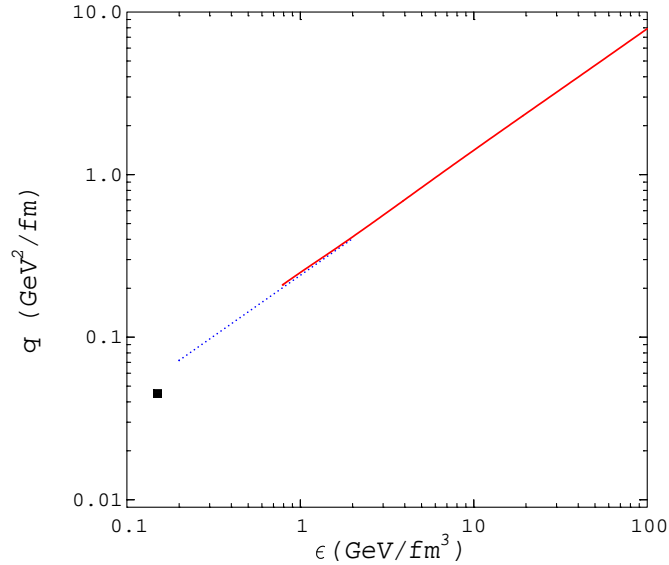}
    \caption{Transport coefficient as a function of energy density for 
             different media: cold (marker), massless hot pion gas 
             (dotted curve) and ideal QGP (solid curve)~\cite{baier}.}
    \label{fig:qtransp}
  \end{center}
\end{figure}

Figure~\ref{fig:qtransp} reports the estimated dependence of $\hat{q}$ on 
the energy density $\varepsilon$ for different equilibrated 
media~\cite{baier}: for a QGP formed at the LHC with 
$\varepsilon\sim 100~\gev/\fm^3$, $\hat{q}$ is expected to be 
of $\sim 10~\gev^2/\fm$. 

In the following examples we consider $L=5~\fm$, which is the typical 
length traveled in the medium for partons produced at mid-rapidity in 
central \PbPb~collisions (we remind that the radius of a $^{208}$Pb 
nucleus is of order $R_{\rm A}\simeq 1.1\,{\rm A}^{1/3}~\fm\simeq 6.5~\fm$)
and $\hat{q}=1~\gev^2/\fm$ in a QGP. 
In Chapter~\ref{CHAP8} we show how a realistic description of the collision
geometry leads to the choice of $\hat{q}\simeq 4~\gev^2/\fm$ for the LHC.

\subsection{Quenching weights}
\label{CHAP2:qw}

The quenching weight is defined as the probability that a hard parton 
radiates an energy $\Delta E$ due to scattering in spatially 
extended QCD matter. In Ref.~\cite{carlosurs}, the weights are 
calculated on the basis of the BDMPS formalism, keeping into account  
(a) the finite in-medium path length $L$ and (b) the dynamic expansion of 
the medium.
The input parameters for the calculation of the quenching weights are
only $L$, the transport coefficient $\hat{q}$ and the parton species
(light quark or gluon).

\begin{figure}[!t]
  \begin{center}
    \includegraphics[width=.6\textwidth]{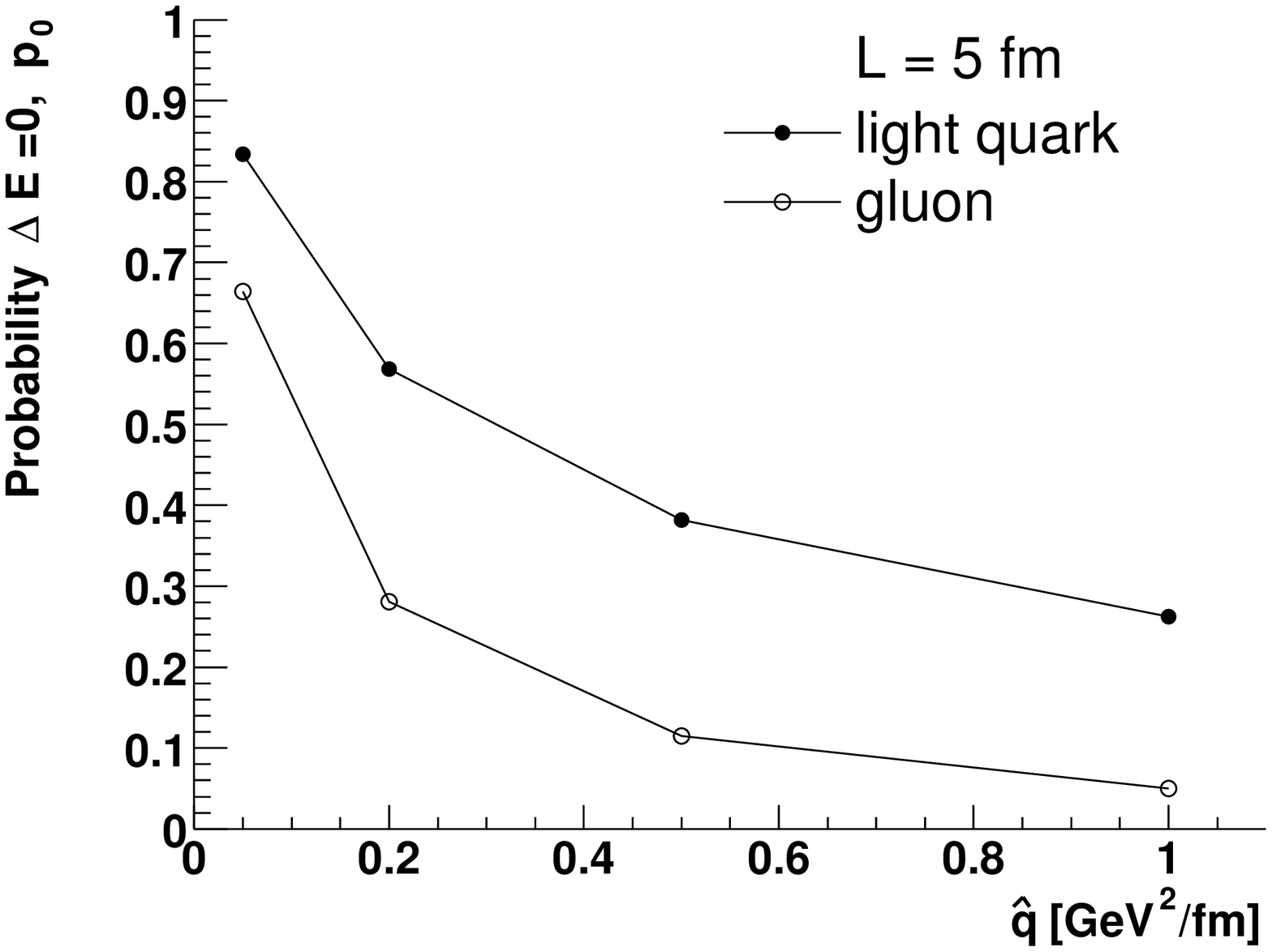}
    \caption{The discrete part of the quenching weight (see text) for 
             $L=5~\fm$ as a function of the transport coefficient.}
    \label{fig:qwDisc}
    \vglue0.2cm
    \includegraphics[width=\textwidth]{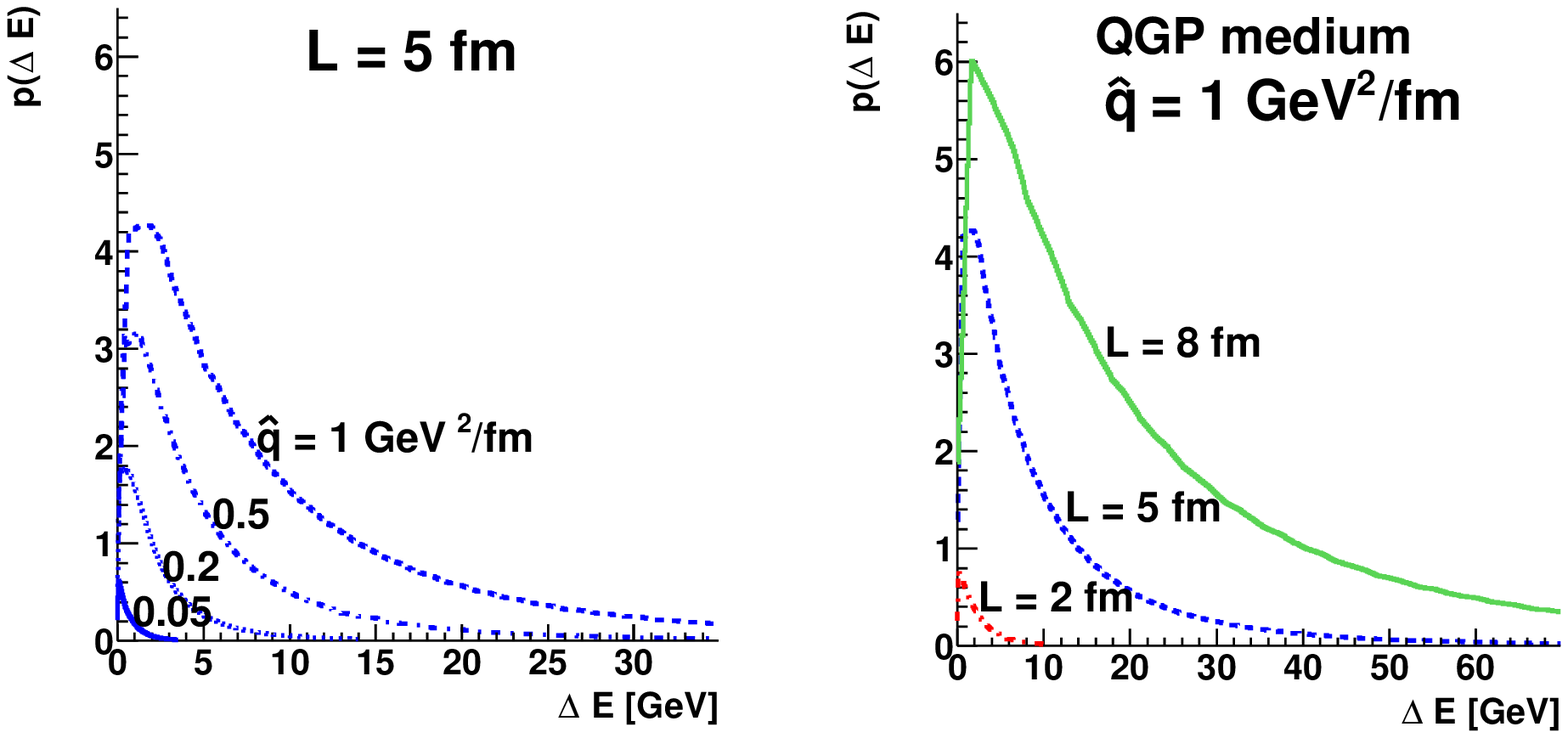}
    \caption{Distribution of the continuous part of the quenching weight 
             (see text) for light quarks in different conditions.}
    \label{fig:qwCont}
  \end{center}
\end{figure}

The probability distribution $P(\Delta E)$ is obtained as the sum of 
a discrete and a continuous part, 
\begin{equation}
  P(\Delta E)=p_0\,\delta(\Delta E)+p(\Delta E).
\end{equation}
The discrete weight $p_0$ is interpreted as the probability that no 
gluon is radiated and hence no in-medium energy loss occurs. The 
continuous weight is the probability to have an energy loss equal 
to $\Delta E$, if at least one gluon is radiated.

Using a numerical routine provided by the authors~\cite{carlosurs}
for $\alpha_s=1/3$, we have plotted the quenching weights as a function 
of the different parameters. 

Figure~\ref{fig:qwDisc} reports the discrete part $p_0$ of the 
weight as a function of $\hat{q}$ for $L=5~\fm$. 
The probability that energy loss does not occur is significantly larger for 
quarks than for gluons, due to their lower QCD coupling, and it decreases
as the density of the medium increases; with $\hat{q}=1~\gev^2/\fm$,
the probability to have energy loss, $1-p_0$, is 75\% for a quark and 
95\% for a gluon. 

\begin{figure}[!t]
  \begin{center}
    \includegraphics[width=.7\textwidth]{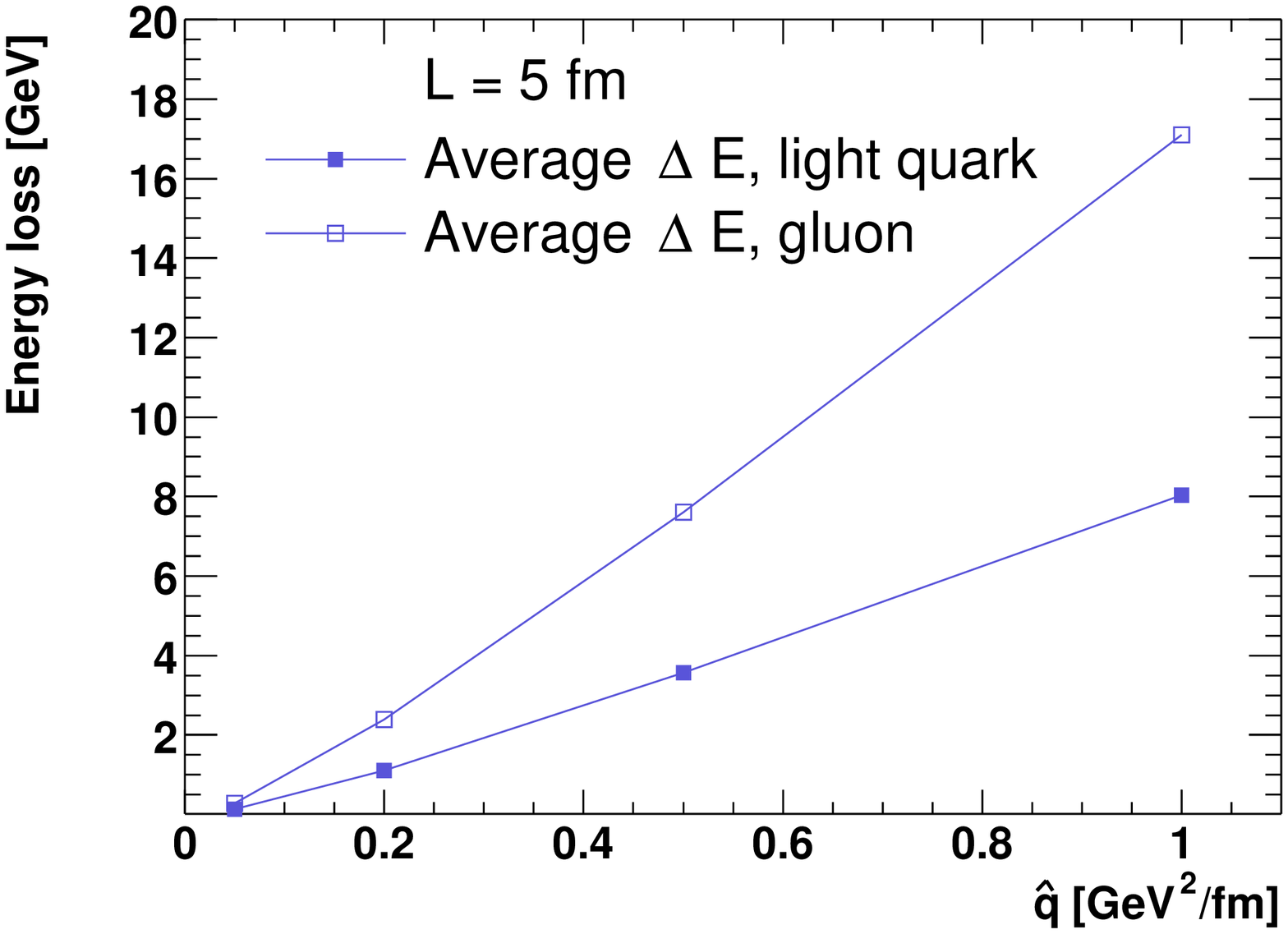}
    \caption{Average energy loss for quarks and for gluons, with $L=5~\fm$, 
             as a function of the transport coefficient.}
    \label{fig:avdEVSqhat_qg}
  \end{center}
\end{figure}

Figure~\ref{fig:qwCont} reports the distribution of the continuous part 
$p(\Delta E)$ of the weight for quarks and for different values of $L$ 
and $\hat{q}$. The average energy loss $\av{\Delta E}$, calculated taking into 
account both the discrete and the continuous parts of the quenching weight,
for $L=5~\fm$ for quarks and 
gluons in shown as a function of $\hat{q}$ in Fig.~\ref{fig:avdEVSqhat_qg}.
As expected (see Eq.~(\ref{eq:avdE})), 
$\av{\Delta E}$ grows approximately linearly with the transport coefficient
and, consequently, with the `maximum' gluon energy $\omega_c$. 
For a quark projectile, we find,
$\av{\Delta E}\approx 0.1\times \omega_c$.  

The average energy loss in a cold medium, 
$\hat{q}=0.05~\gev^2/\fm$, is predicted to be of order $0.1$-$0.2~\mev$.
In a hot medium with $\hat{q}=1~\gev^2/\fm$, the obtained values are
$\av{\Delta E}\simeq 17~\gev$ for gluons and 
$\av{\Delta E}\simeq 8~\gev$ for light quarks. Note that the ratio is 
almost exactly $9/4$. 

These spatially integrated energy losses in a hot medium 
can be `translated' into losses 
per unit path length, $\dEdx_{\rm gluons}\approx 3.5~\gev/\fm$  and
$\dEdx_{\rm quarks}\approx 1.6~\gev/\fm$. 
Such values are one order of magnitude larger than those estimated, 
on the basis of Bjorken's model, for the collisional energy loss.

Given the $L^2$-dependence of the effect, the differential 
energy loss should be given {\sl per unit path length squared}: 
${\rm d}^2E/{\rm d}x^2_{\rm gluons}\approx 0.7~\gev/\fm^2$ and
${\rm d}^2E/{\rm d}x^2_{\rm quarks}\approx 0.3~\gev/\fm^2$.

\subsection{Dead cone effect for heavy quarks}
\label{CHAP2:deadcone}

In Ref.~\cite{dokshitzerkharzeev} Yu.L.~Dokshitzer and D.E.~Kharzeev 
argue that for heavy quarks, 
because of their large mass, the radiative energy loss should be 
lower than for light quarks. The predicted consequence of this effect is
an enhancement of the ratio of D mesons to pions at moderately 
large ($5$-$10~\gev/c$) transverse momenta, with respect to what
observed in the absence of energy loss (\pp~collisions).\\~

Heavy quarks with momenta up to $20$-$30~\gev/c$ propagate with 
a velocity which is smaller than the velocity of light, $c$. As a consequence, 
gluon radiation at angles $\Theta$ smaller than the ratio of their mass to 
their energy $\Theta_0=m_Q/E$ is suppressed by destructive quantum 
interference. In Ref.~\cite{dokshitzerdeadcone} the soft gluon emission 
probability off a heavy quark $Q$ in the vacuum is expressed as:
\begin{equation}
  \d\sigma_{Q\to Q+g}=\frac{\alpha_S\,C_R}{\pi}\,\left(1+\frac{\Theta_0^2}{\Theta^2}\right)^{-2}\,\d\Theta^2\,\frac{\d\omega}{\omega}.
\end{equation}
The relatively depopulated cone around the $Q$ direction with 
$\Theta<\Theta_0$ is indicated as `dead cone'. It is also 
pointed out that the structure of gluon radiation at large angles, 
$\Theta\gg\Theta_0$ appears to be independent of $m_Q/E$ and, thus, 
identical to that for a light quark jet~\cite{dokshitzerdeadcone}. 

A direct consequence of the dead cone effect is the harder fragmentation 
of heavy quarks, with respect to light quarks and gluons, i.e. the fact that 
the `leading' (most energetic) hadron produced by a c or b quark carries 
a larger fraction of the initial quark energy than the leading hadron produced
by a massless parton. In the former case, since less gluons are radiated, 
a larger part of the initial quark energy is available for the leading 
hadron. In Chapter~\ref{CHAP8} we shall further discuss the different 
fragmentation of heavy and light partons.

In Ref.~\cite{dokshitzerkharzeev} the dead cone effect is assumed to 
characterize also in-medium gluon radiation and 
the energy distribution of the radiated gluons (\ref{eq:wdIdw}) is estimated 
to be suppressed by the factor:
\begin{equation}
  \label{eq:FHL}
  \left(1+\frac{\Theta_0^2}{\Theta^2}\right)^{-2}=
  \left[1+\left(\frac{m_Q}{E}\right)^2\sqrt{\frac{\omega^3}{\hat{q}}}\right]^{-2}
  \equiv F_{\rm H/L}(m_Q,E,\hat{q},\omega),
\end{equation} 
where the expression for the characteristic gluon emission 
angle~\cite{dokshitzerkharzeev} 
$\Theta\simeq (\hat{q}/\omega^3)^{1/4}$ has been used. The energy 
distributions radiated off light quark projectiles and heavy quark 
projectiles are related as:
\begin{equation}
  \label{eq:wdIdwHeavy}
  \left(\omega\frac{{\rm d}I}{{\rm d}\omega}\right)_{\rm Heavy}=
  F_{\rm H/L}(m_Q,E,\hat{q},\omega)\times
  \left(\omega\frac{{\rm d}I}{{\rm d}\omega}\right)_{\rm Light}.
\end{equation} 

\begin{figure}[!t]
  \begin{center}
    \includegraphics[width=.6\textwidth]{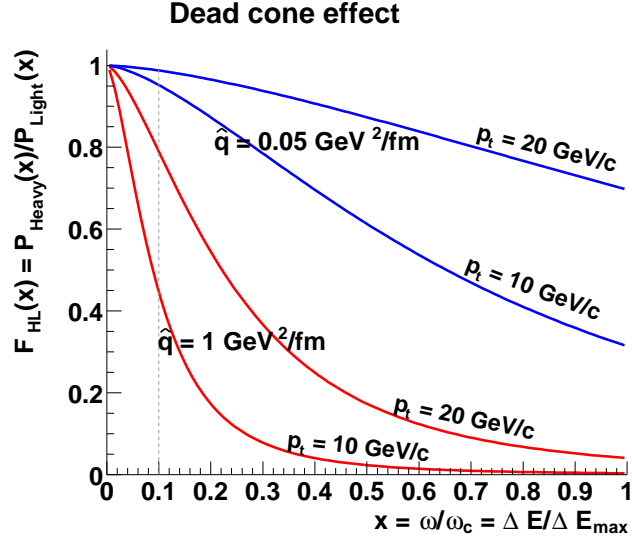}
    \caption{Suppression factor for a charm quark as a function of 
             $x=\omega/\omega_c$ (see text). 
             The in-medium path length considered is $L=5~\fm$.}
    \label{fig:fhl}
  \end{center}
\end{figure}

The heavy-to-light suppression factor $F_{\rm H/L}$ increases 
(less suppression) as the 
heavy quark energy increases (the mass becomes negligible). 
In Fig.~\ref{fig:fhl} we plot the suppression factor for 
charm quarks ($m_{\rm c}=1.2~\gev$) as a function of 
$x=\omega/\omega_c\simeq \Delta E/\Delta E_{\rm max}$. This relative scale
allows to compare directly situations with different transport 
coefficients, i.e. different medium densities. For given $\hat{q}$ and 
$\pt$ of the c quark (we then use $E=\sqrt{\pt^2+m_{\rm c}^2}$ assuming 
production at mid-rapidity), the factor $F_{\rm H/L}(x)$ can be 
interpreted as the decrease of the probability for emitting a 
gluon with energy $x\,\omega_c$. $F_{\rm H/L}$ decreases at large $x$, 
indicating that {\sl the high-energy part of the gluon radiation spectrum 
is drastically suppressed by the dead cone effect}. In a hot medium, 
$\hat{q}=1~\gev^2/\fm$, the probability to radiate a gluon with 
energy $0.1\times \omega_c$ (vertical dashed line), 
which corresponds to the average energy 
loss (see previous section), is reduced by a factor 0.5 for a 
$10~\gev/c$ charm quark and by a factor 0.8 for a $20~\gev/c$ charm quark, 
with respect to a light quark.

\begin{figure}[!t]
  \begin{center}
    \includegraphics[width=.9\textwidth]{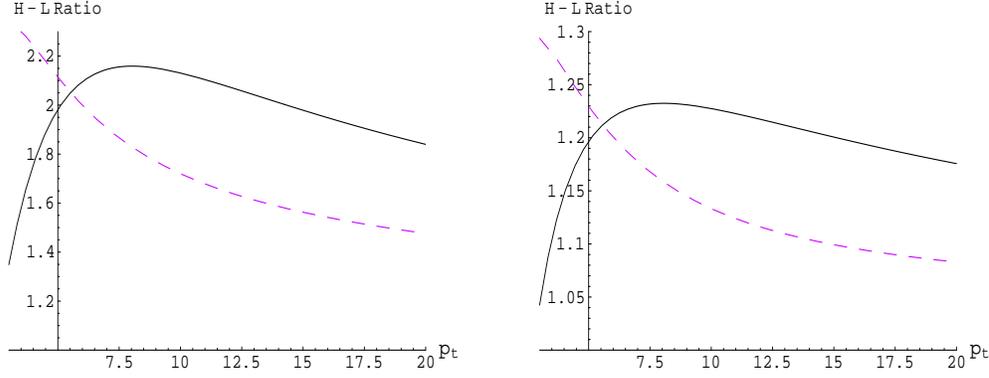}
    \caption{Ratio of the quenching factors for charm (H) and light (L) 
             quarks in hot matter with $\hat{q}=1~\gev^2/\fm$ 
             ($L=5~\fm$, left, and $L=2~\fm$, right), as a function of 
             $\pt~[\gev/c]$. Solid lines correspond
             to unrestricted gluon radiation, while dashed lines are based on 
             the calculation with a cut on the gluon energy 
             $\omega>0.5~\gev$~\cite{dokshitzerkharzeev}.}
    \label{fig:qhql}
  \end{center}
\end{figure}

The energy loss probabilities for heavy quarks are not calculated in 
Ref.~\cite{dokshitzerkharzeev}, as a realistic treatment of nuclear 
geometry and of the time evolution of QCD matter in the final state 
was not included. What is provided is a semi-quantitative illustration of 
the expected consequences of the dead cone effect on the transverse 
momentum dependence of the ratio of hadrons originating from the 
fragmentation of heavy and light quarks in heavy ion collisions; because
of the lower loss of heavy quarks, such ratio should be enhanced with 
respect to what measured in pp collisions.

The ${\rm D}/\pi$ ratio is considered, assuming all pions to originate from 
light quarks. Figure~\ref{fig:qhql} shows this ratio for 
$\hat{q}=1~\gev^2/\fm$, $L=5~\fm$ (left) and $L=2~\fm$ (right), 
with (dashed) and without (solid) a cut on the minimum energy of the 
emitted gluons. For $L=5~\fm$, left panel, a factor $\sim 2$ enhancement is 
expected at $\pt\sim 5$-$10~\gev/c$. The enhancement in the case of a cold 
medium with $\hat{q}=0.05~\gev^{2}/\fm$ is found to be of only 
$\sim 15\%$~\cite{dokshitzerkharzeev}. The conclusion of Dokshitzer and 
Kharzeev is that the ${\rm D}/\pi$ ratio appears to be extremely 
sensitive to the density of colour charges in QCD matter. Also the 
B-meson/D-meson ratio is regarded as specially interesting, because the 
different masses of c and b quarks imply a lower energy loss for the latter.

Concerning the proposed ${\rm D}/\pi$ observable, an important comment has 
to be made. The proton PDF plot in Fig.~\ref{fig:pdf} indicates
that, already at RHIC energies, and even more at LHC energies, 
hadron production come mostly from the fragmentation 
of gluons rather than light quarks (at the LHC, mostly means $\simeq 80\%$, 
as we shall show in Chapter~\ref{CHAP8}), and gluons lose more energy 
than light quarks. On the other hand, D mesons are expected to 
come essentially from the fragmentation of c quarks. If the c 
quark comes from a gluon splitting, the gluon must have a virtuality 
$Q>2\,m_{\rm c}$, meaning that the splitting happens on a spatial scale 
of $\sim 1/(2\,m_{\rm c})\simeq 0.1~\fm$, so that, also in this case, 
the c quark sees the whole medium thickness.

Therefore, {\sl the ${\rm D}/\pi$ (or, more generally, {\rm D}/hadrons) 
ratio is expected to be enhanced both by the different partonic origin 
of {\rm D} mesons and non-heavy-flavour hadrons and by the dead cone effect.}

\mysection{Pre-LHC measurements of open charm production in pA and AA}
\label{CHAP2:data}

\begin{figure}[!t]
  \begin{center}
    \includegraphics[width=.65\textwidth]{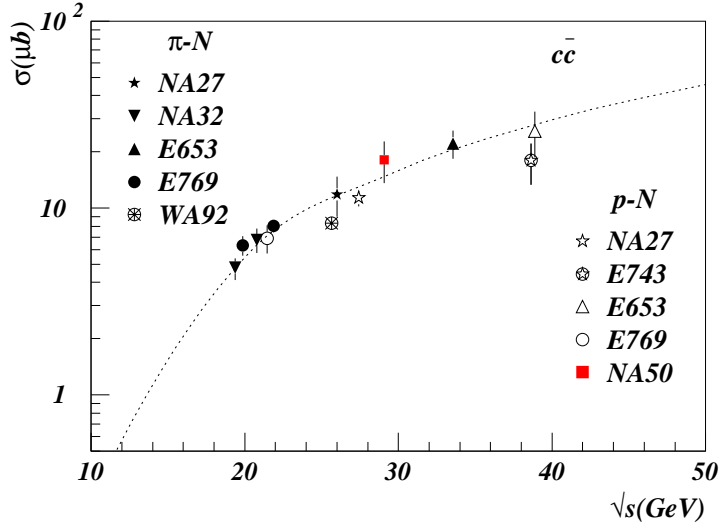}
  \end{center}
  \caption{Compilation of charm hadroproduction cross section 
           measurements~\cite{pbmcharm,Dna50}. 
           The figure is explained in the text.}
  \label{fig:pAcharm}
\end{figure}

Charm production in pion--nucleus and proton--nucleus collisions was 
measured by several experiments in a broad energy range and both the energy 
dependence and the A dependence are well understood and in agreement with 
the binary scaling 
$\sigma^{\scriptstyle {\rm c\overline{c}}}_{\rm pA}={\rm A}\times\sigma^{\scriptstyle {\rm c\overline{c}}}_{\rm pp}$, if no centrality selection is applied.
Figure~\ref{fig:pAcharm} presents a recent 
compilation of charm cross section measurements at different energies: 
the values refer to forward production ($x_F \equiv x_1-x_2>0$) and, for each 
pA or $\pi$A system, the cross section was divided by A to obtain the 
corresponding 
$\sigma^{\scriptstyle {\rm c\overline{c}}}_{\rm pp}$~\cite{pbmcharm,Dna50}.  
The energy dependence is well reproduced by the PYTHIA model~\cite{pythia} 
(dotted line).

On the other hand, the experimental picture on charm production in 
ultra-relativistic nucleus--nucleus reactions is quite unclear: there are
indications of a possible enhanced production in central \mbox{Pb--Pb} 
collisions at SPS energy from the dimuon spectra measured by 
the NA50 experiment~\cite{Dna50}, while no enhancement is observed in 
the first measurement of D meson production in \mbox{Au--Au} collisions 
at RHIC energy by the PHENIX experiment~\cite{Dphenix130}, 
which uses the semi-electronic decay channel. We briefly describe these 
measurements in the next paragraphs.

The NA38 and NA50 experiments have studied muon pair production in 
pA, \mbox{S--U} and \mbox{Pb--Pb} collisions at the SPS. The 
decay of D mesons in the semi-muonic channel allows to indirectly 
measure the production of $\rm D\overline{D}$ meson pairs by an 
analysis of the dimuon invariant mass region between the $\phi$ and 
the J$/\psi$, the so-called intermediate mass region 
($1<M_{\mu^+\mu^-}<3~\gev$). 
The reference process used as a normalization 
is the Drell-Yan (DY), 
which is supposed to be insensitive to 
the nature of the medium produced in the collision and which, therefore, 
scales with the number of binary collisions.

\begin{figure}[!t]
  \begin{center}
    \includegraphics[width=.55\textwidth]{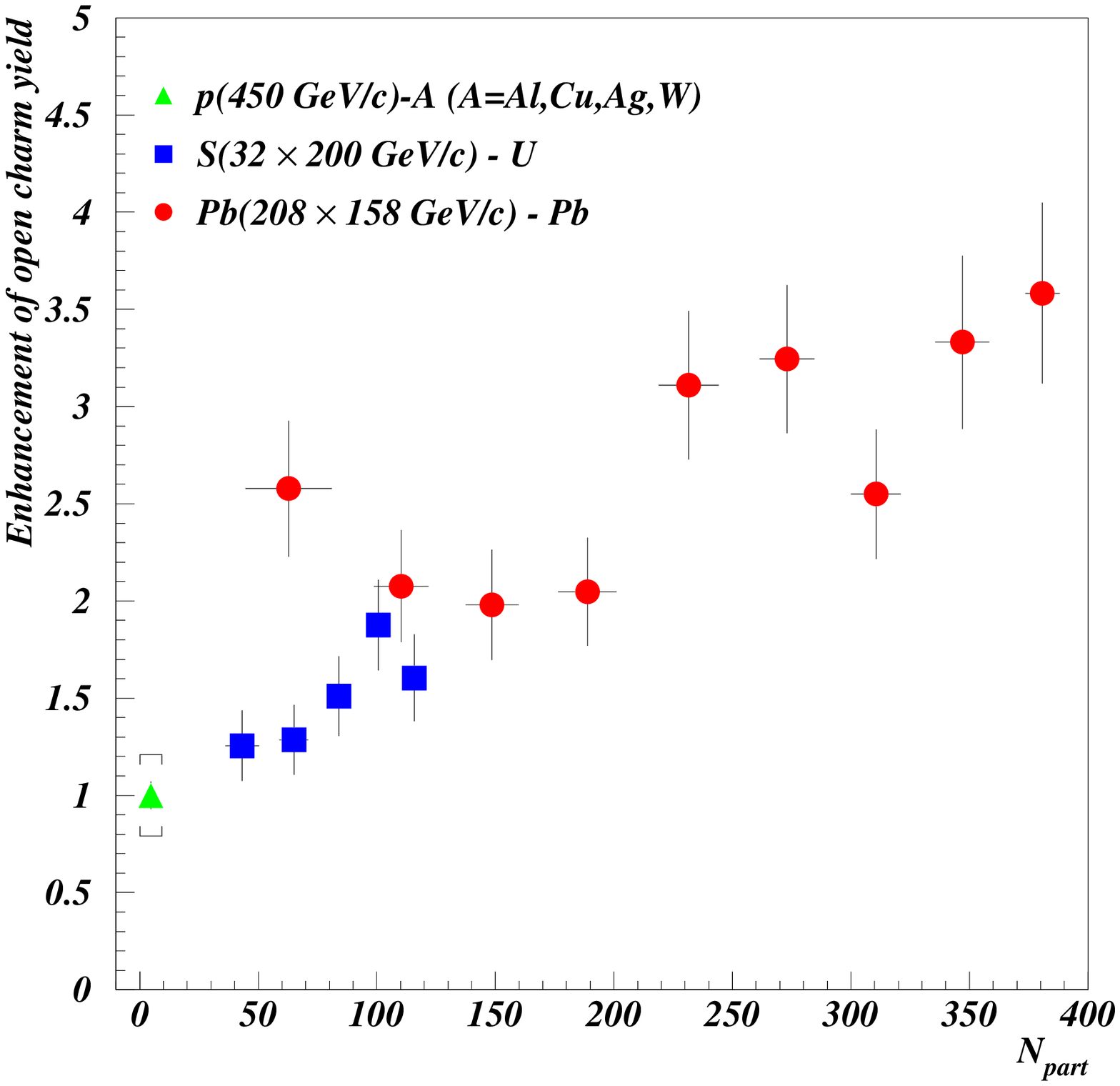}
    \caption{The enhancement $E$ of the charm yield needed to describe
           the dimuon spectrum in the intermediate mass region 
           measured by the NA50 experiment~\cite{Dna50}.}
    \label{fig:Dna50}
    \includegraphics[width=.7\textwidth]{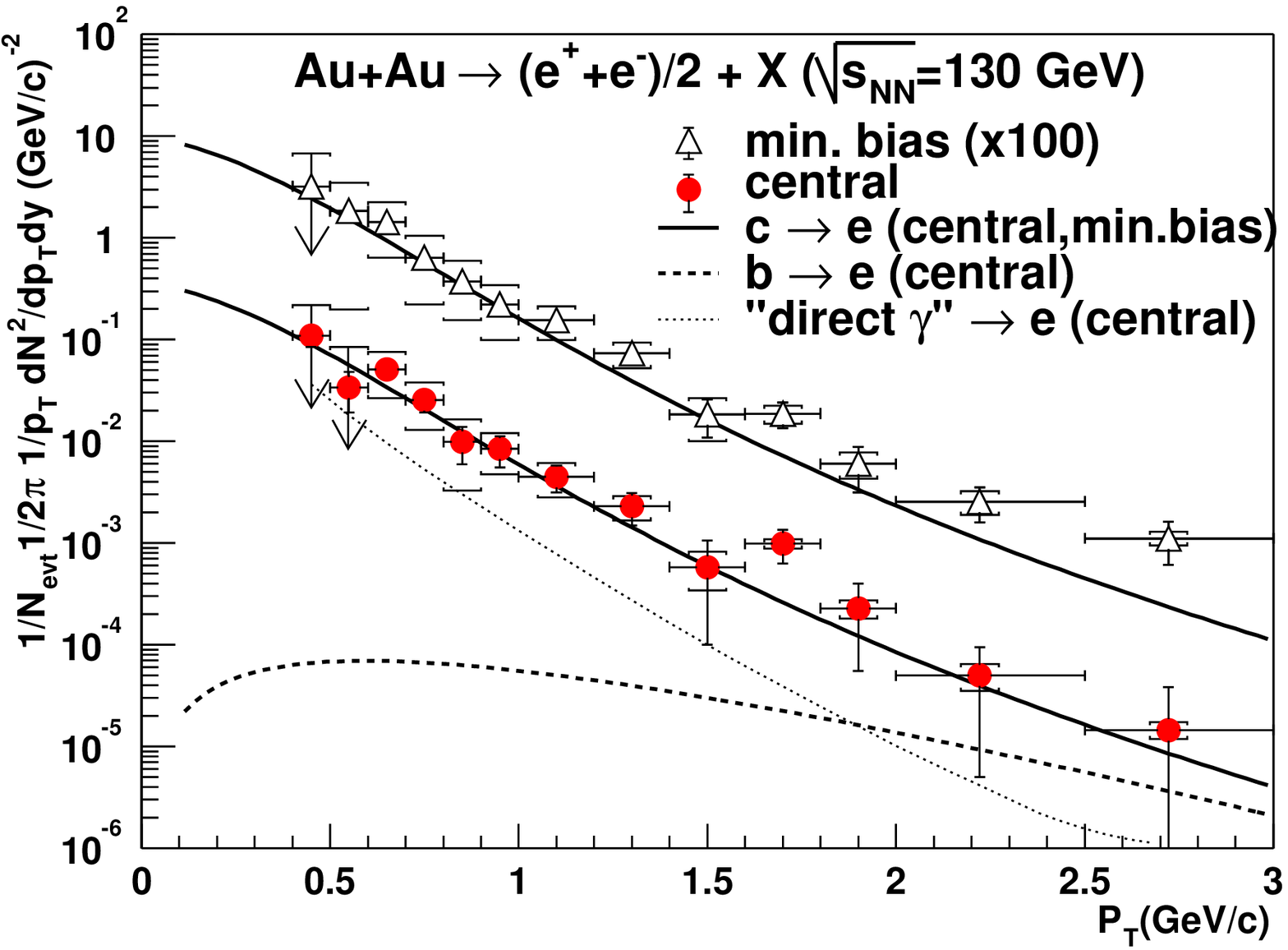}
    \caption{Background-subtracted electron spectra for minimum bias 
           and central collisions, measured by the PHENIX experiment,
           compared to the expected contributions from D decays 
           (PYTHIA). Also shown, for central collisions only, are the 
           expected contribution from beauty decays (dashed line) and the 
           conversion electron spectrum from a direct photon prediction 
           (dotted line)~\cite{Dphenix130}.}
    \label{fig:Dphenix}
  \end{center}
\end{figure}

In Ref.~\cite{Dna50} it is shown that pA data in the intermediate mass 
region can be described as a superposition of DY and $\rm D\overline{D}$ 
dimuons, using PYTHIA to calculate the expected differential spectra of 
the two contributions. When going to nucleus--nucleus collisions, 
a linear extrapolation of the pA sources,
assuming binary scaling, underestimates 
the data by an average factor $\sim 1.27$ for \mbox{S--U}
and $\sim 1.65$ for \mbox{Pb--Pb} collisions.
The expected value of the ratio $({\rm D\overline{D}/DY})$ is calculated 
using PYTHIA and compared to the value obtained from a fit to the data. 
The ratio of the fitted to the expected value is reported in 
Fig.~\ref{fig:Dna50} as a function of the number of participants, 
$N_{\rm part}$: in order to describe the data with a simple superimposition 
of DY plus $\rm D\overline{D}$, the expected charm yield has to be scaled 
up by a factor that increases roughly linearly with $N_{\rm part}$, 
reaching $\sim 3.5$ for central \mbox{Pb--Pb} reactions. 
This result is a bit puzzling, as it is 
very unlikely that additional charm quarks can be thermally 
produced at the SPS energy. We remind once more that this is an indirect 
measurement of charm production.

The PHENIX experiment at RHIC obtained an indirect estimate of 
charm production in \mbox{Au--Au} collisions at $\sqrtsNN=130$ and $200~\gev$ 
from the measurement of single electrons at central rapidity ($|\eta|<0.35$). 
The expected sources of electrons 
are (1) Dalitz and dielectron decays of light hadrons, (2) photon 
conversions, (3) kaon semi-electronic decays and (4) semi-electronic decays 
of D mesons (other contributions, such has beauty decays, are negligible at 
these energies). The contributions (1)-(3) were
 estimated using a simulation tuned to reproduce the $\pi^\pm$ and 
$\pi^0$ measurements by PHENIX and subtracted. 
The background-subtracted electron transverse momentum spectra were
 compared to the 
expected spectra from charm decays using PYTHIA (the event generator was 
tuned in order to describe the charm production data at CERN-SPS and 
at Fermilab 
and the extrapolation from pp to \mbox{Au--Au} was done with a 
scaling according to the number of binary collisions). 
The result at $\sqrtsNN=130~\gev$ is shown in 
Fig.~\ref{fig:Dphenix}~\cite{Dphenix130}. 
The calculated electron spectra show reasonable agreement, within the 
relatively large errors, with the 
data both for the minimum-bias sample and for the central one.
A similar agreement is shown also by the data at $\sqrtsNN=200~\gev$, 
where the contribution of photon conversions was directly estimated 
in a special run with an additional `converter layer' of well defined 
geometry and material thickness~\cite{Dphenix200}.

As we have seen in Section~\ref{CHAP1:rhic}, PHENIX reports
for high-$\pt$ hadrons in central Au--Au collisions a substantial 
suppression relative to binary scaling. Such effect seems not to be present 
(errors are still large) in the single 
electrons from charm. This may be explained by lower charm quark in-medium
energy loss due to the dead cone effect, described 
in Section~\ref{CHAP2:deadcone}.

In the near future, the NA60 experiment will probably clarify the issue of 
charm production in nucleus--nucleus collisions at SPS energy
and more precise measurements, including also the comparison with 
pp and pA collisions, will be performed in the energy 
domain $\sqrtsNN\sim 100$-$200~\gev$ by PHENIX.

\mysection{Probing the QGP with charm at the LHC}
\label{CHAP2:aliceD0}

\subsection{Strategy for the exclusive reconstruction of $\Dz$ mesons
           with ALICE}
\label{CHAP2:D0toKpi}

The investigation of medium-induced effects for charm quarks in the QGP
requires a good sensitivity on the momentum distribution of the 
quarks. Clearly, a direct measurement of the momentum of D mesons
would be more effective to this purpose than the indirect measurement via 
single electrons from the decay \mbox{${\rm D}\to e+X$}.
The exclusive reconstruction of hadronic decays of D mesons is the only way
to directly obtain their $\pt$ distribution. 
 
The mesons $\Dz$ and ${\rm D^+}$ (and antiparticles)
decay through weak processes and have decay lengths of the order of few 
tenths of a millimeter, namely $c\tau = (123.7\pm0.8)\ \mum$ for 
the $\Dz$ and $c\tau = (315.3\pm3.9)\ \mum$ for the ${\rm D^+}$~\cite{pdg}. 
Therefore, the distance between the interaction point (primary vertex) and 
their decay point (secondary vertex) is measurable. 

The selection of a suitable decay channel, which involves only
charged-particle products, allows the direct identification of the charm
states by computing the invariant mass of fully-reconstructed
topologies originating from secondary vertices.

We consider as a benchmark the process $\DtoKpi$
(and ${\rm \overline{D^0}\to K^+\pi^-}$); the fraction of $\Dz$ 
mesons which decay in this channel (branching ratio, $BR$) is 
$(3.80\pm0.09)\%$~\cite{pdg}. 

\begin{figure}[!t]
  \begin{center}
    \includegraphics[width=\textwidth]{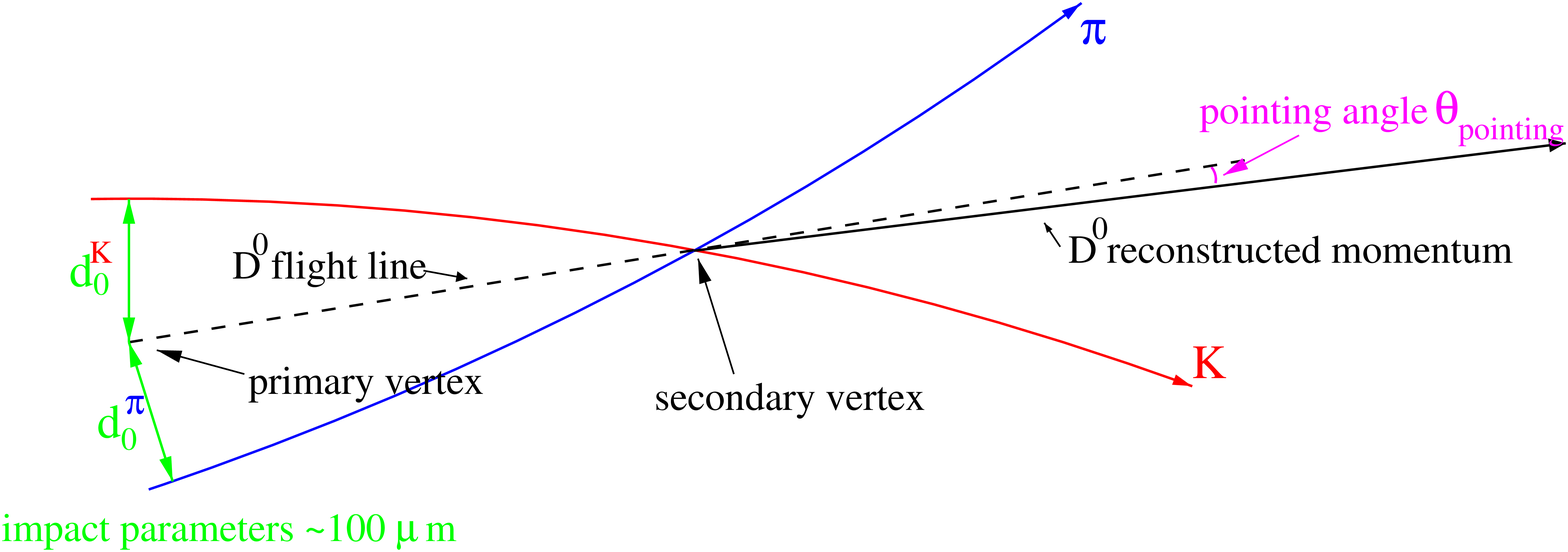}
    \caption{Schematic representation of the $\DtoKpi$ decay with
    the impact parameters ($d_0$) and the pointing angle
    ($\theta_{\rm pointing}$).} 
    \label{fig:D0sketch}
  \end{center}
\end{figure}

A sketch of the decay is shown in Fig.~\ref{fig:D0sketch} (the 
charged tracks are drawn bent by a magnetic field).
The main feature of this topology is the presence of two tracks displaced 
from the primary vertex. The variable that allows to evaluate the 
displacement of a track is the impact parameter, defined 
as the distance of closest approach of the track to the primary 
vertex. Here, we assume the presence of the ALICE solenoidal magnetic field 
and we indicate as $d_0$ the 
projection of the impact parameter on the bending plane, normal to 
the field direction. In Chapter~\ref{CHAP5} a more rigorous
definition of the track impact parameter shall be given.

In Appendix~\ref{App:kine} we show that, in the relativistic limit, the 
impact parameters of the decay products of a particle with mean proper decay 
length $c\tau$ have mean value of order $c\tau$.
Therefore, the decay products of a $\Dz$ particle have typical 
impact parameters of order $100~\mum$.

In this thesis, we study the feasibility for the exclusive reconstruction of 
$\Dz$ decays in heavy ion (and pp) collisions at the LHC by means of the 
invariant mass analysis of $\K^\mp \pi^\pm$ pairs.
Since the background to this decay is combinatorial, the task is a real 
challenge in the scenario of a central \PbPb~collision, where, as we 
have seen in Section~\ref{CHAP1:sqrtsdNdy}, 
up to $3000$-$4000$ charged tracks are expected to be 
produced per unit of rapidity.

In order to extract the signal out of this large background the 
candidate decay tracks 
are required to satisfy the following `secondary vertex criteria':
\begin{enumerate}
\item they are well separated from the interaction vertex, i.e. their impact 
      parameters are of order $100$-$500~\mum$; this is
      mandatory to reject the huge amount of primary tracks produced 
      in a heavy ion collision;
\item the sum of their momenta, which estimates the
      momentum of the $\Dz$ particle,
      points along the reconstructed $\Dz$ flight line; 
      this is realized by requiring the 
      angle $\theta_{\rm pointing}$ between the sum of the momenta and the 
      primary--secondary vertex direction to be small (see sketch in 
      Fig.~\ref{fig:D0sketch}).   
\end{enumerate}
This strategy demands:
\begin{itemize}
\item precise measurement of the momenta to have a good resolution on the
      invariant mass and, thus, reduce the background in the $\Dz$ mass 
      region
      (in Appendix~\ref{App:kine} we show that, in the relativistic 
      approximation, the invariant mass resolution is 
      proportional to the momentum resolution);
\item measurement of the impact parameters with resolution of the order of 
      $50~\mum$ for the decay products of the $\Dz$;
\item particle identification to tag the two decay products and reject 
      $\pi^\mp\pi^\pm$ pairs, which are a large part of the combinatorial 
      background.
\end{itemize}
The ALICE central barrel, described in Chapter~\ref{CHAP4}, 
was designed to fulfill these requirements. It, 
in fact, provides tracking and precise impact parameter measurement, with 
a Time Projection Chamber and a silicon Inner Tracking System, and $\K/\pi$ 
separation, with 
a Time-of-Flight detector, over the pseudorapidity range $-0.9<\eta<0.9$.
The results on the feasibility of the strategy here outlined 
are reported in Chapter~\ref{CHAP6}.

\subsection{Outline for the physics sensitivity studies}
\label{CHAP2:physicsoutline}

As discussed in Section~\ref{CHAP2:charminhic}, the two final state 
effects that can be investigated by means of fully-reconstructed 
$\Dz$ mesons in \PbPb~collisions at the LHC are the
c quark energy loss and the possible additional production of $\ccbar$
pairs in the hot medium.

The sensitivity to thermal charm production goes beyond the scope of this
work and it is an interesting issue for the charm studies in the near 
future. 

Energy loss, or quenching, effects can be addressed with two independent 
approaches, which we have already introduced when presenting the 
recent interesting observations at RHIC (see Section~\ref{CHAP1:rhic}):
\begin{description}
\item[Leading particle analysis:] 
                           study of the medium-induced modification of the 
                           transverse momentum distribution of 
                           identified D mesons.
\item[Correlations analysis:] 
                           study of the event structure, in terms of 
                           angular correlations of the produced particles with 
                           an identified D meson in the event.
\end{description}

In this work we concentrate on the leading particle analysis, as it is more
suitable for comparison of the results with model predictions and 
to disentangle different effects, by looking for example at the 
${\rm D}/hadrons$ ratio to quantify the dead cone suppression of medium-induced
radiation. However, also the study of the structure of 
charm-tagged jets by means of reconstructed $\DtoKpi$ decays is of 
great interest and is currently being investigated.

We consider the {\sl nuclear modification 
factor}, $R_{\rm AA}$, defined as the ratio of the $\pt$ distribution
measured in central AA collisions, divided by the number of 
estimated binary \NN~(NN) collisions, to the $\pt$ distribution measured in pp
collisions, scaled to the same c.m.s. energy:
\begin{equation}
\label{eq:raa}
  R_{\rm AA}(\pt)=
    \frac{{\rm d}\sigma^{\rm AA}/{\rm d}\pt/{\rm binary~NN~collision}}
       {{\rm d}\sigma^{\rm pp}/{\rm d}\pt} =  
    \frac{{\rm d}N^{\rm AA}/{\rm d}\pt}
       {N_{\rm coll}\times{\rm d}N^{\rm pp}/{\rm d}\pt}
\end{equation}

The choice of this ratio allows to reduce the systematic uncertainties, as
the errors which are common to AA and pp data cancel out.

If no nuclear effects were present the nuclear modification factor 
would be 1.
Initial state effects (shadowing and intrinsic $k_{\rm t}$ broadening) and
possible thermal charm production should affect $R_{\rm AA}$ only for 
relatively low transverse momenta ($\pt<5$-$7~\gev/c$), as discussed in 
Section~\ref{CHAP2:charminhic}. Above this region energy loss 
can be reasonably expected to be the only relevant effect. The behaviour 
of $\RAA$ in presence of an energy loss of average magnitude $\Delta E$ 
(for the hadrons) 
can be qualitatively described assuming the $\pt$ distribution to follow 
a power law d$N^{\rm pp}/$d$\pt\propto \pt^{-n}$. In this case, in fact, 
we have 
${\rm d}N^{\rm AA}/{\rm d}\pt/N_{\rm coll}\propto (\pt+\Delta E)^{-n}$ 
and, consequently: 
\begin{equation}
\label{eq:raaVSpt}
\RAA(\pt)\simeq \left(1+\frac{\Delta E}{\pt}\right)^{-n}.
\end{equation}
This ratio is clearly lower than 1 and, if $\Delta E$ does not depend on 
$\pt$, as it is the case for radiative energy loss in the BDMPS formulation, 
Eq.~(\ref{eq:avdE}), $\RAA$ is expected to increase with $\pt$ and reach 
1 for $\pt\gg n\,\Delta E$.
~\\

The sensitivity study (presented in Chapter~\ref{CHAP8}, except where
differently indicated) will be outlined as follows:
\begin{itemize}
\item estimate of the accessible $\pt$ range and of the main experimental 
      uncertainties on the $\pt$ distribution of $\Dz$ mesons, in pp and in 
      \PbPb~(Chapters~\ref{CHAP6} and~\ref{CHAP7});
\item study of a strategy for the extrapolation to $5.5~\tev$ of the 
      $\pt$ distribution measured in pp at $14~\tev$ (Chapter~\ref{CHAP7});
\item study of the effect of energy loss on the $\pt$ distributions of 
      c quarks and D mesons; the quenching weights calculated by Salgado and 
      Wiedemann~\cite{carlosurs} will be used, with a correction that takes 
      into account the dead cone; 
\item estimate of the experimental uncertainties on the nuclear 
      modification factor for $\Dz$ mesons;
\item experimental sensitivity with respect to the predictions 
      of the quenching model, without and with dead cone effect;
\item estimate of the experimental uncertainty on the 
      ${\rm D}/charged~hadrons$ (D/$h$) ratio, defined as 
      \begin{equation}
      \label{eq:RDh}
      R_{{\rm D}/h}(\pt)=R_{\rm AA}^{\rm D}(\pt)/R_{\rm AA}^h(\pt), 
      \end{equation}
      and experimental sensitivity for this ratio, without and with
      dead cone; we point out that, being in practice a double ratio
      Pb--Pb/Pb--Pb$\times {\rm pp/pp}$, $R_{{\rm D}/h}$ 
      is a very sensitive variable since most systematic errors cancel out.   
\end{itemize}

The main goal of this analysis is to assess the capability of ALICE to 
study and {\sl compare the effects induced by the medium on energetic gluons 
and heavy quarks}, which are predicted to be different on the basis of 
well-established properties of quantum chromodynamics, like the colour 
charges of quarks and gluons and the suppression of small-angle radiation off
massive quarks. The capability to experimentally investigate such expected 
differences allows to {\sl test the coherence of our understanding of the
properties of the matter produced in high-energy \AA~collisions.} 

\clearpage
\pagestyle{plain}

\setcounter{chapter}{2}
\mychapter{Charm and beauty production \mbox{at the LHC}}
\label{CHAP3}

\pagestyle{myheadings}

In the previous chapter we have discussed the physics motivations 
for the study of charm (or, more generally, heavy flavour) production
in \AA~collisions at the LHC.  
In this chapter we define the baseline used in the scope of this thesis
for what concerns 
the charm and beauty production cross sections at the LHC and the kinematical 
distributions of the produced heavy quarks. 

The most recent results of next-to-leading order perturbative 
QCD calculations for the cross sections in \NN~collisions at LHC 
energies are presented in Section~\ref{CHAP3:XsecNN}, where also the 
theoretical uncertainties are discussed. In Section~\ref{CHAP3:extrapolation}
the extrapolations to \PbPb~and \pPb~collisions are described.
Some relevant kinematical distributions for c and b quarks are shown 
in Section~\ref{CHAP3:kinematic}.
The PYTHIA event generator was tuned in order to reproduce
the heavy quarks transverse momentum distributions given by the NLO
calculations (Section~\ref{CHAP3:generators}).  
Finally, in Section~\ref{CHAP3:hadr} we report the yields and the
kinematical distributions for charm and beauty mesons.

\mysection{Cross sections in nucleon--nucleon collisions}
\label{CHAP3:XsecNN}

In this section we present the status of the cross section
calculations in nucleon--nucleon collisions and their comparison with
existing data up to a c.m.s. energy of $\simeq 65~\gev$. We then report the
results for LHC energies. The extrapolation to heavy ion
collisions is described in the next section.

\begin{figure}[!t]
  \begin{center}
    \includegraphics[width=0.67\textwidth]{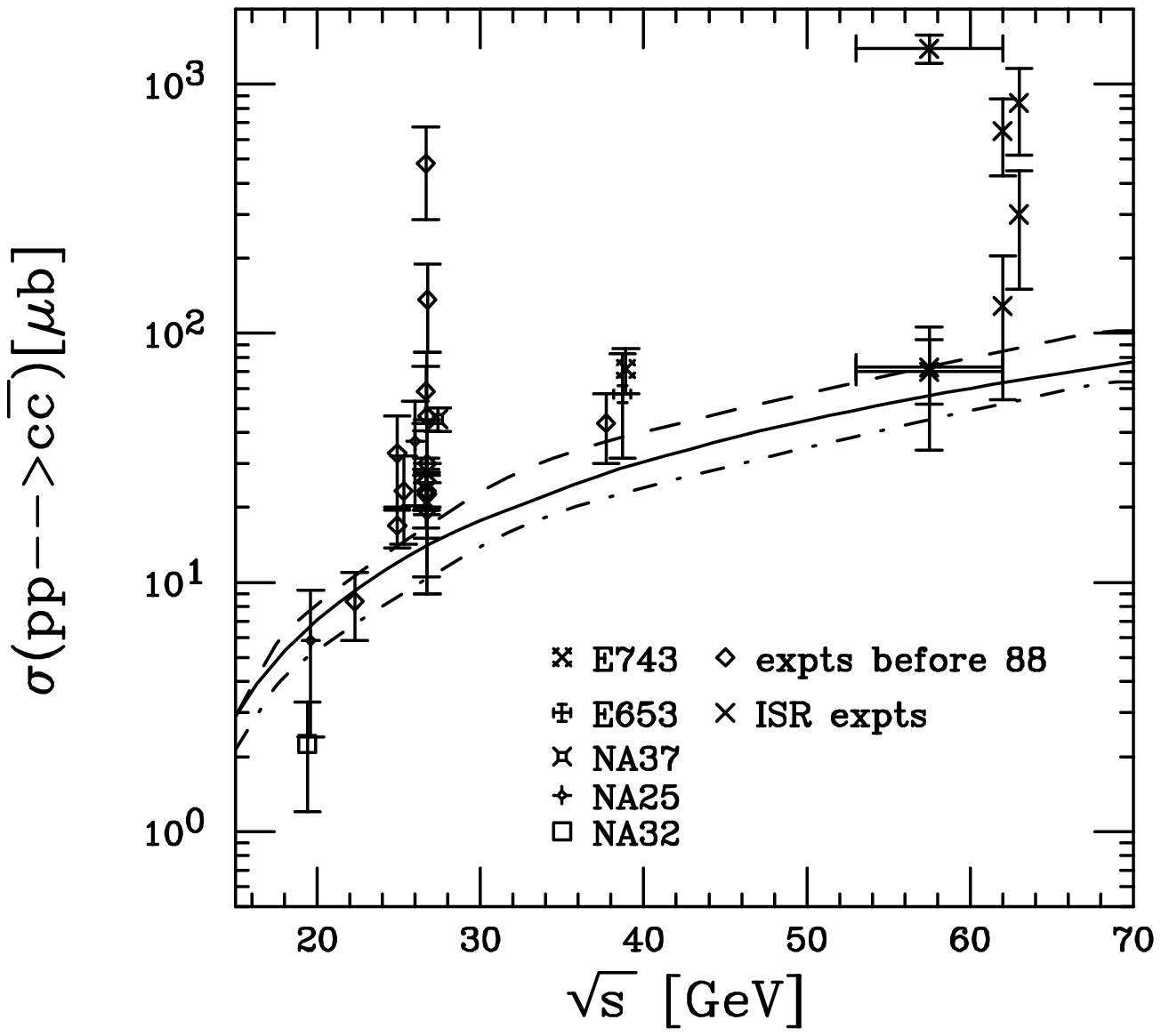}
    \caption{Total charm production cross section from pp and pA
    measurements compared to NLO calculations~\cite{vogtnew} with
    MRS D-' (solid), MRST HO (dashed) and MRST LO (dot-dashed) parton 
    distributions.}
    \label{fig:ramona1}
    \vglue0.5cm
    \includegraphics[angle=270,width=.49\textwidth]{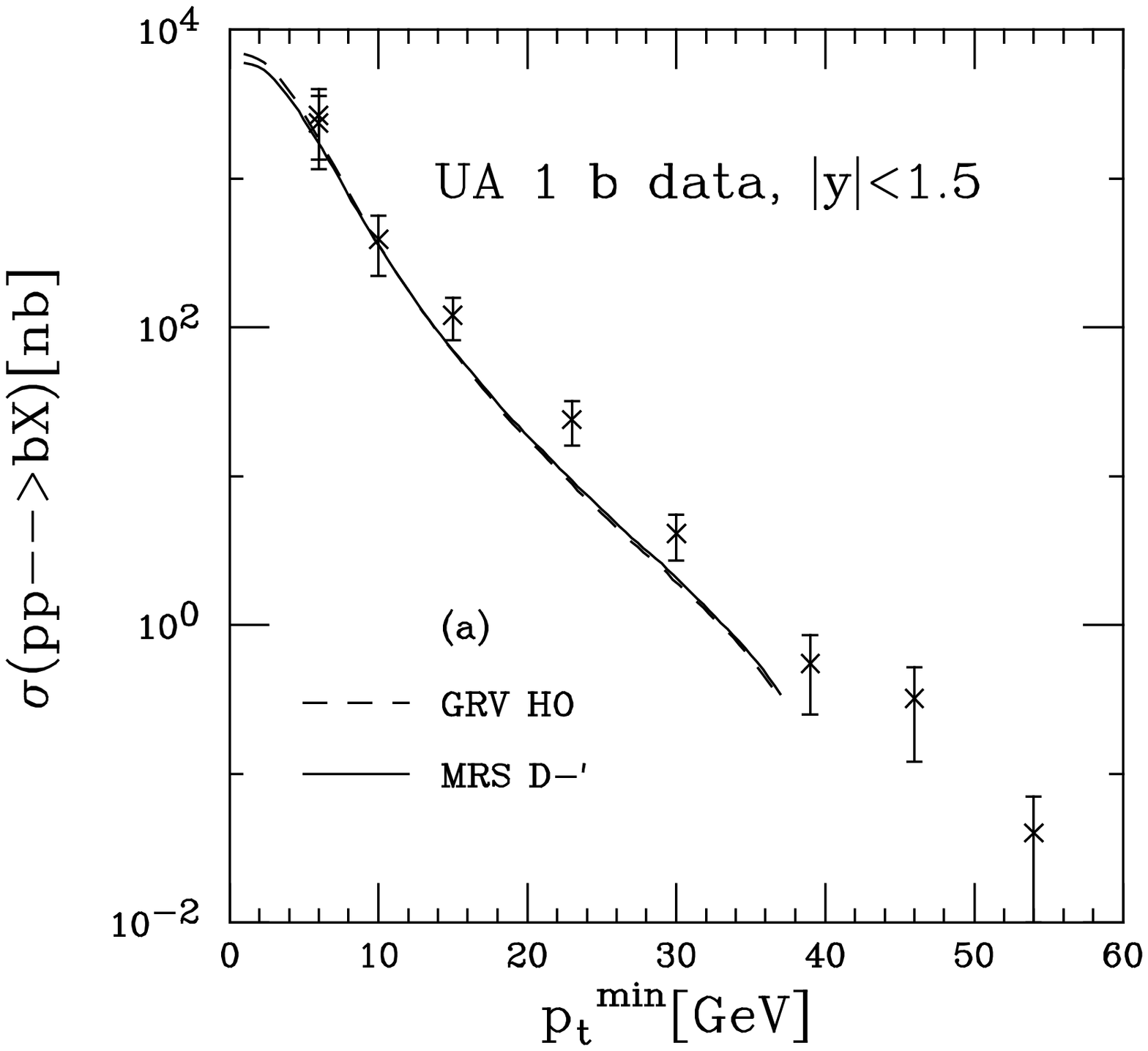}
    \includegraphics[angle=270,width=.49\textwidth]{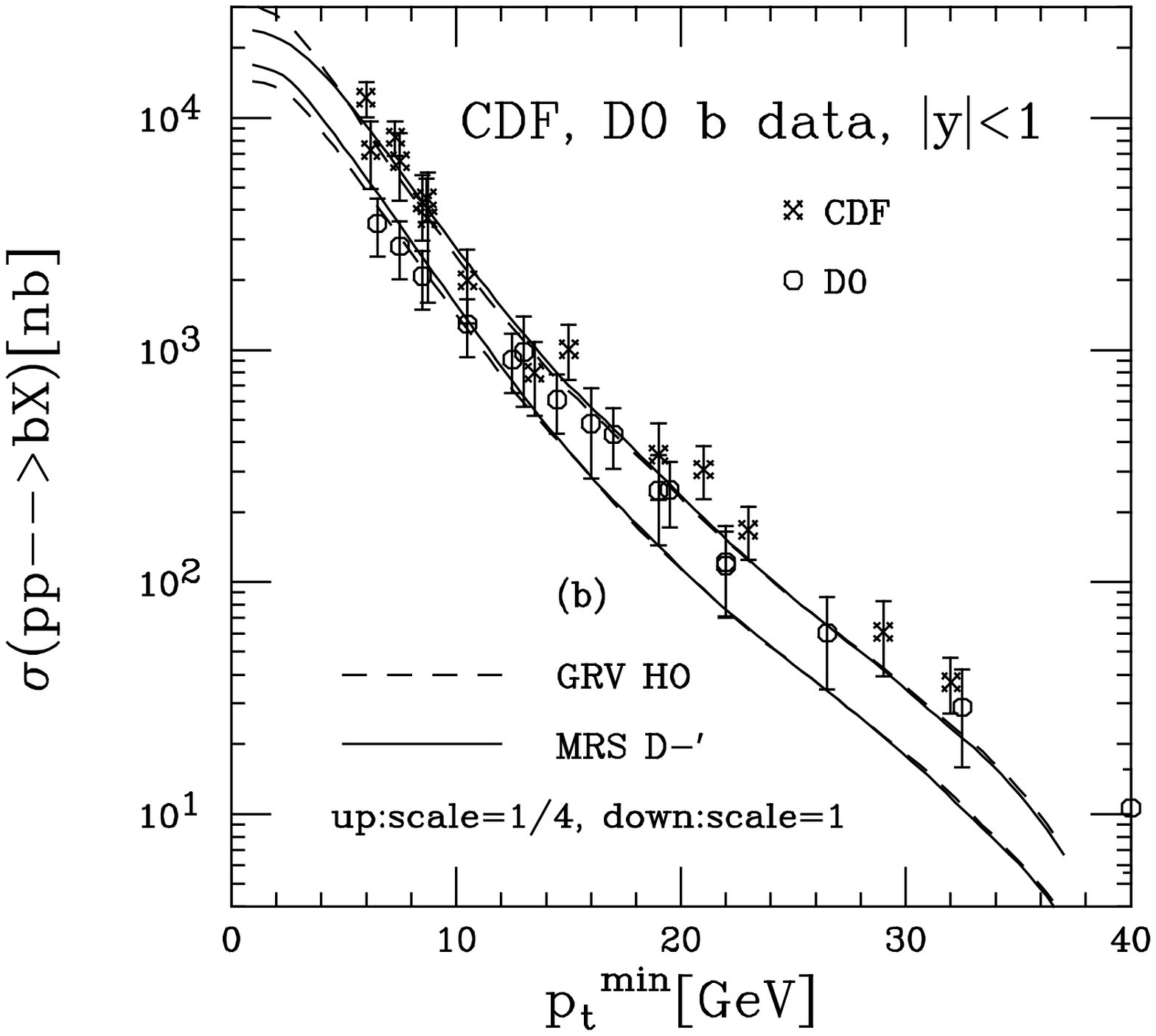}
    \caption{Comparison with b quark production cross section
    integrated over $\pt > \pt^{\rm min}$ from (a)
    UA1~\cite{UA1} and (b) CDF and D0~\cite{CDF}. The NLO calculations
    are with MRS D-' (solid) and GRV HO (dashed) parton
    distributions.}
    \label{fig:ramona2}
  \end{center}
\end{figure}

The existing data on total charm production cross section in pp and pA
collisions\footnote{The pA results were scaled according to the number 
of binary nucleon--nucleon collisions, in order to obtain the 
equivalent cross section in pp.} 
up to ISR energies are compared in Fig.~\ref{fig:ramona1}
with NLO calculations by R.~Vogt~\cite{vogtnew}. In
Fig.~\ref{fig:ramona2} a NLO calculation from Ref.~\cite{ramona94}
is compared to the data in \ppbar~collisions from UA1, CDF
and D0, for which the b quark production cross section integrated
for $\pt > \pt^{\rm min}$ is given. These measurements are taken in the
central rapidity region ($|y| < 1.5$ for UA1, $|y|<1$ for CDF and D0).
All the calculations have been performed using the following values
for the heavy quark masses ($m_{\rm c}$, $m_{\rm b}$) and for the factorization
and renormalization scales ($\mu_F$, $\mu_R$):
\begin{equation}
m_{\rm c} = 1.2~\gev
\,\,\,\,\,\,\,\,\,\,\,\,\,\,\,\,\,\,\,\,\,\,\,\,\,\,\,\,\,\,\,\,\,\,\,\,\,\,\,
\mu_F = \mu_R = 2\,\mu_0
\end{equation}
for charm, and
\begin{equation}
m_{\rm b} = 4.75~\gev
\,\,\,\,\,\,\,\,\,\,\,\,\,\,\,\,\,\,\,\,\,\,\,\,\,\,\,\,\,\,\,\,\,\,\,\,\,\,\,
\mu_F = \mu_R = \mu_0
\end{equation}
for beauty; 
$\mu_0=\sqrt{(p_{\rm t,Q}^2+p_{\rm t,\overline{Q}}^2)/2+m_Q^2}$ 
is approximately equal to the transverse mass of the produced heavy quarks.

For both charm and beauty the theory describes the present data reasonably
well.

\begin{table}[!b]
\caption{NLO calculation~\cite{MNRcode} for the total \ccbar~
  and \bbbar~
  cross sections in pp collisions at 5.5, 8.8 and 14~$\tev$, using the MRST HO
  and CTEQ 5M1 parton distribution functions.}
\label{tab:lhcXsec}
\begin{center}
\begin{tabular}{c|ccc|ccc}
\hline
\hline
 & & $\sigma_{{\rm pp}}^{\scriptstyle{\rm c\overline{c}}} [{\rm mb}]$ & & & $\sigma_{{\rm pp}}^{\scriptstyle{\rm b\overline{b}}} [{\rm mb}]$ & \\
\hline
$\sqrt{s}$ & $5.5~\tev$ & $8.8~\tev$ & $14~\tev$ &  $5.5~\tev$ & $8.8~\tev$ & $14~\tev$ \\
\hline
MRST HO & 5.9 & 8.4 & 10.3 & 0.19 & 0.28 & 0.46 \\
CTEQ 5M1 & 7.4 & 9.6 & 12.1 & 0.22 & 0.31 & 0.55 \\
\hline
Average & 6.6 & 9.0 & 11.2 & 0.21 & 0.30 & 0.51 \\  
\hline
\hline
\end{tabular}
\end{center}
\end{table}

The results for LHC energies ($\sqrt{s}=$ 5.5, 8.8 and 14~$\tev$) 
are reported in Table~\ref{tab:lhcXsec}. 
These values are obtained using the
NLO pQCD calculation implemented in the program
by M.~Mangano, P.~Nason and G.~Ridolfi~\cite{MNRcode} (HVQMNR) and two sets of
parton distribution functions, MRST HO~\cite{mrst} and CTEQ 5M1~\cite{cteq}, 
which include the small-$x$ HERA results. 
The difference due to the choice of the
parton distribution functions is relatively small ($\sim 20$-$25\%$ at
5.5~$\tev$, slightly lower at 14~$\tev$). We chose to use as a baseline 
the average, also reported in the table, of the values obtained 
with these two sets of PDF.

\begin{table}[!b]
\caption{Charm and beauty total cross sections at NLO with different
  choices of the parameters $m_{\rm c}$ ($m_{\rm b}$), $\mu_F$ and
  $\mu_R$~\cite{yelrepHardProbes}. In the last column the ratio 
  of the cross sections at 5.5~TeV and at 14~TeV is reported.}
\label{tab:manganoXsec}
\begin{center}
\begin{tabular}{cc|c|c|c}
\hline
\hline
& parameters & 5.5~$\tev$ & 14~$\tev$ & ratio 5.5~TeV/14~TeV \\
\hline
\hline
& $m_{\rm c} = 1.5~\gev$ & & \\
& $\mu_R=2\mu_0 \,\,\,\,\,\, \mu_F=2\mu_0$ &
\raisebox{1.5ex}[0cm][0cm]{3.7} &
\raisebox{1.5ex}[0cm][0cm]{7.3} &
\raisebox{1.5ex}[0cm][0cm]{0.51} \\
\cline{2-5}
& $m_{\rm c} = 1.2$~GeV & & \\
& $\mu_R=\mu_0 \,\,\,\,\,\, \mu_F=2\mu_0$ &
\raisebox{1.5ex}[0cm][0cm]{9.2} &
\raisebox{1.5ex}[0cm][0cm]{16.7} &
\raisebox{1.5ex}[0cm][0cm]{0.55} \\
\cline{2-5}
& $m_{\rm c} = 1.5$~GeV & & \\
& $\mu_R=\mu_0 \,\,\,\,\,\, \mu_F=2\mu_0$ &
\raisebox{1.5ex}[0cm][0cm]{5.4} &
\raisebox{1.5ex}[0cm][0cm]{10.4} &
\raisebox{1.5ex}[0cm][0cm]{0.52} \\
\cline{2-5}
& $m_{\rm c} = 1.8~\gev$ & & \\
\raisebox{10.5ex}[0cm][0cm]{$\sigma_{{\rm pp}}^{\scriptstyle{\rm c\overline{c}}} [{\rm mb}]$} &
$\mu_R=\mu_0 \,\,\,\,\,\, \mu_F=2\mu_0$ &
\raisebox{1.5ex}[0cm][0cm]{3.4} &
\raisebox{1.5ex}[0cm][0cm]{6.8} &
\raisebox{1.5ex}[0cm][0cm]{0.50} \\
\cline{2-5}
\hline
\hline
& $m_{\rm b} = 4.5~\gev$ & & \\
& $\mu_R=\mu_0 \,\,\,\,\,\, \mu_F=\mu_0$ &
\raisebox{1.5ex}[0cm][0cm]{0.20} &
\raisebox{1.5ex}[0cm][0cm]{0.51} &
\raisebox{1.5ex}[0cm][0cm]{0.39} \\
\cline{2-5}
& $m_{\rm b} = 4.75~\gev$ & & \\
& $\mu_R=\mu_0 \,\,\,\,\,\, \mu_F=\mu_0$ &
\raisebox{1.5ex}[0cm][0cm]{0.17} &
\raisebox{1.5ex}[0cm][0cm]{0.43} &
\raisebox{1.5ex}[0cm][0cm]{0.40} \\
\cline{2-5}
& $m_{\rm b} = 5~\gev$ & & \\
& $\mu_R=\mu_0 \,\,\,\,\,\, \mu_F=\mu_0$ &
\raisebox{1.5ex}[0cm][0cm]{0.15} &
\raisebox{1.5ex}[0cm][0cm]{0.37} &
\raisebox{1.5ex}[0cm][0cm]{0.41} \\
\cline{2-5}
& $m_{\rm b} = 4.75~\gev$ & & \\
& $\mu_R=0.5\mu_0 \,\,\,\,\,\, \mu_F=2\mu_0$ &
\raisebox{1.5ex}[0cm][0cm]{0.26} &
\raisebox{1.5ex}[0cm][0cm]{0.66} &
\raisebox{1.5ex}[0cm][0cm]{0.39} \\
\cline{2-5}
& $m_{\rm b} = 4.75~\gev$ & & \\
\raisebox{13.5ex}[0cm][0cm]{$\sigma_{{\rm pp}}^{\sbbbar} [{\rm mb}]$} &
$\mu_R=2\mu_0 \,\,\,\,\,\, \mu_F=0.5\mu_0$ &
\raisebox{1.5ex}[0cm][0cm]{0.088} &
\raisebox{1.5ex}[0cm][0cm]{0.20} &
\raisebox{1.5ex}[0cm][0cm]{0.44} \\
\hline
\hline
\end{tabular}
\end{center}
\end{table}

The dependence on the PDF set represents only a
part of the error on the theoretical estimate. An evaluation of the
theoretical uncertainties was done by M.~Mangano by varying the
$m_{\rm c}$ ($m_{\rm b}$), $\mu_F$ and $\mu_R$ parameters and is reported in
Table~\ref{tab:manganoXsec}~\cite{yelrepHardProbes}. This table shows that, at
LHC energies, the
theoretical uncertainties span a factor $\sim 2$-$3$ in the total
production cross section of both charm and beauty quarks.
In the last column of the table we report the ratio of the cross section 
at 5.5~TeV to that at 14~TeV. Despite the large spread of the 
absolute values, the ratio is much less dependent on the choice of the 
parameters; its value is $\simeq 0.52$ for charm and $\simeq 0.41$ for beauty. 
This indicates that pQCD can be used to compare the cross 
sections measured in \PbPb~collisions at $\sqrtsNN=5.5~\tev$ to 
those measured in \pPb~at 
$\sqrtsNN=8.8~\tev$ and in pp at $\sqrt{s}=14~\tev$. The uncertainty 
introduced by the energy extrapolation
is estimated in Chapter~\ref{CHAP7}.

\subsubsection*{Yields in \pp~collisions at $\sqrt{s}=14~\tev$}

Using a \pp ~inelastic cross section 
$\sigma^{\rm inel}_{{\rm pp}}=70~{\rm mb}$ 
at 14~$\tev$~\cite{pprCh2} and the average heavy flavour cross sections
in the last row of Table~\ref{tab:lhcXsec}, 
we calculate the yields for the production of $Q\overline{Q}$ pairs as:
\begin{equation}
  N_{{\rm pp}}^{\scriptstyle Q\overline{Q}}=\sigma_{{\rm pp}}^{\scriptstyle Q\overline{Q}}\bigg/\sigma^{\rm inel}_{{\rm pp}}.
\end{equation}
We obtain 0.16~\ccbar~pairs and 0.0072~\bbbar~pairs per event.

\mysection{Extrapolation to heavy ion collisions}
\label{CHAP3:extrapolation}

In this section we derive the extrapolation of the cross sections and yields 
to central \PbPb~collisions first, and then to \pPb~collisions.
We also point out the different weight of the nuclear shadowing effect 
in the two cases, for charm and beauty production.

\subsection{Nucleus--nucleus collisions}
\label{CHAP3:extrapolation2PbPb}

If no nuclear effects are taken into account, a \AA~collision 
can be considered as a superposition of {\sl independent}
\NN~collisions. Thus, the cross section for hard processes in heavy ion 
collisions can be calculated using a simple geometrical extrapolation 
from pp collisions, i.e. assuming that the hard cross section scales 
from pp to \AA~collisions proportionally to the number of 
inelastic \NN~collisions.

Nuclear effects ---such as nuclear shadowing, broadening of the
parton intrinsic transverse momentum ($k_{\rm t}$), energy loss,
as well as possible enhancements due to thermal
production in the medium--- can modify this geometrical scaling from pp to
nucleus--nucleus collisions. Such effects are, indeed, what we want to measure.
We chose to include in the simulation only the nuclear shadowing
and the broadening of the intrinsic $k_{\rm t}$, 
since they are well established
effects. The first effect modifies the 
total hard cross section, while the broadening of the intrinsic $k_{\rm t}$ 
affects only the kinematic distributions of the produced heavy quarks. 
Nuclear shadowing can be accounted for by recalculating the 
hard cross section in elementary \NN~collisions with modified parton 
distribution functions, 
as we have seen in Section~\ref{CHAP1:x}, and extrapolating to 
the \AA~case.

The extrapolation, based on the Glauber model~\cite{glauber,cywong}, 
is derived for the collision of two generic nuclei 
with mass numbers A and B, and numerical examples are given 
for the specific case of \PbPb~reactions at $\sqrtsNN=5.5~\tev$.

We are interested in the cross section for a sample of events in a given 
centrality range, defined by the trigger settings. 
The centrality selection can be assumed to correspond to a cut on the 
impact parameter $b$ of the collision: $0\leq b < b_c$. 
The sample of events defined by this cut contains a fraction of the 
total number of inelastic collisions, i.e. of the total inelastic 
cross section, given by 
\begin{equation}
F(b_c)=\int_0^{b_c}{\rm d}b\,\frac{{\rm d}\sigma^{\rm inel}_{\rm AB}}{{\rm d}b}\bigg/\int_0^{\infty}{\rm d}b\,\frac{{\rm d}\sigma^{\rm inel}_{\rm AB}}{{\rm d}b}.
\end{equation}
The definition of the centrality in terms of fraction of the inelastic 
cross section is more appropriate, since the cross section is directly 
measured, while the impact parameter estimation depends on the model used to 
describe the geometry of the collision.

In the following, we consider the most central class of events that can be 
selected by means of the ALICE Zero Degree Calorimeters (briefly described in 
Chapter~\ref{CHAP4}). This class 
corresponds to 5\% of the total inelastic cross section and
to a maximum impact parameter $b_c$ of about $3.5~{\rm fm}$.

\begin{figure}
  \begin{center}
  \includegraphics[width=0.49\textwidth]{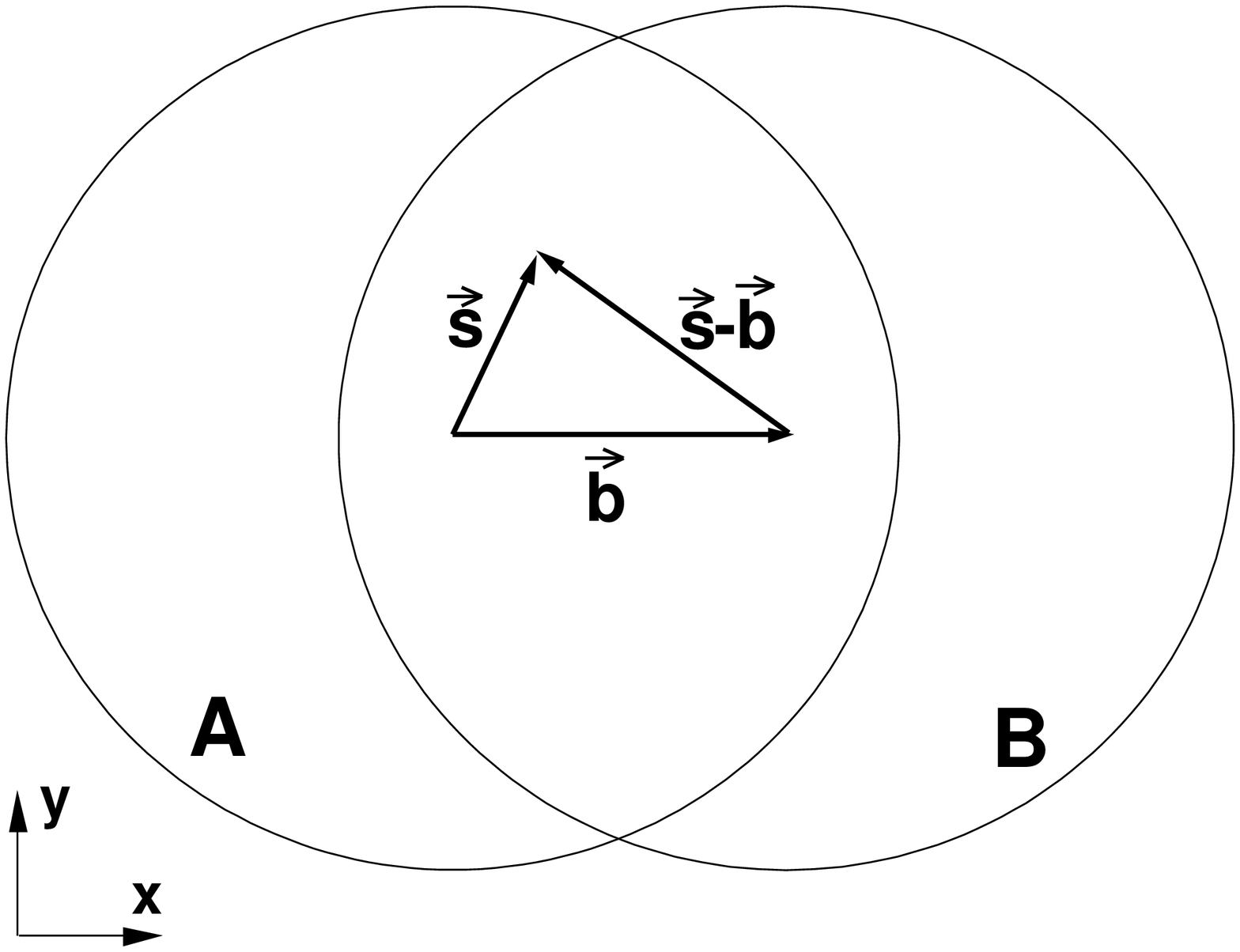}
  \includegraphics[width=0.49\textwidth]{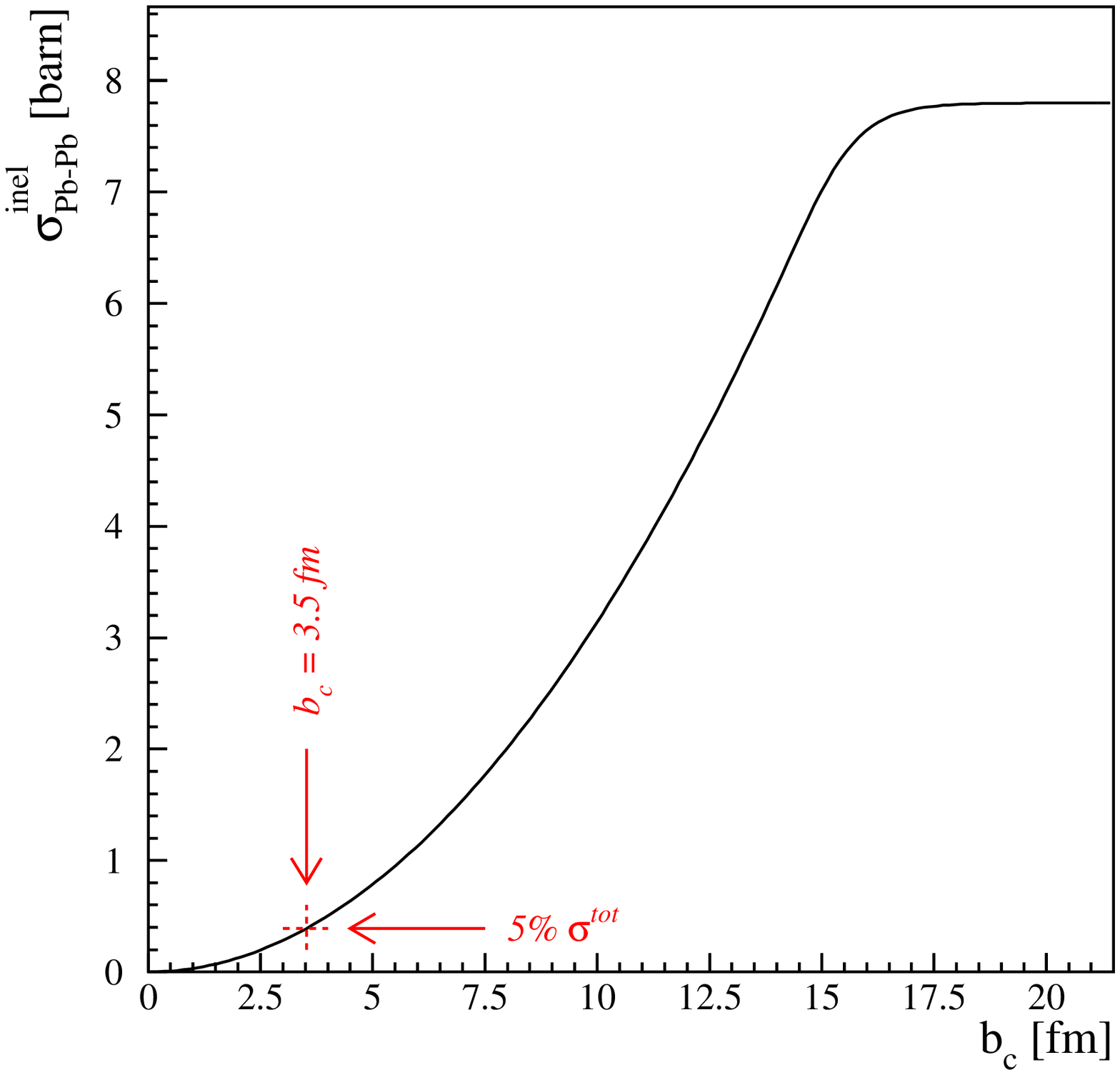}
  \caption{Left: collision geometry in the plane transverse to the beam line.
           Right: inelastic \PbPb~cross section as a function of the impact 
           parameter cut $b<b_c$; the value corresponding to 5\% of the 
           total inelastic cross section is indicated.}
  \label{fig:totinel}
  \end{center}
\end{figure}

The inelastic cross section corresponding to a given centrality
selection is found integrating the interaction probability up to
impact parameter $b_c$:
\begin{equation}
  \sigma_{{\rm AB}}^\mathrm{inel}(b_c) = \int_0^{b_c}{\rm d}b\,\frac{{\rm d}\sigma^{\rm inel}_{\rm AB}}{{\rm d}b} = 2\pi \int_0^{b_c} b\,{\rm d}b\,\, 
  \left\{1 - [1 - \sigma_{\scriptscriptstyle{{\rm NN}}} T_{{\rm AB}}(b)]^{{\rm AB}} \right\}
  \label{eq:totinel}
\end{equation}
where the value $\sigma_{\scriptscriptstyle{{\rm NN}}} = 60~{\rm mb}$ 
was used as the
nucleon--nucleon inelastic cross section at 5.5 TeV~\cite{pdg}, 
and the total thickness function $T_{{\rm AB}}$
\begin{equation}
  T_{{\rm AB}}(b) = \int {\rm d}^2s\, T_{\rm A}(\vec{s})\,T_{\rm B}(\vec{s}-\vec{b})
\end{equation}
(vectors defined as in Fig.~\ref{fig:totinel}, left) 
is expressed in terms of the thickness function of the nucleus
$T_i(\vec{s}) = \int {\rm d}z\, \rho_i(z,\vec{s})$ for $i={\rm A,\,B}$, 
where $\rho_i$ is the Wood-Saxon nuclear density profile~\cite{woodsaxon} 
---the thickness function is 
normalized to unity: $\int {\rm d^2}s\,T_i(\vec{s})=1$. 
In Fig.~\ref{fig:totinel} (right) the
inelastic cross section (\ref{eq:totinel}) is shown as a function of $b_c$. 

The average number of inelastic collisions for a given impact 
parameter $b$ is:
\begin{equation}
\label{eq:collisions}
  \sigma_{\scriptscriptstyle{{\rm NN}}}\cdot {\rm AB}\,\,T_{{\rm AB}}(b).
\end{equation}
By replacing the inelastic \NN~cross section 
$\sigma_{\scriptscriptstyle{{\rm NN}}}$ with the 
elementary cross section for a given hard process 
$\sigma_{\rm pp}^{\rm hard}$, we obtain 
the average number of inelastic collisions that yield the considered hard 
process:
\begin{equation}
\label{eq:hardcollisions}
  \sigma_{\rm pp}^{\rm hard}\cdot {\rm AB}\,\,T_{{\rm AB}}(b), 
\end{equation}
and the cross section for hard processes for $0\leq b<b_c$:
\begin{equation}
\label{eq:sigmaABhard}
  \sigma_{{\rm AB}}^\mathrm{hard}(b_c) = \sigma_{{\rm pp}}^\mathrm{hard}\cdot 2\pi \,\int_0^{b_c} b\,{\rm d}b\,\,{\rm AB}\,\,
  T_{{\rm AB}}(b).
\end{equation}
For minimum-bias collisions ($b_c=+\infty$), we have:
\begin{equation}
\label{eq:proptoAB}
  \sigma_{{\rm AB}}^\mathrm{hard} = \sigma_{{\rm pp}}^\mathrm{hard}\,{\rm AB}.
\end{equation}

\begin{figure}
  \begin{center}
    \includegraphics[width=0.49\textwidth]{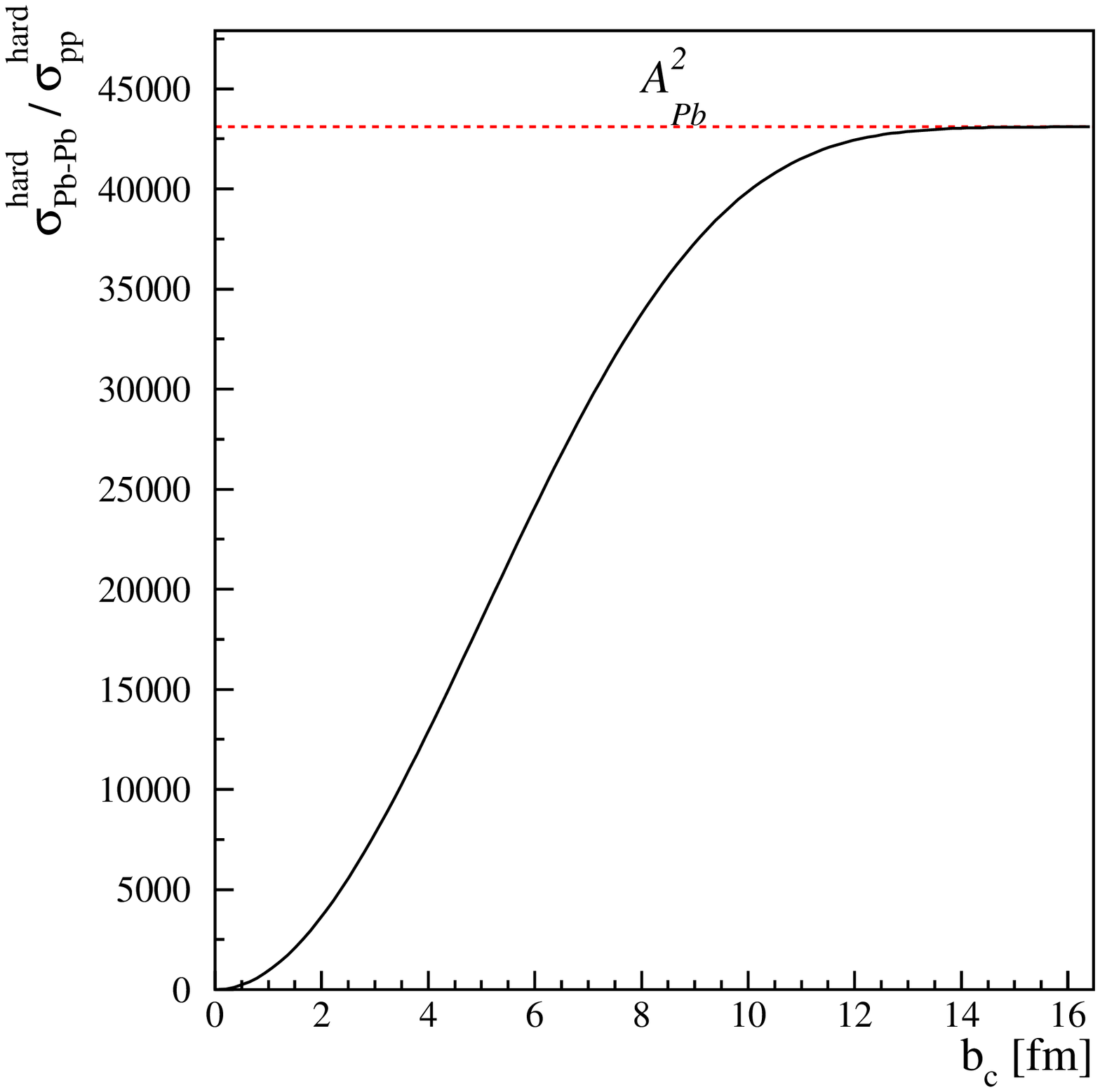}
    \includegraphics[width=0.49\textwidth]{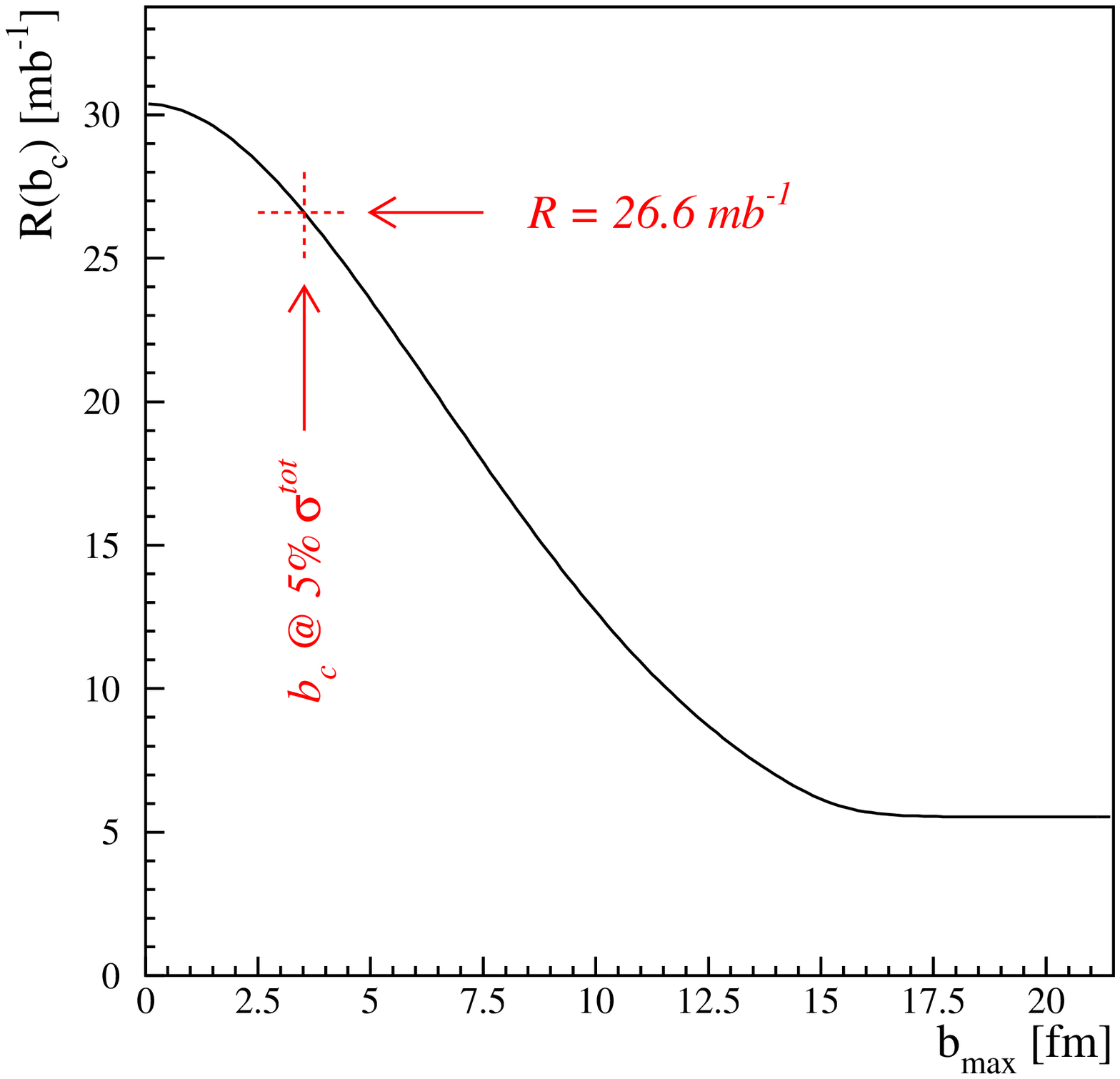}
  \caption{Left: cross section for a hard process in \mbox{Pb--Pb} collisions
    relative to the one in nucleon-nucleon collisions as a function of
    the impact parameter cut $b<b_c$. Right: yield of the hard process in 
    \mbox{Pb--Pb} collisions
    relative to the cross section in nucleon--nucleon collisions as a
    function of the impact parameter cut $b<b_c$.}
  \label{fig:fhardANDrbc}
  \end{center}
\end{figure}

The ratio of the hard cross section in nucleus--nucleus collisions, with a
centrality cut $b<b_c$, relative to the cross section in nucleon--nucleon
interactions is (see Fig.~\ref{fig:fhardANDrbc}, left):
\begin{equation}
  f^\mathrm{hard}(b_c) =
  \frac{\sigma_{{\rm AB}}^\mathrm{hard}(b_c)}{\sigma_{{\rm pp}}^\mathrm{hard}} =
  2\pi \int_0^{b_c} b\,{\rm d}b\,\, {\rm AB}\,\, T_{{\rm AB}}(b).
\end{equation}

The number (yield) of hard processes per triggered event is:
\begin{equation}
  N_{{\rm AB}}^\mathrm{hard}(b_c) = \frac{\sigma_{{\rm AB}}^\mathrm{hard}(b_c)}{\sigma_{{\rm AB}}^\mathrm{inel}(b_c)} = R(b_c) \cdot \sigma_{{\rm pp}}^\mathrm{hard}
\end{equation}
where (Fig.~\ref{fig:fhardANDrbc}, right)
\begin{equation}
\label{eq:rbc}
  R(b_c) = \frac{\int_0^{b_c} b\,{\rm d}b\,\,{\rm AB}\,\,T_{{\rm AB}}(b)}
     {\int_0^{b_c} b\,{\rm d}b\,\, 
  \left\{1 - [1 - \sigma_{\scriptstyle{{\rm NN}}} T_{{\rm AB}}(b)]^{{\rm AB}} \right\}}.
\end{equation}

For a 5\% centrality cut in \mbox{Pb--Pb} collisions, the yield 
$N_{\rm AB}^{\rm hard}$ is obtained by multiplying the elementary cross 
sections by $26.6~{\rm mb}^{-1}$.

\subsubsection*{Cross sections and yields in \PbPb~collisions at 
$\sqrtsNN=5.5~\tev$}

We used the EKS98 parameterization~\cite{EKS} of nuclear shadowing
(introduced in Section~\ref{CHAP1:x})
to recalculate the elementary charm and beauty production cross sections.
The reduction of the cross section due to shadowing amounts
to about 35\% for $\ccbar$ pairs, while it amounts only to
about 15\% for $\bbbar$ pairs, since, as pointed out in 
Section~\ref{CHAP1:x}, beauty production corresponds to
larger values of Bjorken $x$ (due to the larger mass of the b quark), 
that are less affected by the shadowing suppression 
(Fig.~\ref{fig:shadowingmodels}). In Section~\ref{CHAP3:kinematic} we will 
show how nuclear shadowing modifies the heavy quark kinematical distributions.

The values of the parton intrinsic $k_{\rm t}$ used in the simulation
were taken from Ref.~\cite{vogtnew}. 

Table~\ref{tab:xsecpb} summarizes the charm and beauty total cross
sections and yields in pp (with and without shadowing) and
\mbox{Pb--Pb} collisions (5\% centrality selection) 
at $\sqrtsNN=5.5~\tev$. 
The values shown correspond to the average of the
results obtained with MRST HO and CTEQ 5M1 parton distribution functions.

\begin{table}
\caption{Total cross sections and yields for charm and beauty
  production in pp and \mbox{Pb--Pb} collisions at 
  \mbox{$\sqrtsNN=5.5~\tev$}. 
  The effect of shadowing
  is shown as the ratio $C_{\rm shad}$ of the cross section calculated
  with and without the modification of the parton distribution
  functions.
  For the \mbox{Pb--Pb} case the centrality selection corresponds to 5\% 
  of the total inelastic cross section.}
\label{tab:xsecpb}
\begin{center}
\begin{tabular}{cc|c|c}
\hline
\hline
& & Charm & Beauty \\
\hline
& {\rm w/o~shadowing} & 6.64 & 0.21 \\
\cline{2-4}
\raisebox{1.5ex}[0cm][0cm]{$\sigma^{\scriptstyle Q\overline{Q}}_{{\rm pp}}\, [{\rm mb}]$} & {\rm w/~shadowing} & 4.32 & 0.18 \\
\hline
$C_{\rm shad}$ & & 0.65 & 0.84 \\
\hline
$\sigma^{\scriptstyle Q\overline{Q}}_{\rm Pb-Pb} [{\rm b}]$ & $5\%~\sigma^{\rm tot}$ & 45.0 & 1.79 \\
\hline
$N^{\scriptstyle Q\overline{Q}}_{\rm Pb-Pb}$ & $5\%~\sigma^{\rm tot}$ & 115 & 4.56 \\
\hline
\hline
\end{tabular}
\end{center}
\end{table}

\subsection{Proton--nucleus collisions}
\label{CHAP3:extrapolation2pPb}

For the extrapolation to \pA~collisions we use the geometrical 
Glauber-based method already described for the case of \AA~collisions. 
If we consider minimum-bias 
collisions (with no centrality selection), and we use $\rm B=1$ and 
$T_{\rm B}(\vec{s})=\delta(\vec{s})$ for the 
proton\footnote{The proton is assumed to be point-like.}, 
the total cross section for hard processes (\ref{eq:sigmaABhard}) 
becomes:
\begin{equation}
  \sigma_{{\rm pA}}^\mathrm{hard} = \sigma_{{\rm pp}}^\mathrm{hard}\cdot 2\pi \,\int_0^{\infty} b\,{\rm d}b\,\,{\rm A}\,\,T_{{\rm A}}(b) = {\rm A}\,\sigma_{{\rm pp}}^\mathrm{hard}.
\end{equation}

The number of hard processes per minimum-bias pA collision is:
\begin{equation}
  N^{\rm hard}_{\rm pA} = {\rm A}\,\sigma^{\rm hard}_{\rm pp}\bigg/\sigma^{\rm inel}_{\rm pA}.
\end{equation}

\subsubsection*{Cross sections and yields in \mbox{p--Pb} collisions 
  at $\sqrtsNN=8.8~\tev$}

We consider, as the reference proton--nucleus system,
\mbox{p--Pb} (and \mbox{Pb--p})\footnote{When we write \mbox{p--Pb}, 
we mean that the proton moves with $p_{\rm z}>0$; when we write \mbox{Pb--p}, 
we mean that the proton moves 
with $p_{\rm z}<0$.}. In this case we have $\sqrtsNN=8.8~\tev$ and 
the rapidity shift is $\Delta y=+0.47$ ($-0.47$ for \mbox{Pb--p}).

Using $\rm A=208$ and $\sigma_{\rm p-Pb}^{\rm inel}=1.9$~barn~\cite{pprCh2},
the yield of $Q\overline{Q}$ pairs per minimum-bias collision is:
\begin{equation}
  N^{\scriptstyle Q\overline{Q}}_{\rm p-Pb} = \sigma^{\scriptstyle Q\overline{Q}}_{\rm pp}\cdot 0.109~{\rm mb^{-1}}.
\end{equation}

As for the \PbPb~case, the effect of nuclear shadowing was accounted 
for by using the EKS98 parameterization~\cite{EKS}. Clearly, the effect is
lower for \mbox{p--Pb}, since one of the colliding nuclei is a proton: 
the reduction of the cross sections due to nuclear shadowing is 
20\% for charm and 10\% for beauty.

The cross sections and yields for charm and beauty production in pp 
(with and without shadowing) and minimum-bias \mbox{p--Pb} collisions at 
$\sqrtsNN=8.8~\tev$ are reported in Table~\ref{tab:xsecpPb}.
The values shown correspond to the average of the
results obtained with MRST HO and CTEQ 5M1 parton distribution functions.

A summary of the production yields and of the average magnitude of nuclear
shadowing in the three considered colliding systems is presented in 
Table~\ref{tab:summarytable}.

\begin{table}
\caption{Total cross sections and yields for charm and beauty
  production in pp and \mbox{p--Pb} collisions at 
  \mbox{$\sqrtsNN=8.8~\tev$}. 
  The effect of shadowing
  is shown as the ratio $C_{\rm shad}$ of the cross section calculated
  with and without the modification of the parton distribution
  functions.} 
\label{tab:xsecpPb}
\begin{center}
\begin{tabular}{cc|c|c}
\hline
\hline
& & Charm & Beauty \\
\hline
& {\rm w/o~shadowing} & 9.00 & 0.30 \\
\cline{2-4}
\raisebox{1.5ex}[0cm][0cm]{$\sigma^{\scriptstyle Q\overline{Q}}_{{\rm pp}}\, [{\rm mb}]$} & {\rm w/~shadowing} & 7.16 & 0.27 \\
\hline
$C_{\rm shad}$ & & 0.80 & 0.90 \\
\hline
$\sigma^{\scriptstyle Q\overline{Q}}_{\rm p-Pb} [{\rm b}]$ & & 1.49 & 0.056 \\
\hline
$N^{\scriptstyle Q\overline{Q}}_{\rm p-Pb}$ & & 0.78 & 0.029 \\
\hline
\hline
\end{tabular}
\end{center}
\end{table}

\begin{table}
\begin{center}
\caption{Summary table of the production yields and of the average 
         magnitude of nuclear shadowing in pp, \mbox{p--Pb} and \mbox{Pb--Pb}.}
\label{tab:summarytable}
\begin{tabular}{c|ccc|ccc}
\hline
\hline
 & & Charm & & & Beauty & \\
\hline
system & pp & p--Pb & Pb--Pb & pp & p--Pb & Pb--Pb \\
centrality & min.-bias & min.-bias & centr. (5\%) & min.-bias & min.-bias & centr. (5\%) \\ 
$\sqrtsNN$ & $14~\tev$ & $8.8~\tev$ & $5.5~\tev$ &  $14~\tev$ & $8.8~\tev$ & $5.5~\tev$ \\
\hline
$N^{\scriptstyle Q\overline{Q}}$/ev & 0.16 & 0.78 & 115 & 0.0072 & 0.029 & 4.56 \\
$C_{\rm shad}$ & 1 & 0.80 & 0.65 & 1 & 0.90 & 0.84 \\
\hline
\hline
\end{tabular}
\end{center}
\end{table}

\mysection{Heavy quark kinematical distributions}
\label{CHAP3:kinematic}

\begin{figure}
  \begin{center}
    \includegraphics[width=\textwidth]{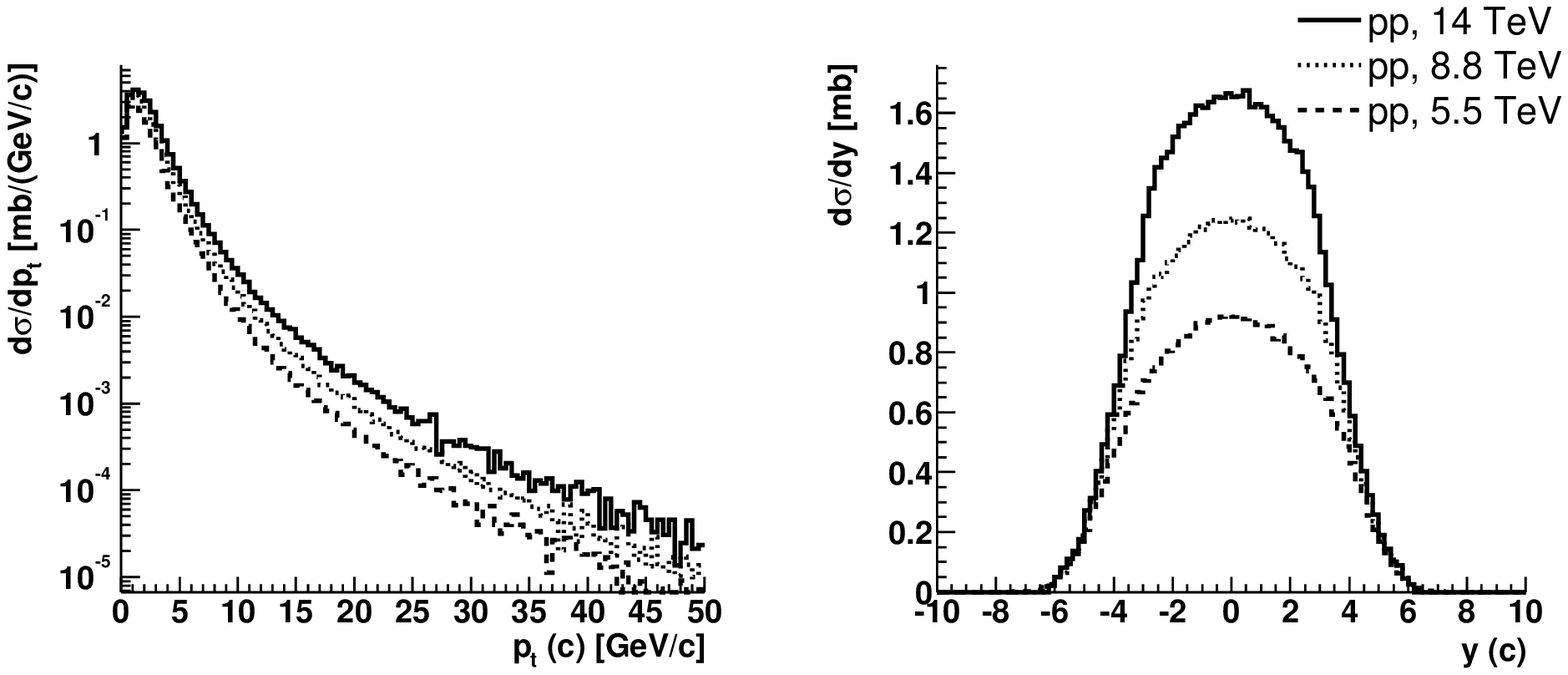}
    \includegraphics[width=\textwidth]{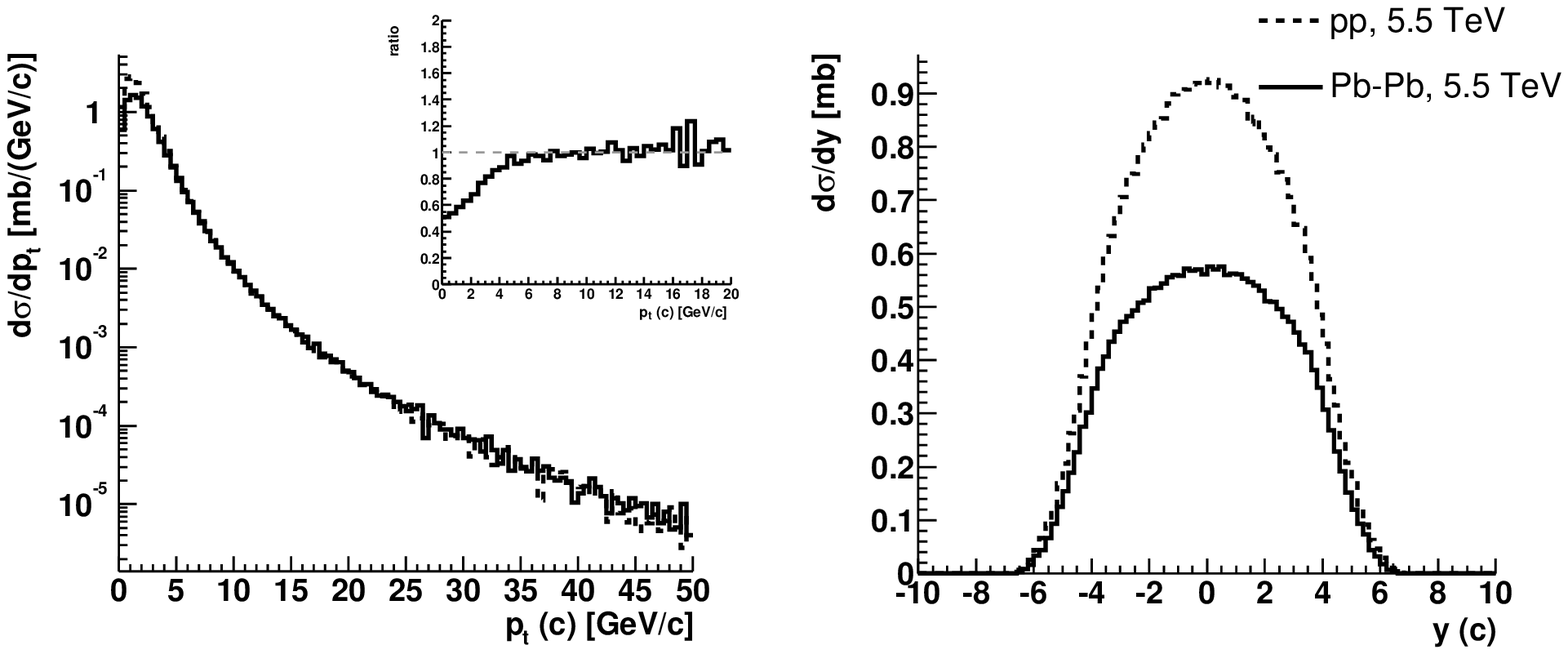}
    \includegraphics[width=\textwidth]{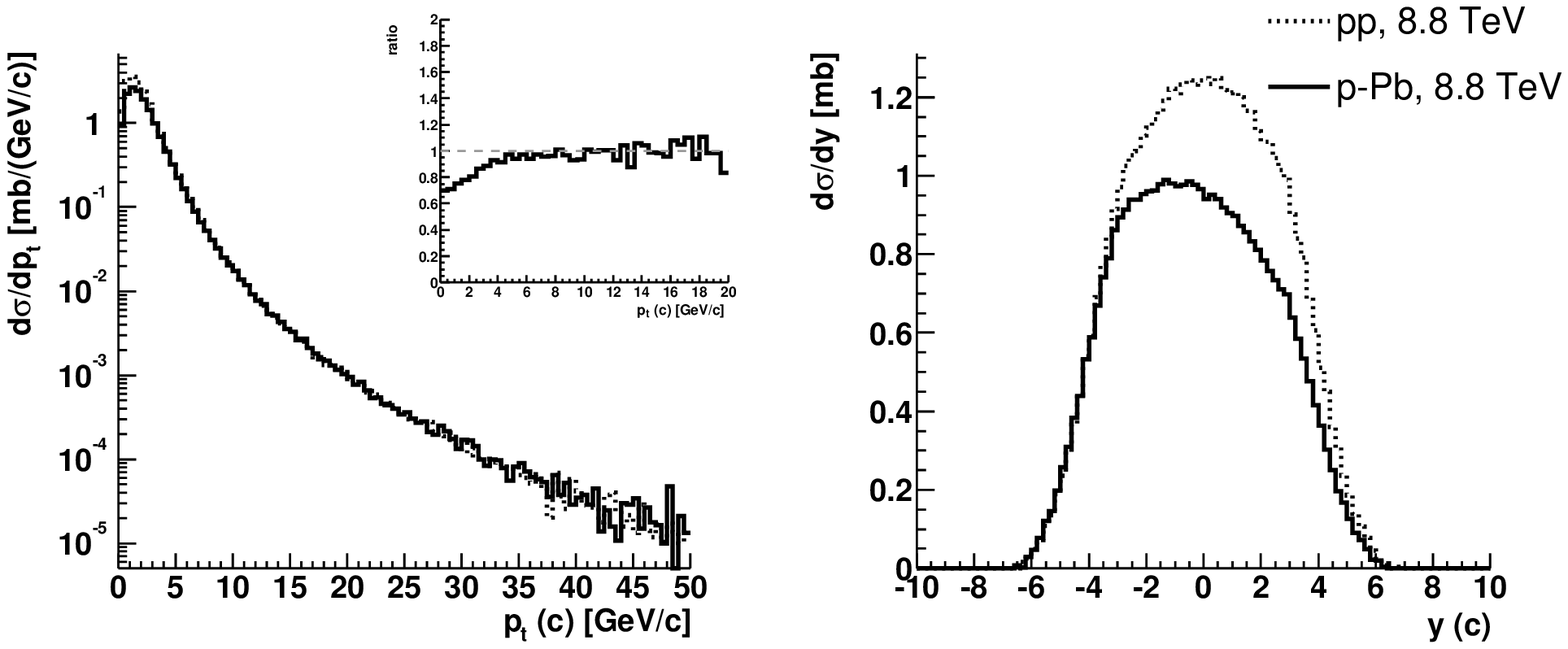}
  \end{center}
  \caption{Inclusive c quark $\pt$ and rapidity distributions obtained 
           from the HVQMNR program, using the CTEQ 5M1 set of PDF. 
           The distributions for \PbPb~and \pPb~are 
           normalized to the cross sections per \NN~collision and they 
           include the effects of nuclear shadowing and intrinsic 
           $k_{\rm t}$ broadening.}
  \label{fig:ckine}
\end{figure}

\begin{figure}
  \begin{center}
    \includegraphics[width=\textwidth]{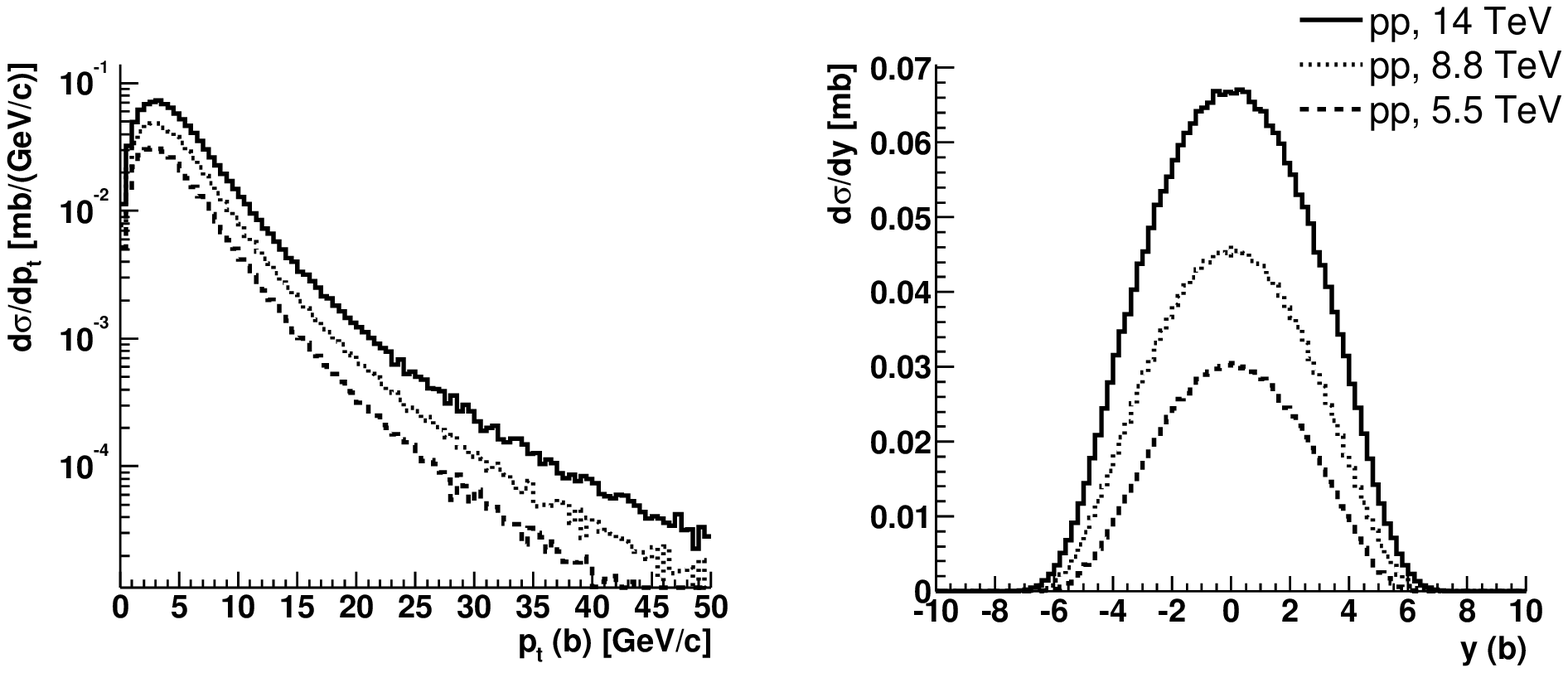}
    \includegraphics[width=\textwidth]{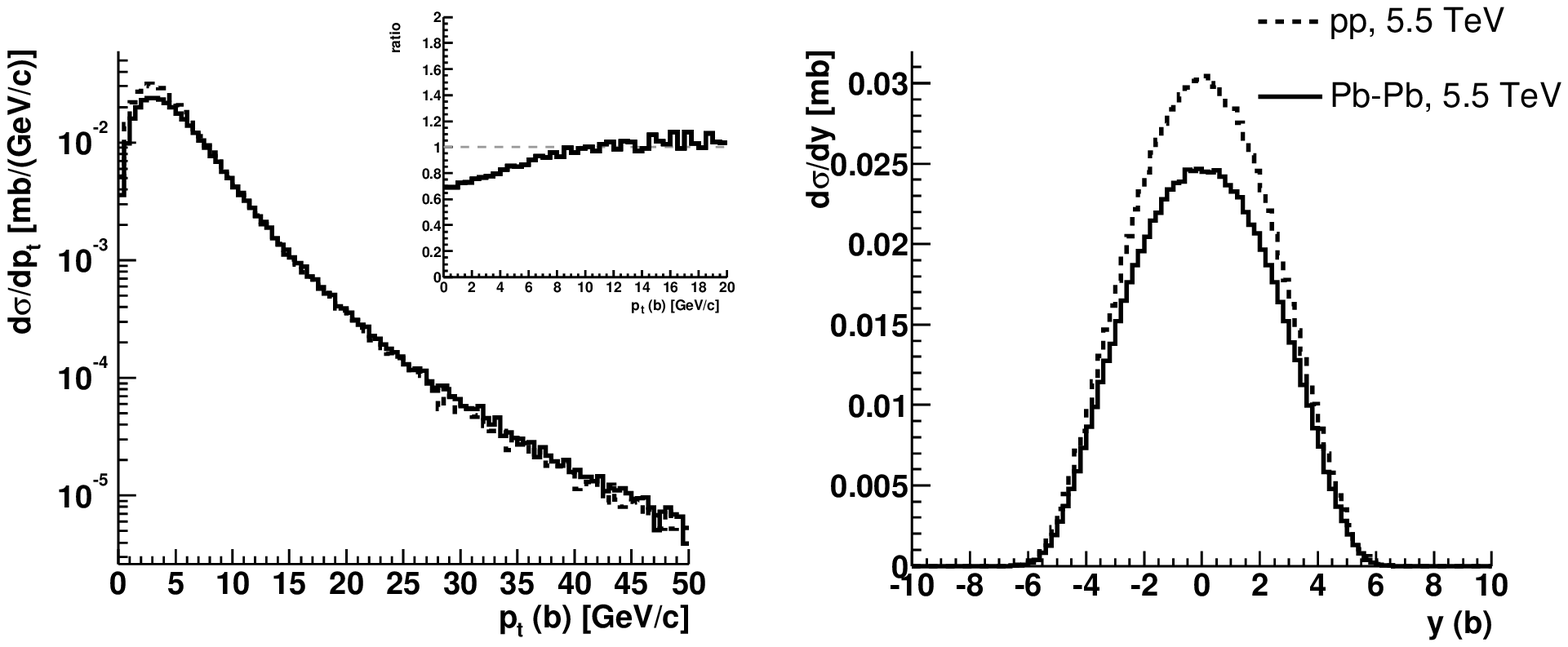}
    \includegraphics[width=\textwidth]{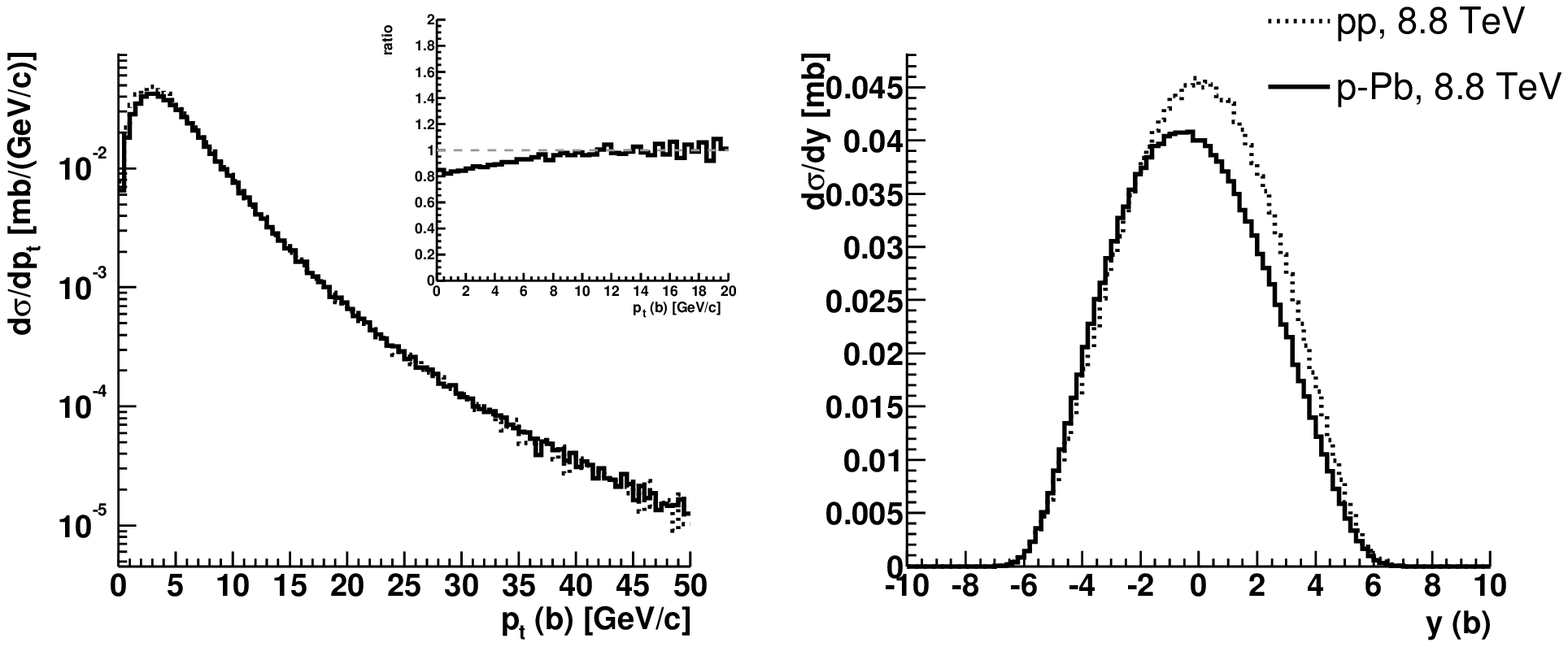}
  \end{center}
  \caption{Inclusive b quark $\pt$ and rapidity distributions obtained 
           from the HVQMNR program, using the CTEQ 5M1 set of PDF. 
           The distributions for \PbPb~and \pPb~are 
           normalized to the cross sections per \NN~collision and they 
           include the effects of nuclear shadowing and intrinsic 
           $k_{\rm t}$ broadening.}
  \label{fig:bkine}
\end{figure}

Figures~\ref{fig:ckine} and~\ref{fig:bkine} present the transverse momentum 
and rapidity distributions, obtained using the NLO pQCD program 
HVQMNR, for c and b quarks, respectively.
The distributions for \PbPb~and \pPb~are normalized to the cross 
section per \NN~collision. Nuclear shadowing is included via the EKS98 
parameterization and intrinsic $k_{\rm t}$ broadening is included as well.
In the case of \pPb~events the rapidity distribution in the centre-of-mass 
frame is plotted; the rapidity distribution in the laboratory frame 
would be shifted by $\Delta y=0.47$.

We notice that the $\pt$ distributions for pp collisions at 5.5, 8.8 and 14~TeV
 (top-left panel) have essentially the same shape.

The comparison of the $\pt$ distributions for pp and \PbPb~(and for 
pp and \pPb) at the same centre-of-mass energy shows that, as expected,
nuclear shadowing affects heavy quark production only for relatively 
low transverse momenta ($\pt<5$-$6~\gev/c$ with EKS98), 
where the $\QQbar$ pairs
are produced by low-$x$ gluons. This is clearly seen in the ratios of the 
distributions, reported in the insets.

A relevant feature of $\QQbar$ production in \pPb~collisions
is a depletion in the forward region (where the proton goes) of the rapidity 
distributions. This effect is due to the shadowing which is biased toward
forward rapidities, where the smallest $x$ values in the Pb nucleus 
are probed. 

\mysection{Heavy quark production in Monte Carlo event generators}
\label{CHAP3:generators}

The program used for the NLO calculations reported in
the previous sections is
not well suited to be included in a simulation, since it is not
an event generator and it does not provide parton kinematics.
On the other hand, widely used event
generators, like PYTHIA~\cite{pythia} and HERWIG~\cite{herwig}, 
are exact only at leading order, when only the pair production processes
($q\overline{q}\to Q\overline{Q}$ and $gg\to Q\overline{Q}$) are included. 
Higher-order contributions are included in these generators
in the parton shower approach (see e.g. Ref.~\cite{norrbin}). 
This model is not exact at next-to-leading order, but it reproduces 
some aspects of the multiple-parton-emission phenomenon.
In the following we will concentrate on the PYTHIA event generator; 
the version we have used is PYTHIA 6.150. We have also investigated 
heavy quark production in HERWIG, observing an incorrect behaviour
in the final kinematical distributions of both c and b quarks. We, therefore, 
concluded that HERWIG is not suitable for heavy quark simulations at LHC
energies. More details can be found in Ref.~\cite{yelrepHardProbes}.  


In PYTHIA, the processes giving rise to contributions above
leading order, see Section~\ref{CHAP2:pQCD}, 
like flavour excitation ($qQ\rightarrow qQ$ and
$gQ\rightarrow gQ$) and gluon splitting ($g\rightarrow
Q\overline{Q}$), are calculated using a massless matrix element. As a
consequence the cross sections for these processes diverge as
$\pt^{\rm hard}$ vanishes\footnote{$\pt^{\rm hard}$ is defined as the
transverse momentum of the outgoing quarks in the rest frame of the
hard interaction.}. These divergences are regularized by putting a lower
cut-off on $\pt^{\rm hard}$. The value of the minimum
$\pt^{\rm hard}$ cut has a large influence on the heavy flavour cross
section at low $\pt$, a region of prime interest for ALICE physics and
covered by the ALICE acceptance. Our approach was, therefore, to
tune the PYTHIA parameters in order to reproduce as well as possible the NLO
predictions (HVQMNR). 
We used PYTHIA with the option MSEL=1, that allows to switch 
on one by one the different processes (see Appendix~\ref{App:pythiahvq} 
for more details). 
The main parameter we tuned is the lower $\pt^{\rm hard}$
limit. In this procedure we compared the following distributions of the
bare quarks:
\begin{itemize}
\item inclusive $\pt$ and rapidity distributions of the quark (antiquark);
\item mass of the pair: $M(Q\overline{Q}) =
  \sqrt{(E_Q+E_{\overline{Q}})^2-(\vec{p}_Q+\vec{p}_{\overline{Q}})^2}$,
  where $E_Q = \sqrt{m_Q^2 + p_Q^2}$ is the quark energy;
\item $\pt$ of the pair, defined as the projection on the plane normal
  to the beam axis of the $Q\overline{Q}$ total momentum;
\item angle $\Delta\phi$ between the quark and the antiquark in the
  plane normal to the beam axis.
\end{itemize}

In the simulations for \mbox{Pb--Pb} collisions at $\sqrtsNN=5.5~\tev$ 
the parton distribution functions used are the
CTEQ 4L, modified for nuclear shadowing using the EKS98~\cite{EKS}
parameterization. We verified that the results given by the CTEQ 4 set 
lie in between the ones obtained with the more recent CTEQ 5 and MRST sets 
for all the relevant kinematical quantities.

The overall normalization was set to the value of the 
cross sections obtained for \pp~with shadowing  
(second row of Table~\ref{tab:xsecpb}).

\begin{figure}
  \begin{center}
    \includegraphics[width=0.9\textwidth]{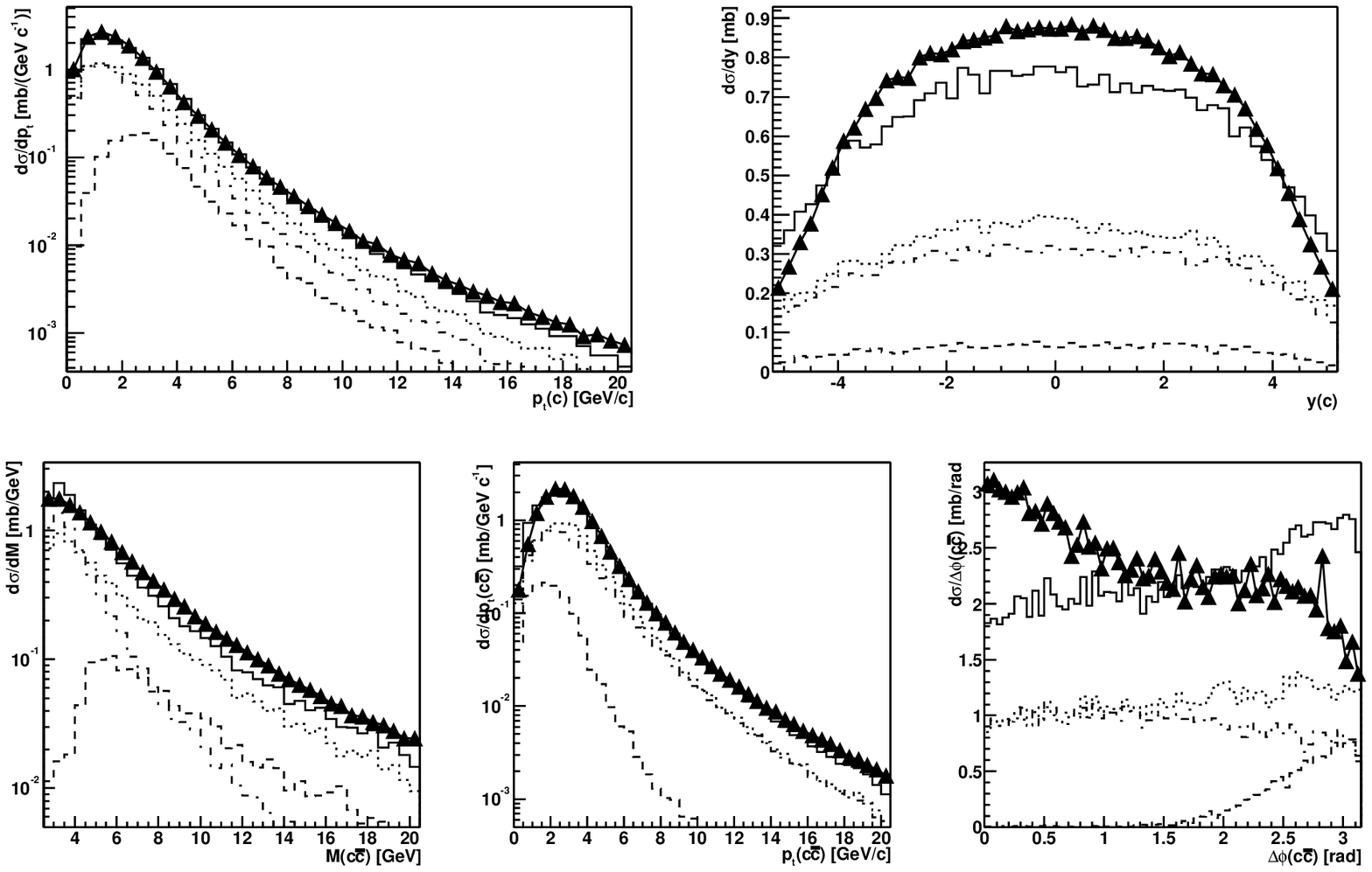}
  \caption{Comparison between charm production in \mbox{Pb--Pb} collisions at 
    $\sqrtsNN=5.5~\tev$ in the NLO calculation
    by Mangano, Nason, Ridolfi and in PYTHIA with parameters tuned as
    described in the text. The triangles show the NLO calculation, the
    solid histogram corresponds to the PYTHIA total production. The
    individual PYTHIA contributions are pair production (dashed),
    flavour excitation (dotted) and gluon splitting (dot-dashed).}
  \label{fig:charmPbPbPyMNR}
    \includegraphics[width=0.9\textwidth]{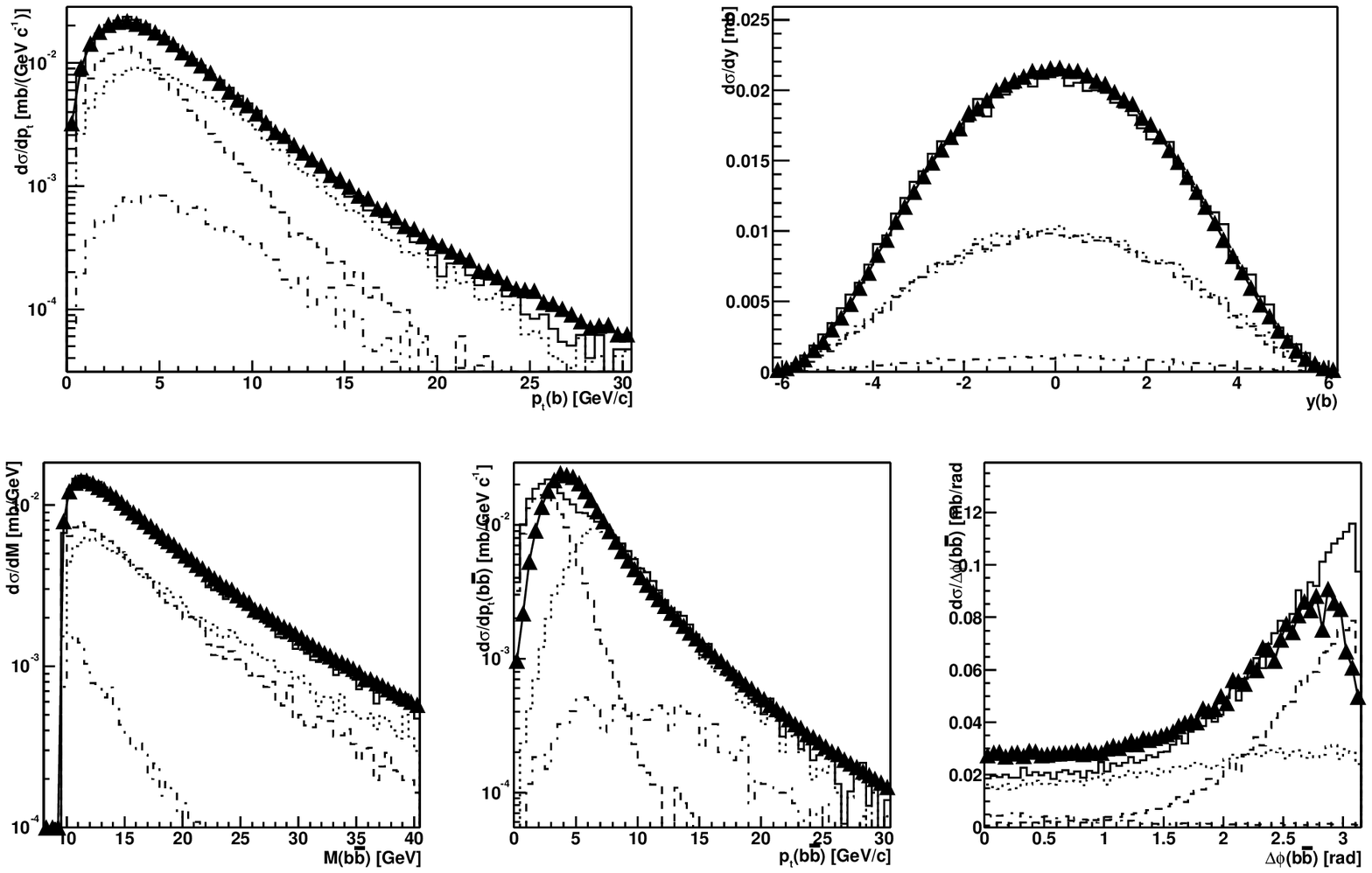}
  \caption{Equivalent of Fig.~\ref{fig:charmPbPbPyMNR} for beauty production.}
  \label{fig:beautyPbPbPyMNR}
  \end{center}
\end{figure}

\begin{figure}
  \begin{center}
    \includegraphics[width=0.9\textwidth,height=8cm]{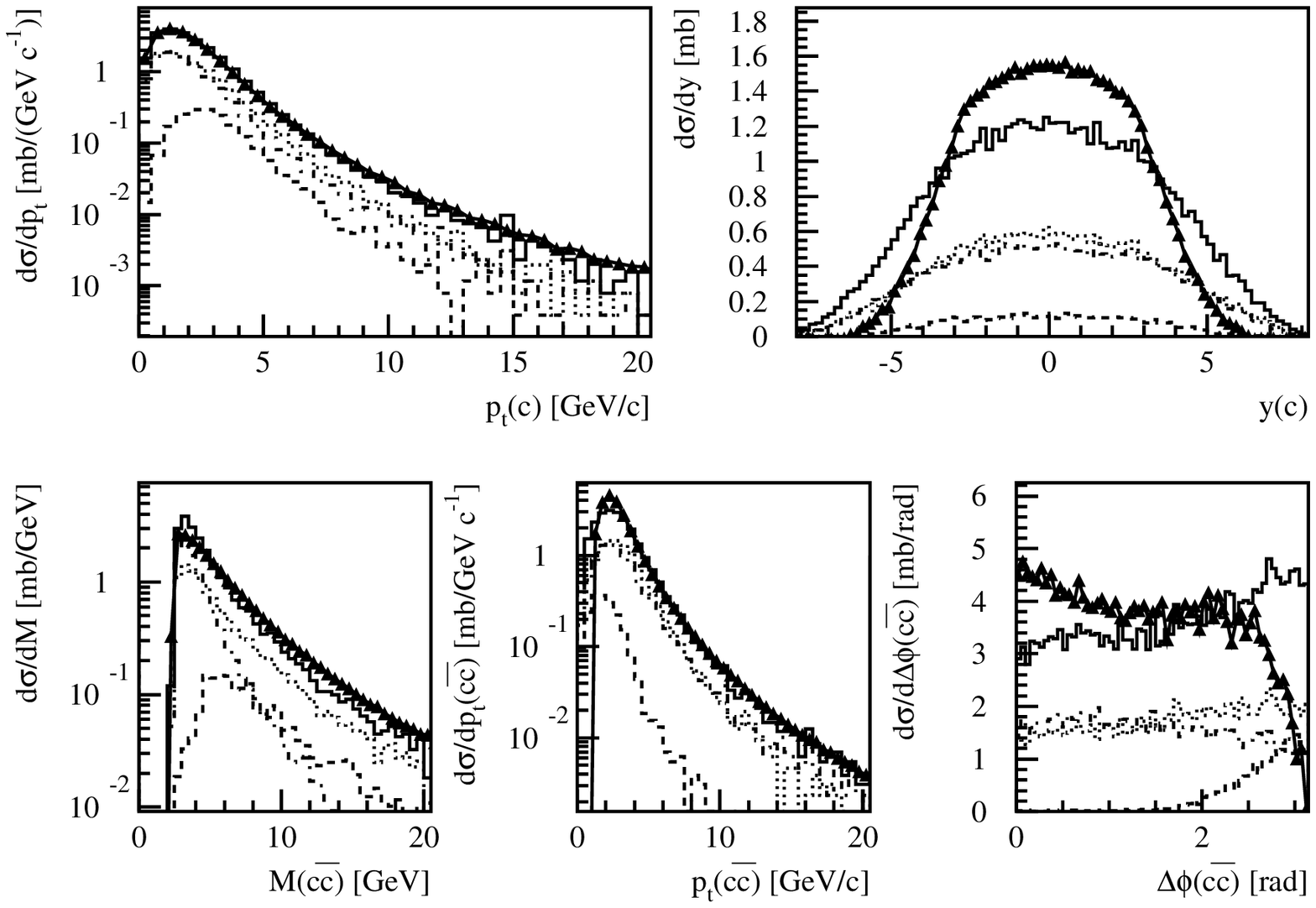}
  \caption{Comparison between charm production in pp collisions at 
    $\sqrt{s}=14~\tev$ in the NLO calculation
    by Mangano, Nason, Ridolfi and in PYTHIA with parameters tuned as
    described in the text. The triangles show the NLO calculation, the
    solid histogram corresponds to the PYTHIA total production. The
    individual PYTHIA contributions are pair production (dashed),
    flavour excitation (dotted) and gluon splitting (dot-dashed).}
  \label{fig:charmPpPyMNR}
    \includegraphics[width=0.9\textwidth,height=8cm]{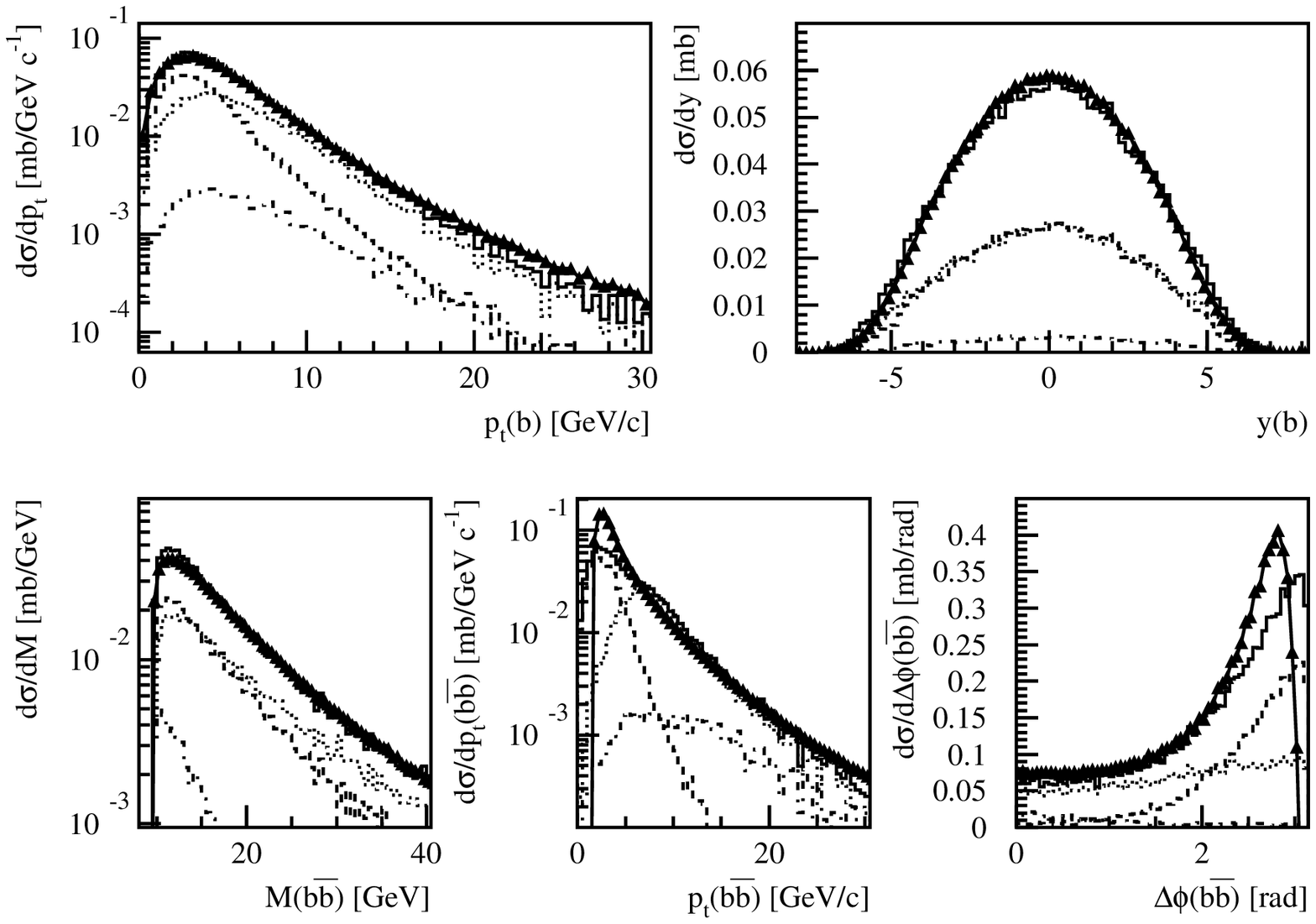}
  \caption{Equivalent of Fig.~\ref{fig:charmPpPyMNR} for beauty production.}
  \label{fig:beautyPpPyMNR}
  \end{center}
\end{figure}

The results of the tuning are shown in Figs.~\ref{fig:charmPbPbPyMNR}
and~\ref{fig:beautyPbPbPyMNR}, where the distributions from PYTHIA and the
NLO calculation are compared.  The PYTHIA results are scaled to give
the same total cross section as the NLO calculation. Despite the
fundamental differences between the two models, the agreement is
relatively good. However, significant discrepancies are present,
especially in the $\Delta\phi$ distribution for \ccbar~pairs.

A similar tuning of the PYTHIA event generator was done also for the 
production of $\ccbar$ and $\bbbar$ pairs in pp collisions at 
$\sqrt{s}=14~\tev$.
The same set of parton distribution functions (CTEQ 4L) was used, 
without the modification for nuclear shadowing. 
Results are shown in 
Figs.~\ref{fig:charmPpPyMNR} and~\ref{fig:beautyPpPyMNR}. The largest 
difference with the results obtained for the \mbox{Pb--Pb} case is a worse 
description of the rapidity distribution of charm quarks. This is due 
to a feature of the parameterizations of the parton distribution functions:
most of them, including CTEQ 4, are valid only
down to $x=10^{-5}$; below this value the behaviour depends on 
the implementation of the specific parameterization but has no physical 
meaning (e.g. for the CTEQ 4 the gluon density $g(x)$ is kept constant at 
$g(10^{-5})$). The rapidity range in which the evolution of the parton 
distribution functions is reliable depends on the c.m.s. energy; for 
charm production at $\sqrt{s}=5.5~\tev$ (14~$\tev$) this range is 
found to be $|y|<4.3$ ($|y|<3.4$), using equation~(\ref{eq:yx1x2}) with 
$x_1>10^{-5}$ and $x_2>10^{-5}$.

The values of the PYTHIA parameters obtained from the tuning are reported 
in Appendix~\ref{App:pythiahvq}.

\mysection{Hadron yields and distributions}
\label{CHAP3:hadr}

For the hadronization of heavy quarks we used the default Lund string
fragmentation model~\cite{norrbin} included in PYTHIA (JETSET package).
The total yield and the rapidity density d$N$/d$y$ in the central region 
for hadrons with open 
charm and beauty in \mbox{Pb--Pb} at 5.5~$\tev$~($5\%~\sigma^{\rm tot}$ 
centrality selection), pp at 14~$\tev$ and \mbox{p--Pb} at 
$8.8~\tev$ are summarized 
in Tables~\ref{tab:hadyieldsPbPb},~\ref{tab:hadyieldspp} 
and~\ref{tab:hadyieldspPb}, respectively.
The rapidity densities are calculated in $-1<y_{\rm lab}<1$, corresponding to 
$-1.47<y_{\rm c.m.s.}<0.53$ for \mbox{p--Pb} 
and $-0.53<y_{\rm c.m.s.}<1.47$ for \mbox{Pb--p}. 
No dependence of the relative hadron abundances on the centre-of-mass 
energy is observed. 

\begin{table}
  \caption{Total yield and average rapidity density for $|y|<1$ for
    hadrons with charm and beauty in \mbox{Pb--Pb} collisions at 
    $\sqrtsNN=5.5~\tev$. The values reported correspond to a centrality
    selection of $5\%~\sigma^{\rm tot}$.}
  \label{tab:hadyieldsPbPb}
  \begin{center}
  \begin{tabular}{ccc|ccc}
\hline
\hline
  Particle & Yield & $\langle$d$N$/d$y\rangle_{|y|<1}$ & Particle & Yield &
    $\langle$d$N$/d$y\rangle_{|y|<1}$ \\
\hline
${\rm D^0}$&   68.9&   6.87&${\rm B^0}$&   1.86&  0.273\\
${\rm \overline{D}^0}$&   71.9&   6.83&${\rm \overline{B}^0}$&   1.79&  0.262\\
${\rm D^+}$&   22.4&   2.12&${\rm B^+}$&   1.82&  0.251\\
${\rm D^-}$&   22.2&   2.00&${\rm B^-}$&   1.83&  0.270\\
${\rm D_s^+}$&   14.1&   1.30&${\rm B_s^0}$&   0.53&  0.077\\
${\rm D_s^-}$&   12.7&   1.22&${\rm \overline{B}_s^0}$&   0.53&  0.082\\
${\rm \Lambda_c^+}$&    9.7&   1.18&${\rm \Lambda_b^0}$&   0.36&  0.050\\
${\rm \overline{\Lambda_c}^-}$&    8.2&   0.85&${\rm \overline{\Lambda_b}^0}$&
    0.31&  0.047\\
\hline
\hline
  \end{tabular}
  \end{center}

  \caption{Total yield and average rapidity density for $|y|<1$ for
    hadrons with charm and beauty in pp collisions at 
    $\sqrt{s}=14~\tev$.}
  \label{tab:hadyieldspp}
  \begin{center}
  \begin{tabular}{ccc|ccc}
\hline
\hline
  Particle & Yield & $\langle$d$N$/d$y\rangle_{|y|<1}$ & Particle & Yield &
    $\langle$d$N$/d$y\rangle_{|y|<1}$ \\
\hline
${\rm D^0}$ &   0.0938 & 0.0098 & ${\rm B^0}$ & 0.00294 & 0.00043\\
${\rm \overline{D}^0}$ & 0.0970 & 0.0098 & ${\rm \overline{B}^0}$ & 0.00283 & 0.00041\\
${\rm D^+}$ & 0.0297 & 0.0029&${\rm B^+}$ & 0.00287 & 0.00040\\
${\rm D^-}$ & 0.0290 & 0.0029&${\rm B^-}$ & 0.00289 & 0.00043\\
${\rm D_s^+}$ & 0.0186 & 0.0018&${\rm B_s^0}$ & 0.00084 & 0.00012\\
${\rm D_s^-}$ & 0.0176 & 0.0020&${\rm \overline{B}_s^0}$ & 0.00084 & 0.00013\\
${\rm \Lambda_c^+}$ & 0.0113 & 0.0013 & ${\rm \Lambda_b^0}$ & 0.00057 & 0.00008\\
${\rm \overline{\Lambda_c}^-}$ &    0.0110 & 0.0013 & ${\rm \overline{\Lambda_b}^0}$ & 0.00049 & 0.00008\\
\hline
\hline
  \end{tabular}
  \end{center}

  \caption{Total yield and average rapidity density for $|y_{\rm lab}|<1$ for
    hadrons with charm and beauty in \mbox{p--Pb} collisions at 
    $\sqrtsNN=8.8~\tev$.}
  \label{tab:hadyieldspPb}
  \begin{center}
  \begin{tabular}{ccc|ccc}
\hline
\hline
  Particle & Yield & $\langle$d$N$/d$y\rangle_{|y_{\rm lab}|<1}$ & 
  Particle & Yield & $\langle$d$N$/d$y\rangle_{|y_{\rm lab}|<1}$ \\
\hline
${\rm D^0+\overline{D}^0}$ &   0.926 & 0.096 & ${\rm B^0+\overline{B}^0}$ & 0.0221 & 0.0030\\
${\rm D^++D^-}$ & 0.293 & 0.030 & ${\rm B^++B^-}$ & 0.0221 & 0.0030\\
${\rm D_s^++D_s^-}$ & 0.176 & 0.018&${\rm B_s^0+\overline{B}_s^0}$ & 0.0064 & 0.0009\\
${\rm \Lambda_c^++\overline{\Lambda_c}^-}$ & 0.118 & 0.012 & ${\rm \Lambda_b^0+\overline{\Lambda_b}^0}$ & 0.0041 & 0.0005\\
\hline
\hline
  \end{tabular}
  \end{center}
\end{table}

It is interesting to notice the large ratio of the neutral-to-charged 
D meson yields: $N({\rm D^0})/N({\rm D^+})\simeq 3.1$. 
In PYTHIA, charm quarks are 
assumed to fragment to D (spin singlets: $J=0$) and D$^*$ (spin triplets: 
$J=1$) mesons according to the number of available spin states; therefore,
 $N({\rm D^0}):N({\rm D^+}):N({\rm D^{*0}}):N({\rm D^{*+}})=1:1:3:3$.
Then, the resonances D$^*$ are decayed to D mesons according to the 
branching ratios. The difference between neutral and charged D mesons arises 
here: due to the slightly larger ($\approx 4~\mev$) mass of the D$^+$, the 
D$^{*+}$ decays preferably to $\Dz$ and the D$^{*0}$ decays exclusively to 
$\Dz$. We have~\cite{pdg}:
\begin{equation}
\begin{array}{rcl}
\displaystyle
\frac{N({\rm D^0})}{N({\rm D^+})}&=&\displaystyle\frac{N({\rm D^0_{primary}})+N({\rm D^{*+}})\times BR({\rm D^{*+}\to D^0})+N({\rm D^{*0}})\times BR({\rm D^{*0}\to D^0})}{N({\rm D^+_{primary}})+N({\rm D^{*+}})\times BR({\rm D^{*+}\to D^+})+N({\rm D^{*0}})\times BR({\rm D^{*0}\to D^+})}\\
&=&\displaystyle\frac{1+3\times 0.68+3\times 1}{1+3\times 0.32+3\times 0}\\
&=& 3.08.
\end{array}
\end{equation}  

We chose to use the relative abundances given by PYTHIA, although, 
experimentally, the fraction $\Dz/{\rm D^+}$ was found to be lower than 3.
The value measured in $e^+e^-$ collisions at LEP by the ALEPH Collaboration 
is $\approx 2.4$~\cite{alephD}. This would reduce by about 6\% the 
expected yield for the $\Dz$ mesons.

\begin{figure}[!t]
  \begin{center}
    \includegraphics[width=0.49\textwidth]{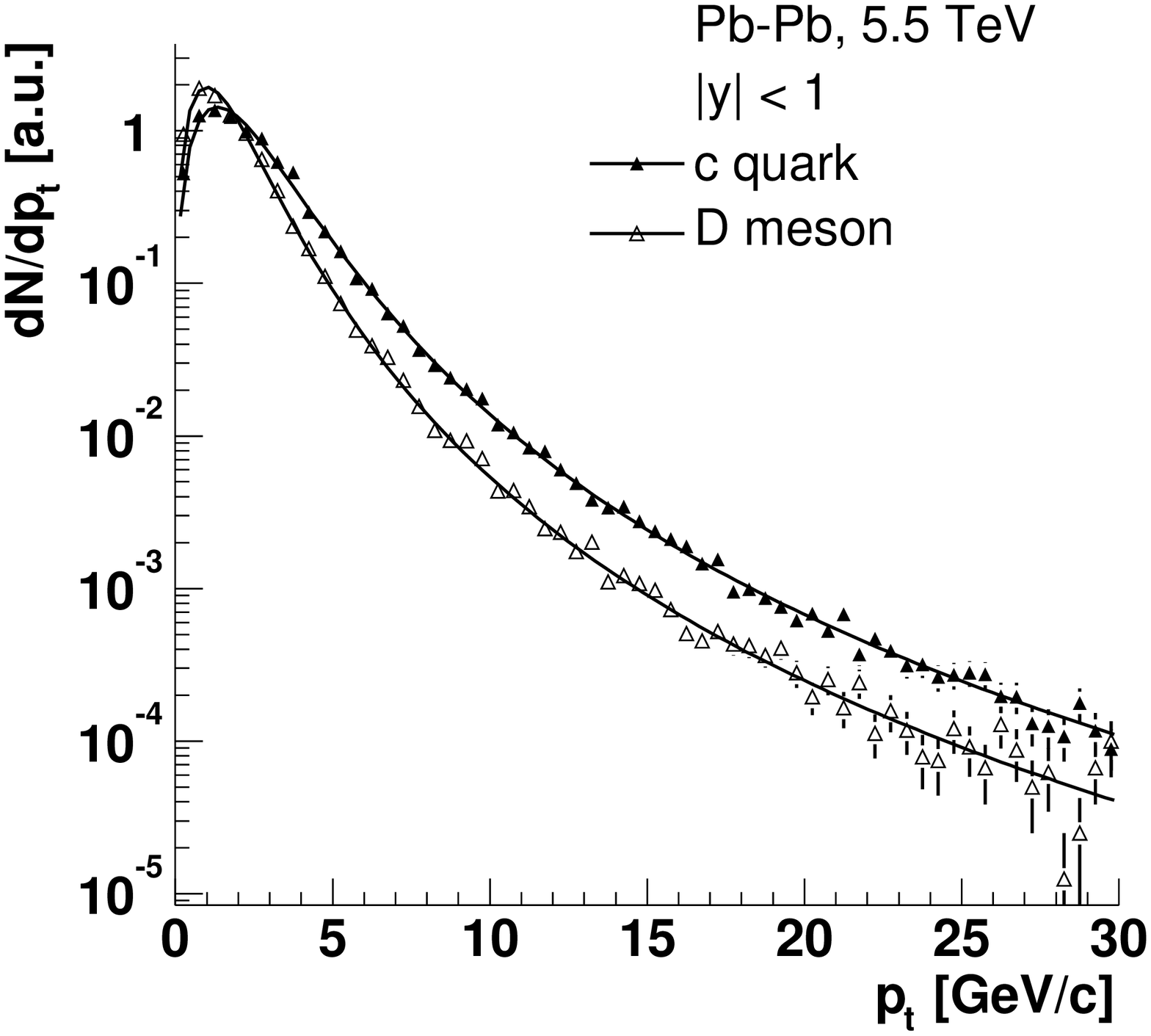}
    \includegraphics[width=0.49\textwidth]{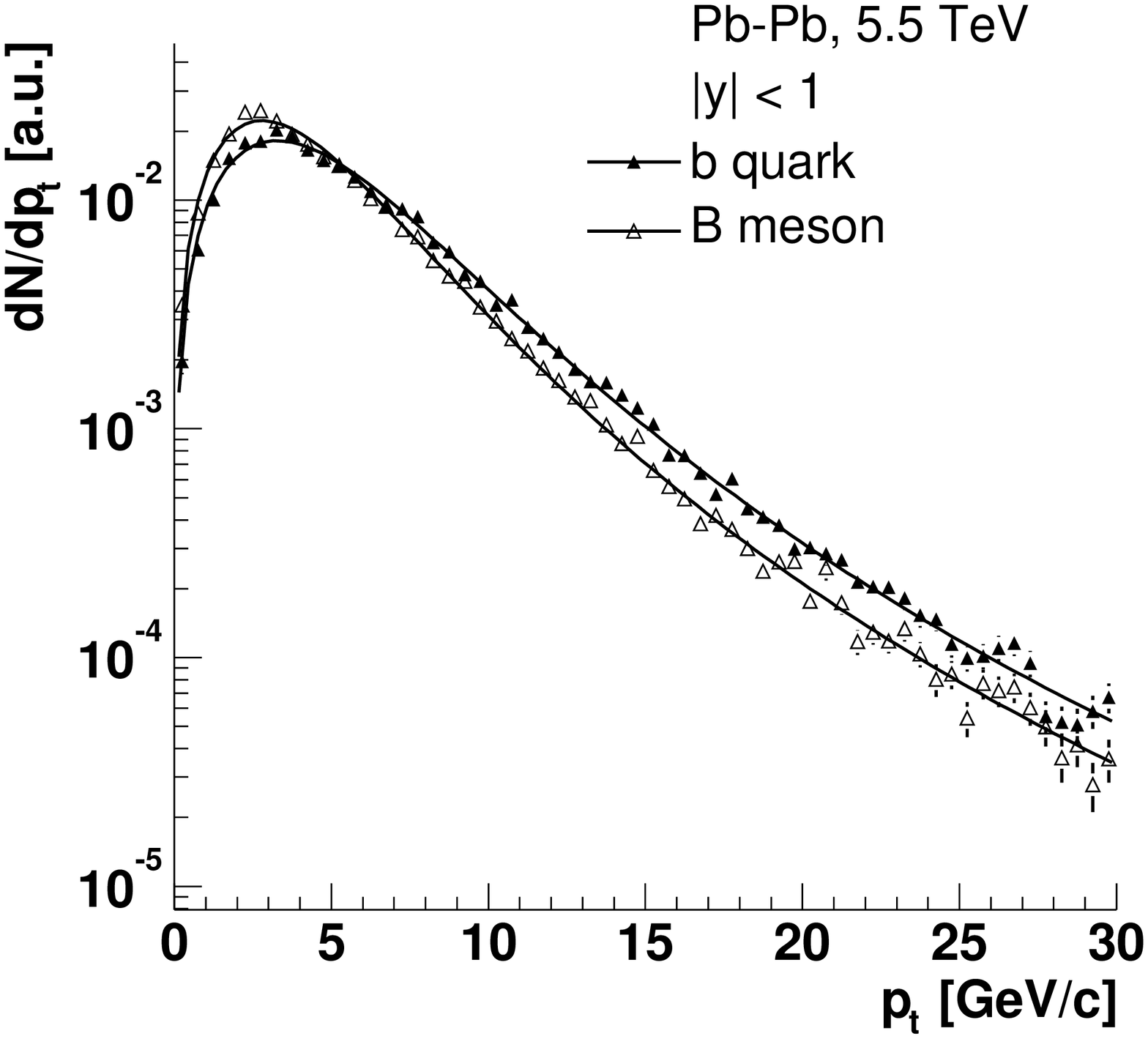}
  \end{center}
  \caption{Transverse momentum distributions at mid-rapidity for heavy quarks 
           and mesons in \mbox{Pb--Pb} at 5.5 TeV. The distributions are 
           normalized to the same integral in order to compare their
           shapes.}
  \label{fig:ptDcBb}
\end{figure}
 
Figure~\ref{fig:ptDcBb} presents the transverse
momentum distributions at mid-rapidity ($|y|<1$) 
for c quarks and for D mesons (left panel) and  
for b quarks and B mesons (right panel), in \mbox{Pb--Pb} at 5.5~$\tev$. 
For $\pt>0$ and $|y|<1$, we have, on average, 
$\pt^{\rm D}\simeq 0.75\,\pt^{\rm c}$
and $\pt^{\rm B}\simeq 0.85\,\pt^{\rm b}$. 
The shape of the transverse momentum distributions for D and B mesons 
was fitted to the following expression:
\begin{equation}
  \label{eq:fitMesons}
  \frac{1}{\pt}\frac{{\rm d}N}{{\rm d}\pt}\propto\left[1+\left(\frac{\pt}{\pt^0}\right)^2\right]^{-n}.
\end{equation}
The $\pt$ distributions were studied also for pp at 
14~TeV and for \pPb~at 8.8~TeV.  
The results of the fits are reported in Table~\ref{tab:fitDBpt}, 
together with the average $\pt$ of D and B mesons in the different 
conditions. The average $\pt$ does not depend strongly on the 
colliding system and on the energy in the centre of mass. On the other 
hand, we remark that $\av{\pt}$ is larger by $\approx 10\%$ 
at mid-rapidity than 
in the forward region ($2.5<y<4$). These two regions correspond 
to the acceptance of the ALICE detector: barrel, $|\eta|<0.9$, 
and forward muon arm, $2.4<\eta<4$ (more details on the structure of the
detector are given in Chapter~\ref{CHAP4}).
 
\begin{figure}[!t]
  \begin{center}
    \includegraphics[width=.8\textwidth]{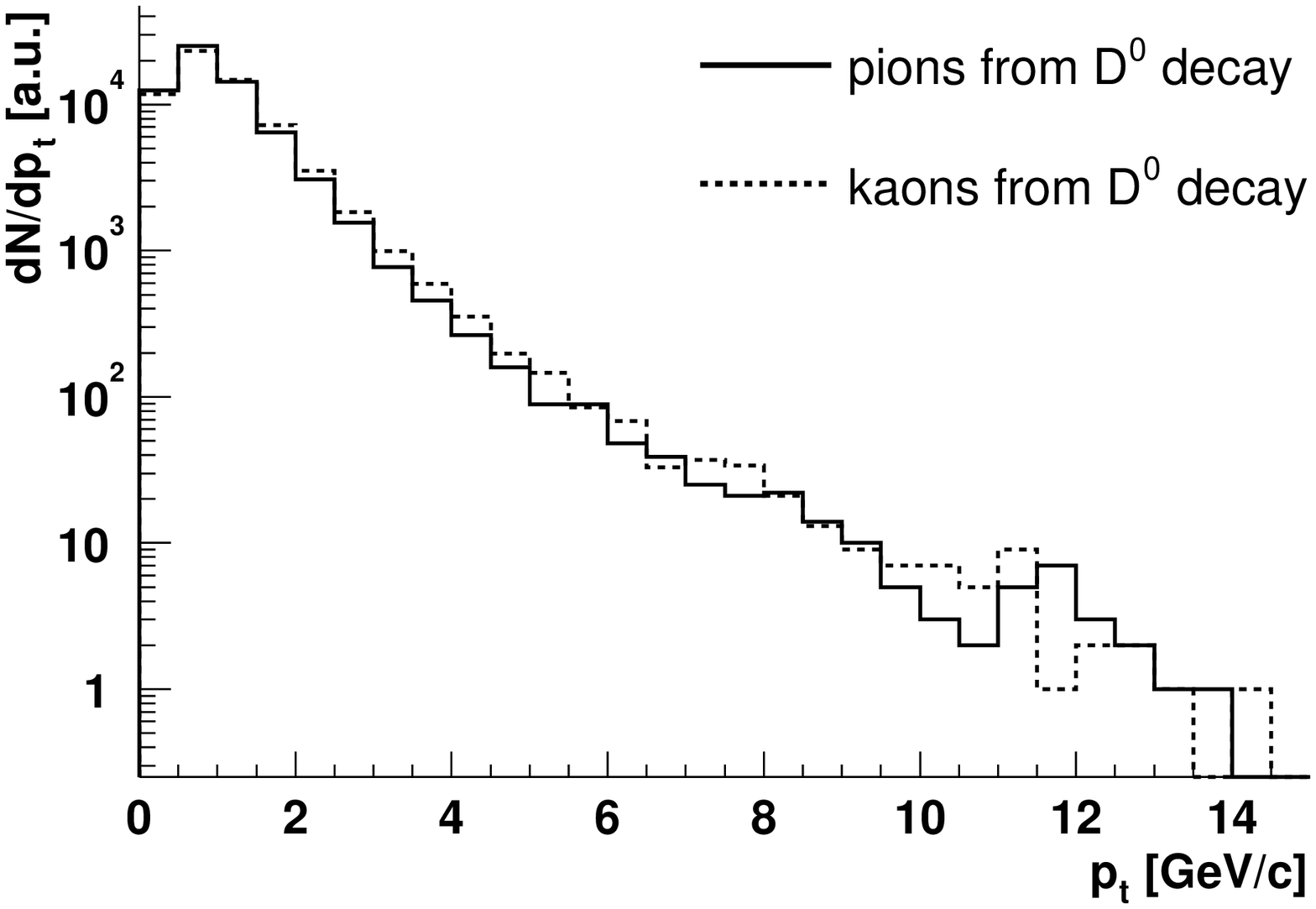}
  \end{center}
  \caption{Transverse momentum distributions for kaons and pions 
           from $\Dz$ meson decays in \mbox{Pb--Pb} at 5.5 TeV.}
  \label{fig:ptKpi}
\end{figure}

Finally, we show in Fig.~\ref{fig:ptKpi} the $\pt$ distributions
of kaons and pions from the decay $\DtoKpi$ (and charge 
conjugate) in \PbPb~at 
$5.5~\tev$: the average $\pt$ is $1.2~\gev/c$ for the kaons and 
$1.1~\gev/c$ for the pions. For kaons and pions in $|\eta|<0.9$ (ALICE
barrel acceptance) the average momenta are $\av{p_{\rm K}}=1.5~\gev/c$
and $\av{p_\pi}=1.3~\gev/c$; thus, they are with good approximation
relativistic, since $\av{E_{\rm K}}/\av{p_{\rm K}}\simeq 1.05$ and 
$\av{E_\pi}/\av{p_\pi}\simeq 1.01$.

\begin{table}[!t]
\footnotesize
  \caption{Parameters derived from the fit of the $\pt$ distributions of 
           D and B mesons to the expression~(\ref{eq:fitMesons}) and average 
           value of $\pt$ for these particles.}
  \label{tab:fitDBpt}
  \begin{center}
  \begin{tabular}{c|cc|c|c|c}
\hline
\hline
  Particle & System & $\sqrtsNN$ [TeV] & $\pt^0~[\gev/c]$ & $n$ & $\av{\pt} [\gev/c]$\\
\hline
\hline
                    & pp     & 14  & 2.04 & 2.65 & 1.85 \\
                  D & p--Pb  & 8.8 & 2.09 & 2.72 & 1.83 \\
($|y_{\rm lab}|<1$) & Pb--Pb & 5.5 & 2.12 & 2.78 & 1.81 \\
\hline
                      & pp     & 14  & 2.18 & 3.04 & 1.67 \\
                    D & p--Pb  & 8.8 & 2.22 & 3.11 & 1.66 \\
($2.5<y_{\rm lab}<4$) & Pb--Pb & 5.5 & 2.25 & 3.17 & 1.64 \\
\hline
\hline
                    & pp     & 14  & 6.04 & 2.88 & 4.90 \\
                  B & p--Pb  & 8.8 & 6.08 & 2.90 & 4.89 \\
($|y_{\rm lab}|<1$) & Pb--Pb & 5.5 & 6.14 & 2.93 & 4.89 \\
\hline
                      & pp     & 14  & 6.45 & 3.54 & 4.24 \\
                    B & p--Pb  & 8.8 & 6.49 & 3.56 & 4.24 \\
($2.5<y_{\rm lab}<4$) & Pb--Pb & 5.5 & 6.53 & 3.59 & 4.24 \\
\hline
\hline
  \end{tabular}
  \end{center}
\end{table}

\clearpage
\pagestyle{plain}

\setcounter{chapter}{3}
\mychapter{The ALICE experiment at the LHC}
\label{CHAP4}

\pagestyle{myheadings}

This chapter is devoted to the description of ALICE, 
the dedicated heavy ion experiment at the Large Hadron Collider. 
We start by presenting the ALICE layout and its 
different sub-systems, with particular emphasis on the detectors of 
the barrel, in the central rapidity region, which are employed for 
the measurement of hadronic observables (Section~\ref{CHAP4:detector}).
In Section~\ref{CHAP4:reco} we describe event simulation and 
reconstruction in ALICE. After a brief overview of the 
employed event generators, of the
detector response simulation and of the track 
reconstruction strategy, we concentrate on the 
tracking performance, in terms of efficiency and momentum resolution, 
in different multiplicity environments. 
Finally, we discuss some 
parameters of the LHC heavy ion and proton beams which are relevant for 
the study presented in this thesis (Section~\ref{CHAP4:lhc}).
 
\mysection{The ALICE detector}
\label{CHAP4:detector}

ALICE is conceived as a general-purpose detector, in which most of the 
hadrons, leptons
and photons produced in the interaction can be measured and identified.

The requirement of the combined capability to track and identify particles
of very low up to fairly high $\pt$, and to reconstruct the decays of
hyperons and D and B mesons in an environment with 
charged particle multiplicities
up to 8000/unit rapidity at mid-rapidity, has led to a unique design, with a
very different optimization with respect to the pp-dedicated experiments 
at the LHC. 



\subsection{Detector layout}
\label{CHAP4:detector_layout}

\begin{figure}[!ht]
  \includegraphics[clip,width=\textwidth]{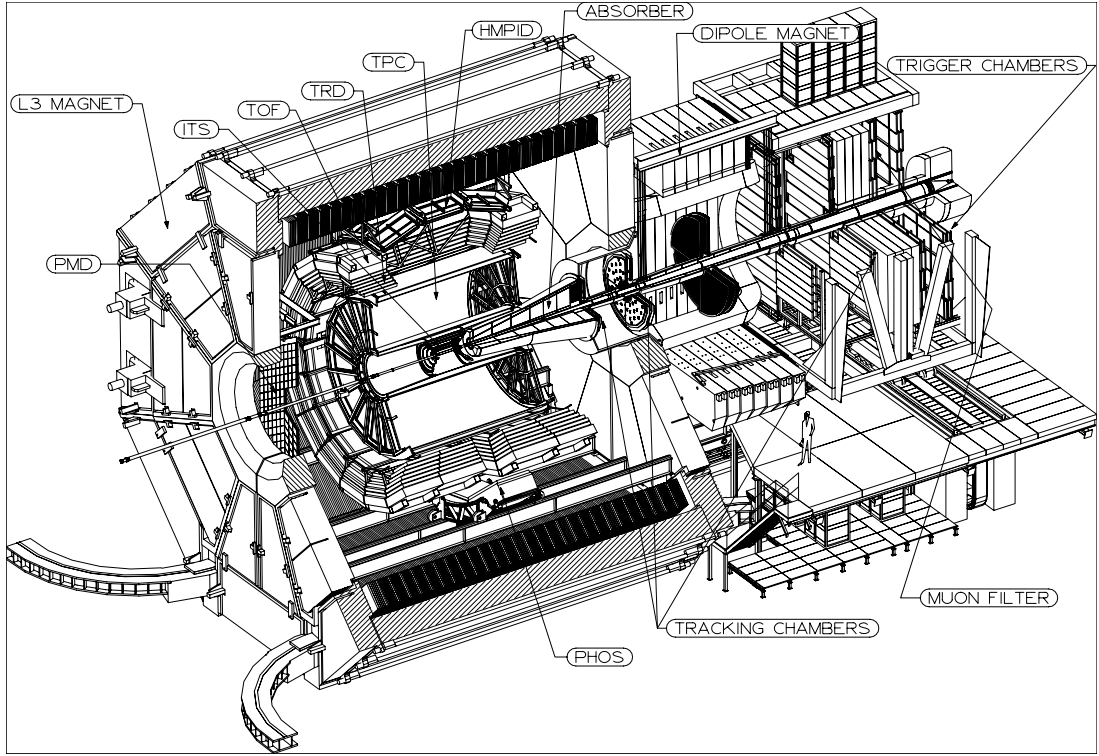}
  \caption{Layout of the ALICE detector.}
  \label{fig:Alice}
\end{figure}

The general ALICE layout is shown in Fig.~\ref{fig:Alice}.

It consists of a central detector ($|\eta|<0.9$) covering the full azimuth,
where hadrons, electrons and photons are measured, and a forward
muon arm ($2.4<\eta<4$)~\cite{tpalice,tpmu,tptrd}. We define here the ALICE
global reference frame: it has $z$ axis parallel to the beam direction 
and pointing towards the muon arm, $x$ and $y$ axes in the plane transverse
to the beam direction. 

The central detector is embedded in a large solenoidal magnet with a 
weak field of $<0.5$~T, parallel to $z$, and it consists 
of the Inner Tracking System (ITS) with six layers of high-resolution
silicon detectors, the cylindrical Time Projection Chamber (TPC), 
a Transition Radiation Detector (TRD) for electron identification,
a barrel Time of Flight (TOF), a small-area ring imaging Cherenkov detector at 
large distance for the identification of high-momentum particles 
(High Momentum Particle Identifier - HMPID), and a single-arm electromagnetic
calorimeter of high-density crystals (Photon Spectrometer - PHOS). 

The design of the tracking system was primarily driven by the
requirement for safe and robust track finding.
It uses mostly three-dimensional hit information and
dense tracking with many points (TPC).
The detection of hyperons, and even more of D and B mesons,
requires in addition a high-resolution vertex detector close to the beam 
pipe (ITS).

The field strength is a compromise between
momentum resolution and low momentum acceptance.
The momentum cut-off should be as low as possible
($\simeq 100~\mev/c$), in order to detect the decay products of
low-$\pt$ hyperons.
At high $\pt$ the magnetic field determines the momentum
resolution, which is essential for the study of jet quenching
and  high-$\pt$ leptons.
The ideal choice for hadronic physics, maximizing reconstruction
efficiency, would be around $0.2$ T, while for the high-$\pt$
observables the maximum field the magnet can produce, $0.5$ T, would be
the best choice.
Since the high-$\pt$ observables are
limited by statistics, ALICE will run mostly with the higher field option.

The beam pipe has the smallest possible thickness 
(in terms of the radiation length, $X_0$) to minimize the 
multiple scattering undergone by the particles produced in the collision. 
It is built in beryllium (usually chosen for its 
low atomic number, i.e. low radiation length) and it has an outer
radius of 3~cm and a thickness of 0.8~mm, corresponding to 0.3\% of 
$X_0$. 

The muon spectrometer is designed to measure the production of the 
complete spectrum of heavy quark resonances, namely
J/$\psi$ and $\psi'$, $\Upsilon$, $\Upsilon'$ and $\Upsilon''$.
It is constructed of an absorber very close to the vertex followed by a 
spectrometer with a dipole magnet and, finally, an iron wall to 
select the muons.

The set-up is completed by a forward
photon counting detector (Photon Multiplicity Detector - PMD) and
a multiplicity detector (Forward Multiplicity Detector - FMD) 
covering the forward rapidity region, that, in conjunction with the 
ITS allows the measurement of the charged multiplicity in 
the range $-3.4<\eta<5.1$. 

A system of scintillators (V0 detector) and quartz counters (T0 detector) 
provide fast trigger signals. 

The collision centrality is determined by measuring the energy (and, thus, 
the number) of spectator nucleons, that lay outside the transverse 
superposition region of the two colliding nuclei and continue to propagate 
along the beam direction. Owing to their different Z/A values, it is possible 
to separate in space the neutron and proton spectators and the beam particles
(Z/A~$\simeq 0.4$ for Pb beams) by means of the first LHC dipole.
Therefore, the neutron and proton spectators are detected in two distinct
calorimeters (Zero Degrees Calorimeters - ZDC), made respectively of 
tantalum and brass with embedded quartz 
fibers, located on both sides of the interaction region $\approx 90~\m$ 
downstream in the machine tunnel.

In the following, we describe the sub-systems of ALICE which are employed 
for the measurement of open charm particles in hadronic decay channels:
the TPC and the ITS, that allow tracking and reconstruction of the 
interaction vertex and of secondary vertices; the TOF detector, that 
provides particle identification for hadrons over the full geometrical 
acceptance of the central barrel.

\subsection{Inner Tracking System (ITS)}
\label{CHAP4:its}

The task of the inner tracker is to provide: 
\begin{itemize}
\item primary and secondary vertex reconstruction with the high 
resolution that is required for the 
detection of hyperons and particles with open charm and open beauty;
\item tracking and identification of low-$\pt$ particles which are strongly 
bent by the magnetic field and do not reach the TPC;
\item improved momentum resolution for the higher-$\pt$ particles
which also traverse the TPC.
\end{itemize}

\begin{figure}[!t]
  \centering
  \includegraphics[width=\textwidth]{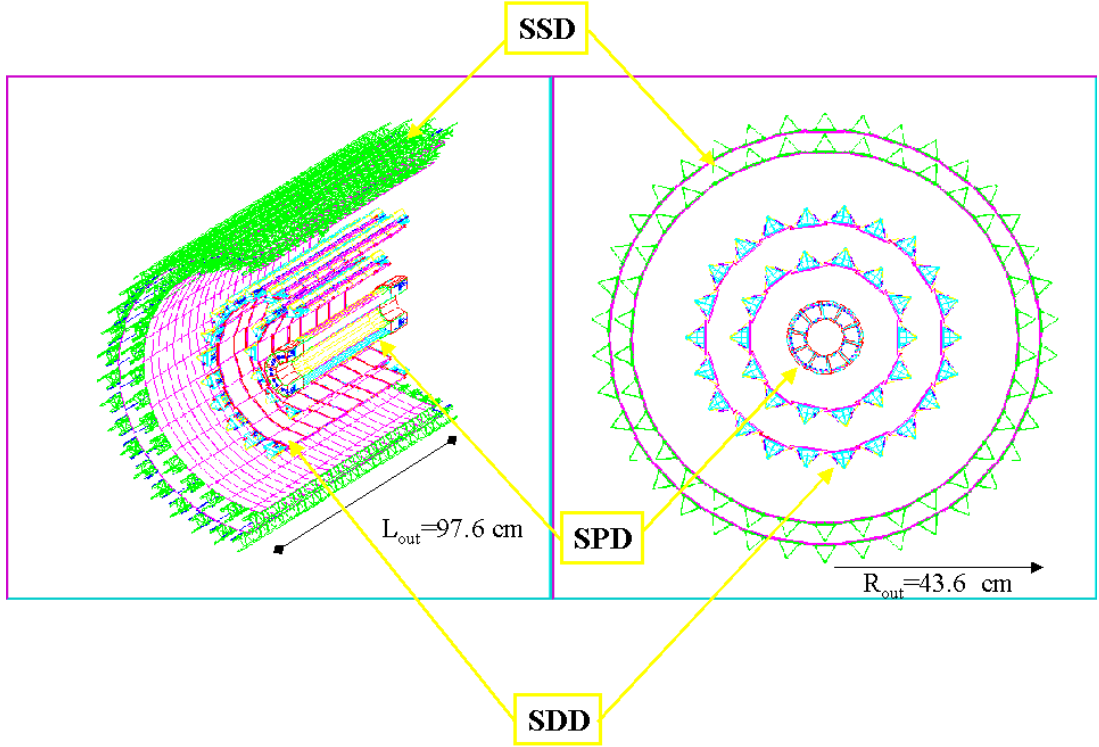}
  \caption{Layout of the ITS.}
  \label{fig:its}
\end{figure}

These goals are achieved with a silicon detector structured in 
six cylindrical layers, from the inside to the outside: 
two layers of Silicon Pixel Detectors (SPD), located at $r=4$ and $7~\cm$;
two layers of Silicon Drift Detectors (SDD), $r=14$ and $24~\cm$;
two layers of Silicon Strip Detectors (SSD), $r=39$ and $44~\cm$.
A general view of the ITS is shown in Fig.~\ref{fig:its} and the main 
parameters of each of the three detector types are reported in 
Table~\ref{tab:its}.

\begin{table}[!b]
\begin{center}
\caption{Parameters of the various detector types of the ITS. 
  The occupancy is calculated 
  for the maximum expected particle density in \PbPb~collisions at the LHC 
  ($\dNdy=8000$).}
\label{tab:its}
\begin{tabular}{llccc} 
\hline
\hline
Parameter && Silicon Pixel & Silicon Drift & Silicon Strip \\ 
\hline
Radius (inner layer) & [cm] & 4 & 14 & 39 \\ 
Radius (outer layer) & [cm] & 7 & 24 & 44 \\ 
Cell size ($r\phi \times z$) & [$\mum^2$] & $50 \times 425$  & $150 \times 
300$ & $95 \times 40000$ \\
Spatial precision ($r\phi\times z$) & [$\mum^2$] & $12\times 120$ & 
$38\times 28$ & $20 \times 830$ \\
Readout channels & [k] & 9835 & 133 & 2719 \\
Av. occup. (inner layer) & [\%] & 2.1 & 2.5 & 4 \\
Av. occup. (outer layer) & [\%] & 0.6 & 1.0 & 3.3 \\ 
Thickness per layer & [\% of $X_0$] & 1.24 & 0.95 & 0.90 \\ 
\hline
\hline
\end{tabular}
\end{center}
\end{table}

The high particle density expected in heavy ion collisions 
and the requirement for optimal reconstruction of secondary vertices, 
have dictated the choice of silicon detectors with high granularity and
true two-dimensional readout for the four innermost planes.
  
The pixel detectors, in layers 1 and 2, with a cell (pixel) size of 
$50(r\phi)\times 425(z)~\mum^2$, allow excellent position resolution in an 
environment where the track density may exceed 50~tracks/$\cm^2$.
For the two intermediate layers, 3 and 4, silicon drift detectors have been 
selected, since they couple a very good multi-track capability
to the information on the specific energy loss
(with the two-dimensional analog readout).

At larger radii, layers 5 and 6, 
the requirements in terms of granularity are less 
stringent, therefore double-sided silicon micro-strip detectors are used. 
Double-sided micro-strips have been selected rather than single-sided ones
because they offer the possibility to correlate the pulse height 
read out from the two sides, thus helping 
to resolve ambiguities inherent in the use of detectors with 
one-dimensional readout. This aspect is very important for the connection 
of tracks from the TPC to the ITS.

With the drift and the strip detectors, the four outer layers have analog
readout. This allows to apply a truncated-mean method (requiring 
at least four measurements) for the estimate of the d$E$/d$x$ and gives 
the ITS a stand-alone capability as low-momentum particle spectrometer,
in the 1/$\beta^{2}$ region of the Bethe-Bloch curve~\cite{pdg}.

The pseudorapidity coverage of the ITS is $|\eta|<0.9$ 
for collisions with vertex located within the length of the 
interaction diamond, i.e. $-5.3<z<5.3~\cm$ along the beam  
direction.           
The first layer of pixel detectors has a more extended coverage 
($|\eta|<1.98$) to provide, together with the forward multiplicity 
detectors, a continuous coverage in rapidity for the measurement of 
charged multiplicity.

The track momentum and position resolutions for particles with 
small transverse momenta are dominated by multiple scattering 
effects.
Therefore, the minimization of the material thickness is an absolute 
priority in the ITS, which is the first detector crossed by the particles 
produced in the collision.
In the two innermost layers, both the pixel sensor and 
the electronics chip are $200~\mum$ thick, for a total silicon budget
of $400~\mum$ per layer. Including also the carbon-fiber supports 
and the cooling system, the average material per layer traversed by 
a straight track perpendicular to the beam line corresponds to 
$1.2\%$  of $X_0$. 
Also the drift and strip layers have a similar material budget, so that 
the total thickness of the ITS corresponds to $\approx~6\%$ of $X_0$.

The improvement of the momentum measurement, obtained when the tracks 
reconstructed in the TPC are prolonged in the ITS,  
is discussed in Section~\ref{CHAP4:ptres}. The performance of the ITS
for the issues related to the reconstruction of secondary vertices 
(namely, track position resolution and primary vertex reconstruction)
was studied in detail in the scope of this thesis and it is 
described in a dedicated chapter (Chapter~\ref{CHAP5}).

\subsection{Time Projection Chamber (TPC)}
\label{CHAP4:tpc}

The TPC is the main tracking detector in ALICE: 
it provides track finding, 
momentum measurement and particle identification via d$E$/d$x$. 

\begin{figure}[!t]
  \centering
  \includegraphics[width=0.58\textwidth]{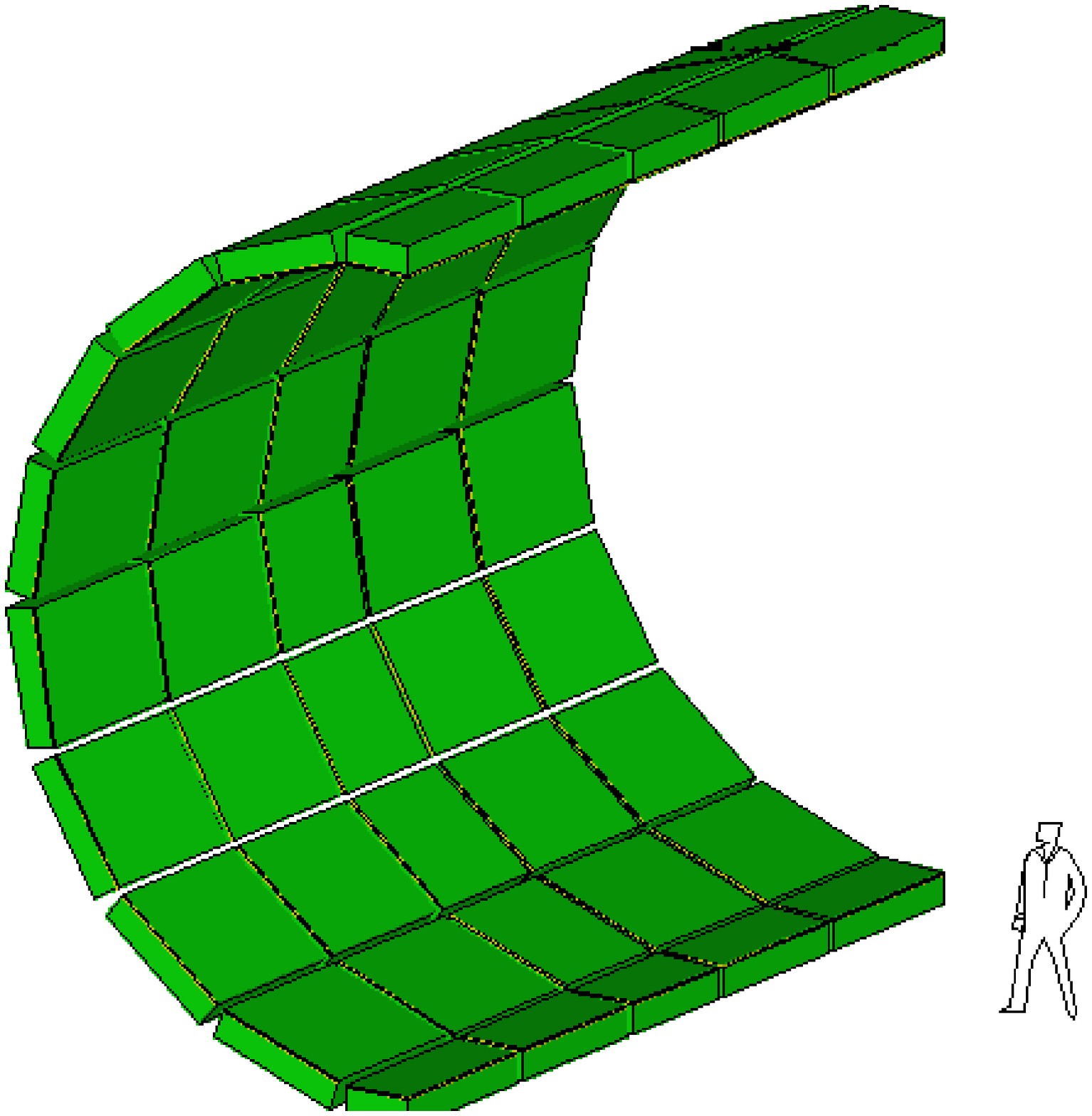}
  \includegraphics[width=0.40\textwidth]{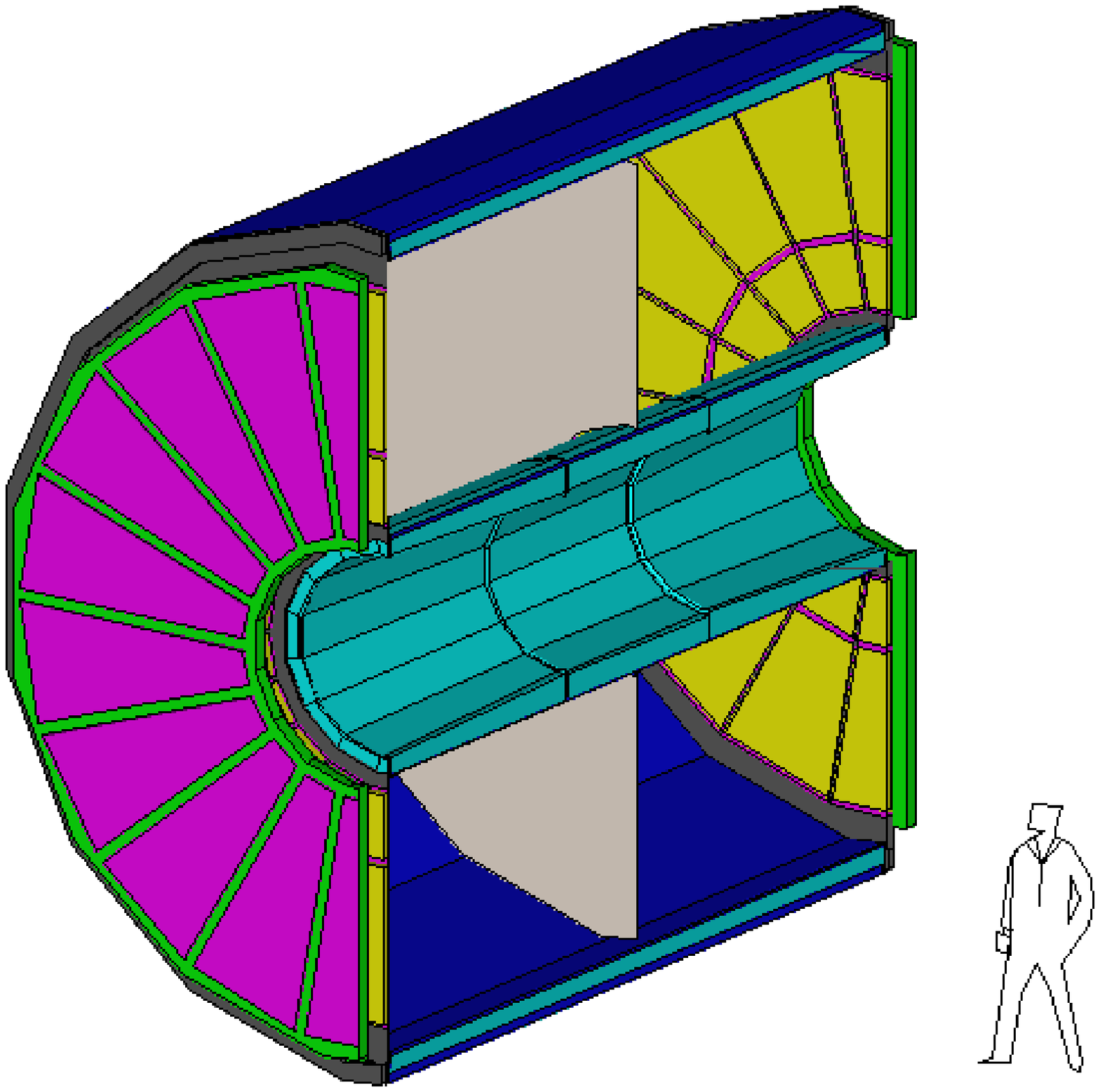}
  \caption{A view of the ALICE Time of Flight (left) and Time Projection 
           Chamber (right) how they are described in the detector 
           simulation framework. For graphical clearness only half 
           of the azimuthal coverage of the detectors is shown.}
  \label{fig:tpc}
\end{figure}

A view of the detector is shown in Fig.~\ref{fig:tpc} (right).
The TPC has an inner radius of 80~cm, given by the maximum
acceptable hit density (0.1~cm$^{-2}$), and an outer radius of 250~cm,
given by the length required for a d$E$/d$x$ resolution better than $10\%$, 
necessary for particle identification. The total active length  
of 500~cm allows the acceptance in the pseudorapidity range $|\eta|<0.9$.

The gas mixture Ne/CO$_2$ (90\%/10\%) is optimized for drift velocity, 
low electron diffusion and low radiation length. 

The TPC readout chambers are multi-wire proportional chambers with 
cath\-ode-pad readout. The readout planes at the two ends of the large 
drift volume ($88~\m^3$) are azimuthally segmented in 18 sectors, 
each covering 
an angle of $20^\circ$. The non-active region between two adjacent sectors 
is 2.7~cm wide, implying an azimuthal acceptance of $\approx 90\%$ for 
straight tracks originating from the interaction point. 
The radial thickness of
the detector is of $3.5\%$ of $X_0$ at central rapidity and grows to 
$\approx 40\%$ towards the acceptance edges. 

Track reconstruction strategy and performance in the TPC in the 
high-multi\-plicity environment of heavy ion collisions and in pp 
collisions will be described in detail in Section~\ref{CHAP4:tracking}.

\subsection{Particle identification system}

One of the distinctive features of ALICE is the particle identification
capability, which is realized using a number of different techniques. 

Charged hadron identification is provided over the full barrel acceptance 
($|\eta|<0.9$) by the combination of (a) d$E$/d$x$ measurement in 
the four outer layers of the ITS and in the TPC, 
for momenta up to $\simeq 0.5~\gev/c$,
with (b) a barrel Time of Flight at $r=370~\cm$, in the range
$0.5<p<2.5~\gev/c$.
Electron are separated from pions for $\pt>1~\gev/c$ by means of a 
dedicated Transition Radiation Detector and by exploiting the 
relativistic rise of the specific energy loss measured in the TPC. 
A smaller-area ring imaging Cherenkov detector (HMPID), 
covering about $15\%$ of the acceptance of the ALICE central detectors, 
allows the separation of hadrons up to higher
momenta ($\pi$/K up to $3~\gev/c$ and K/p up to $5~\gev/c$).

Photons and neutral pions are identified in the small-acceptance
electromagnetic calorimeter PHOS.

Here we describe in detail only the TOF detector. For the other 
detectors see Refs.~\cite{tpalice,tptrd}.

%
%
%

\subsubsection{Time of Flight detector (TOF)}
\label{CHAP4:tof}

The TOF barrel is positioned outside the TRD and has an internal radius
of $370~\cm$ and an external radius of $399~\cm$. Its task is to 
provide hadron separation in the momentum range from $0.5~\gev/c$, where 
the d$E$/d$x$ technique is no longer effective, to 
about $2.5~\gev/c$. PID in this momentum range allows the study of the 
kinematical distributions of the different particle types on an 
event-by-event basis in heavy ion collisions. Moreover, given the large mass 
of the charm quark, the decay products of D mesons have typical 
momenta of the order of $1$-$2~\gev/c$; therefore, the Time of Flight, 
with K/$\pi$ separation up to $2.5~\gev/c$, is very effective for the 
reconstruction of exclusive decays of D mesons in hadronic channels.  
 
The time-of-flight is measured using the technology of the 
Multi-gap Resistive Plate Chambers (MRPC). 
The RPC is a gaseous detector with resistive electrodes, which 
quench the streamers so that they do not initiate a spark breakdown. 
 
The TOF MRPC design consists of a double stack with $2\times 5$ gaps. 
The basic unit is a MRPC pad of size $3.5\times 2.5~\cm^2$;
the pads are organized in large modules and the full barrel counts 
18~(in~$r\phi$)~$\times$~5~(along~$z$) modules, for a total active 
area of $\approx 140~\m^2$. A view of the detector is shown in 
Fig.~\ref{fig:tpc} (left).

The MRPC resolution has been measured to be in the $50$-$60$~ps
range, with efficiency above $99\%$. Including the other sources
of timing errors, the overall resolution is estimated to be 
$\approx 120$ ps. 

In Chapter~\ref{CHAP6} the PID performance of the detector will 
be presented, along with the 
TOF PID strategy optimized for the detection of charm mesons in \PbPb~and 
in pp collisions.

\mysection{Event simulation and reconstruction}
\label{CHAP4:reco}

\begin{figure}[!b]
  \centering
  \includegraphics[width=0.7\textwidth]{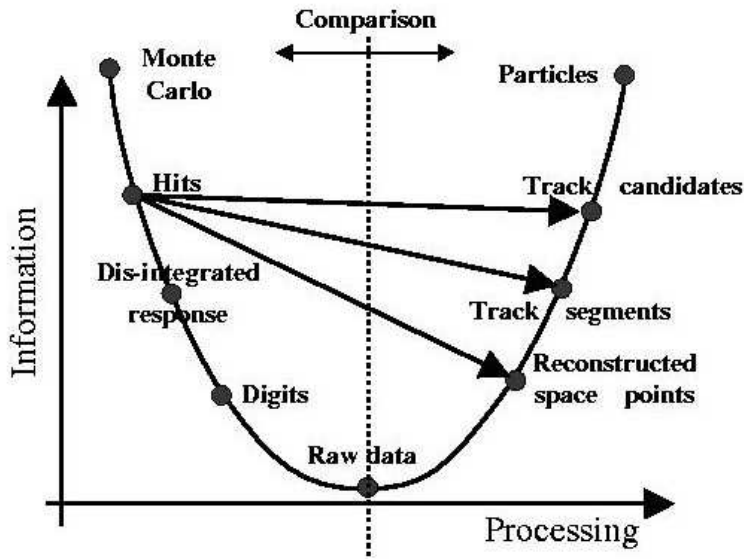}
  \caption{Schematic representation of the data processing chain.}
  \label{fig:aliroot}
\end{figure}

The ALICE off-line framework, AliRoot~\cite{alisoft}, is described in 
detail in Ref.~\cite{pprCh4}. This framework, based on the Object 
Oriented / C++ environment of ROOT~\cite{root}, allows to reconstruct and 
analyze physics data coming from simulations and real interactions.
The role of the framework can be graphically represented as shown in 
Fig.~\ref{fig:aliroot}. Events are generated via Monte Carlo simulation 
programs, generators and detector simulation, and are then transformed 
into the format produced by the detector ({\sl raw data}). Here 
we have a minimum of the physics information. At this point, the 
reconstruction and analysis chain is used to evaluate the detector 
and the physics performance, and most of the initial information on 
the generated event can be retrieved (e.g. particle ID and kinematics, 
event topology). In the next paragraphs we will 
follow from the left to the right the parabola in Fig.~\ref{fig:aliroot} 
and detail the aspects which are relevant to the studies reported 
in this thesis.

\subsection{Event generators}
\label{CHAP4:generators}

We briefly describe here some relevant features of the Monte Carlo event 
generators that were used
for the simulation of pp minimum-bias collisions (PYTHIA) and of \PbPb~central 
collisions (HIJING) at LHC energies.

\subsubsection{PYTHIA}
\label{CHAP4:pythia}

The PYTHIA~\cite{pythia} event generator, was introduced in 
Section~\ref{CHAP3:generators}, where we have reported how its parameters were 
tuned in order to reproduce the $\pt$ distributions for heavy quarks 
as predicted by NLO pQCD.

PYTHIA was also employed for the production of a large sample of pp 
non-diffractive interactions\footnote{The single-diffractive 
processes, $A+B\to A+X$ and $A+B\to X+B$, and the 
double-diffractive processes, $A+B\to A+B+X$, are excluded.}
at $\sqrt{s}=14~\tev$ that served as 
background (or, more appropriately, {\sl underlying}) events for the 
study on the detection of D mesons in pp. To this purpose, the 
version 6.150 of the 
program was used with the CTEQ 4L set of PDF and with the parameters 
tuned in order to reproduce the multiplicity of all available collider 
data~\cite{pythiamb}. As detailed in Ref.~\cite{pythiamb}, the main 
parameter tuned is the low-$\pt$ cut-off, 
$\pt^{\rm min}$, introduced in the model to regularize the dominant 
$2\to 2$ QCD cross sections, which diverge as $\pt\to 0$ and drop 
rapidly at high $\pt$. In order to reproduce the data, a monotonic 
increase of $\pt^{\rm min}$ with $\sqrt{s}$ has to be introduced.

The predicted average charged particle rapidity density in non-diffractive
inelastic pp collisions
at $\sqrt{s}=14~\tev$ is
$\dNdy\approx 6$. In PYTHIA, at this energy the ratio
of the non-diffractive inelastic cross section to the total inelastic 
cross section is $\simeq 0.7$.

For events with 
charm production we can expect a multiplicity higher than the average, 
due to the fact that heavy quarks, having a large mass, yield more particles,
in their fragmentation, than light quarks and gluons.
We shall discuss in detail this point in Chapter~\ref{CHAP6}, where 
we will compare the charged particle multiplicity and the mean $\pt$ 
predicted by PYTHIA for events with and without charm production.  


\subsubsection{HIJING}
\label{CHAP4:hijing}

HIJING~\cite{hijing} (Heavy Ion Jet INteraction Generator) 
combines a QCD-inspired
model of jet production with the Lund string
model~\cite{norrbin} for jet fragmentation.
Binary scaling with Glauber geometry 
is used to extrapolate to proton--nucleus and 
nucleus--nucleus collisions.

Nuclear shadowing and parton energy loss are included in the HIJING model 
and they can be selected by the user.
 
The charged particle rapidity density at mid-rapidity and the total number 
of charged particles in the ALICE barrel acceptance ($|\eta|<0.9$) 
given by HIJING for central (impact parameter $b<2~\fm$) \PbPb~collisions at 
$\sqrtsNN=5.5~\tev$, with and without jet quenching, are 
reported in Table~\ref{tab:hijing}. Taking into account energy loss 
(quenching) leads to a factor of 2 increase in multiplicity at mid-rapidity.


\begin{table}[!h]
  \caption{Charged particle multiplicity and total number of charged particles
           in the ALICE barrel acceptance given by HIJING 
           for central~\PbPb~($b<2~\fm$) at the LHC.}
  \label{tab:hijing}
\begin{center}
  \begin{tabular}{l|c|c}
  \hline
  \hline
  Setting & $\dNdy$ at $y=0$ & $N_{\rm ch}$ in $|\eta|<0.9$  \\
  \hline
   Energy loss on & $\simeq$ 6200 & $\simeq$ 10800 \\
   Energy loss off & $\simeq$ 2900  & $\simeq$ 5200 \\
  \hline
  \hline
  \end{tabular}
\end{center}
\end{table}

For our background studies in \PbPb~we used HIJING events 
with $b<2~\fm$ and with energy loss. 
This is a conservative choice, since the obtained multiplicity 
$\dNdy\approx 6000$ is close to the most pessimistic expectation 
for the LHC, $\dNdy=8000$, and about a factor 2 larger than the 
value predicted by the most recent analyses of RHIC results, 
$\dNdy\approx 2500$ (see Section~\ref{CHAP1:sqrtsdNdy}).

Finally, we remark that the use of HIJING allows to keep into 
account all sources of physical backgrounds (e.g. decays of 
strange particles, that can fake the topology of a charm particle 
decay vertex). We will give more details on this point in 
Chapter~\ref{CHAP6}.


\subsection{Simulation of the detector response}
\label{CHAP4:detresponse}

After event generation, in order to produce the equivalent of the 
{\sl raw data} (minimum of the parabola in Fig.~\ref{fig:aliroot})
the following steps are necessary:
\begin{itemize}
\item {\sl Particle transport}. The particles
  are transported in the material of the
  detector, simulating their interaction with it and the energy
  deposition that generates the detector response ({\sl hits}).
  To this purpose the program GEANT3.21~\cite{geant3}, interfaced to    
  AliRoot, is used. All physical processes are taken into account, 
  including a complete description of Coulomb scattering of 
  charged particles on atomic nuclei. The ALICE detector is 
  described in great detail, including support structures, beam pipe 
  and services.
\item {\sl Signal generation and detector response}. During this phase the
  detector response is generated from the energy deposition of the
  particles traversing it. This is the `ideal' detector response,
  before the conversion to digital signal and the formatting of the
  front-end electronics are applied.
\item {\sl Digitization}. The detector response is digitized and formatted
  according to the output of the front-end electronics and the Data
  Acquisition System. The results should resemble closely the real
  data that are produced by the detector.
\end{itemize}

\subsection{Track reconstruction}
\label{CHAP4:tracking}

The event reconstruction procedure includes:
\begin{enumerate}
  \item cluster finding;
  \item track reconstruction;
  \item reconstruction of the position of the interaction vertex.
\end{enumerate}
We shall describe the reconstruction of the interaction (or primary) vertex
in Chapter~\ref{CHAP5},
which is dedicated to the items concerning displaced vertices identification.
Here, after a brief note on cluster finding, we focus on track 
reconstruction and, particularly, on tracking performance.    

\subsubsection{Cluster finding}

During cluster finding, the information 
given by the detector electronics ({\sl digits}) is converted to 
space points, interpreted as (a) the crossing points between the tracks and
the centres of the pad rows in the readout chambers, 
in the case of the TPC, and (b) the crossing points
between the tracks and the silicon sensitive volumes, in the case of the ITS.
Another important piece of information provided by the 
cluster finder, is the estimate of the errors of the 
reconstructed space points. At present, a procedure for parallel clustering 
and tracking in the TPC is being tested. In the high-multiplicity 
scenario of \PbPb~collisions clusters from different tracks may overlap
and a preliminary knowledge of the track parameters is very helpful 
in the cluster deconvolution.  

The possibility to use a fast simulation of the detector response 
is implemented for many sub-systems of ALICE. The clusters are obtained 
directly from the hits via a parameterization of the response, 
in terms of efficiency and spatial resolution. 
The dramatic reduction in computing time (e.g. a factor $\simeq 25$ in the 
case of the ITS) allows the use of very high statistics in simulation studies.
The clusters obtained via the
fast simulation are called {\sl fast points}, while those obtained from the
detailed detector response are called {\sl slow points}.   

\subsubsection{Track reconstruction in TPC--ITS}

Due to the expected charged particle multiplicity,
track finding in ALICE is a very challenging task. In the most 
pessimistic case, the occupancy (defined as the ratio of the number of
read-out channels over threshold to the total number of channels) 
in the inner part of the TPC may reach 40\%. 


The track finding procedure developed for the barrel 
(ITS, TPC, TRD, TOF) is based on the Kalman filtering algorithm~\cite{kalman}, 
widely used in high-energy physics experiments. The Kalman filter 
is a method for simultaneous track recognition and reconstruction 
(or, in other words, track finding and fitting) and its main property is that, 
being a local method, at any given point along the track it provides 
the optimal estimate of the track geometrical parameters at that point.
For this reason it is a natural way to find the extrapolation of a track 
from a detector to another (for example from the TPC to the ITS or TRD).
As we will explain, in the Kalman filter energy loss and multiple 
scattering are accounted for in a direct and simple way. 

The complete chain of track reconstruction in the ALICE barrel 
foresees the following steps: (a) track finding in the TPC, inward 
(i.e. from the outer to the inner part); (b) matching to the ITS 
outer layer and track finding inward down to the innermost pixel layer;
(c) back-propagation and refit of the track outward in ITS and TPC, 
up to the outer radius of the TPC; (d) matching to the TRD and track finding 
(outward) in the TRD; (e) matching to the TOF detector, for 
PID.

Here we describe only steps (a) and (b), since this is the 
part of the chain which was employed in the studies performed for this work.  

{\small 
In the Kalman filter procedure, as implemented in ALICE, a track in 
the magnetic field of the barrel is locally (i.e. at a certain 
radial position in the barrel) parameterized as an helix, 
identified by a state vector of 5 parameters. Two parameters describe the 
track geometry in the beam direction ($z$) and three in the plane 
transverse to the beam (also referred to as {\sl bending} plane).
The description of the 
track state is completed by the $5\times 5$ covariance matrix of the 
parameters, which, at any given point, contains the best estimate of the 
errors on the parameters and of their correlations. In the TPC 
tracking, the procedure starts from the searching of track {\sl seeds} 
in the outermost pad rows of the detector, where the occupancy is lower.
All pairs of points, the first on the outermost pad row and the second on 
the pad row which is $n$ rows closer to the interaction point, are 
considered. For each pair, using the two points and the 
primary vertex position,
a first estimate of the state vector at the outermost pad row is
obtained.  Then, track points in the next $n$ rows are searched, and, 
if at least $n/2$ points are found, the candidate is saved as a 
track seed. Subsequently, a second seed-finding step is performed using 
another pair of rows. At this point the Kalman filter through the TPC 
starts, beginning with the tracks with lower curvature (i.e. higher 
$\pt$) that are found more easily, because the effect of multiple 
scattering is inversely proportional to the track momentum. 
The algorithm proceeds with an iteration of three steps:
\begin{enumerate}
  \item Prediction: given the state $j-1$ of the track at a certain layer, a 
        prediction of the state $j$ is obtained by propagating the 
        track-helix to the next layer. In this prolongation
        the track curvature is modified to take into account the energy loss
        and the covariance matrix is updated according to the multiple 
        scattering in the material encountered by the track.
  \item Filtering: after the extrapolation to the state $j$, all clusters 
        whose coordinates are inside a suitable `road' are 
        considered, the road being defined by the 
        track covariance matrix and by the spatial resolution of the 
        present detector layer. For each cluster the state vector is 
        updated and a $\chi^2$-increment is calculated. Then, all the 
        possible prolongations are `filtered' and the cluster that gives 
        the minimum $\chi^2$-increment is assigned to the track, provided 
        that the increment is lower than a given $\chi^2_{\rm max}$.  
  \item Update: finally the state $j$ of the track is updated 
        using the information of the assigned cluster. 
\end{enumerate} 

In the ITS implementation, the Kalman procedure was modified towards a more 
global approach. The tracks found in the TPC are used as seeds, again 
beginning with higher-$\pt$ tracks. In order to find the prolongation 
of a TPC track inside the ITS, for each seed all clusters on the 
outer ITS layer which are located in the fiducial road are considered.
For each of them a new candidate track is defined and propagated to the 
next layer, without applying any filtering. 
In this way, a track-tree with many candidates 
is built from a single TPC track, and only when the inner pixel layer 
is reached the filtering is applied and the candidate with the lowest 
$\chi^2$ per assigned cluster is selected. The assigned clusters are 
`removed' and the next TPC track is considered. In this way the finding 
of the low-$\pt$ tracks is facilitated, since all the clusters from the 
previously found tracks are not considered.  

Two track finding steps are used in the ITS: the first with a constraint on 
the position of the primary vertex, measured by the pixel layers 
(see Chapter~\ref{CHAP5}), to increase the efficiency for the 
tracks originating from the primary vertex (primary tracks); in the
second step this constraint is removed in order to allow the finding 
of tracks coming from displaced vertices (secondary tracks, e.g. 
decay products of strange particles). Decay products of charm (and beauty)
mesons can be considered as primary, from the point of view of track 
finding, since their displacement is usually lower than $1~\mm$.  
Afterwards, all found tracks are refitted without vertex constraint, 
in order to get an unbiased estimate of the distance 
from the interaction point. } 

In the next sections we shall present the relevant performance parameters
of the track reconstruction in the TPC and in TPC--ITS. 
The two extremes of the multiplicity scale are considered: central 
\PbPb~($\dNdy=6000$, HIJING) and pp collisions ($\dNdy=6$, PYTHIA).
The value of the magnetic field is 0.4 Tesla.

\subsubsection{Tracking efficiency in pp and in \PbPb}
\label{CHAP4:trackingeff}

\begin{figure}[!ht]
  \centering
  \includegraphics[width=\textwidth]{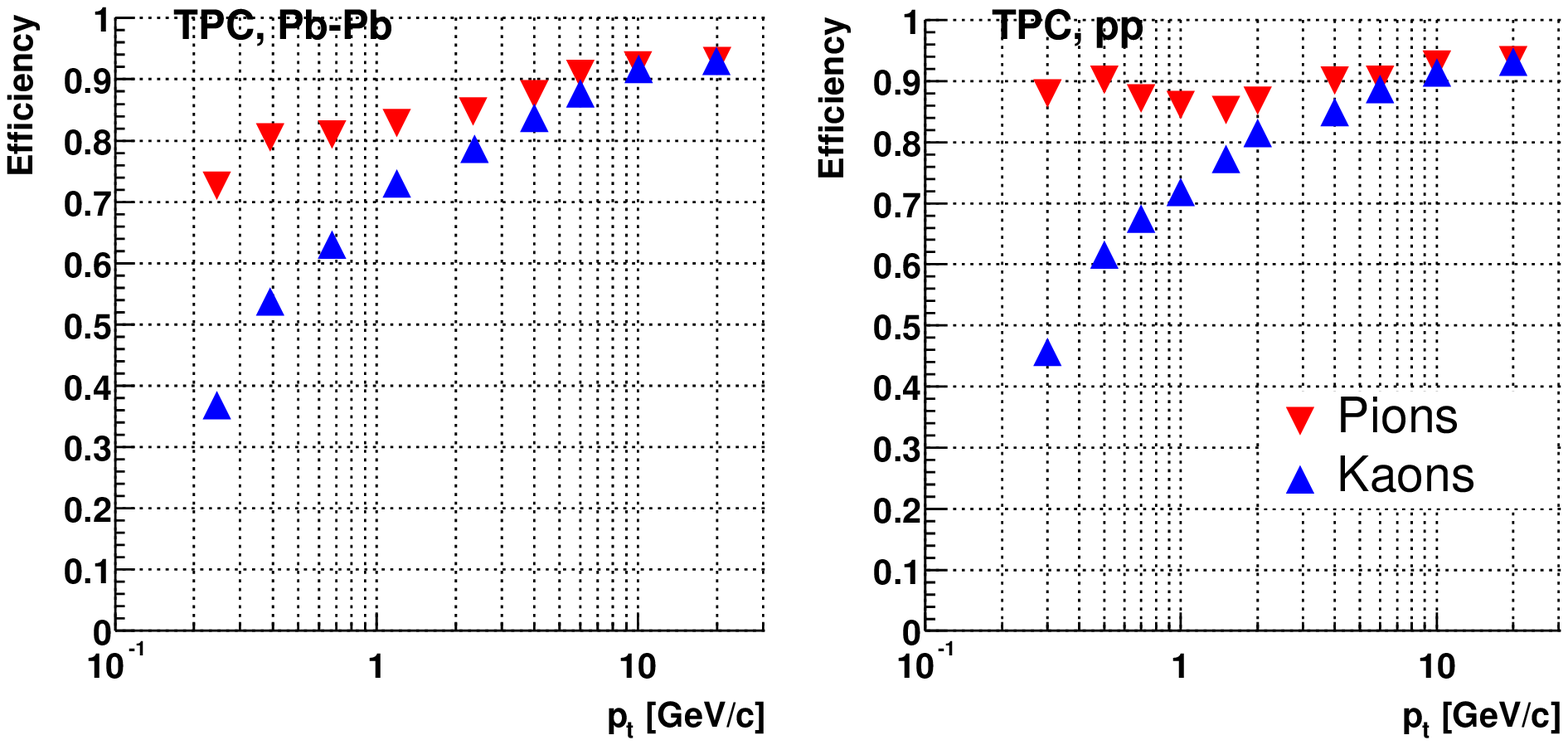}
  \caption{Tracking efficiency in the TPC, as a function of the transverse
           momentum, for charged pions and kaons in \PbPb~(left) and pp 
           (right) events. 
           The efficiency is defined as the ratio (in $|\eta|<0.9$)
           of the number of found primary tracks to the number of generated 
           primary tracks.}
  \label{fig:effTPC}
\vglue0.5cm
  \includegraphics[width=\textwidth]{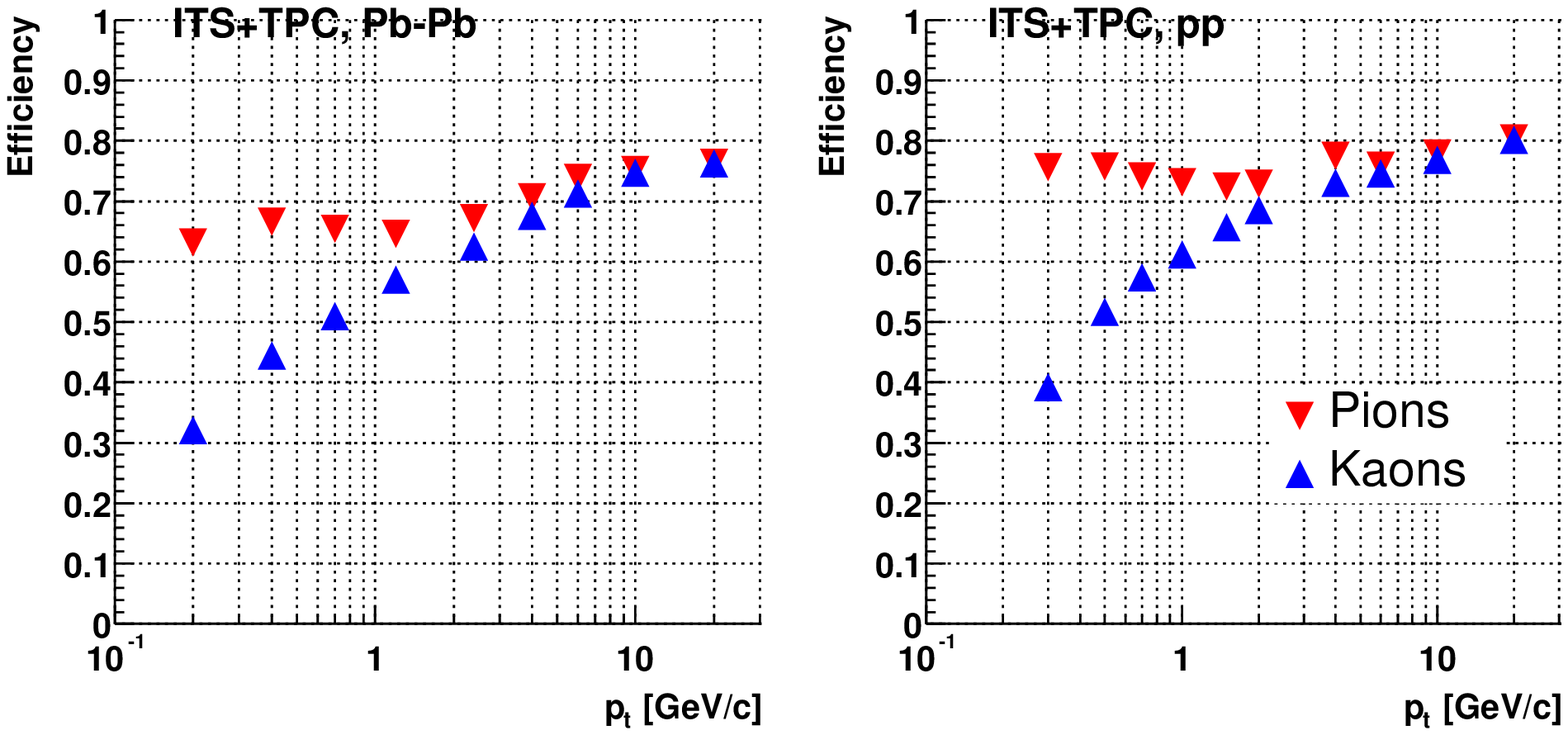}
  \caption{Same as the previous figure, but for TPC--ITS. 
           Six assigned clusters per track in the ITS are required.}
  \label{fig:effTPCITS}
\end{figure}

Figure~\ref{fig:effTPC} presents the tracking efficiency for 
primary tracks in the TPC, defined as a the ratio of the number of 
reconstructed tracks to the number of generated tracks
(the acceptance cut of the barrel, $|\eta|<0.9$, 
is applied to both numerator and denominator). The left panel corresponds to 
\PbPb, the right one to pp. The efficiency is 
shown as a function of the transverse momentum for charged pions
and charged kaons separately. The large difference between kaons 
and pions is due to the lower mean decay length of the former
---$7.5~\m\cdot p(\gev/c)$ for kaons and $55.7~\m\cdot p(\gev/c)$ for pions. 
In pp, the lower occupancy allows 
a higher track finding efficiency. At very 
high $\pt$ (straight tracks) the inefficiency of $\sim 10\%$ is due to the 
non-active regions of the TPC (separation between different sectors).

\begin{figure}[!ht]
  \centering
  \includegraphics[width=.6\textwidth]{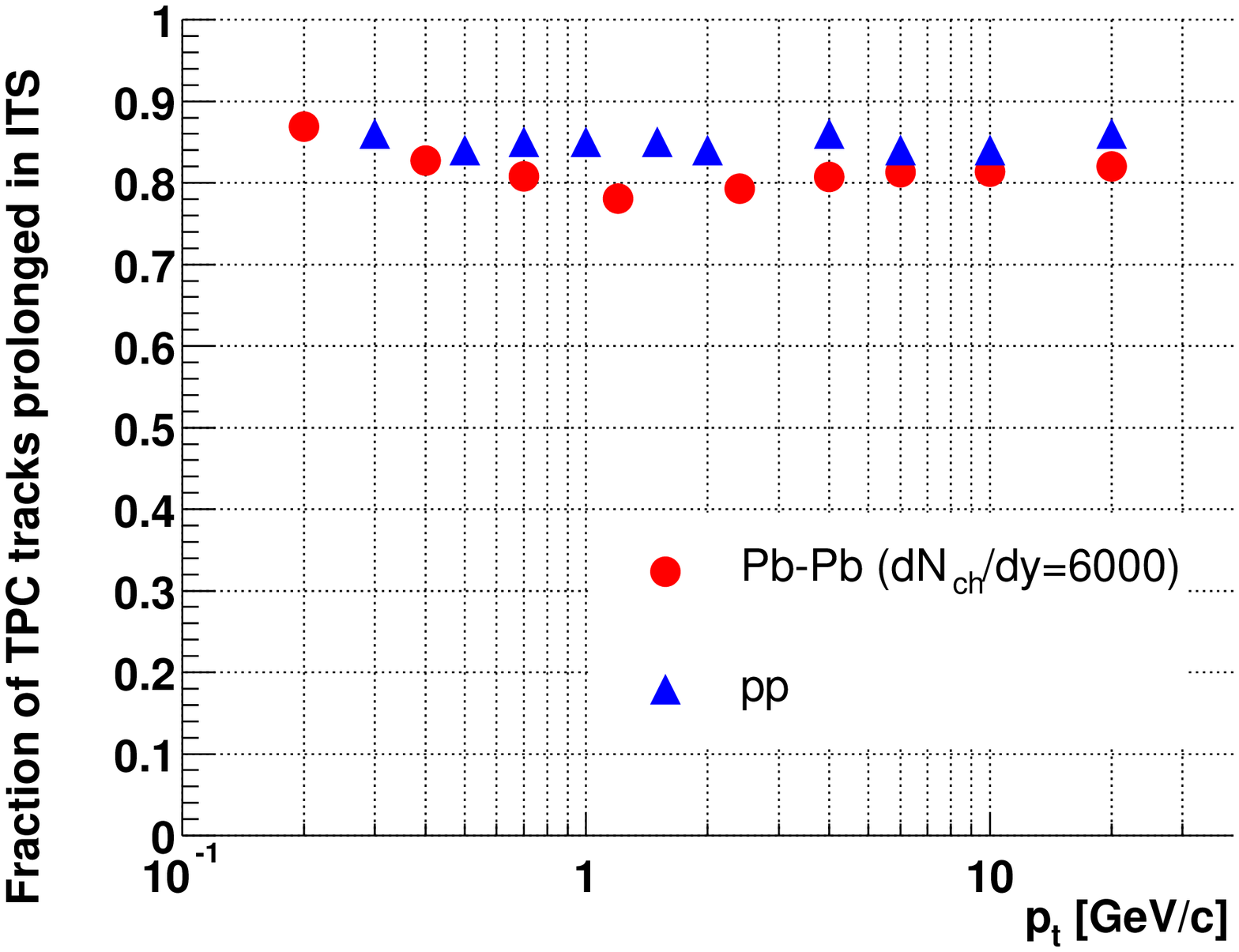}
  \includegraphics[width=.6\textwidth]{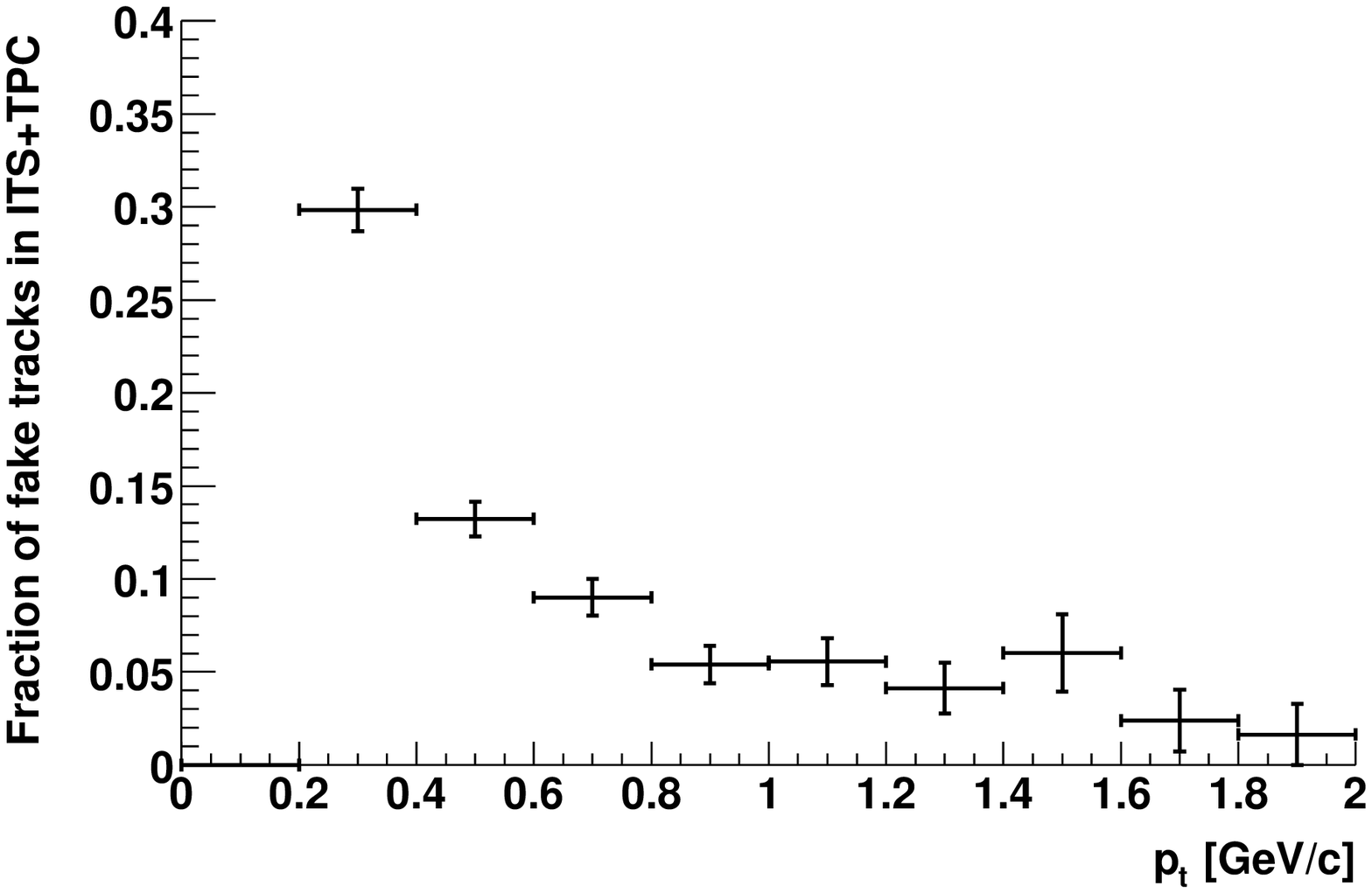}
  \caption{Fraction of TPC tracks prolonged to the ITS, with 6 clusters, 
           in pp and in \PbPb~(top) and, for \PbPb, fraction of ITS tracks 
           with at least one misassigned cluster in the ITS (bottom).}
  \label{fig:ITStoTPCandFakes}
\end{figure}

Figure~\ref{fig:effTPCITS} shows the efficiency, defined as for the 
previous figure, for the system TPC--ITS. In the ITS an
assigned cluster per layer is required (6 clusters in total). 
The average efficiency, taken into account the $\pt$ distribution 
of the particles generated by PYTHIA and HIJING, is 
$\simeq 65\%$ ($\simeq 75\%$) for pions and 
$\simeq 45\%$ ($\simeq 50\%$) for kaons in \PbPb~(pp).  

The fraction of TPC tracks prolonged in the ITS is roughly the 
same in pp and \PbPb~(Fig.~\ref{fig:ITStoTPCandFakes}, upper panel). However, 
in \PbPb, due to the high occupancy in the ITS layers, there is 
a non-negligible fraction of tracks that have at least one 
incorrectly-assigned cluster ({\sl fake} tracks). As shown in 
Fig.~\ref{fig:ITStoTPCandFakes} (lower panel), this effect is mainly 
restricted to low-$\pt$ tracks ($\pt<1~\gev/c$). We will compare the 
quality of correctly-reconstructed and fake tracks, in terms of track 
position resolution, in Chapter~\ref{CHAP5}.

\subsubsection{Momentum resolution in pp and in \PbPb}
\label{CHAP4:ptres}

The momentum resolution is a very important parameter for the selection 
of rare signals and for the measurement of their $\pt$ distribution. 
In particular, we have seen (Section~\ref{CHAP2:D0toKpi} and 
Appendix~\ref{App:kine}) that in the case of invariant mass analyses,
if the decay particles are relativistic (as in the 
$\DtoKpi$ decay at LHC energies) the mass resolution is 
proportional to the momentum resolution. Therefore, the 
signal-to-background ratio is inversely proportional to the momentum 
resolution.

Given the importance of this parameter, we cross-checked, in a 
simple case, the result obtained using the detector simulation 
in AliRoot. The transverse momentum resolution
can be approximated as a quadratic sum of a contribution due 
to the multiple scattering and a contribution due to the detector 
resolution. If the relative resolution is considered, the 
first contribution is a constant and the second is proportional to 
the transverse momentum itself~\cite{pdg}:
\begin{equation}
  \frac{\sigma(\pt)}{\pt} = K_{\rm scatt} \oplus K_{\rm meas}\cdot\pt
\equiv \sqrt{K_{\rm scatt}^2 + K_{\rm meas}^2\cdot\pt^2}.
\label{eq:ptres1}
\end{equation}
If $\pt$ is measured in a solenoidal magnetic field $B$ using 
uniformly distributed points with the same spatial resolution, 
the term $K_{\rm meas}$ should be~\cite{pdg}:
\begin{equation}
  K_{\rm meas} = \frac{\sigma_{r\phi}}{0.3\,B\,\ell^2}\sqrt{\frac{720}{N+2}}
\label{eq:ptres2}
\end{equation}
where $N$ is the number of points, $\sigma_{r\phi}$ is their 
spatial resolution in the transverse plane and $\ell$ is the lever 
arm of the arc between the first and the last measured point.

\begin{figure}[!t]
  \centering
  \includegraphics[width=.7\textwidth]{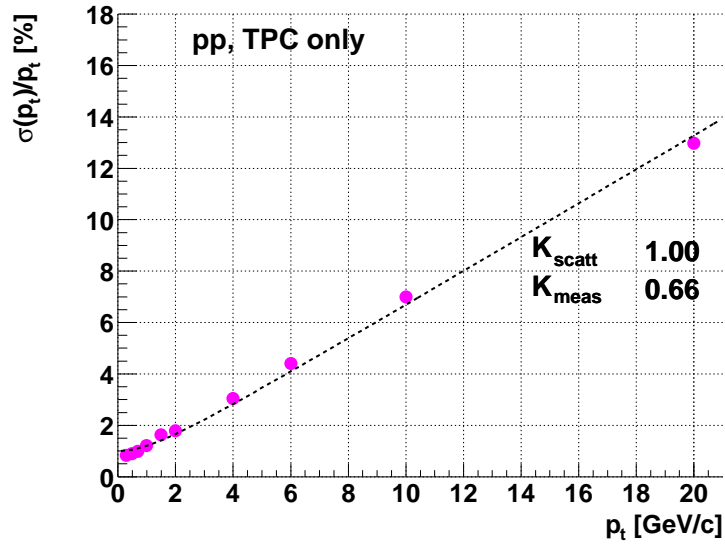}
  \caption{Relative $\pt$ resolution in the TPC for pp events with 
           $B=0.4$~T. 
           The resolution is fitted to the expression 
           $\sigma(\pt)/\pt=K_{\rm scatt}\oplus K_{\rm meas}\cdot\pt$.}
  \label{fig:TPCptresFit}
\end{figure}

In the TPC the spatial resolution of the space points can be assumed
as constant only in the case of pp collisions; in fact, in the 
high-multiplicity environment of heavy ion collisions, 
the clusters overlap in the inner part 
of the detector and the resolution is worse than in the outer part.   
The fit of the relative $\pt$ resolution, in pp collisions, in the TPC 
(see Fig.~\ref{fig:TPCptresFit}) to the expression (\ref{eq:ptres1}) 
gives:
\begin{equation}
  \frac{\sigma(\pt)}{\pt} \simeq 1.00\% \oplus 0.66\% \cdot \pt(\gev/c).
\end{equation}
The obtained value for $K_{\rm meas}$ is in good agreement with the 
value 0.65\%, 
calculated using $\sigma_{r\phi}=0.8~\mm$~\cite{tpctdr}, 
$N=100$~\cite{yura}, $\ell=1.6~\m$ and $B=0.4$~T.

\begin{figure}[!t]
  \centering
  \includegraphics[width=.49\textwidth]{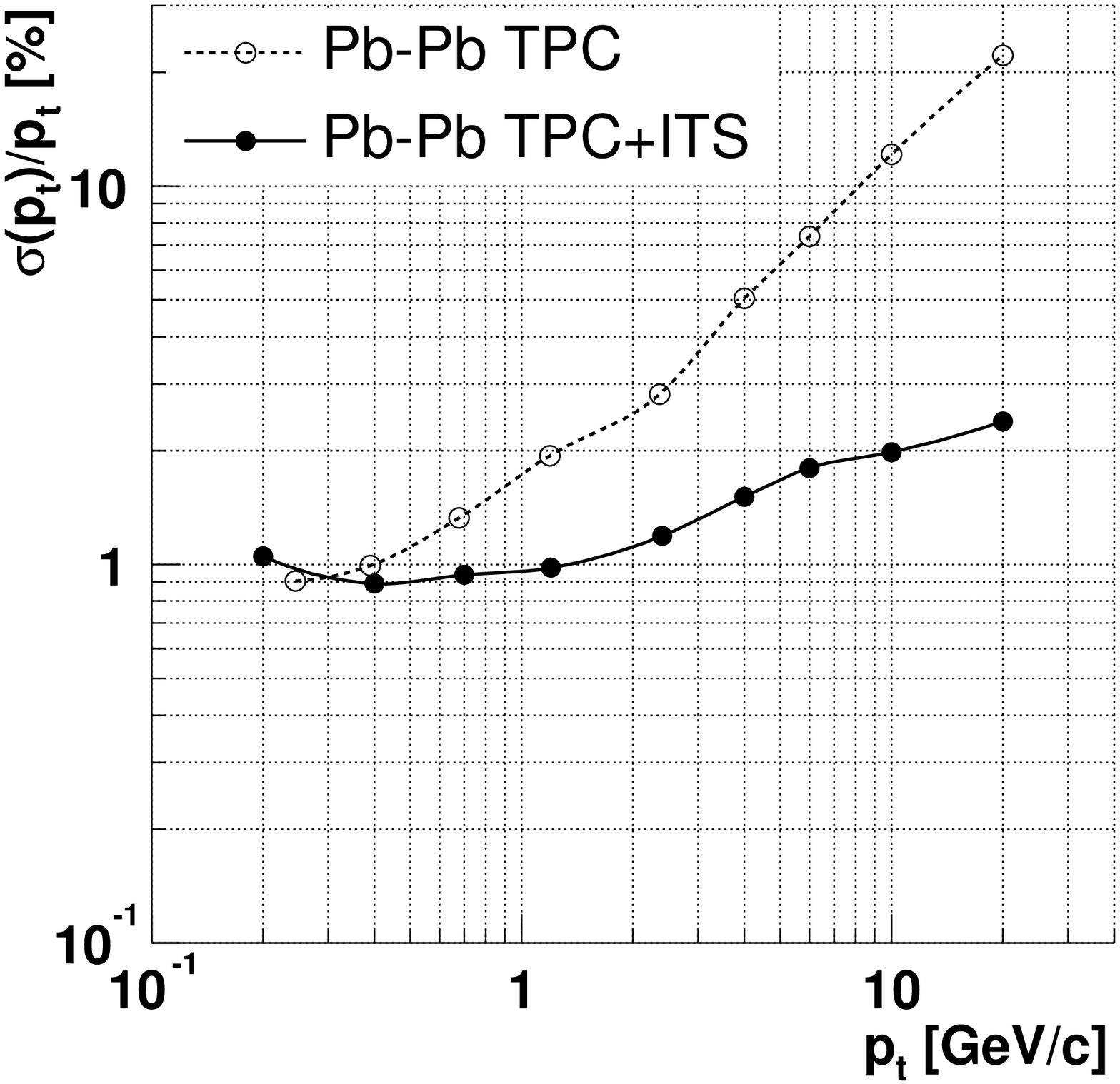}
  \includegraphics[width=.49\textwidth]{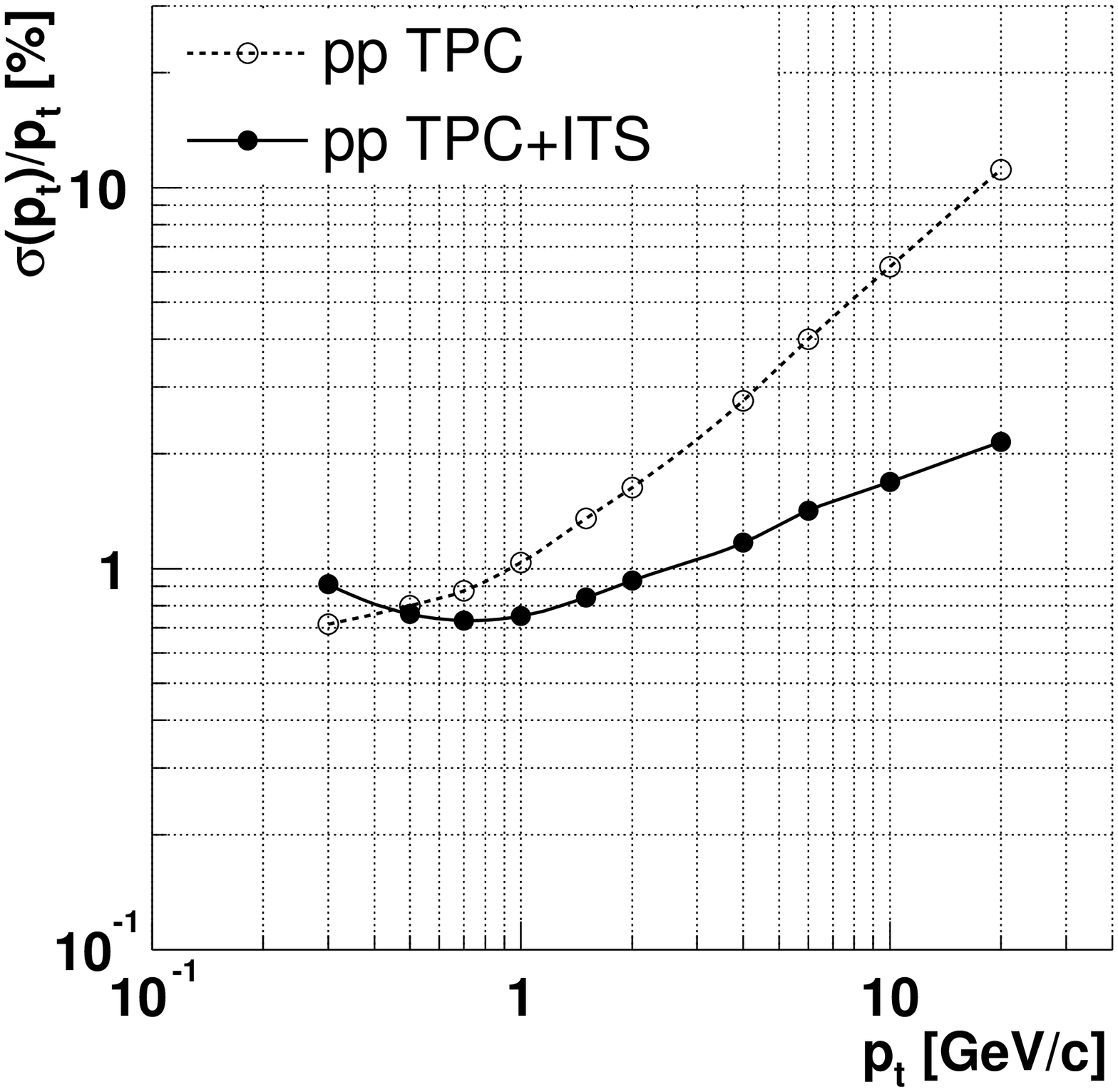}
  \caption{Relative $\pt$ resolution in the TPC and in TPC--ITS, 
           for \PbPb~(left) and for pp events (right).
           The value of the magnetic field is $B=0.4$~T.}
  \label{fig:ptres}
\end{figure}

The relative transverse momentum resolution, as a function of the 
transverse momentum, is shown in Fig.~\ref{fig:ptres}. The resolutions 
achieved in the TPC and in TPC--ITS are compared on the same plot, 
for \PbPb~(left panel) and for pp events (right panel). 
For the definition of the resolution, the reconstructed $\pt$ is 
compared to the `true' $\pt$ at the entrance of the TPC, for the TPC,
and to the `true' $\pt$ at the interaction point, for TPC--ITS.
This is the reason why the resolution at very low $\pt$ is better
for the TPC alone: at low $\pt$ the resolution is dominated 
by multiple scattering, which is much lower in the TPC gas than in
the layers of the ITS.  
But the most striking feature is the improvement of about 
one order of magnitude 
for high-$\pt$ tracks when including the ITS. This effect is partially
due to the increased lever arm ($\ell_{\rm TPC}\simeq 1.6~\m$ 
in the TPC and $\ell_{\rm TPC-ITS}\simeq 2.5~\m$ in TPC--ITS), 
which accounts for an improvement of a factor 
$(\ell_{\rm TPC-ITS}/\ell_{\rm TPC})^2\simeq 2.5$. The remaining 
difference is due to the much better spatial resolution of the 
points in the ITS, with respect to those in the TPC.
For TPC--ITS the $\pt$ resolution in pp is $\approx 15$-$20\%$ better 
than in \PbPb.

\subsubsection{Parameterization of TPC tracking response: a tool 
for high-statistics simulation studies}
\label{CHAP4:tpcparam}

The computing time and disk space requirements for the detailed simulation of 
the TPC are very large and they correspond to more than 90\% of the total 
requirements for the tracking dedicated detectors of the ALICE 
barrel (ITS and TPC). The size of 
the file containing the hits and the digits (in addition to the 
kinematic information on the generated particles) for 
1 central \PbPb~event produced with HIJING is $0.02$~Gbytes if only the ITS
is included and 1.2~Gbytes if also the detailed description of the TPC is 
included. Similar proportions are maintained also for pp events.
This implies that {\sl the complete simulation of the TPC 
cannot be used in the studies of physics signals requiring a large 
number of events}. 
The study for the detection of open charm via hadronic decays is 
a typical example. In fact, as we will see in Chapter~\ref{CHAP6}, 
the selection strategy to extract the D meson signal out of the large 
combinatorial background has to be optimized on a very large 
sample of events ($\sim 10^4$-$10^5$ for \PbPb~and $\sim 10^6$-$10^7$ for pp). 
The required statistics is similar also for the studies 
on semi-electronic charm and beauty decays and on hyperon 
decays ($\Xi$ and $\Omega$, in particular). 
In all these cases, the performance of the apparatus is determined mainly 
by the ITS, which is dedicated to the reconstruction of secondary vertices. 
Therefore, {\sl the optimal solution is to use a detailed 
simulation of the ITS (and of the beam pipe material) and a 
parameterization of the response of the tracking algorithm in the TPC}.

The design and implementation of this parameterization, 
developed specifically in the scope of this thesis work, 
is described in an ALICE Internal Note~\cite{notetpcparam}. 
We remark that this work allowed also a systematic study of the response 
of TPC tracking and a fruitful analysis of the error handling in the 
tracking algorithm (see Ref.~\cite{notetpcparam} for details).

In the following we outline the idea on which the tool is based.
The tests that were performed in order to 
validate the tool for the use in physics studies are reported in 
Appendix~\ref{App:tpc}.

After the Kalman filter reconstruction in the TPC, the tracks that have 
been found are described by the state vector and its
covariance matrix, 
defined at a reference plane corresponding to the radial position of the inner 
pad row of the TPC.
At this stage all the information from the TPC is `summarized' 
in the state vector and the covariance matrix.
Such tracks are taken as seeds for the track finding algorithm in the ITS.

The idea is to parameterize the response of the TPC at this level and 
create the TPC reconstructed tracks taking as a starting point the 
generated tracks.
Since for every track entering the TPC in a sensitive region the position and 
the momentum in the entering point are stored ({\sl hit}), it is 
possible to determine the exact state vector of the track at the reference 
plane. 
Then, the following steps are performed:
\begin{enumerate}
  \item Apply a first selection based on the geometrical acceptance of the TPC.
  \item Assign a covariance matrix to the track. This is done using look-up
        tables. 
  \item Smear the track parameters according to the Kalman filter resolutions,
        by means of the covariance matrix.
  \item Assign a value of d$E/$d$x$ to the track. This is important 
        because the d$E/$d$x$ from the 
        TPC is used to make a mass hypothesis for the 
        estimation of multiple scattering and energy loss during 
        tracking in the ITS. 
  \item Apply a second selection based on the efficiency of the TPC 
        (which accounts for decays, detector and tracking algorithm 
        efficiency).
\end{enumerate}
In all these steps different track kinematics and particle types 
are appropriately taken into account.  
The tracks `built' with this procedure can be used to seed the standard 
reconstruction with the Kalman filter in the ITS. Since only the 
information at the inner radius of the TPC is required, the transport 
done by GEANT can be stopped at $r\simeq 90$~cm 
and the digitization of the TPC is not needed. This reduces the simulation 
requirements in terms of disk space and computing time by a factor 
$\simeq 35$. Is was verified that the performance of the ITS tracking, 
in terms of tracking efficiency and resolutions on the track parameters, is 
not altered by the use of this parameterized tracking in the TPC
(see Appendix~\ref{App:tpc}). The parameterization was realized for 
\PbPb~events with $\dNdy=6000$ and for pp events, at $B=0.4~$T.

\mysection{LHC beams and interaction region}
\label{CHAP4:lhc}

The running strategy for the heavy ion program at the LHC, along 
with the dependence  on the colliding system of the available 
energy in the centre of mass were discussed in Section~\ref{CHAP1:sqrtsdNdy}.
Here we focus on the features of the particle beams at the ALICE 
intersection point (IP) and, consequently, of the interaction region, 
the region in space that contains all the possible collision 
vertices. These parameters are, 
indeed, very important for the reconstruction of secondary vertices, 
as we will see in the next chapters.

In Section~\ref{CHAP4:luminosity} we will define the luminosity and show
how it is related 
to the size of the interaction region. Then, we will consider the 
specific cases of \PbPb~(Section~\ref{CHAP4:lhcPbPb}) and pp 
(Section~\ref{CHAP4:lhcpp}) running.

\subsection{Luminosity and beam size}
\label{CHAP4:luminosity}

The event rate $R$ in a collider is proportional to the interaction cross 
section $\sigma_{\rm int}$ and the factor of proportionality is called 
{\sl luminosity} (${\cal L}$):
\begin{equation}
R={\cal L}\,\sigma_{\rm int}.
\end{equation}
The luminosity, as we will show, is entirely defined by the characteristics 
of the colliding beams at the interaction point, and it is
an important issue, because there are limitations
on the maximum event rate coming from both the ALICE detector 
and the accelerator. As we will detail in the next sections, in the 
case of \PbPb~events the limitations of the detector and of the machine 
coincide, while in the case of pp events the maximum event rate is 
determined by the detector and the machine parameters have to be tuned 
in order not to exceed this limit.

Let us consider two intersecting bunches, labelled 1 and 2. To a good 
approximation, the particles in each of them 
will be distributed according to Gaussians in the three perpendicular 
directions~\cite{pdg}:
\begin{equation}
N_i(x,y,z)=N_i\,G(x,\overline{x}_i,\sigma_{x,i})\,G(y,\overline{y}_i,\sigma_{y,i})\,G(z,\overline{z}_i,\sigma_{z,i})~~~~~~~~i=1,~2
\end{equation}
where $N_i$ is the total number of particles in the bunch $i$ and
\begin{equation}
G(q,\overline{q},\sigma_q)=\frac{1}{\sqrt{2\pi}\sigma_q}\exp\left[-\frac{(q-\overline{q})^2}{2\sigma_q^2}\right]~~~~~~~~~~~~~~~~~q=x,~y,~z.
\end{equation}
If $f$ is the revolution frequency and $N_{\rm b}$ is the number of bunches, 
the luminosity is obtained as:
\begin{equation}
{\cal L}=f\,N_{\rm b} \times \int \d x\,\d y\,\d z\,N_1(x,y,z)\,N_2(x,y,z).
\end{equation} 
Here, the integration along the beam direction, $z$, gives 1, since the two 
particle bunches cross each other and, therefore, their distributions 
along $z$ are equivalent to delta functions from the point of view of 
the interaction probability.

At the LHC the bunches in the two beams have the same number of particles
($N$) and the same dispersions; moreover the dispersion is the same for 
the two directions transverse to the beam axis 
($\sigma_x=\sigma_y=\sigma_{x,y}$). Thus:
\begin{equation}
\label{eq:lumi}
{\cal L}=f\,N_{\rm b}\,\frac{N^2}{4\pi\sigma_{x,y}^2}\exp\left[-\frac{d^2}{4\sigma_{x,y}^2}\right]
\end{equation} 
where $d^2=(\Delta \overline{x})^2+(\Delta \overline{y})^2$ is the square
of the distance between the centres of the two beams. Normally, the two 
beams are centred and $d=0$.

On the other hand, the interaction region is defined as the convolution 
of the two particle distributions in the two intersecting bunches: 
the interaction vertex lies in a `diamond' with `dimensions'
\begin{equation}
\sigma^{\rm vertex}_q=\sigma_q/\sqrt{2}~~~~~~~~~q=x,~y,~z
\end{equation}
that do not depend on the value of $d$, if the distributions are gaussian. 

The size of the bunches at the IP depends on the {\sl transverse emittance} 
$\epsilon$ (a beam quality parameter) and on the value of the {\sl 
amplitude function} $\beta$ at the IP, indicated as $\beta^\star$, which is 
determined by the accelerator magnets configuration. We have~\cite{pdg}:
\begin{equation}
\sigma_q=\sqrt{\frac{\epsilon_q\,\beta^\star}{\pi}}~~~~~~~~~~q=x,~y,~z.
\end{equation}
From the last three equations we see that the luminosity can be decreased 
locally (i.e. only at the ALICE IP) by increasing $\beta^\star$, but this 
increases the transverse size of the interaction region.
In Table~\ref{tab:lhcbetastar} we report the LHC machine nominal parameters
at the ALICE IP for pp and \PbPb~runs~\cite{pprCh2}. We will discuss them 
in the next sections.

\begin{table}[!t]
  \caption{LHC parameters for pp and \PbPb~runs at the ALICE IP~\cite{pprCh2}.}
  \label{tab:lhcbetastar}
\begin{center}
  \begin{tabular}{ll|c|c}
  \hline
  \hline
  Parameter & & pp & \PbPb \\ 
  \hline
  Energy per nucleon & [TeV] & 7 & 2.76 \\
  $\beta^\star$ & [m] & 10 & 0.5 \\
  $\sigma_{x,y}$ & $[\mum]$ & 71 & 16 \\
  $\sigma_{z}$ & $[\cm]$ & 7.5 & 7.5 \\
  $\sigma^{\rm vertex}_{x,y}$ & $[\mum]$ & 50 & 11 \\
  $\sigma^{\rm vertex}_{z}$ & $[\cm]$ & 5.3 & 5.3 \\
  Luminosity & $[\cm^{-2}\s^{-1}]$ & $5\times 10^{32}$ & $10^{27}$ \\
  \hline
  \hline
  \end{tabular}
\end{center}
\end{table}

\subsection{Interaction region in \PbPb~collisions}
\label{CHAP4:lhcPbPb}

When taking data with \PbPb~beams the ALICE detector is limited to 
a maximum luminosity of $10^{27}~\cm^{-2}\s^{-1}$ by the 
drift time in the Time Projection Chamber. Due to the maximum allowed
heating in the beam pipe\footnote{The energy deposition in the beam pipe 
comes from Pb ions `emitted' by the beam because of electromagnetic 
interactions.} it is currently assumed that the 
maximum \PbPb~luminosity at the LHC is also limited to 
\mbox{$0.5$-$1\times 10^{27}~\cm^{-2}\s^{-1}$}.

The parameters of the Pb beams ($N,~N_{\rm b},~f$) are, therefore, tuned 
in order to have $\mathcal{L}\simeq 10^{27}~\cm^{-2}\s^{-1}$ 
while running with a low value 
of $\beta^\star$ ($0.5~\m$), which allows
to focus very well the beams in the transverse plane. 
The transverse size of the beams at the ALICE IP 
is $\sigma_{x,y}\simeq 16~\mum$ and 
the transverse size of the interaction region is 
$\sigma_{x,y}/\sqrt{2}\simeq 11~\mum$. 
Since the position of the centres of the beams is stable for 
a given run (of a duration of about 4 hours), 
the mean position of the interaction point during each run 
is measured with very high precision, by integration over a long
time interval. Therefore, the uncertainty on the vertex position in 
the transverse plane can be assumed to be given by the size of the interaction 
region: $\sigma_{\rm vertex}\simeq 11~\mum$. 

Along the $z$ direction, the interaction point is distributed 
according to a Gaussian with a dispersion of $\simeq 5.3~\cm$ and the ALICE
interaction trigger selects the events with vertex located in the 
fiducial region $-5.3<z<5.3~\cm$.
Clearly, the position of the vertex in $z$ has to be reconstructed on an 
event-by-event basis. This task is achieved exploiting the correlation 
between clusters in the two silicon pixel layers of the 
ITS~\cite{vtxPbPb1,vtxPbPb2}; the results of this method will be briefly 
described in Chapter~\ref{CHAP5}.

\subsection{Interaction region in pp collisions}
\label{CHAP4:lhcpp}

For the pp runs ALICE will take data in parallel with the 
pp-dedicated experiments, which will exploit the maximum design 
luminosity of the LHC for pp, ${\cal L}\simeq 10^{34}~\cm^{-2}~\s^{-1}$.
The nominal luminosity at the ALICE IP, reduced using a larger value of 
$\beta^\star$ with respect to the other experiments, 
is $5\times 10^{32}~\cm^{-2}\s^{-1}$.
Nevertheless, this nominal luminosity 
has to be reduced to $<3\times 10^{30}~\cm^{-2}\s^{-1}$, in order to 
limit the pile-up of events in the TPC
 and in the Silicon Drift Detectors (SDD)\footnote{A maximum pile-up of 
$\simeq 20$ pp events can be disentangled by the High Level Trigger, 
exploiting the fact that the tracks from different events `point' to 
different interaction points along $z$.}.
 
As we can see from Eq.~(\ref{eq:lumi}), such reduction can be achieved in 
two ways: 
\begin{itemize}
\item by further increasing the value of $\beta^\star$ and/or
\item by displacing the 
      two beams in the transverse plane ($d>0$) to make a collision between 
      the tails of the particle distributions. 
\end{itemize}

If the first option is chosen, $\beta^\star$ might be increased 
up to $100~\m$; this would broaden of a factor $\sqrt{100~\m/10~\m}\simeq 3$ 
the transverse size of 
the interaction `diamond', with respect to the nominal value reported 
in Table~\ref{tab:lhcbetastar}, up to $\simeq 150~\mum$. 

If the second option 
is necessary, the beams might be displaced to a distance of 
$\simeq 4$-$5~\sigma_{x,y}\simeq 300~\mum$ and the collisions will occur in the 
tails at $4$-$5~\sigma$ from the centre of the beams: these tails will 
much likely be non-gaussian and the size of the interaction `diamond'
may be even larger than $150~\mum$.

For what concerns the $z$ direction, the situation will be the same
as for the heavy ion running.

Given that the uncertainty on the vertex position in the transverse plane 
might be even larger than the mean decay length of neutral charm mesons
($c\tau\simeq 123~\mum$), 
we conclude that the position of the interaction vertex in 
pp collisions has to be reconstructed in all three dimensions, and not only
along $z$ as in \PbPb, on an event-by-event basis. A large part of the 
next chapter is dedicated to this topic. In Chapter~\ref{CHAP6} we
demonstrate that the uncertainty on the position of the interaction vertex 
is the main limiting factor for the performance of ALICE in the exclusive 
reconstruction of charm particles in \pp~events. 

\clearpage
\pagestyle{plain}

\setcounter{chapter}{4}
\mychapter{Identification of heavy flavour \mbox{decay vertices}:
experimental issues}
\label{CHAP5}

\pagestyle{myheadings}

Secondary vertices are the signature of the 
(weak) decay of particles containing strangeness, charm or beauty.
The identification of these decays is particularly challenging   
in the case of open charm and open beauty hadrons that have mean proper 
decay lengths of $\sim 100$-$500~\mum$, namely D$^0$ ($c\tau\simeq~123~\mum$), 
D$^+$ ($c\tau\simeq~315~\mum$) and B mesons ($c\tau\sim~500~\mum$)~\cite{pdg}.

The most effective constraint for the selection of such particles 
is the presence of one or more tracks that are displaced from the
interaction (primary) vertex. 
The variable allowing to evaluate the displacement of a track is its 
{\sl impact parameter}, which was defined in Chapter~\ref{CHAP2} as the 
distance of closest approach of the reconstructed particle trajectory 
to the primary vertex. 

Now that the procedure for track reconstruction in the 
ALICE barrel has been described (Section~\ref{CHAP4:tracking}), a more 
precise definition of the two projections of the impact parameter, 
in the transverse plane and along the beam direction, can be given. 
After the reconstruction, the state vector of the track is given at the 
radial position corresponding to the inner pixel layer ($r\simeq 4~\cm$),
where the last cluster assigned to the track lies. From this point
the state vector is propagated to the radius of the beam pipe
($r\simeq 3~\cm$) and a correction is applied to the track curvature 
in order to account for the energy loss in the material ($0.8~\mm$ of 
beryllium). The impact parameter projection in the transverse (bending) 
plane, $d_0(r\phi)$, is defined as:
\begin{equation}
  d_0(r\phi)\equiv q\cdot \left[R-\sqrt{(x_V-x_C)^2+(y_V-y_C)^2}\right],
\end{equation}
where $q$ is the sign of the particle charge, $R$ and $(x_C,y_C)$ are 
the radius and the centre of the track projection in the transverse plane 
(which is a circle) and  
$(x_V,y_V)$ is the position of the primary vertex in the transverse plane. 
In this way, the impact parameter has also a sign; this is very useful 
for the identification of specific topologies, in particular for the 
$\Dz\to\K^-\pi^+$ decay, as we will see in the next chapter. 
The $z$ projection of the impact parameter, $d_0(z)$, is defined as:
\begin{equation}
  d_0(z)\equiv z_{\rm track}-z_V,
\end{equation}
where $z_{\rm track}$ is the $z$ position of the track after it has been 
propagated to the distance of closest approach in the bending plane,
and $z_V$ is the position of the primary vertex along the beam direction.

Clearly, for both the $r\phi$ and $z$ 
projections, the impact parameter resolution 
has a contribution due to the track position resolution and a contribution 
due to the uncertainty on the primary vertex position:
\begin{equation}
\label{eq:d0res}
  \sigma(d_0) = \sigma_{\rm track} \oplus \sigma_{\rm vertex}.
\end{equation}
{\sl Since the $r\phi$ impact parameter of the decay products of $\Dz$ mesons 
is of the order of $100~\mum$, as shown in Appendix~\ref{App:kine}, 
it is crucial to achieve a very good resolution ($<50$-$60~\mum$) 
not only on the track 
position at the vertex, but also on the position in $r\phi$ of the primary 
vertex itself.} 
~\\

We consider separately the two cases of \PbPb~(or, more generally, heavy ion)
events and pp events. 

We have seen in Section~\ref{CHAP4:lhc}
that, in the former case, the transverse beam size is very small and the
primary vertex position is know for a given run with an uncertainty of 
only $\simeq 10~\mum$. 
This uncertainty is negligible and, therefore, the resolution on $d_0(r\phi)$ 
coincides with the resolution on the track position.
In Section~\ref{CHAP5:vtxPbPb} we show how the pixel detector 
provides, before track reconstruction, a very precise estimate of the 
position along $z$ of the interaction point. Then, we report on
the performance of ALICE for the measurement of the track
impact parameter in  heavy ion collisions (Section~\ref{CHAP5:d0PbPb}). 
The achieved resolution is studied as a function
of particle kinematics and particle type, and the effect of missing
or misassigned clusters in the ITS is discussed.

In the case of pp running, since the beams 
have to be defocused and/or displaced to reduce the luminosity,
the  {\sl a priori} information on the vertex position 
might be extremely poor ($\sigma\sim150~\mum$) and an event-by-event 
reconstruction of all the three 
coordinates is mandatory in order to fulfill the resolution requirements
stated above. A method to accomplish this task is presented  
in Section~\ref{CHAP5:vtxpp}. The achieved resolution on the 
the vertex is eventually combined with that on the track position 
to obtain the impact parameter resolution in pp events 
(Section~\ref{CHAP5:d0pp}).  

Finally, we describe the reconstruction of the secondary vertex position
in the case of the two-body decay $\Dz\to\K^-\pi^+$ 
(Section~\ref{CHAP5:secondary}). We show the 
achieved spatial resolution, which is important for the precise 
determination of the $\Dz$ flight line and, consequently, 
of the pointing angle, the other key variable for the identification 
of displaced vertex topologies.

\mysection{Primary vertex reconstruction in \PbPb}
\label{CHAP5:vtxPbPb}

Along the $z$ direction, the interaction points are distributed according 
to a Gaussian with $\sigma=5.3~\cm$, both in heavy ion and in pp runs 
(Section~\ref{CHAP4:lhc}). The $z$ coordinate of the primary vertex in 
heavy ion collisions can be measured very precisely, before track 
reconstruction, using the correlation between clusters
in the two pixel layers at $r=4~\cm$ and $r=7~\cm$~\cite{vtxPbPb1,vtxPbPb2}. 
A {\sl tracklet} (line segment), built from two clusters 
generated by the same track, points to the primary vertex and gives an estimate
of its position along $z$. All pairs of clusters in the two layers
are considered and the background from uncorrelated pairs is partially 
removed requiring the two clusters to lie within the same (small) azimuthal
window $\Delta \phi$. The achieved resolution on $z_V$ is
proportional to $1/\sqrt{N_{\rm tracklets}}$, where $N_{\rm tracklets}$ is 
the number of tracklets from 
correlated clusters, which is proportional to the 
event multiplicity $\dNdy$. The resolution in heavy ion collisions
was parameterized as~\cite{vtxPbPb1}:
\begin{equation}
\sigma_z(\dNdy)=\left(2 + \frac{292}{\sqrt{\dNdy}}\right)~\mum,
\end{equation} 
where the constant additive term is due to the residual background of 
uncorrelated clusters.
The resolution is $\simeq 5~\mum$ for $\dNdy=6000$ and $\simeq 7~\mum$ for 
a lower multiplicity, $\dNdy=3000$. Therefore, in \PbPb~also the 
uncertainty on $z_V$ is negligible and the resolution on the $z$ 
projection of the impact parameter is given only by the track position 
resolution. 

{\small 
We adapted and tuned this method for the estimate of $z_V$ also in 
pp collisions~\cite{vtxpp}. The resolution was studied as a function of 
the multiplicity in pp events simulated with PYTHIA. 
The result is~\cite{vtxpp}:  
\begin{equation}
\label{eq:zpixelspp}
\sigma_z(\dNdy)=\left(42 + \frac{290}{\sqrt{\dNdy}}\right)~\mum.
\end{equation} 
The value corresponding to the average multiplicity predicted by PYTHIA 
for pp non-diffractive events at $\sqrt{s}=14~\tev$ ($\dNdy=6$) 
is $\simeq 160~\mum$. Although
not very precise, it is important to have an estimate of the vertex 
position along $z$ before track reconstruction; this information is, 
indeed, helpful to improve the track finding efficiency in the ITS.
} 

\mysection{Track impact parameter resolution in \PbPb}
\label{CHAP5:d0PbPb}

\subsection{Transverse momentum dependence}
\label{CHAP5:d0VSpt}

For the estimation of the $d_0$ $r\phi$- and 
$z$-resolution as a function of the 
transverse momentum, we superimposed samples of 500 particles, 
50\% positively charged and 50\% negatively charged,
with a given $\pt$ and homogeneously distributed in $|\eta|<0.9$, on top of 
standard \PbPb~central events, with collision impact parameter $b<2~\fm$, 
generated with HIJING. For 
each considered $\pt$ value 10 such combined events were used. 
The magnetic field was set to 0.4~T.
In the ITS the 
clusters were obtained using a realistic simulation of the detector 
response ({\sl slow points}) and
track reconstruction in TPC--ITS was performed as described in 
Chapter~\ref{CHAP4}.


At first we consider primary 
charged pions reconstructed in the TPC and in the ITS, 
with a cluster in each ITS layer; 
this is very important, because the impact parameter resolution is 
strongly dependent on the number of clusters associated to the 
track in the ITS, 
and, in particular, on the presence of the clusters in the two pixel layers, 
as we shall detail in the following.

\begin{figure}[!t]
  \begin{center}
    \leavevmode
    \includegraphics[width=.7\textwidth]{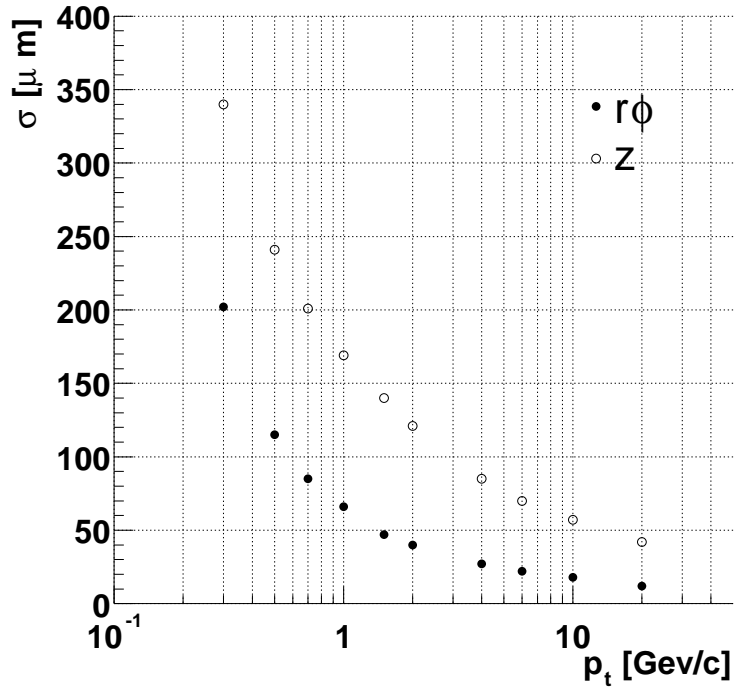}
    \caption{Impact parameter resolutions for primary charged pions 
             reconstructed in the TPC and in the ITS (with 
             6 clusters in the ITS) in central 
             \PbPb~collisions ($\dNdy=6000$).} 
    \label{fig:d0resMain}
  \end{center}
\end{figure}

The resolutions, obtained from a gaussian fit to the $d_0$
distributions, are presented in Fig.~\ref{fig:d0resMain} as a function of the 
track transverse momentum. 
The resolution in $r\phi$ ($z$) is $65~(170)~\mum$ at $\pt=1~\gev/c$ and
 $12~(40)~\mum$ at $\pt=20~\gev/c$. The large difference between the 
two projections reflects the different 
spatial resolutions in the $r\phi$ and $z$ directions of the detectors in 
the ITS (see Table~\ref{tab:its}). A more detailed analysis of the 
$\pt$-dependence of the resolutions and of their value at high $\pt$ can be 
found in Appendix~\ref{App:d0}, along with a study of their dependence on the 
thickness of the two pixel layers.

A resolution on $d_0(r\phi)$ better than $65~\mum$ ($\approx c\tau(\Dz)/2$) 
for $\pt>1~\gev/c$ is, in principle, sufficient to separate from the 
background the tracks from a $\Dz$ decay and, therefore, the Inner Tracking 
System with pixel detectors fulfills its most severe design requirement.

\subsection{Effect of missing and misassigned clusters}
\label{CHAP5:d0VSclusters}

\begin{figure}[!t]
  \begin{center}
    \leavevmode
    \includegraphics[width=\textwidth]{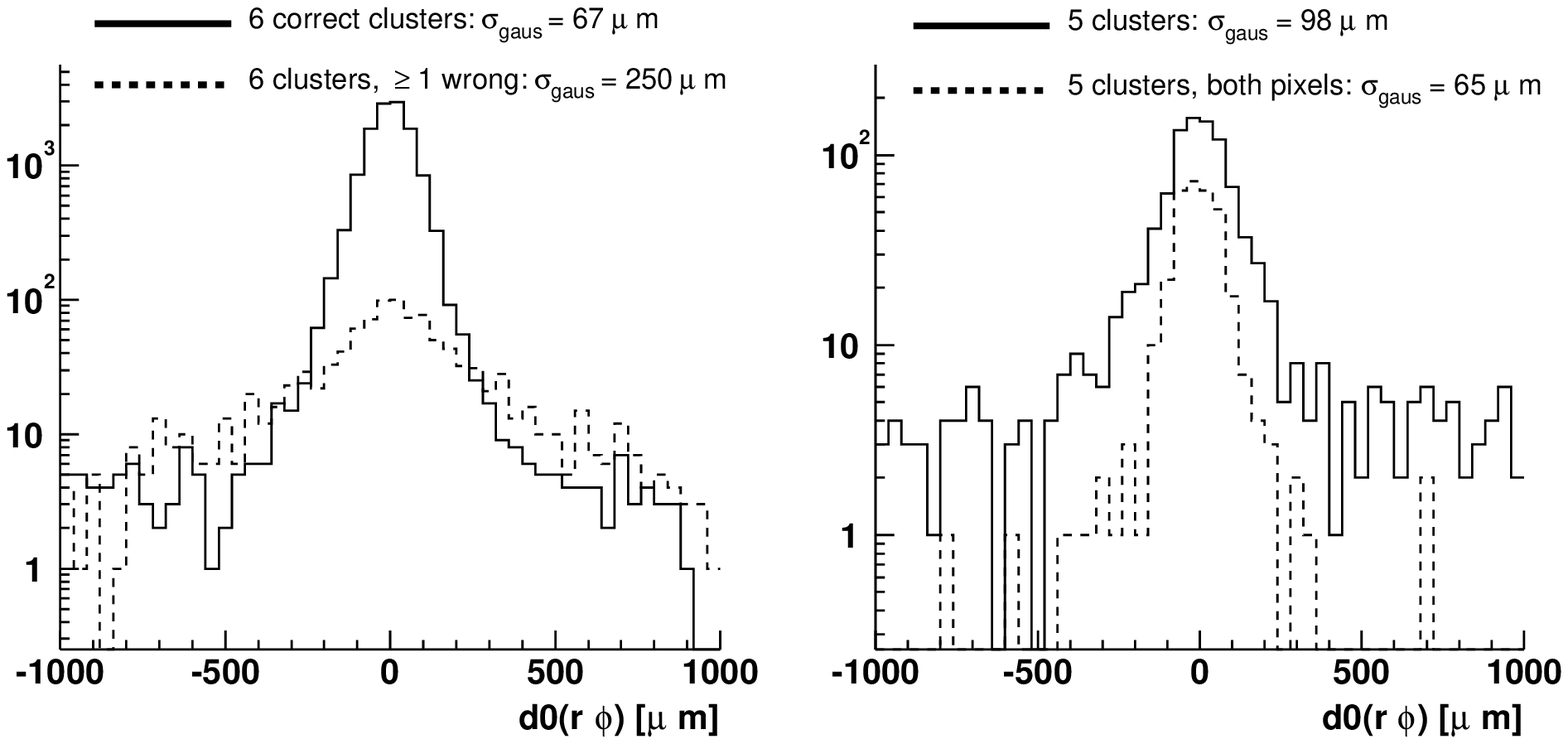}
    \caption{Distributions of the $r\phi$ impact parameter for 
             primary pions with $\pt\approx 1~\gev/c$. On the left, 
             tracks with 6 correctly assigned ITS clusters (solid) and 
             {\sl fake} tracks, with at least 1 misassigned cluster (dashed).
             On the right, tracks with 5 ITS clusters (solid) and tracks 
             with 5 clusters but with clusters in the pixel layers (dashed).
             We report the resolutions estimated with a gaussian fit.} 
    \label{fig:clusters}
    \vglue0.5cm
    \includegraphics[width=.6\textwidth]{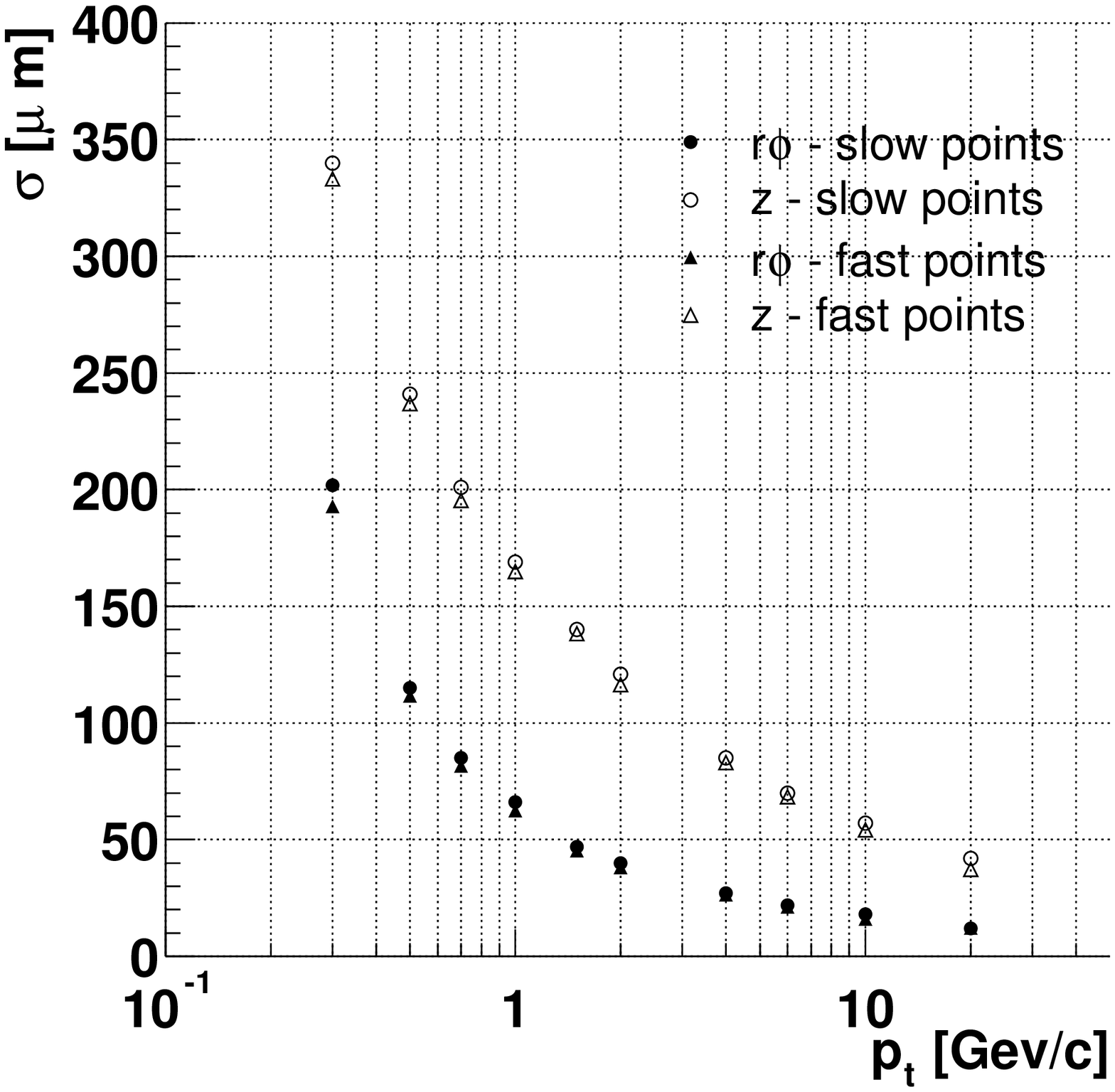}
    \caption{Comparison of the resolutions obtained with detailed 
             (circles) and with fast (triangles) response of the ITS.} 
    \label{fig:spVSfp}
  \end{center}
\end{figure}

We now consider the effect on the impact parameter resolution 
of misassigned or missing clusters in the ITS layers. We focus on 
the $r\phi$ impact parameter for pions with $\pt\approx 1~\gev/c$.
Figure~\ref{fig:clusters} (left) shows that the distribution of the impact 
parameters for {\sl fake} tracks (at least 1 misassigned cluster) is 
much broader than that for tracks with 6 correctly-assigned clusters.
However, since the fraction of fake tracks is $\simeq 5\%$ at $\pt=1~\gev/c$
and it vanishes very rapidly as $\pt$ increases (as 
shown in Section~\ref{CHAP4:trackingeff}---Fig.~\ref{fig:ITStoTPCandFakes}), 
for relatively large $\pt$ tracks, the 
effect due to misassigned clusters is very small, if 6 ITS clusters are 
required. 

As it can be seen in Fig.~\ref{fig:clusters} (right), 
for tracks with only 5 clusters in the ITS, the resolution is still 
good if there is a cluster in each of the pixel layers. Therefore, 
for the physics studies requiring an optimal impact parameter resolution, 
the loosest track quality condition is to have at least 5 clusters in the ITS
and two of these in the pixel layers. This increases the total number of 
`tracks for physics' by 8\%, with respect to requiring always all 6 points 
in the ITS; the increase is 5\% for $\pt>0.5~\gev/c$
and 4\% for $\pt>0.8~\gev/c$.  

\subsection{Comparison of detailed and fast ITS simulation}
\label{CHAP5:d0slowfast}

\begin{figure}[!t]
  \begin{center}
    \leavevmode
    \includegraphics[width=.69\textwidth]{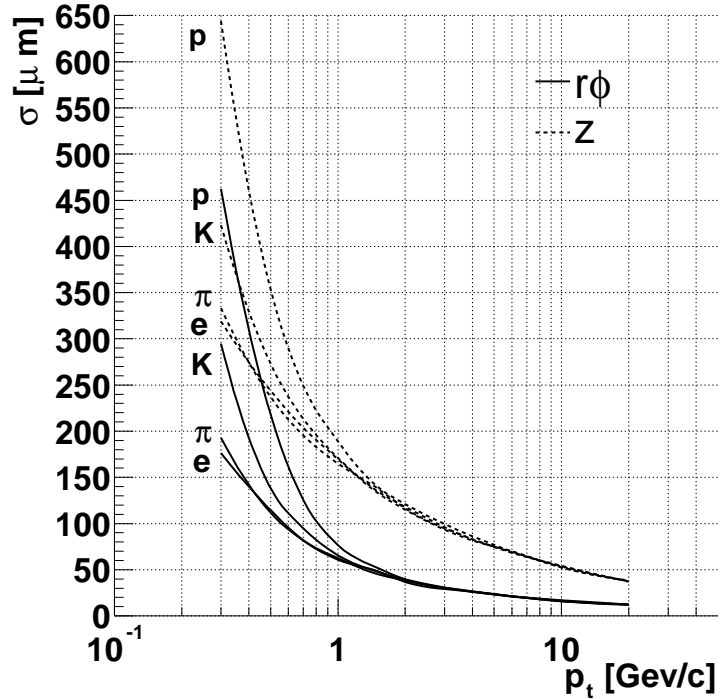}
    \caption{Impact parameter resolutions for electrons, 
             pions, kaons and protons 
             as a function of the transverse momentum.} 
    \label{fig:d0ElPiKaPr}
  \end{center}
\end{figure}

Physics simulation studies requiring very large statistics 
(e.g. open charm and open beauty feasibility studies) use a simplified 
and faster version of the detector 
response of the the ITS ({\sl fast points}). 
In this fast simulation approach 
the position of the clusters in each layer is directly obtained 
from the position of the hits (the points where the particles crossed the 
sensitive volume) applying a gaussian smearing in $r\phi$ and in $z$ 
that accounts for the detector spatial precision (the precisions used 
for the 6 layers are those reported in Table~\ref{tab:its}). 
It is, therefore, very important to observe, see Fig.~\ref{fig:spVSfp}, that 
the impact parameter resolution is essentially not altered by the use 
of the fast response of the ITS.

\subsection{Particle type dependence}
\label{CHAP5:d0VSpart}

For low momenta, the main contribution to the impact parameter resolution is 
due to the multiple scattering, which depends on $1/\beta$~\cite{pdg}.
Consequently, for a given $\pt$, the resolution itself is worse for  
heavier particles, that have lower velocity $\beta$.
Figure~\ref{fig:d0ElPiKaPr}
presents the resolutions for electrons ($e^\pm$), pions ($\pi^\pm$), 
kaons ($\rm K^\pm$) and protons ($\rm p$ and $\rm\overline{p}$).
For $\pt>1~(1.5)~\gev/c$ the resolutions for kaons (protons) 
are the same as for pions. 

The separation at low $\pt$ between pions and electrons is not well defined, 
because the latter can suffer from energy loss due to the bremsstrahlung
process; even if the probability is quite low ($\sim 1\%$ at 
$\pt=1~\gev/c$), this spoils both the momentum and the impact parameter 
resolution.

\mysection{Primary vertex reconstruction in pp}
\label{CHAP5:vtxpp}

\subsection{Outline of the method} 

After track reconstruction in TPC and ITS, 
all tracks are propagated to the nominal position of the interaction vertex, 
which, during the proton--proton running, will be given by the machine 
with a resolution of $\sim 100$-$200~\mum$ (Section~\ref{CHAP4:lhcpp}). 

The reconstruction in three dimensions of the primary vertex position 
by means of the reconstructed tracks is performed in two steps:
\begin{enumerate}
  \item {\sl Vertex finding}: a first estimate of the vertex position is 
        obtained using track pairs.
  \item {\sl Vertex fitting}: tracks are propagated to the position 
        estimated in the first step and the optimal estimate of the vertex 
        position, as well as the vertex covariance matrix and a `quality 
        parameter' ($\chi^2$), are obtained via a fast fitting algorithm.
        In this step a cut on the maximum contribution to the total $\chi^2$
        is applied in order to remove secondary tracks from the fit. 
\end{enumerate}

The algorithm was tested on a sample of 5000 minimum-bias pp events 
at $\sqrt{s}=14~\tev$ (without single- and 
double-diffractive topologies) generated with PYTHIA, as described 
in Section~\ref{CHAP4:pythia}.
The generated vertex position was sampled along $z$ from a Gaussian centred
 at $z=0$ with $\sigma_z =5.3~\cm$. The sampling was limited in the region 
$-\sigma_z < z < +\sigma_z$. 
The coordinates in the bending plane were sampled from 2 Gaussians with 
dispersions $\sigma_x = \sigma_y = 150~\mum$. 
In order to simulate a beam-offset condition, 
the two Gaussians were centred at $x_0=y_0=5~\mm$.

\subsection{Expected resolutions}
\label{CHAP5:expected}

The attainable resolution on the vertex position
can be estimated on the basis of the resolutions on the track 
position in the bending plane and in the longitudinal direction, which 
are shown in Fig.~\ref{fig:d0resMain}.

The average number of reconstructed tracks in pp non-diffractive events 
is $\av{N}\simeq 7$ and the average value of the transverse momentum 
for these tracks is $\av{\pt}\simeq 0.6~\gev/c$.
Therefore, the expected 
resolutions on the $x$ (for $y$ it is the same) and on the $z$ 
coordinate of the vertex are:
\begin{equation}
\begin{array}{l}
  \sigma_{\rm vertex}(x) = \sigma_{\rm track}(r\phi)_{\pt=0.6~\gev/c}\bigg/\sqrt{\av{N}/2} \simeq 110~\mum\bigg/\sqrt{3.5} = 60~\mum \\
  \sigma_{\rm vertex}(z) = \sigma_{\rm track}(z)_{\pt=0.6~\gev/c}\bigg/\sqrt{\av{N}} \simeq 240~\mum\bigg/\sqrt{7} = 90~\mum.
\end{array}
\end{equation}
For the $x$ (and $y$) coordinate the number of contributing tracks is taken 
equal to $\av{N}/2$, since, in the transverse plane, the information from 
$\av{N}$ tracks is used to determine 2 coordinates.  
  
\subsection{Vertex finding}
\label{CHAP5:vertfinder}

The reconstructed tracks are propagated to the 
radial position corresponding to the nominal vertex 
position in the bending plane. Such position is expected to be known 
with a precision of the order of $100$-$200~\mum$, however the vertex finding 
procedure should succeed even though the nominal position is known with a 
poorer accuracy.
The nominal position used in the test is $(x_{\rm nom}=0,y_{\rm nom}=0)$, 
i.e. $\approx 7~\mm$ far from the generated position,
in order to check the performance of the vertex recognition 
algorithm in a condition far worse than expected.

A loose cut ($3~\cm$) on the transverse impact parameter of the tracks is 
applied in order to exclude particles originating from secondary vertices 
well displaced with respect to the primary vertex. 

Each track is approximated with the straight line tangent to the 
reconstructed helix at the nominal vertex position. Then, all possible 
track pairs $(i,j)$ are considered and, for each pair, the centre 
$C(i,j)\equiv(x_{ij},y_{ij},z_{ij})$ of the segment of minimum approach 
between the two lines\footnote{i.e. the segment perpendicular to 
both lines.} is found. The coordinates of the primary vertex 
are determined as:
\begin{equation}
q_{\rm found}={1\over N_{\rm pairs}}\sum_{i,j}q_{ij}~~~~~~~~~~~~~~~~q = x,~y,~z
\end{equation}
where $N_{\rm pairs}$ is the number of track pairs.

The obtained resolution on the position of the vertex is of order $100~\mum$
for all three coordinates~\cite{vtxpp}. 
The results are actually independent of the value used 
for the nominal position 
of the primary vertex, when this is in the range of few millimeters with 
respect to the true position.

\subsection{Vertex fitting}

The vertex finding algorithm provides, as described, a first estimate of the 
vertex position and propagates track parameters to this position. 
The task of the vertex fitting algorithm is to determine the best fit 
coordinates of the vertex and the vertex covariance matrix.
We have implemented this step on the basis of the fast vertex fitting 
method described in Ref.~\cite{cmsvtxnote}.

\subsubsection{Vertex fitting algorithm}

Since the measurements of different tracks are independent of each other, 
the $\chi^2$ function to be minimized can be written as a sum over tracks.
In Ref.~\cite{cmsvtxnote} it is shown that, if the tracks can be approximated 
to straight lines in the vicinity of the vertex position, the $\chi^2$ is:
\begin{equation}
  \label{eq:chi2}
  \chi^2(\vec{r}_{\rm vertex})=\sum_i (\vec{r}_{\rm vertex}-\vec{r}_i)^T\,{\bf V}_i^{-1}(\vec{r}_{\rm vertex}-\vec{r}_i).
\end{equation}
In this expression, $\vec{r}_i$ is the current position of the 
track $i$ (i.e. the position given by the vertex finder) 
and ${\bf V}_i$ is the covariance matrix of the vector $\vec{r}_i$, 
obtained from the covariance matrix of the track.

The approximation of the track to a straight line allows to neglect, 
in the covariance matrix of $\vec{r}$, the contribution of the elements 
of the track covariance matrix relative to the curvature and direction
parameters. This simplifies the calculation, as detailed in 
Refs.~\cite{vtxpp,cmsvtxnote}. 
We will now verify that this approximation holds in our 
particular case.

{\small
Since the tracks are propagated by the vertex finder to the first estimate 
of the vertex position, which is determined with a resolution 
$\sigma\sim 100~\mum$ in the bending plane, the length over which we 
neglect the curvature and the changes in the direction parameters is 
of the same order of magnitude; however we consider a safety factor of 10 
and we estimate the effects of the linear approximation over a length 
$\ell\sim 1~\mm$. The sagitta of the arc with cord $\ell$ and with radius of 
curvature $R=1~\m$ is $\ell^2/8R\simeq 0.125~\mum$. In ALICE tracks with 
$R<1.5~\m$, that do not cross the whole TPC volume, are reconstructed with 
lower quality and will be excluded from vertexing studies.
}

Given that the matrix ${\bf V}$ is independent of $\vec{r}_{\rm vertex}$, 
the expression (\ref{eq:chi2}) is a linear function of $\vec{r}_{\rm vertex}$. 
The solution 
for the vertex coordinates which minimize (\ref{eq:chi2}) reads then:
\begin{equation}
  \vec{r}_{\rm vertex}=\left(\sum_i {\bf W}_i\right)^{-1}\,\sum_i {\bf W}_i\,\vec{r}_i
\end{equation} 
with ${\bf W}_i={\bf V}_i^{-1}$ and the covariance matrix of 
$\vec{r}_{\rm vertex}$ is
\begin{equation}
  {\bf C}_{\rm vertex}=\left(\sum_i {\bf W}_i\right)^{-1}.
\end{equation} 

\subsubsection{Optimization of the algorithm: rejection of tracks with 
  large contribution to the total $\chi^2$}

\begin{figure}[!ht]
  \begin{center}
    \leavevmode
    \includegraphics[width=0.65\textwidth]{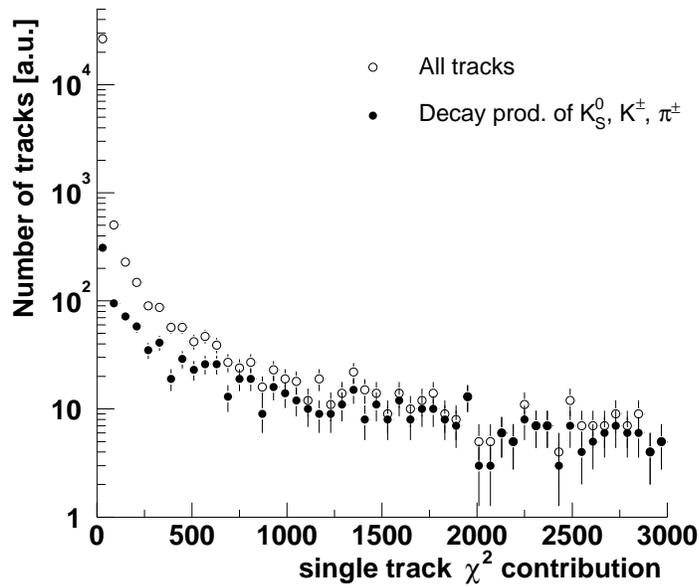}
    \caption{Distribution of single-track $\chi^2$-contributions, for
             all tracks (open circles) and tracks coming from decays 
             of $\rm K^0_S,~K^\pm,~\pi^\pm$ (full circles).} 
    \label{fig:chi2perTrack}
  \end{center}
\end{figure}

The resolution of the fitted position of the primary vertex might be 
spoiled by the presence of tracks that had large scatterings in the material
or tracks coming from decays, in particular decays of strange particles, 
which have relatively large life-times (e.g. $c\tau\simeq 2.7~\cm$ for 
$\rm K^0_S$). Typically, these tracks bring large contributions to the global 
$\chi^2$ of the vertex; in Fig.~\ref{fig:chi2perTrack} we show the 
distribution of single-track $\chi^2$-contributions for all tracks 
and for tracks coming from decays; the latter are clearly dominating 
in the large-$\chi^2$ region.

In order to minimize the loss of resolution due to these effects,
we apply a cut on the maximum contribution of a single track to the total 
$\chi^2$ of the fit, calculated as in (\ref{eq:chi2}). We proceed as follows:
the fit is initially performed using all tracks and the $\chi^2$-contribution 
of each track w.r.t. the result of the fit is calculated; 
then the tracks whose contribution exceeds some value $\xi$ are removed 
and the fit is repeated. The process is iterated until no further tracks 
are removed or less than 3 tracks are left. 

\begin{figure}[!t]
  \begin{center}
    \leavevmode
    \includegraphics[width=\textwidth]{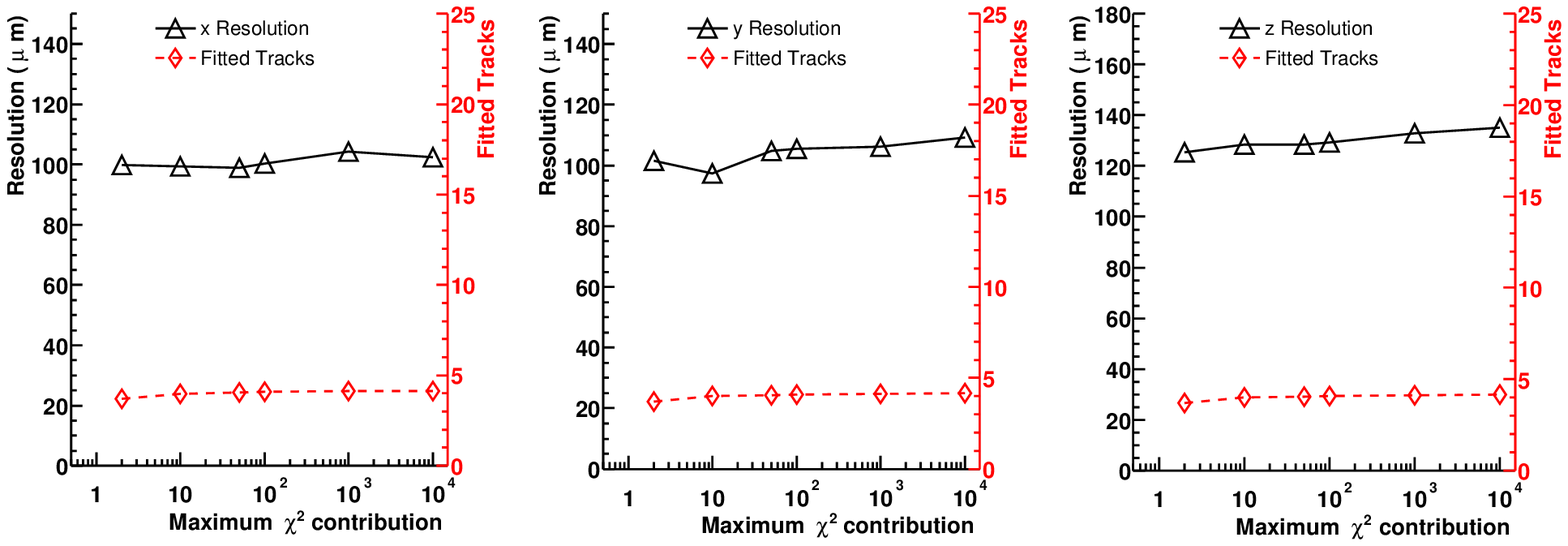}
    \vglue0.5cm
    \includegraphics[width=\textwidth]{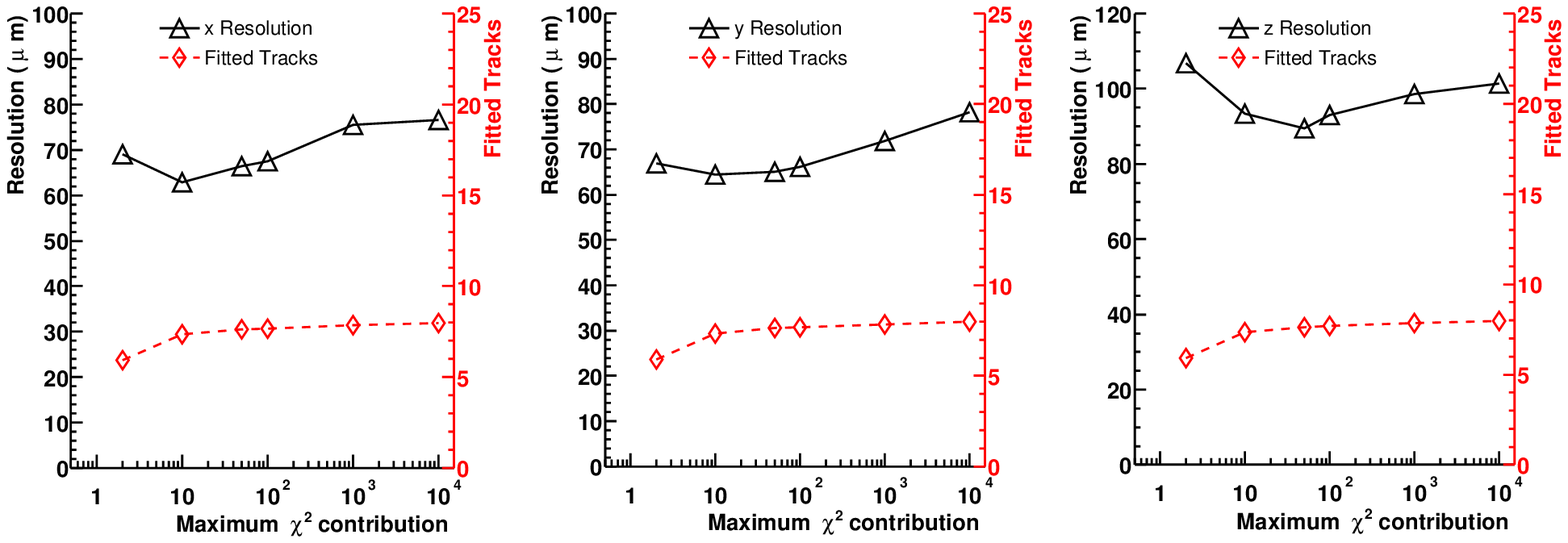}
    \vglue0.5cm
    \includegraphics[width=\textwidth]{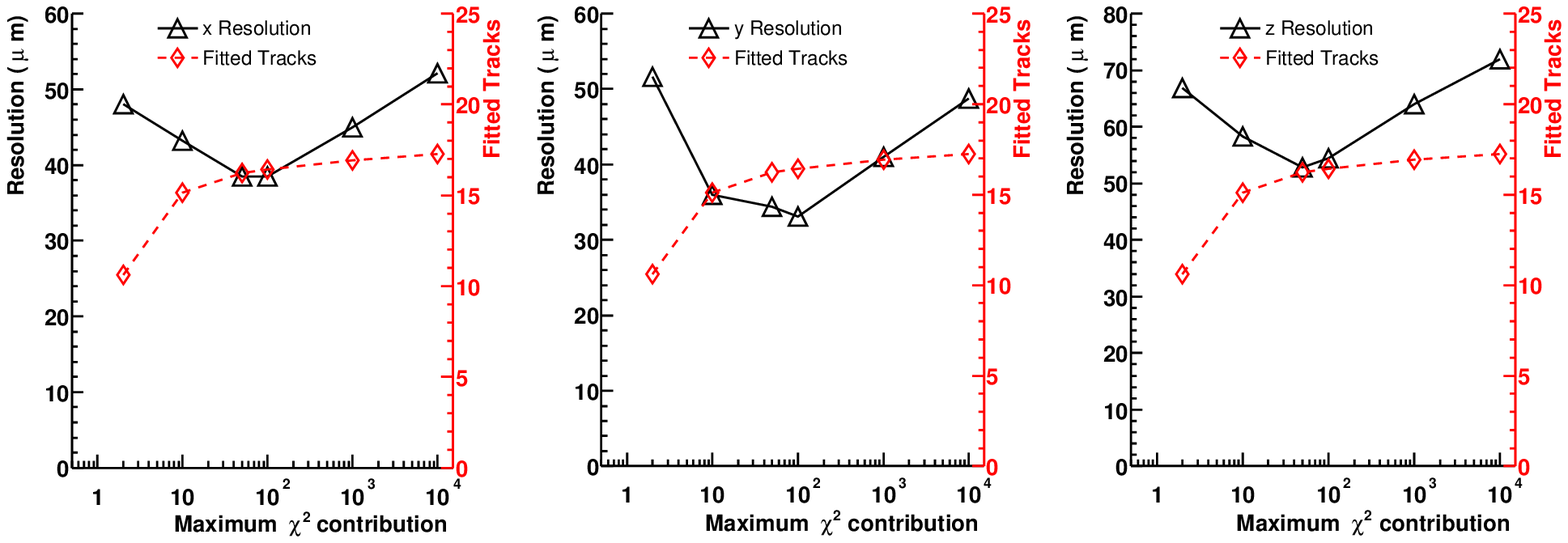}
    \caption{Tuning of the cut on the maximum $\chi^2$-contribution. Each 
             plot presents the resolution on one of the three coordinates 
             of the vertex ($x$, $y$ and $z$ from the left to the right)  
             and the mean number of tracks included in the fit 
             as a function of the cut. Top row: low-multiplicity events;
             central row: medium-multiplicity events; bottom row:
             high-multiplicity events.} 
    \label{fig:chi2cut}
  \end{center}
\end{figure}

{\small
The cut on $\xi$ has to be optimized in order to maximize the resolution 
of the fit. Indeed, if the cut is too loose, secondary or scattered 
tracks enter the fit and the resolution degrades; if the cut is too tight, 
primary tracks are removed and statistical precision is lost. 
The tuning has been done as a function of the number of reconstructed 
tracks in the event. The results are displayed in Fig.~\ref{fig:chi2cut},
where we present the resolutions on the three coordinates of the vertex 
together with the mean number of tracks used in the fit, as a function of the 
cut on $\xi$; as it can be seen from the number of tracks, in the top row 
we select low-multiplicity events, in the central row medium-multiplicity 
events and in the bottom row high-multiplicity events. At low 
multiplicity there is no dependence of the resolutions on the value of the 
cut, since there are essentially no secondary tracks in such events; 
while as the multiplicity increases, we observe the expected trend and 
the resolutions show a clear minimum.    
}

\subsubsection{Results for 3D vertex reconstruction}

We tested the algorithm on the sample of pp events already described.  
Requiring a minimum of 3 tracks in 
the vertex fit, the vertex was reconstructed in about 65\% of the events 
of the sample. It was verified that essentially all the events in 
which the vertex was not reconstructed had less than 3 found tracks
(see Ref.~\cite{vtxpp}). 
The efficiency for vertex reconstruction as a function of 
event multiplicity will be detailed in the following.

\begin{figure}[!t]
  \begin{center}
    \leavevmode
    \includegraphics[width=\textwidth]{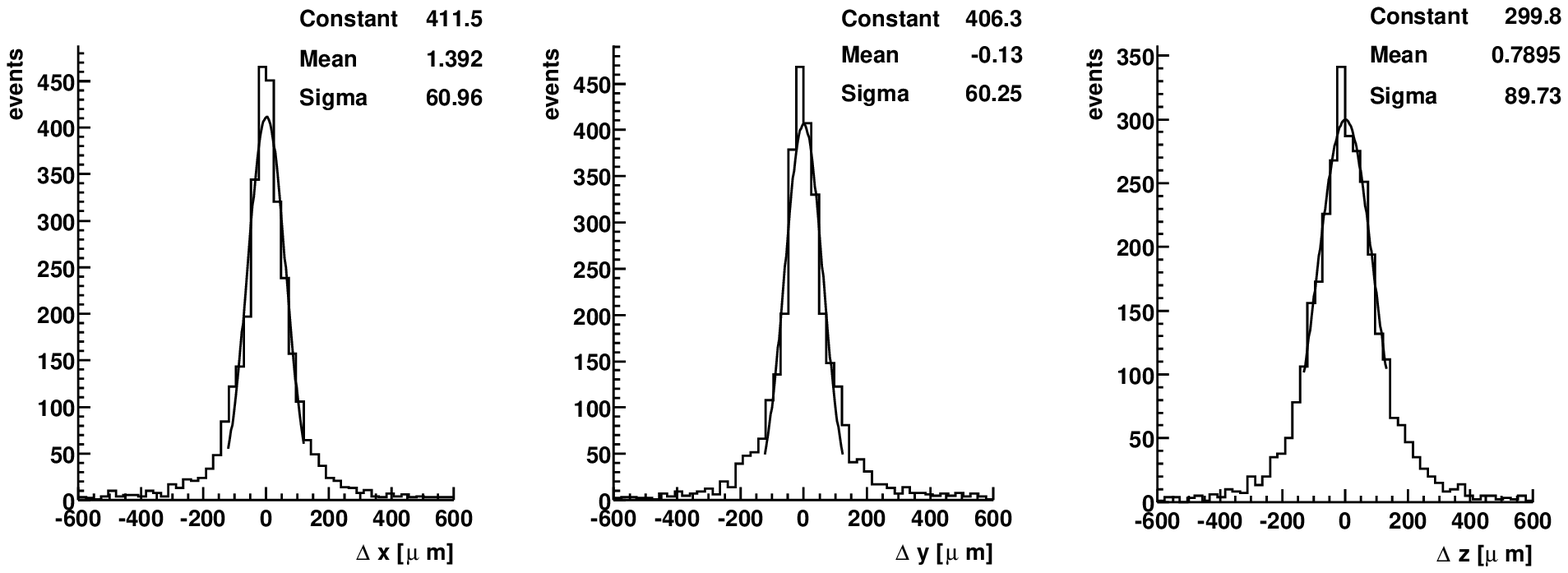}
    \caption{Distributions of the residuals: 
             $\Delta q = q_{\rm measured}-q_{\rm true}$, for $q=x,\,y,\,z$.} 
    \label{fig:resXYZ}
    \vglue0.2cm
    \includegraphics[width=\textwidth]{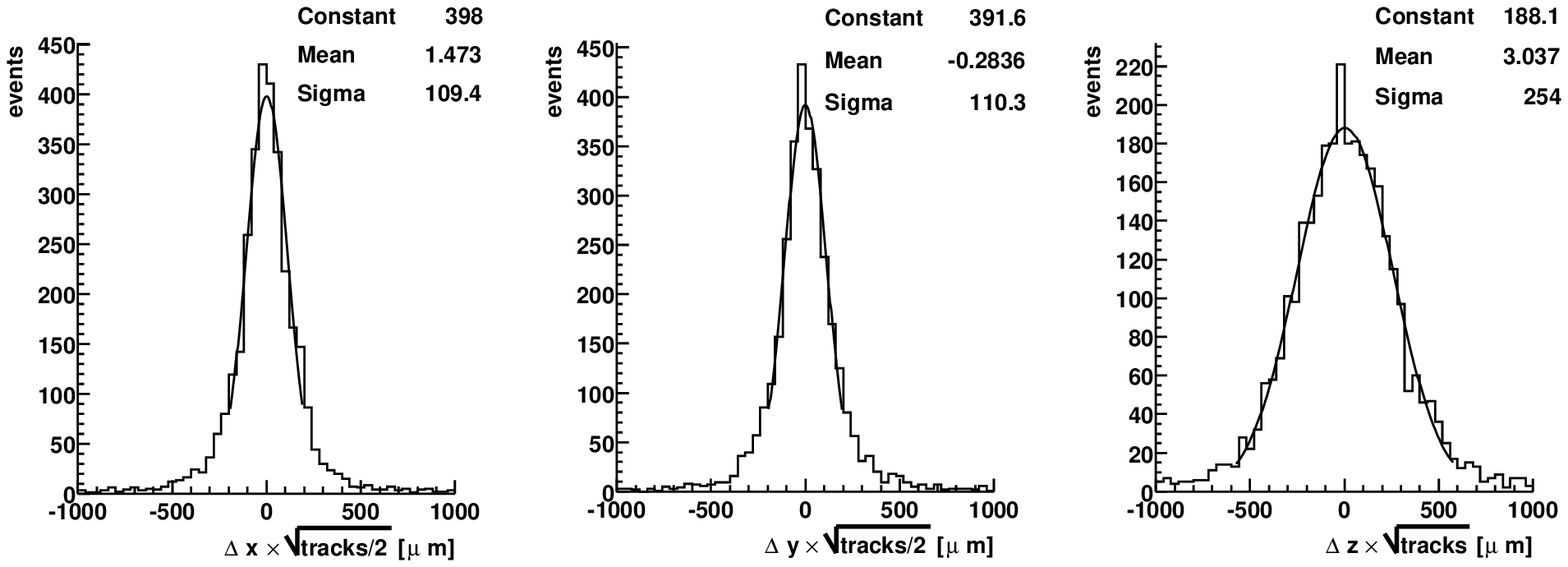}
    \caption{Distributions of $\Delta x\times\sqrt{N_{\rm tracks}/2}$,
             $\Delta y\times\sqrt{N_{\rm tracks}/2}$ 
             and $\Delta z\times\sqrt{N_{\rm tracks}}$. The width of these
             distributions gives the ``resolution corresponding to 1 track''.} 
    \label{fig:res1trackXYZ}
    \vglue0.2cm
    \includegraphics[width=\textwidth]{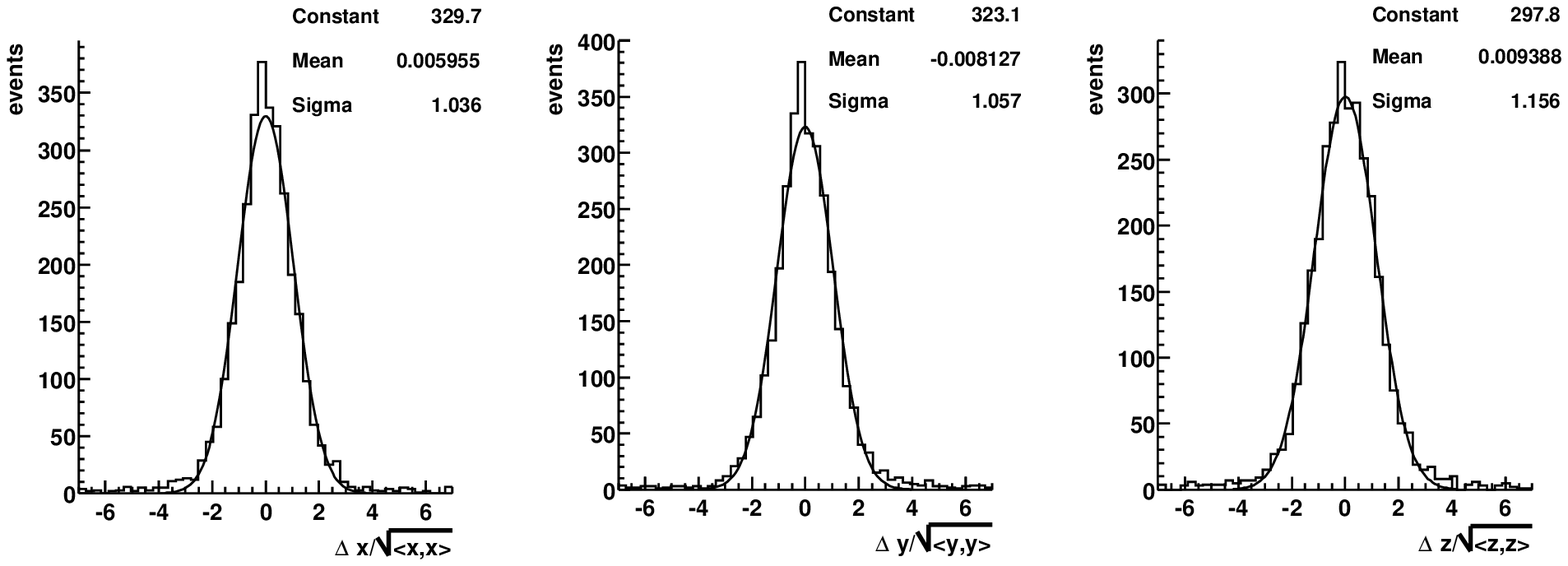}
    \caption{Distributions of the standardized residuals.} 
    \label{fig:pullsXYZ}
  \end{center}
\end{figure}

Figure~\ref{fig:resXYZ} presents the residuals $\Delta q$ for the three vertex 
coordinates ($q=x,~y,~z$), integrated over the full statistics. 
The distributions 
are clearly non-gaussian, since they are the convolution of many gaussian 
distributions with different dispersions, depending on the number of 
tracks used in each event for the fit. However, we fitted with a Gaussian 
the central part of the distributions, in order to quantify the global 
resolution. We obtain $\sigma_x \simeq\sigma_y\simeq 60~\mum$ and 
$\sigma_z\simeq 90~\mum$, in fair agreement with the expected values
(Section~\ref{CHAP5:expected}).
Such agreement emerges also from the fit of the distributions of 
$\Delta x\times\sqrt{N_{\rm tracks}/2}$ and 
$\Delta z\times\sqrt{N_{\rm tracks}}$, 
which should give the `resolution corresponding to 1 track' 
(Fig.~\ref{fig:res1trackXYZ}).

We checked the reliability of the estimated errors on the vertex position 
with the test of the pulls. The distribution of the standardized residuals, 
defined as $residual/\sqrt{variance}=\Delta q/\sqrt{\av{q,q}}$ 
for $q=x,\,y,\,z$, are normal (mean $\simeq 0$, $\sigma\simeq 1$), 
as shown in Fig.~\ref{fig:pullsXYZ}; 
therefore, we conclude that the errors given in the 
vertex covariance matrix describe correctly 
the resolution on the vertex estimate.

These global results are summarized in Table~\ref{tab:globresults}.

\begin{table}[!h]
  \caption{Summary table of the results on the vertex fit, integrated over all 
           events.}
\begin{center}
  \begin{tabular}{ll|ccc}
  \hline
  \hline
  Parameter && $x$ & $y$ & $z$ \\ 
  \hline
  Resolution & [$\mum$] & 61 & 60 & 90 \\
  Resolution per track & [$\mum$] & 109 & 110 & 254 \\
  Pull && 1.04 & 1.06 & 1.16 \\
  \hline
  \hline
  \end{tabular}
  \label{tab:globresults}
\end{center}
\end{table}

The sample was subdivided into five multiplicity ($\dNdy$) 
classes in order to study the 
resolution and the efficiency of the method as a function of event 
multiplicity. The multiplicity was 
computed in the central pseudorapidity unit counting charged pions, 
charged kaons, protons, electrons and muons originating within 
$100~\mum$ from the primary vertex.

\begin{figure}[!t]
  \begin{center}
    \leavevmode
    \includegraphics[width=0.7\textwidth]{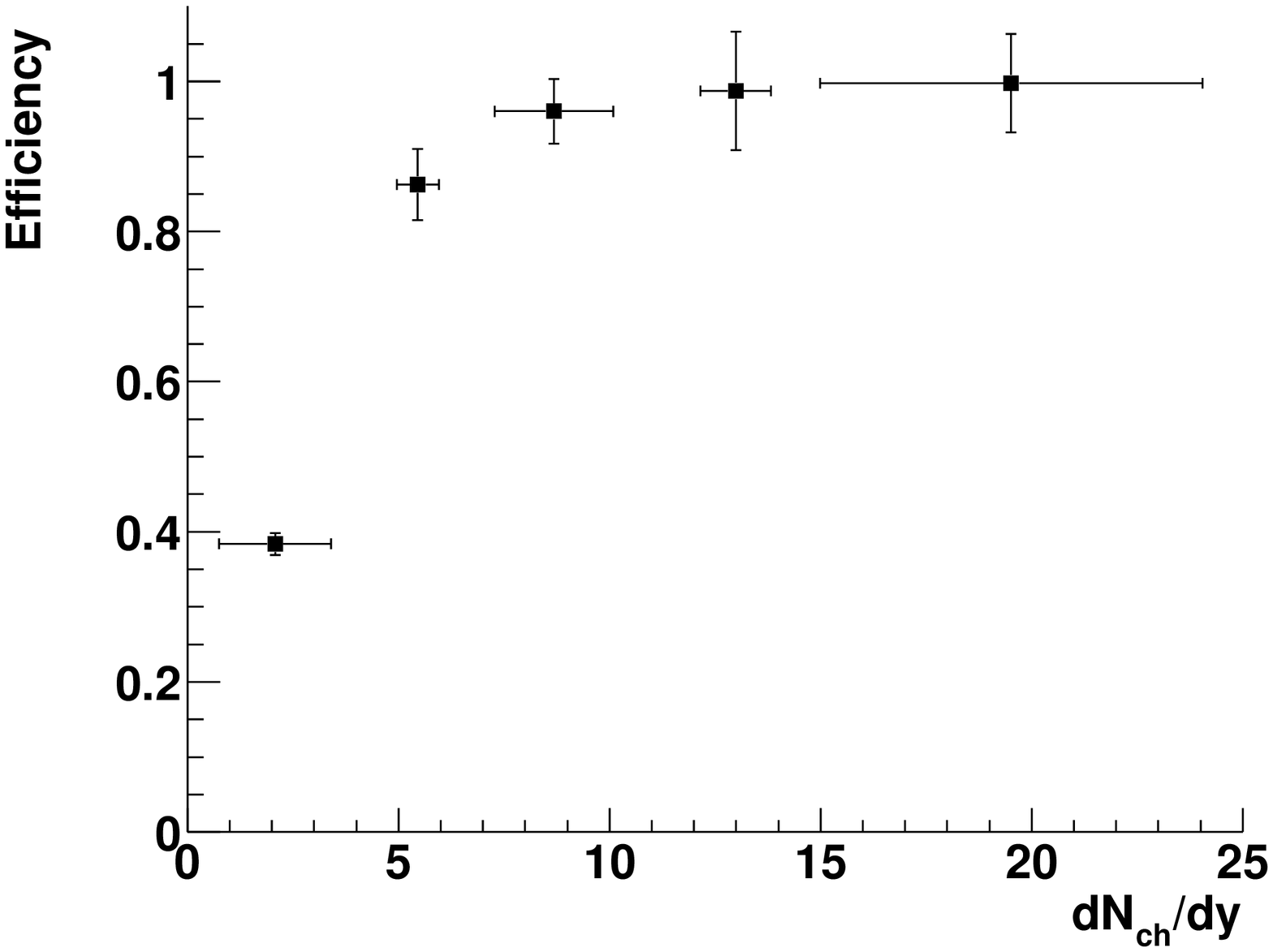}
    \caption{Fraction of events for which the vertex is reconstructed 
             from the tracks, as a function of event multiplicity.} 
    \label{fig:probXYZ}
    \vglue0.5cm
    \includegraphics[width=0.8\textwidth]{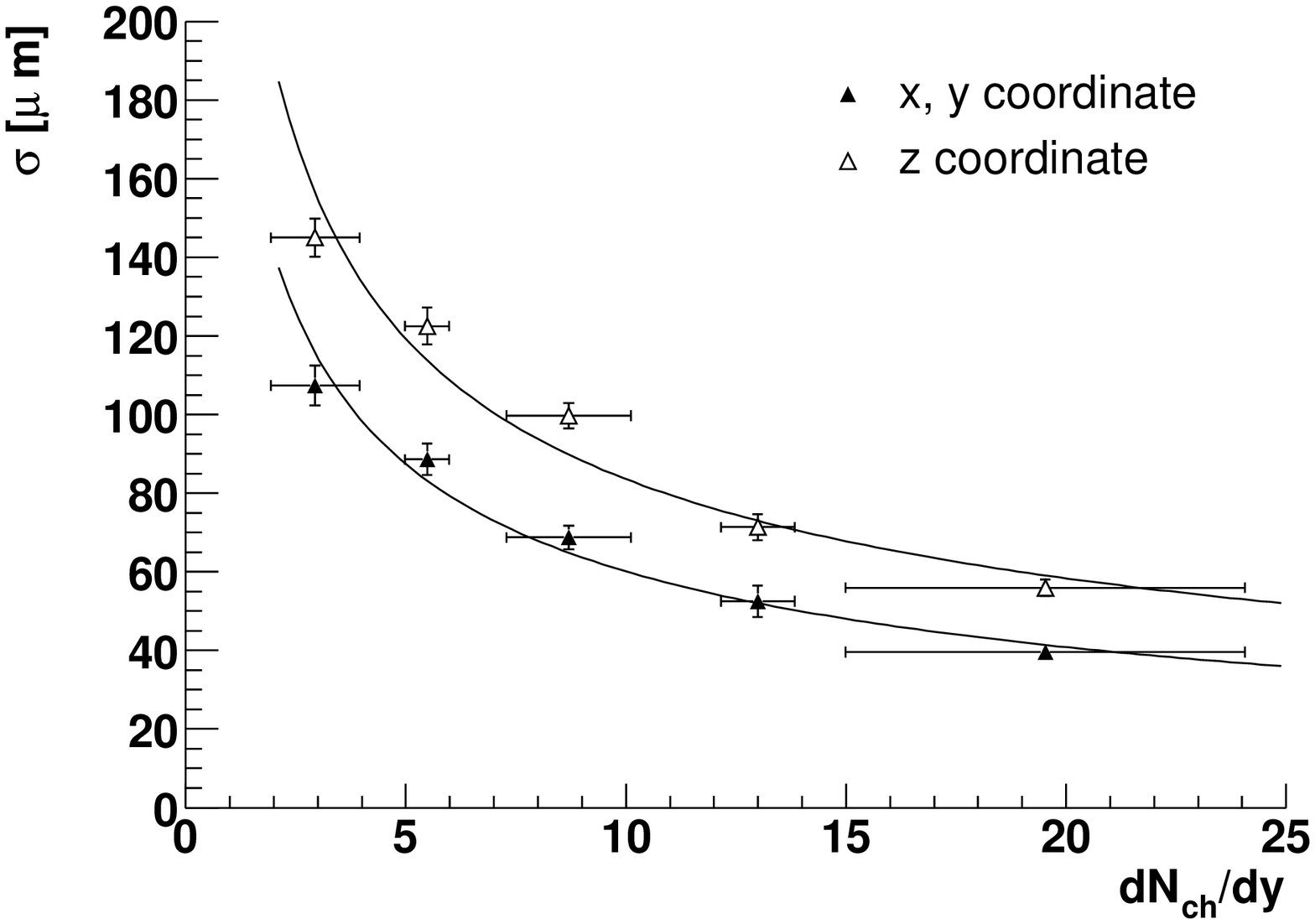}
    \caption{Resolutions on vertex position in $x$ and $z$ as a function 
             of event multiplicity. The trends have been fitted to the 
             expression $a+b/\sqrt{\dNdy}$ and the fitted parameters are 
             reported in Table~\ref{tab:fitXYZresults}.} 
    \label{fig:resXYZfit}
  \end{center}
\end{figure}

The efficiency for vertex reconstruction 
(defined as the ratio of the number of events with vertex found to 
the total number of events, in each bin) grows as a function of the 
multiplicity and it is saturated at 1 for $\dNdy>10$ (Fig.~\ref{fig:probXYZ}).



The resolutions as a function of the event multiplicity 
are shown in Fig.~\ref{fig:resXYZfit}. As done for the $z$ measurement 
with the pixels, the resolutions were fitted to the 
expression:
\begin{equation}
  \sigma(\dNdy)=a+b\bigg/\sqrt{\dNdy}.
\end{equation}  
The results of the fit are reported in Table~\ref{tab:fitXYZresults}.  

\begin{table}[!t]
  \caption{Results of the fit of the multiplicity dependence of the 
           resolutions to the expression $a+b/\sqrt{\dNdy}$.} 
\begin{center}
  \begin{tabular}{cc|cc}
  \hline
  \hline
  Parameter && $x,~y$ & $z$ \\ 
  \hline
  $a$ & $[\mum]$ & $-6\pm 4$ & $-3\pm 4$ \\
  $b$ & $[\mum]$ & $208\pm 13$ & $272\pm 13$ \\
  \hline
  \hline
  \end{tabular}
  \label{tab:fitXYZresults}
\end{center}
\end{table}

\begin{figure}[!t]
  \begin{center}
    \leavevmode
    \includegraphics[width=0.9\textwidth]{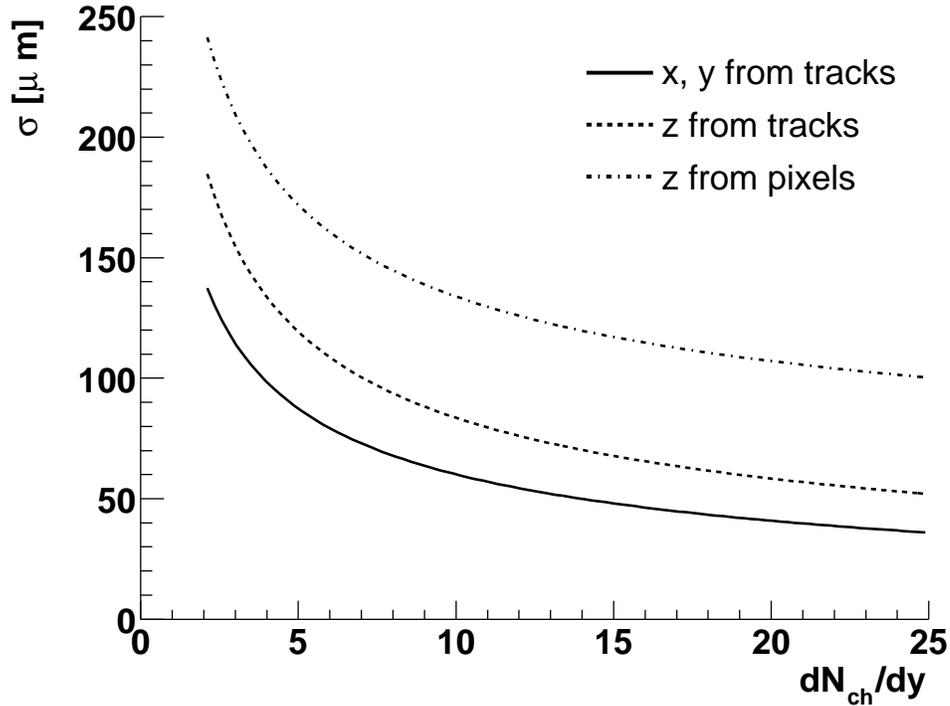}
    \caption{Summary plot for the achieved resolutions as a function of 
             multiplicity.} 
    \label{fig:summary}
  \end{center}
\end{figure}

We summarize the results for the reconstruction of the interaction vertex
in pp collisions in Fig.~\ref{fig:summary}, where the resolutions achieved 
for the $z$ coordinate with pixels, for the $z$ and $x$ ($y$) coordinates 
with tracks are displayed as a function of event multiplicity. The 
improvement on the $z$ resolution when using the tracks is of 
$\simeq 50~\mum$, but it increases at high $\dNdy$ as 
the average $\pt$ of the reconstructed tracks increases with the multiplicity, 
and hence their average position resolution improves.  

\mysection{Track impact parameter resolution in pp}
\label{CHAP5:d0pp}

Given the results presented in the former section, 
in proton--proton the impact parameter resolution has
a significant contribution from the uncertainty on the primary vertex
position, which is, on average, about one order of magnitude larger than in 
the \PbPb~case.

\begin{figure}[!t]
  \begin{center}
    \includegraphics[width=.55\textwidth]{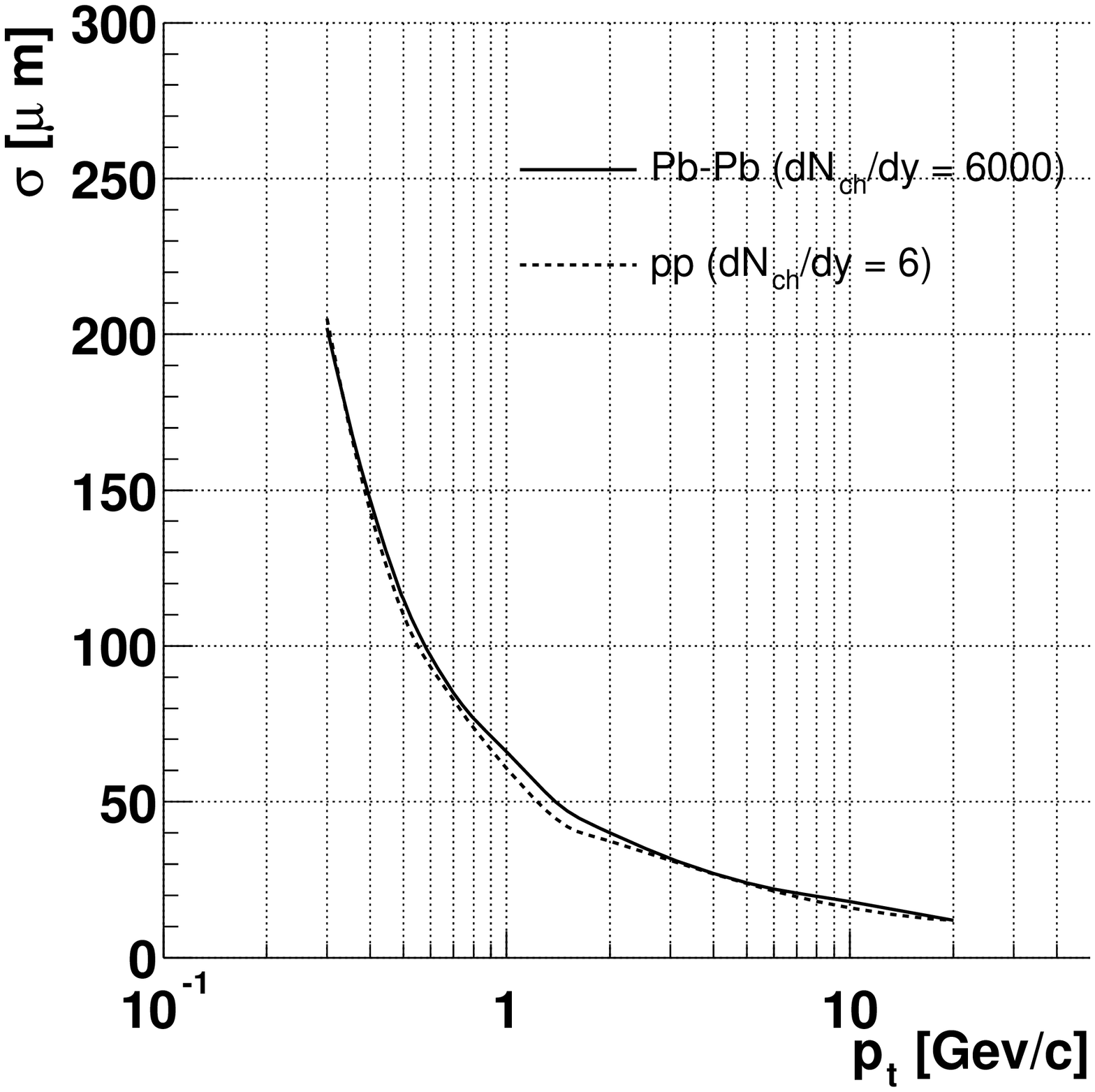}
    \caption{Resolution on the track position in the transverse plane 
             in \PbPb~and in pp for pions.} 
    \label{fig:d0ppPbPb}
    \includegraphics[width=0.8\textwidth]{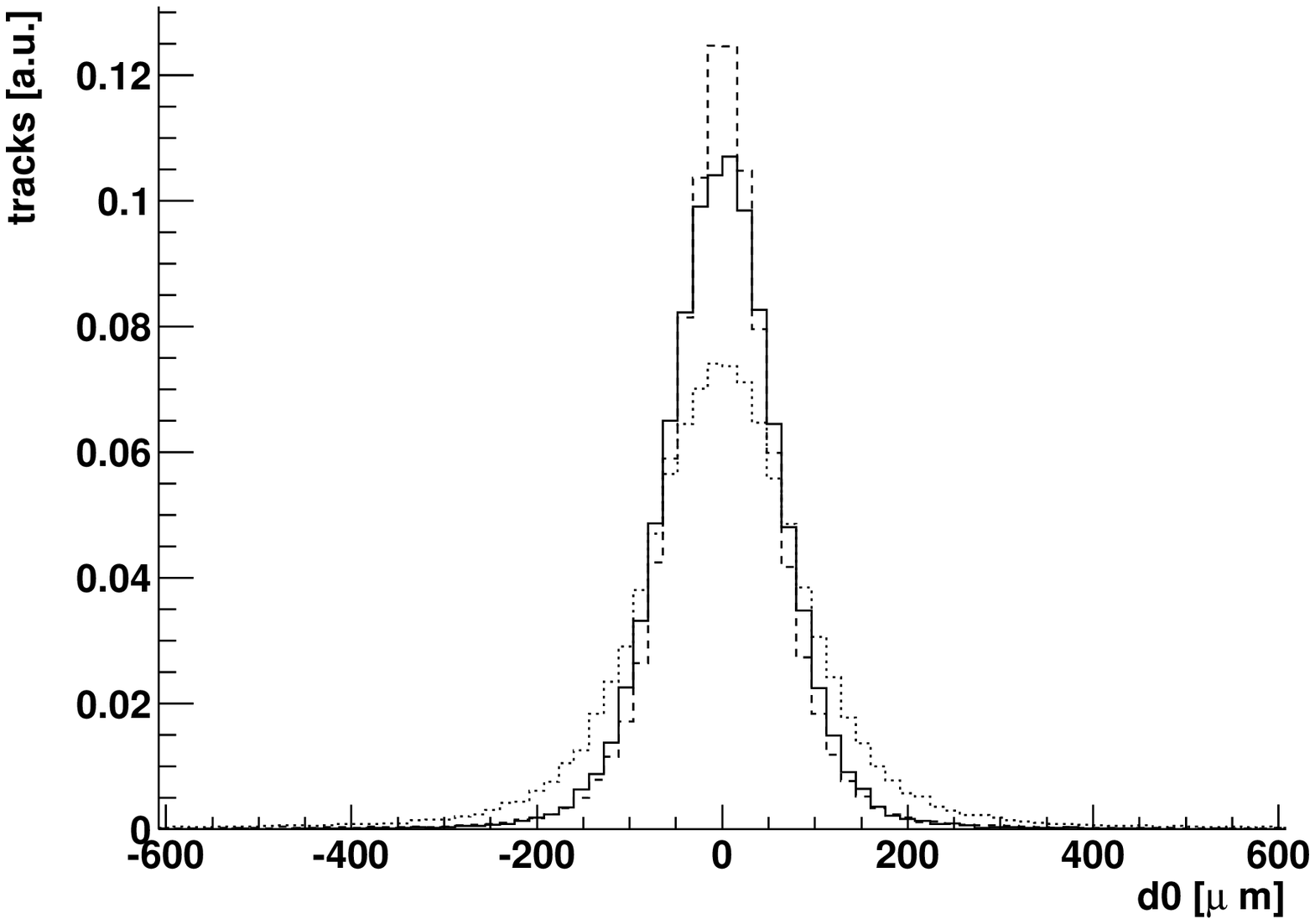}
    \caption{Distributions of the impact parameters for primary pions 
             with \mbox{$\pt\approx 1~\gev/c$} obtained: 
            using true vertex position 
            (solid), using vertex estimated from all tracks in the event 
            (dashed), using vertex estimated from all tracks but the 
            current one (dotted).} 
    \label{fig:d0shrink}
  \end{center}
\end{figure}

The resolution on the track position in the transverse plane
---the main contribution to the impact parameter resolution--- is 
essentially the same in pp and in \PbPb~if 6 ITS clusters are required 
(Fig.~\ref{fig:d0ppPbPb}).
This is not surprising, since we have already verified that the 
effect due to the presence of fake tracks in \PbPb~is quite small
(Section~\ref{CHAP5:d0VSclusters}). 

Let us now concentrate on the impact parameter resolution in the transverse 
plane ($r\phi$).
The aim of the measurement of the tracks impact parameter is 
the identification of one or more displaced tracks with respect to the 
interaction vertex. Therefore, in pp collisions, we adopt the 
following strategy for the measurement of the impact parameter:
the impact parameter of a given track $j$ is estimated as the distance of 
closest approach of the track $j$ to the vertex position obtained by 
excluding the track $j$ from the vertex reconstruction. In fact, 
if the track $j$ was included in the reconstruction, it would bias 
the vertex position, leading to a systematic underestimation of the 
impact parameter. This effect is shown in Fig.~\ref{fig:d0shrink}, 
where the following distributions are compared for primary pions 
with $\pt\approx 1~\gev/c$: impact parameter using true vertex position 
(solid), impact parameter using vertex estimated from all tracks (dashed), 
impact parameter using vertex estimated from all tracks but $j$ 
(dotted). In the case of the $\Dz\to\K^-\pi^+$ vertex reconstruction, 
as we will detail in Chapter~\ref{CHAP6}, when we consider a given 
{\sl pair} of tracks as a $\Dz$ candidate, we exclude {\sl both} of them 
from the fit of the primary vertex.

\begin{figure}[!t]
  \begin{center}
    \leavevmode
    \includegraphics[width=.49\textwidth]{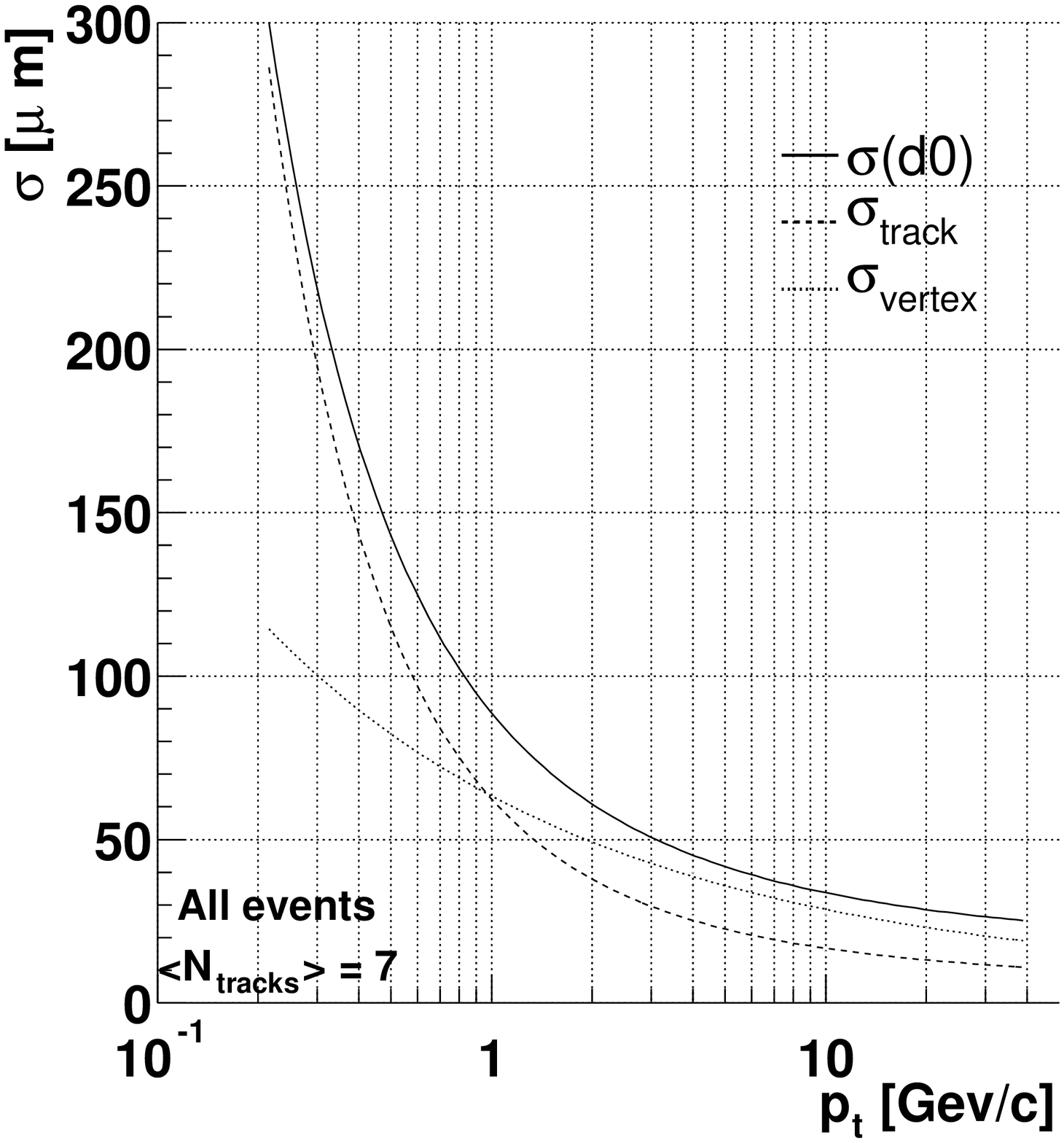}
    \includegraphics[width=.49\textwidth]{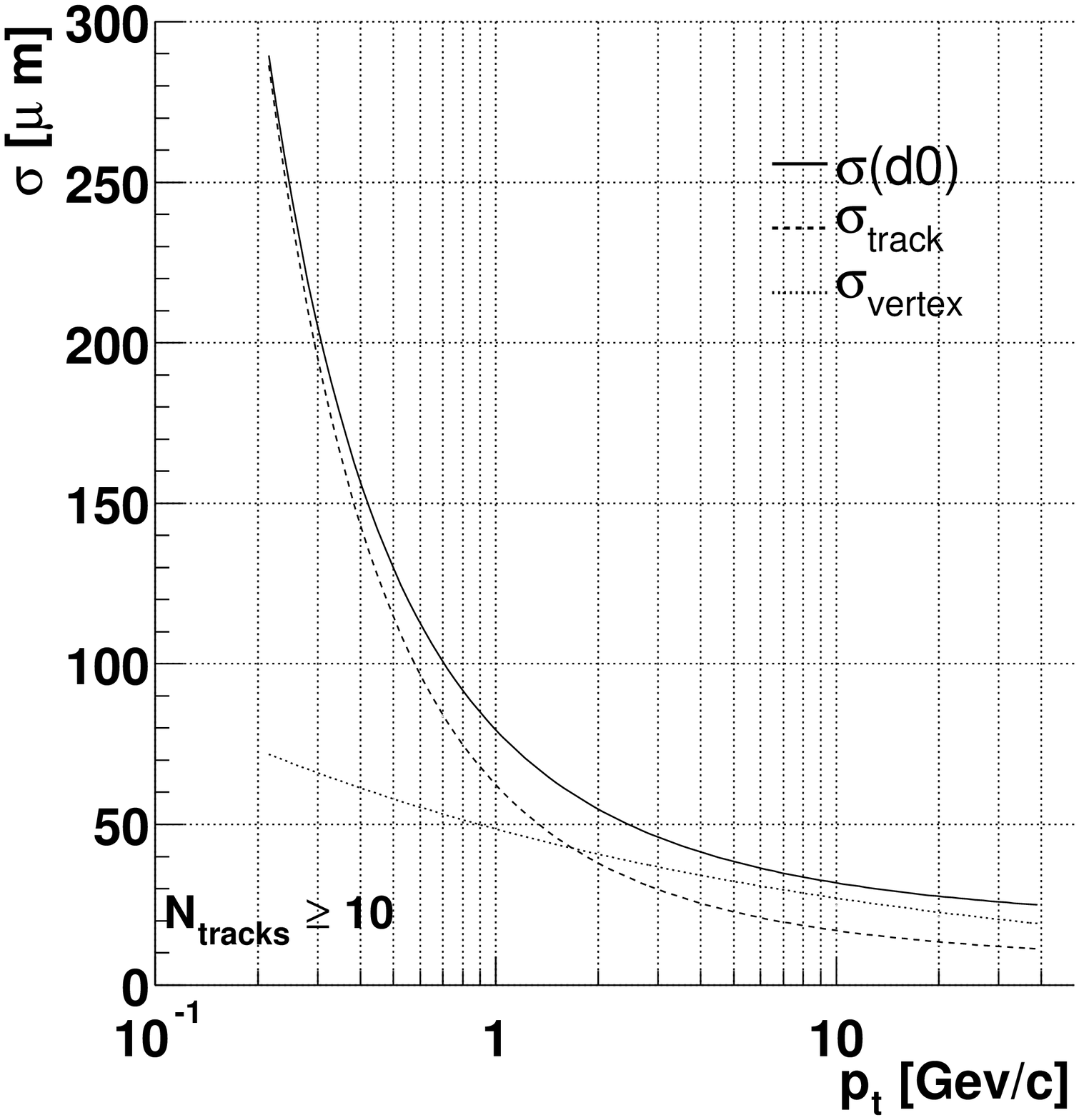}
    \includegraphics[width=.49\textwidth]{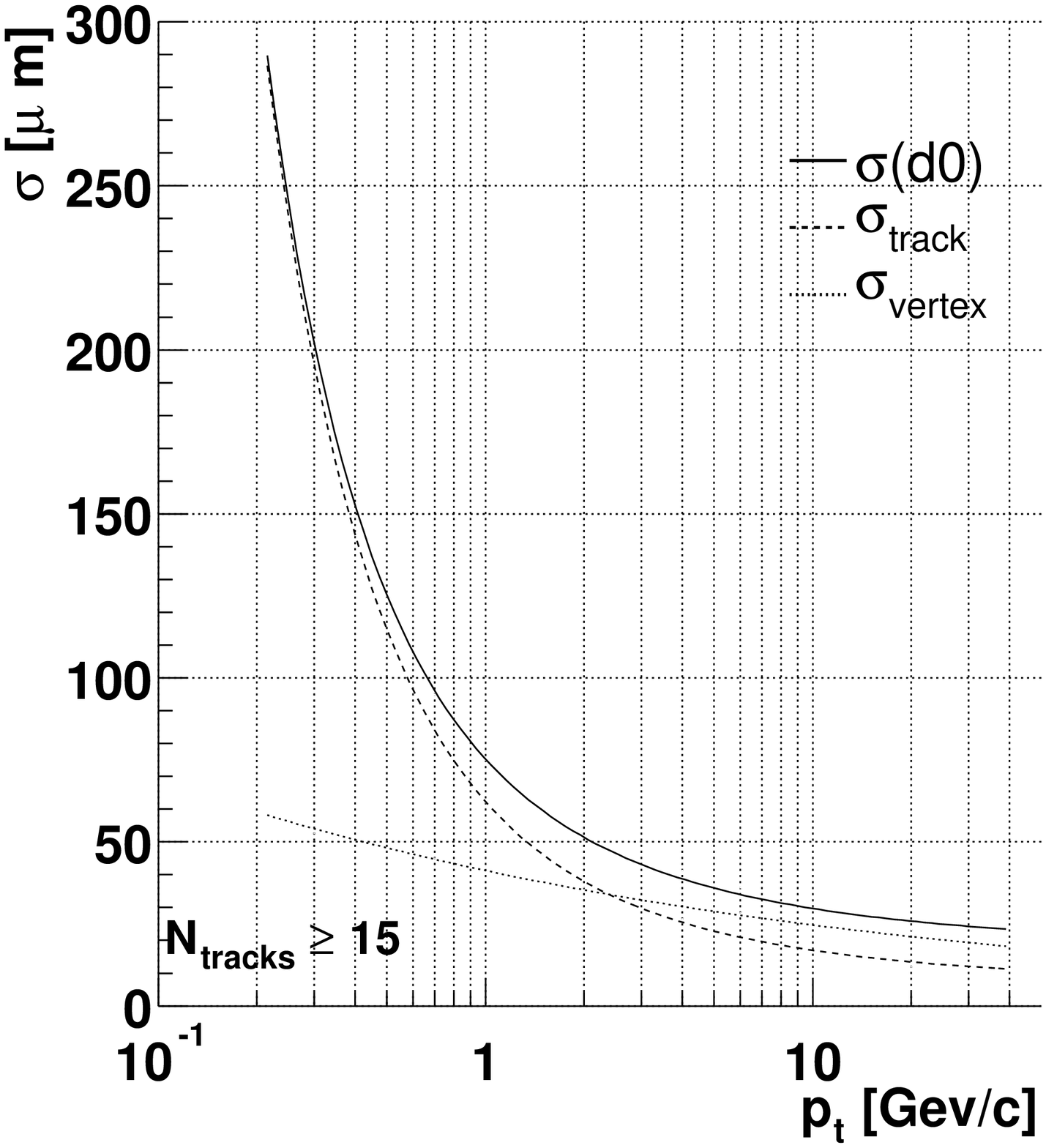}
    \includegraphics[width=.49\textwidth]{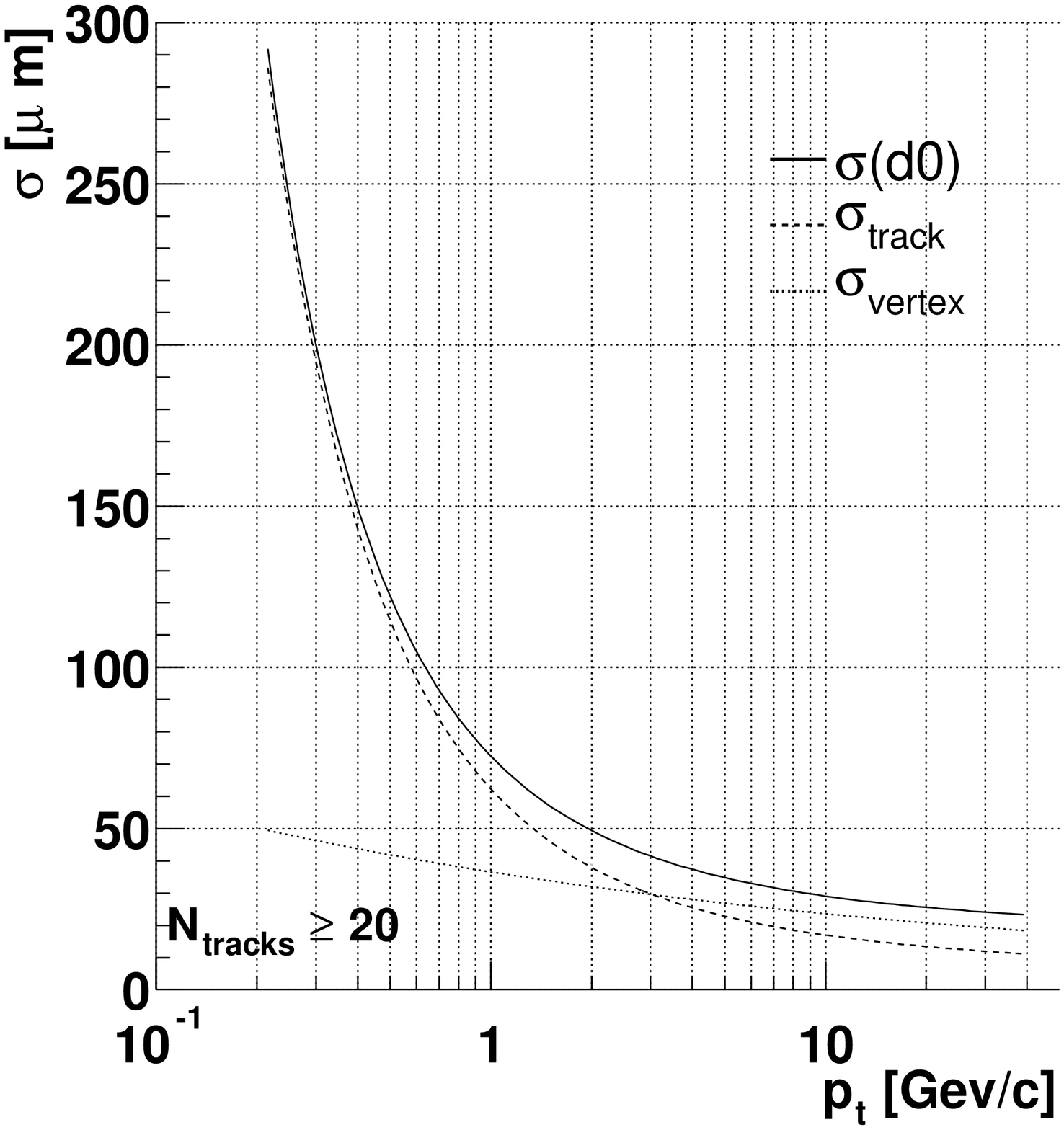}
    \caption{Impact parameter resolution in the bending plane as 
             function of the transverse momentum in pp collisions.
             The panels differ in the minimum number of reconstructed 
             tracks required (see text).}   
    \label{fig:d0full}
  \end{center}
\end{figure}

Figure~\ref{fig:d0full} presents the impact parameter resolution
as a function of the transverse momentum obtained with this strategy. 
We considered primary pions reconstructed in TPC and ITS, with 6 points
in the ITS layers. We used a sample of $7.5\cdot 10^{6}$ PYTHIA events, in
order to have sufficient statistics at high transverse momentum. 
In the top-left panel, we show the resolution integrated 
over all events (for this sample the average number of reconstructed 
tracks is $\av{N_{\rm tracks}}=7$). 
Then, we consider only events with $N_{\rm tracks}\geq N_{\rm min}$, with 
$N_{\rm min}=10,\,15,\,20$. We plot the resolution $\sigma(d_0)$ on 
the impact parameter 
(solid line), the resolution $\sigma_{\rm track}$ 
on the track position (dashed line) 
and an `equivalent resolution' 
on the vertex position (dotted line), calculated, 
`inverting' the expression (\ref{eq:d0res}), as:
\begin{equation}
  \sigma_{\rm vertex} = \sqrt{\sigma(d_0)^2 - \sigma_{\rm track}^2} 
\end{equation}
(for all quantities we consider the $r\phi$ component).

From the analysis of these plots, we observe the following:
\begin{itemize}
  \item the worsening in the $d_0$ resolution w.r.t. the case of perfect 
        knowledge of the vertex position, 
        $[\sigma(d_0)-\sigma_{\rm track}]/\sigma_{\rm track}$, 
        is negligible for 
        very-low-$\pt$ tracks, of the order of $30\%$ for $\pt=1~\gev/c$ 
        and of the order of $50\%$ for $\pt=10~\gev/c$;
  \item the impact of the uncertainty on the vertex position is 
        not too dramatic for medium- and high-momentum tracks, since 
        these tracks are always produced in events with large multiplicity, 
        in which the vertex can be reconstructed quite precisely;
        this is clearly shown by the strong $\pt$-dependence 
        of the equivalent vertex resolution, $\sigma_{\rm vertex}$;
  \item as the number of reconstructed tracks (i.e.
        the multiplicity of the 
        event) increases, the impact of the uncertainty on the vertex position 
        becomes smaller: for a track with $\pt=1~\gev/c$ it becomes 
        $20\%$ if the total number of tracks is $\geq 10$ and $15\%$ if 
        the total number of tracks is $\geq 15$. 
\end{itemize}

We, therefore, conclude that the algorithm for vertex reconstruction 
in pp collisions
allows the measurement of the impact parameter projection 
in the bending plane with a resolution that is not substantially worse than 
the track position resolution for low and medium transverse 
momentum tracks, in particular for tracks produced in high-multiplicity 
events. To this respect, we observe that, indeed, events with heavy flavour 
production have a multiplicity which is larger than the mean 
multiplicity in pp minimum-bias events. We will discuss and quantify this 
observation in Chapter~\ref{CHAP6}.
For high-momentum tracks 
the achieved impact parameter resolution is roughly twice the track position 
resolution; however, this is not dramatic, since, at high $\pt$, the 
background to heavy flavour particles is almost negligible and, therefore,
the selection based on the impact parameter is not as crucial as it is 
for low-momentum particles.

\mysection{Secondary vertex reconstruction}
\label{CHAP5:secondary}

The knowledge of the position of the secondary vertex allows the 
complete reconstruction of the momentum of the particle that has 
decayed. This provides an effective selection handle, because it 
allows to require the pointing of the momentum to the main interaction 
vertex. Moreover, a direct measurement of the $\pt$ distribution 
(of charm particles, for example) is possible. 

For a given pair of opposite-sign tracks the secondary vertex is 
reconstructed by a minimization of the distance in space between the 
two helices representing the tracks. Once the `minimum segment' between 
the tracks is found, the position of the vertex on this segment 
is defined keeping into account the different position precision of the two 
tracks. This information can be retrieved from the track covariance matrix.
As we have seen in the section on the impact parameter resolution, 
the track with the largest momentum has a better precision and, therefore, 
the vertex is usually estimated to be closer to this track than to the 
lower-momentum one.  
This method was originally developed 
for the reconstruction of strange particle decays (namely, ${\rm K^0_S}$, 
$\Lambda$, $\Xi$ and $\Omega$) and it was adopted also 
for the reconstruction of $\Dz$ decays.

\begin{figure}[!t]
  \begin{center}
    \leavevmode
    \includegraphics[width=\textwidth]{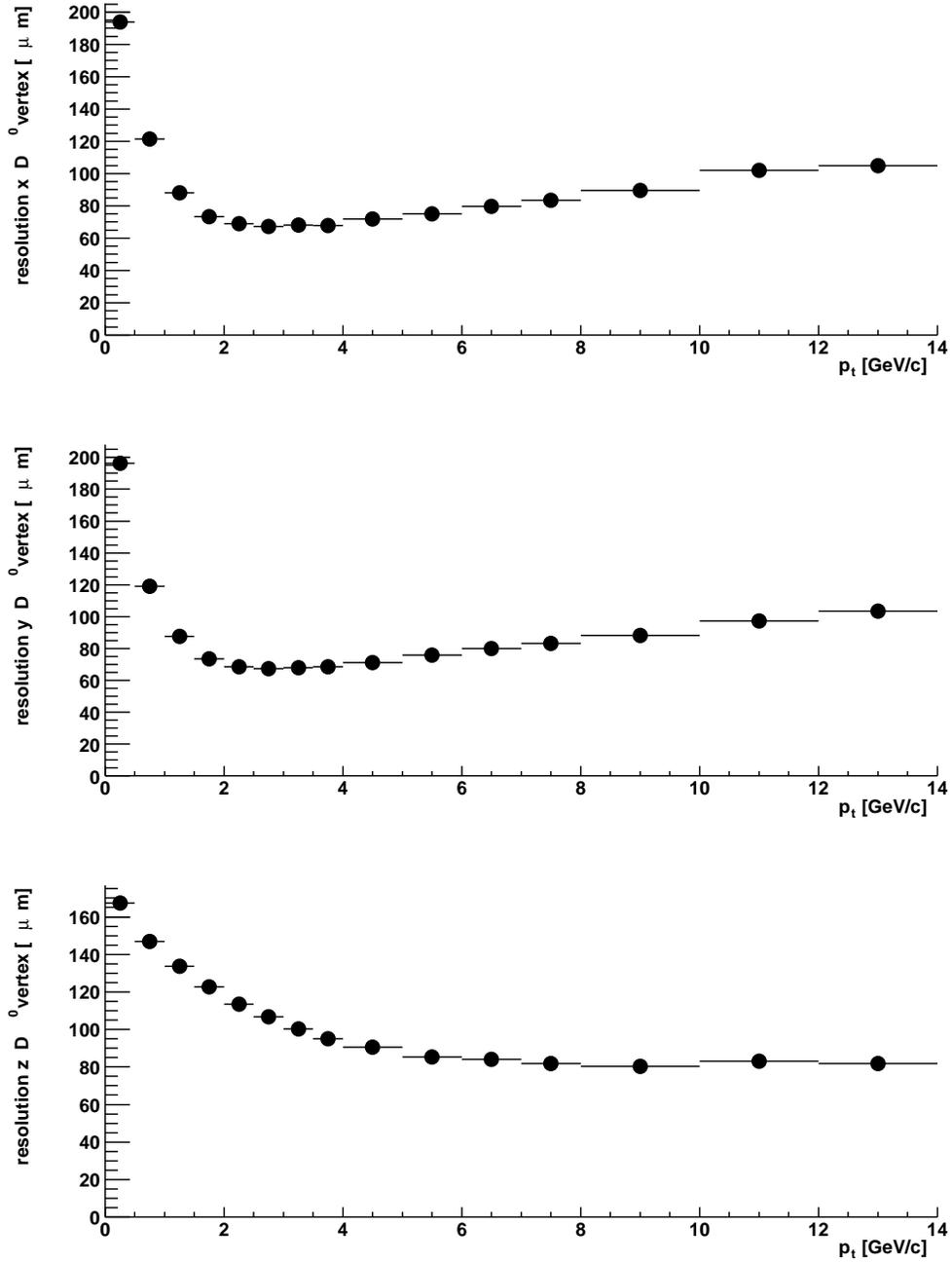}
    \caption{Resolution on the three coordinates 
             ($x,~y,~z,$ from top to bottom) 
             of the position of the $\Dz$ decay vertex, as a function of the 
             transverse momentum of the $\Dz$.}   
    \label{fig:vtx2res}
  \end{center}
\end{figure}

Figure~\ref{fig:vtx2res} reports the resolutions on the three coordinates
of the secondary vertex, as a function of the transverse momentum 
of the $\Dz$. There is clearly a strong correlation with the impact 
parameter resolution: the resolution is better in the bending plane 
($x$ and $y$) than in $z$ and, for $\pt<2$-$3~\gev/c$, it improves as 
$\pt$ increases. We will comment later on the worsening observed at high 
$\pt$ for the $x$ and $y$ coordinates. We can first try to understand 
quantitatively the obtained resolutions. 

We concentrate on the values at 
$\pt\simeq 2~\gev/c$, which is the average $\pt$ of the $\Dz$ particles 
produced at the LHC, as we have seen in Section~\ref{CHAP3:hadr}.
The average $\pt$ of the decay products is $\simeq 1~\gev/c$.
If the opening angle between the two decay tracks is not too small, we can 
consider each track to `measure' one of two perpendicular coordinates
in the bending plane; in this way, we expect a resolution of the order of the 
impact parameter resolution in $r\phi$ at $\pt\simeq 1~\gev/c$, i.e. 
$\simeq 50$-$70~\mum$. Along the $z$ direction, both tracks contribute to the 
measurement of the secondary vertex position; thus, we expect a resolution
of $\simeq \sigma(d_0(z))_{\pt\simeq 1~\gev/c}/\sqrt{2}\simeq 100$-$120~\mum$. 
The observed resolutions at $\pt=2~\gev/c$ agree with these simple
estimates.

At higher $\pt$ the resolution in the bending plane 
worsens as $\pt$ increases: this is due to the fact that the angle 
between the two decay tracks becomes smaller as the $\pt$ of the $\Dz$ 
increases and, thus, the two coordinates $x$ and $y$ are measured 
using essentially only 1 track. The expected resolution per coordinate 
in this case is $\simeq \sigma(d_0(r\phi))_{\pt\simeq 1~\gev/c}/\sqrt{0.5}\simeq 80$-$90~\mum$. 
~\\

In the first detailed study for the $\Dz$ detection~\cite{noteD0} a different 
method for the reconstruction of the secondary vertex was used.
The two tracks were approximated to straight lines starting from 
the intersection point of the circles obtained projecting the 
tracks on the bending plane. The secondary vertex was defined as the 
middle-point of the shortest segment joining the two lines.

\begin{figure}[!t]
  \begin{center}
    \leavevmode
    \includegraphics[width=.81\textwidth]{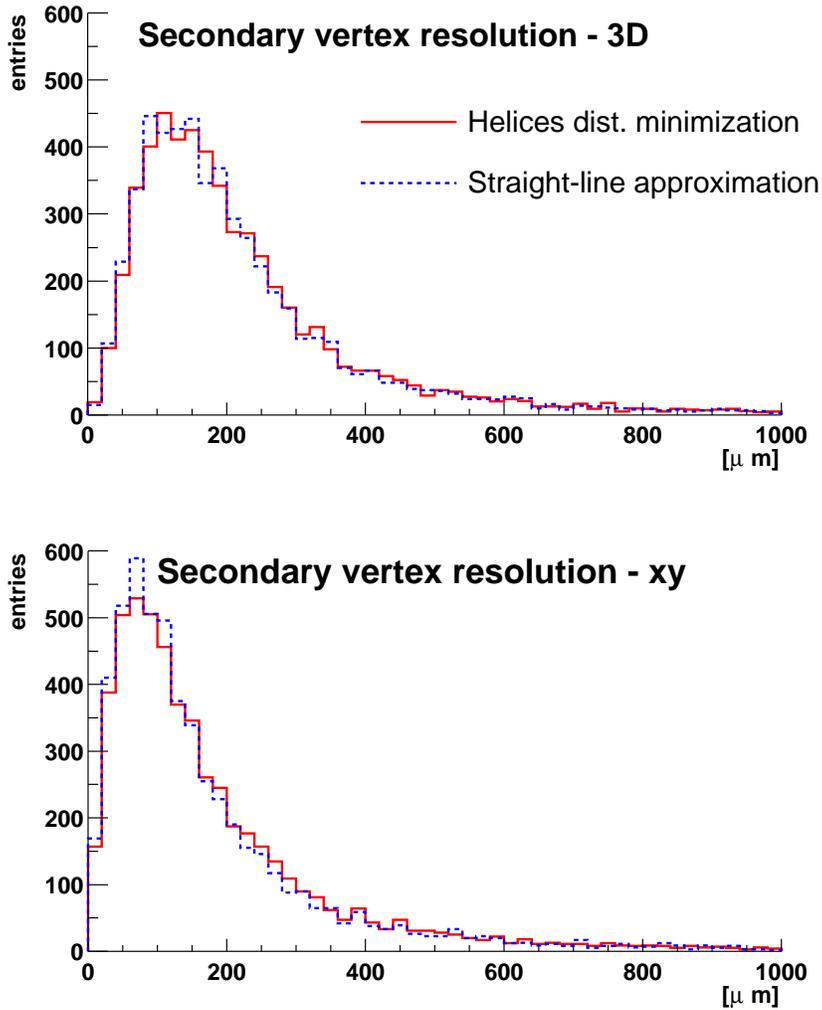}
    \caption{Distance between the reconstructed and the true position 
             of the secondary vertex for the two reconstruction methods: 
             minimization of the distance between helices (solid) and 
             straight-line approximation of the tracks (dashed). Both the 
             distance in space (top) and in the bending plane (bottom) are 
             shown.}   
    \label{fig:vtx2cmp}
  \end{center}
\end{figure}

It is interesting to compare the results, in terms of position 
resolution, given by the two different methods. In 
Fig.~\ref{fig:vtx2cmp} we report, for the two methods,
the distances in space and
in the bending plane between the reconstructed and the true vertex.
The achieved resolutions are very similar in the two cases, allowing to 
conclude that there is no much room for further optimization. 

\clearpage
\pagestyle{plain}

\setcounter{chapter}{5}
\mychapter{Exclusive reconstruction of $\Dz$ particles}
\label{CHAP6}

\pagestyle{myheadings}

In Chapter~\ref{CHAP2} we have shown that the exclusive reconstruction 
of charm mesons is a very effective tool to study the production 
of charm quarks in \AA~collisions at the LHC 
and we have outlined a detection strategy 
for the decay $\DtoKpi$. In Chapters~\ref{CHAP4} and~\ref{CHAP5} we have 
described the ALICE experimental apparatus and shown how its capabilities 
for tracking and vertexing allow to pursue this measurement.
This chapter presents the details and results of the feasibility study, that 
was carried out for both the \PbPb~and the pp case. The study done for 
\PbPb~collisions is described in the first part of the chapter 
(Section~\ref{CHAP6:D0recoPbPb}). Then, the \pp~case is addressed 
(Section~\ref{CHAP6:D0recopp}) with 
particular emphasis on the aspects that are distinctive of the low
multiplicity of pp events. The results obtained for \PbPb~and pp 
are compared at various stages of the reconstruction and selection chain 
in order to check their consistency and understand their differences.  

\mysection{Feasibility study for \PbPb~collisions}
\label{CHAP6:D0recoPbPb}

As discussed in Chapter~\ref{CHAP2}, in \AA~collisions the 
signal of the $\Dz$ hadronic decay has to be selected out of a very 
large combinatorial background (proportional to the square 
of the charged particle rapidity density $\dNdy$) given by pairs of 
uncorrelated tracks with large impact parameters. These can be:
\begin{enumerate} 
\item primary tracks which acquire an impact parameter due to scatterings 
      in the material of the beam pipe and of the innermost pixel layer;
\item tracks from the decays of hyperons ($\Lambda$, $\Xi$, $\Omega$) 
      and kaons;
\item tracks from undetected charm decays;
\item pions produced by annihilations of primary anti-protons 
      ($\rm \overline{p}$) and anti-neutrons ($\rm \overline{n}$) in the beam 
      pipe and in the innermost pixel layer. 
\end{enumerate}

As we will see, we have $S/B\sim 10^{-6}$ in the mass 
range $M_{\rm D^0}\pm 3~\sigma$, 
before selections; therefore, in order to extract the charm signal with 
good significance, one has to apply cuts selective enough to reduce 
the background by $6$-$7$ orders of magnitude.
In view of this severe selection procedure, the required statistics 
for the study of the background and of the signal is very large
($\sim 10^5$ events  for the background). Thus, 
we adopted a number of fast simulation techniques; namely:
\begin{itemize}
\item separate generation of signal and background events;
\item event mixing technique to increase the statistics of the background;
\item parameterization of track reconstruction in the TPC 
      (this tool, described in Section~\ref{CHAP4:tracking} 
      and in Ref.~\cite{notetpcparam}, was specifically 
      developed in the scope of the charm studies presented here);
\item fast simulation of the response of the ITS detectors
      ({\sl fast points});
\item parameterization of the particle identification (PID) 
      in the TOF detector.
\end{itemize} 

\subsection{Background and signal generation}
\label{CHAP6:generPbPb}

\subsubsection{Background}

We chose to use the HIJING event generator, introduced in 
Section~\ref{CHAP4:generators}, for the simulation of \PbPb~events, 
because it includes all the background sources listed above.
This is not true, for example, for some parameterized generators, where 
the event multiplicity can be adjusted by the user but only primary 
pion and kaon tracks are provided. 
We have seen in Section~\ref{CHAP4:generators} that the multiplicity 
given by HIJING when parton energy loss is included is $\dNdy\simeq 6000$, 
for central \PbPb~collisions. Since the recent 
extrapolations from the RHIC measurements give values of the order 
of $2000$-$3000$ charged particles per unit of rapidity
(Section~\ref{CHAP1:sqrtsdNdy}), the number obtained in 
HIJING can be considered conservative. An extrapolation of the 
results to the case of lower multiplicity is presented in the last part of 
this section. 

The collision impact parameter $b$ was sampled from the physical distribution 
d$N/$d$b\propto b$ and the condition $b<2~\fm$ was applied in order to generate
central collisions\footnote{The impact parameter
range used for the generation of the background events is $b<2~\fm$, 
while it was $b<3.5~\fm$ for the estimate of the charm production rate 
(Section~\ref{CHAP3:extrapolation2PbPb}). 
This choice is due to technical reasons; however, it is a conservative one, 
since $b<2~\fm$ gives a larger multiplicity for background tracks than 
$b<3.5~\fm$.}. 

Our background sample consists of $2\cdot 10^4$ such events, that were
generated in 1000 sub-samples of 20 events each, the events of a sub-sample
having the same values for the impact parameter $b$ and for the three 
coordinates of the primary vertex. Since the background is combinatorial, 
400 equivalent events were obtained out of each sub-sample by combining 
each positive track of the sub-sample with all negative tracks of the 
sub-sample. In this way, $4\cdot 10^5$ equivalent events were obtained, 
increasing the background statistics by a factor 20. 

\subsubsection{Signal: ${\rm c}\to \DtoKpi$ and ${\rm b}\to{\rm B}\to \DtoKpi$}

At LHC energies the ratio of the production cross sections for beauty and 
for charm 
is of the order of 5\% (see Section~\ref{CHAP3:XsecNN}). Considering also
that the average inclusive branching ratio of B mesons to $\Dz$ is 
$\simeq 65\%$~\cite{pdg}, we conclude that a significant fraction of 
all produced $\Dz$ particles comes from b quarks
(${\rm b}\to{\rm B}\to \Dz$).
The ratio ($\Dz$ from b)/($\Dz$ from c) can be calculated as:
\begin{eqnarray*}
\label{eq:bD02cD0}
\frac{\d N({\rm b}\to{\rm B}\to \Dz)/\d y}{\d N({\rm c}\to \Dz)/\d y}
= \frac{\d N({\rm b}\to {\rm B^0,B^+})/\d y \times BR({\rm B^0,B^+\to\Dz})}{\d N({\rm c}\to \Dz)/\d y}= \\
= \left\{ \begin{array}{ll}
0.049 & {\rm for}~\PbPb~{\rm at}~\sqrtsNN=5.5~\tev \\
0.054 & {\rm for}~{\rm pp}~{\rm at}~\sqrt{s}=14~\tev
\end{array} \right.
\end{eqnarray*}
where the rapidity densities are taken from Tables~\ref{tab:hadyieldsPbPb} 
and~\ref{tab:hadyieldspp}.
This `B contribution' was included in the study presented here, since it is 
important to understand how the ratio of secondary (from b) to primary 
(from c) $\Dz$ is affected by the selections that we apply. 
After the selections this contribution has to be corrected for by 
subtracting, in bins of transverse momentum, the estimated number of 
secondary $\Dz$. 
As we will discuss in the next chapter, the uncertainty on this number, 
which is proportional to the uncertainty on the beauty cross section and
to the fraction of secondary-to-primary $\Dz$, is one of the main contributions
to the final systematic error. It is, therefore, essential to keep under 
control and, possibly, low this fraction. This is also motivated by the fact 
that,  since the $\pt$ 
distribution of b quarks is harder than that of c quarks and the selections 
will naturally tend to be more efficient for larger momenta, we 
expect the fraction secondary/primary $\Dz$ to increase after the selections.  

The signal was generated using PYTHIA, tuned to reproduce the 
$\pt$ distribution of charm and beauty quarks given by the NLO calculations 
by Mangano, Nason and Ridolfi, as explained in Section~\ref{CHAP3:generators}.
Many $\Dz/\overline{\Dz}$ 
mesons, with decay forced in a charged $\K\pi$ pair, were superimposed 
in special `signal events'. 
The number of $\Dz$ per event (13000 in $|y|<2$) was tuned in order to 
have the same track multiplicity as in a central HIJING event.
In this way, the $\Dz$ decay products are reconstructed with the same 
efficiency as if they were produced in a central \PbPb~collision. It was 
verified that the different momentum and impact parameter distributions
of these `signal events' with respect to central HIJING events
do not affect significantly the reconstruction efficiency
(more details are given Section~\ref{CHAP6:recoPbPb}).

\begin{figure}[!t]
  \begin{center}
    \includegraphics[width=.8\textwidth]{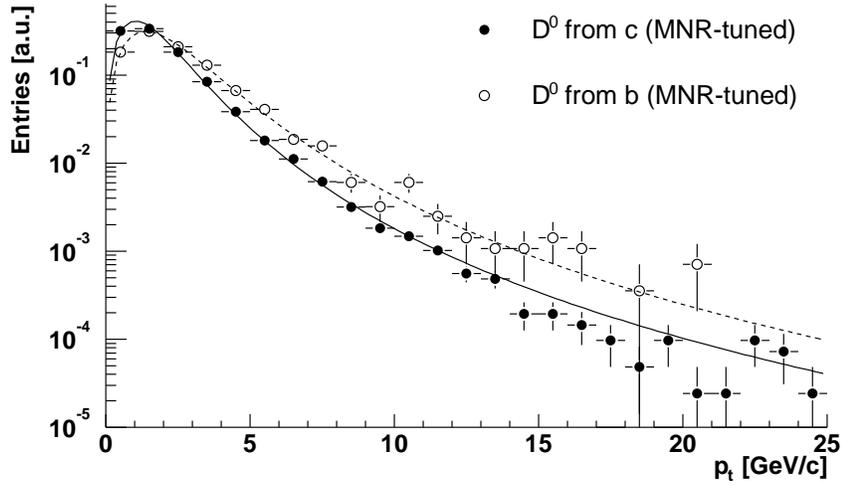}
    \caption{Transverse momentum distributions for primary and secondary 
             (from B meson decays) $\Dz$ mesons, in \PbPb~collisions at 
             $5.5~\tev$.}
    \label{fig:ptD0cb}
  \end{center}
\end{figure}

A total of 1000 such `signal events' were generated with primary $\Dz$
particles. 
Using our present rate estimate (from Table~\ref{tab:hadyieldsPbPb}) 
and a branching ratio of $3.8\%$~\cite{pdg}, such a number of $\DtoKpi$ decays 
corresponds to $\simeq 6.1\cdot 10^6$ central \PbPb~events. 
In addition, we have generated 49 similar events, but with 
secondary $\Dz$ from the decay of B mesons. In this case we set PYTHIA in 
order to reproduce the NLO pQCD results for the $\pt$ distribution of 
b quarks. We did not transport through the detector the other decay 
products of the b quarks and of the B mesons, but only the kaons and pions 
from the $\Dz$ decays. Figure~\ref{fig:ptD0cb} shows the $\pt$ distributions 
for primary and secondary $\Dz$ mesons: as expected the latter have a harder
spectrum.

All the results presented here are scaled to $10^7$ central \PbPb~events, 
expected to be collected in the 1-month heavy ion run of 1 LHC year.

\subsubsection{Interaction vertex}

For both signal and background events, the $x$ and $y$ coordinates of the 
vertex were sampled from two gaussian distributions with $\sigma=15~\mum$
and centred at $(0,0)$, 
while the $z$ position was sampled from a Gaussian with $\sigma = 5.3~\cm$, 
according to the expected beam configuration during the heavy ion running 
(Section~\ref{CHAP4:lhcPbPb}). A cut at $\pm 1~\sigma$ on the $z$ position 
of the vertex was applied; this corresponds to the expected width of 
the fiducial interaction region. 

\subsection{Detector simulation and event reconstruction}
\label{CHAP6:recoPbPb}

The detector simulation was done in the framework of AliRoot, as 
described in Section~\ref{CHAP4:detresponse}, using a detailed 
description of the materials for the beam pipe and the ITS, which are
instrumental in determining track impact parameters and 
secondary vertices positions, the crucial quantities to extract the 
charm signal. The response of the TPC was parameterized and 
in Appendix~\ref{App:tpc} we show that this does not affect the 
momentum and impact parameter resolutions obtained after completing the
track reconstruction in the ITS. In order to contain the CPU time and storage 
space within reasonable limits, the {\sl fast points} were used in the ITS
(Section~\ref{CHAP4:tracking}); in Chapter~\ref{CHAP5}, 
Fig.~\ref{fig:spVSfp}, we have shown 
that this approximation does not affect the impact parameter resolution.  
Also the particle identification response of the Time Of Flight detector
was parameterized, as we shall detail in the next section. 

The value of the magnetic field used in the simulation was $B=0.4$~T, which 
is close to the maximum value that can be provided by the ALICE magnet.
This kind of physics studies yields better performance 
with relatively large values of the magnetic field, since (a) the 
invariant mass resolution is better with larger fields and (b) the acceptance
at very low $\pt$ ($<500~\mev/c$) is not crucial. 
The extrapolation of the results for lower 
values of the magnetic field is straight-forward and will be eventually 
discussed. 

The $z$ position of primary vertex was reconstructed for each event by means 
of the Silicon Pixel Detectors using the method briefly described in 
Section~\ref{CHAP5:vtxPbPb} 
(see Refs.~\cite{vtxPbPb1,vtxPbPb2} for the details). We remind that 
a resolution of $\simeq 6~\mum$ is obtained with $\dNdy=6000$. 
The vertex position in the transverse plane was taken at the nominal value 
$(0,0)$, i.e. it was assumed to be known within the uncertainty given by the 
size of the interaction region ($\simeq 15~\mum$).

Track reconstruction in the ITS was performed as described in 
Section~\ref{CHAP4:tracking}, using as input for the standard Kalman 
filter the parameterized tracks of the TPC. At least 5 clusters per track 
were required in the ITS, including the clusters in the two pixel layers;
as shown in Section~\ref{CHAP5:d0VSclusters}, 
in this way the impact parameter resolution is still optimal. 
   
\begin{figure}[!t]
  \begin{center}
    \includegraphics[width=.6\textwidth]{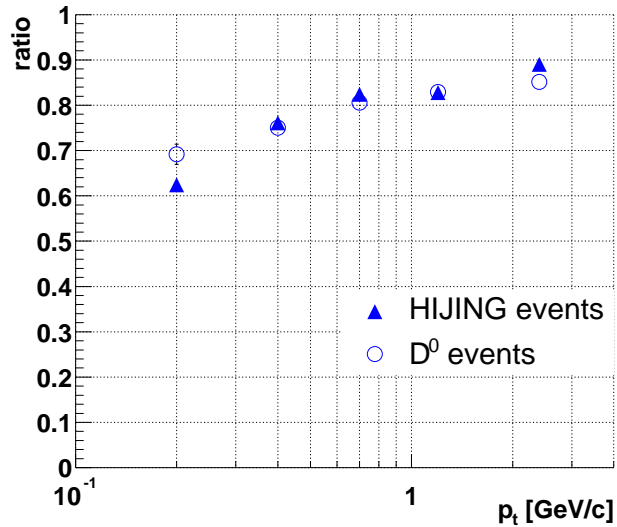}
    \caption{ITS tracking efficiency, defined as the ratio between the
    number of tracks reconstructed in the ITS with respect to the
    number of tracks reconstructed in the TPC, as a function of $\pt$
    for central HIJING events and for `signal events'
    containing only $\Dz/\overline{\Dz}$, decaying in $\K\pi$, with the same 
    track multiplicity as a central HIJING event.}
    \label{fig:SgnBkgITSEffic}
  \end{center}
\end{figure}

The reconstruction was performed in the same way for background and 
`signal events'. Figure~\ref{fig:SgnBkgITSEffic} 
demonstrates that the tracking efficiency as a function of $\pt$
in the ITS for the `signal events' is the same as for the background events.

\subsection{Particle identification in the TOF detector}
\label{CHAP6:tofPbPb}

\begin{figure}[!t]
  \begin{center}
    \leavevmode
    \includegraphics[clip,width=.7\textwidth]{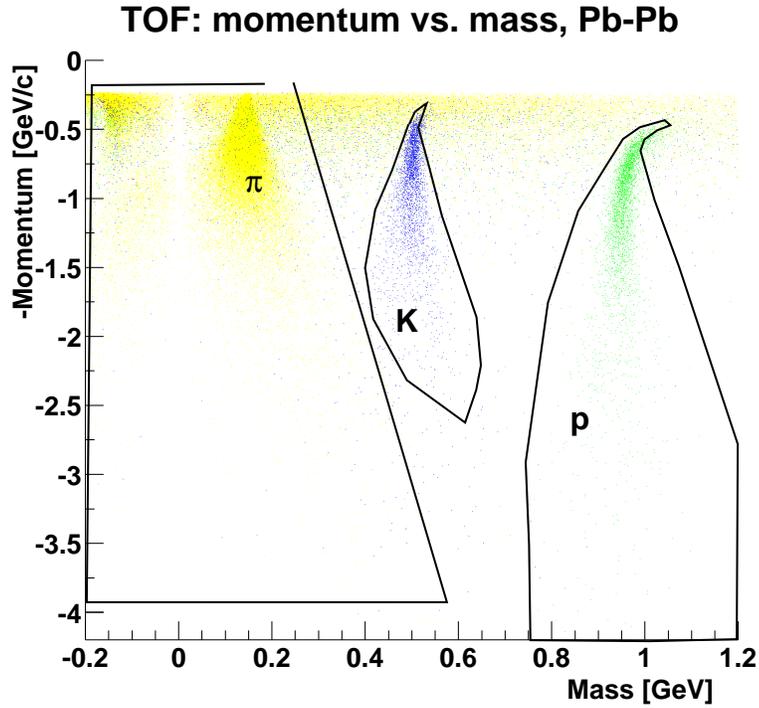}
    \caption{Momentum versus mass calculated from TOF for a sample of
    HIJING \PbPb~events. The lines correspond to the chosen
    graphical cuts relative to the selection of pions, kaons and protons.
    Negative values of the mass are assigned when the argument of the 
    square root in equation~\ref{eq:mtof} is lower than 0.}
    \label{fig:TOFpVSmPbPb}
  \end{center}
\end{figure}

In the momentum range of interest for charm analysis 
($p\simeq 0.5$-$2~\gev/c$) the particle identification capability of ALICE 
is determined mainly by the TOF detector. The measurement of the 
time-of-flight $t$ across a known distance $L$ for a track with momentum 
$p$ (measured in the TPC and in the ITS) allows an estimation of the 
particle mass as:
\begin{equation}
\label{eq:mtof}
m = p \cdot \sqrt{\frac{t^2}{L^2}-1}.
\end{equation}
Figure~\ref{fig:TOFpVSmPbPb} presents a scatter plot of the 
measured momenta versus the estimated masses 
for the particles produced in central \PbPb~collisions
generated with HIJING. The points corresponding to electrons, pions, 
kaons and protons are coloured in red, yellow, blue and green, respectively.
The figure, from the TOF Technical Design Report (TDR)~\cite{toftdr}, 
is obtained for $B=0.4$~T assuming an overall time resolution\footnote{
Since the TOF TDR, the design of the Multigap Resistive Plate Chambers has 
been improved and an overall time resolution of $\approx 120$~ps can now be
achieved. The analysis presented here uses the slightly worse resolution 
that was expected at the time of the TOF TDR.} of $150$~ps. 
The association of the time-of-flight and, hence, of the mass to a specific 
reconstructed track is obtained by means of a matching algorithm, that 
propagates the track from the outer radius of the TPC to the TOF detector 
and matches it with one of the $\approx 3\times 3~\cm^2$ pads of the 
detector~\cite{toftdr}. 
Tracks matched with a non-active region of the detector 
or with a non-fired or multi-fired\footnote{A pad is non-fired if it does 
not have hits; it is single-fired if it has only one hit; it is multi-fired 
if it has more than one hit.} pad are not assigned a mass.
The TRD, which lies between the TPC and the TOF, will provide a `bridge' 
between the two detectors and improve the matching procedure. Since 
the extension of track reconstruction to the TRD is still under development,  
Fig.~\ref{fig:TOFpVSmPbPb} 
was obtained from a simulation that does not include 
the material of the TRD. Indeed, it would be too pessimistic to consider the 
TRD as a layer of inactive material.

For $0.5<p<2$-$2.5~\gev/c$ there is a good mass separation of pions, kaons 
and protons. For lower momenta the matching tends to fail because of 
multiple scattering and energy loss, while for $p>2$-$2.5~\gev/c$ the
separation vanishes, especially between pions and kaons, as they 
become relativistic.

The association of the particle type to a track (tagging) is determined 
by applying cuts on the momentum-versus-mass plane.
The values of these cuts determine the identification efficiency and the 
contamination of the sample. The identification efficiency for a 
particle type $i$ is defined as the ratio of the number of tracks of
type $i$ correctly tagged as $i$ to the total number of tracks of type $i$;
the contamination is defined as the ratio of the number of tracks 
incorrectly tagged as $i$ to the total number of tracks tagged as $i$.
The optimal level of contamination and efficiency depends on
the specific physics case under study.

We divide our set of reconstructed
tracks into four samples: those identified as pions ($\pi_{\rm tag}$), 
as kaons ($\K_{\rm tag}$), 
as protons (${\rm p}_{\rm tag}$) and non-identified ($?_{\rm tag}$). A
$\DtoKpi$ decay for which both the pion and the
kaon tracks are reconstructed corresponds to a pair of tracks 
of opposite charge $(-,+)$. According to their PID, the pair can fall in one
of the following samples:
\begin{description}
\item[Sample A] ($\K_{\rm tag}$, $\pi_{\rm tag}$) + 
                ($\K_{\rm tag}$, $?_{\rm tag}$): the
kaon is identified while the other track can be identified as pion or
non-identified;
\item[Sample B] ($?_{\rm tag}$, $\pi_{\rm tag}$): only the positive track is
identified as pion;
\item[Sample C] ($?_{\rm tag}$, $?_{\rm tag}$): both tracks are not
identified; in this sample each pair is counted twice: once as a $\Dz$
candidate and once as a $\overline{\Dz}$ candidate.
\item[Sample D] All other combinations, like e.g. ($\pi_{\rm tag}$, 
$\pi_{\rm tag}$). These pairs are rejected.
\end{description}

If the pion from a $\Dz$ decay is correctly identified, but the kaon is 
misidentified as a pion, the candidate falls in sample D and is lost.
Therefore, for open charm detection, the PID strategy has to be optimized
in order to minimize the number of kaons tagged as pions, while tagging 
correctly a large fraction of the pions.

On the basis of this guideline the PID tags have been defined in the
following way:
\begin{itemize}
\item any track not matched with a single-fired TOF pad 
is tagged as $?_{\rm tag}$;
\item tracks matched with a single-fired TOF pad are tagged according
  to the graphical cuts shown in Fig.~\ref{fig:TOFpVSmPbPb}; if
  a track falls outside all graphical cuts it is tagged as
  $?_{\rm tag}$. 
\end{itemize}
The graphical cuts were optimized in order to minimize the probability 
to tag a kaon as a pion, i.e. to minimize the loss of signal. 

\begin{figure}[!t]
  \begin{center}
    \leavevmode
    \includegraphics[clip,width=\textwidth]{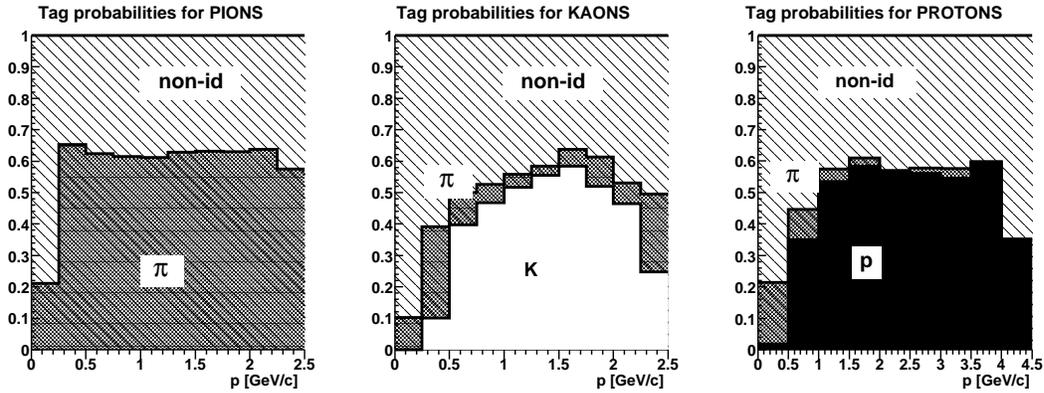}
    \caption{PID tag probabilities for reconstructed pions, kaons and protons
             in \PbPb~collisions with the TOF detector.}
    \label{fig:TOFtagprobPbPb}
  \end{center}
\end{figure}

In this way, for every particle type, we can compute the probabilities
to be tagged as pion, kaon, proton or non-identified. These
probabilities are shown in Fig.~\ref{fig:TOFtagprobPbPb} as a function of
the total momentum. We give an example of how these figures should be read:
for a reconstructed track (in TPC and ITS), known from the simulation to 
be a kaon, with $p=1~\gev/c$, the probability to be tagged as kaon is 45\%, 
the probability to tagged as pion is 8\% and the probability to be tagged 
as non-identified is the remaining 47\% (see central panel). 

In our study the TOF detector was not included in the simulation of all 
the events and the three samples A, B and C were populated with 
$\Dz$ candidates according to the tabulated probabilities from the figure,
both for the signal and for the background. The PID information was used 
for $p<2~\gev/c$ for pions and kaons and for $p<4~\gev/c$ for protons;
for larger momenta all tracks were tagged as $?_{\rm tag}$ (non-id).
The fraction of signal lost because the kaon is tagged as a pion is 
10\%.

\subsection{Analysis}
\label{CHAP6:analysisPbPb}

\begin{figure}[!t]
  \begin{center}
    \leavevmode
    \includegraphics[clip,width=.7\textwidth]{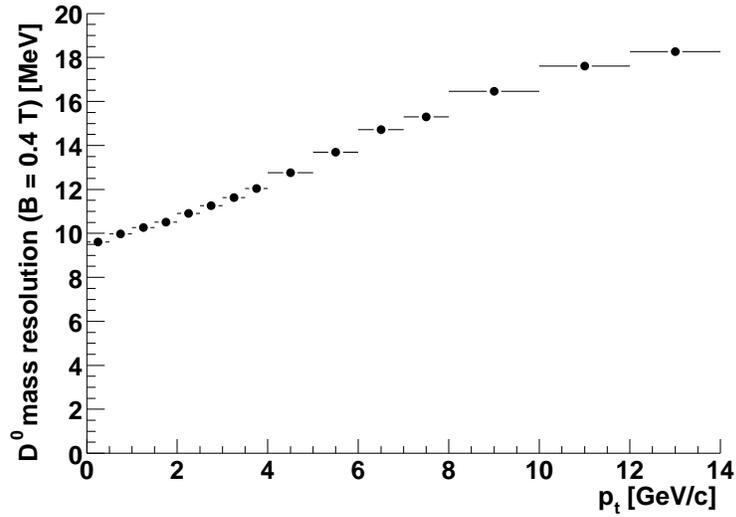}
    \caption{$\Dz$ invariant mass resolution as a function of $\pt$, for 
             $B=0.4$~T.}
    \label{fig:D0massres}
  \end{center}
\end{figure}

For each $\Dz$ candidate (opposite-charge tracks pair) 
the position of the decay vertex is computed, 
as explained in Section~\ref{CHAP5:secondary}, by a minimization of the 
distance in space between the two helices representing the particle 
trajectories. The momentum of the $\Dz$ candidate is calculated as the sum 
of the momenta of the kaon and of the pion at the position of closest
approach between the two tracks. 
The invariant mass of the pair is calculated as 
$M=\sqrt{(E_++E_-)^2-(\vec{p}_++\vec{p}_-)^2}$ and, for the 
signal, the resolution $\sigma$ for $B=0.4$~T is reported in 
Fig.~\ref{fig:D0massres} as a function of the $\Dz$ transverse momentum. 
The average resolution is $12~\mev$ and the $\pt$ dependence reflects 
the trend of the momentum resolution (Fig.~\ref{fig:ptres}). 

In Table~\ref{tab:initialPbPb} we present the signal-to-background ratios for
the three samples A, B and C in the invariant mass range 
$|M_{\rm K\pi}-M_{\rm D^0}| <3~\sigma$, 
before any geometrical or kinematical selection. Due to the 
small fraction of kaons in the background, sample A (kaon identification 
required) shows the highest $S/B$ ratio ($\sim 2\cdot 10^{-5}$). 
However, Fig.~\ref{fig:TOFtagprobPbPb} (central panel) shows that the
identification probability decreases rapidly for kaons with 
momentum larger than $1.5~\gev/c$;  
therefore, for $\Dz$ momenta larger than $\sim 2$-$3~\gev/c$, 
the fraction of signal that populates sample A becomes
marginal. For this reason, we consider as our standard sample the sum 
of the three samples A, B and C (called `Total' in 
Table~\ref{tab:initialPbPb});
this corresponds to the rejection of ($\pi_{\rm tag}$, $\pi_{\rm tag}$) and 
($\K_{\rm tag}$, $\K_{\rm tag}$) pairs. 
In the low-$\pt$ region, it will be eventually
convenient to restrict the PID selection to sample A only.

\begin{table}[!t]
  \caption{Initial values of $S/B$ in the invariant mass range
  $M_{\rm D^0}\pm 3~\sigma$, before selection.}
  \begin{center}
  \begin{tabular}{c|ccc}
    \hline
    \hline 
    Sample & $S$/event & $B$/event        & $S/B$                \\
    \hline
    A      & $0.054$ & $2.5\cdot10^3$ & $2.16\cdot10^{-5}$ \\
    B      & $0.041$ & $1.4\cdot10^4$ & $2.98\cdot10^{-6}$ \\
    C      & $0.031$ & $1.2\cdot10^4$ & $2.69\cdot10^{-6}$ \\
    \hline
    Total  & $0.126$ & $2.8\cdot10^4$ & $4.53\cdot10^{-6}$ \\
    \hline
    \hline      
  \end{tabular}
  \label{tab:initialPbPb}
\end{center}
\end{table}

Several selection cuts are applied in order to increase the $S/B$ ratio to
the level needed to extract the signal. Their definition is presented in
the following paragraphs.

Pairs for which the distance of closest approach ($dca$) between 
the tracks is larger than $dca_{\rm max}$ ($300$-$400~\mum$, depending 
on the transverse momentum of the $\Dz$ candidate) are rejected. 

Since the transverse momentum distributions for the signal are harder
than those for the background, it is convenient to apply a cut on the
minimum $\pt$ for $\K$ and $\pi$ ($\pt>800~\mev/c$). 

\begin{figure}[!t]
  \begin{center}
    \includegraphics[width=.38\textwidth]{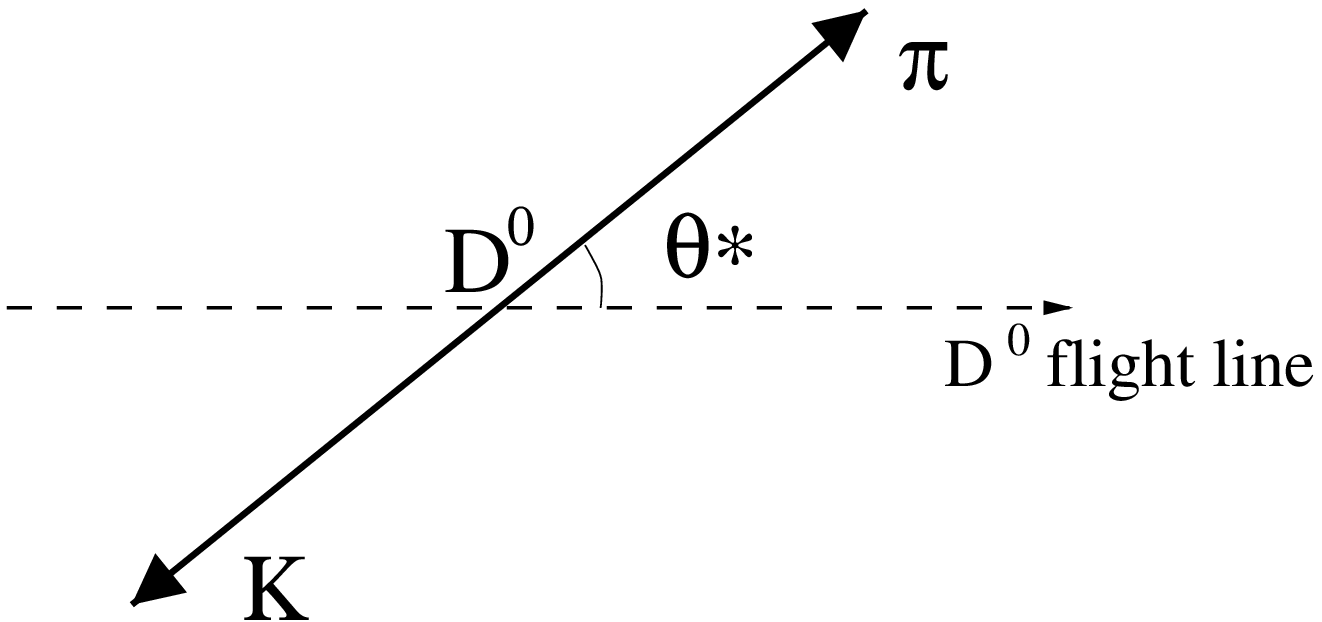}
    \includegraphics[width=.6\textwidth]{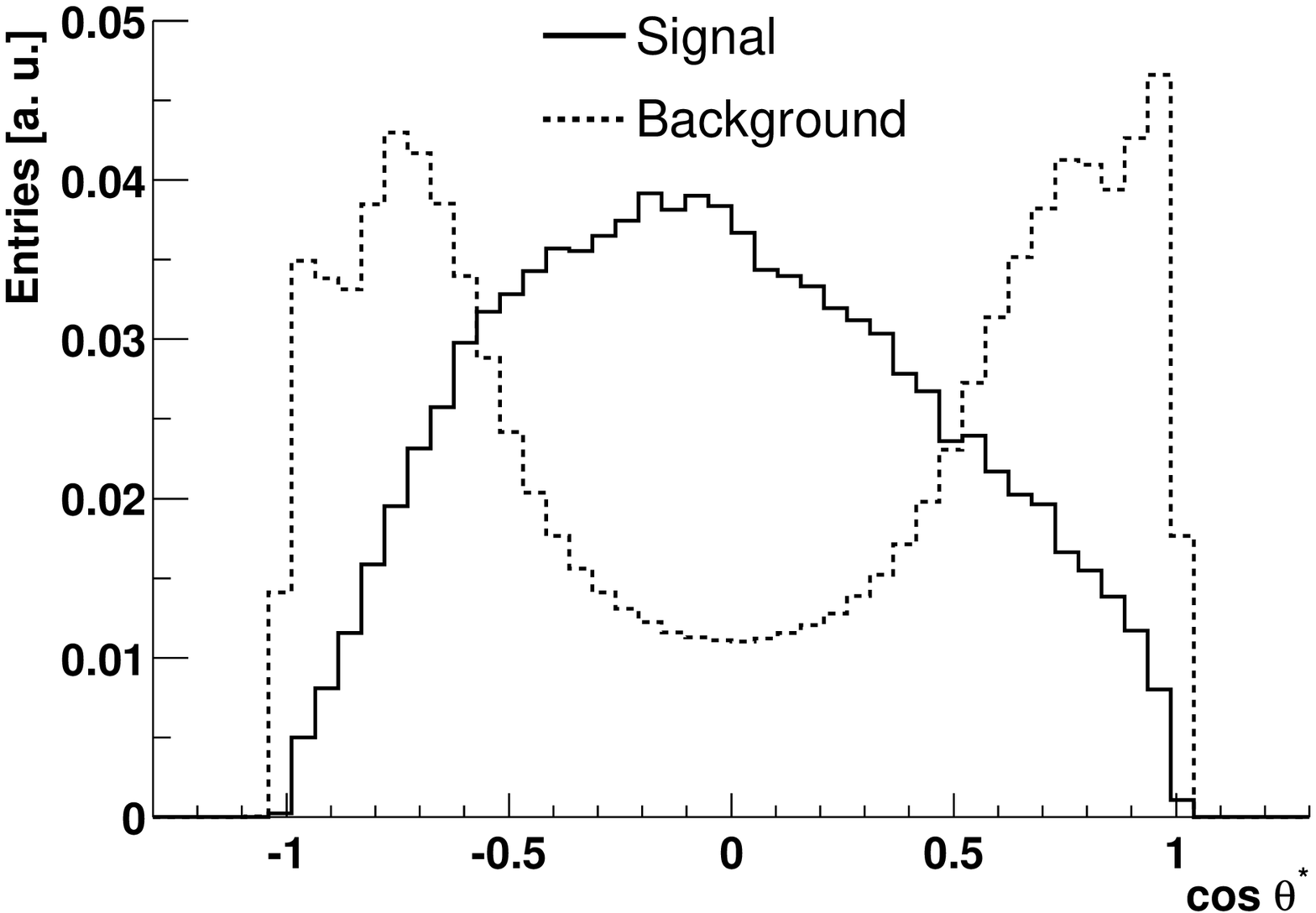}
    \caption{Definition of the decay angle $\theta^\star$ in the $\Dz$
    reference frame (left). On the right, distribution of $\cos\theta^\star$ 
    for the $\Dz$ signal (solid line) and for the background (dashed line). 
    The histograms are normalized to the same integral.} 
    \label{fig:thstar}
  \end{center}
\end{figure}

In the reference frame of the
decaying $\Dz$, we define $\theta^\star$ as the angle between the pion
momentum and the $\Dz$ flight line (see sketch in Fig.~\ref{fig:thstar}). 
As shown in Fig.~\ref{fig:thstar} (right),
the background accumulates at $\cos\theta^\star = \pm 1$. The distribution
for the signal is not uniform due to the other cuts applied,
in particular the cut on the minimum $\pt$ of the pion and of the
kaon. The slight asymmetry reflects the different masses of the kaon and 
the pion. Only pairs with $|\cos\theta^\star|<cth_{\rm max}$ ($\simeq 0.6$)
are kept.

With these cuts the signal-to-background ratio increases by a factor 
$\sim 100$. Further improvement can be obtained by applying the displaced
vertex identification procedure, outlined in Section~\ref{CHAP2:D0toKpi},
based on the impact parameter and on the requirement that the 
reconstructed $\Dz$ points back to the primary vertex.

\begin{figure}[!t]  
  \begin{center}
    \includegraphics[width=.85\textwidth]{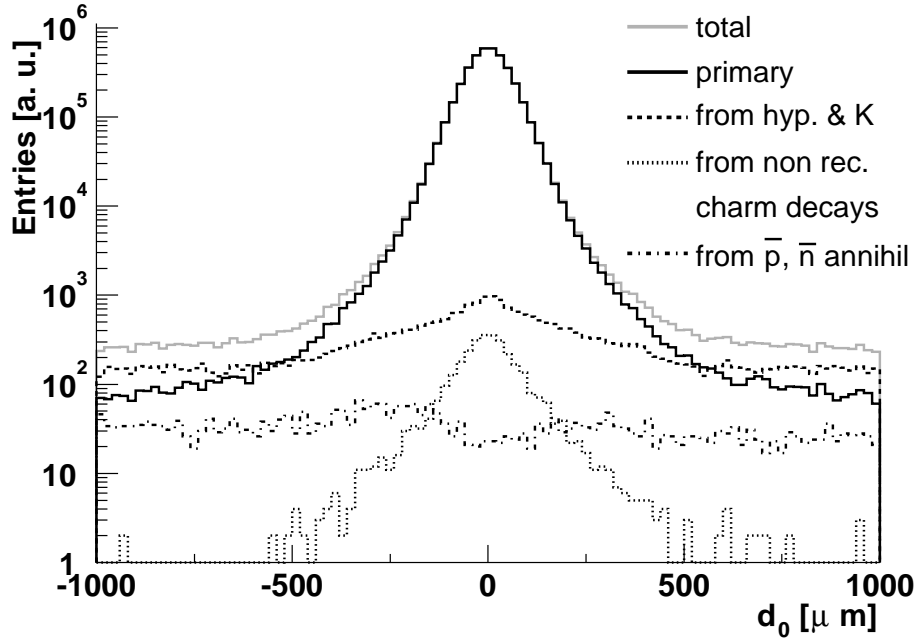}
    \caption{Impact parameter distribution for
             pions coming from the different background sources. 
             The analysis cut of $\pt>800~\mev/c$ is applied.} 
    \label{fig:d0bkg}
  \end{center}
\end{figure}

We consider only the impact parameter projection on the transverse
plane ($d_0(r\phi)$, simply indicated as $d_0$ in the following) 
since it is measured much more precisely than that along the $z$
direction (see Section~\ref{CHAP5:d0VSpt}). 
The impact parameter distribution for the various sources of
background listed at the beginning of this chapter is shown in 
Fig.~\ref{fig:d0bkg}. It can be seen that for large impact parameters 
($|d_0|> 500~\mum$) the dominant background comes from the decay of hyperons
and kaons. Indeed, we have found that an upper cut on $d_0$ 
gives an efficient rejection of this background contribution.

\begin{figure}
  \begin{center}
    \leavevmode
    \includegraphics[width=.95\textwidth]{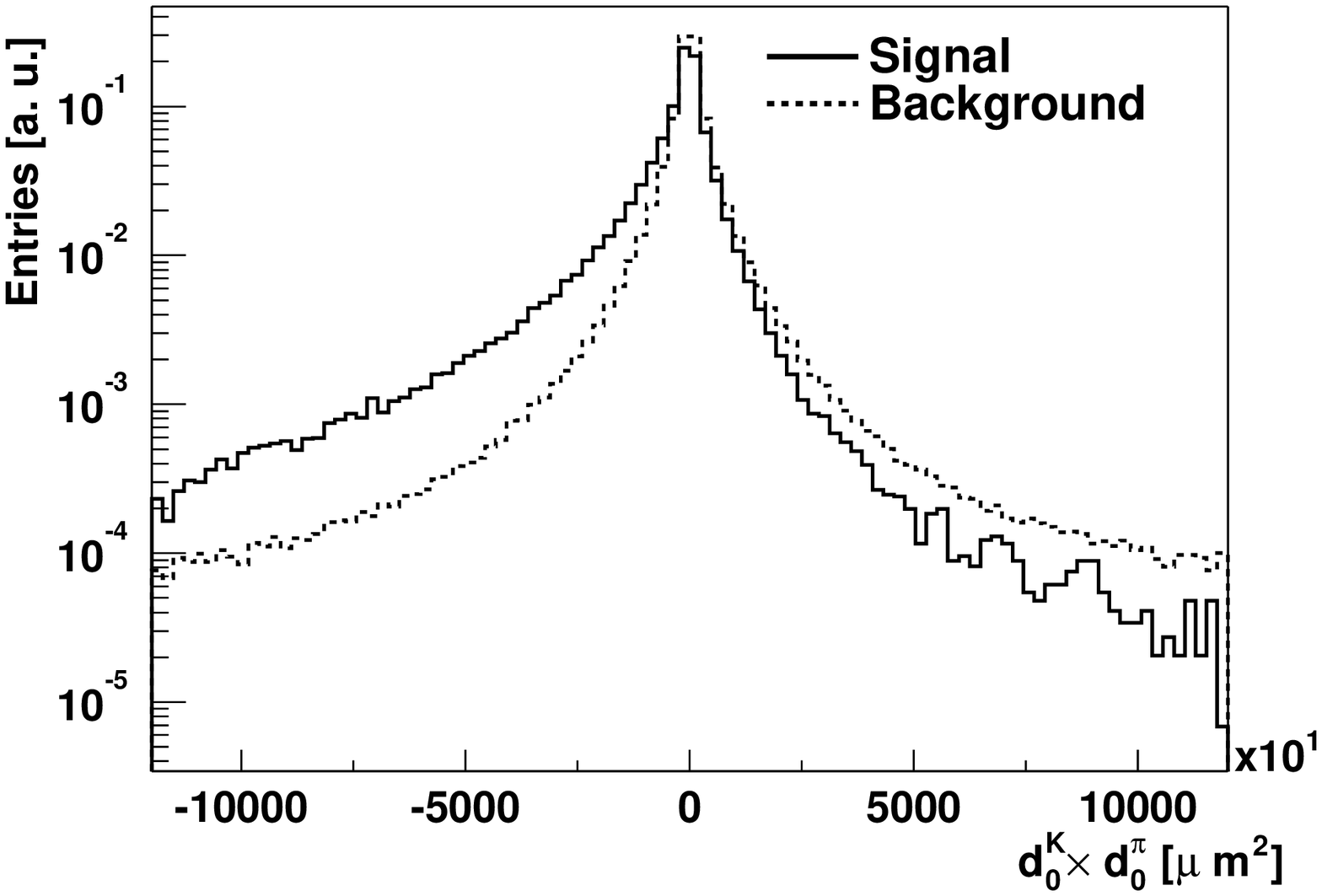}
    \caption{Product of the pion and kaon $r\phi$ impact parameters
    for signal and background combinations. The two distributions are
    normalized to the same integral.}  
    \label{fig:d0d0PbPb}  
    \includegraphics[width=.9\textwidth]{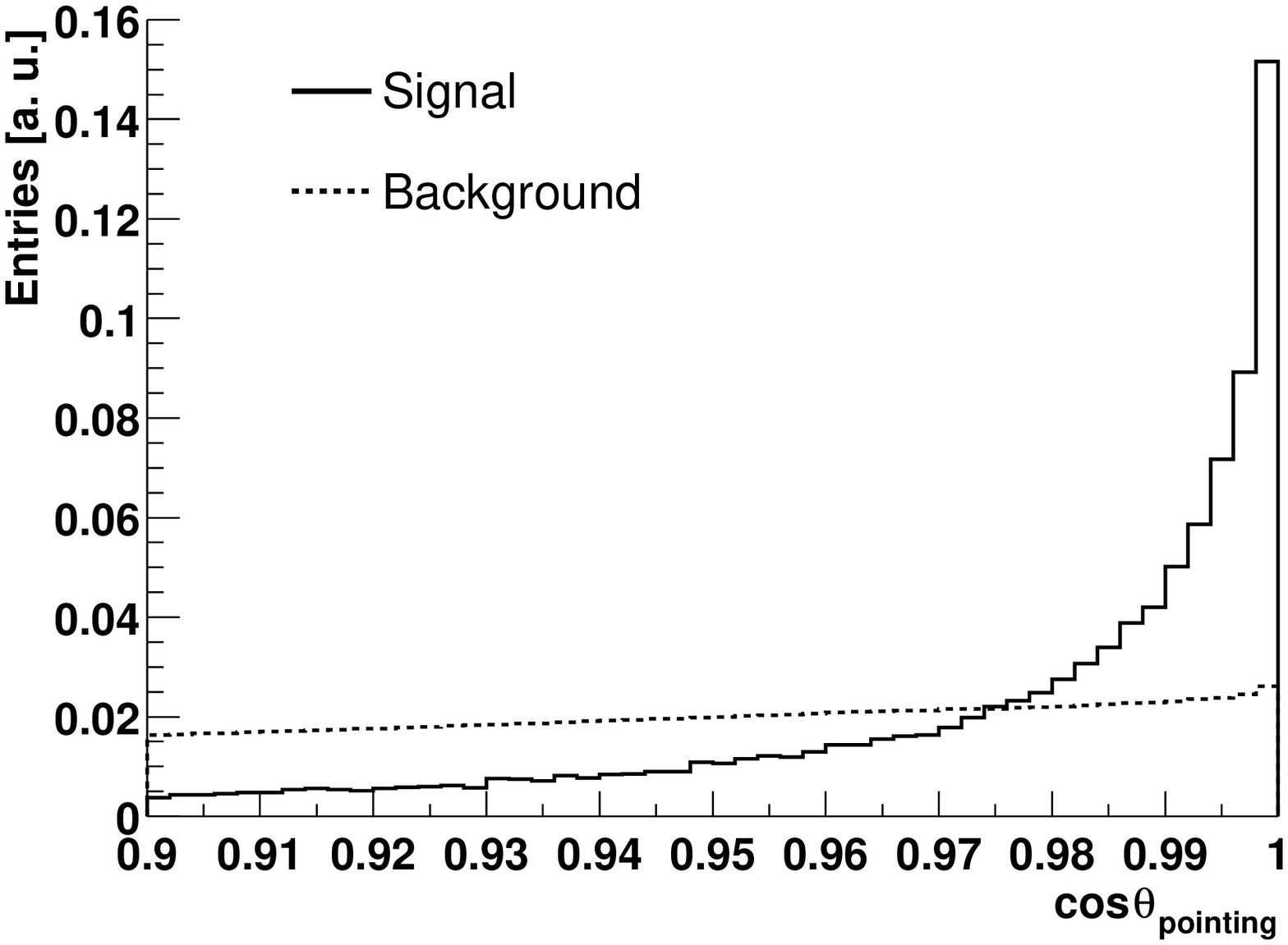}
    \caption{Cosine of the pointing angle for signal and background
    combinations. The two distributions are normalized to the same
    integral.}  
    \label{fig:thetapPbPb}
  \end{center}
\end{figure}

The projection of the tracks on the bending plane allows us to define
a sign for the impact parameter. This sign is positive or negative 
according to the position of the track projection with respect to the 
primary vertex
(the orientation is given by the direction of the track momentum). The
tracks of opposite charge originating from a $\Dz$ decaying far from
the primary vertex will then have impact parameters of opposite signs
and large in absolute value. A very appropriate variable for selection is the
product of the two transverse projections of the impact parameters.
For true decays this quantity should tend to be negative and large in absolute
value. In Fig.~\ref{fig:d0d0PbPb} we plot the distribution of the product
of impact parameters for signal and background, normalized to the same
integral. The cut $d_0^{\rm K}\times d_0^{\pi}<-40000~\mum^2$ 
improves the $S/B$ ratio by a factor $\simeq 10$.

The condition for the $\Dz$ to point back to the primary vertex is
imposed by a cut on the angle between the momentum vector of the $\Dz$
candidate and the line connecting the primary and the secondary vertex
(pointing angle $\theta_{\rm pointing}$). The cosine of
$\theta_{\rm pointing}$ peaks at $+1$ for the signal, and is almost uniformly
distributed for the background, as shown
in Fig.~\ref{fig:thetapPbPb}. Requiring to have 
$\cos\theta_{\rm pointing}>0.98$ 
would also give, by itself, a background rejection of about one order 
of magnitude.

\begin{figure}[!t]
  \begin{center}
    \includegraphics[width=.9\textwidth]{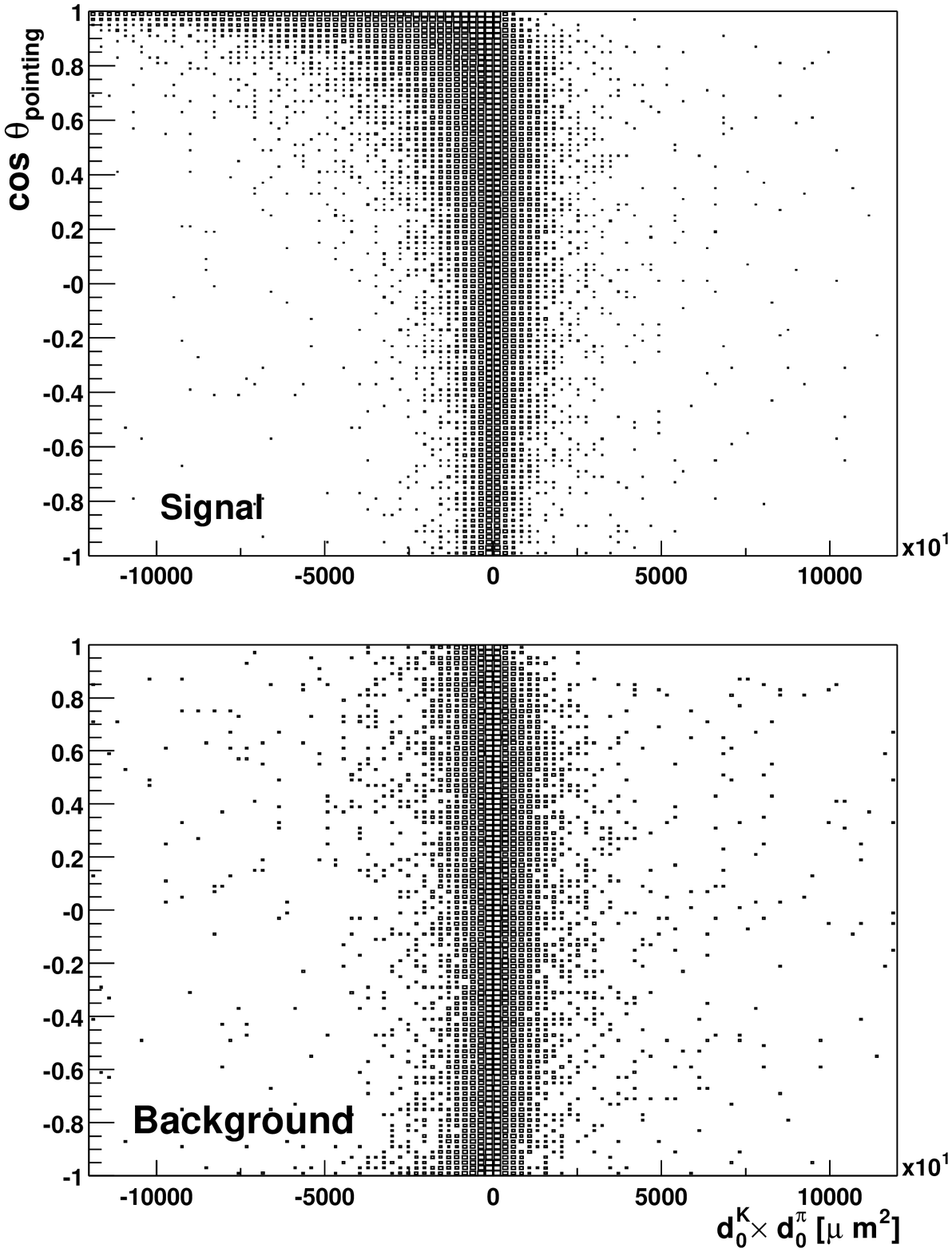}
    \caption{Cosine of the pointing angle versus product of the impact
    parameters for signal and background combinations.}  
    \label{fig:bidimPbPb}
    \end{center}
\end{figure}

A much larger rejection factor can be obtained by combining these two
cuts. In fact, if the secondary vertex is well separated from the primary one,
the impact parameters are large and the pointing angle is small, since the 
$\Dz$ flight direction is measured with a better resolution. Therefore, 
the two variables are strongly correlated in the signal, while this 
correlation is absent in the background. 
 This can be seen in Fig.~\ref{fig:bidimPbPb}, which shows the
bidimensional plot of $\cos\theta_{\rm pointing}$ versus the product of
impact parameters. The improvement in the signal-to-background ratio 
obtained by applying the combined cut is about a factor $10^3$. 

\begin{figure}[!t]
  \begin{center}
  \begin{minipage}{0.5\textwidth}
    \begin{center}
      \includegraphics[width=\textwidth]{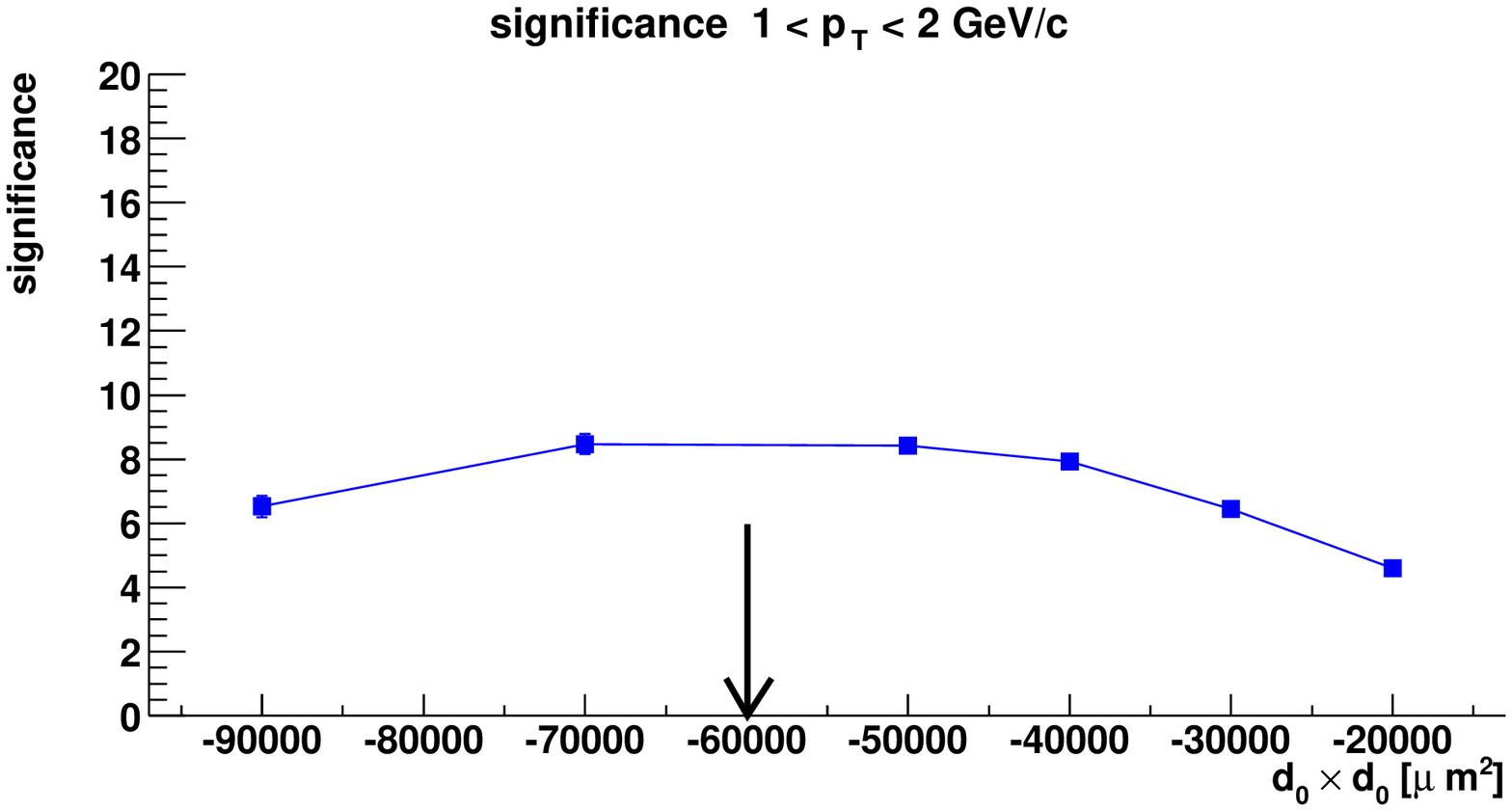}
    \end{center}
  \end{minipage}%
  \begin{minipage}{0.5\textwidth}
    \begin{center}
      \includegraphics[width=\textwidth]{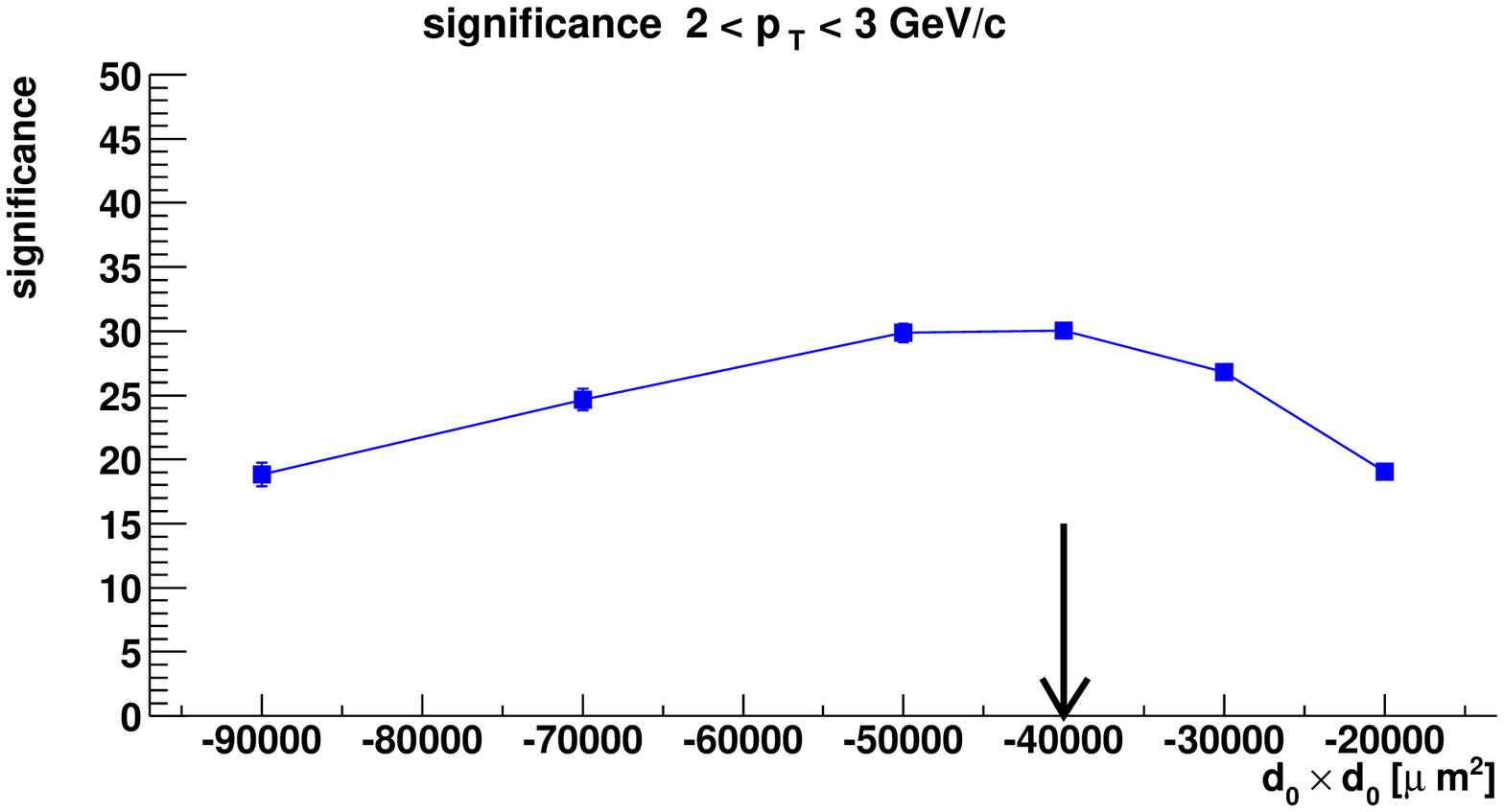}
    \end{center}
  \end{minipage} \\
  \begin{minipage}{0.5\textwidth}
    \begin{center}
      \includegraphics[width=\textwidth]{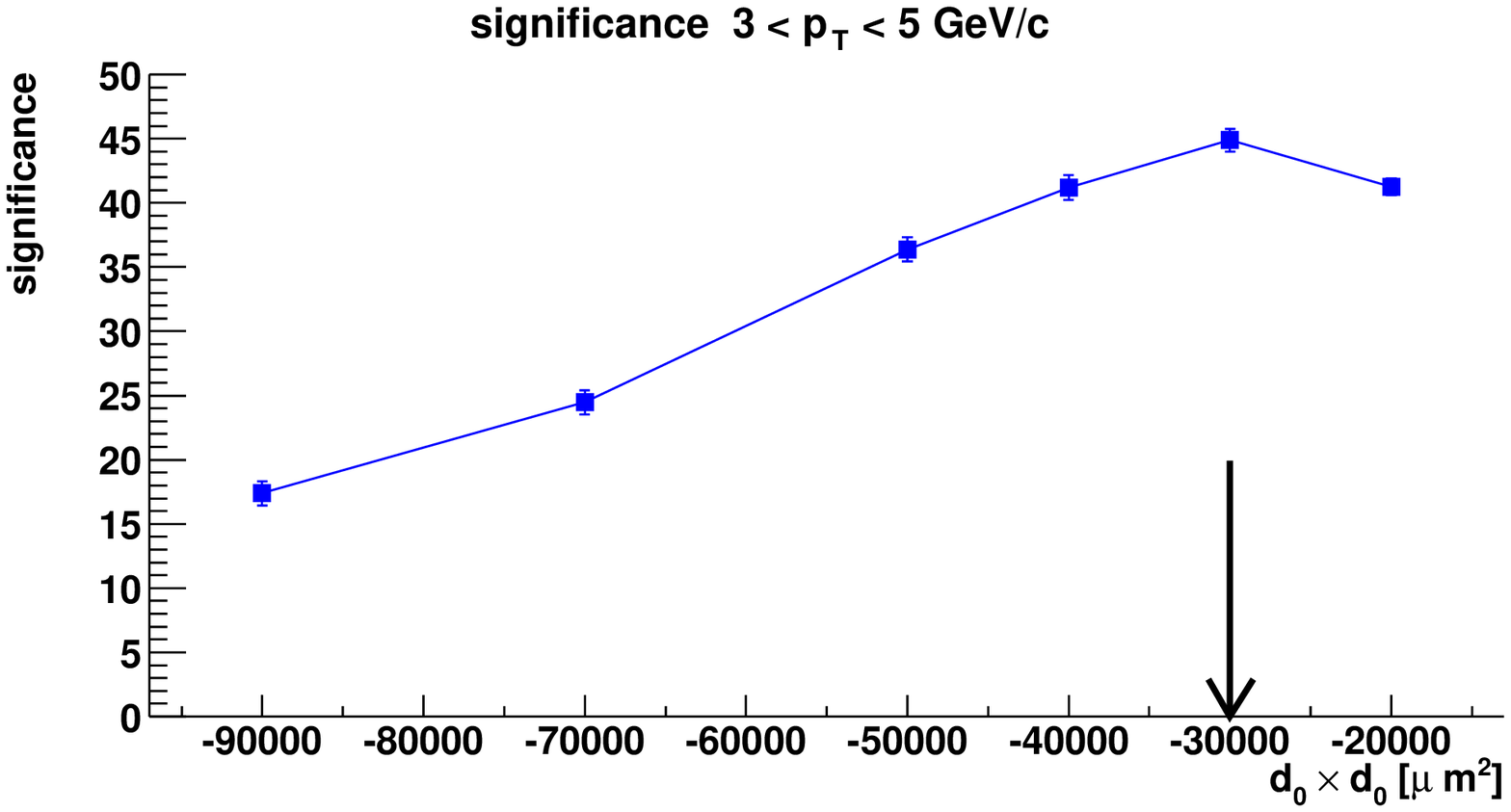}
    \end{center}
  \end{minipage}%
  \begin{minipage}{0.5\textwidth}
    \begin{center}
      \includegraphics[width=\textwidth]{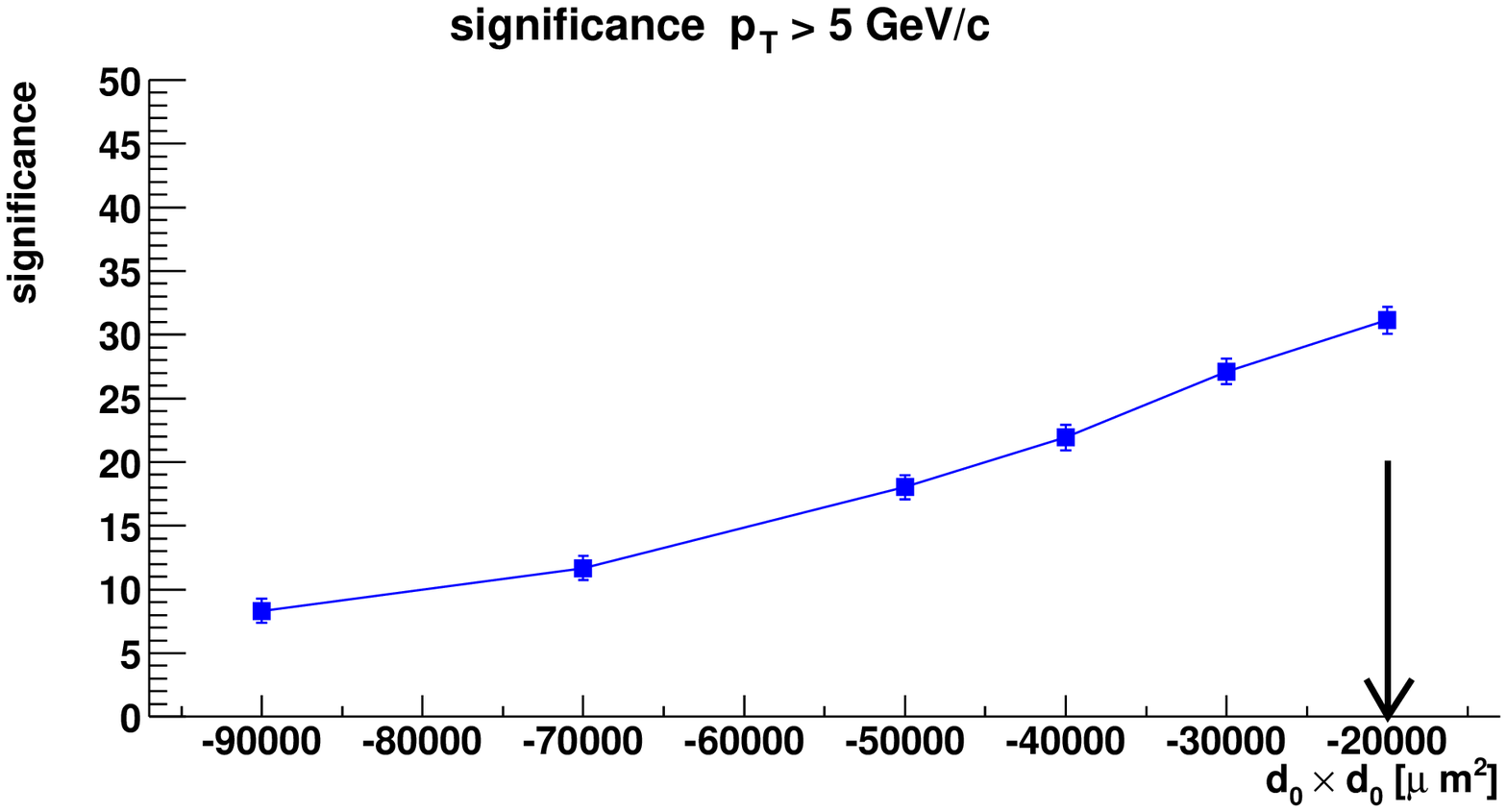}
    \end{center}
  \end{minipage}
  \end{center}
  \caption{Tuning of the $d_0^{\rm K}\times d_0^{\pi}$ cut for the different
  $\pt^{\rm D^0}$ bins. The arrows mark the values chosen for the cut.}
  \label{fig:d0d0tuning}
\end{figure}

Each cut was studied in order to maximize the 
statistical significance $S/\sqrt{S+B}$, calculated for $10^7$ central 
Pb-Pb events. Also, the optimization of the cuts was 
done separately for the following
bins in the $\pt$ of the $\Dz$: $1<\pt<2\ \gev/c$, $2<\pt<3\ \gev/c$,
$3<\pt<5\ \gev/c$, $\pt>5\ \gev/c$. The optimization
procedure consists in varying one cut at a time while the others are
kept constant and selecting the value of the cut which maximizes 
the significance. 
As an example, 
Fig.~\ref{fig:d0d0tuning} shows the tuning of the
$d_0^{\rm K}\times d_0^{\pi}$ cut for the different $\pt$ bins. The
significance is plotted as a function of the value of the cut. For
larger momenta the maximum of the significance is found at higher
values of the cut, since the impact parameter resolution improves as
$\pt$ increases. 

\begin{table}
  \caption{Final value of the cuts in the different $\pt$ bins.}
\begin{center}
\scriptsize
  \begin{tabular}{c|c|c|c|c}
    \hline
    \hline
    Cut name &  $1<\pt<2\ \gev/c$ & $2<\pt<3\ \gev/c$ &
$3<\pt<5\ \gev/c$ & $\pt>5\ \gev/c$ \\
\hline
distance of & & & & \\
closest approach & & & & \\
$(dca)$ & $<400\ \mum$ & $<300\ \mum$ & $<300\ \mum$ &
  $<300\ \mum$ \\
\hline
decay angle &&&&\\
$|\cos\theta^\star|$ & $<0.6$ &
    $<0.6$ & $<0.6$ &
    $<0.6$\\
\hline
K, $\pi$ $\pt$ & $>800~\mev/c$ &
$>800~\mev/c$ & $>800~\mev/c$ &
$>800~\mev/c$\\
\hline
K, $\pi$ $|d_0|$ & $<700\ \mum$ & $<500\ \mum$ &
$<500\ \mum$ & $<500\ \mum$ \\
\hline
$\Pi d_0=$&&&&\\
$d_0^{\rm K}\times d_0^\pi$ & $< -60000\ \mum^2$ & $<
-40000\ \mum^2$ & $< -30000\ \mum^2$ & $<
-20000\ \mum^2$ \\
\hline
pointing angle&&&&\\
$\cos\theta_{\rm pointing}$ & $>0.95$ & $> 0.98$ &
$> 0.98$ & $> 0.98$ \\
    \hline
    \hline
  \end{tabular}
  \label{tab:finalCut}
\end{center}
\end{table}

\begin{figure}[!t]
  \begin{center} 
    \includegraphics[width=.8\textwidth]{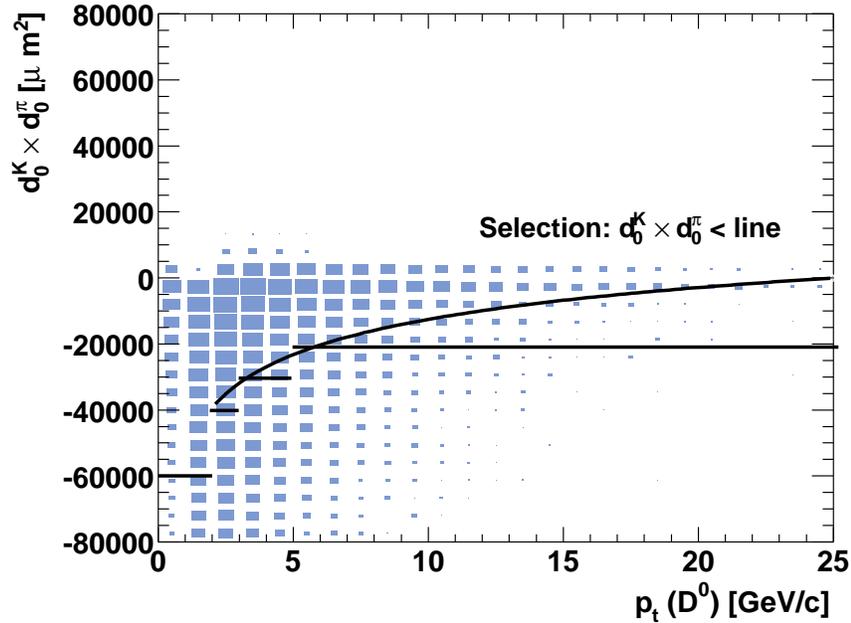}
    \caption{Product of the impact parameters as a function of the $\Dz$ $\pt$
             for signal candidates. The step-like cut obtained by the 
             cut-tuning procedure is shown. The line starting from $2~\gev/c$
             represents the smooth cut chosen in order to avoid the loss 
             of signal at high $\pt$.}
    \label{fig:d0d0VSpt}
    \end{center}
\end{figure}

In Table~\ref{tab:finalCut} the final values of the 
cuts are reported. The most sensible cut is the one on the product of the 
impact parameters, since it selects the tail of the distribution 
in order to exploit the different shapes of signal and background
(as shown in Fig.~\ref{fig:d0d0PbPb}). Therefore, we studied more carefully 
the $\pt$ dependence of this cut. Figure~\ref{fig:d0d0VSpt} shows 
the bidimensional plot of $d_0^{\rm K}\times d_0^{\pi}$ versus $\pt^{\rm D^0}$ 
for the signal candidates, after all 
other cuts have been applied, as reported in Table~\ref{tab:finalCut}. 
The distribution becomes very narrow at high $\pt$ as a consequence of 
the strong $\pt$ dependence of the impact parameter resolution.
The step-like cut obtained by the tuning procedure 
described before is shown. 
From the shape of the distribution it is clear that 
using this step-like cut would determine the loss of most of the 
signal for $\pt>8$-$10~\gev/c$; this high-$\pt$ region is extremely 
important for the parton energy loss studies, as we will see in 
Chapter~\ref{CHAP8}.
It is, therefore, mandatory to have a `really' $\pt$-dependent cut, 
for $\pt>2~\gev/c$, such as the one indicated by the line in the figure. 

\subsection{Results}
\label{CHAP6:resultsPbPb}

\begin{figure}[!b]
  \begin{center} 
    \includegraphics[width=.8\textwidth]{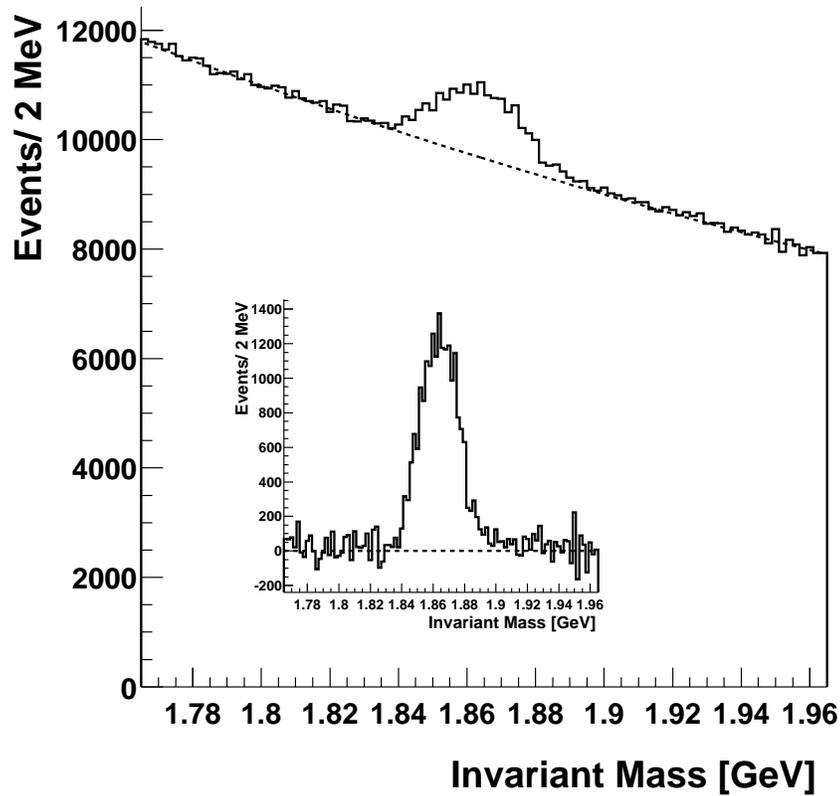}
    \caption{$\K\pi$ invariant mass distribution for $10^7$ events. The
    same distribution after background subtraction is shown in the
    inset.}
    \label{fig:minvtotPbPb}
    \end{center}
\end{figure}

Figure~\ref{fig:minvtotPbPb} shows the invariant mass distribution for
the sum of samples A, B and C after selection, corresponding to $10^7$
events.

In Table~\ref{tab:finalPbPb} the values for $S/$event, $B/$event and $S/B$ ,
in the invariant mass range $|M_{\rm K\pi}-M_{\rm D^0}|<1~\sigma$, are
presented. In the same table,
we report also the statistical significance $S/\sqrt{S+B}$ for $10^7$ 
central \PbPb~events and the relative error $\sigma_S/S$ on the 
estimation of the 
number $S$ of detected $\Dz$ mesons. As we shall demonstrate in 
Chapter~\ref{CHAP7} this relative error is $\sqrt{S+B}/S$, i.e. the 
inverse of the significance.

\begin{figure}[!t]
  \begin{center}
    \includegraphics[width=\textwidth]{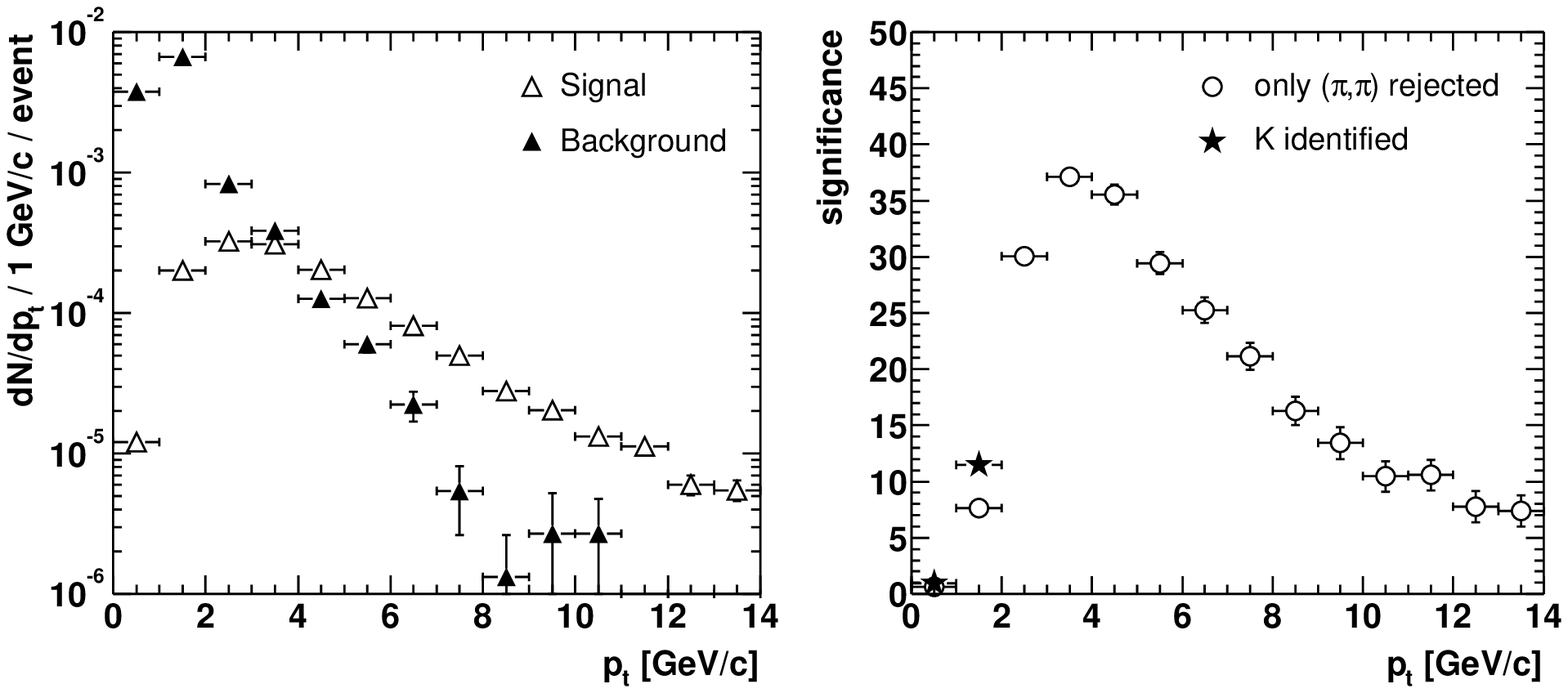}
    \caption{Transverse momentum distribution for the signal and for the 
    background after selection (left);  
    the normalization corresponds to 1 central \PbPb~event. Corresponding
    significance for $10^7$ events as a function of $\pt$ (right). The full 
    markers
    shows the significance obtained for $\pt<2~\gev/c$ requiring 
    the identification of the kaon in the Time of Flight.}
    \label{fig:resultsPt_PbPb}
  \end{center}
\end{figure}

Considering the sum of the three samples A, B and C, the 
$\pt$-integrated significance is 37. Figure~\ref{fig:resultsPt_PbPb} 
shows the $\pt$ distribution of the signal and of the background absolutely
normalized and the significance as a function of $\pt$, in bins of $1~\gev/c$.
With $10^7$ events the significance is larger than 10 up to about 
$10~\gev/c$ of $\pt$. For $\pt>4~\gev/c$, the $S/B$ ratio grows but the 
significance decreases due to the decrease in the signal statistics.   

\begin{figure}[!b]
  \begin{center}
    \includegraphics[width=.75\textwidth]{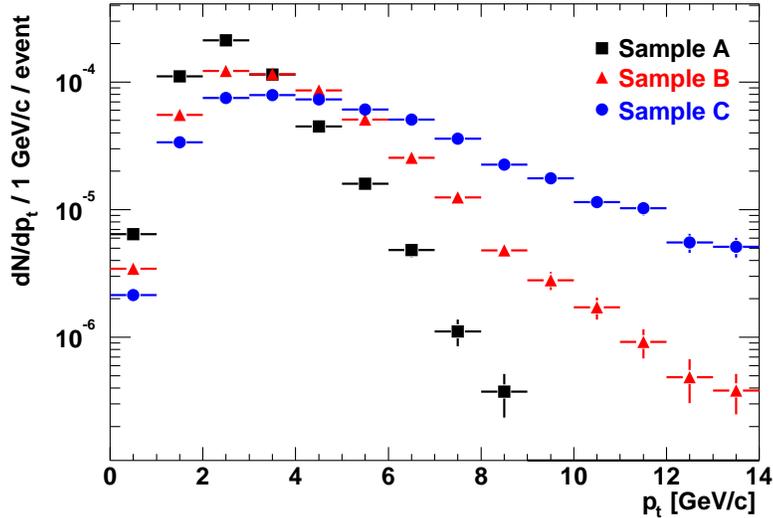}
    \caption{Transverse momentum distributions for the signal in the 
    three TOF-PID samples.}
    \label{fig:samplesVSptPbPb}
  \end{center}
\end{figure}

In Fig.~\ref{fig:samplesVSptPbPb} we show how the signal is distributed
in the three PID classes, as a function of the transverse momentum: sample 
A covers the low-$\pt$ region, where the kaon can be efficiently identified 
in the TOF detector; samples B and C, even if their integrated significances 
are quite low, are essential to cover the high-$\pt$ region above $5~\gev/c$.

\begin{figure}[!t]
  \begin{center}
    \includegraphics[width=\textwidth]{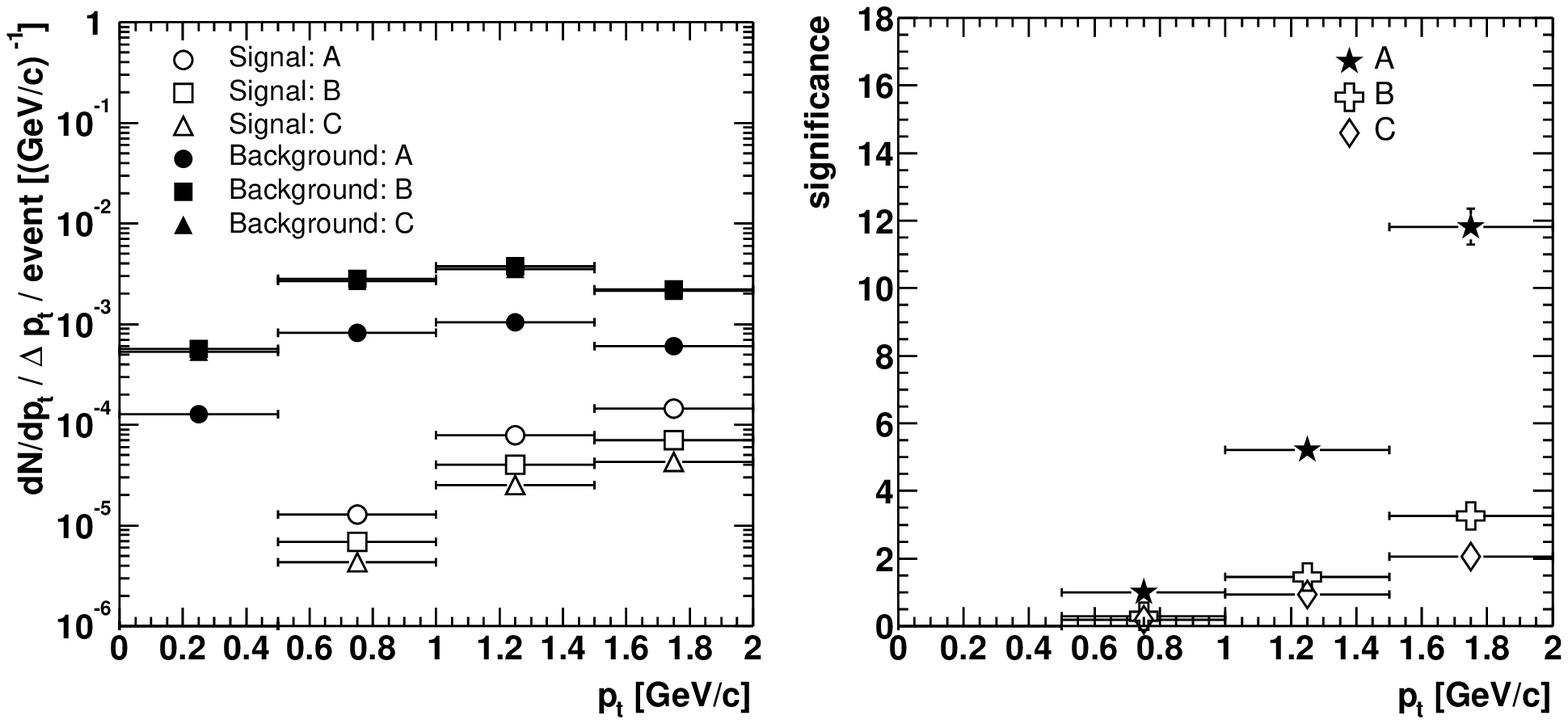}
    \caption{Transverse momentum distributions for the signal and for the 
    background after selection for $\pt<2~\gev/c$ (left);  
    the normalization corresponds to 1 central \PbPb~event. Corresponding
    significance for $10^7$ events as a function of $\pt$ (right).}
    \label{fig:lowptPbPb}
    \vglue0.3cm
    \includegraphics[width=.6\textwidth]{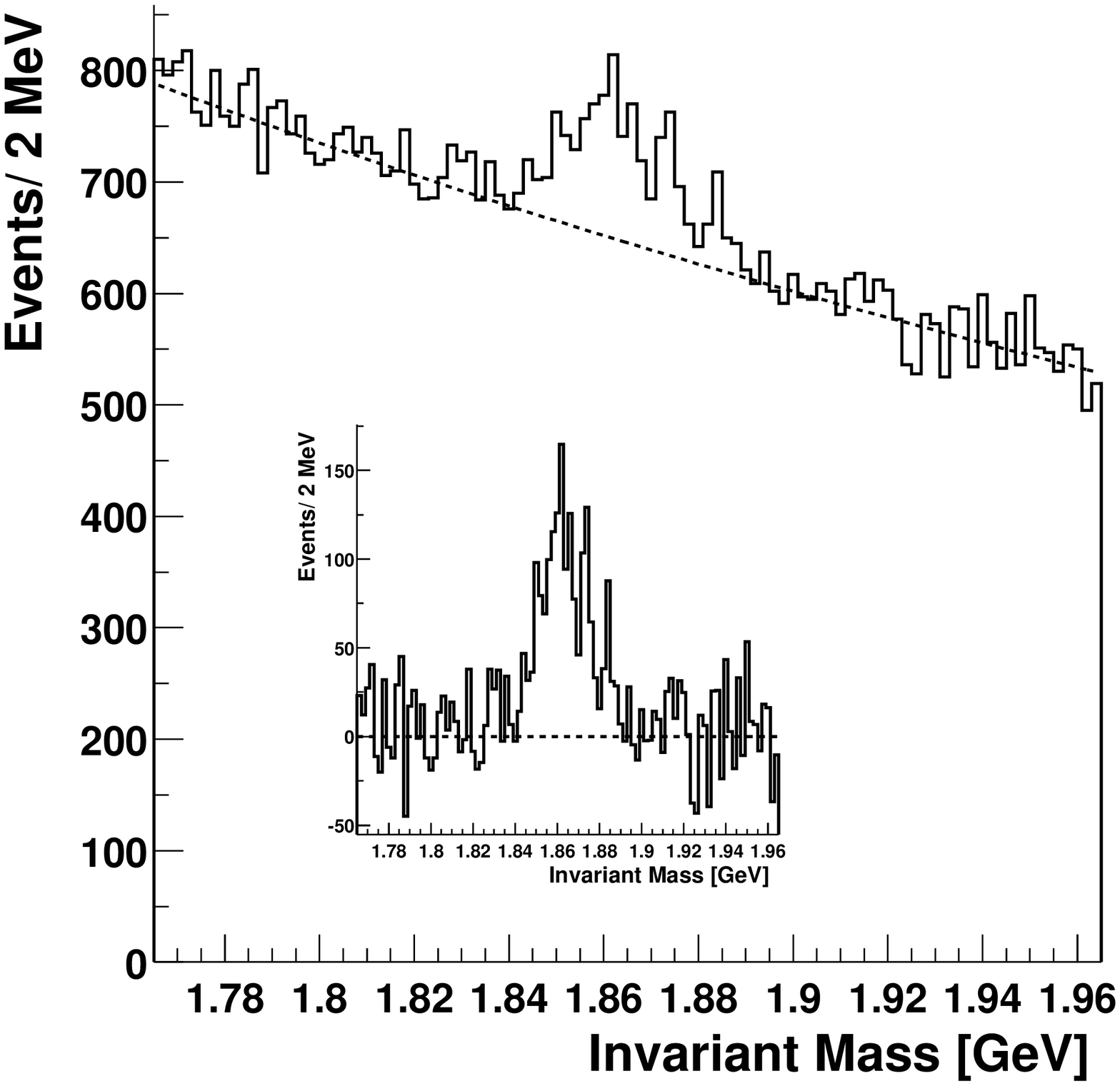}
    \caption{$\K\pi$ invariant mass distribution in the bin $1<\pt<2~\gev/c$ 
    for the sample of candidates with kaon identified in the 
    Time of Flight ($10^7$ events).}
    \label{fig:minvbin1PbPb}
    \end{center}
\end{figure}

In general, it is convenient to merge the three samples, which essentially
corresponds to rejecting only $(\pi_{\rm tag},\,\pi_{\rm tag})$
pairs\footnote{$(\rm K_{\rm tag},\,K_{\rm tag})$ pairs and pairs with 
an identified proton are a small fraction of the total background.}
using the TOF information.
However, this strategy gives the quite marginal significance of 8 in the bin 
$1<\pt<2~\gev/c$.
Since for $\pt<2~\gev/c$ sample A contains 
most of the signal and only a small fraction of the background 
(see Fig.~\ref{fig:lowptPbPb}),
considering only candidates with the kaon identified 
(sample A) yields a significance of 12 in $1<\pt<2~\gev/c$ (as showed by the 
star markers in the right panel of Fig.~\ref{fig:resultsPt_PbPb}). 
Figure~\ref{fig:minvbin1PbPb} shows the invariant mass distribution for this 
sample in $1<\pt<2~\gev/c$; even for this low-$\pt$ bin, the signal is well 
visible over the background.

\begin{table}[!t]
  \caption{Final values of $S/B$ and $S/\sqrt{S+B}$ for $10^7$ \PbPb~events.}
  \begin{center}  \begin{tabular}{cccccc}
    \hline 
    \hline
    Sample & $S/$event & $B/$event & $S/B$ & $S/\sqrt{S+B}$ ($10^7$
  events) & $\sigma_S/S$ \\
    \hline
    A & $4.4\cdot10^{-4}$ & $1.4\cdot10^{-3}$ & $32\%$ &
  $33$ & $3\%$ \\
    B & $4.3\cdot10^{-4}$ & $5.2\cdot10^{-3}$ & $8\%$ &
  $8$ & $13\%$ \\
    C & $4.6\cdot10^{-4}$ & $5.0\cdot10^{-3}$ & $9\%$ &
  $9$ & $11\%$ \\
    \hline
    Total & $1.3\cdot10^{-3}$ & $1.16\cdot10^{-2}$ & $11\%$ &
  $37$ & $3\%$ \\
    \hline
    \hline
  \end{tabular}
  \label{tab:finalPbPb}
\end{center}
\end{table}

With the choice of parameters we have used for the generation of the signal,
the fraction of the transverse momentum distribution for which we have
sensitivity ($\pt>1~\gev/c$) corresponds to about $70\%$ of the 
total $\Dz$  production cross section, at mid-rapidity.

\begin{figure}[!t]
  \begin{center}
    \includegraphics[width=.49\textwidth]{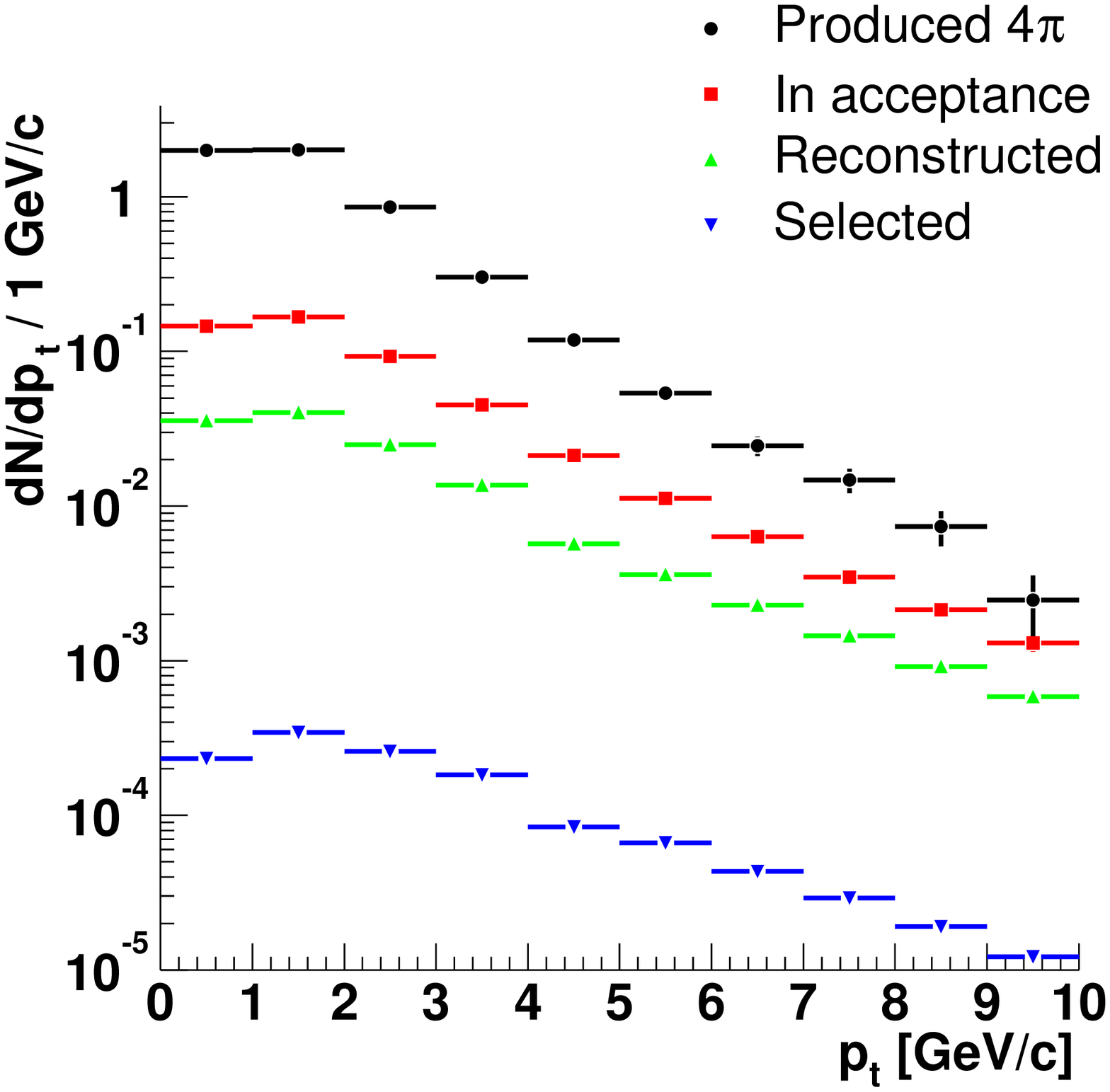}
    \includegraphics[width=.49\textwidth]{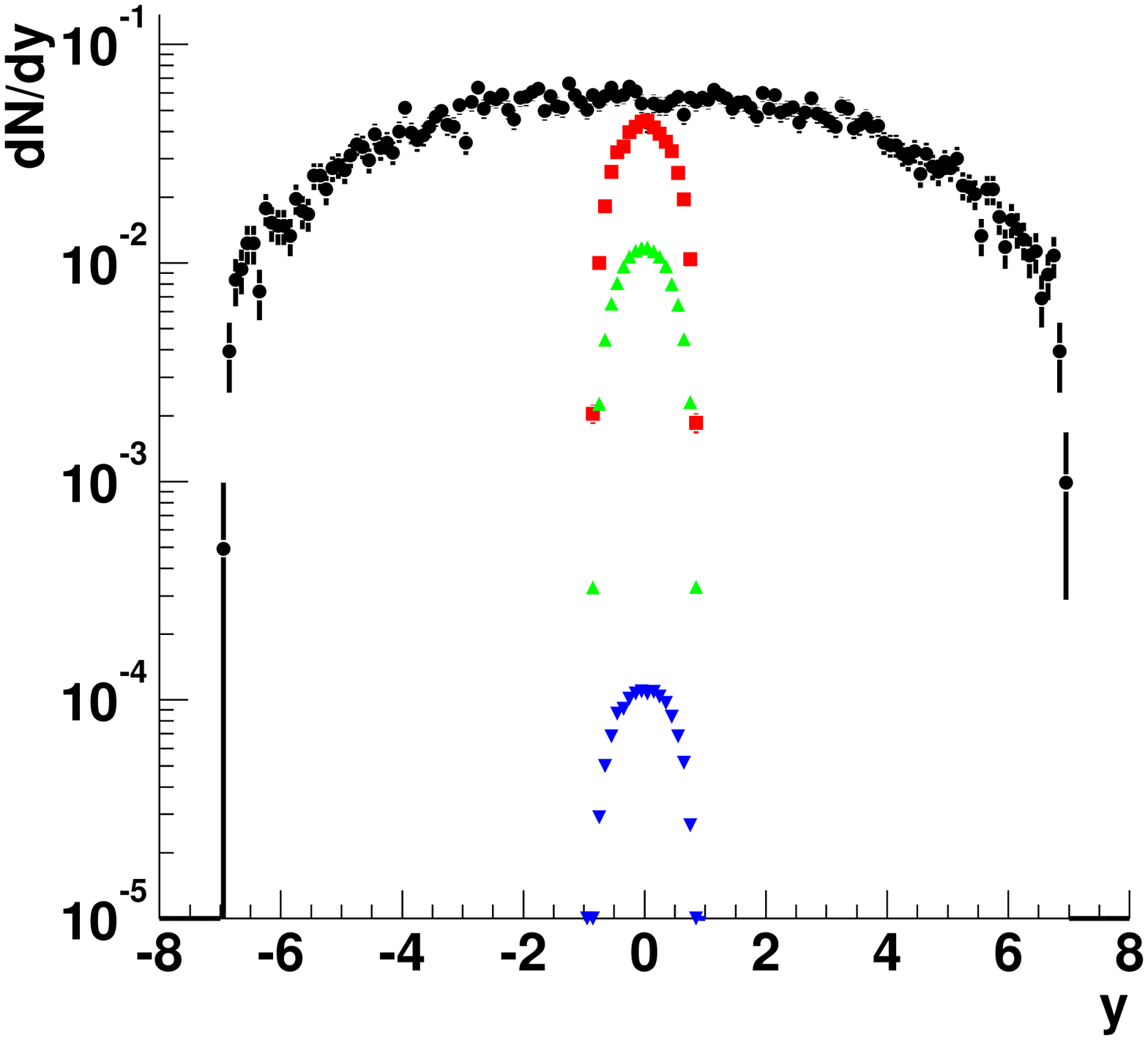}
    \vglue0.5cm
    \includegraphics[width=.49\textwidth]{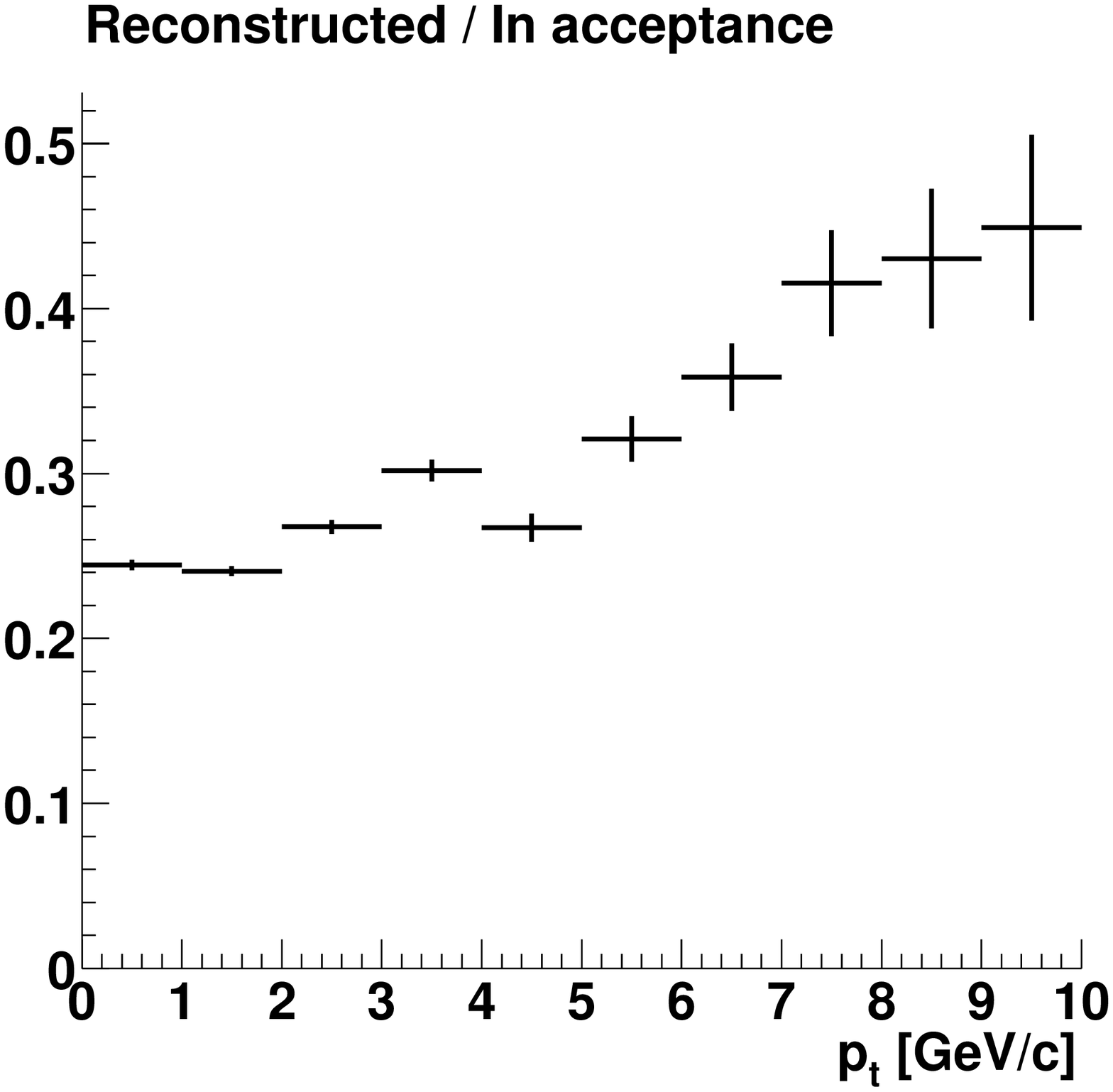}
    \includegraphics[width=.49\textwidth]{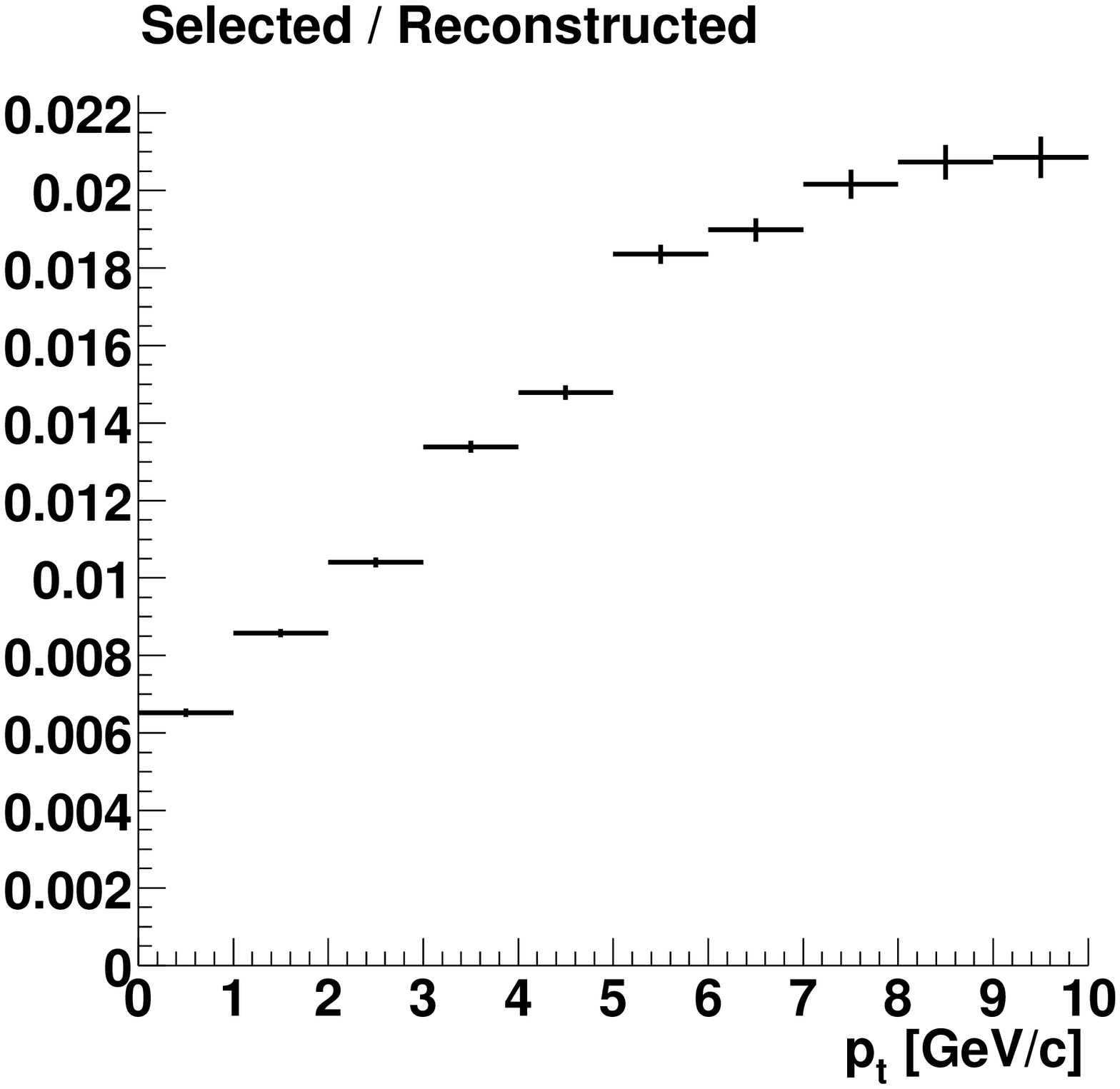}
    \caption{In the upper row, transverse momentum and rapidity 
             distributions for the $\DtoKpi$ 
             signal: produced per event, 
             with $\K$ and $\pi$ in the acceptance of the 
             barrel ($|\eta|<0.9$), reconstructed and selected. In the lower 
             row: reconstruction and selection efficiencies as a function of 
             $\pt$.}
    \label{fig:acceffPbPb}
    \end{center}
\end{figure}

\begin{figure}[!t]
  \begin{center}
    \includegraphics[width=.49\textwidth]{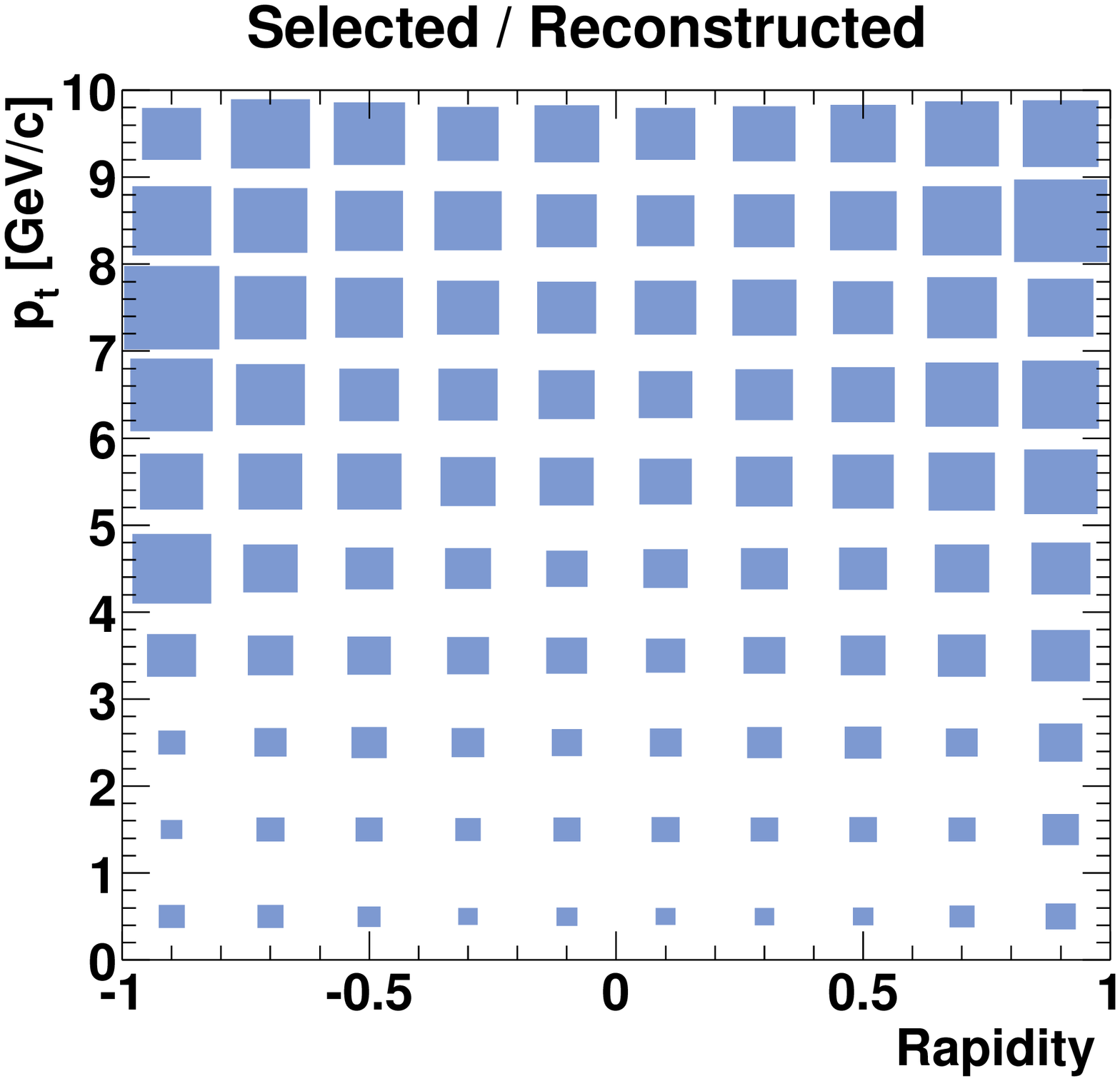}
    \includegraphics[width=.49\textwidth]{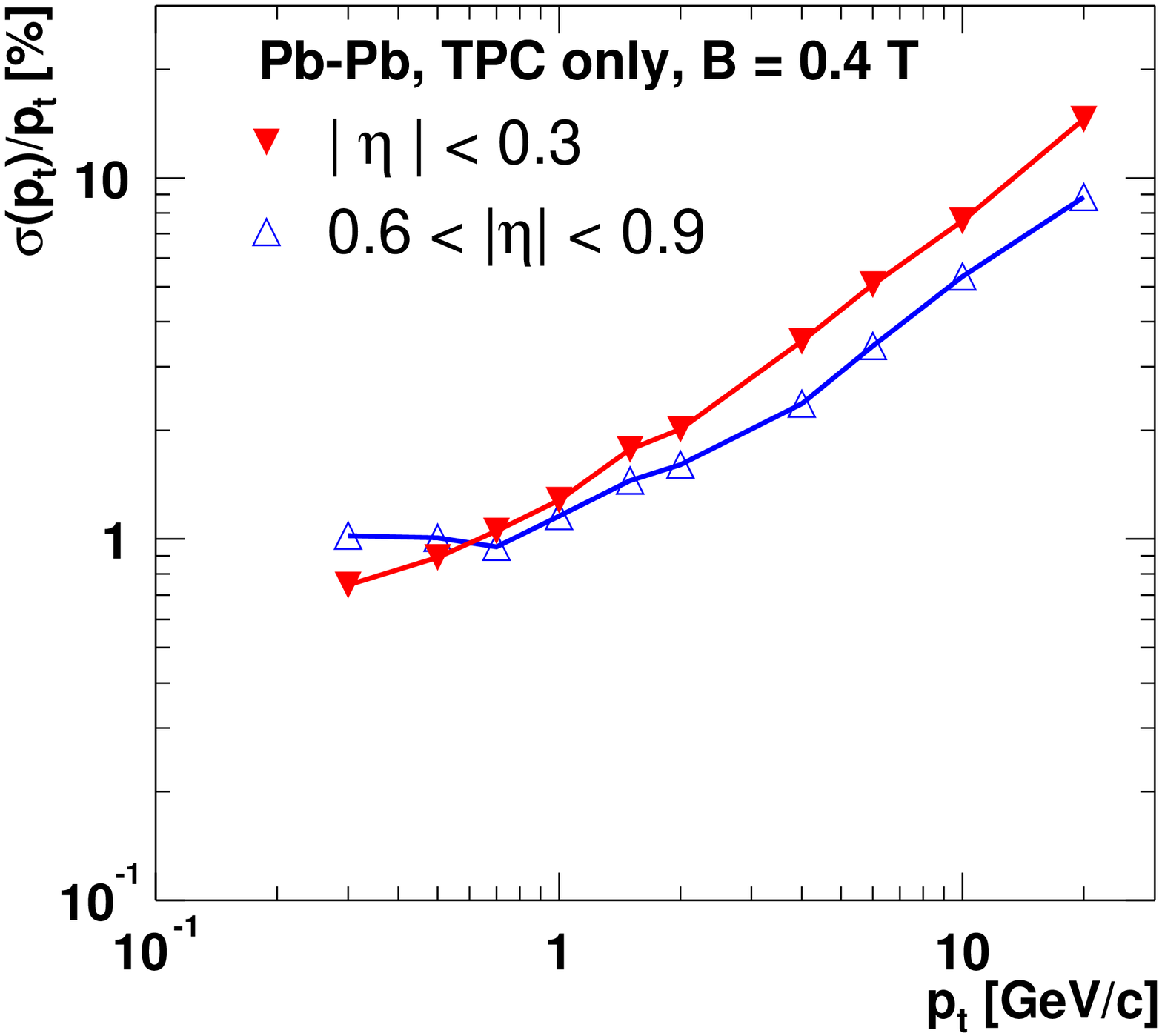}
    \caption{Left: selection efficiency as a function of $\pt$ and $y$.
             Right: relative $\pt$ resolution in the TPC 
             as a function of $\pt$ for 
             tracks in $|\eta|<0.3$ and $0.6<|\eta|<0.9$.}
    \label{fig:effVSpty}
    \end{center}
\end{figure}

The cuts applied so far, including also the $\pm 1~\sigma$ 
cut on the invariant mass, reduce the
background by a factor $4\cdot 10^{-7}$ and select $\simeq 1\%$ of the 
signal we had after track reconstruction. 
In Table~\ref{tab:historyPbPb} we summarize the `history' of the signal, 
showing the effects of acceptance, reconstruction efficiency and selection
efficiency. These effects are illustrated as a function of transverse 
momentum and rapidity in Fig.~\ref{fig:acceffPbPb}. The double-differential 
plot of the ratio of selected-to-reconstructed $\Dz$ as a function of 
$\pt$ and rapidity (Fig.~\ref{fig:effVSpty}, left) 
shows an interesting feature: at a given $\pt$, the signal is 
selected with higher efficiency at the edges of the rapidity acceptance 
($|y|\simeq 1$) than in the very central region. This is due to the fact 
that, at a given $\pt$, the $\pt$ resolution in the TPC is better 
for tracks with large $|\eta|$ (Fig.~\ref{fig:effVSpty}, right): 
for these tracks the drift length in the TPC is on average shorter than 
for tracks close to $\eta=0$ and, therefore, the diffusion effects 
are smaller and the clusters have better position resolution; only at very
low $\pt$, where multiple scattering is significant, the resolution is better
at small $|\eta|$, because less material is crossed.

\begin{table}[!t]
  \caption{`History' of the $\Dz/\overline{\Dz}$ signal in \PbPb~events.}
  \begin{center}  
   \begin{tabular}{lc}
     & $S/$event \\
    \hline
    \hline
    Total produced ($4\pi$) & 141 \\
    Decaying to $\K^{\mp}\pi^{\pm}$ & 5.4 \\
    With $\K$ and $\pi$ in $|\eta|<0.9$ & 0.5 \\
    With $\K$ and $\pi$ reconstructed & 0.14 \\
    After $(\pi_{\rm tag},\,\pi_{\rm tag})$ rejection & 0.13 \\
    After selection cuts (including $\pm 1~\sigma$ mass cut) & 0.0013 \\
    \hline
    \hline
  \end{tabular}
  \label{tab:historyPbPb}
\end{center}
\end{table}

\subsection{Results scaled to a lower-multiplicity scenario}
\label{CHAP6:multdep}

The present analysis assumes a charged particle rapidity density 
$\dNdy=6000$ for the underlying events. According to recent 
extrapolations of the RHIC data, the multiplicity is more likely to be
of the order of $\dNdy=3000$ (Section~\ref{CHAP1:sqrtsdNdy}). 
We have therefore estimated how the 
results on signal-to-background ratio and significance scale in this 
scenario.

If $\dNdy$ decreases, the number of background pairs decreases 
as $(\dNdy)^2$. Therefore, $S/B$ is proportional to $(\dNdy)^{-2}$.
The significance is proportional to $(\dNdy)^{-1}$ if $S\ll B$, so 
that $S/\sqrt{S+B}\simeq S/\sqrt{B}$. This condition holds for our 
\mbox{$\pt$-integrated} significance, since we have $S\simeq B/10$; 
for the \mbox{$\pt$-dependent} significance such a scaling can be applied 
only up to $\pt\simeq 3~\gev/c$, as for larger transverse momenta 
the significance is dominated by the statistics of the signal,
$S/\sqrt{S+B}\simeq \sqrt{S}$.
In Table~\ref{tab:multiplicityscaling} 
we just scale the \mbox{$\pt$-integrated} 
results, according to the proportionalities mentioned above, 
to a case of lower multiplicity. The $\ccbar$ production rate is not 
rescaled because it is not clear how it is correlated with the total 
multiplicity and because we have used the average of the values given by 
the different PDFs, which is already a conservative estimate.
In addition, with a lower multiplicity the tracking efficiency 
would improve and a further improvement can be expected from a refinement 
of the cuts.

\begin{table}[!h]
  \caption{$S/B$ and $S/\sqrt{S+B}$ as from this analysis ($\dNdy=6000$) 
           and scaled for a lower multiplicity ($\dNdy=3000$).}
  \begin{center}
  \begin{tabular}{ccc}
    \hline
    \hline 
    $\dNdy$ & $S/B$ & $S/\sqrt{S+B}$ ($10^7$ events) \\
    \hline
    6000 & $11\%$ & $37$ \\
    3000 & $44\%$ & $74$ \\
    \hline
    \hline	
  \end{tabular}
  \label{tab:multiplicityscaling}
\end{center}
\end{table}

\subsection{Feed-down from beauty}
\label{CHAP6:beautyPbPb}

In Section~\ref{CHAP6:generPbPb} we pointed out that, for \PbPb~collisions at 
$\sqrtsNN=5.5~\tev$, about 5\% of all the produced $\Dz$ mesons come 
from the decay of B mesons. After the described selections, the ratio of 
secondary-to-primary $\Dz$ increases to $\simeq 12\%$. Such result does not 
match the expectation that the pointing requirement
should suppress the $\Dz$ from beauty, as they point to the decay vertex 
of the B meson and not to the primary vertex.
In the following we clarify this picture by analyzing how the main 
selections affect secondary $\Dz$ particles.

\begin{figure}[!b]
  \begin{center}
    \includegraphics[width=.7\textwidth]{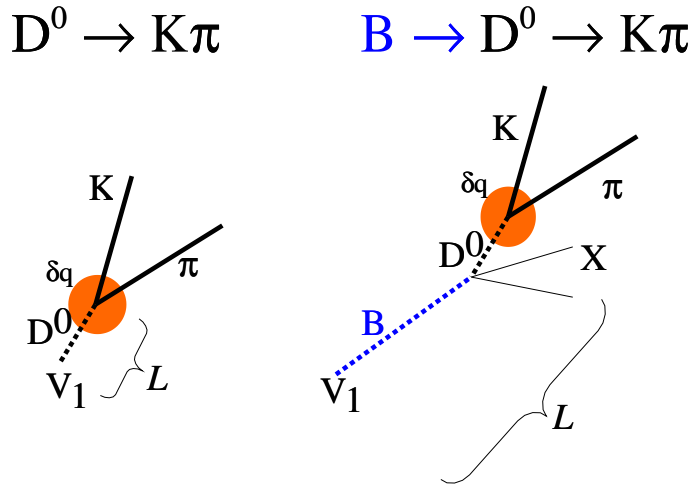}
    \caption{Sketch of the decay topologies of primary and 
             secondary $\Dz$ mesons. The shaded circle represents the 
             uncertainty $\delta q$ on the reconstructed position of 
             the secondary vertex (see text).}
    \label{fig:cptasketch}
    \end{center}
\end{figure}

\begin{figure}[!th]
  \begin{center}
    \includegraphics[width=.64\textwidth]{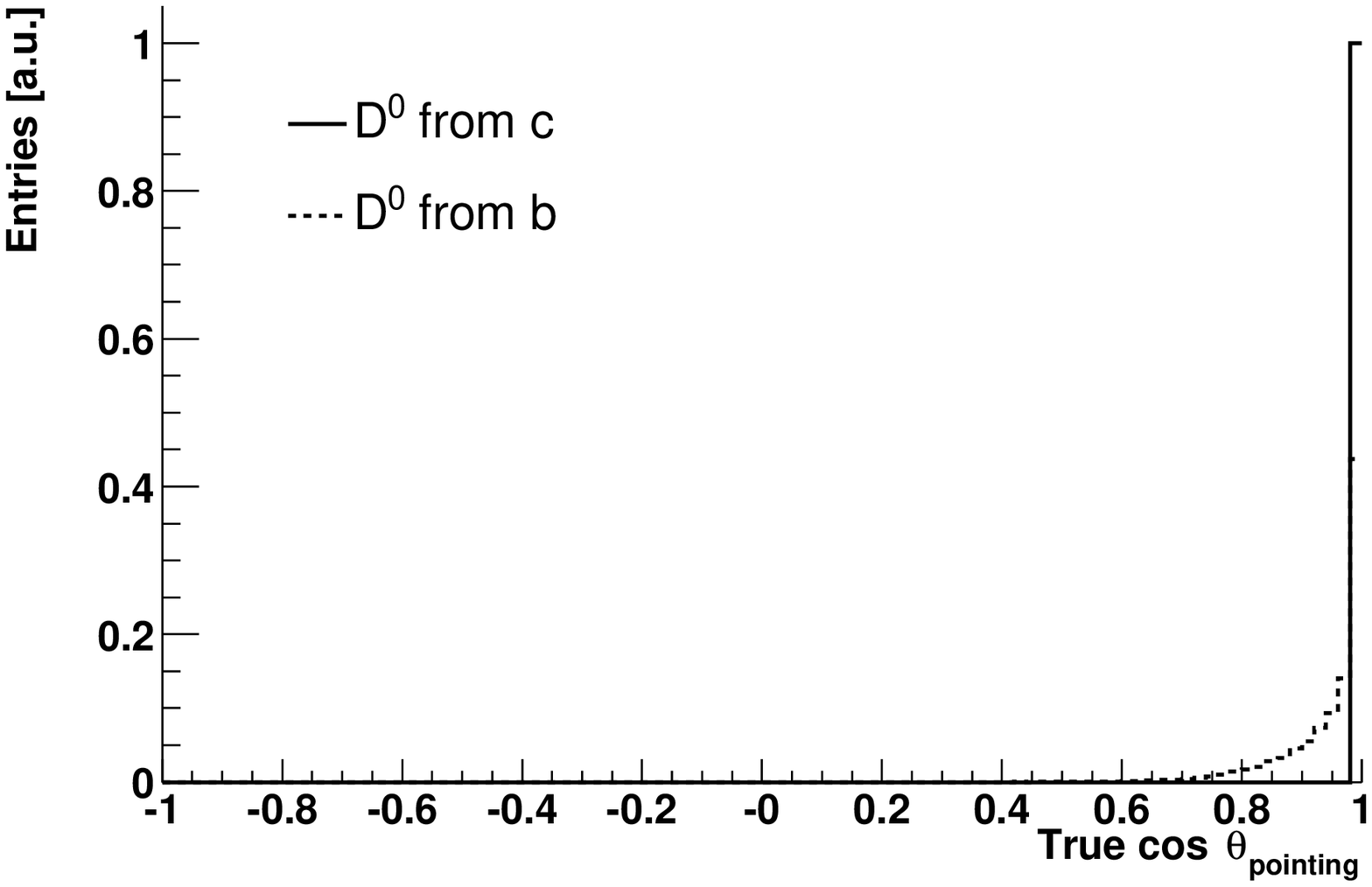}
    \includegraphics[width=.64\textwidth]{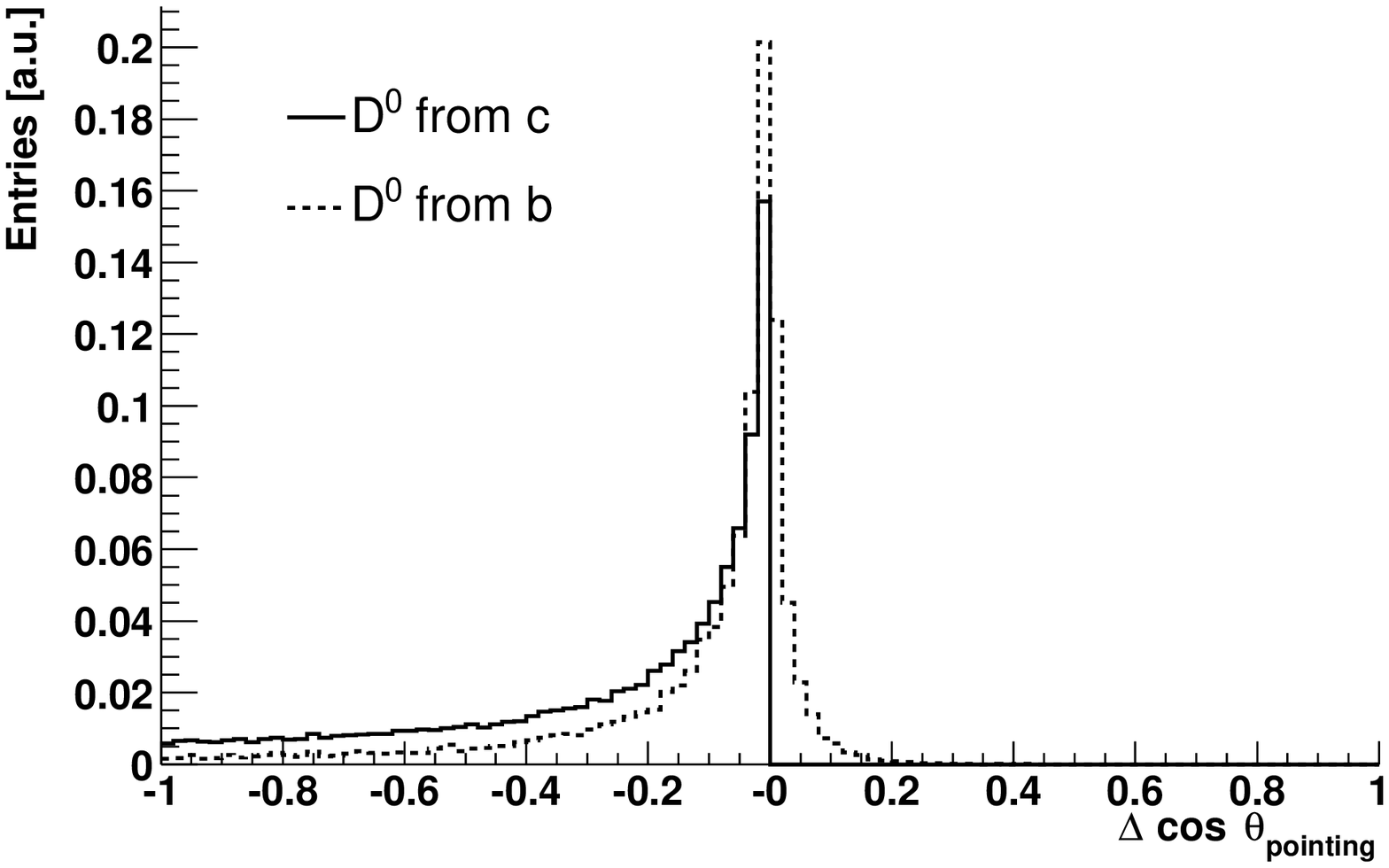}
    \includegraphics[width=.64\textwidth]{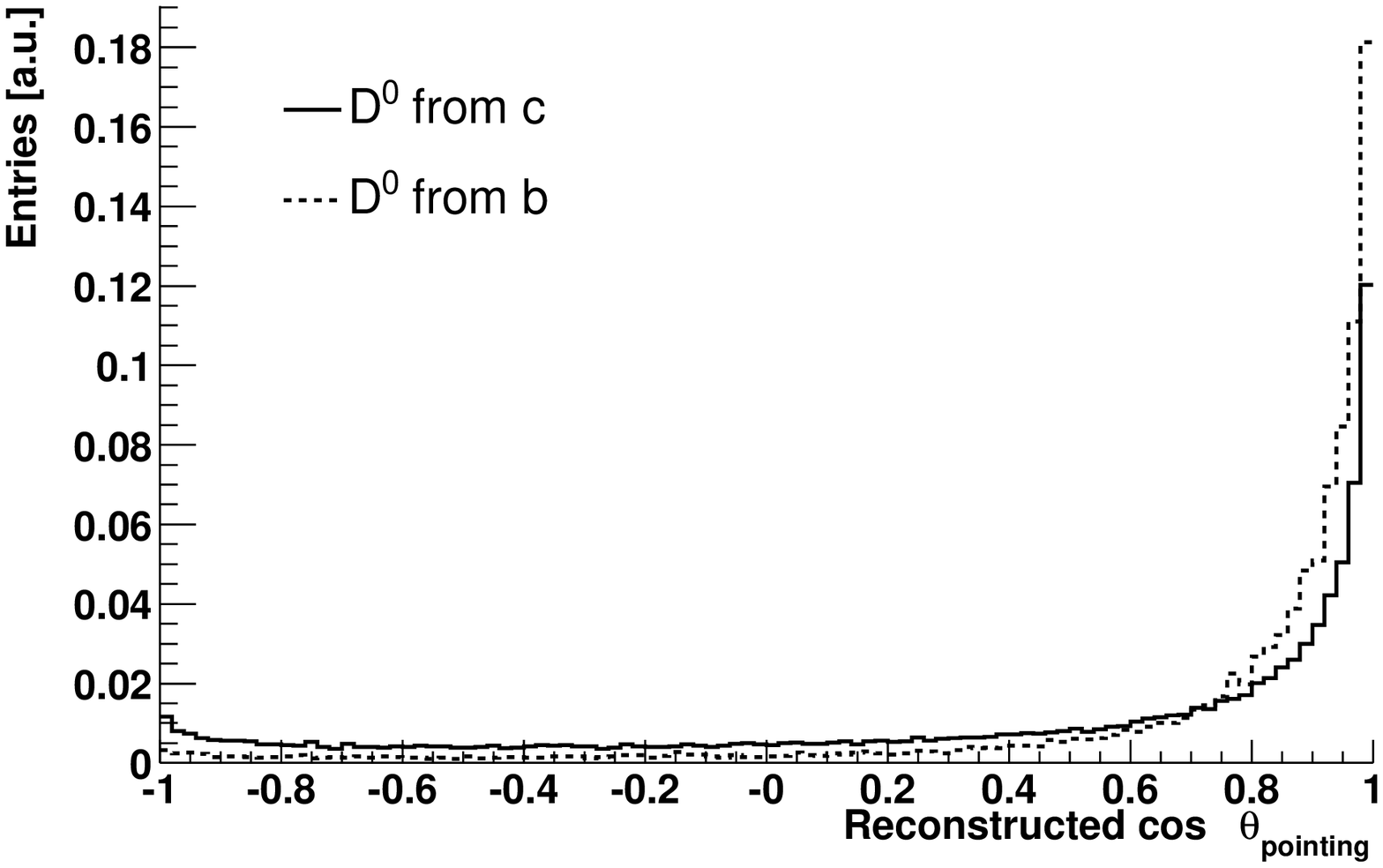}
    \caption{Comparison between primary (from c) and secondary (from b) 
             $\Dz$ mesons for: true distribution of 
             $\cos\theta_{\rm pointing}$ 
             (top), resolution on $\cos\theta_{\rm pointing}$ (middle) and 
             reconstructed distribution of $\cos\theta_{\rm pointing}$
             (bottom). The cut $2<\pt<3~\gev/c$ is applied in order to
             compare the two signals at the same $\pt$.}
    \label{fig:DfromBcpta}
    \end{center}
\end{figure}

\begin{figure}[!t]
  \begin{center}
    \includegraphics[width=.65\textwidth]{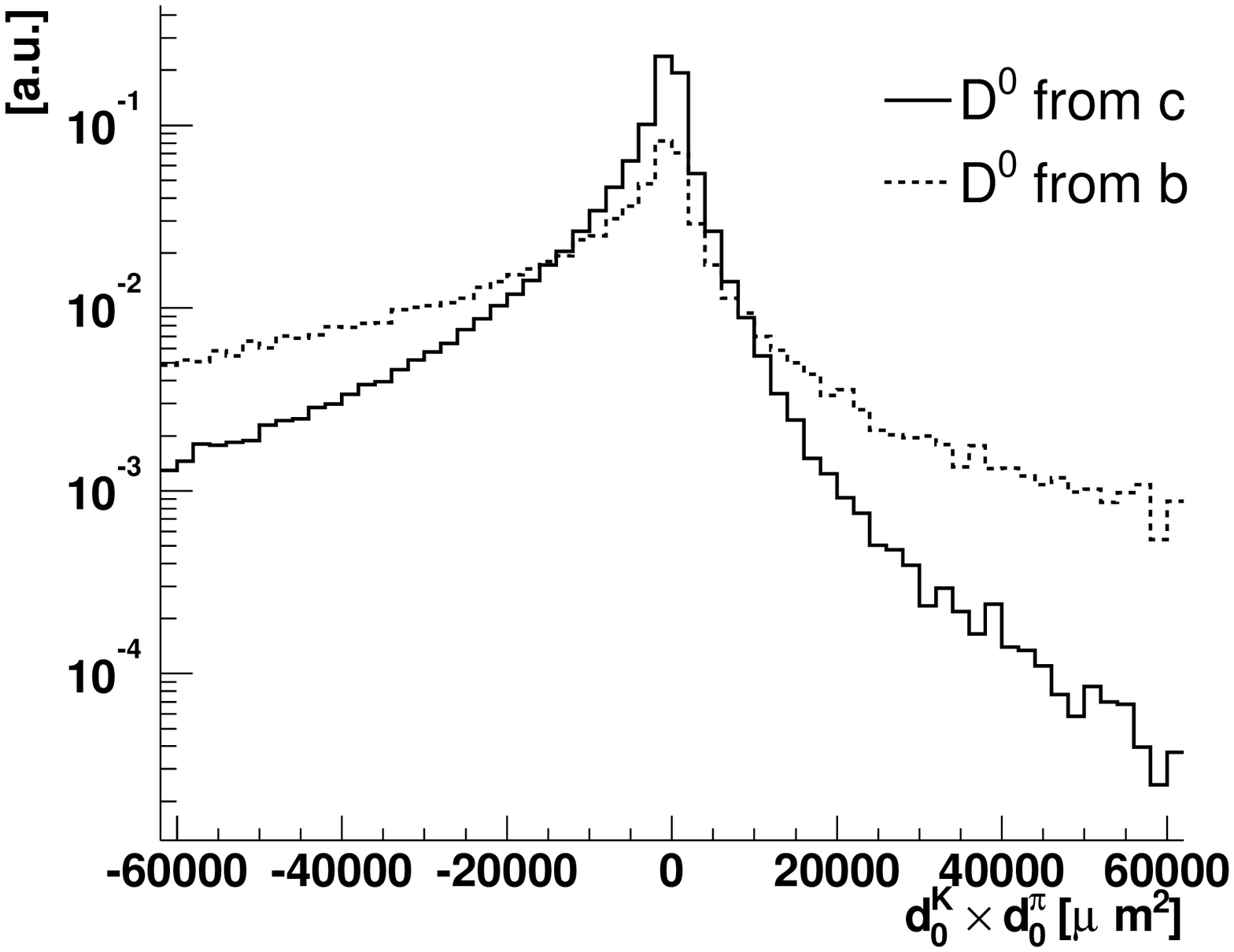}
    \caption{Distributions of the product of the impact parameters for 
             primary and secondary $\Dz$ mesons (normalized to the same 
             integral).}
    \label{fig:DfromBd0d0}
\vglue0.5cm
    \includegraphics[width=.8\textwidth]{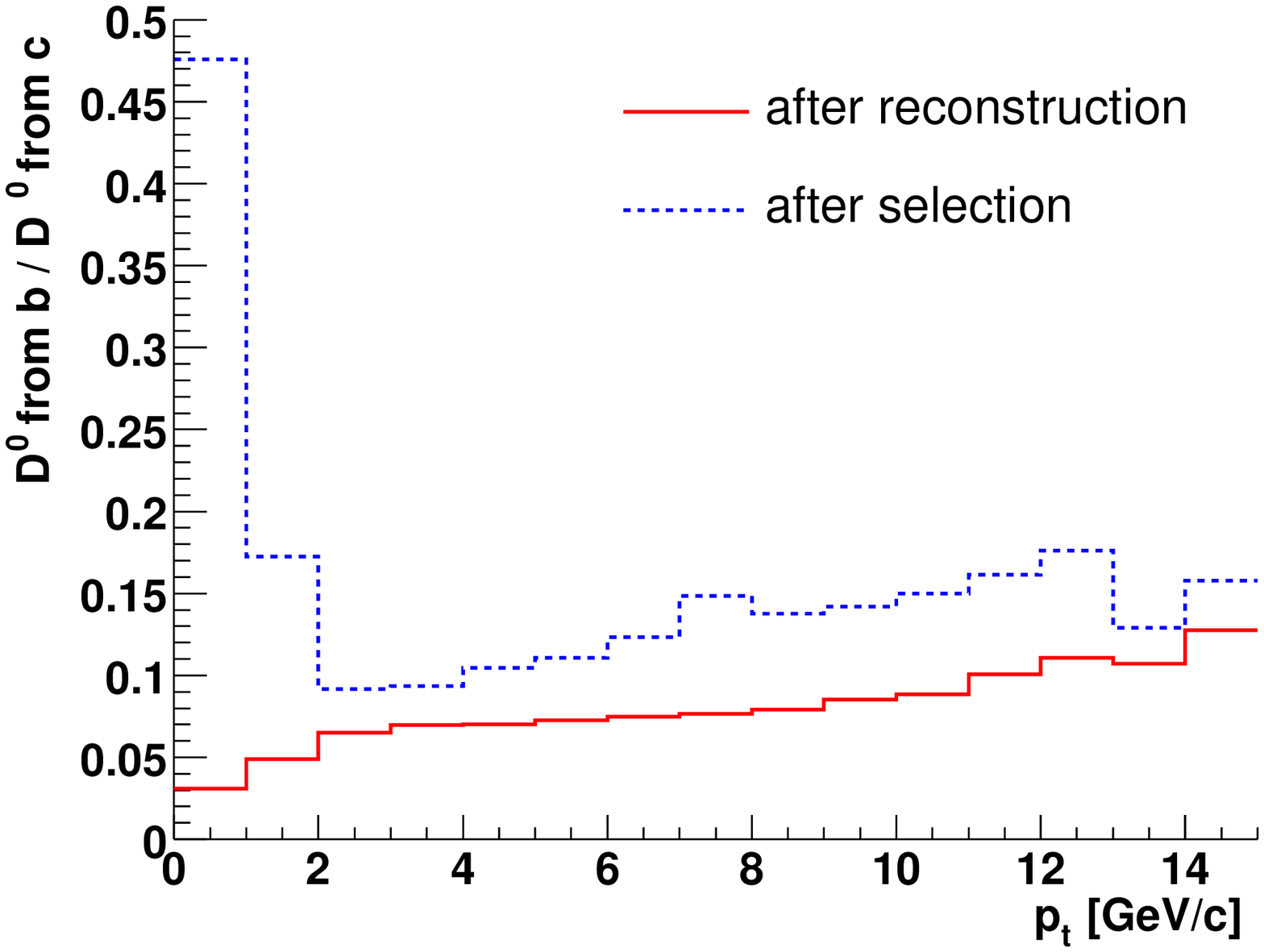}
    \caption{Ratio of secondary-to-primary $\Dz$ mesons after track 
             reconstruction and after selections, as a function of 
             $\pt$.}
    \label{fig:bD0tocD0PbPb}
    \end{center}
\end{figure}

\begin{description}
\item[Pointing angle:] this cut has the effect to {\sl enhance} 
  the fraction of secondary 
  $\Dz$. In fact, in the decay ${\rm B}\to{\rm D^0}+X$, the $\Dz$ takes most 
  of the momentum of the B meson and, thus, it approximately follows  
  its flight line. Figure~\ref{fig:DfromBcpta} (top panel) 
  shows that 
  the `true' distribution of $\cos\theta_{\rm pointing}$ (generator level)
  for secondary $\Dz$ is accumulated in the
  region $\cos\theta_{\rm pointing}>0.9$. Moreover, the resolution on this 
  variable is better in the case of secondary $\Dz$ particles 
  (Fig.~\ref{fig:DfromBcpta}, middle panel, for $2<\pt<3~\gev/c$). 
  This is explained by the 
  sketch in Fig.~\ref{fig:cptasketch}: at the same value of $\pt$ 
  the resolution $\delta q$ on the position 
  of the secondary vertex is the same for primary and secondary $\Dz$ 
  particles, but the resolution on the pointing angle is proportional 
  to $\delta q/L$, where $L$ is the distance of the secondary vertex 
  from the interaction point, which is larger for $\Dz$ from B decays.
  As a consequence, the reconstructed distribution of 
  $\cos\theta_{\rm pointing}$ is more accumulated at 1 for secondary than 
  for primary $\Dz$ particles (Fig.~\ref{fig:DfromBcpta}, bottom panel).    
\item[Product of the impact parameters:] the tracks from the decay 
  ${\rm B}\to\DtoKpi$
  have larger impact parameters than those from $\DtoKpi$ and the selection 
  $d_0^{\rm K}\times d_0^\pi< -40000~\mum^2$ 
  {\sl enhances} the fraction of the former in the final sample 
  (Fig.~\ref{fig:DfromBd0d0}). 
\item[Upper cut on $d_0$:] the analysis cut $|d_0|<500~\mum$, introduced 
  to reject the background from strange-particle decays, is effective also 
  to {\sl reduce} by $\simeq 30\%$ the fraction of $\Dz$ from B mesons.
\end{description}

The ratio ($\Dz$ from b)/($\Dz$ from c) is reported 
in Fig.~\ref{fig:bD0tocD0PbPb} as a function of $\pt$, after 
track reconstruction and after selections. The ratio
grows with $\pt$ due to the harder spectrum of the $\Dz$ from beauty.
The increase in the fraction of secondary $\Dz$ after selections is 
very large at low $\pt$ where tighter displaced vertex requirements have to  
be applied in order to reject the combinatorial background. 

\clearpage

\mysection{Feasibility study for pp collisions}
\label{CHAP6:D0recopp}

The study for the detection of $\DtoKpi$ decays in pp events followed 
the same general lines as that for the \PbPb~case. In particular, the 
same selection strategy was adopted, with cuts on the product of impact 
parameters and on the pointing angle. 

The significance of the extracted signal should be much higher in pp 
than in \PbPb, because:
\begin{enumerate}
\item The detector performance is better in the very-low-multiplicity 
      environment of pp collisions. As an example the tracking efficiency 
      is larger by about 10\% (see Fig.~\ref{fig:effTPCITS}).
\item Without taking into account the improved efficiency, the initial 
      $S/B$ ratio is proportional to 
      $N^{\scriptstyle{\rm c\overline{c}}}/(\dNdy)^2$; 
      the charm production yield is lower by a factor about 700
      in pp with respect to \PbPb~(see Table~\ref{tab:summarytable}), 
      but the multiplicity of the 
      background event is lower by a factor 1000 
      ($\av{\dNdy}=6$ in pp with PYTHIA); therefore, the initial $S/B$ is 
      larger by a factor $\simeq 1500$ in pp collisions.  
\end{enumerate}
However, the larger uncertainty on the position of the interaction vertex, 
extensively discussed in Chapters~\ref{CHAP4} and~\ref{CHAP5}, is a clear and 
important disadvantage for the displaced vertex selection in the pp case.
In the following, along with the results for the realistic scenario 
(indicated as ``vertex reconstructed''), we present also the results for a
scenario of perfect knowledge of the vertex position (``vertex known'').
This is done in order to (a) have a situation (``vertex known'') 
more directly comparable to the
\PbPb~one and (b) quantify and understand the weight of the larger 
uncertainty on the vertex position. 

\subsection{Background and signal generation}
\label{CHAP6:generpp}

The magnetic field was set to same value as for \PbPb, $0.4$~T. The same
settings were used also for the generation of the position of the 
interaction vertex.

\subsubsection{Background}

Proton--proton minimum-bias events at a centre-of-mass energy 
$\sqrt{s}=14~\tev$ were generated using PYTHIA, as described in 
Section~\ref{CHAP4:pythia}, excluding diffractive topologies. 
The average charged particle rapidity density is $\av{\dNdy}=6$.
A total of $8.5\cdot 10^6$ events were used. These events correspond to 
$12.1\cdot 10^6$ minimum-bias events 
using $\sigma_{\rm pp}^{\rm non-diffr.}/\sigma_{\rm pp}^{\rm inel}=0.7$ 
from PYTHIA.
Half of the statistics was 
produced and analyzed using the distributed computing facilities at 
CERN and in about 10 sites in Europe. Large productions, for 
a total of \mbox{$\approx$~300,000} CPU hours, were managed by 
means of AliEn~\cite{alien}, the ALICE interface to the grid computing
network.

All the results are given for $10^9$ pp minimum-bias events, corresponding
to a run of 1 month.

\subsubsection{Signal}

For the generation of the $\Dz$ signal we did not use the same method as 
in the \PbPb~case, i.e. generating many $\DtoKpi$ decays in special 
signal events. In the case of pp collisions, it is essential to have 
the signal with `its own pp event' for two main reasons:
\begin{enumerate}
  \item The primary vertex has to be reconstructed 
        event-by-event using the tracks; therefore, the signal 
        events must be pp events with charm, as we will 
        show that, in PYTHIA, events with charm have 
        different properties, in terms of particle production, 
        with respect to events without charm.
  \item In pp, medium- and high-multiplicity events are characterized 
        by the presence of jets of particles emitted in narrow cones in 
        the fragmentation of 
        energetic partons; there might be a significant background 
        originated from the association of one track from the $\Dz$ decay 
        with one of the tracks in the same jet.
\end{enumerate}
Therefore, 
we generated standard pp events with PYTHIA, with the same settings as
for the generation of the background, and we selected events that contained 
a $\Dz$ in $|y|<1$ (the decay in the $\K\pi$ channel was forced). 
The part of these $\Dz$ mesons coming from beauty feed-down was weighted 
in order to match the correct ratio secondary/primary 
(Section~\ref{CHAP6:generPbPb}).   
We used $2\cdot 10^6$ such events, 
corresponding to $1.7\cdot 10^9$ pp minimum-bias events, using 
the yields in Table~\ref{tab:hadyieldspp}.

The drawback of this method is that the produced $\Dz$ mesons have to be
reweighted according to their $\pt$ in order to reproduce the
distributions given by the NLO pQCD calculations. In fact, the 
settings of PYTHIA necessary to reproduce these distributions cannot 
be used to generate standard pp events. The $\Dz$ from charm and from 
beauty were reweighted separately, in order to keep into account their 
different $\pt$ spectra.

\subsubsection{Properties of pp events with charm production}

\begin{figure}[!b]
  \begin{center}
    \includegraphics[width=.49\textwidth]{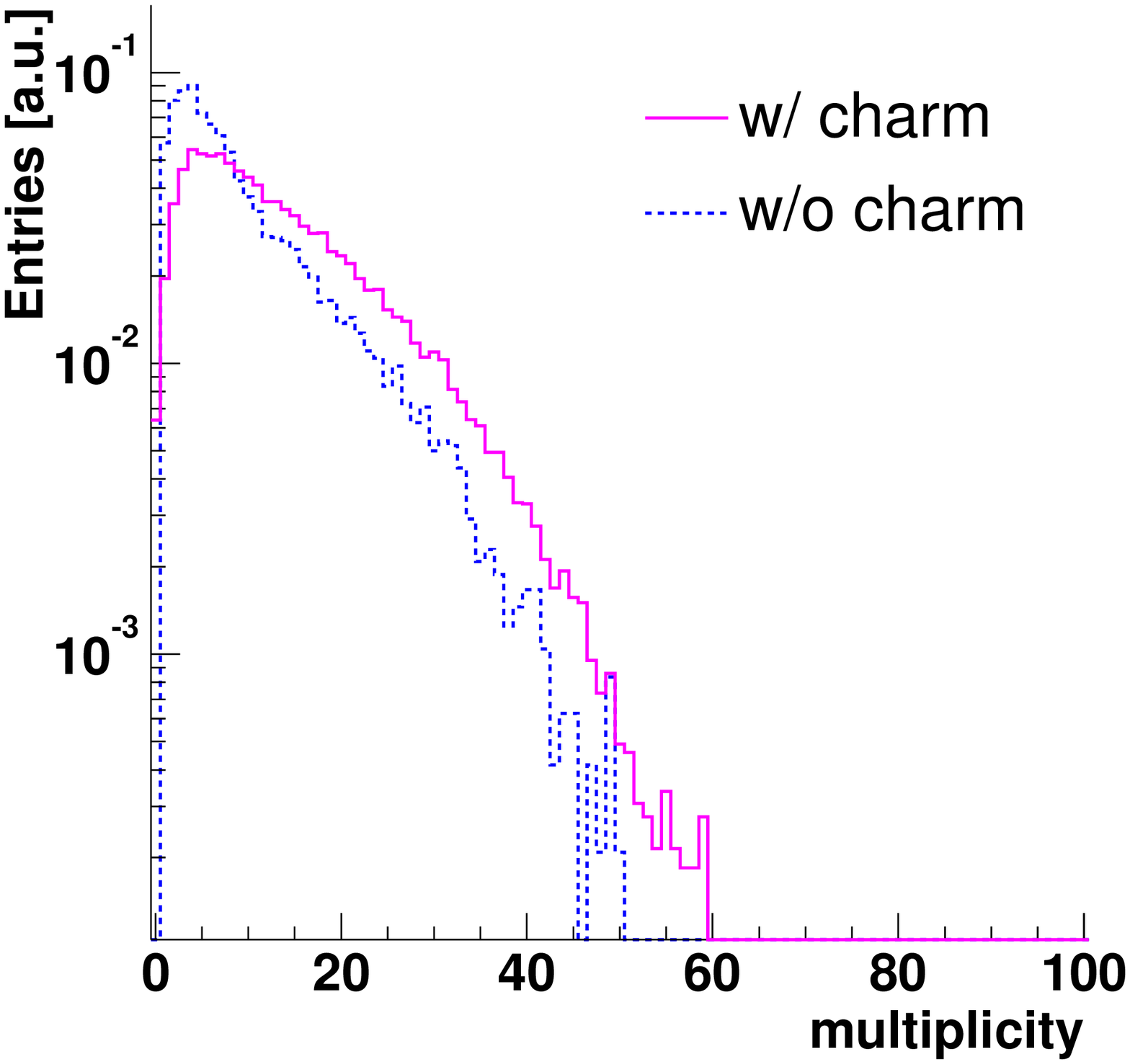}
    \includegraphics[width=.49\textwidth]{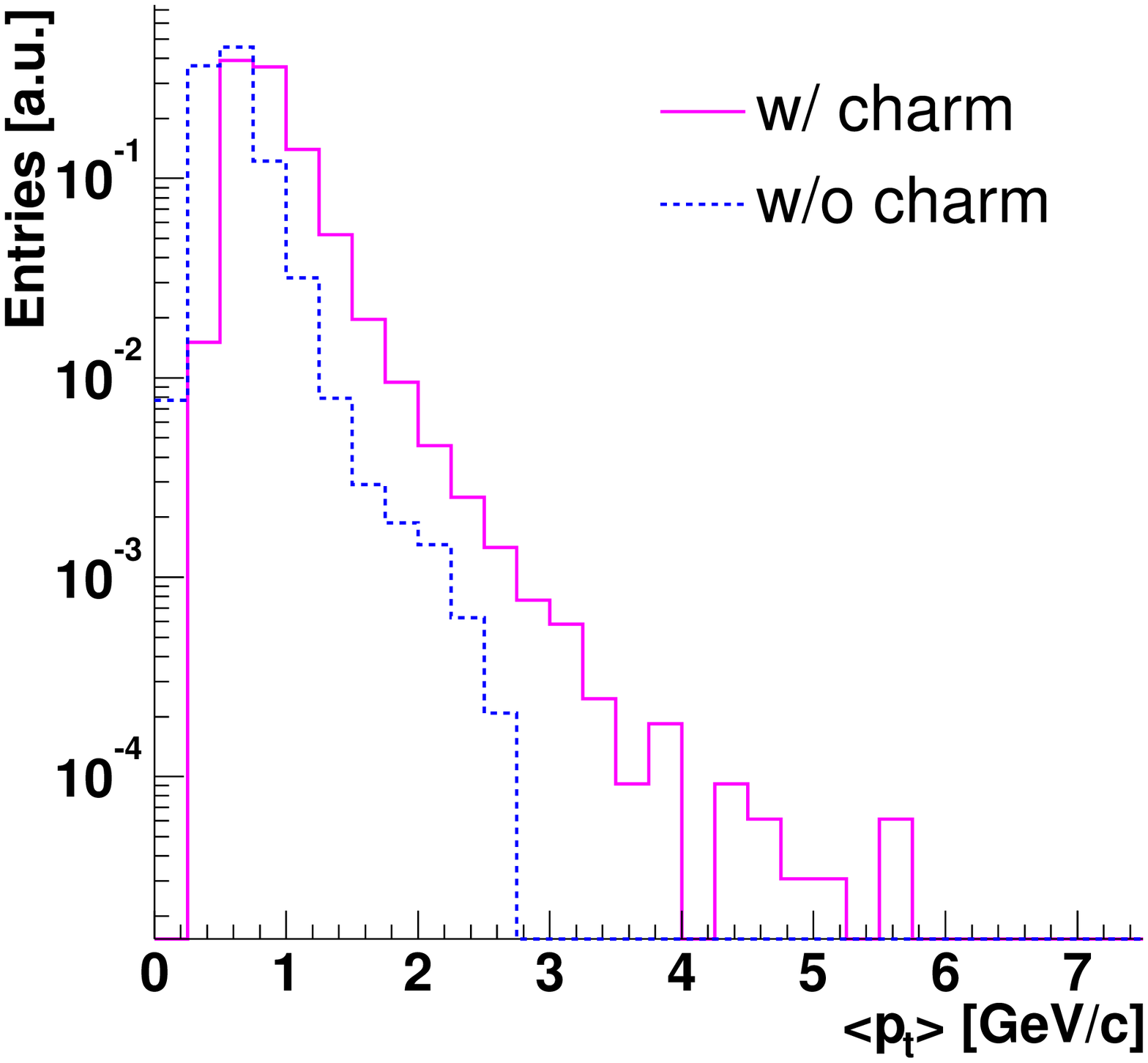}
    \caption{Charged multiplicity (left) and mean $\pt$ 
             (right) in $|\eta|<0.9$ for PYTHIA events with and 
             without charm production. Histograms are normalized to the
             same integral.}
    \label{fig:cpythia1}
    \end{center}
\end{figure}

\begin{figure}[!ht]
  \begin{center}
    \includegraphics[width=.85\textwidth]{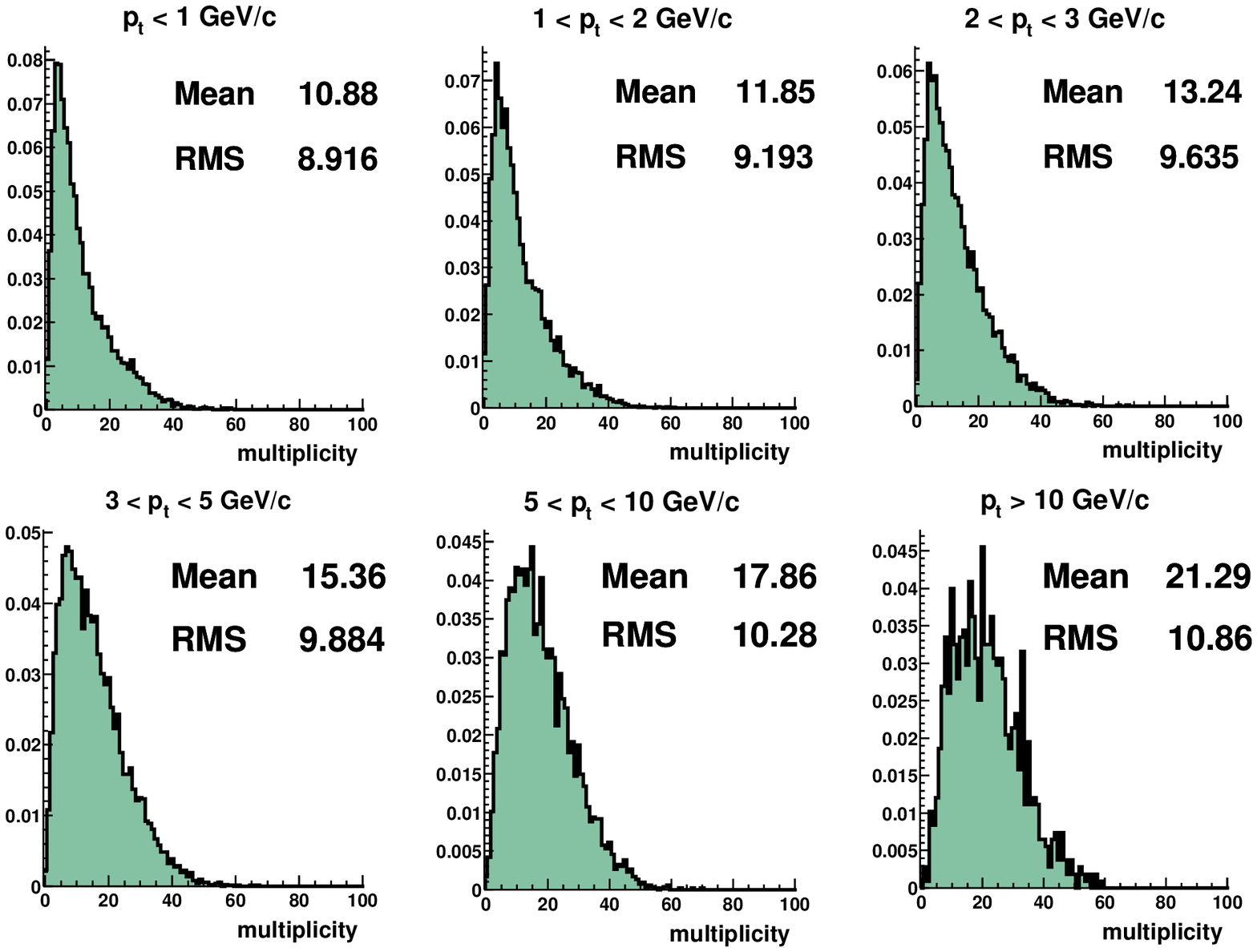}
    \includegraphics[width=.85\textwidth]{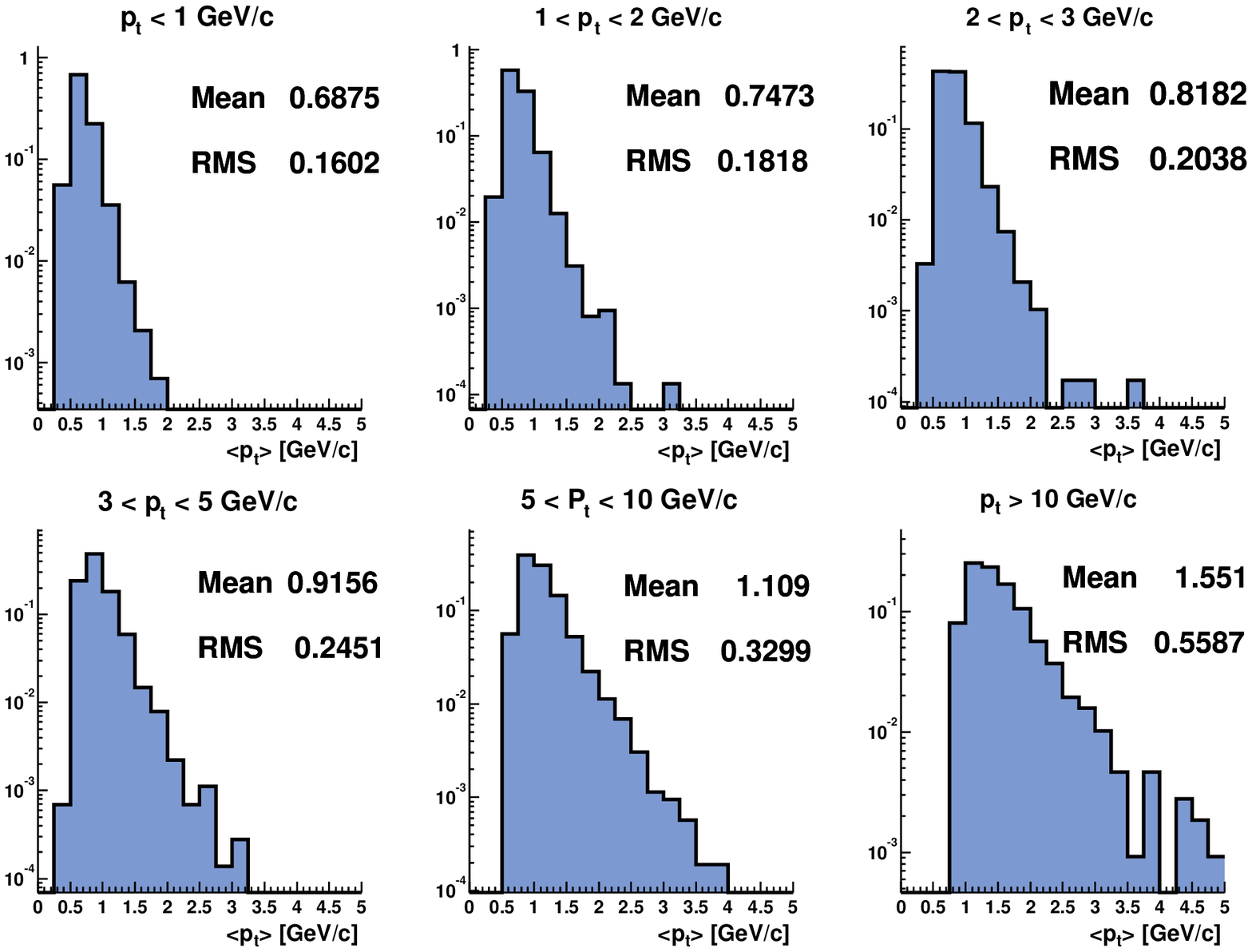}
    \caption{Charged multiplicity (top) and mean $\pt$ 
             (bottom) in $|\eta|<0.9$ for PYTHIA events with charm production,
             in bins of $\pt$ of a D meson produced in the event.}
    \label{fig:cpythia2}
    \end{center}
\end{figure}

In Chapter~\ref{CHAP5} we have shown that the efficiency and resolution 
for the reconstruction of the primary vertex in pp events depends 
on the number of produced particles and on their 
average transverse momentum. 

Due to the large mass of heavy quarks, one expects the events 
with charm (or beauty) to yield more particles than other events, and 
with larger mean energy. This is a very important aspect, because it 
would imply that the vertex information is quantitatively better in 
events containing charm particles.

At lower energies, a difference in the mean multiplicity of events with and 
without charm was observed by the NA27 Collaboration in pp fixed-target 
interactions with a beam energy of $400~\gev$~\cite{na27}. The average 
charged multiplicity was found to be $\av{N_{\rm ch}}=11$ and 
$\av{N_{\rm ch}}=9$ for 
events with and without charm, respectively. In the former case, 
two completely reconstructed charm mesons were required; each of them 
contributed for one unit to the count of the multiplicity and their 
decay products were not counted.
~\\   

Comparing the properties, in terms of multiplicity and mean 
transverse momentum, of PYTHIA pp events with and without charm production,
we found that:
\begin{enumerate}
  \item Events with charm production have significantly larger 
        multiplicity and mean $\pt$.
  \item With respect to a D meson produced in the collisions, the quantities
        $\av{\rm multiplicity}$ and $\av{\pt}$ increase as the 
        transverse momentum of the D meson increases.
\end{enumerate} 

Figure~\ref{fig:cpythia1} (left) presents the distribution of the charged 
multiplicity in $|\eta|<0.9$ (ALICE barrel acceptance) for 
events with and without charm (for events with charm we used the same
counting rules as in Ref.~\cite{na27}): 
the averages are 14 and 10, respectively.
The mean $\pt$ is shown on the right panel of the same figure:
the average values are $850~\mev/c$ and $600~\mev/c$, respectively.
In Fig.~\ref{fig:cpythia2} the same distributions, for events with charm 
production, are shown in bins of $\pt$ of a D meson produced in the 
event.

\subsection{Event reconstruction and particle identification}
\label{CHAP6:recopp}

Track reconstruction was performed in the same way as for the \PbPb~case,
using the parameterization of the tracking response in the TPC,
with a pp-specific tuning, and the Kalman filter in the ITS, 
where at least 5 clusters (2 in the pixel layers) were required.

After the tracking, the interaction vertex position was determined 
by means of the reconstructed tracks, as described in 
Section~\ref{CHAP5:vtxpp}. In order not to bias the measurement of the 
impact parameters of the two $\Dz$ decay tracks, for each $\Dz$ candidate 
the vertex was reconstructed excluding the two tracks belonging to the 
candidate. 
 
\begin{figure}[!t]
  \begin{center}
    \leavevmode
    \includegraphics[clip,width=\textwidth]{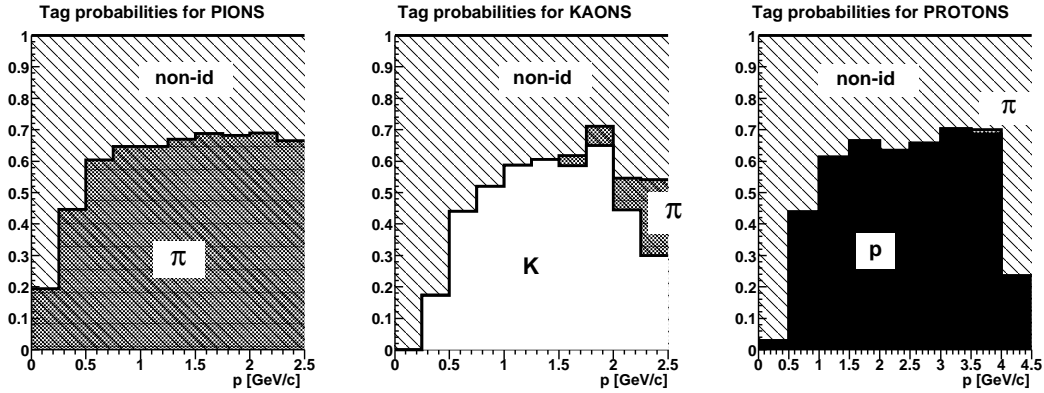}
    \caption{PID tag probabilities for reconstructed pions, kaons and protons
             in pp collisions with the TOF detector.}
    \label{fig:TOFtagprobpp}
  \end{center}
\end{figure}

In the low-multiplicity environment of \pp~collisions, the particle 
identification in the Time of Flight is more efficient, because the 
probability to match incorrectly the tracks with the TOF pads is much lower.
In Fig.~\ref{fig:TOFtagprobpp} we show the tagging probabilities, defined 
as for \PbPb~and obtained by optimizing for pp collisions the 
graphical cuts applied on the momentum-versus-mass plane. We remark that in 
this 
case the contamination of kaons in the pion sample, i.e. the probability 
to tag a kaon as pion, is almost negligible. Consequently, the 
fraction of $\Dz$ signal lost for this reason goes down to 2\% in pp
(it is about 10\% in \PbPb). 

\subsection{Analysis and results}
\label{CHAP6:resultspp}

\begin{table}[!b]
  \caption{Initial statistics for signal and background (in 
           $M_{\rm D^0} \pm 3~\sigma$), for pp
           and \PbPb. Sample A+B+C. The ratio \PbPb/pp is also reported.}
  \begin{center}
  \begin{tabular}{cccc}
    \hline
    \hline 
    & $S/$event & $B/$event & $S/B$ \\
    \hline
    pp    & $2.4\cdot 10^{-4}$ & $1.1\cdot 10^{-1}$ & $2.3\cdot 10^{-3}$ \\
    \PbPb & $1.3\cdot 10^{-1}$ & $2.8\cdot 10^{4}$ & $4.5\cdot 10^{-6}$ \\
\hline
    ratio \PbPb/pp & 520 & $2.6\cdot 10^{5}$ & $2\cdot 10^{-3}$ \\
    \hline
    \hline	
  \end{tabular}
  \label{tab:initialpp}
\end{center}
\end{table}

The statistics for signal and background after track reconstruction 
are reported in Table~\ref{tab:initialpp}. Here and in the following we 
consider the sum of the three samples A, B and C, where not differently 
specified. Only a $\pm 3~\sigma$ cut 
on the invariant mass is applied (the invariant mass resolution 
for the $\Dz$ is 10\% better in pp than in \PbPb, due to the improved
momentum resolution). In the table, along with the values for pp, 
also the same values for \PbPb~and the ratio between the two are quoted, 
for comparison purposes. The initial signal-to-background ratio is much 
larger (about three orders of magnitude) in pp than in \PbPb. This is due to 
the fact that, when going from pp to \PbPb, the background, which is 
combinatorial, increases much more than the signal. It is important to 
understand in a semi-quantitative way the changes in $S$ and $B$:
\begin{itemize}
\item The charm yield increases by a factor 720 from pp to 
      \PbPb~(Table~\ref{tab:summarytable}), but, as tracking 
      and PID efficiencies are worse in \PbPb~by $\approx 10\%$ each, 
      this increase is reduced to 520 after tracking and TOF-PID:
      $720\times (\epsilon_{\rm PID}^{\rm Pb-Pb}/\epsilon_{\rm PID}^{\rm pp})\times (\epsilon_{\rm track}^{\rm Pb-Pb}/\epsilon_{\rm track}^{\rm pp})^2=720\times 0.9 \times (0.9)^2\approx 520$. 
\item For the background, the average number of reconstructed tracks per 
      event is 4600 for \PbPb~and 6 for pp; therefore, the increase in the 
      number of track pairs should be $(4600/6)^2\approx 5.9\cdot 10^5$.
      The obtained ratio is a factor 2 lower (see Table~\ref{tab:initialpp}); 
      this is due to the 
      fact that in pp the number of reconstructed tracks has large 
      event-by-event fluctuations, and thus the combinatorial background is 
      larger than $\av{N_{\rm tracks}}^2$.
\end{itemize}

The background with 1 track from a D meson and the other from the underlying
event is a negligible fraction (0.1\%) of the total background and it is 
about a factor 2 lower than the signal. We will show in the following 
that it is almost completely suppressed by the selection cuts.

Selection and results are shown for the case ``primary vertex known'' first, 
and then for the more realistic case ``primary vertex reconstructed''.

\subsubsection{Scenario 1: ``primary vertex known''}

In this scenario we assume the $(x,y)$ position of the interaction 
point to be 
known as precisely as it is in the case of \PbPb~collisions. 
The position along $z$ is measured using the reconstructed tracks 
with a resolution of $\sim 100~\mum$.
We remark that 
this might be the case if the machine luminosity is below the 
nominal value so that beam defocusing or displacements are not necessary at 
the ALICE IP (see Section~\ref{CHAP4:lhcpp}). This is likely to happen 
during the start-up runs at the LHC.

\begin{figure}[!t]
  \begin{center}
    \includegraphics[width=.49\textwidth]{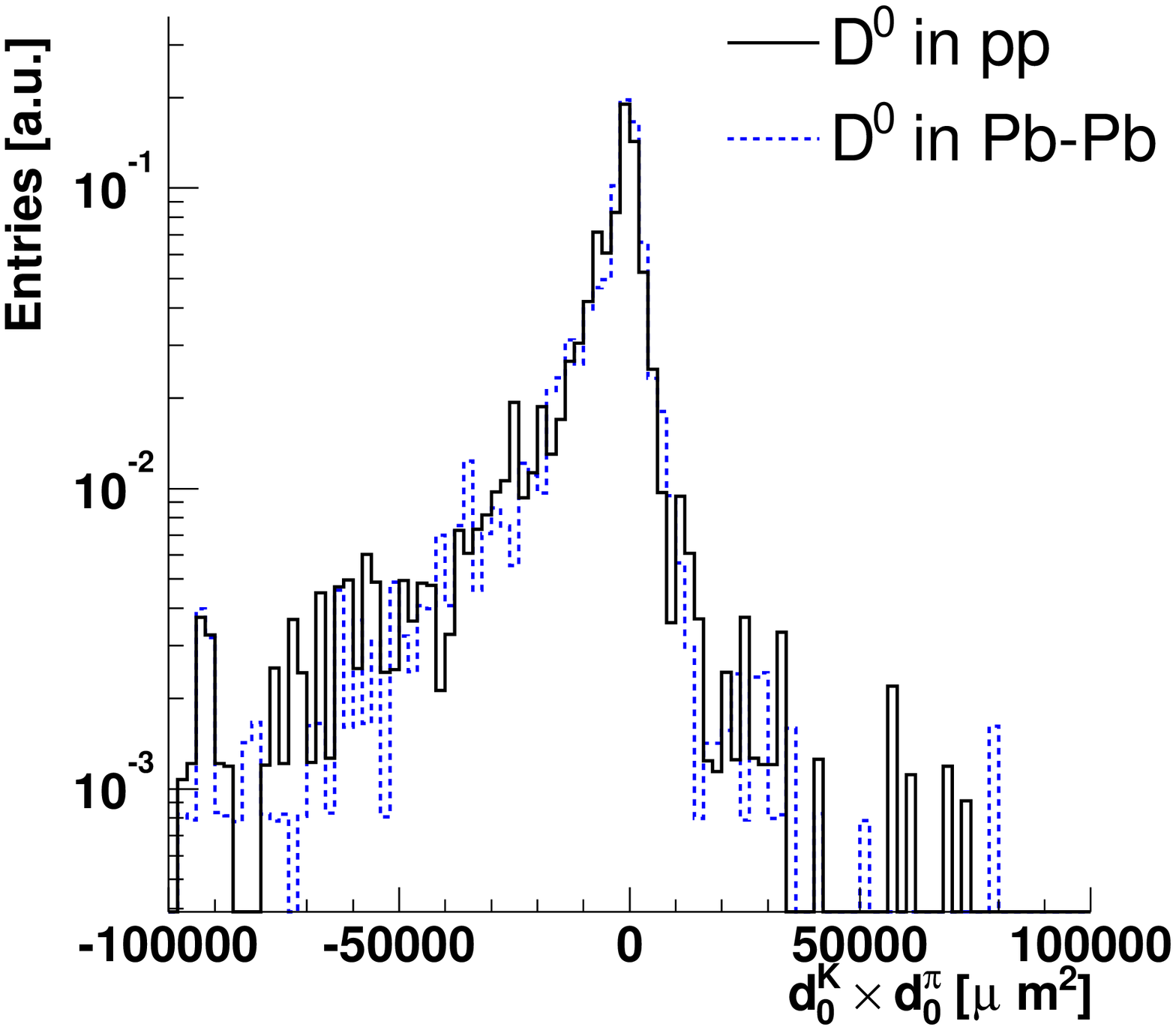}
    \includegraphics[width=.49\textwidth]{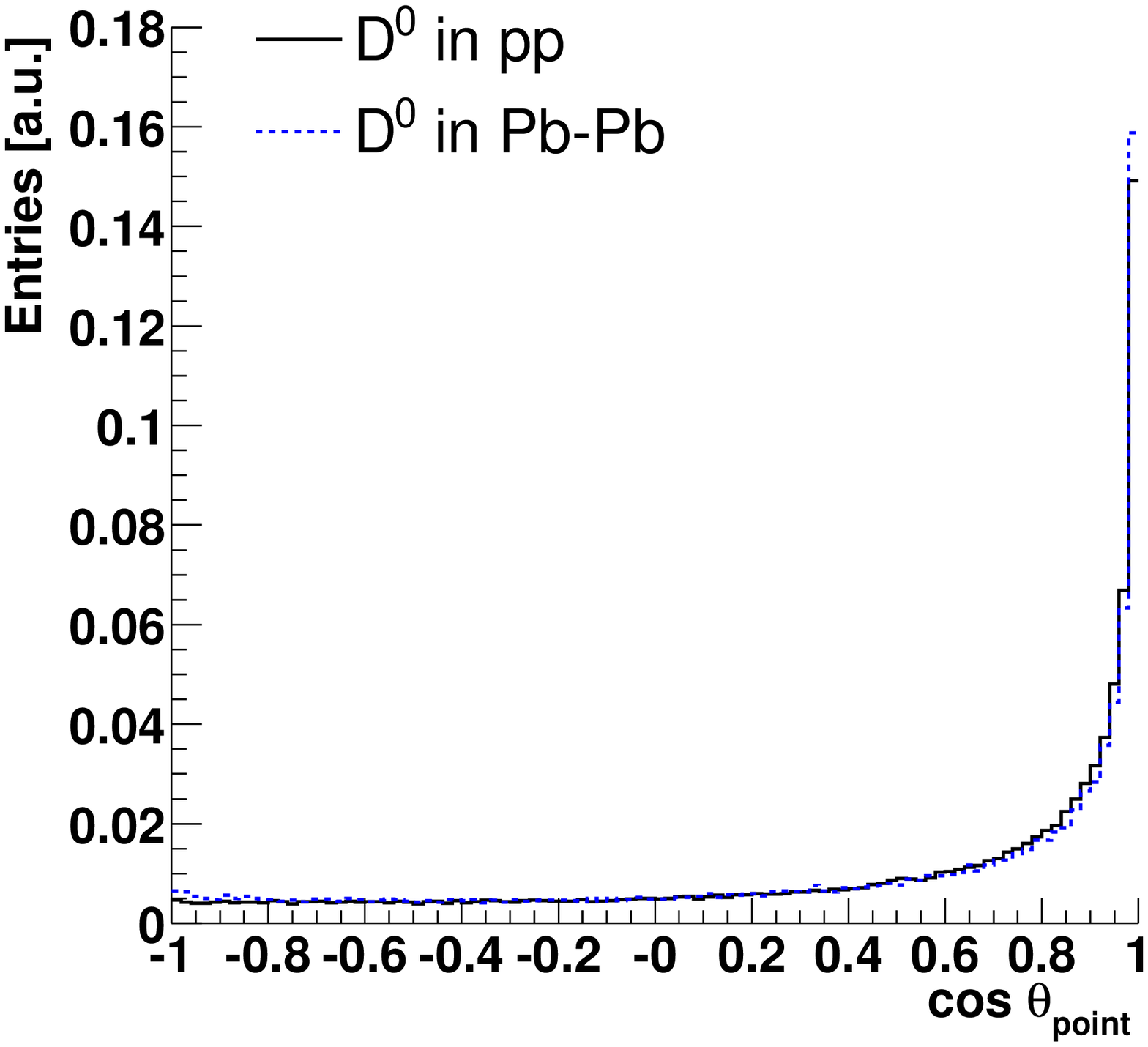}
    \caption{Distribution of the product of impact parameters (left) and
             of the cosine of the pointing angle (right) for $\Dz$ mesons
             in \PbPb~and pp (case ``vertex known'') events.}
    \label{fig:varcmpPbPbpp}
  \end{center}
\end{figure}

In Fig.~\ref{fig:varcmpPbPbpp} we report the distributions of the two 
main variables for the selection, $d_0^{\rm K}\times d_0^{\pi}$ and
$\cos\theta_{\rm pointing}$, for the signal in \PbPb~and in pp. Both 
distributions have similar shape in the two cases. This is not 
surprising for the product of the impact parameters, since it was shown in 
Chapter~\ref{CHAP5} that the track position resolution is the same 
for \PbPb~and pp events and we are assuming the same resolution on the 
primary vertex in the bending plane. In the case of the pointing angle 
one may expect the resolution in pp to be spoiled by the fact that the 
$z$ position of the primary vertex is known with a resolution of $100~\mum$,
with respect to $6~\mum$ in \PbPb. However, this difference does not 
affect too much the pointing angle resolution, since the dominant 
contributions to it come from the position resolution of the secondary vertex,
which is of $\sim 70\times 70\times 120~\mum^3$ in the three perpendicular
directions for a $\Dz$ with $\pt\simeq 2~\gev/c$. 

\begin{table}[!b]
\begin{center}
  \caption{Final value of the cuts in the different $\pt$ bins for pp 
           (case ``vertex known'').}
\scriptsize
  \begin{tabular}{c|c|c|c|c|c}
    \hline
    \hline
    Cut name &  $\pt<1~\gev/c$ & $1<\pt<2\ \gev/c$ & $2<\pt<3\ \gev/c$ &
$3<\pt<5\ \gev/c$ & $\pt>5\ \gev/c$ \\
\hline
distance of & & & & \\
closest approach & & & & \\
$(dca)$ & $<400\ \mum$ & $<300\ \mum$ & $<200\ \mum$ &
  $<200\ \mum$ & $<200\ \mum$ \\
\hline
decay angle &&&&&\\
$|\cos\theta^\star|$ & $<0.8$ &
    $<0.8$ & $<0.8$ & 
    $<0.8$ & $<0.8$\\
\hline
K, $\pi$ $\pt$ & $>500~\mev/c$ &
$>600~\mev/c$ & $>700~\mev/c$ & $>700~\mev/c$ &
$>700~\mev/c$\\
\hline
K, $\pi$ $|d_0|$ & $<\infty$ & $<\infty$ &
$<\infty$ & $<\infty$ & $<\infty$ \\
\hline
$\Pi d_0=$&&&&\\
$d_0^{\rm K}\times d_0^\pi$ & $< -20000\ \mum^2$ & $<
-20000\ \mum^2$ & $< -20000\ \mum^2$ & $<
-10000\ \mum^2$ & $<-5000\ \mum^2$ \\
\hline
pointing angle&&&&\\
$\cos\theta_{\rm pointing}$ & $>0.7$ & $> 0.7$ & $> 0.7$ &
$> 0.7$ & $> 0.7$ \\
    \hline
    \hline
  \end{tabular}
  \label{tab:cutsppvtxknown}
\end{center}
\end{table}

The values of the cuts, tuned in order to maximize the significance, in 
different $\pt$ bins, are reported in Table~\ref{tab:cutsppvtxknown}.
We notice that, as a consequence of the better initial $S/B$ ratio, 
the cuts on impact parameters and pointing angle are much less tight than in 
the \PbPb~case. In addition, the cut on the maximum absolute value of 
$d_0$ is not necessary. As in \PbPb, a smooth $\pt$-dependent cut for 
the product of the impact parameters was used. 

The final statistics are shown in 
Table~\ref{tab:finalppvtxknown}: the integrated $S/B$ ratio is 50\% and 
the significance for $10^9$ pp minimum-bias events is 84. The lower 
$\pt$ limit is $\simeq 0$, with a significance of 17 for 
$0<\pt<1~\gev/c$ if the K identification in the TOF is required. 

\begin{table}[!t]
  \caption{Final values of $S/B$ and significance for pp 
           (case ``vertex known''). The results for \PbPb~are reported for 
           comparison.}
  \begin{center}  
  \begin{tabular}{cccccc}
    \hline 
    \hline
    System & $S/$event & $B/$event & $S/B$ & $S/\sqrt{S+B}$ & $\sigma_S/S$ \\
    \hline
    pp ($10^9$ events) & $2.1\cdot10^{-5}$ & $4.1\cdot10^{-5}$ & $50\%$ &
  $84$ & $1\%$ \\
    \PbPb ($10^7$ events) & $1.3\cdot10^{-3}$ & $1.16\cdot10^{-2}$ & $11\%$ &
  $37$ & $3\%$ \\
    \hline
    \hline
  \end{tabular}
  \label{tab:finalppvtxknown}
\end{center}
\end{table}

\subsubsection{Scenario 2: ``primary vertex reconstructed''}

\begin{figure}[!t]
  \begin{center}
    \includegraphics[width=\textwidth]{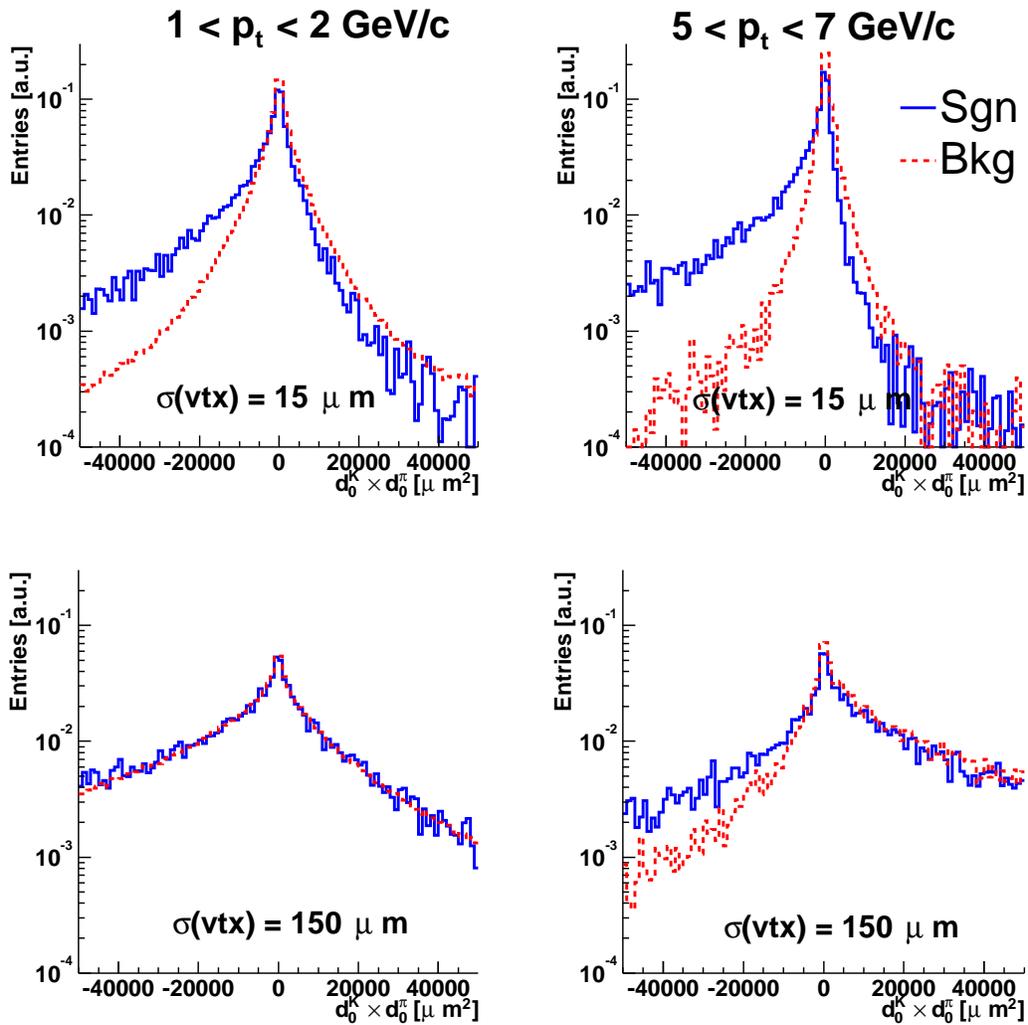}
    \caption{Distribution of the product of impact parameters for 
             ``vertex known'' (top) and ``vertex unknown'' (bottom).}
    \label{fig:d0d0pp15VS150}
  \end{center}
\end{figure}

We now consider the scenario in which the information on the vertex position 
in the transverse plane given by the position and size of the proton 
beams is very poor ($\sim 150~\mum$). Since this uncertainty is larger than 
the track position resolution given by the pixels and the mean impact 
parameter of the decay products of $\Dz$ mesons is $\sim 100~\mum$, it is
clear that, without a primary vertex reconstruction, it is impossible 
to separate the decay vertex from the interaction point. 
Figure~\ref{fig:d0d0pp15VS150} shows the distribution of the product 
of impact parameters ($\Pi d_0=d^{\rm K}_0\times d_0^\pi$) in the two 
cases ``vertex known'' (labelled $\sigma({\rm vtx})=15~\mum$) and 
``vertex unknown''
(labelled $\sigma({\rm vtx})=150~\mum$); 
two $\pt$ bins are considered: $1$-$2~\gev/c$ (left)
and $5$-$7~\gev/c$ (right). In the case ``vertex unknown'', 
at low $\pt$ the distributions of signal and background have exactly the 
same shape and even at high $\pt$, where the effect of multiple scattering
is negligible, the difference is very tiny.

\begin{figure}[!t]
  \begin{center}
    \includegraphics[width=\textwidth]{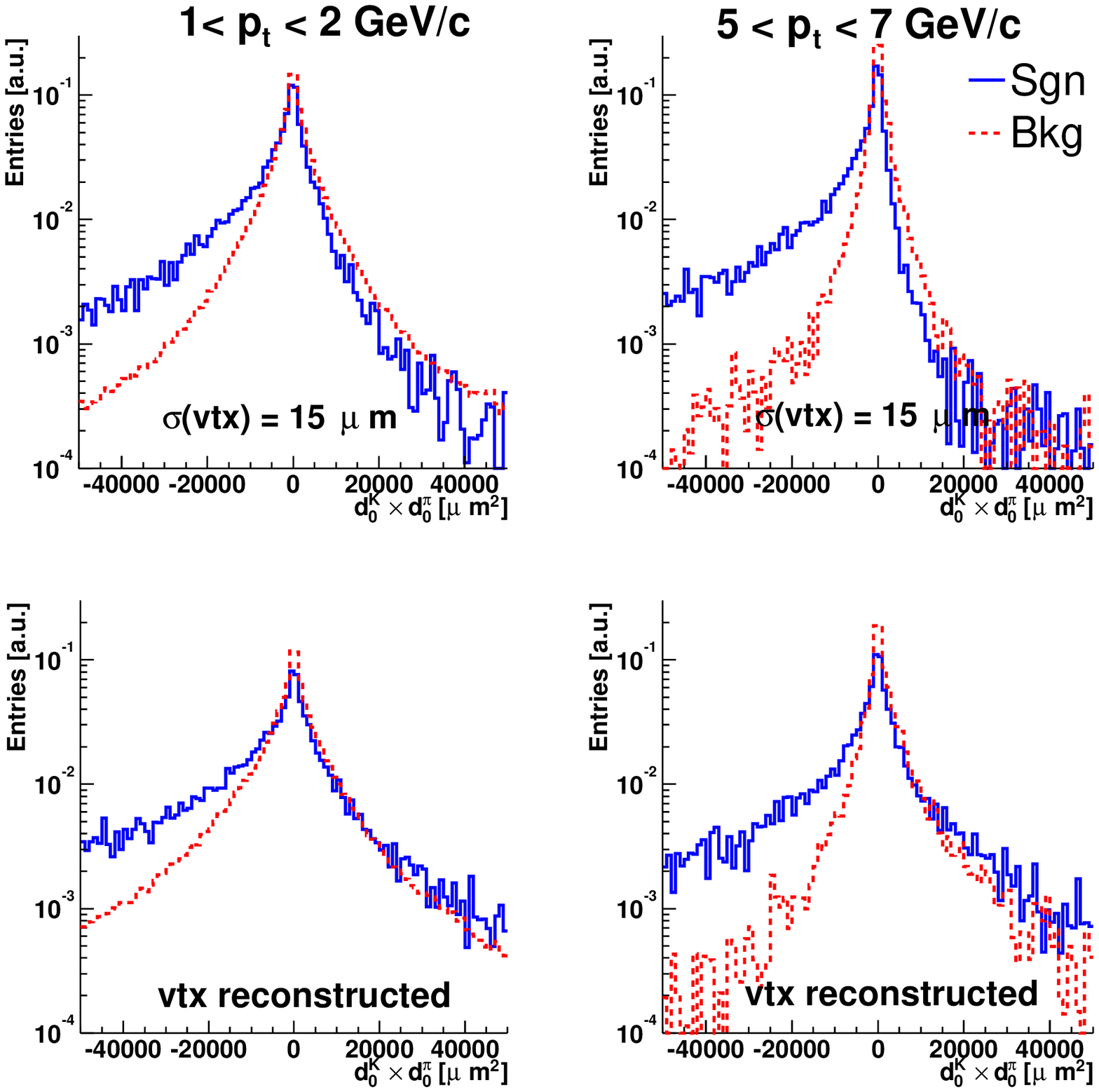}
    \caption{Distribution of the product of impact parameters for 
             ``vertex known'' (top) and ``vertex reconstructed'' 
             using the tracks (bottom).}
    \label{fig:d0d0pp15VStracks}
  \end{center}
\end{figure}

After the reconstruction of the vertex using the method, specifically
developed to this purpose, described in Chapter~\ref{CHAP5}, we obtain
the distributions reported in Fig.~\ref{fig:d0d0pp15VStracks}.
Now a cut at $\Pi d_0<-20000~\mum^2$ allows to improve the $S/B$ ratio 
even at very low $\pt$. 

\begin{figure}[!t]
  \begin{center}
    \includegraphics[width=.6\textwidth]{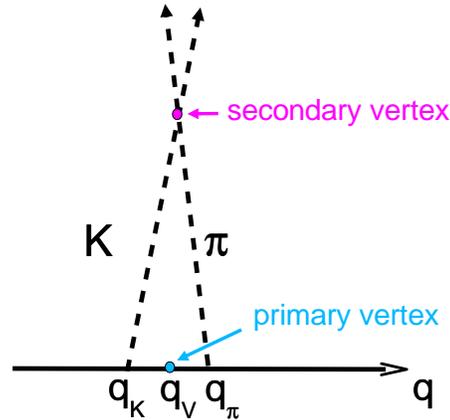}
    \caption{Sketch of the decay of a high-$\pt$ $\Dz$. 
             The impact parameters of the two decay tracks are
             $q_{\rm K}-q_V$ and $q_\pi-q_V$.}
    \label{fig:D0highptsketch}
  \end{center}
\end{figure}

By comparing the distributions for the case 
``vertex reconstructed'' (top) to those for the case ``vertex known''
(bottom) in Fig.~\ref{fig:d0d0pp15VStracks}, 
we notice that, at low $\pt$ (left), 
the distributions of signal and background are both broader, while
at higher $\pt$ (right) they are almost unchanged. This is, partially, due 
the fact that a high-$\pt$ $\Dz$ candidate usually belongs to an event 
with a relatively high multiplicity, where the vertex is reconstructed more
precisely. The other interesting reason is 
that high-$\pt$ $\Dz$ candidates are less affected by the uncertainty on the
vertex position, in spite of the fact that for high-$\pt$ tracks the 
impact parameter resolution is more affected by this uncertainty than 
for low-$\pt$ tracks (as shown in Fig.~\ref{fig:d0full}). 
We explain this point by means of a sketch of a high-$\pt$
$\Dz$ decay (Fig.~\ref{fig:D0highptsketch}). The two decay tracks form 
a small angle and the primary vertex lies between them. If we
consider the direction $q$ approximately perpendicular to the two tracks,
the product of the impact parameters is (see figure):
\begin{equation}
\Pi d_0 = d_0^{\rm K}\times d_0^\pi \approx
(q_{\rm K}-q_V) \times (q_\pi-q_V).
\end{equation}
The error on $\Pi d_0$, is function of the errors on the positions of the 
two tracks, $\sigma(q_{\rm K})$ and $\sigma(q_\pi)$, and of the error 
on the primary vertex position, $\sigma(q_V)$. We obtain:
\begin{equation}
\label{eq:d0d0err}
\sigma^2(\Pi d_0)=(d_0^{\rm K})^2\cdot \sigma^2(q_\pi)+(d_0^\pi)^2\cdot \sigma^2(q_{\rm K})+(d_0^{\rm K}+d_0^\pi)^2\cdot\sigma^2(q_V). 
\end{equation}
The error on the vertex position, in the last term of (\ref{eq:d0d0err}), 
is ``weighted'' by 
the square of the sum of the impact parameters. Now, since the impact 
parameters have preferably opposite signs (as in the example in 
Fig.~\ref{fig:D0highptsketch}), we have:
\begin{equation}
(d_0^{\rm K}+d_0^\pi)^2 \ll (d_0^{\rm K})^2~~~~{\rm and}~~~~(d_0^{\rm K}+d_0^\pi)^2 \ll (d_0^\pi)^2. 
\end{equation}
Thus, at high-$\pt$, the uncertainty on $\Pi d_0$ is dominated by the 
errors on the track positions (first two terms of equation~(\ref{eq:d0d0err})) 
and not by the error on the primary vertex position.

All the cuts were re-optimized and are shown in 
Table~\ref{tab:cutsppvtxunknown}. Also, the cut $|d_0|<500~\mum$ used 
in \PbPb~was introduced in order to reduce the feed-down from B meson decays
(more details on this are given in the next section).
The final statistics for the realistic scenario with interaction vertex 
reconstruction are given in Table~\ref{tab:finalppvtxunknown}. In the
table we have reported as a separate entry the values obtained after 
applying the additional cut on $|d_0|$, in order to point out 
its effect.

\begin{table}
\begin{center}
  \caption{Final value of the cuts in the different $\pt$ bins for pp 
           (case ``vertex reconstructed'').}
\scriptsize
  \begin{tabular}{c|c|c|c|c|c}
    \hline
    \hline
    Cut name &  $\pt<1~\gev/c$ & $1<\pt<2\ \gev/c$ & $2<\pt<3\ \gev/c$ &
$3<\pt<5\ \gev/c$ & $\pt>5\ \gev/c$ \\
\hline
distance of & & & & \\
closest approach & & & & \\
$(dca)$ & $<400\ \mum$ & $<300\ \mum$ & $<200\ \mum$ &
  $<200\ \mum$ & $<200\ \mum$ \\
\hline
decay angle &&&&&\\
$|\cos\theta^\star|$ & $<0.8$ &
    $<0.8$ & $<0.8$ & 
    $<0.8$ & $<0.8$\\
\hline
K, $\pi$ $\pt$ & $>500~\mev/c$ &
$>600~\mev/c$ & $>700~\mev/c$ & $>700~\mev/c$ &
$>700~\mev/c$\\
\hline
K, $\pi$ $|d_0|$ & $<500~\mum$ & $<500~\mum$ &
$<500~\mum$ & $<500~\mum$ & $<500~\mum$ \\
\hline
$\Pi d_0=$&&&&\\
$d_0^{\rm K}\times d_0^\pi$ & $< -20000\ \mum^2$ & $<
-20000\ \mum^2$ & $< -20000\ \mum^2$ & $<
-10000\ \mum^2$ & $<-5000\ \mum^2$ \\
\hline
pointing angle&&&&\\
$\cos\theta_{\rm pointing}$ & $>0.5$ & $> 0.6$ & $> 0.8$ &
$> 0.8$ & $> 0.8$ \\
    \hline
    \hline
  \end{tabular}
  \label{tab:cutsppvtxunknown}
\end{center}
\end{table}

\begin{table}[!t]
  \caption{Final values of $S/B$ and significance for pp (bold) and \PbPb.}
  \begin{center}  
  \begin{tabular}{cccccc}
    \hline 
    \hline
    System & $S/$event & $B/$event & $S/B$ & $S/\sqrt{S+B}$ & $\sigma_S/S$ \\
    \hline
    pp ($10^9$ events) &&&&&\\
    vertex known & $2.1\cdot10^{-5}$ & $4.1\cdot10^{-5}$ & $50\%$ &
  $84$ & $1\%$ \\
    \hline
    pp ($10^9$ events) &&&&&\\
    vertex reconstructed & $1.9\cdot10^{-5}$ & $17.3\cdot10^{-5}$ & $11\%$ &
  $44$ & $2\%$ \\
    \hline
    {\bf pp (${\bf 10^9}$ events)} &&&&&\\
    vertex reconstructed & ${\bf 1.5\cdot10^{-5}}$ & ${\bf 12.4\cdot10^{-5}}$ & ${\bf 12\%}$ &
  ${\bf 39}$ & ${\bf 3\%}$ \\
    cut $|d_0|<500~\mum$\\
    \hline
    {\bf \PbPb~(${\bf 10^7}$ events)} & ${\bf 1.3\cdot10^{-3}}$ & ${\bf 1.16\cdot10^{-2}}$ & ${\bf 11\%}$ &
  ${\bf 37}$ & ${\bf 3\%}$ \\
    \hline
    \hline
  \end{tabular}
  \label{tab:finalppvtxunknown}
\end{center}
\end{table}

The larger uncertainty on the position of the interaction point in the case 
``vertex reconstructed'' has a quite dramatic effect on the 
performance for the detection of $\DtoKpi$ decays: the background increases
by a factor about 4 and, consequently, the $S/B$ ratio and the 
significance go down by factors 4 and 2, respectively, with 
respect to the case with small and well-defined interaction region.
The additional cut on $|d_0|$ reduces the signal by 20\% 
(10\% if only sample A is considered for $\pt<2~\gev/c$) and the background by 
30\%; as we will detail in the next section, is reduces the 
feed-down from beauty by $30\%$.

\begin{figure}[!t]
  \begin{center}
    \includegraphics[width=\textwidth]{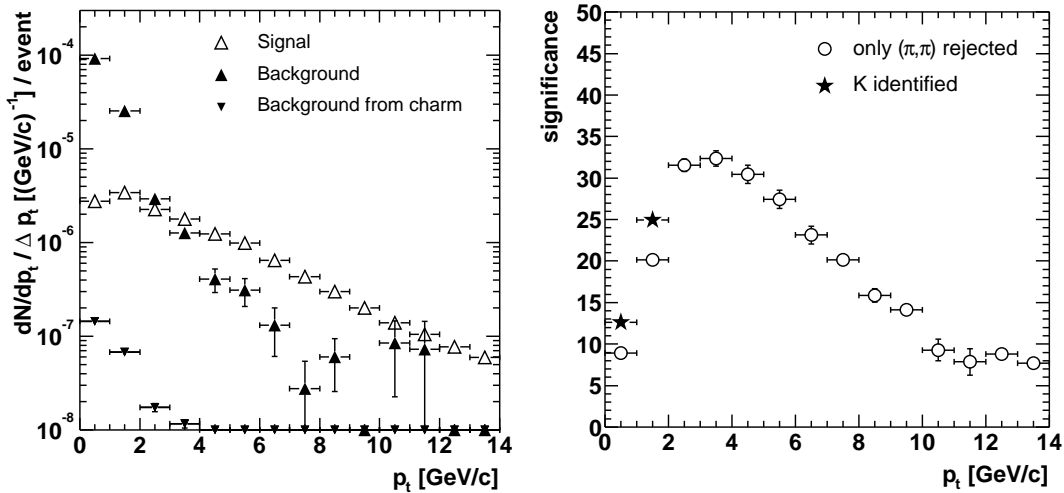}
    \caption{Transverse momentum distribution for the signal and for the 
    background after selection (left);  
    the normalization corresponds to 1 pp minimum-bias event. Corresponding
    significance for $10^9$ events as a function of $\pt$ (right). The full 
    markers
    show the significance obtained for $\pt<2~\gev/c$ requiring 
    the identification of the kaon in the Time of Flight.}
    \label{fig:resultsPt_pp}
  \end{center}
\end{figure}

In Fig.~\ref{fig:resultsPt_pp} (left) we report the transverse momentum 
distributions for signal and background. The background with a 
track from charm, also shown, is negligible. The right panel of the 
same figure reports the significance as a function of $\pt$: 
as it can be seen, the method developed for the vertex reconstruction 
allows to maintain the capability of ALICE to measure charm mesons 
down to essentially 0 in $\pt$: the significance is 14 for 
$0<\pt<1~\gev/c$ if the kaon identification in the TOF is required.
For larger transverse momenta, the significances are very close to 
those obtained in the \PbPb~case.

The `history' of the signal is summarized in Table~\ref{tab:historypp}. 
The cuts applied, including a $\pm 1~\sigma$ cut on the invariant mass, 
select 6\% of the signal.

\begin{table}[!t]
  \caption{`History' of the $\Dz/\overline{\Dz}$ signal in pp events.}
  \begin{center}  
   \begin{tabular}{lc}
     & $S/$event \\
    \hline
    \hline
    Total produced ($4\pi$) & 0.2 \\
    Decaying to $\K^{\mp}\pi^{\pm}$ & $7.5\cdot 10^{-3}$ \\
    With $\K$ and $\pi$ in $|\eta|<0.9$ & $7.0\cdot 10^{-4}$ \\
    With $\K$ and $\pi$ reconstructed & $2.5\cdot 10^{-4}$ \\
    After $(\pi_{\rm tag},\,\pi_{\rm tag})$ rejection & $2.4\cdot 10^{-4}$ \\
    After selection cuts (including $\pm 1~\sigma$ mass cut) & $1.5\cdot 10^{-5}$ \\
    \hline
    \hline
  \end{tabular}
  \label{tab:historypp}
\end{center}
\end{table}

\subsection{Feed-down from beauty}
\label{CHAP6:beautypp}

\begin{figure}[!h]
  \begin{center}
    \includegraphics[width=.65\textwidth]{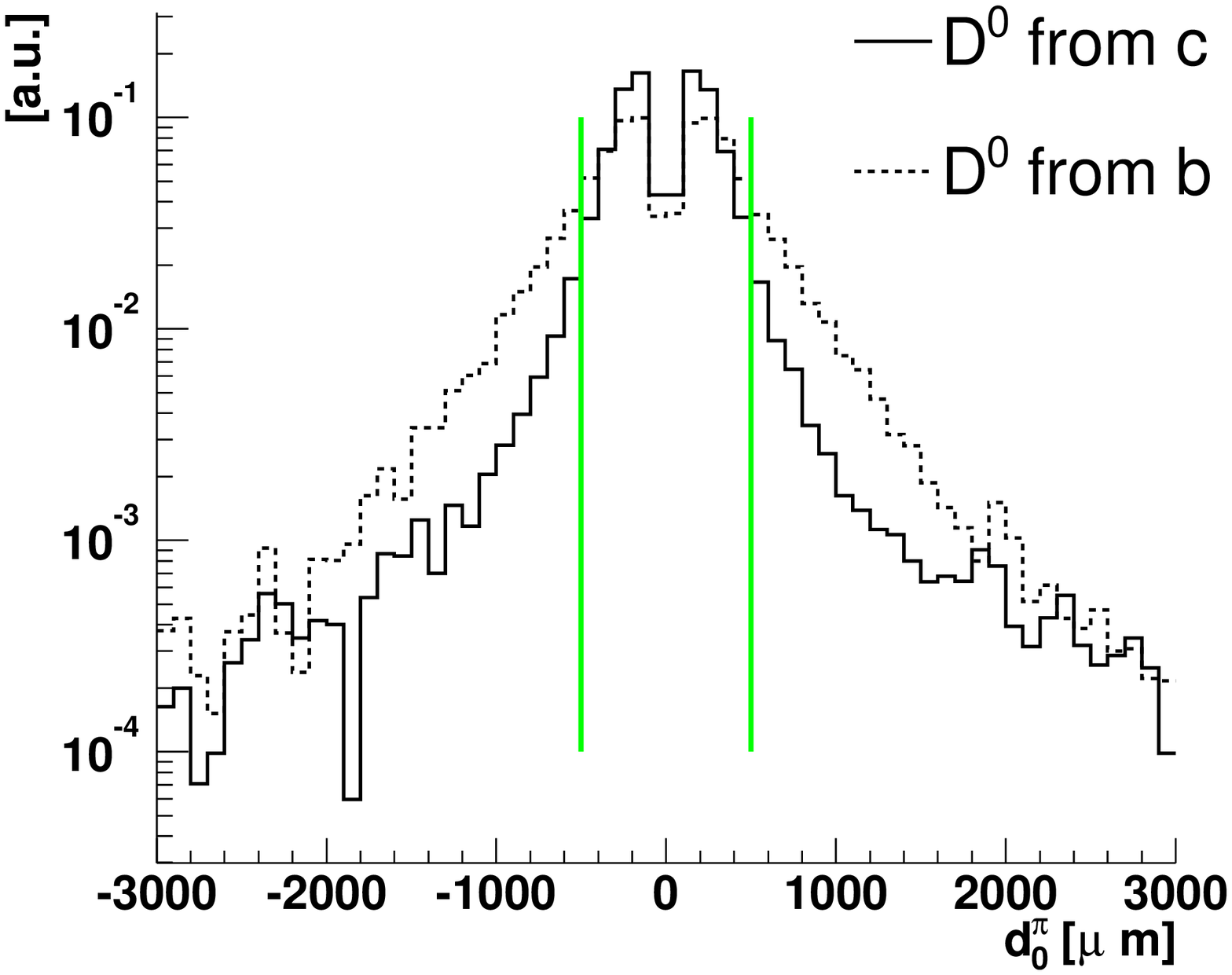}
    \caption{Distribution of the pion impact parameter after all 
             selections for primary and secondary $\Dz$ mesons.
             The lines show the cut applied in order to reduce the latter
             contribution.}
    \label{fig:d0cbD0}
\vglue0.5cm
    \includegraphics[width=.65\textwidth]{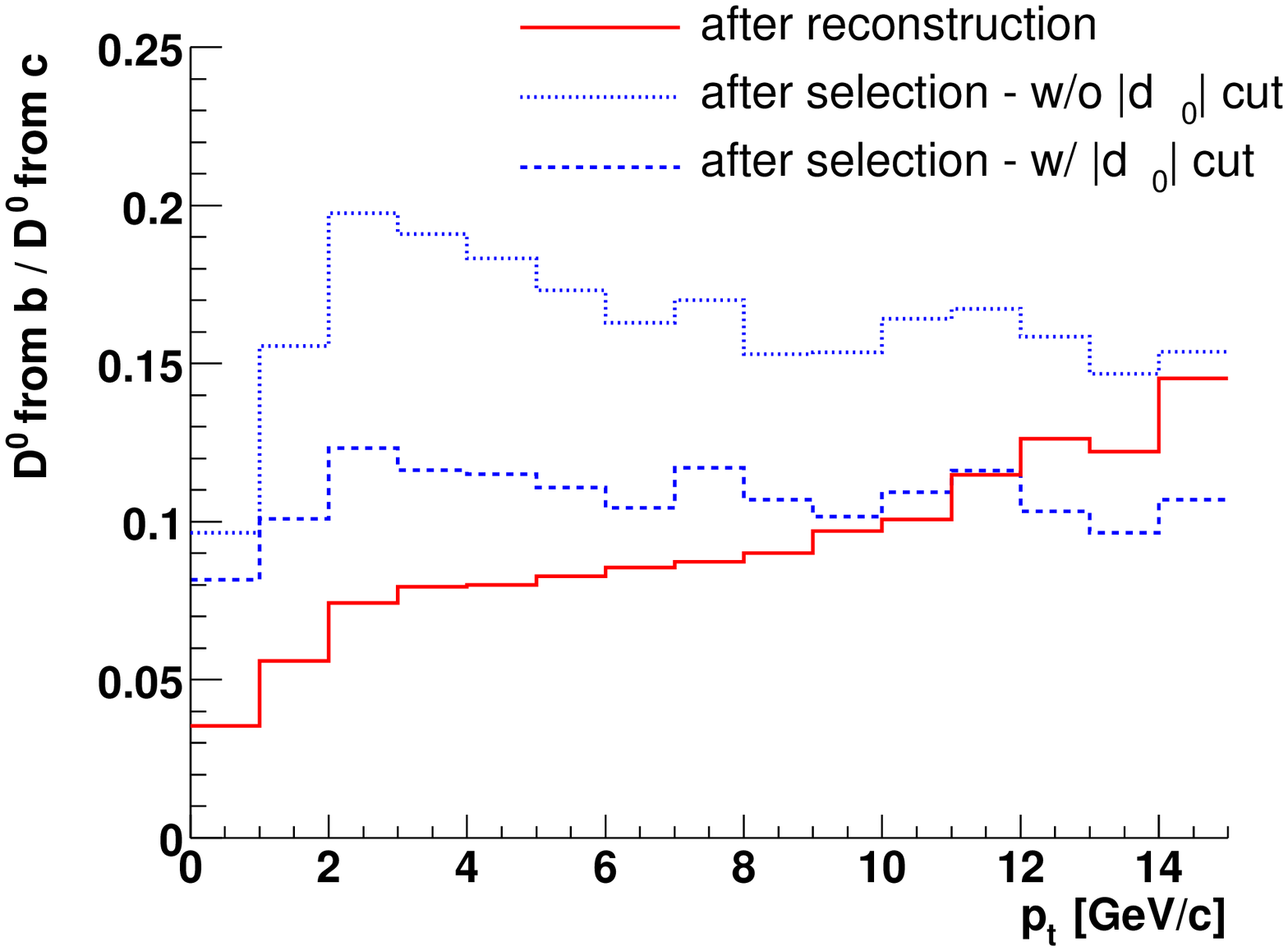}
    \caption{Ratio of secondary-to-primary $\Dz$ mesons after track 
             reconstruction and after selections, as a function of 
             $\pt$.}
    \label{fig:bD0tocD0pp}
    \end{center}
\end{figure}

As in the \PbPb~case, also in pp we observe that the selection cuts have the 
effect to increase the ratio of secondary-to-primary $\Dz$ mesons. The
ratio, which is 5.5\% after track reconstruction, becomes $\simeq 16\%$ 
after selections, if we do not apply any cut on the maximum absolute value 
of the impact parameter, $|d_0|$. We remind that the systematic error due to 
the correction for the feed-down is proportional to this ratio. 
In order to reduce it, we introduce also for pp the cut $|d_0|<500~\mum$.
Figure~\ref{fig:d0cbD0} shows the distribution of $d_0^\pi$ for the two 
contributions, after all other selections are applied. The cut, marked 
by the vertical lines, selects 90\% of the $\Dz$ from c and 70\% 
of those from b, thus allowing to reduce their ratio from $\simeq 16\%$ 
to $\simeq 11\%$.
The ratio after reconstruction and after selection, without and with the 
$|d_0|$ cut, is reported in Fig.~\ref{fig:bD0tocD0pp} as a function of $\pt$.

\mysection{Expected results for p--Pb collisions}
\label{CHAP6:pPb}

Since the study of the $\Dz$ production in pA collisions is a very important
tool to disentangle cold- and hot-medium
effects in \AA~collisions, 
in particular to investigate nuclear shadowing in the low-$\pt$ region
(see Section~\ref{CHAP1:hardprobes}), 
we give here an estimate of the results for the \pPb~system at 
$\sqrtsNN=8.8~\tev$. We consider minimum-bias collisions, 
because the precision of a centrality selection for pA collisions is 
not yet clear. 

The average multiplicity given by HIJING for this case, $\dNdy\simeq20$, is 
quite close to that obtained for pp with PYTHIA, $\dNdy=6$. Therefore, 
the detector performance, in terms of tracking and PID, can be assumed to 
be the same as for \pp. For what concerns the interaction vertex, the beams 
will be focused to the same transverse size as for the \PbPb~runs, given 
that \pPb~is a dedicated heavy ion run, optimized for ALICE. On the basis 
of these considerations, we conclude that a reliable extrapolation can 
be obtained starting from the results for pp in the scenario with 
``vertex known''.
  
In Table~\ref{tab:pppPb} we report for the charm yield, the multiplicity 
and the multiplicity squared (which is proportional to the combinatorial 
background) the values in pp and \pPb, and the ratio between the two.
From pp to \pPb, the charm yield increases by a factor 4.9, 
while the multiplicity increases by a factor 3.5.

\begin{table}[!b]
  \caption{Charm yield, multiplicity and square of the multiplicity for 
           pp and \pPb~minimum-bias collisions at LHC energy. 
           The ratio of the values for \pPb~and pp is also reported.}
  \begin{center}  
   \begin{tabular}{lccc}
    \hline
    \hline
    Parameter & pp & \pPb & \pPb/pp \\
    \hline
    $N^{\scriptstyle \ccbar}$/event~$(\propto S)$ & 0.16 & 0.78 & 4.9 \\
    $\av{\dNdy}$ & 6 & 20 & 3.5 \\
    $\av{\dNdy}^2~(\propto B)$ & 36 & 400 & 11 \\
    \hline
    \hline
  \end{tabular}
  \label{tab:pppPb}
\end{center}
\end{table}

If we assume to use the same values for all the selection cuts as in pp,
also the ratios $S_{\rm SELECTED}/S_{\rm INITIAL}$ and 
$B_{\rm SELECTED}/B_{\rm INITIAL}$ are the same. Thus, $S_{\rm SELECTED}$
and $B_{\rm SELECTED}$ are larger in \pPb~with respect to pp 
by factors 4.9 and 11, respectively, using Table~\ref{tab:pppPb}.

For the signal-to-background ratio we have:
\begin{equation}
{\bf (S/B)_{\rm p-Pb}=}(S/B)_{\rm pp}\times (4.9/11)\simeq 50\%\times 0.5 = 
{\bf 25\%}.
\end{equation}

For the $\pt$-integrated significance, 
keeping into account also that the expected number of 
collected events is $10^9$ for pp and $10^8$ for \pPb:
\begin{equation}
\begin{array}{ccl}
{\bf (S/\sqrt{S+B})_{\rm p-Pb}} & \simeq & (S/\sqrt{B})_{\rm p-Pb}=(S/\sqrt{B})_{\rm pp}\times 4.9/\sqrt{11}\times \sqrt{10^8/10^9}\\
& \simeq & 84\times 1.5 \times \sqrt{0.1}={\bf 40}. 
\end{array}
\end{equation}

\mysection{Results at lower magnetic field: $B=0.2~{\rm T}$}

We pointed out in Chapter~\ref{CHAP2} 
that the width of the mass peak of the $\Dz$ is
proportional to the momentum resolution and, consequently, the 
integral $B$ of the background under the peak is also proportional to it.
As the momentum resolution goes like $1/B_{\rm MAG}$~\cite{pdg} (we use here 
the notation $B_{\rm MAG}$ for the magnetic field to distinguish it 
from the background, $B$), the extrapolation of the results to 
$B_{\rm MAG}=0.2$~T is straight-forward: the background would be larger by 
a factor 2.

The $S/B$ ratio would be lower by a factor 2: $\simeq 5\%$ for \PbPb~and pp,
and $\simeq 12\%$ for \pPb. The $\pt$-integrated significance 
($\simeq S/\sqrt{B}$) would be lower by a factor $\sqrt{2}$: $\simeq 30$ 
for \PbPb, pp and \pPb. The low-$\pt$ limit would not be much affected:
a significance of 10 for the bin $1$-$2~\gev/c$ in \PbPb~and for the bin 
$0$-$1~\gev/c$ in pp gives a statistical error of 10\%, which may still
be acceptable. At high-$\pt$ the significance does not depend on the magnetic 
field, since the background is negligible.

\mysection{Summary}
\label{CHAP6:summary}

The results of the feasibility studies for the detection of 
$\DtoKpi$ decays in lead--lead and \pp~collisions, and the estimated 
results for proton--lead collisions, are summarized in 
Table~\ref{tab:D0summary}. 

The results are quantitatively comparable for the three considered systems: 
the $\Dz$ production cross section can be measured with a statistical 
error of the order of 3\% (in the next chapter we shall address with 
more detail the estimation of statistical and systematic errors).
In \pp, the uncertainty on the interaction vertex position is the main 
limitation to the performance and a precise vertex reconstruction in 
three dimensions is a crucial issue.

In terms of performance, the only difference among the three systems 
is the lower $\pt$ limit: $\simeq 1~\gev/c$ for \PbPb~and $\simeq 0$ 
for pp and \pPb. The upper 
$\pt$ limit is of the order of $15~\gev/c$, assuming the $\pt$ distributions
predicted by our baseline NLO pQCD estimates.

If the multiplicity in \PbPb~collisions at the LHC
turns out to match the value extrapolated from RHIC 
energies ($\dNdy\sim 3000$), the expected significance for this system will 
be larger by a factor 2.  

In the previous section, we have shown that, if the lower-field option (0.2~T)
was chosen, the results are not expected to change qualitatively, 
as the lower and upper $\pt$ limits are not dramatically affected.

\begin{table}[!t]
\caption{Summary table of the expected results for $\DtoKpi$ detection in pp 
         and \mbox{Pb--Pb} collisions and extrapolated results for 
         \pPb~collisions (using the same cuts as in pp).}
\label{tab:D0summary}
\begin{center}
\large
\begin{tabular}{l|ccc}
\hline
\hline
System & pp & p--Pb & Pb--Pb \\
Centrality & min.-bias & min.-bias & central (5\% $\sigma^{\rm tot}$) \\ 
$\sqrtsNN$ & $14~\tev$ & $8.8~\tev$ & $5.5~\tev$ \\
Number of events & $10^9$ & $10^8$ & $10^7$ \\
\hline
$S$ (total) & 15,000 & 7,500 & 13,000 \\
$S/B$ & 12\% & 25\% & 11\% \\
$S/\sqrt{S+B}$ & 39 & 40 & 37 \\
Lower $\pt$ limit & $\simeq 0$ & $\simeq 0$ & $\simeq 1~\gev/c$ \\
Upper $\pt$ limit & $\simeq 15~\gev/c$ & $\simeq 15~\gev/c$ & $\simeq 15~\gev/c$ \\
\hline
\hline
\end{tabular}
\end{center}
\end{table}

\clearpage
\pagestyle{plain}

\setcounter{chapter}{6}
\mychapter{Performance for the measurement \mbox{of $\Dz$ production}}
\label{CHAP7}

\pagestyle{myheadings}

The results of the feasibility study presented in Chapter~\ref{CHAP6}
are here used to derive an estimate of the sensitivity for the 
measurement of the $\Dz$ production cross sections and of the transverse
momentum distributions in \PbPb~and pp collisions. Statistical 
errors are estimated in Section~\ref{CHAP7:stat} and the main systematic 
uncertainties are discussed in Section~\ref{CHAP7:syst}. Eventually, 
errors are combined and the expected sensitivity is compared, for the
pp case, to the theoretical uncertainty in pQCD calculations
(Sections~\ref{CHAP7:errsummary} and~\ref{CHAP7:cmppQCD}). 
The extrapolation of the pp results from $\sqrt{s}=14~\tev$
to $\sqrt{s}=5.5~\tev$, required to compute the nuclear modification 
factor $\RAA$, is discussed in Section~\ref{CHAP7:ppextrapolation}.  

In the last part of the chapter we present a preliminary study on the 
possibility to separate 
primary (from c) and secondary (from b) $\Dz$ mesons by means of the 
impact parameter of the $\Dz$ itself to the interaction vertex 
(Section~\ref{CHAP7:D0d0}). Such analysis would be of great interest in 
order to directly estimate the amount of feed-down from B decays.

\mysection{Estimation of the statistical uncertainty}
\label{CHAP7:stat}

The relative statistical error, $\sigma_S/S$, on the number of 
reconstructed $\Dz$ candidates, in a given $\pt$-bin and for a given 
number of events, is equal, as we demonstrate in the following, to the 
inverse $\sqrt{S+B}/S$ of the statistical significance.

The signal $S$ is obtained as $S=T-B$, where $T$ is the total number 
($T=S+B$) of 
candidates in an invariant mass window entirely containing the $\Dz$ peak  
(e.g. $|M-M_{\rm D^0}|<4~\sigma\approx 40$-$60~\mev$) and $B$ is the number 
of background 
candidates in the same window, estimated using, for example, a fit on the 
side-bands of the invariant mass distribution 
($4~\sigma<|M-M_{\rm D^0}|<10~\sigma\approx 100$-$200~\mev$). 
The error on $S$ is, therefore,
\begin{equation}
\label{eq:sigmaS}
\sigma_S=\sqrt{\sigma_T^2+\sigma_B^2}\approx \sigma_T=\sqrt{T}=\sqrt{S+B}.
\end{equation}
The error on $B$, $\sigma_B$, can be neglected; it is, in fact, much lower 
than the error on $T$, $\sigma_T$, since $B$ is estimated on a larger 
invariant mass range and, thus, with better precision. From 
Eq.~(\ref{eq:sigmaS}) we derive trivially $\sigma_S/S=1/{\rm significance}$.

In the next section we optimize the $\pt$-binning in order to have 
a significance larger than 10, i.e. statistical error lower than 10\%,
up to $14~\gev/c$ and we calculate $S/B$ and significance as a 
function of $\pt$.
In Section~\ref{CHAP7:massfit} we test the expression derived above 
for $\sigma_S/S$ by a fit of Monte Carlo invariant mass distributions.

\subsection{$S/B$ and $S/\sqrt{S+B}$ with optimized $\pt$-binning}
\label{CHAP7:binning}

\begin{figure}[!b]
  \begin{center}
    \includegraphics[width=.85\textwidth]{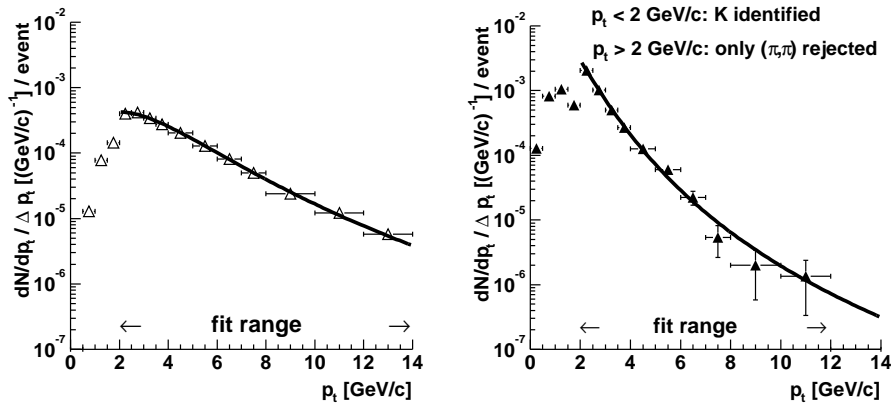}
    \caption{Transverse momentum distribution for the signal (left) and for the
             background (right) in \PbPb~collisions. The fit in the 
             indicated ranges is performed using the expression introduced in 
             Eq.~(\ref{eq:fitMesons}).} 
    \label{fig:resultsPt_PbPb_fit}
  \end{center}
\end{figure}

We fitted the reconstructed $\pt$ distributions for signal and background,
presented in Figs.~\ref{fig:resultsPt_PbPb} and~\ref{fig:resultsPt_pp},
in order to remove the large fluctuations at high $\pt$ due to the limited
statistics of the background in the simulations. The fit is shown, 
for the \PbPb~case, in Fig.~\ref{fig:resultsPt_PbPb_fit}.

\begin{figure}[!t]
  \begin{center}
    \includegraphics[width=\textwidth]{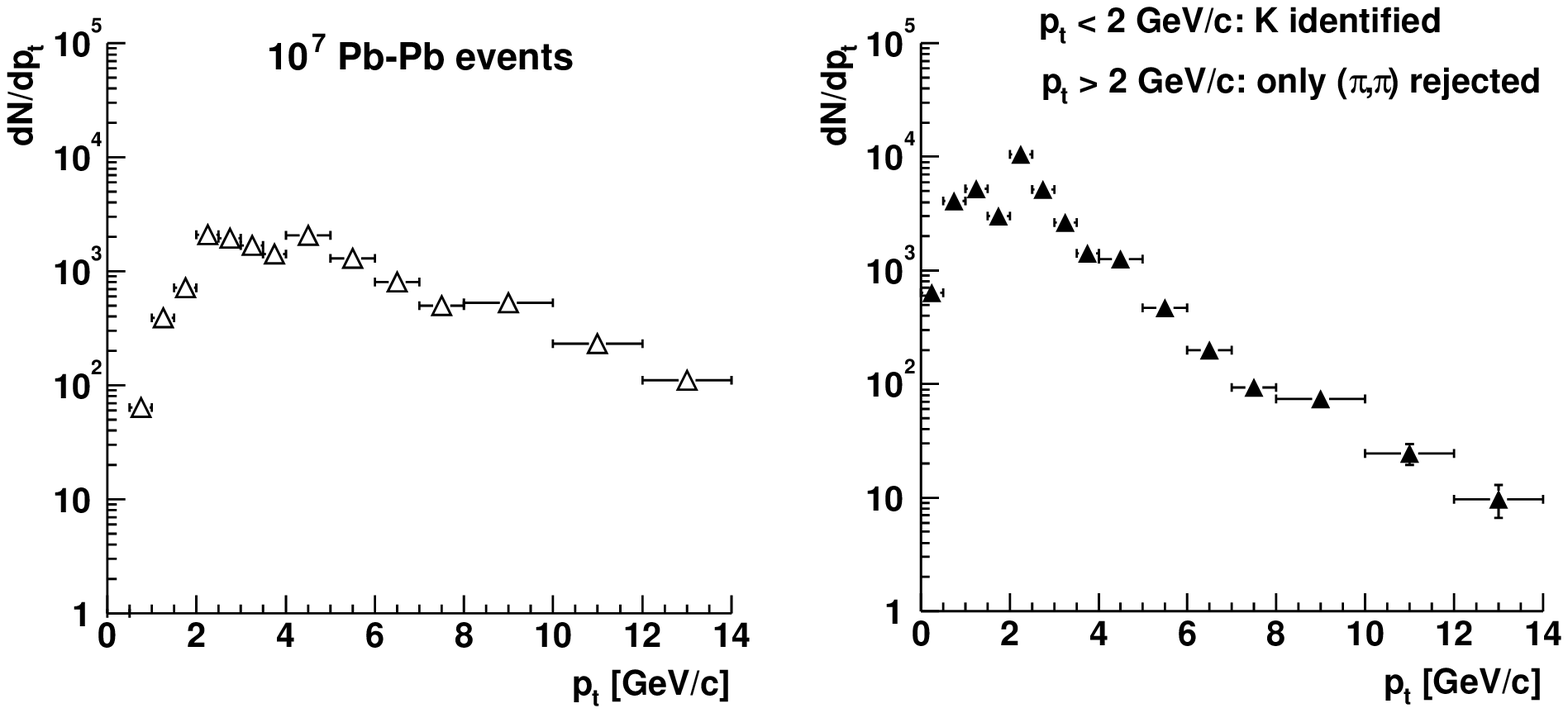}
    \includegraphics[width=\textwidth]{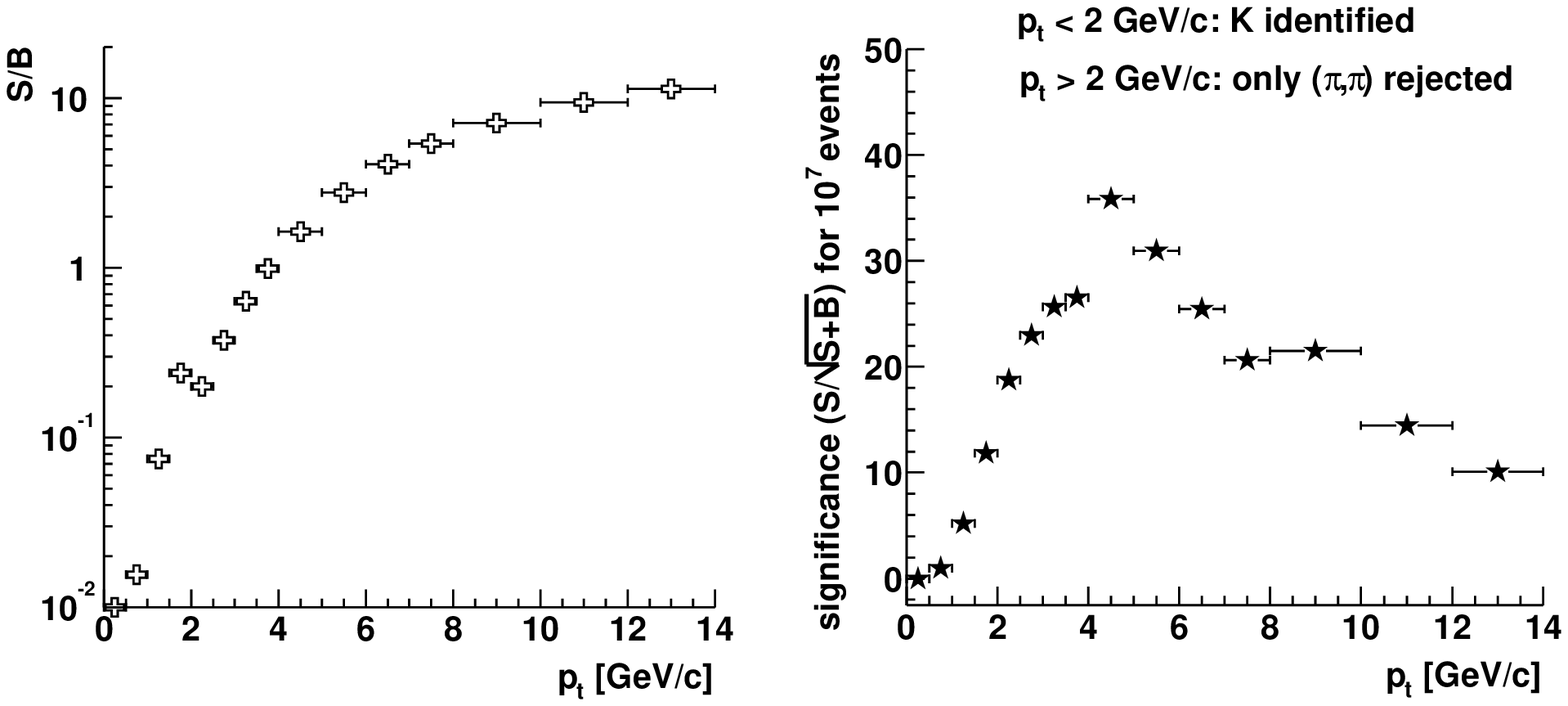}
    \caption{Statistics as a function of $\pt$ for signal (top-left) and 
             background (top-right) in $|M-M_{\rm D^0}|<1~\sigma$,
             for $10^7$ \PbPb~events. $S/B$ ratio (bottom-left) and 
             significance (bottom-right).} 
    \label{fig:stat_PbPb}
  \end{center}
\end{figure}

\begin{figure}[!t]
  \begin{center}
    \includegraphics[width=\textwidth]{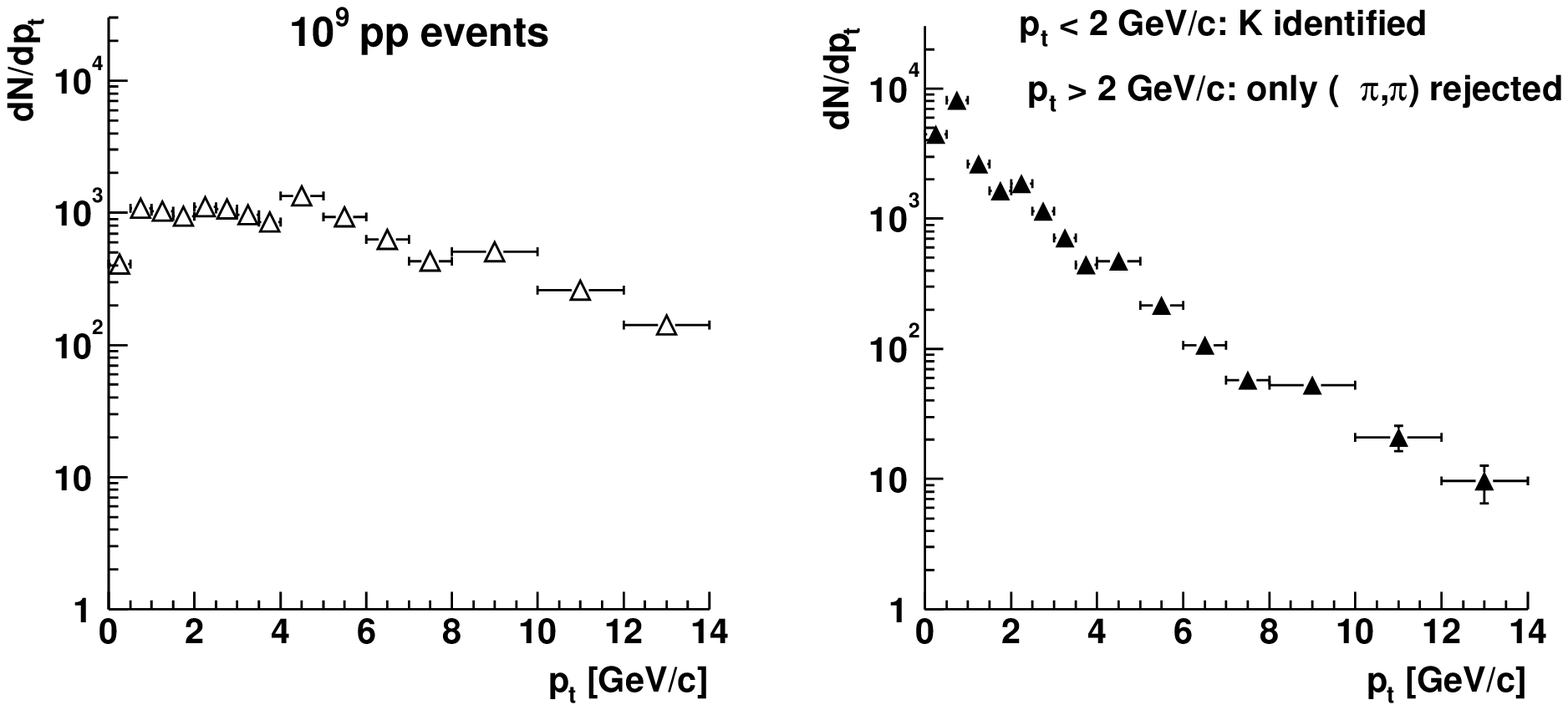}
    \includegraphics[width=\textwidth]{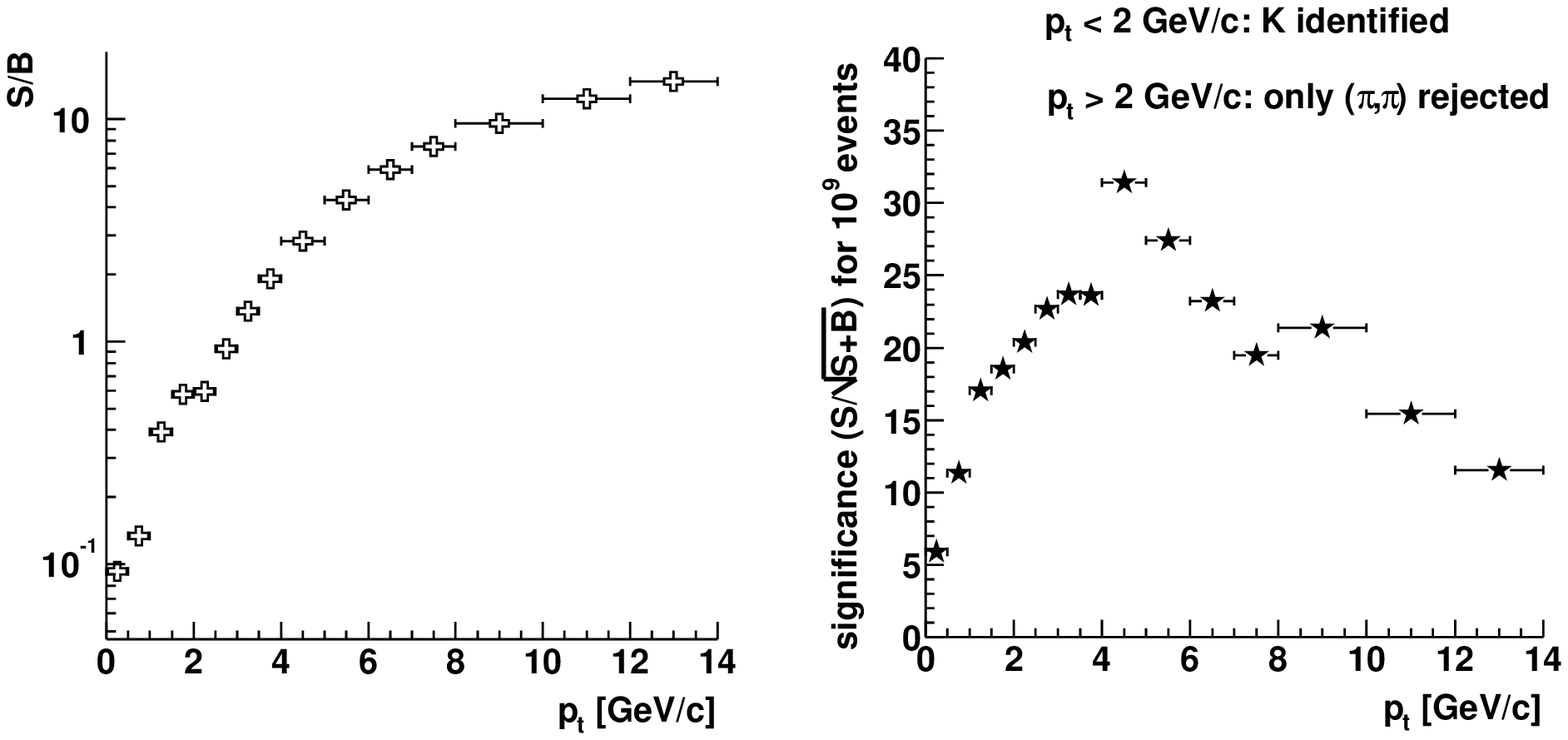}
    \caption{Statistics as a function of $\pt$ for signal (top-left) and 
             background (top-right) in $|M-M_{\rm D^0}|<1~\sigma$, 
             for $10^9$ pp events. 
             $S/B$ ratio (bottom-left) and significance 
             (bottom-right).} 
    \label{fig:stat_pp}
  \end{center}
\end{figure}

At high $\pt$ the background is negligible and the statistical error 
is determined by the statistics of the signal: $\sigma_S/S\simeq 1/\sqrt{S}$.
In order to compensate the decrease in signal statistics, the width of the 
$\pt$-bins is usually increased as $\pt$ increases. We chose the following
binning:
\begin{itemize}
\item $0<\pt<4~\gev/c$: 8 bins with width $\Delta\pt=0.5~\gev/c$;
\item $4<\pt<8~\gev/c$: 4 bins with width $\Delta\pt=1~\gev/c$;
\item $8<\pt<14~\gev/c$: 3 bins with width $\Delta\pt=2~\gev/c$.
\end{itemize}

The total statistics for signal and background, the signal-to-background
ratio and the significance obtained with this binning are presented in 
Fig.~\ref{fig:stat_PbPb} for \PbPb~($10^7$ events) and in 
Fig.~\ref{fig:stat_pp}
for pp ($10^9$ events). The invariant mass window $|M-M_{\rm D^0}|<1~\sigma$
is considered. The $\pt$ distributions were not divided by the bin 
width\footnote{We show here d$N/$d$\pt$, not d$N/$d$\pt/\Delta\pt$.} 
in order to directly show the expected number of signal $\Dz$
in each bin.

In both cases, \PbPb~and pp, the number of selected candidates is of 
order $\sim 10^3$/bin for $\pt<8$-$10~\gev/c$ and of order $\sim 10^2$/bin
for larger $\pt$, up to $14~\gev/c$. The $S/B$ ratio increases with $\pt$ 
and the significance is larger than 10 (i.e. $\sigma_S/S<10\%$) up to 
$14~\gev/c$. We consider here $14~\gev/c$ as the upper $\pt$ limit, 
but we remark that such limit can probably be extended to $17$-$18~\gev/c$ 
with one or two more bins of $\Delta\pt=2$-$3~\gev/c$.

\subsection{Fit of the invariant mass distribution}
\label{CHAP7:massfit}

Before showing the distribution of the relative statistical errors as a
function of $\pt$, we report the results of a test on the fit of the 
invariant mass distribution. The test was meant to check:
\begin{enumerate}
\item the relation $\sigma_S/S=\sqrt{S+B}/S$;
\item the reliability of the statistical error on $S$ given by 
      standard fit algorithms (CERNLIB MINUIT package~\cite{minuit});
\item that the determination of $S$ by means of a fit does not
      introduce systematic errors (systematic underestimation or 
      overestimation of $S$).
\end{enumerate} 
The `true' shape (without statistical fluctuations) of the invariant mass 
distribution in 
$|M-M_{\rm D^0}|<200~\mev/c$ 
was parameterized, for each $\pt$-bin, as an exponential (background) plus a 
Gaussian (signal). The 
slope of the exponential and the width of the Gaussian were obtained,
as a function of $\pt$, from the full simulation; the signal and background  
contributions were normalized in order to have integrals in 
$|M-M_{\rm D^0}|<1~\sigma$ according to the values reported in 
the top panels of Figs.~\ref{fig:stat_PbPb} and~\ref{fig:stat_pp}.

\begin{figure}[!t]
  \begin{center}
    \includegraphics[width=.49\textwidth]{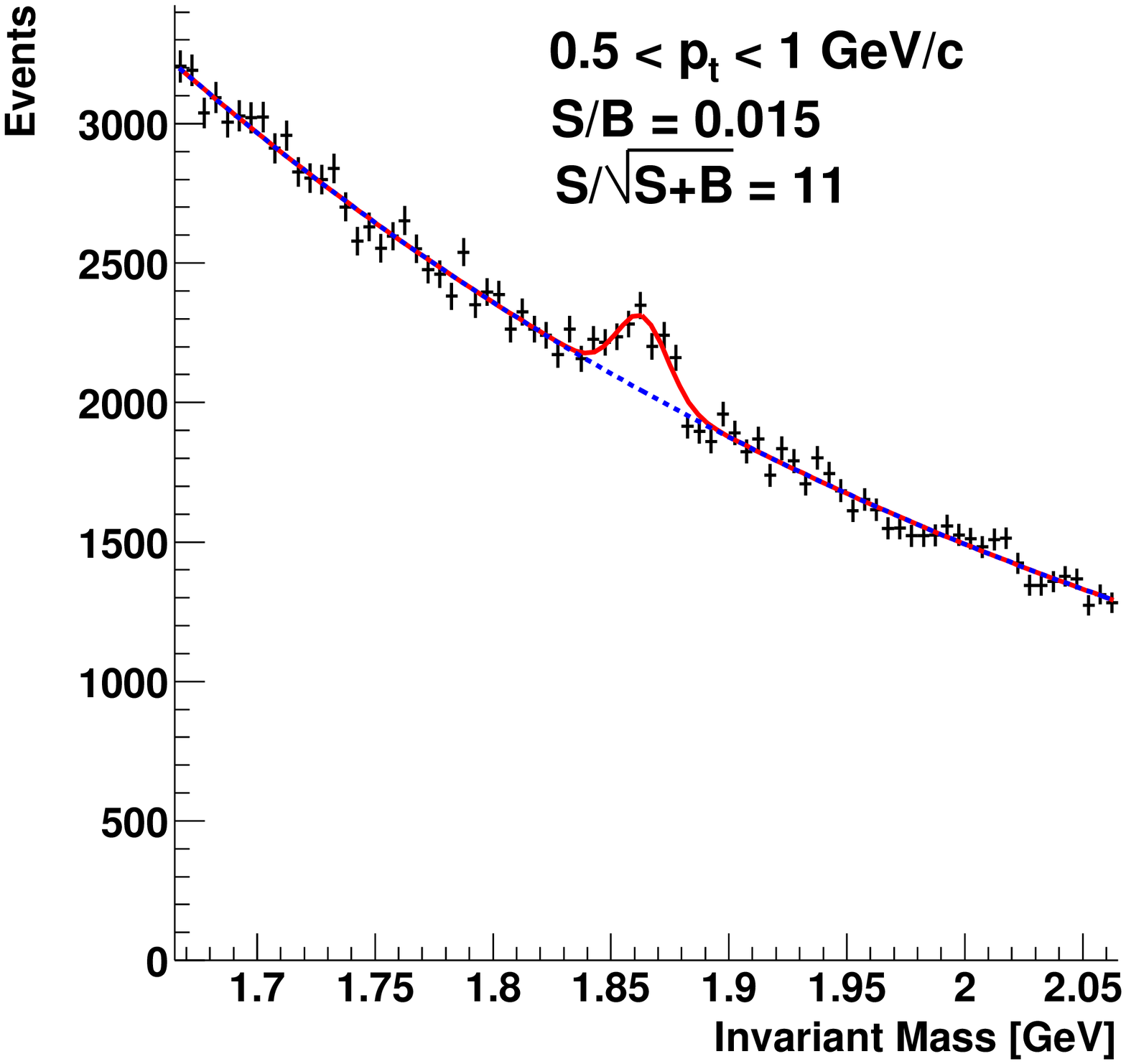}
    \includegraphics[width=.49\textwidth]{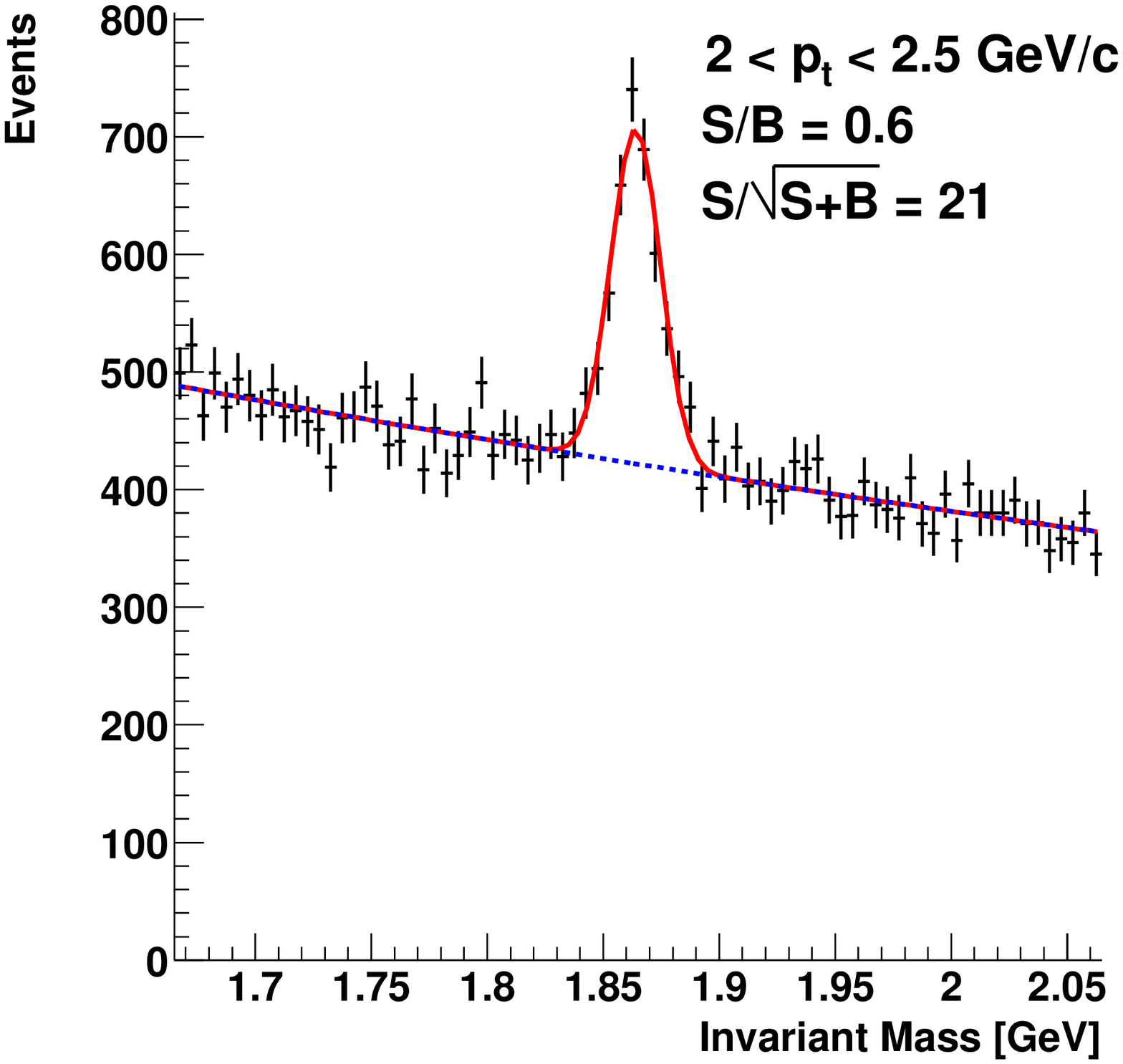}
    \includegraphics[width=.49\textwidth]{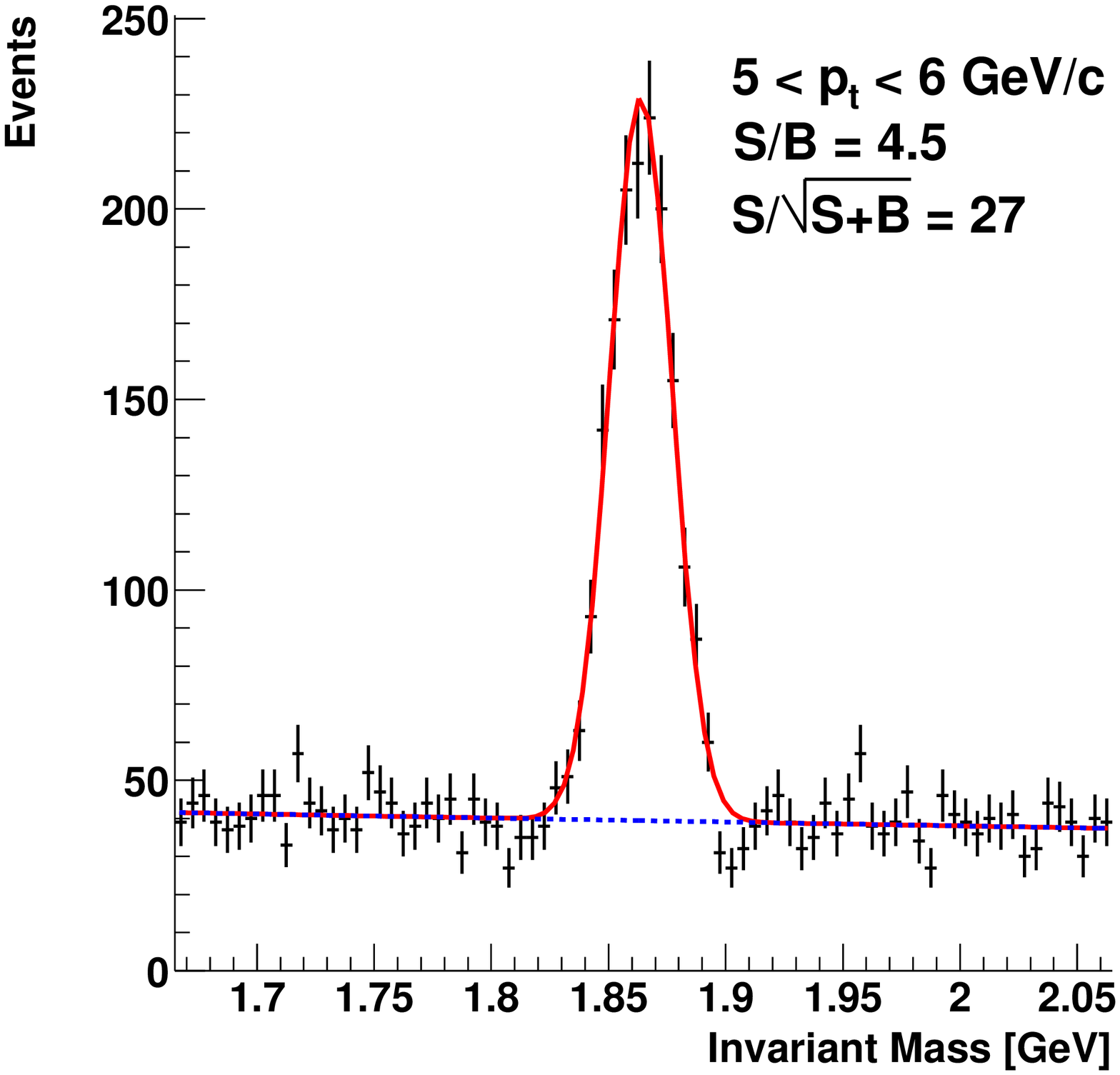}
    \includegraphics[width=.49\textwidth]{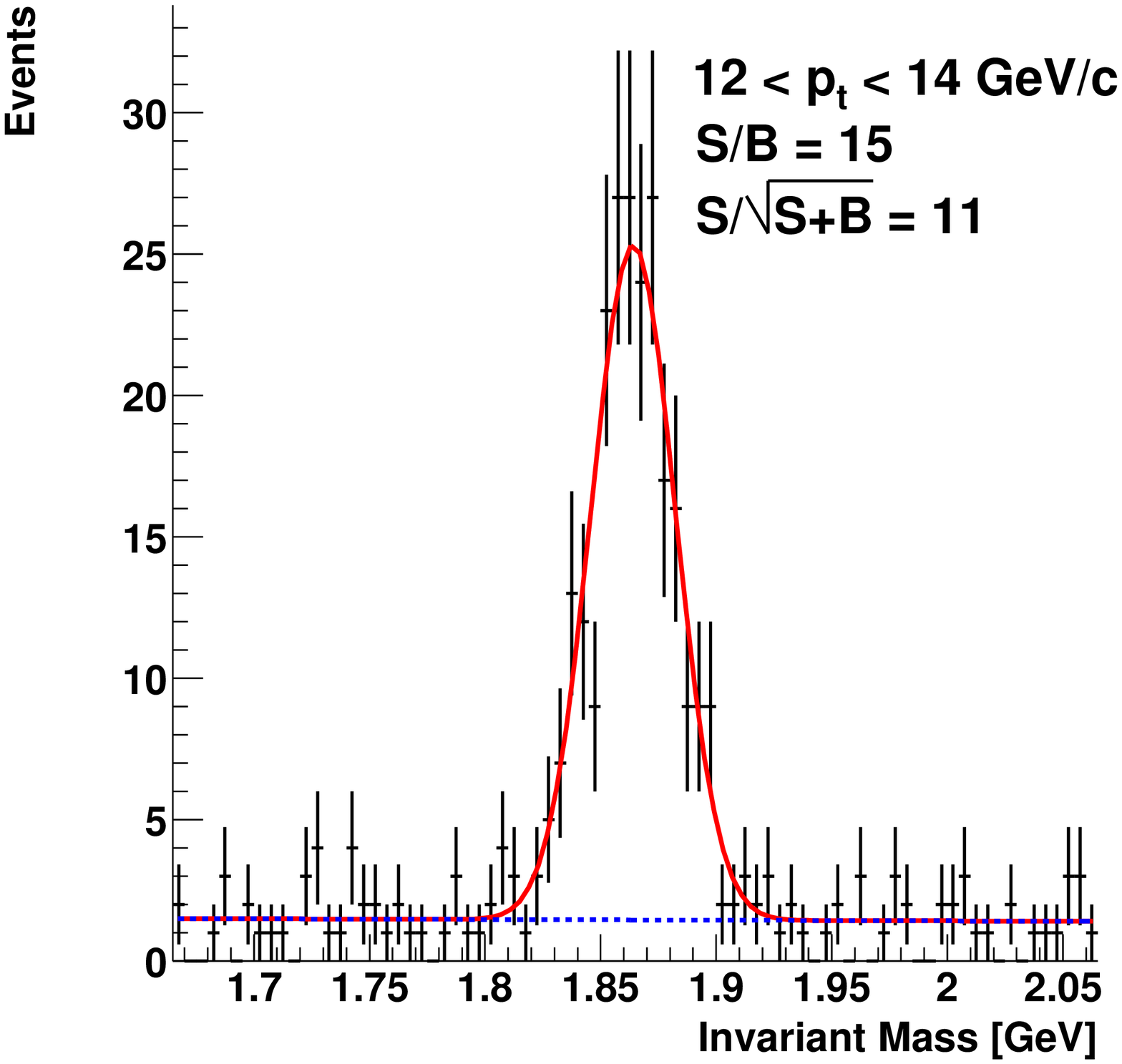}
    \caption{Invariant mass distributions in \pp~collisions fitted to an 
             exponential plus a Gaussian. The statistics correspond to 
             $10^9$ pp events.} 
    \label{fig:massfits}
  \end{center}
\end{figure}

From the `true' invariant mass distribution an histogram (with a 5~MeV bin 
width) was filled and statistical fluctuations were generated by smearing 
the content of each bin according to a Poisson distribution. The histogram
was then fitted to the expression
\begin{equation}
\label{eq:massfit}
f(M)=\frac{P_1 (P_0-P_2)}{\exp(-P_1 M_{\rm min})-\exp(-P_1 M_{\rm max})}\exp(-P_1 M)+\frac{P_2}{\sqrt{2\pi}P_4}\exp\left[-\frac{(M-P_3)^2}{2 P_4^2}\right],
\end{equation}
where $[M_{\rm min},M_{\rm max}]$ is the fit range, $P_0$ is the integral 
of the distribution in the fit range, which is known (sum of bin contents) 
and fixed, $P_1$ is
the slope of the exponential background and $P_2$, $P_3$, $P_4$ are, 
respectively, the integral ($S$), the mean and the $\sigma$ of the Gaussian
that represents the signal. The fit was performed in two steps: (1) fit 
of the side-bands with an exponential to determine a first approximation
of the slope parameter, $P_1$; (2) fit of the whole distribution
to determine $P_1$ and the three parameters of the Gaussian together. Examples
of fitted invariant mass distributions for different $\pt$-bins in 
pp collisions are shown in Fig.~\ref{fig:massfits}.

\begin{figure}[!t]
  \begin{center}
    \includegraphics[width=.72\textwidth]{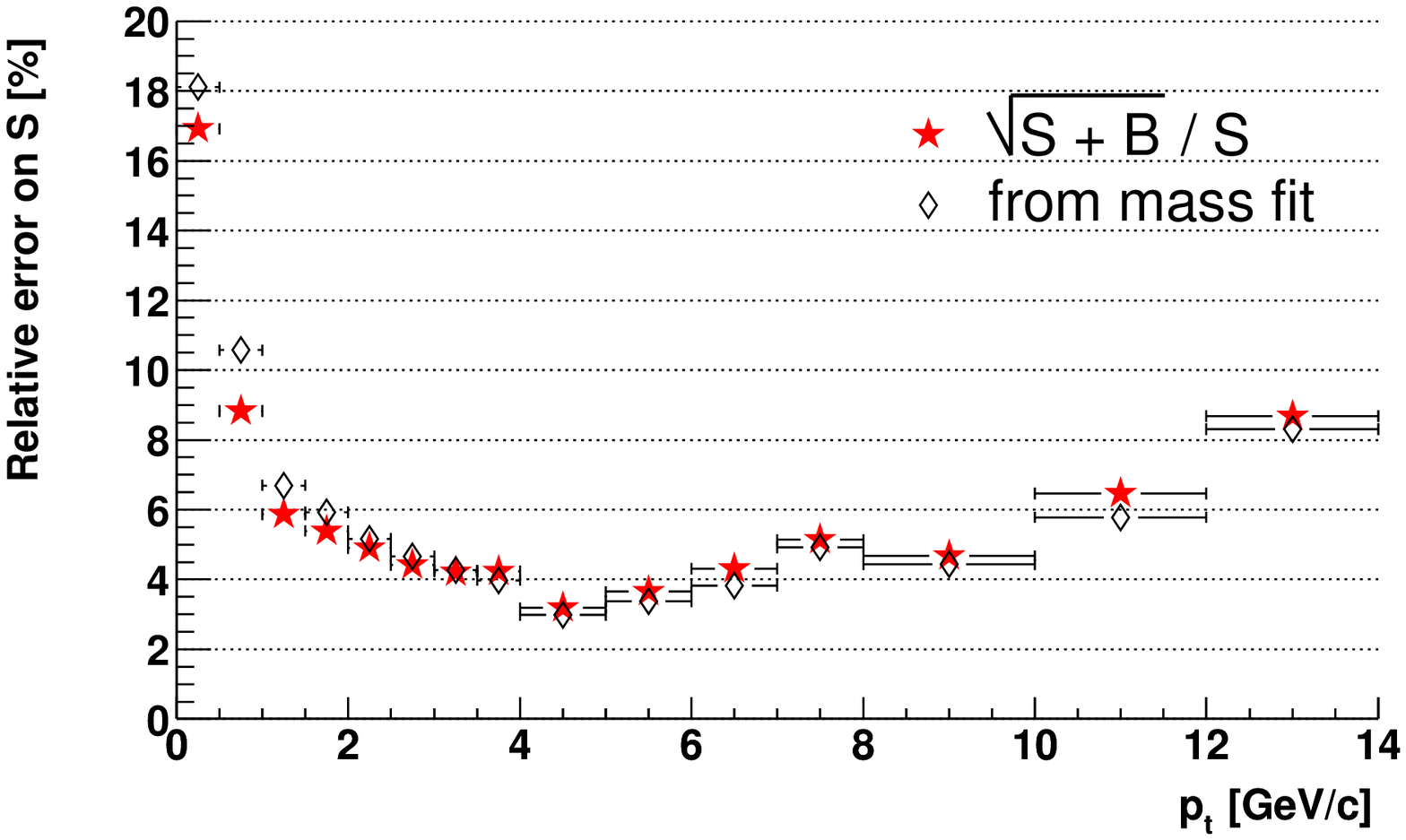}
    \includegraphics[width=.72\textwidth]{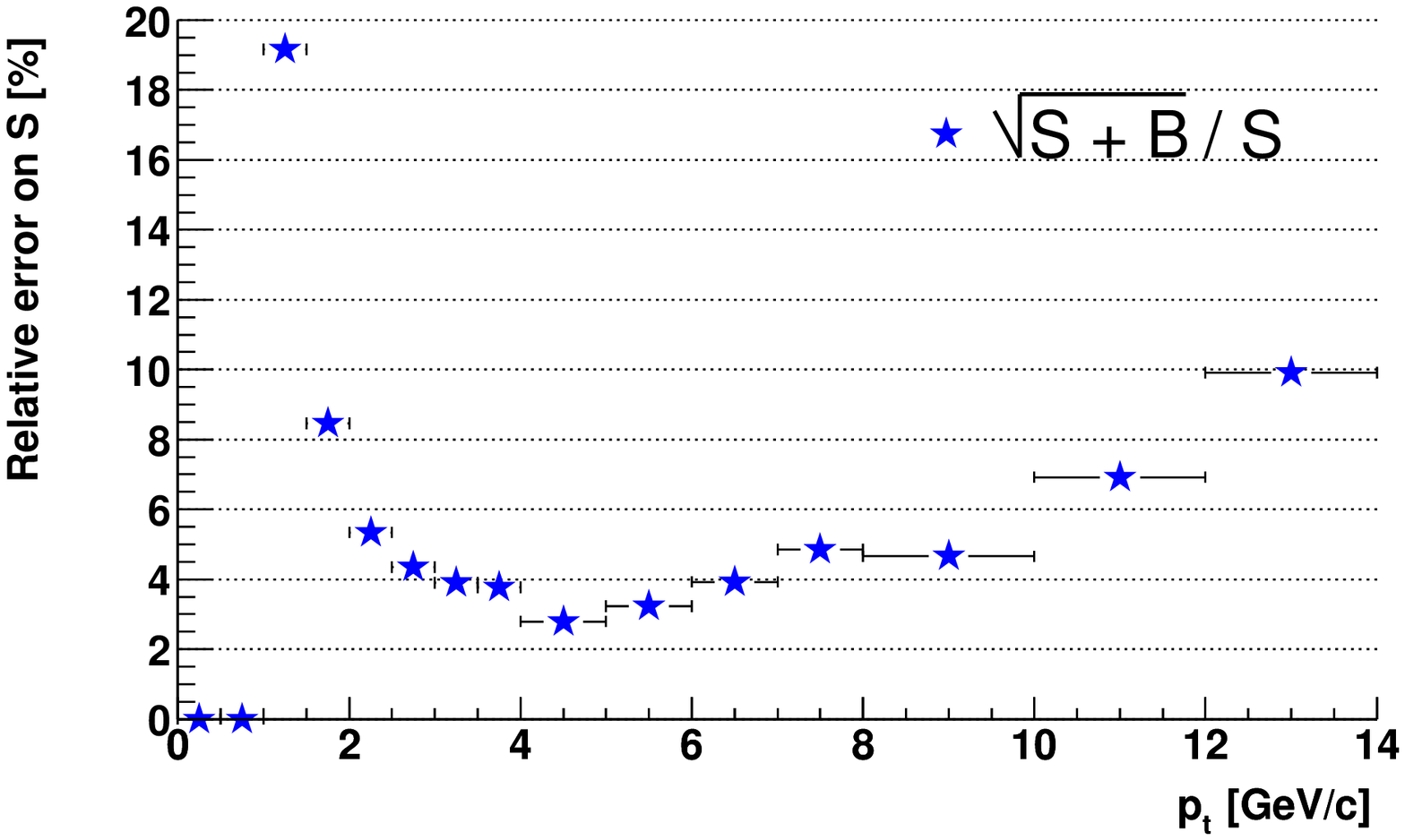}
    \caption{Relative statistical errors estimated using the fit test and
             as the inverse of the significance for $10^9$ pp events (top).
             Relative statistical errors as the inverse of the significance
             for $10^7$ central \PbPb~events (bottom); 
             for $\pt<1~\gev/c$ the error
             is larger than 50\% and was put to 0 only for graphical
             representation; these first two bins are not considered in the 
             following analyses.} 
    \label{fig:estat}
  \end{center}
\end{figure}

The procedure of smearing of the histogram bin contents and fit was 
iterated 1,000 times for each of the 15 $\pt$-bins. For a given bin 
the distribution of the 1,000 residuals of the signal 
integral, $P_2^{\rm fit}-P_2^{\rm true}$, is gaussian and its sigma
gives the statistical error on $S$. We report in Fig.~\ref{fig:estat}
(top panel) the relative statistical errors $\sigma_S/S$ for pp as obtained 
from the described fit test, compared to the expected values, 
$\sqrt{S+B}/S$: the agreement is satisfactory. We use, therefore, 
the inverse of the significance as relative statistical error. 
In the bottom panel of the same figure we show this error for the \PbPb~case.

The relative statistical error is larger at low $\pt$ ($\simeq 20\%$ at
$\pt\simeq 0$-$0.5~\gev/c$  in pp and at $\pt\simeq 1~\gev/c$ in \PbPb), 
where the background accumulates; then it goes down to $\simeq 3\%$
at $\simeq 4~\gev/c$ and it increases again at high $\pt$, where the signal 
statistics decreases.

\begin{figure}[!t]
  \begin{center}
    \includegraphics[width=.85\textwidth]{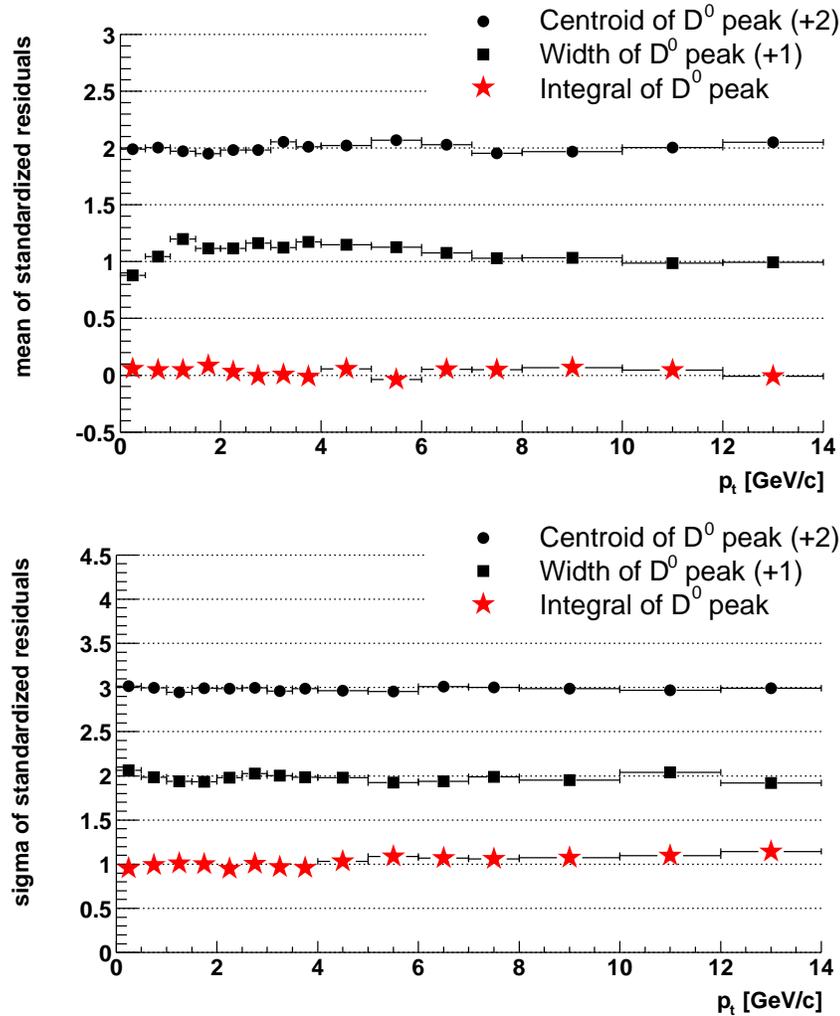}
    \caption{Means (top) and sigmas (bottom) 
             of the distributions of the standardized
             residuals for the three parameters of the invariant mass
             distribution that describe the signal peak. The case of pp 
             collisions is considered.} 
    \label{fig:massfitpulls}
  \end{center}
\end{figure}

The fit algorithm gives also an estimate of the errors, 
$\delta P_i$ for $i=1,2,3,4$, on the 4 parameters,
based on the statistical errors on the bin contents of the invariant mass 
distribution (these errors are visible in Fig.~\ref{fig:massfits}). From the
1,000 iterations we can study the distributions of the standardized residuals,
already introduced in Chapter~\ref{CHAP5}, and defined as 
$(P_i^{\rm fit}-P_i^{\rm true})/\delta P_i$. 
If the fit does not introduce systematic shifts in the estimate of the 
parameters and if the errors given by the fit,
$\delta P_i$, estimate correctly the statistical errors on the parameters,
the distributions of the standardized residuals should be Gaussians with 
mean equal to 0 (no systematic shift) and $\sigma$ equal to 1 (correct error
estimation). Figure~\ref{fig:massfitpulls} shows, for pp, the means (top) and 
the sigmas (bottom) of the distributions of the standardized residuals for the 
parameters $P_2$ (integral of the $\Dz$ peak), $P_3$ (centroid of the 
$\Dz$ peak) and $P_4$ (width of the $\Dz$ peak). Over the whole transverse 
momentum range the means are 0 and the sigmas are 1; we can, therefore, 
conclude that {\sl it is possible to extract the number of selected 
signal $\Dz$ candidates, and its statistical error, by means of a fit on 
the invariant mass distribution, without introducing additional systematic 
errors}. 

\mysection{Estimation of systematic uncertainties}
\label{CHAP7:syst}

The number $S$ of selected signal $\Dz$, estimated from the fit, will have to 
be corrected for efficiency and acceptance in order to obtain the total
and $\pt$-differential yields, or the cross sections, for $\Dz$ production 
per event.
A number of corrections are applied and, in principle, each of them 
introduces a systematic error. A correction consists, essentially, 
in multiplying 
$S$ by a certain factor: $S({\rm corrected})=C\times S($non-corrected$)$;
the systematic error introduced is $\delta C\times S($non-corrected$)$, 
where $\delta C$  is the error on the correction factor $C$. 
In Table~\ref{tab:systerr} we list the main corrections and the 
expected systematic errors that are introduced.

\begin{table}[!t]
  \caption{Main corrections and related systematic errors.} 
  \begin{center}
  \begin{tabular}{rl|l}
  \hline
  \hline
  & {\sl Correction} & {\sl Systematic error} \\
  \hline
  1) & Extrapolation from TOF PID & Matching and PID efficiencies \\
  & to perfect PID &  and contaminations in the TOF \\
  \hline 
  2) & Feed-down from beauty & Uncertainty on $\bbbar$ production at LHC \\
  \hline
  3) & Reconstruction efficiency & Tracking efficiencies and resolutions \\
  \hline
  4) & Acceptance & Geometrical detector acceptance \\
  \hline 
  5) & From $\DtoKpi$ to $\Dz\to X$ & Error on branching ratio $\DtoKpi$\\
  \hline 
  6) & Cross section normalization & \PbPb: error on centrality selection \\
  && and number of binary collisions \\
  && pp: error on inelastic cross section \\
  \hline
  \hline
  \end{tabular}
  \label{tab:systerr}
  \end{center}
\end{table}

The corrections for tracking and PID efficiency and for acceptance
are usually done by means of the Monte Carlo simulation of detector 
geometry and response. The non-perfect description in the simulation 
of the geometry and of the physics processes that determine the detector 
response introduces systematic uncertainties (entries 1, 3 and 4 in 
Table~\ref{tab:systerr}). 
It is reasonable to assume that these uncertainties will initially amount to 
about 10\%. However, we remark that experience from other experiments
tells that this kind of systematic error can be reduced after few years of 
running as the understanding of the detector response improves.

The error on the branching ratio of the $\Dz$ to the ${\rm K^-}\pi^+$ 
channel (entry 5 in Table~\ref{tab:systerr}) is 
quite small, 2.4\%~\cite{pdg}, and it is essentially
negligible with respect to the errors from other sources. However, it 
is included for completeness. 

The other errors listed in the table are considered in the next paragraphs.
 
\subsection{Correction for feed-down from beauty}
\label{CHAP7:systbeauty}

After the selection described in Chapter~\ref{CHAP6}, 
the number of $\Dz$ from c quarks will be determined as 
$N({\rm c\to D^0})=N(\Dz)-N({\rm b\to B \to D^0})$, where $N(\Dz)$ is the 
total number of selected $\Dz$ and $N({\rm b\to B \to D^0})$ is the
amount of feed-down from beauty, that will be estimated via Monte Carlo.

The systematic error introduced by this correction is equal to the 
error on the estimated number of $\Dz$ mesons from beauty that pass the 
selection cuts, $N({\rm b\to B \to D^0})$. The relative error on 
$N({\rm b\to B \to D^0})$ is essentially equal to the relative error on 
the $\bbbar$ production cross section, 
$\sigma^{\scriptstyle {\rm b\overline{b}}}$, at LHC energy and, thus, the 
relative error on $N({\rm c\to D^0})$ is equal to the relative error 
on $\sigma^{\scriptstyle {\rm b\overline{b}}}$ multiplied by the ratio of 
secondary to primary $\Dz$, after selections. Such ratio, as obtained with 
the present baseline on charm and beauty production, has been shown
in Figs.~\ref{fig:bD0tocD0PbPb} and~\ref{fig:bD0tocD0pp}, for 
\PbPb~and pp respectively, and it amounts to about 10\% on average.
At present, the $\bbbar$ cross section at LHC energies is estimated by pQCD
calculations at NLO with a very large theoretical uncertainty, $\simeq 80\%$, 
as we have reported in Chapter~\ref{CHAP3}. We assume here this large relative
error. This is probably an overestimate, since B meson production will be 
measured by ALICE, in the semi-electronic decay 
channel~\cite{notebeauty}, and also by ATLAS and CMS. However, it is not 
yet clear how precise these measurements can be, especially at low $\pt$. 

\begin{figure}[!t]
  \begin{center}
    \includegraphics[width=.85\textwidth]{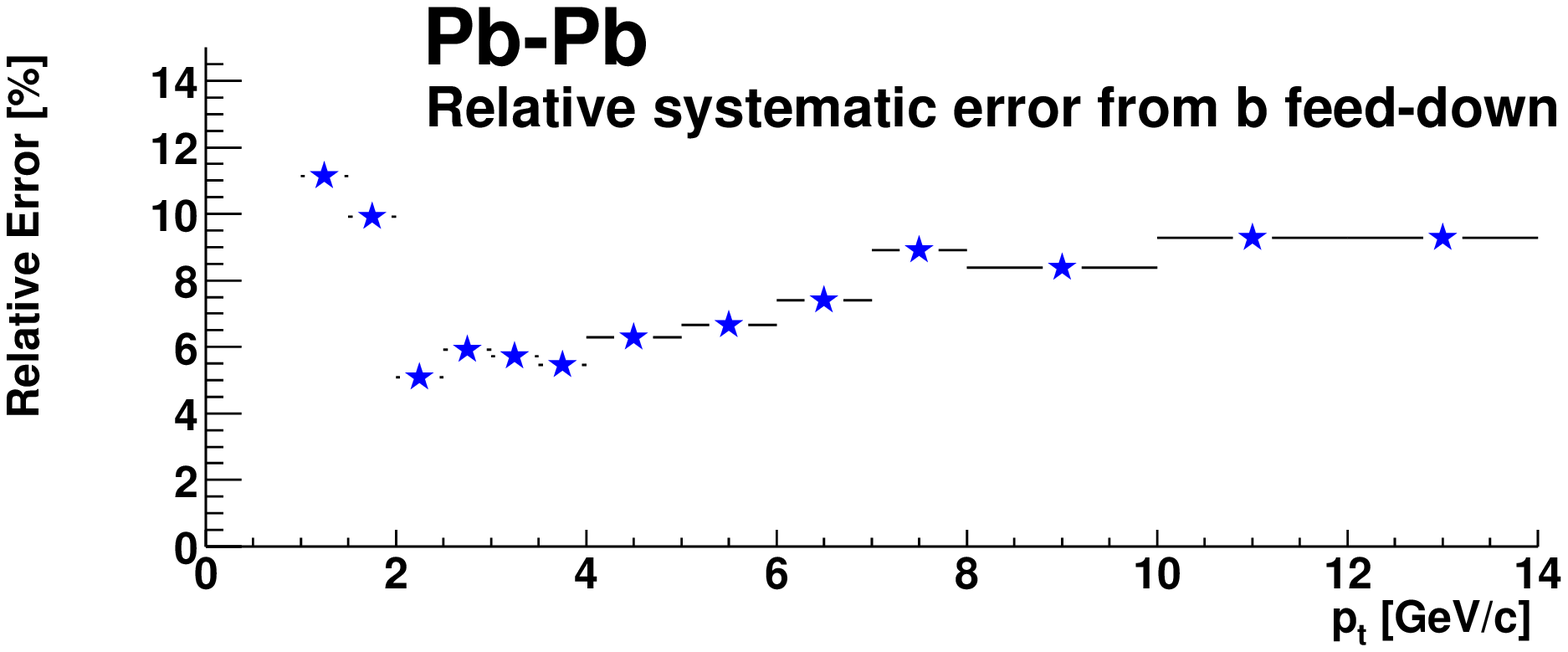}
    \includegraphics[width=.85\textwidth]{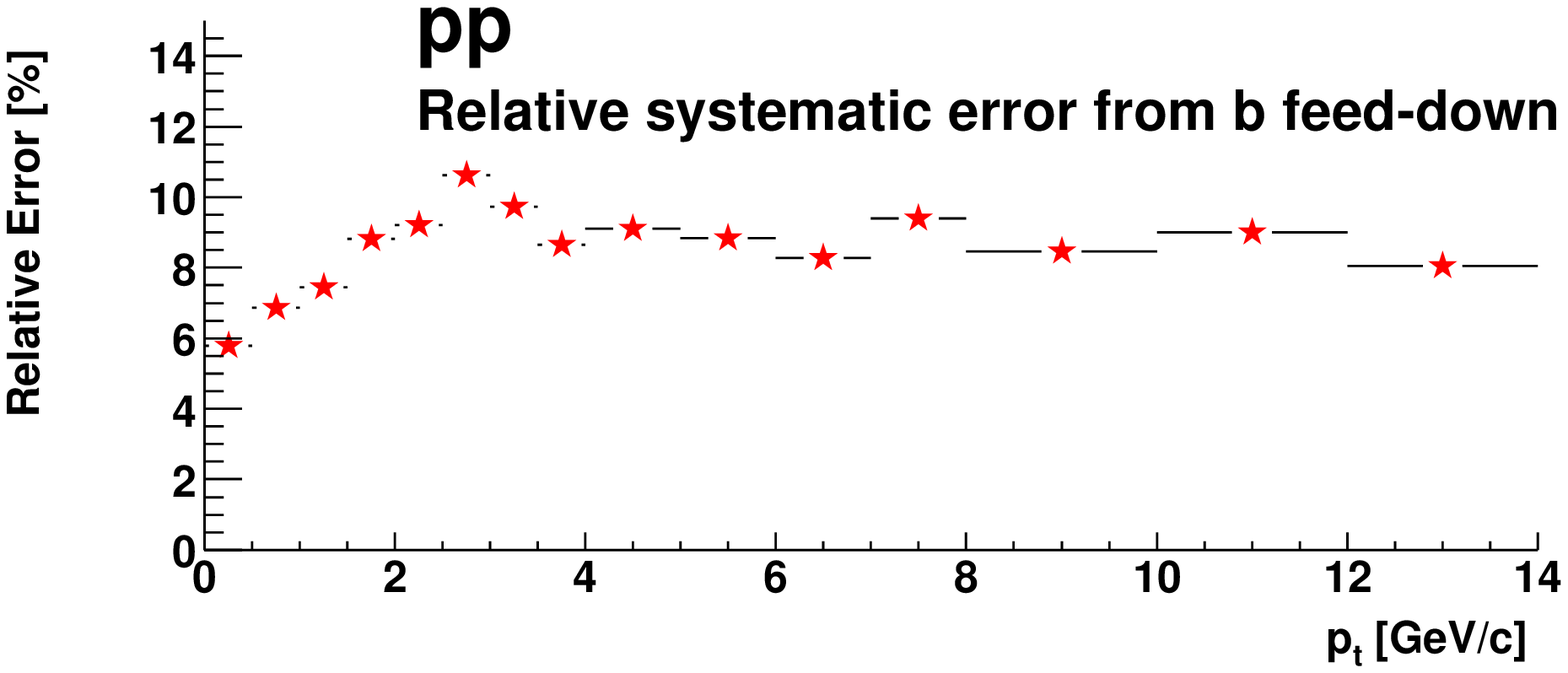}
    \caption{Relative systematic error from the correction for feed-down 
             from beauty, in \PbPb~(top) and pp (bottom).} 
    \label{fig:errfromb}
  \end{center}
\end{figure}

Multiplying the two factors, 10\% (ratio of selected secondary to primary 
$\Dz$) and 80\% (relative uncertainty on 
$\sigma^{\scriptstyle {\rm b\overline{b}}}$), we obtain an average relative 
error arising from the correction for beauty feed-down of about 8\%.
We report in Fig.~\ref{fig:errfromb} this relative error as a function of 
$\pt$ for \PbPb~(top) and for pp (bottom). The error has different trends 
as a function of $\pt$ in the two cases because different selection cuts
are applied in \PbPb~and in pp. 

Recently, the CDF Collaboration has directly estimated the fraction of 
primary $\Dz$ using the distance of the reconstructed $\Dz$ 
flight line to the interaction vertex as a variable to separate primary 
and secondary $\Dz$~\cite{CDFd0D0}. This technique allows them to 
correct for the feed-down with a systematic error as low as $3$-$5\%$.
A preliminary study on the possibility to use the same technique in 
ALICE was carried out and will be described at the end of this chapter.  

\subsection{Cross section normalization}
\label{CHAP7:systxsec}

In \pp~collisions, the cross section for $\Dz$ production, which is 
necessary for example for the comparison with pQCD calculations, 
can be determined 
multiplying the estimated number of (primary) $\Dz$ produced per inelastic 
event by the inelastic pp cross section at $\sqrt{s}=14~\tev$:
\begin{equation}
\sigma_{\rm pp}(\Dz) = N_{\rm pp}(\Dz)/{\rm inel.~event} \times \sigma^{\rm inel}_{\rm pp}.
\end{equation}
The pp cross section will be measured at the LHC by the TOTEM 
experiment~\cite{totem} with an expected precision of about 5\%. Therefore, 
the normalization of $\Dz$ production to pp inelastic collisions will 
contribute a systematic error of $\simeq 5\%$, of course independent of $\pt$.

In the case of central \AA~collisions, the $\Dz$ production cross section per 
binary NN collision, which enters in the calculation of the nuclear 
modification factor
(Section~\ref{CHAP2:physicsoutline}), can be derived as:
\begin{equation}
\sigma_{\scriptscriptstyle{\rm NN}}(\Dz) = N_{\rm AA}(\Dz)/{\rm event}/ R(b_c),
\end{equation} 
where $R(b_c)$, defined in Eq.~(\ref{eq:rbc}), is essentially the 
average number of binary NN collisions in an AA collision with impact 
parameter $b<b_c$, divided by the inelastic AA cross section corresponding 
to the same impact parameter range.
Three sources contribute to the error on $R(b_c)$: 
\begin{enumerate}
\item {\sl Error on the centrality selection}, i.e. on the determination of 
      the upper limit $b_c$ ($\simeq 3.5~\fm$) in impact parameter for the 
      class of most 
      central \PbPb~collisions (5\% of the total cross section). 
      The impact parameter, as mentioned in 
      Chapter~\ref{CHAP4}, is measured in ALICE by means of the Zero Degree 
      Calorimeters, which are expected to provide a relative precision 
      $\delta b/b\simeq 30\%$ for $b<3$-$4~\fm$~\cite{pprCh6-1}. 
      The upper limit  
      will be, therefore, determined as $b_c =(3.5\pm 1.0)~\fm$, which gives 
      $R(b_c)=(27\pm 2)~\mb^{-1}$ (from Fig.~\ref{fig:fhardANDrbc}, right 
      panel). The relative error on $R(b_c)$ and on 
      $\sigma_{\scriptscriptstyle{\rm NN}}(\Dz)$
      from this source is then $2/27 \simeq 8\%$.
\item {\sl Error on the NN inelastic cross section} at $\sqrt{s}=5.5~\tev$.
      Once the cross section will be measured at $\sqrt{s}=14~\tev$ by 
      TOTEM, the 
      extrapolation to lower energy should not introduce large additional
      uncertainties. We, therefore, assume a 5\% precision also at 5.5~TeV.
\item {\sl Uncertainty on the parameters of the Wood-Saxon nuclear density 
      profile.} These uncertainties are of order 5\%~\cite{woodsaxon}.   
\end{enumerate}
Combining these three contributions we obtain an overall normalization 
error of about 11\% for central \PbPb~collisions.

\mysection{Errors on ${\rm d}^2\sigma(\Dz)/{\rm d}\pt{\rm d}y$ and 
           ${\rm d}\sigma(\Dz)/{\rm d}y$}
\label{CHAP7:errsummary}

\begin{figure}[!b]
  \begin{center}
    \includegraphics[width=.85\textwidth]{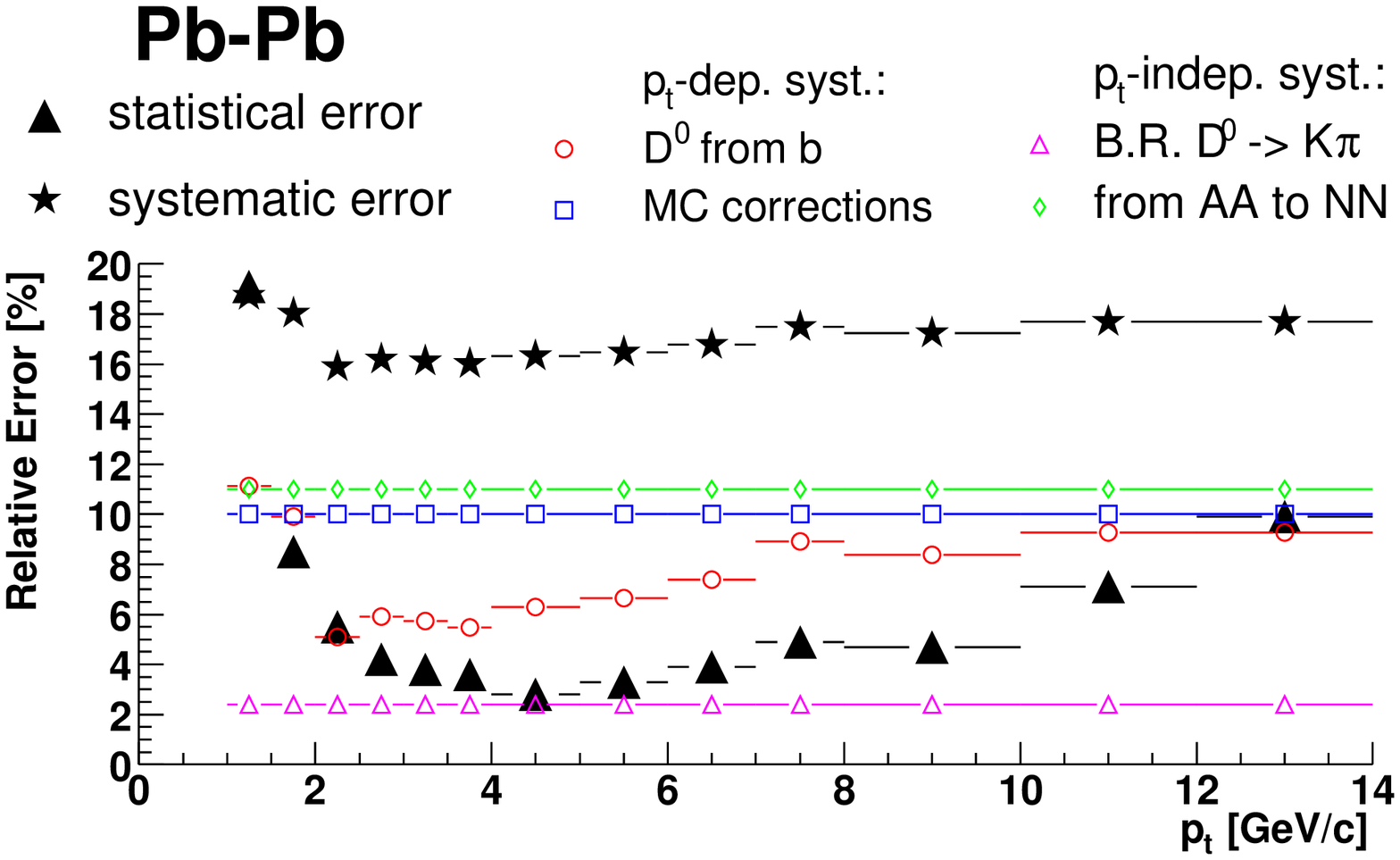}
    \includegraphics[width=.85\textwidth]{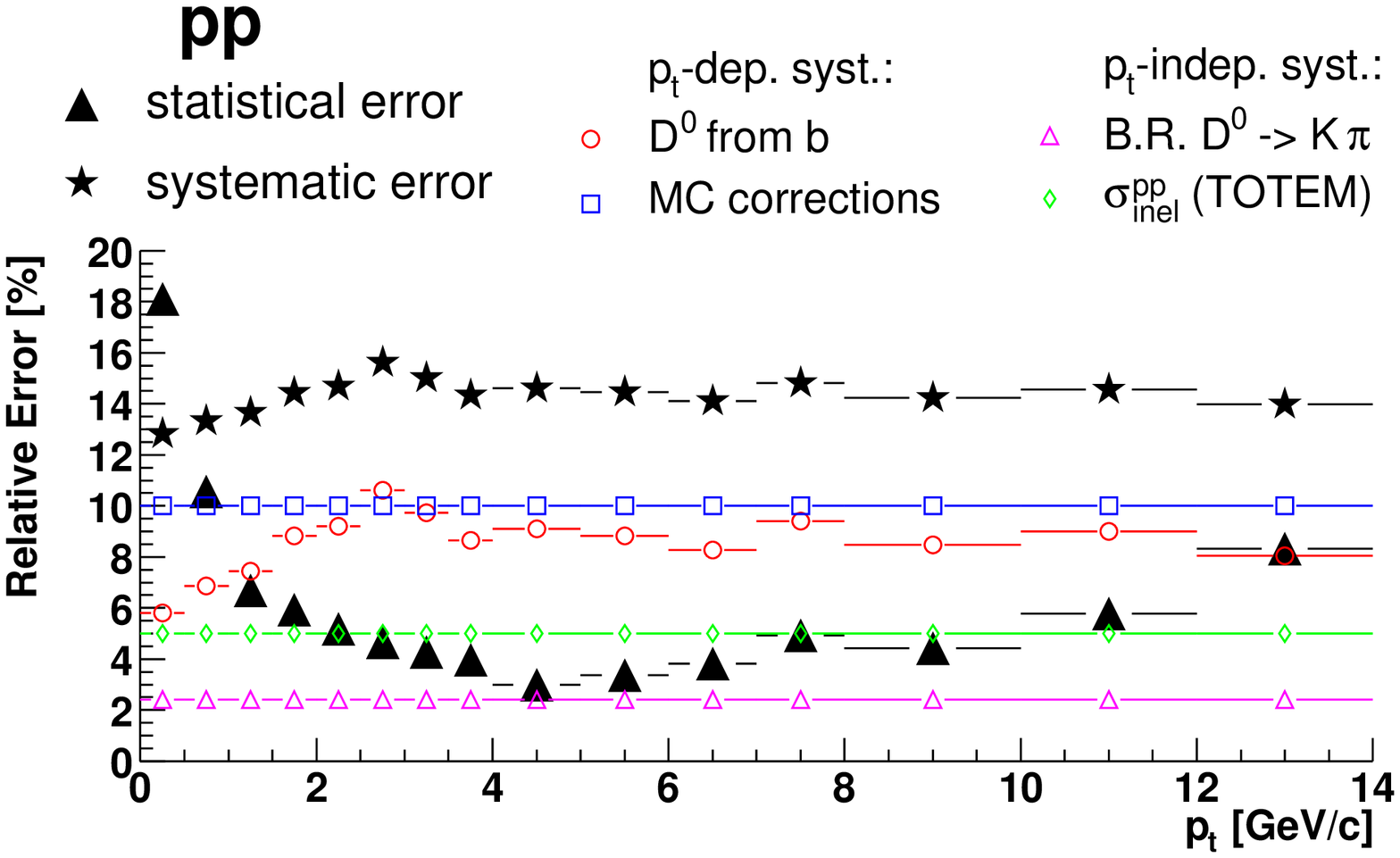}
    \caption{Summary of relative errors in \PbPb~and pp collisions.} 
    \label{fig:allerrors}
  \end{center}
\end{figure}

The relative statistical errors and the different contributions to the
relative systematic errors are summarized in Fig.~\ref{fig:allerrors}.
The total systematic error, obtained as a quadratic sum of the single 
contributions, amounts to 16-17\% for the \PbPb~case and 
14-15\% for the pp case. However, we remark that (a) some of the systematic 
errors do not affect the shape of the $\pt$ distributions (uncertainty on 
branching ratio and normalization errors) and (b) many of them are common 
to \PbPb~and pp and will cancel in the ratio $\RAA$ (correction for b 
feed-down, branching ratio, uncertainty on NN cross section and, 
partially, Monte Carlo corrections, e.g. acceptance).

\begin{figure}[!b]
  \begin{center}
    \includegraphics[width=.75\textwidth]{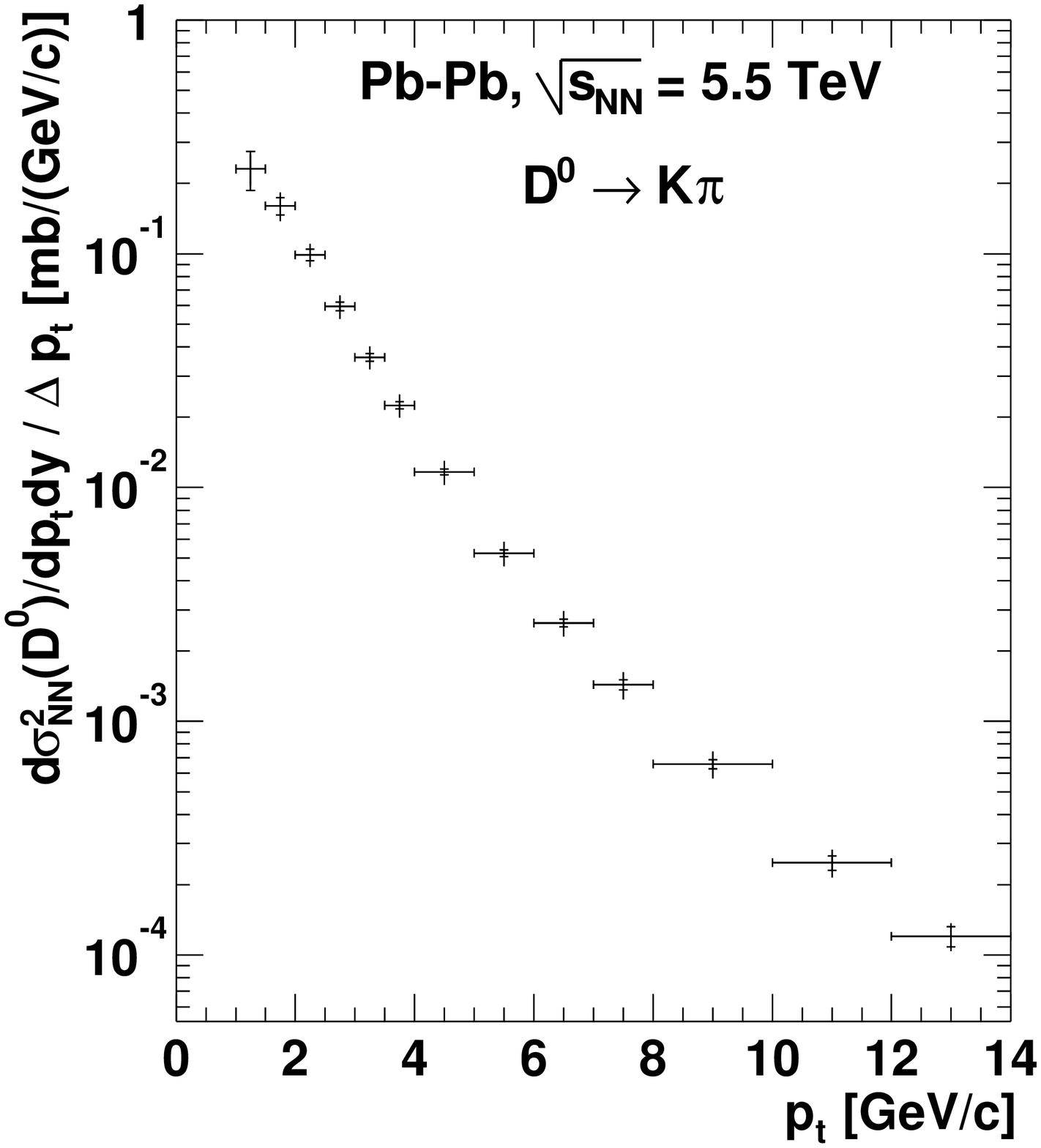}
    \caption{Double differential cross section per \NN~collision 
             for $\Dz$ production as a 
             function of $\pt$, as it can be measured with $10^7$ central  
             \PbPb~events. Statistical (inner bars) and $\pt$-dependent 
             systematic errors (outer bars) are shown. A normalization error
             of 11\% is not shown.} 
    \label{fig:dsigmadpt_PbPb}
  \end{center}
\end{figure}

\begin{figure}[!b]
  \begin{center}
    \includegraphics[width=.75\textwidth]{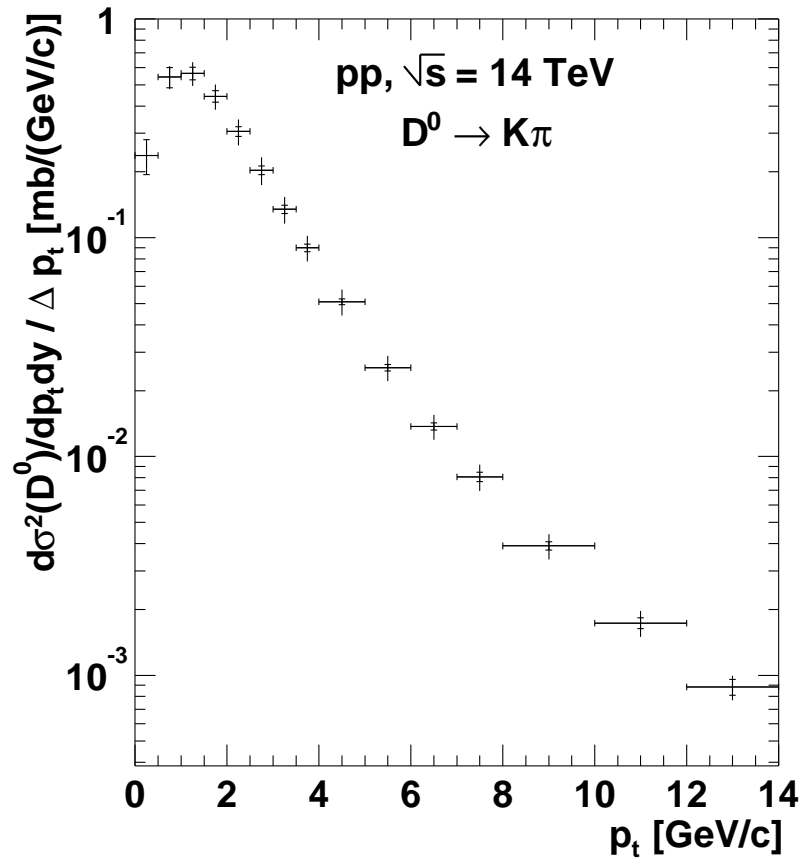}
    \caption{Double differential cross section for $\Dz$ production as a 
             function of $\pt$, as it can be measured with $10^9$ pp  
             events. Statistical (inner bars) and $\pt$-dependent 
             systematic errors (outer bars) are shown. A normalization error
             of 5\% is not shown.} 
    \label{fig:dsigmadpt_pp}
  \end{center}
\end{figure}

Figures~\ref{fig:dsigmadpt_PbPb} and~\ref{fig:dsigmadpt_pp} present
the distributions of d$^2\sigma(\Dz)/$d$\pt$d$y$ in $|y|<1$ with the estimated 
statistical (inner) and $\pt$-dependent systematic (outer) error bars. 
A normalization error of 11\% for \PbPb~and 5\% for pp is not included in 
the error bars, as it will not affect the shape of the transverse momentum 
distribution. 

The expected relative uncertainties for the measurement of the 
$\Dz$ production cross section per unit of rapidity, integrated over 
$\pt>\pt^{\rm min}=1~\gev/c$ for \PbPb~and $\pt>\pt^{\rm min}=0.5~\gev/c$ 
for pp, are reported in Table~\ref{tab:errtotxsec}. 
The statistical uncertainty was obtained as a 
quadratic sum of the statistical errors of the $\pt$-bins for 
$\pt>\pt^{\rm min}$. The single contributions to the systematic uncertainty 
were obtained as a linear sum over the $\pt$-bins and they were then 
added in quadrature to get the total systematic uncertainty.

Concerning the pp result, we remark that, with statistical and systematic 
errors of 3\% and 14\%, and $\pt^{\rm min}=0.5~\gev/c$,
the reconstruction of $\DtoKpi$ decays in ALICE will provide the only 
precise measurement of charm production cross section at LHC energy. 
For comparison, the CDF Collaboration has recently measured $\Dz$ production 
in ${\rm p\overline{p}}$ collisions at the Tevatron, $\sqrt{s}=1.96~\tev$, 
with similar uncertainties, 1.5\% statistical and 11\% systematic, 
but with a much higher low-$\pt$ cut-off, 
$\pt^{\rm min}=5.5~\gev/c$~\cite{CDFd0D0}. 

\begin{table}[!t]
  \caption{Expected relative uncertainties for the measurement of 
           d$\sigma(\Dz)/$d$y$ 
           in $|y|<1$ and $\pt>\pt^{\rm min}$.} 
  \begin{center}
  \begin{tabular}{l|c|c}
  \hline
  \hline
  System & \PbPb & pp \\
  & $\pt^{\rm min}=1~\gev/c$ & $\pt^{\rm min}=0.5~\gev/c$ \\
  \hline
  {\bf Statistical error} & {\bf 7\%} & {\bf 3\%} \\
  {\bf Systematic error} & {\bf 17\%} & {\bf 14\%} \\
  ~~~~~~~Correction for b feed-down & 9\% & 8\% \\
  ~~~~~~~Monte Carlo corrections & 10\% & 10\% \\
  ~~~~~~~Branching ratio & 2\% & 2\% \\
  ~~~~~~~Cross section normalization & 11\% & 5\% \\
  \hline
  \hline
  \end{tabular}
  \label{tab:errtotxsec}
  \end{center}
\end{table}

\mysection{Comparison with pQCD predictions}
\label{CHAP7:cmppQCD}

In Section~\ref{CHAP3:XsecNN} we have seen that the results of perturbative 
QCD calculations for $\ccbar$ (and $\bbbar$) production at the LHC have a 
strong dependence on the choice of the heavy quark masses and of the 
factorization and renormalization scales, $\mu_F$ and $\mu_R$.
We compare here to this theoretical uncertainty the sensitivity of ALICE 
for the measurement of the total and $\pt$-differential cross section for 
$\Dz$ production in pp collisions at $14~\tev$. We used the program by 
Mangano, Nason and Ridolfi (HVQMNR)~\cite{MNRcode} to calculate 
the cross sections for different sets of parameters. The $\pt$ distributions 
for D mesons were obtained from those for c quarks using the 
PYTHIA fragmentation model (more details on the procedure are given 
in the next chapter, Section~\ref{CHAP8:charmenergyloss}). 

\begin{figure}[!t]
  \begin{center}
    \includegraphics[width=.85\textwidth]{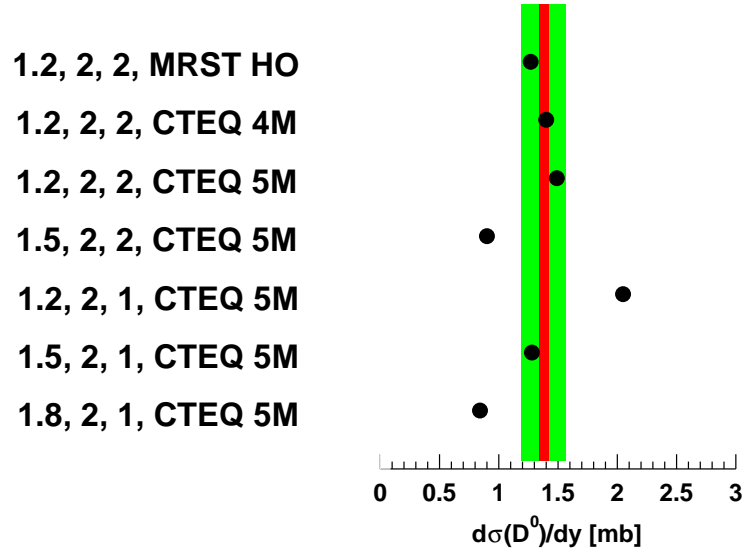}
    \caption{ALICE sensitivity on d$\sigma(\Dz)/$d$y$ integrated for
             $\pt>0.5~\gev/c$, in pp at 14~TeV, compared to the 
             pQCD predictions obtained with different sets of the input 
             parameters $m_{\rm c}$ [GeV], 
             $\mu_F/\mu_0$, $\mu_R/\mu_0$ and
             PDF set ($\mu_0$ is defined in the text). The narrower band 
             represents the statistical error, the broader band the systematic 
             error, including all normalization errors.} 
    \label{fig:cmptotxsec}
  \end{center}
\end{figure}

\begin{figure}[!t]
  \begin{center}
    \includegraphics[width=.49\textwidth]{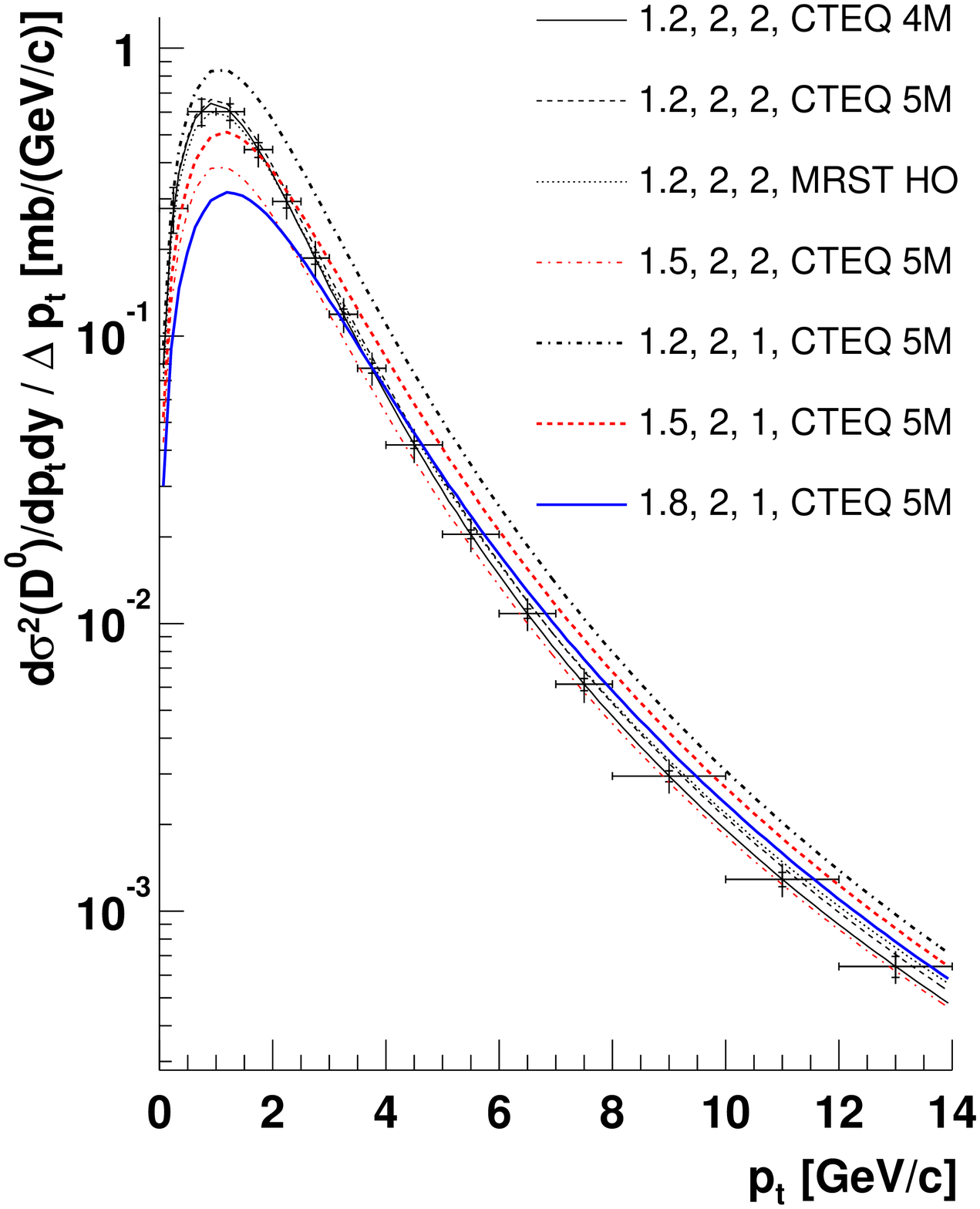}
    \includegraphics[width=.49\textwidth]{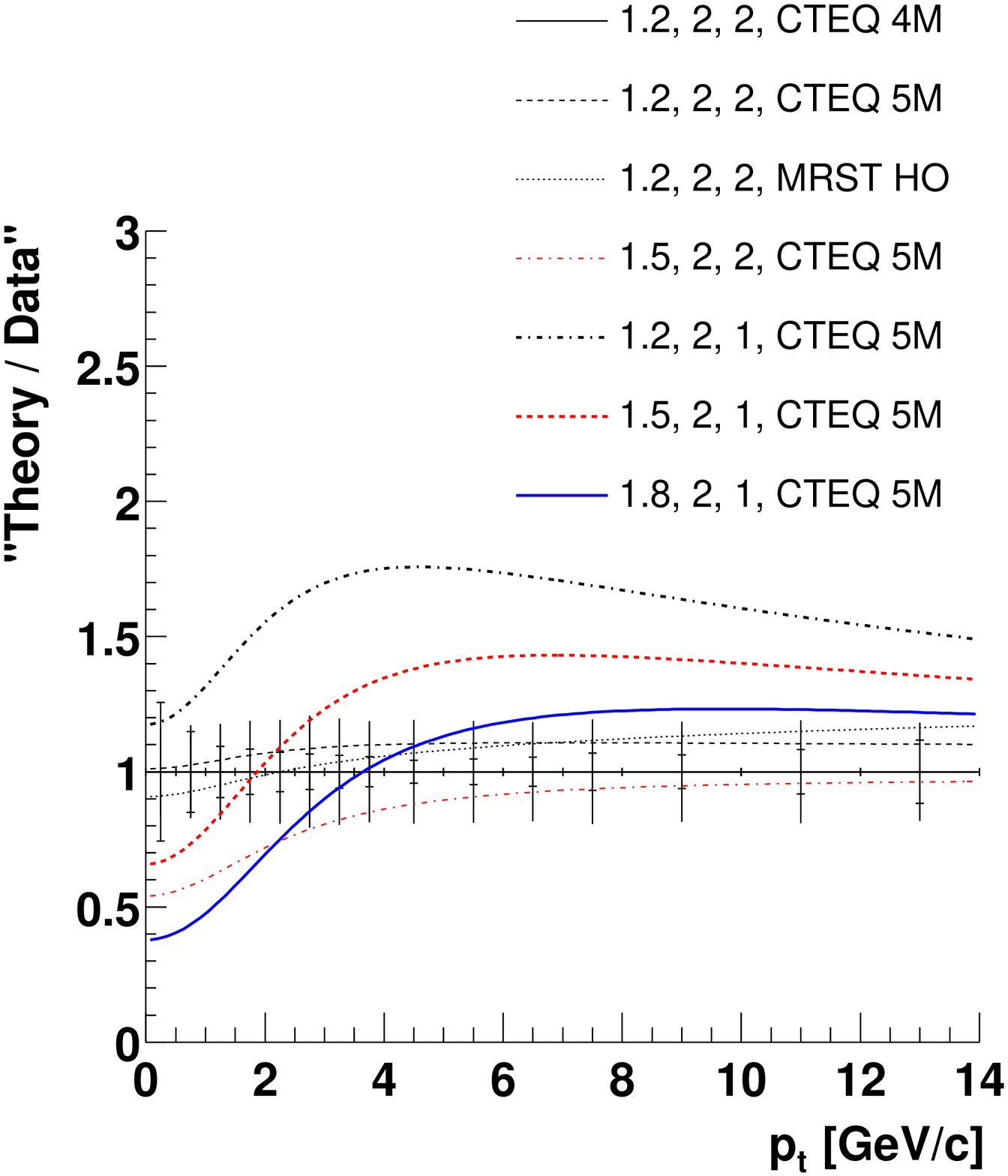}
    \caption{ALICE sensitivity on d$^2\sigma(\Dz)/$d$\pt$d$y$,
             in pp at 14~TeV, compared to the 
             pQCD predictions obtained with different sets of the input 
             parameters $m_{\rm c}$ [GeV], $\mu_F/\mu_0$, $\mu_R/\mu_0$ and
             PDF set ($\mu_0$ is defined in the text). The inner bars 
             represent the statistical error, the outer bars the 
             $\pt$-dependent systematic error. A normalization error of 
             5\% is not shown. The panel on the right shows how a `theory/data'
             plot could look like.} 
    \label{fig:cmpptxsec}
  \end{center}
\end{figure}

Figure~\ref{fig:cmptotxsec} shows the comparison for d$\sigma(\Dz)/$d$y$, 
integrated for $\pt>0.5~\gev/c$. Statistical (narrower) and systematic 
(broader) error bands are reported; the latter include all normalization 
errors. The error bars are hereafter applied to the value obtained with the 
set of parameters used in our simulations (`default parameters'): 
$m_{\rm c}=1.2~\gev$, 
$\mu_F=\mu_R=2\mu_0=2\sqrt{(p_{\rm t,c}^2+p_{\rm t,\overline{c}}^2)/2+m_{\rm c}^2}$
and PDF set = CTEQ 4M.
The comparison for the $\pt$-differential cross section is presented 
in Fig.~\ref{fig:cmpptxsec} along with the ratio `theory/data' 
(`theory parameters/default parameters') which better allows
to compare the different $\pt$-shapes obtained by changing the input 
`theory parameters' and to illustrate the sensitivity of the 
ALICE measurement. 

\mysection{Energy extrapolation of the pp result}
\label{CHAP7:ppextrapolation}

The comparison plots presented in the previous section show that the 
measurement of $\Dz$ production in pp collisions at 14~TeV will be 
accurate enough to allow some kind of `tuning' of the pQCD calculations
by choosing the set/sets of input parameters that better reproduce the 
experimental total cross section and $\pt$ distribution.  

After such parameter selection, perturbative QCD can be used for the 
extrapolation of pp results from 14 to 5.5~TeV. We remind that this step 
is necessary in order to calculate the nuclear modification factor $\RAA$. 

\begin{figure}[!t]
  \begin{center}
    \includegraphics[width=.6\textwidth]{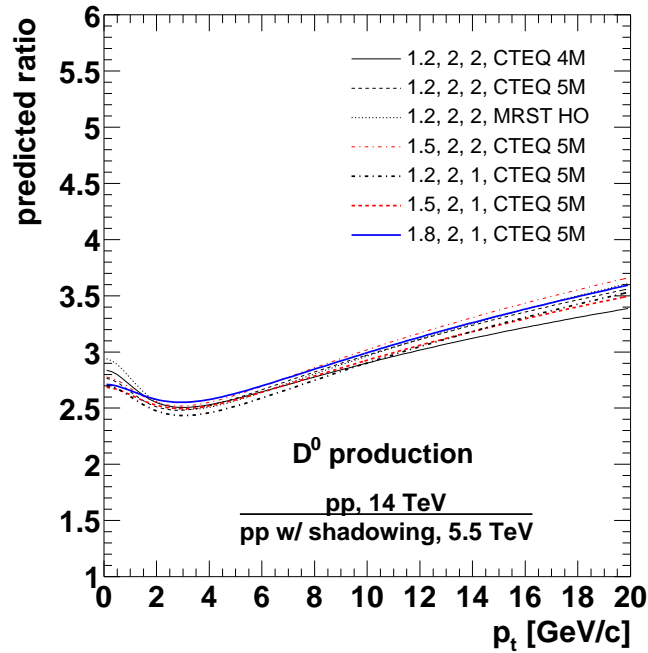}
    \caption{Ratio of the $\pt$-differential $\Dz$ cross section 
             given by pQCD at 14 and at 5.5~TeV for different sets of the 
             input parameters: $m_{\rm c}$ [GeV], $\mu_F/\mu_0$, $\mu_R/\mu_0$
             and PDF set.} 
    \label{fig:energyextrapolation}
  \end{center}
\end{figure}

In Chapter~\ref{CHAP3} (Table~\ref{tab:manganoXsec}) 
we showed that the ratio of the
total $\ccbar$ cross section at different energies is almost independent 
of the input parameters used. We consider here this point more in detail 
by comparing the ratios of the $\pt$-differential $\Dz$ cross sections 
at 14 and at 5.5~TeV for different sets of parameters. The result is 
reported in Fig.~\ref{fig:energyextrapolation}: the ratio is independent 
of the input parameters within 10\% up to $\pt=20~\gev/c$. Therefore,
the `tuning' of the parameters has only a marginal importance, since the 
$\pt$ distribution measured at 14~TeV allows to predict the $\pt$ 
distribution at 5.5~TeV. The spread in the ratio displayed in 
Fig.~\ref{fig:energyextrapolation} can be taken as an estimate of the 
error introduced by the energy extrapolation; such error shall be accounted 
for in the evaluation of the sensitivity on $\RAA$ in Chapter~\ref{CHAP8}.

\mysection{Perspectives for the measurement of\\ 
$N({\rm b\to B\to D^0})/N({\rm c\to D^0})$}
\label{CHAP7:D0d0}

Primary and secondary $\Dz$ mesons can, in principle, be separated on the 
basis of their impact parameter to the interaction vertex. For a $\Dz$
with reconstructed decay vertex $(V_x,V_y,V_z)$ and reconstructed momentum
at the decay vertex 
$(p_x,p_y,p_z)$, the impact parameter in the transverse plane
is defined as the distance of the straight line parameterized as 
$(V_x,V_y)+k\,(p_x,p_y)$ to the reconstructed position of the interaction 
vertex in the transverse plane. For $\Dz$ mesons coming from c quarks,
the impact parameter should be 0, within the experimental resolution,
while for $\Dz$ mesons coming from the decay of B mesons, it should 
differ significantly from 0, as B mesons have a mean decay length of 
order $500~\mum$. However, the separation may not be very 
clear because, in the B decay, the $\Dz$ is the heavier particle and 
it tends to carry most of the momentum of the B and, consequently, 
to follow its flight direction, thus having a small impact parameter.

\begin{figure}[!t]
  \begin{center}
    \includegraphics[width=.7\textwidth]{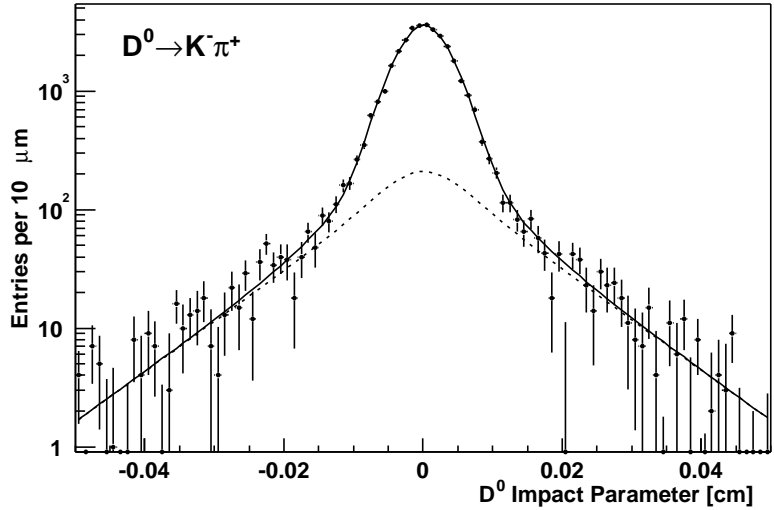}
    \caption{CDF data: the impact parameter distribution of $\Dz$ mesons,
             measured from the $\pm 2~\sigma$ signal region of the 
             invariant mass distribution and corrected for combinatorial 
             background measured in the invariant mass side-bands. The 
             dashed curve shows the contribution of secondary $\Dz$ from B
             decays~\cite{CDFd0D0}.} 
    \label{fig:CDFd0D0}
  \end{center}
\end{figure}

\begin{figure}[!t]
  \begin{center}
    \includegraphics[width=.9\textwidth]{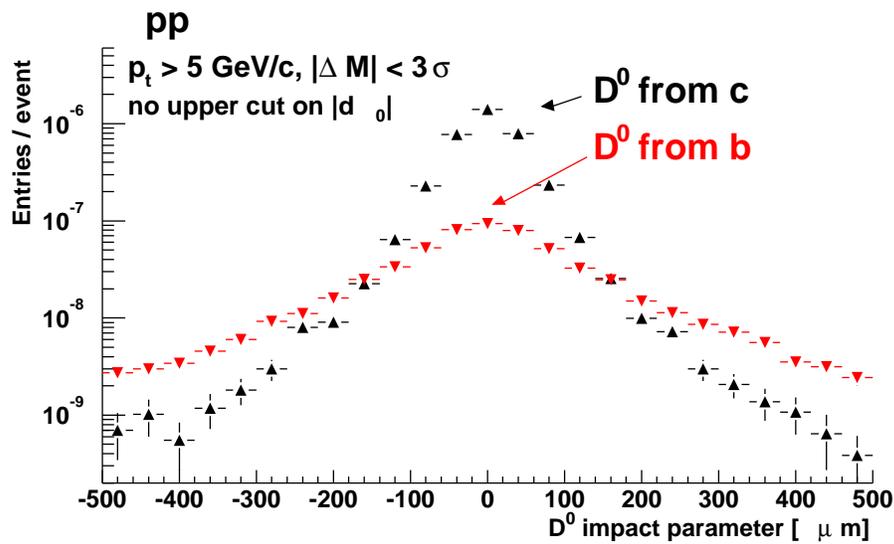}
    \caption{Impact parameter distribution for primary and secondary 
             $\Dz$ mesons in pp collisions. All detector and reconstruction 
             effects are taken into account, including the reconstruction 
             of the interaction vertex position using tracks.} 
    \label{fig:d0D0_pp}
  \end{center}
\end{figure}

As mentioned, the analysis of the $\Dz$ impact parameter has been done 
by the CDF Collaboration on the data collected in Run II at the Tevatron. 
Figure~\ref{fig:CDFd0D0} shows the distribution of this variable for 
signal $\Dz$ candidates with invariant mass in the range 
$M_{\rm D^0}\pm 2~\sigma$ and with $\pt>5.5~\gev/c$~\cite{CDFd0D0}. 
The contribution of the combinatorial background has been subtracted. 
This can be done because the integral of the background 
is known from the fit of the invariant mass distribution and 
its shape is obtained from the impact parameter distribution of the 
$\Dz$ candidates that populate the side-bands of the invariant mass 
($|M-M_{\rm D^0}| > 4~\sigma$). The true impact parameter
of primary $\Dz$ is 0, while the shape of the true impact parameter 
distribution of secondary $\Dz$ is derived from a generator-level
Monte Carlo simulation of B meson production and decay. Both components are
smeared with a resolution function (Gaussian + exponential tails) obtained
from a sample of ${\rm K^0_S}\to \pi^+\pi^-$ (${\rm K^0_S}$ are almost 
exclusively primary, thus having true impact parameter equal to 0). 
In this way the shapes of the two distributions for primary and 
secondary $\Dz$ are known and their relative integral 
is obtained from a fit. Averaged over $\pt>5.5~\gev/c$, the fraction 
of primary $\Dz$ is estimated to be 
$(86.6\pm 0.4({\rm stat})\pm 3.0({\rm syst}))\%$~\cite{CDFd0D0}.

We reported a detailed description of the analysis performed in CDF, because 
the idea is to apply the same strategy in ALICE, not only in pp but also in 
\PbPb~events. A preliminary study was carried out and the results, shown in 
the following, are quite promising.

The impact parameter distribution for selected $\Dz$ candidates in pp events 
is presented in Fig.~\ref{fig:d0D0_pp}. The contributions of signal $\Dz$ 
from c and from b are plotted separately. 
The cuts reported in 
Table~\ref{tab:cutsppvtxunknown} are applied; only the cut on the maximum 
value of the impact parameter of the decay tracks, which would suppress the 
component of $\Dz$ from b, is removed.
The additional cuts  $|M-M_{\rm D^0}|< 3~\sigma$ and $\pt>5~\gev/c$ are 
applied. All detector and reconstruction effects are taken into account, 
including the reconstruction of the interaction vertex position using tracks.
The distributions of primary and secondary $\Dz$ have significantly 
different shapes, similarly to what observed by CDF.

\begin{figure}[!b]
  \begin{center}
    \includegraphics[width=.9\textwidth]{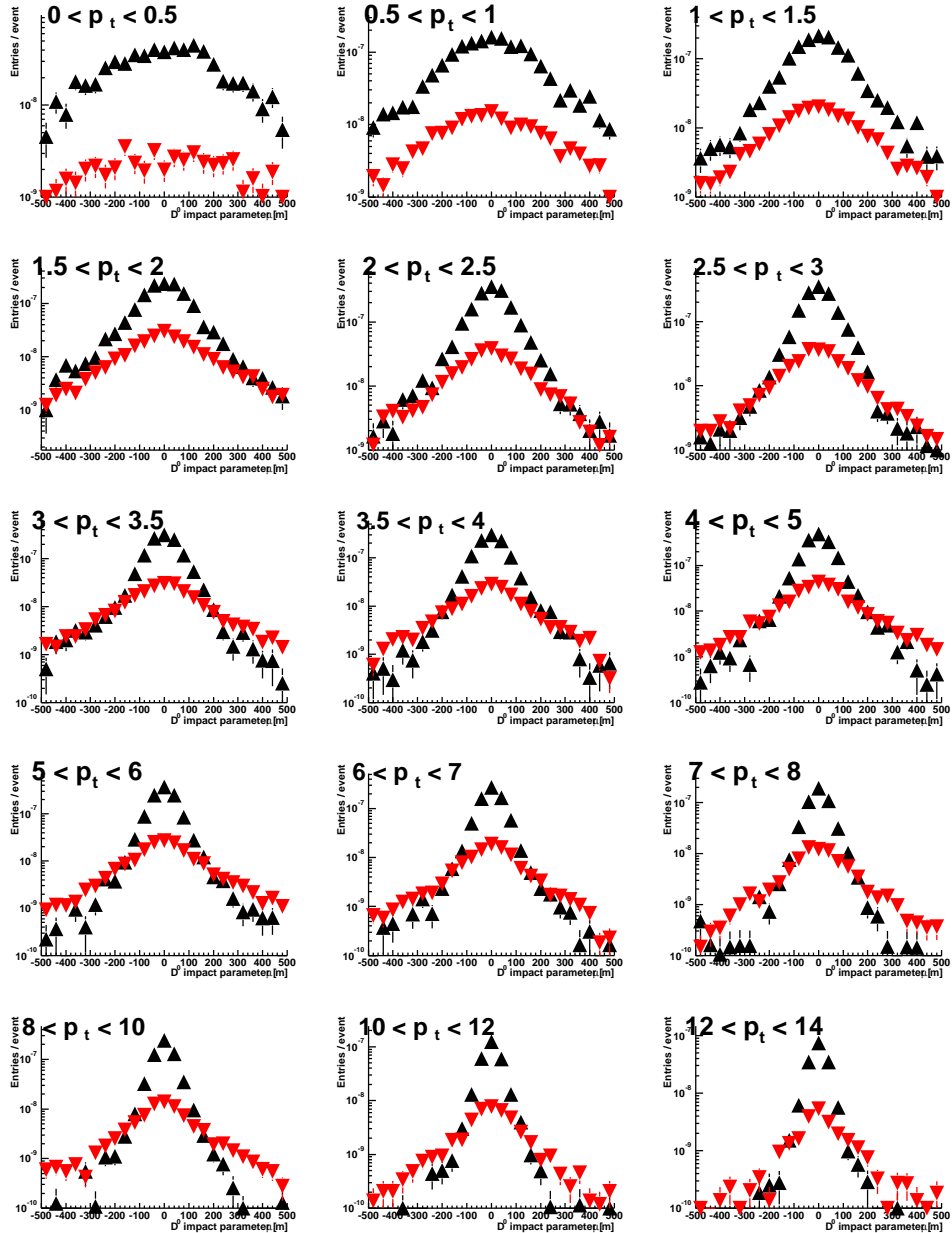}
    \caption{Impact parameter distribution for primary 
             (triangle up) and secondary 
             (triangle down)  
             $\Dz$ mesons in pp collisions, in bins of the $\Dz$ transverse
             momentum.} 
    \label{fig:d0D0_pp_VSpt}
  \end{center}
\end{figure}

\begin{figure}[!b]
  \begin{center}
    \includegraphics[width=.9\textwidth]{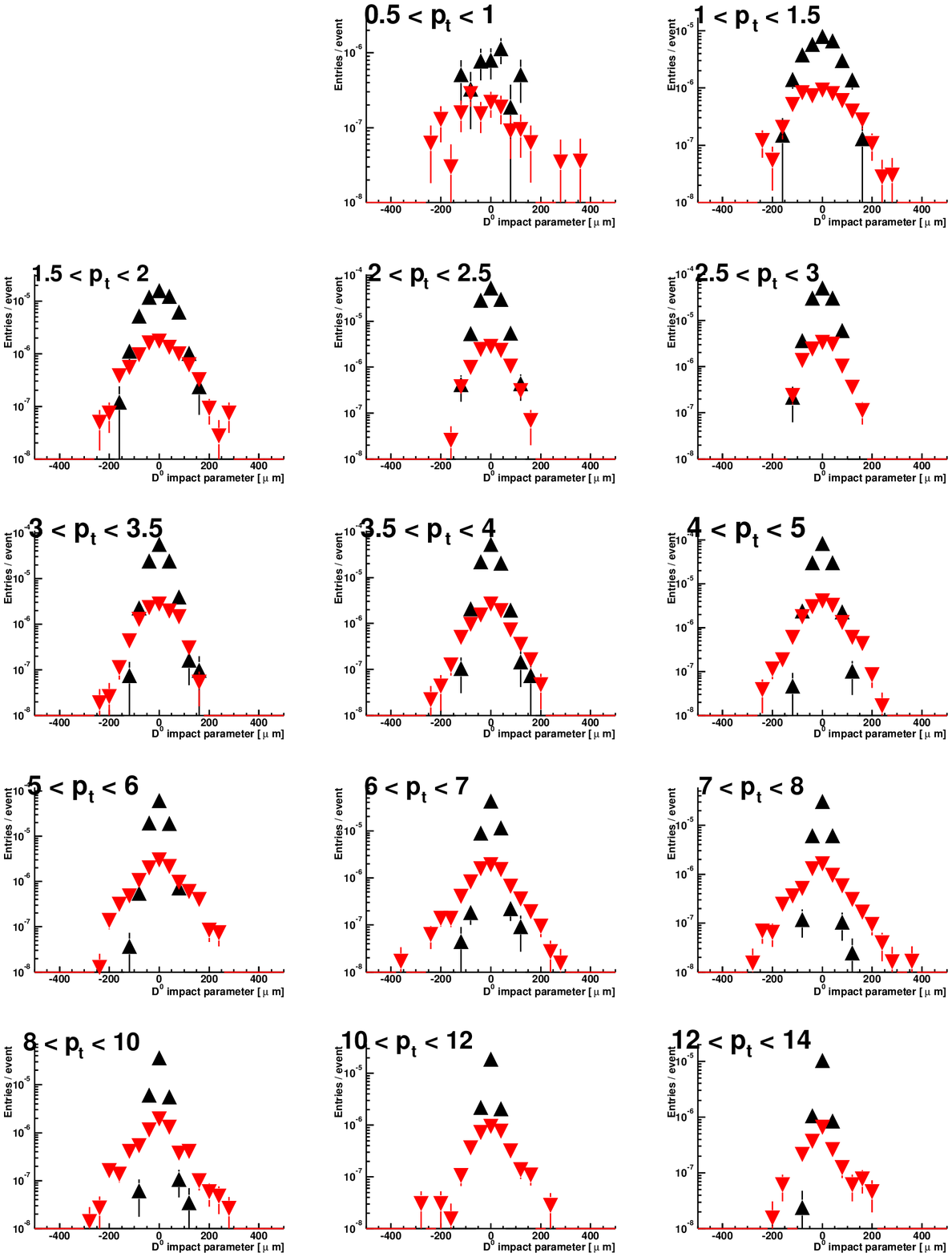}
    \caption{Impact parameter distribution for primary (triangle up)
             and secondary (triangle down) 
             $\Dz$ mesons in \PbPb~collisions, in bins of the $\Dz$ transverse
             momentum.} 
    \label{fig:d0D0_PbPb_VSpt}
  \end{center}
\end{figure}

The same distributions are shown in bins of $\pt$ in 
Figs.~\ref{fig:d0D0_pp_VSpt} and~\ref{fig:d0D0_PbPb_VSpt} for pp
and \PbPb, respectively. In both cases, the difference in shape between the 
two contributions is visible down to $\pt\simeq 4$-$5~\gev/c$; for 
lower $\pt$ the resolution on the position of the 
$\Dz$ decay vertex is quite poor due to multiple scattering effects
and the impact parameter distribution of primary $\Dz$ mesons becomes 
very broad. We point out that the separation is sharper 
in \PbPb~than in pp collisions, since the resolution on the transverse 
position of the interaction vertex is better (see Chapter~\ref{CHAP5}).  

More detailed investigations are necessary for a clear assessment of the 
feasibility of an analysis `\`a la CDF'. In particular, the effect of the
cut on the pointing angle should be evaluated and the strategy for background
subtraction and fit of the two signal components should be studied. 
However, there are good indications of the possibility to measure the ratio of 
primary to secondary $\Dz$ mesons as a function of $\pt$, 
down to \mbox{$4$-$5~\gev/c$}, in pp, \pPb~and \PbPb~collisions. This would
allow a direct correction of the $\Dz$ yield for feed-down from B decays 
and might provide information on the ratio of B/D mesons as a function of 
$\pt$, which is regarded as particularly interesting to 
investigate the mass dependence of medium-induced 
energy loss for heavy quarks (see Section~\ref{CHAP2:deadcone}).

\clearpage
\pagestyle{plain}

\setcounter{chapter}{7}
\mychapter{Quenching of open charm mesons}
\label{CHAP8}

\pagestyle{myheadings}

The role of a comparative analysis of the quenching of charm mesons and 
charged hadrons in the scope of a systematic study of the properties 
of QCD matter produced in \PbPb~collisions at the LHC was emphasized 
in Chapter~\ref{CHAP2}.
In this chapter we evaluate the effect of medium-induced energy loss
on the transverse momentum distributions of c quarks and D mesons 
and we discuss the potential of ALICE for carrying out such comparative 
analysis. 

The choice of the input parameters for the simulation of energy loss is 
considered in 
Section~\ref{CHAP8:ellandqhat}. We derive the distribution of in-medium 
path lengths using a detailed description of the collision geometry and 
we estimate a value for the transport coefficient of the medium at the LHC,
keeping into account the pion suppression observed in Au--Au collisions at 
RHIC (see Section~\ref{CHAP1:rhic}). 
In Section~\ref{CHAP8:charmenergyloss} we describe how the SW 
(Salgado-Wiedemann) weights (Section~\ref{CHAP2:qw}) are used to 
quench charm quarks and D mesons generated with PYTHIA and we
introduce a correction to the weights that accounts for the dead cone effect 
predicted for heavy quarks. The nuclear modification factor of D mesons and 
the D/$hadrons$ ratio for different quenching scenarios (different transport 
coefficients, dead cone option) are presented in 
Sections~\ref{CHAP8:RAA} and~\ref{CHAP8:RDh}. The attainable precision in the
measurement of these observables is derived from the uncertainties on the 
$\pt$ distributions estimated in Chapter~\ref{CHAP7}.

\mysection{Medium parameters: path length and transport coefficient}
\label{CHAP8:ellandqhat}

\begin{figure}[!t]
  \begin{center}
    \includegraphics[width=.5\textwidth]{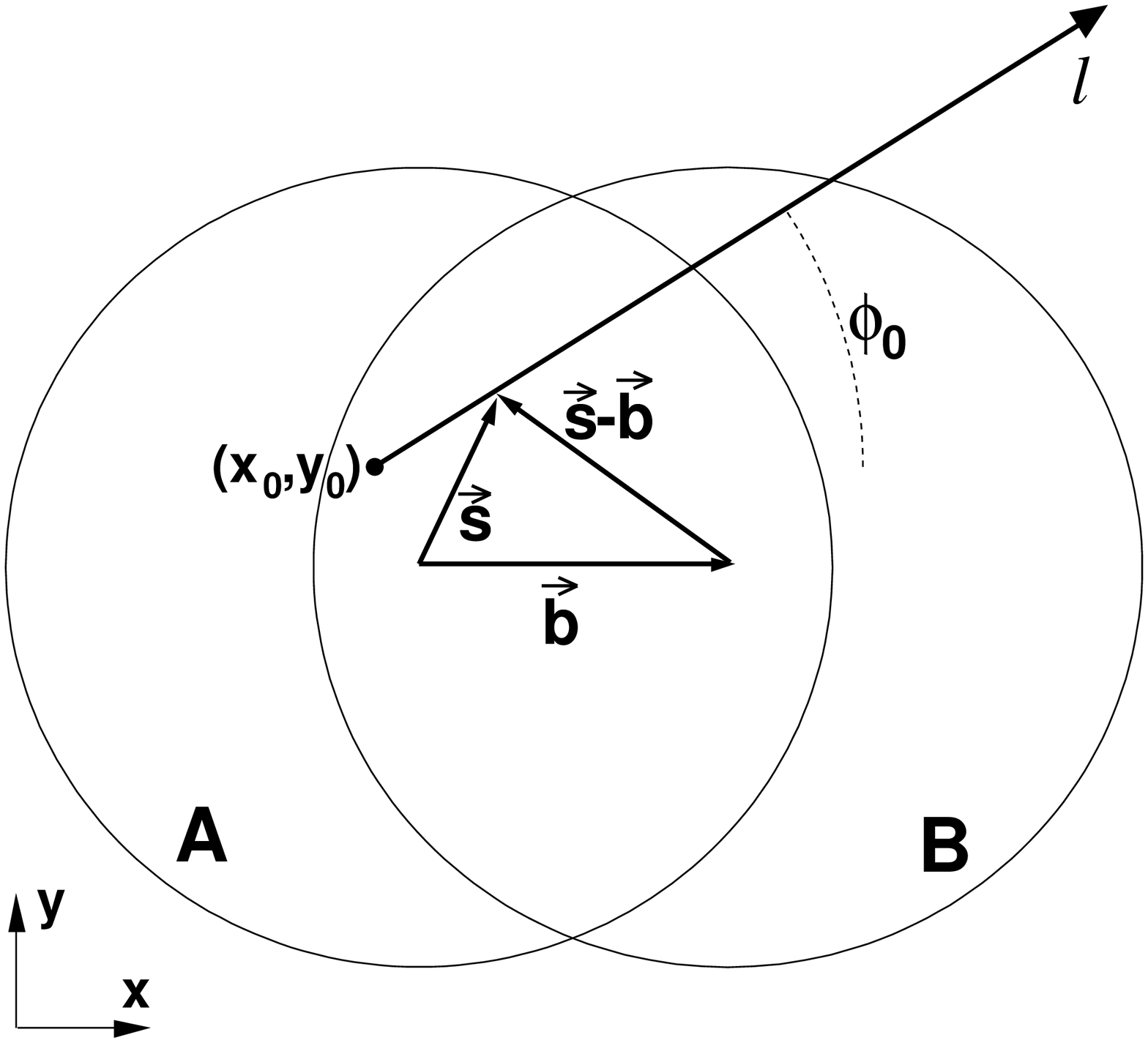}
    \caption{Geometry of the collision in the transverse plane and 
             definition of the in-medium path length.} 
    \label{fig:elldef}
  \end{center}
\vglue1cm
  \begin{center}
    \includegraphics[width=.8\textwidth]{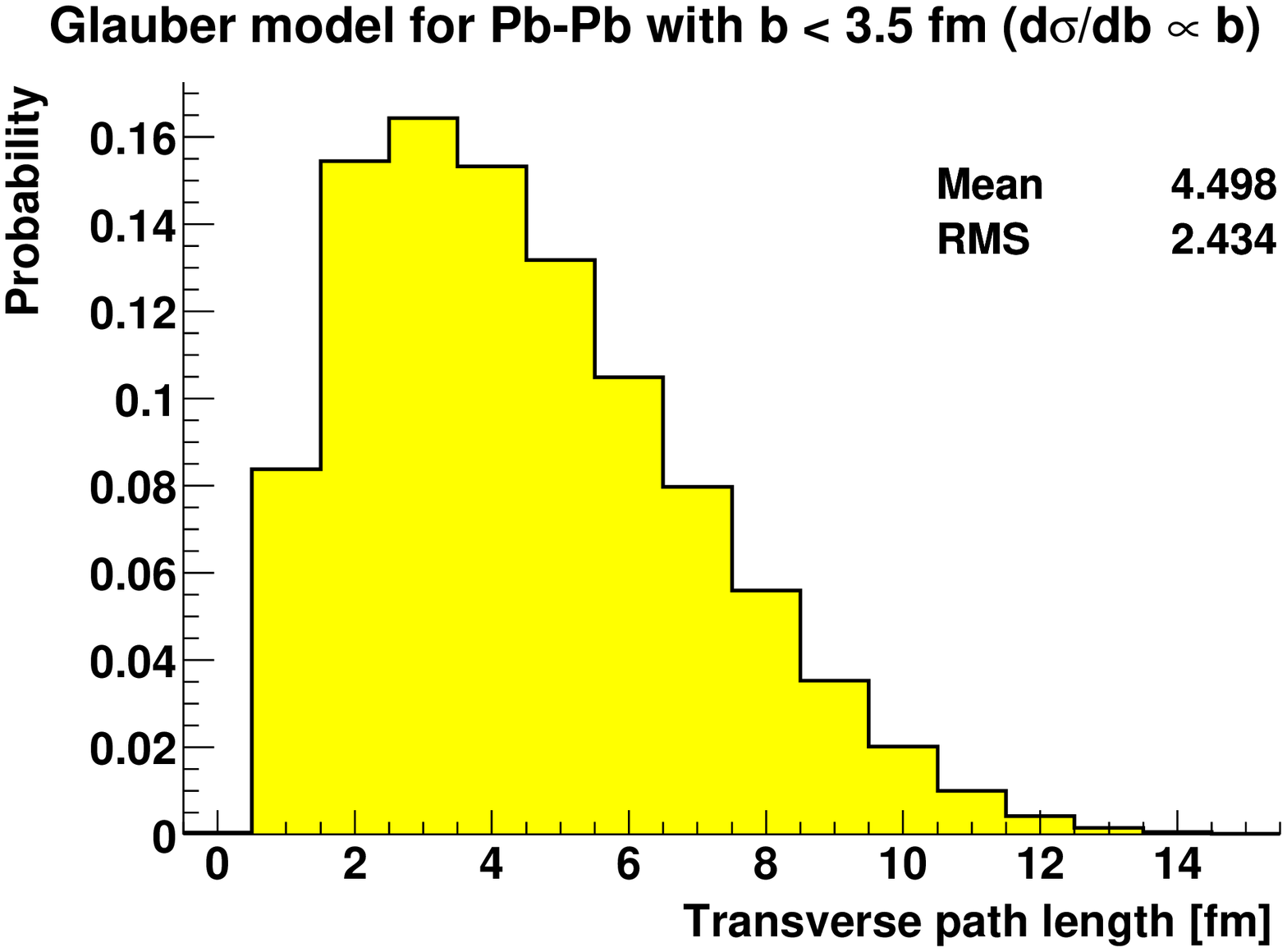}
    \caption{Distribution of the path lengths in the transverse plane for 
             partons produced in \PbPb~collisions with $b<3.5~\fm$ 
             (5\% of the inelastic cross section).} 
    \label{fig:Ldistr}
  \end{center}
\end{figure}

The distribution of the in-medium path length in the transverse 
plane\footnote{Partons produced at central rapidities 
propagate in the plane transverse to the beam line.}, 
$L$, for central
\PbPb~collisions (impact parameter $b<3.5~\fm$) is calculated in the framework
of the Glauber model of the collision geometry. We refer to 
the nomenclature introduced in Section~\ref{CHAP3:extrapolation2PbPb}. 
For fixed impact parameter, the density of binary NN collisions in the 
transverse plane is obtained as the product of the thickness functions of 
the two colliding Pb nuclei, 
$T_{\rm A}(\vec{s})\,T_{\rm B}(\vec{s}-\vec{b})$; 
parton production points are sampled according to this density and their 
azimuthal propagation directions are sampled uniformly in the range 
$[0,2\pi]$. For a parton with given production point $(x_0,y_0)$ 
and azimuthal direction  $\phi_0$ (see Fig.~\ref{fig:elldef}), 
the path length $L$ is defined as the line integral 
from the production point to `well outside the superposition region of the two 
nuclei' weighted by the product of the thickness functions and normalized with 
its integral:
\begin{equation}
\label{eq:ell}
  L = \frac{\int_0^\infty {\rm d}l\, l\, T_{\rm A}(\vec{s}(l))\,T_{\rm B}(\vec{s}(l)-\vec{b})}
           {0.5\,\int_0^\infty {\rm d}l\, T_{\rm A}(\vec{s}(l))\,T_{\rm B}(\vec{s}(l)-\vec{b})}.
\end{equation} 

Many sampling iterations are performed varying the impact 
parameter $b$ from $0.25~\fm$ to $3.25~\fm$ in steps of $0.5~\fm$. 
The obtained distributions are given a weight $b$, since we verified that 
d$\sigma^{\rm hard}/$d$b\propto b$ for $b<3.5~\fm$, and added together. 
The result is shown in 
Fig.~\ref{fig:Ldistr}. The average length is $4.5~\fm$, corresponding to 
about 70\% of the radius of a Pb nucleus and 
the distribution is significantly accumulated towards low values of $L$
because a large fraction of the partons are produced in the periphery
of the superposition region of the two nuclei (`corona' effect).
~\\

For the estimation of the value of the other input parameter, the 
transport coefficient $\hat{q}$, we consider that it is 
reasonable to require for AA collisions at the LHC a quenching of
hard partons of the same magnitude as that observed at RHIC.
We, therefore, derive the nuclear modification factor $\RAA$ for charged 
hadrons produced at the LHC and we choose the transport coefficient in 
order to obtain $\RAA\simeq 0.2$-$0.3$ in the range $\pt=5$-$10~\gev/c$
(PHENIX results were reported in Fig.~\ref{fig:phenixRAA}). 

The transverse momentum distributions, for $\pt>5~\gev/c$, of charged hadrons 
are generated by means of the chain:
\begin{enumerate}
\item generation of a parton, quark or gluon, with $\pt>5~\gev/c$;
\item sampling of an energy loss $\Delta E$ and calculation of the quenched 
      transverse momentum of the parton, $\pt'=\pt-\Delta E$ (if 
      $\Delta E>\pt$, $\pt'$ is set to 0);
\item (independent) fragmentation of the parton to a hadron. 
\end{enumerate}
Quenched and unquenched $\pt$ distributions are obtained including or 
excluding the second step of the chain. We shall now detail each single step.

\begin{figure}[!t]
  \begin{center}
    \includegraphics[width=.49\textwidth]{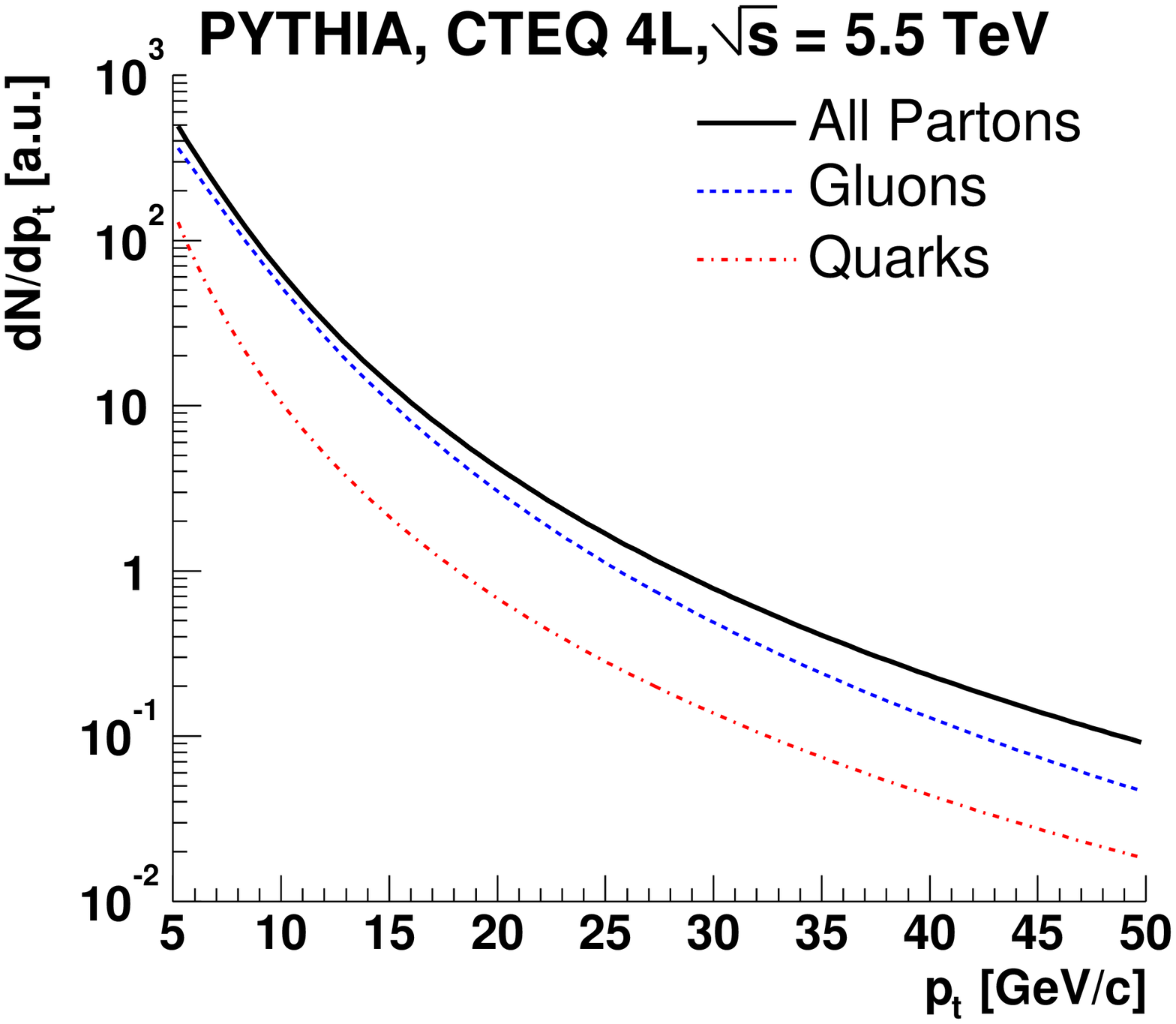}
    \includegraphics[width=.49\textwidth]{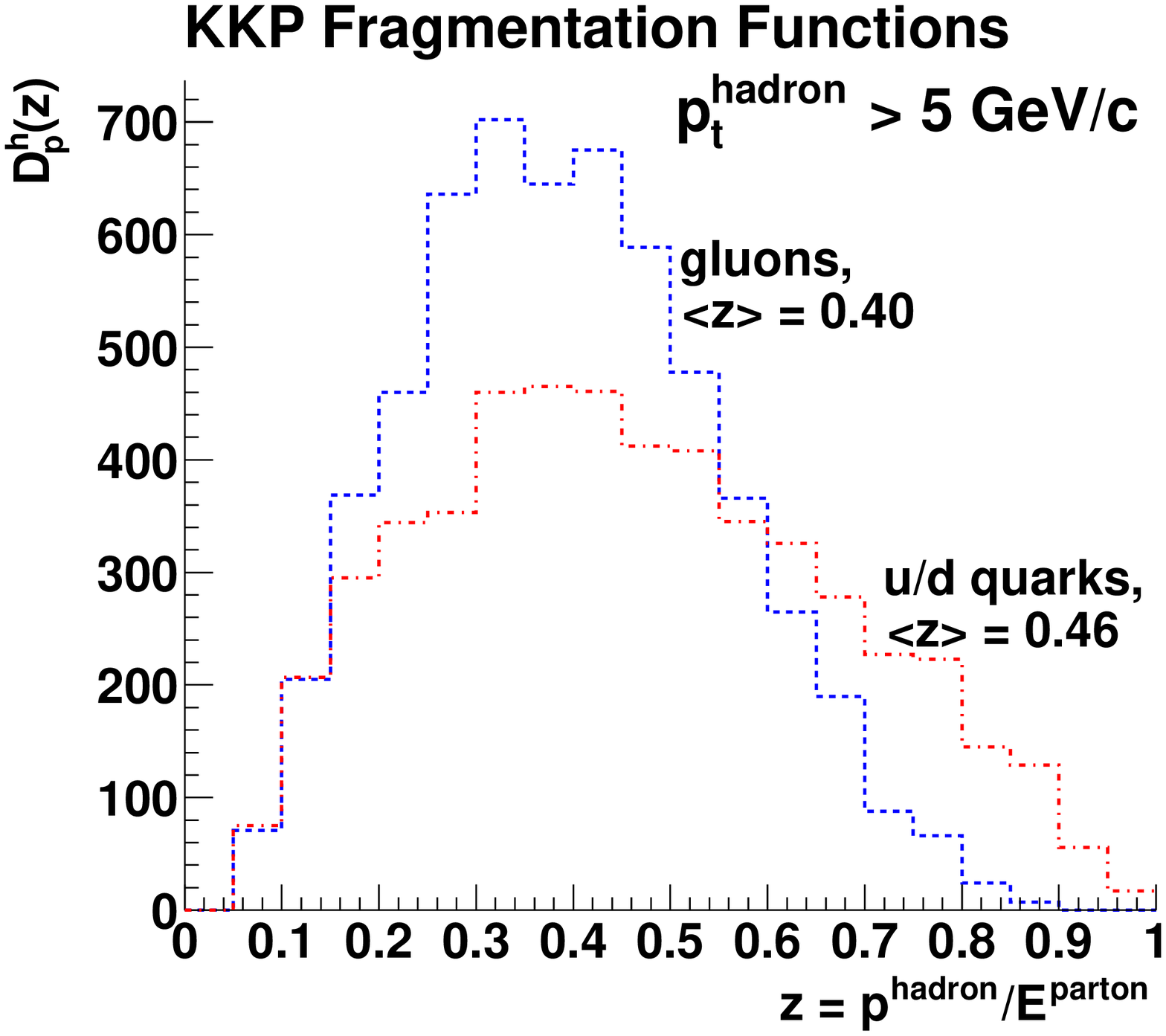}
    \caption{Left: fitted transverse momentum distributions for quarks and 
             gluons generated with PYTHIA; the sum of quarks and gluons is 
             shown as well. Right: KKP fragmentation functions for quarks and 
             gluons into hadrons, with a cut $\pt^{\rm hadron}>5~\gev/c$.} 
    \label{fig:PYTHIApartonsFF}
  \end{center}
\end{figure}

Input $\pt$ distributions of quarks and gluons are generated in PYTHIA
with the same settings used for the generation of pp events at 
$\sqrt{s}=14~\tev$ (Section~\ref{CHAP4:generators}), except the c.m.s. 
energy, which was set to $5.5~\tev$. The CTEQ 4L set of parton distribution 
functions was used. Figure~\ref{fig:PYTHIApartonsFF} (left) reports the result
of a fit on the $\pt$ distributions of quarks and gluons for $\pt>5~\gev/c$.
With this $\pt$ cut, the parton composition is found to be 78\% gluons 
and 22\% quarks, with the ratio quarks/gluons increasing with $\pt$
(i.e. with Bjorken $x$), as the contribution of the valence quarks of the 
incoming protons becomes more and more important (see parton distributions in 
Fig.~\ref{fig:pdf}).

For the fragmentation (step 3) we use the leading order (LO) 
Kniehl-Kramer-P\"otter (KKP) fragmentation functions~\cite{kkp}. This kind
of fragmentation procedure is called independent, as each parton fragments 
to a single hadron with transverse momentum 
$\pt^{\rm hadron}=z\,E^{\rm parton}\simeq z\,\pt^{\rm parton}$
(see Fig.~\ref{fig:PYTHIApartonsFF}, right). We choose 
such approach, rather than using the standard string fragmentation of
PYTHIA, because this would require the implementation of partonic 
energy loss in PYTHIA, which is highly non-trivial and goes beyond the 
scope of this work. In the next section we will show that a simple Charm 
quantum number
conservation argument allows to use the PYTHIA string fragmentation in the 
calculations for the quenching of D mesons.

The quenching procedure is as follows.
As described in Section~\ref{CHAP2:qw}, the SW quenching weight has a discrete
part $p_0$ ($\equiv$ probability to have no gluon radiation) and a continuous 
part $p(\Delta E)$
($\equiv$ probability to radiate an energy $\Delta E$, if at least one gluon 
is radiated). Both parts of the weight have to be included in the quenching 
procedure. In addition, the Glauber-based distribution of path lengths 
$L$ has to be accounted for. 

\begin{figure}[!t]
  \begin{center}
    \includegraphics[width=.95\textwidth]{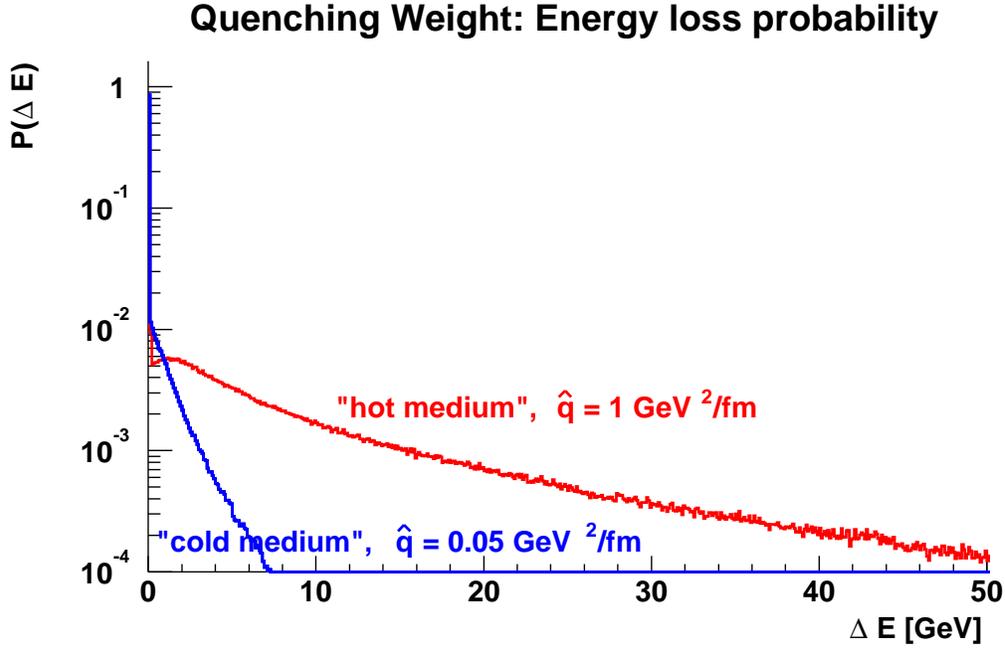}
    \caption{Quark energy loss probability for $\hat{q}=0.05~\gev^2/\fm$ and
             for $\hat{q}=1~\gev^2/\fm$. The discrete and the continuous 
             parts of the quenching weight are included and 
             the distribution of in-medium path lengths is taken into 
             account.} 
    \label{fig:QWcoldhot}
  \end{center}
\end{figure}

For a given value of the transport coefficient $\hat{q}$ and a given parton 
species (quark or gluon), we use the routine 
provided by the authors~\cite{carlosurs} to get $p_0$ and to histogram the 
distribution $p(\Delta E)$ for all integer values of $L$ up to 
$15~\fm$ ($1~\fm$, $2~\fm$, \dots, $15~\fm$). Since $p_0$ is 
essentially a probability to have $\Delta E=0$ (no radiated gluons),
we add the value of $p_0$ to the bin corresponding to $\Delta E=0$ in the 
histogram of the continuous part of the weight $p(\Delta E)$. In this 
way the complete distribution of energy loss probability, 
including both parts of the quenching weight and properly normalized, 
is obtained for each value of $L$. Finally, these 15 distributions are 
weighted according to the path length distribution in Fig.~\ref{fig:Ldistr}
and added together. The resulting energy loss probability distributions 
$P(\Delta E)$ for 
quarks in media with $\hat{q}=0.05~\gev^2/\fm$ (cold medium) and 
$\hat{q}=1~\gev^2/\fm$ (hot medium) are reported in Fig.~\ref{fig:QWcoldhot}
(the `peak' at $\Delta E=0$ represents the discrete part of the quenching 
weight). The energy loss to be used in the second step of the chain 
reported above can be directly sampled from the $P(\Delta E)$ distribution 
corresponding to the chosen $\hat{q}$ and to the correct parton species.

\begin{figure}[!t]
  \begin{center}
    \includegraphics[width=\textwidth]{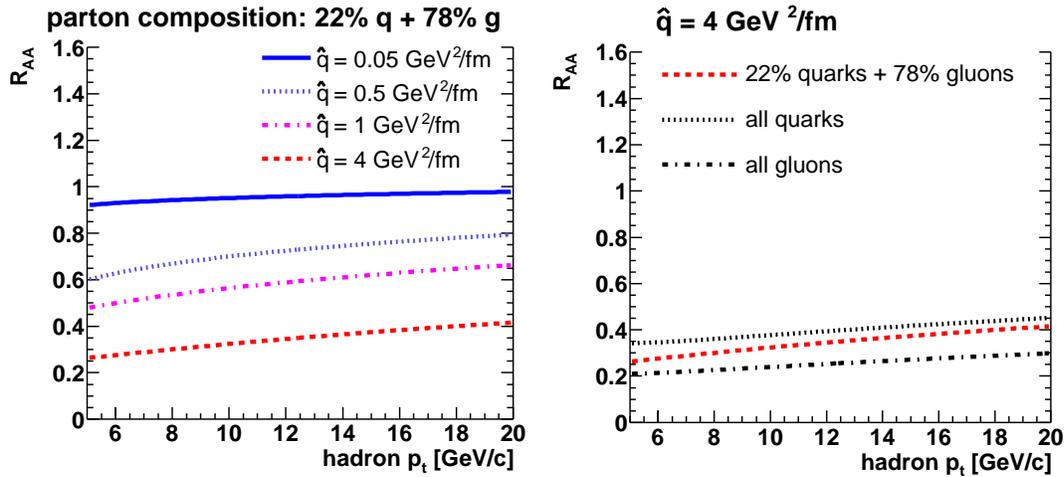}
    \caption{Left: nuclear modification factor for charged hadrons 
             for different values of $\hat{q}$. Right: comparison 
             of quark and gluon quenching for $\hat{q}=4~\gev^2/\fm$.} 
    \label{fig:RAA-hadrons}
  \end{center}
\end{figure}

Experimentally, the nuclear modification factor will be obtained as the ratio 
of the $\pt$ distribution measured in central \PbPb~collisions at 
$\sqrtsNN=5.5~\tev$, divided by the estimated number of binary collisions, to 
the $\pt$ distribution measured in pp collisions at $\sqrt{s}=14~\tev$, 
extrapolated to $\sqrt{s}=5.5~\tev$ by means of pQCD.
Here, for the moment, we go directly to $\RAA$ (the uncertainties 
involved in the experimental estimate will be discussed in the last 
part of this chapter). 
Figure~\ref{fig:RAA-hadrons} shows $\RAA$ for 
hadrons, calculated as the ratio of the $\pt$ distribution with quenching 
to the $\pt$ distribution without quenching. Different values of $\hat{q}$
are considered in the left panel of the figure: a value as large as 
$4~\gev^2/\fm$ is necessary to have $\RAA\simeq 0.25$-$0.3$ in 
$5<\pt<10~\gev/c$. In the right panel, for $\hat{q}=4~\gev^2/\fm$, we 
compare the results obtained considering all partons as gluons or all 
partons as quarks, in order to remark and quantify the larger quenching 
of gluons with respect to quarks.

Since the transport coefficient determines the size of the energy loss
effect, we shortly discuss the choice of $\hat{q}=4~\gev^2/\fm$.
This value corresponds, using the plot in 
Fig.~\ref{fig:qtransp}, to an energy density 
$\varepsilon\simeq 40$-$50~\gev/\fm^3$, which is about a factor 2 lower 
than the maximum energy density expected for central \PbPb~collisions at the
LHC (see Table~\ref{tab:spsrhiclhc}). The value looks, therefore, reasonable.

\begin{figure}[!t]
  \begin{center}
    \includegraphics[width=\textwidth]{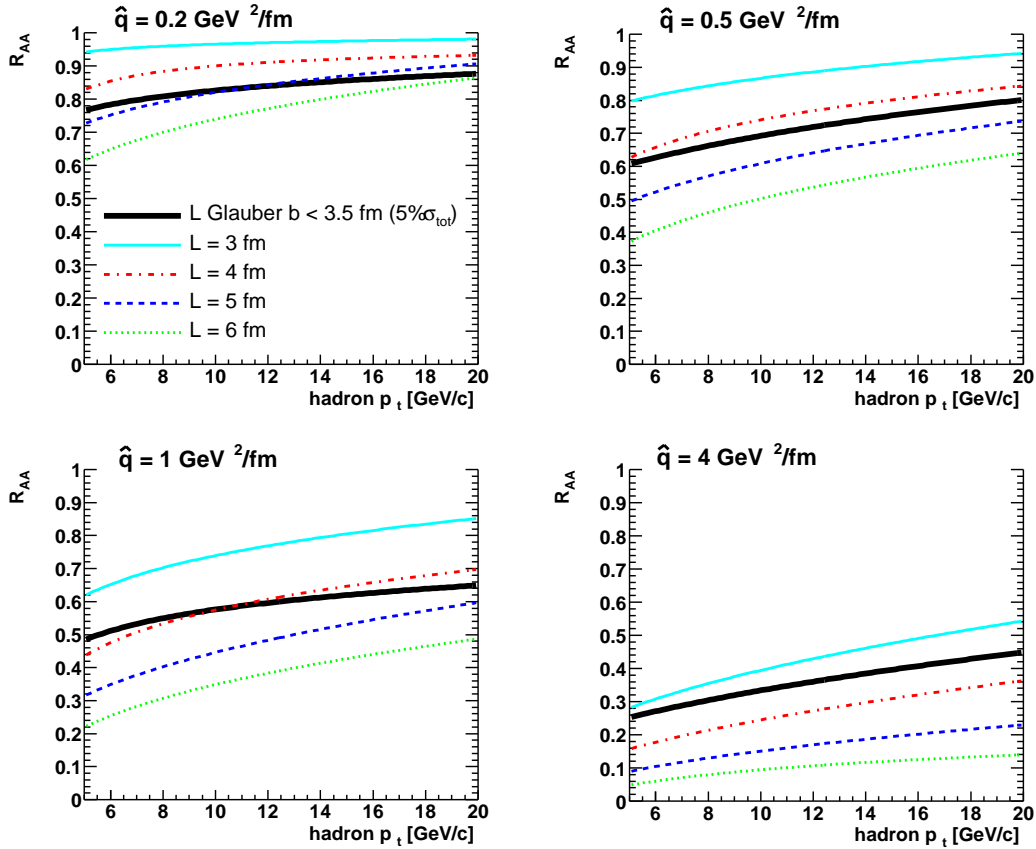}
    \caption{Nuclear modification factor for charged hadrons 
             for different values of $\hat{q}$ as obtained using the complete
             Glauber-based distribution or constant values of $L$.} 
    \label{fig:RAA-hadrons-cmpL}
  \end{center}
\end{figure}

Using the same quenching weights, 
Salgado and Wiedemann reproduce the suppression observed at RHIC with 
the much lower value $\hat{q}=0.75~\gev^2/\fm$~\cite{carlosurs}. However, 
they use the constant length $L=6~\fm$ rather than a realistic distribution 
of lengths, thus obtaining a significantly stronger quenching. 

The use of a constant length of the order of the nuclear radius or even 
the use of the average length from a detailed distribution can produce quite 
different results with respect to those obtained keeping into account 
the complete distribution. This is demonstrated in 
Fig.~\ref{fig:RAA-hadrons-cmpL}: for $\hat{q}=0.5$-$1~\gev^2/\fm$, 
$L=6~\fm$ gives almost a factor 2 difference in $\RAA$ at $\pt\sim 10~\gev/c$
and the complete $L$ distribution is equivalent to a constant length which 
decreases as $\hat{q}$ increases, 5, 4.5, 4, 3.5~fm for 
$\hat{q}=0.2,~0.5,~1,~4~\gev^{2}/\fm$. This behaviour is clearly due to 
an upper `cut-off' of the length distribution: large lengths correspond to 
very high values of $\Delta E$, but, since $\Delta E$ cannot be higher than 
the initial parton energy, large lengths are not `fully exploited'; 
this corresponds to a cut-off; e.g. for many partons of moderate energy a 
length of $8~\fm$ is equivalent to a length of $4~\fm$, because after
propagating for $4~\fm$ they have lost all their initial energy. 
As a consequence,
the length distribution corresponds to an average effective length lower than
its arithmetic average. The cut-off moves towards lower lengths as
$\hat{q}$ increases and, thus, the average effective length decreases.
Another important observation revealed by Fig.~\ref{fig:RAA-hadrons-cmpL} 
is the fact that the use of the complete $L$ distribution reduces 
the increase of $\RAA$ with $\pt$, which is not observed in RHIC data.
This happens because higher energy partons can exploit the large-$L$ tail more
than lower energy partons and, consequently, for them the cut-off is shifted 
towards larger lengths.

Finally, we consider that the transport coefficient can be related
to the initial gluon rapidity density via the approximated 
formula~\cite{carlosurs,carlosurs2}:
\begin{equation}
\label{eq:qhatdNgdy} 
\hat{q}=\frac{2}{R_{\rm A}^2\,L}\times\frac{{\rm d}N_{\rm gluons}}{{\rm d}y}
\end{equation}
where $R_{\rm A}$ is the nuclear radius. For $\hat{q}=4~\gev^2/\fm$, 
$R_{\rm A}=6~\fm$ and $L=3.5~\fm$ (see bottom-right panel in 
Fig.~\ref{fig:RAA-hadrons-cmpL}) we obtain d$N_{\rm gluons}/$d$y=6300$. 
Given the large uncertainties~\cite{carlosurs2} in the relation 
(\ref{eq:qhatdNgdy}),
this value is in reasonable agreement with the estimated~\cite{eskola1} 
d$N_{\rm gluons}/$d$y\simeq 5000$ for LHC, reported in 
Table~\ref{tab:spsrhiclhc}.

In summary, we use the in-medium path length distribution shown in 
Fig.~\ref{fig:Ldistr} for \PbPb~collisions with $b<3.5~\fm$ and a transport 
coefficient $\hat{q}=4~\gev^2/\fm$. For comparison, we will as well present 
the results for lower values of $\hat{q}$.

\mysection{Charm energy loss with quenching weights}
\label{CHAP8:charmenergyloss}

Energy loss for charm quarks is simulated following a slightly different 
procedure with respect to that for light quarks and gluons, because (a) 
the total number of $\ccbar$
pairs per event has to be conserved and (b) the string fragmentation 
implemented in PYTHIA is more reliable for heavy quarks than the independent
fragmentation.

The starting point is the generation with PYTHIA of charm events, each 
one containing a $\ccbar$ pair. The parameters obtained from the tuning 
described in Section~\ref{CHAP3:generators} are used for PYTHIA and 
charm quarks are fragmented to hadrons via the default string model.
Since c and $\overline{\rm c}$ quarks fragment to D 
and $\overline{\rm D}$ mesons, respectively, in each event related 
${\rm (c,D)}$ and ${\rm (\overline{c},\overline{D})}$ pairs can be 
easily identified\footnote{Events containing charm baryons were rejected.}. 

\begin{figure}[!b]
  \begin{center}
    \includegraphics[width=.7\textwidth]{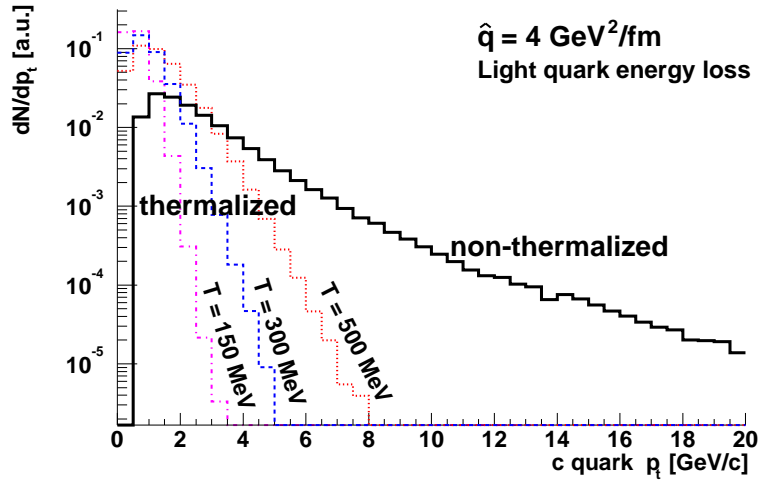}
    \caption{$\pt$ distributions of non-thermalized 
             ($\pt^{\rm c}-\Delta E>1.5\,T$) and thermalized
             ($\pt^{\rm c}-\Delta E<1.5\,T$) c quarks for 
             $\hat{q}=4~\gev^2/\fm$ and $T=150,\,300,\,500~\mev$
             (for clearness the non-thermalized component is shown only for 
             $T=500~\mev$). The quenching 
             corresponding to light quarks was applied.} 
    \label{fig:ptc_thermal}
  \end{center}
\end{figure}

The second step consists in applying the quenching to the $\pt$ distribution 
of charm quarks. The procedure is close to that used for light quarks and 
gluons: for each quark an energy loss $\Delta E$ is sampled from 
a distribution $P(\Delta E)$ (similar to those shown in 
Fig.~\ref{fig:QWcoldhot}) 
and the transverse momentum is modified to  
$(\pt^{\rm c})'=\pt^{\rm c}-\Delta E$. In Ref.~\cite{linvogt} it is pointed 
out that if the energy loss is close to the transverse momentum of the c 
quark, or larger than it, the quark should be considered as thermalized with 
the dense medium. We use $T=300~\mev$ as the temperature 
of the medium and, in the cases when $\pt^{\rm c}-\Delta E<1.5\,T$, we 
assign to the quark a thermal transverse momentum according the 
distribution d$N/$d$m_{\rm t}\propto m_{\rm t}\cdot \exp(-m_{\rm t}/T)$, being
$m_{\rm t} = \sqrt{m_{\rm c}^2+\pt^2}$ the transverse mass of the quark.
This procedure allows to conserve the number of $\ccbar$ pairs. 
In Fig.~\ref{fig:ptc_thermal} we show that the thermalized component 
accumulates in the region $\pt<5$-$6~\gev/c$ for temperatures up to 
$T=500~\mev$, which is a factor 3 larger than the expected thermalization 
temperature of $170~\mev$. As mentioned in Chapter~\ref{CHAP2} and as we 
shall detail in the next section, the quenching of D mesons can be better 
studied for $\pt>6$-$7~\gev/c$. We, then, conclude that the choice of $T$ 
is not critical in our energy loss simulation.

\begin{figure}[!t]
  \begin{center}
    \includegraphics[width=.47\textwidth]{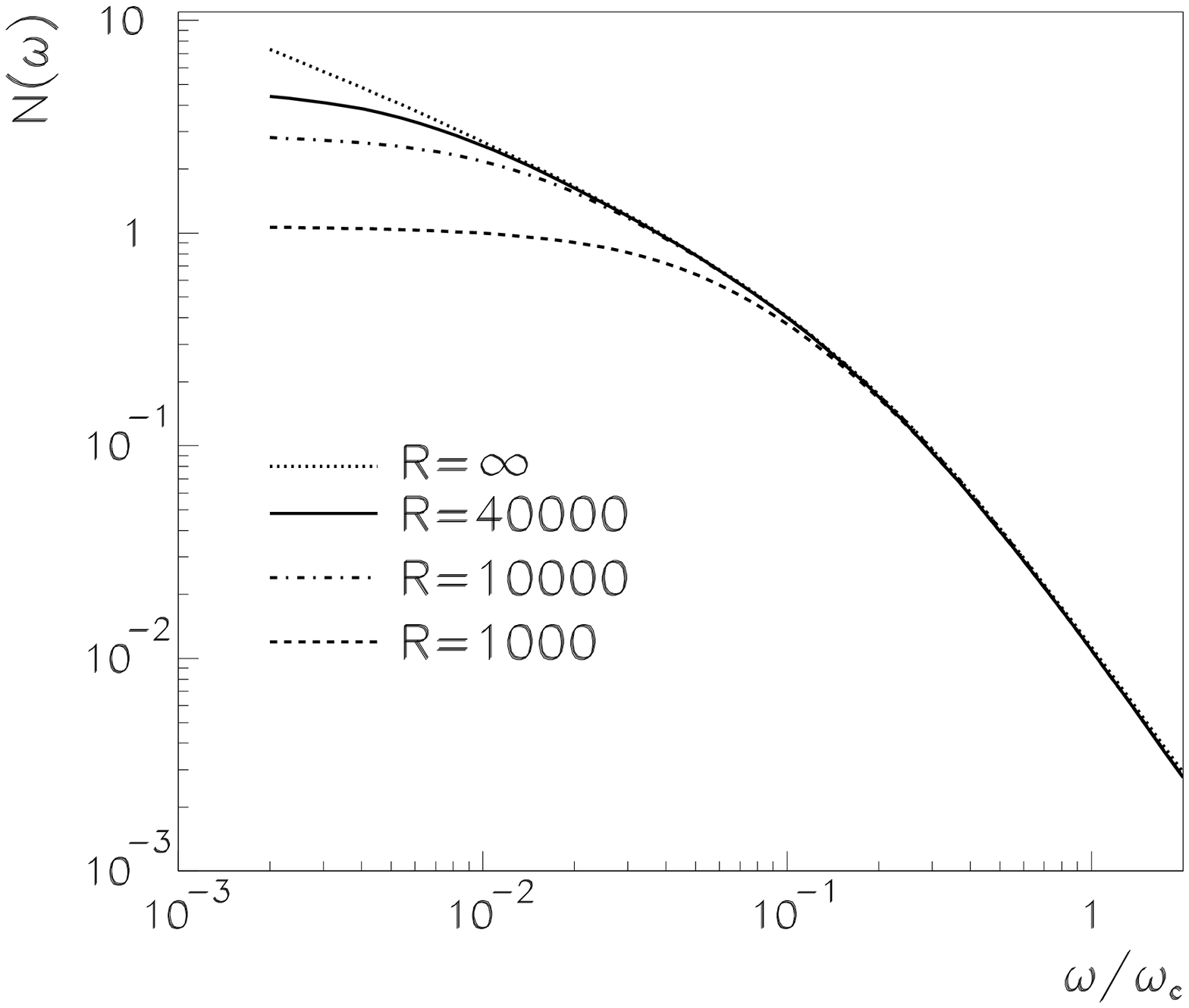}
    \includegraphics[width=.51\textwidth]{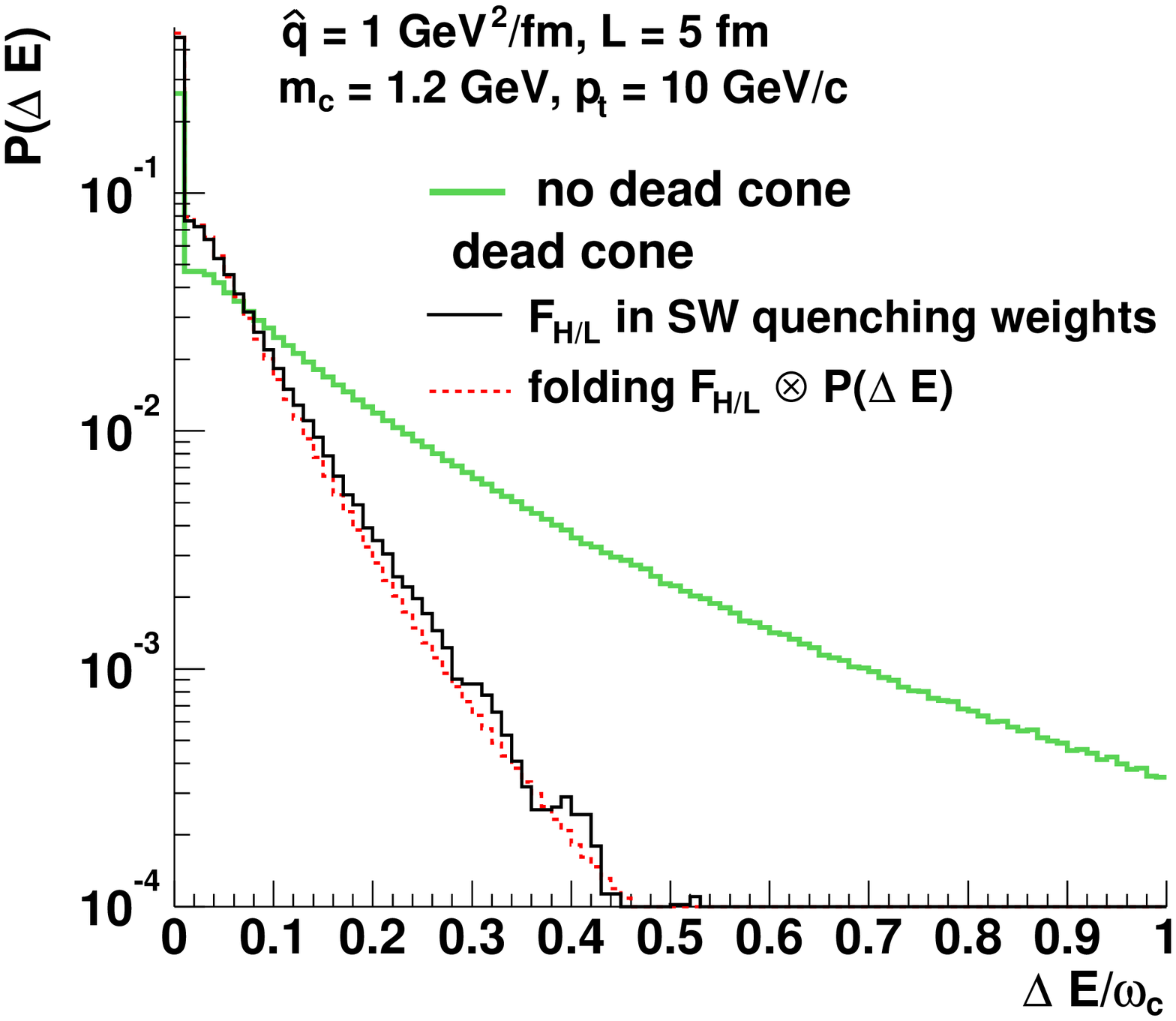}
    \caption{Left: multiplicity of medium-induced gluons radiated with energy 
             larger than $\omega$, from Ref.~\cite{carlosurs}.
             Right: energy loss probability distribution without dead cone
             and with the two different implementations of the dead cone 
             effect as described in the text.} 
    \label{fig:deadconecorr}
  \end{center}
\end{figure}

As a last step, starting from the unquenched, $N^{\rm vacuum}_{\rm c}(\pt)$, 
and quenched, $N^{\rm medium}_{\rm c}(\pt)$, $\pt$ distributions for 
charm quarks, the quenched $\pt$ distribution for D mesons was obtained 
by applying to each of the mesons generated with PYTHIA the weight: 
\begin{equation}
\mathcal{W}(\pt^{\rm c}) = \frac{N^{\rm medium}_{\rm c}(\pt^{\rm c})}
                                {N^{\rm vacuum}_{\rm c}(\pt^{\rm c})},
\end{equation}
where $\pt^{\rm c}$ is the original transverse momentum of the parent quark.

{\sl The energy loss probability distributions $P(\Delta E)$ for charm quarks 
were corrected to account for the dead cone effect}, described in 
Section~\ref{CHAP2:deadcone}. This was done {\sl by folding the distribution 
$P(\Delta E)$ for light quarks with the dead cone suppression factor} 
introduced in Ref.~\cite{dokshitzerkharzeev} 
and reported in Section~\ref{CHAP2:deadcone}. In the following we 
detail and justify this solution.

Figure~\ref{fig:deadconecorr} (left) 
from Ref.~\cite{carlosurs} shows that for 
$R\equiv 0.5\,\hat{q}\,L^3\simeq 2100$ (with $\hat{q}=4~\gev^2/\fm$ and 
$L=3.5~\fm$) the average number $N(\omega=0)$ of radiated gluons per parton, 
when at least 1 gluon is radiated, is approximately 1. If the number of 
radiated gluons was always 1, the distribution of the continuous part 
$p(\Delta E)$ of the quenching weight would coincide with the energy 
distribution d$I/$d$\omega$ of radiated gluons (see Eq.~(\ref{eq:wdIdw})).
Then, the dead cone suppression factor 
$F_{\rm H/L}(m_{\rm c},E_{\rm c},\hat{q},\omega)$ 
in Eq.~(\ref{eq:FHL}), which multiplies d$I/$d$\omega$, could be applied 
directly to $p(\Delta E)$ with $\omega=\Delta E$. The complete energy loss 
probability distribution $P(\Delta E)$ is the weighted sum over the 
different path lengths $L$ of the continuous weight distributions 
$p(\Delta E)$ with the discrete weight $p_0$ added to the first bin 
($\Delta E=0$). But $F_{\rm H/L}$ does not depend on $L$ and it is 1 for 
$\omega=\Delta E=0$. Therefore, in the approximation of single-gluon emission,
the dead cone effect can be included by folding the energy loss probability 
distributions $P(\Delta E)$ with the dead cone factor $F_{\rm H/L}$.
Since $F_{\rm H/L}$ depends on the heavy quark energy ($\approx\pt$), this
folding has to be done for each c quark or, more conveniently, in bins of 
$\pt$.

\begin{figure}[!t]
  \begin{center}
    \includegraphics[width=.49\textwidth]{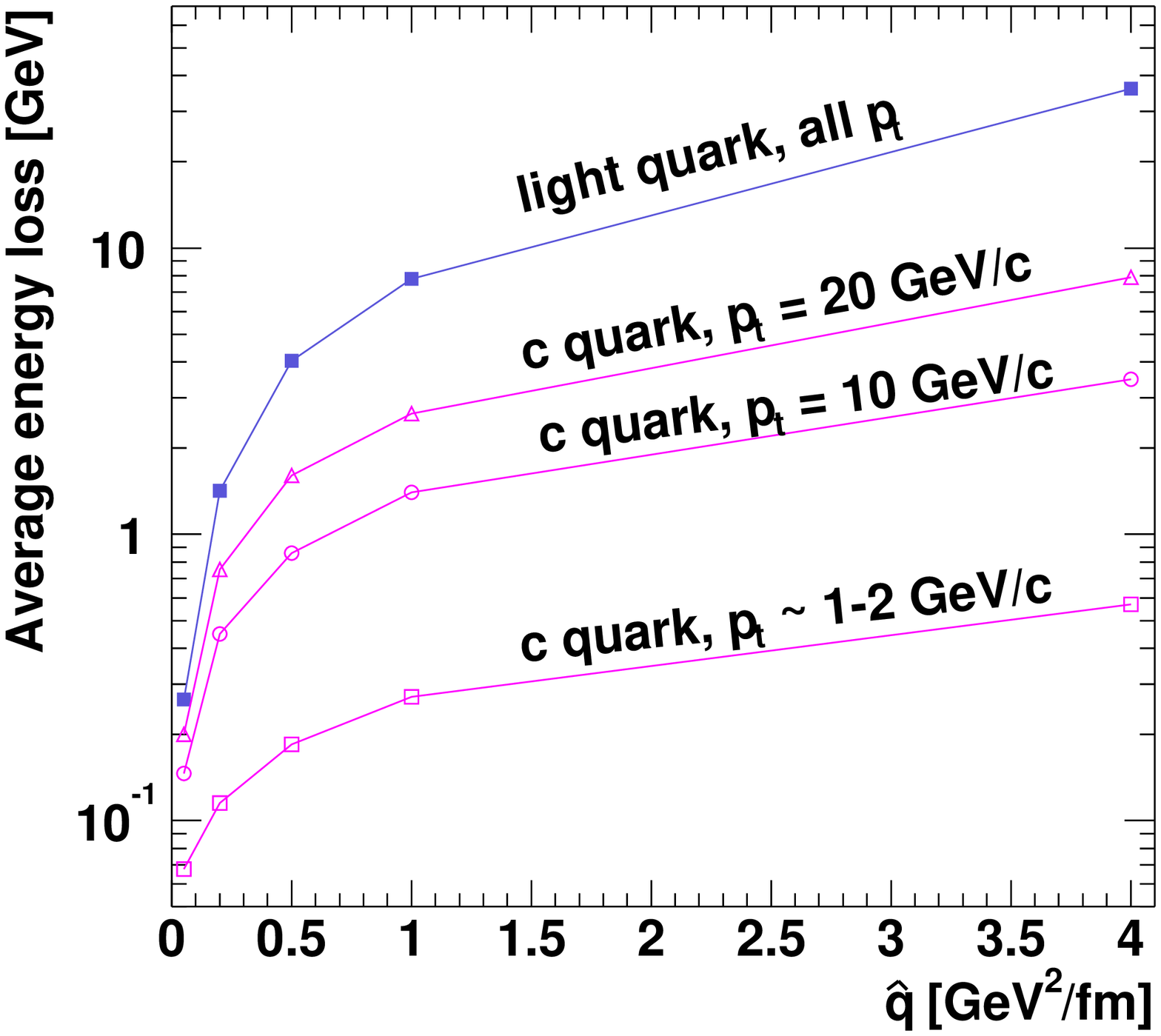}
    \includegraphics[width=.49\textwidth]{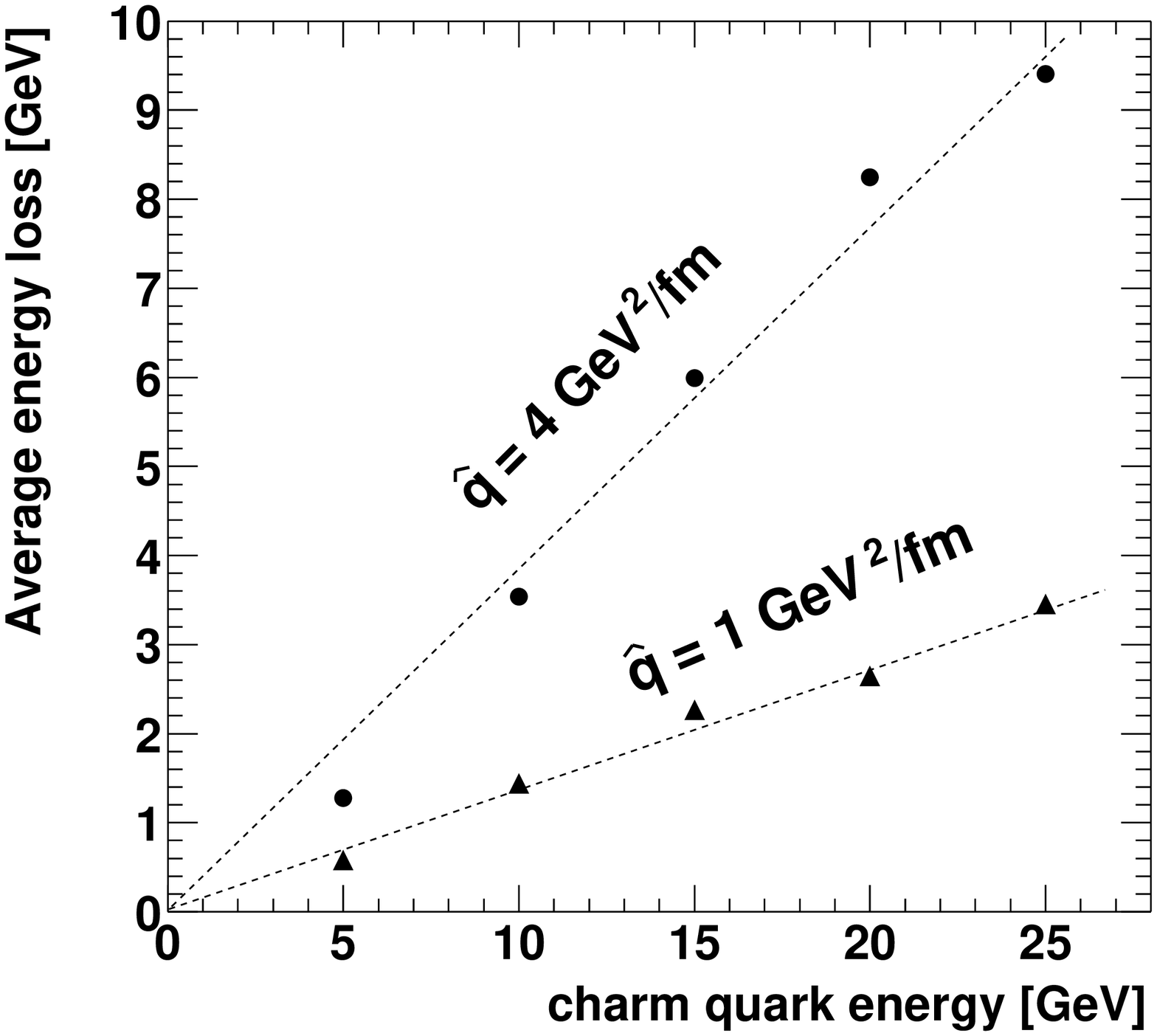}
    \caption{Left: average energy loss as function of the transport 
             coefficient for light (massless) quarks and for charm 
             ($m_{\rm c}=1.2~\gev$) quarks with different transverse momenta.
             Right: average energy loss for charm quarks as a function of 
             their energy; the dashed lines correspond to 
             $\av{\Delta E}\propto E$.} 
    \label{fig:avdE-LightCharm}
  \end{center}
\end{figure}

\begin{figure}[!t]
  \begin{center}
    \includegraphics[width=.49\textwidth]{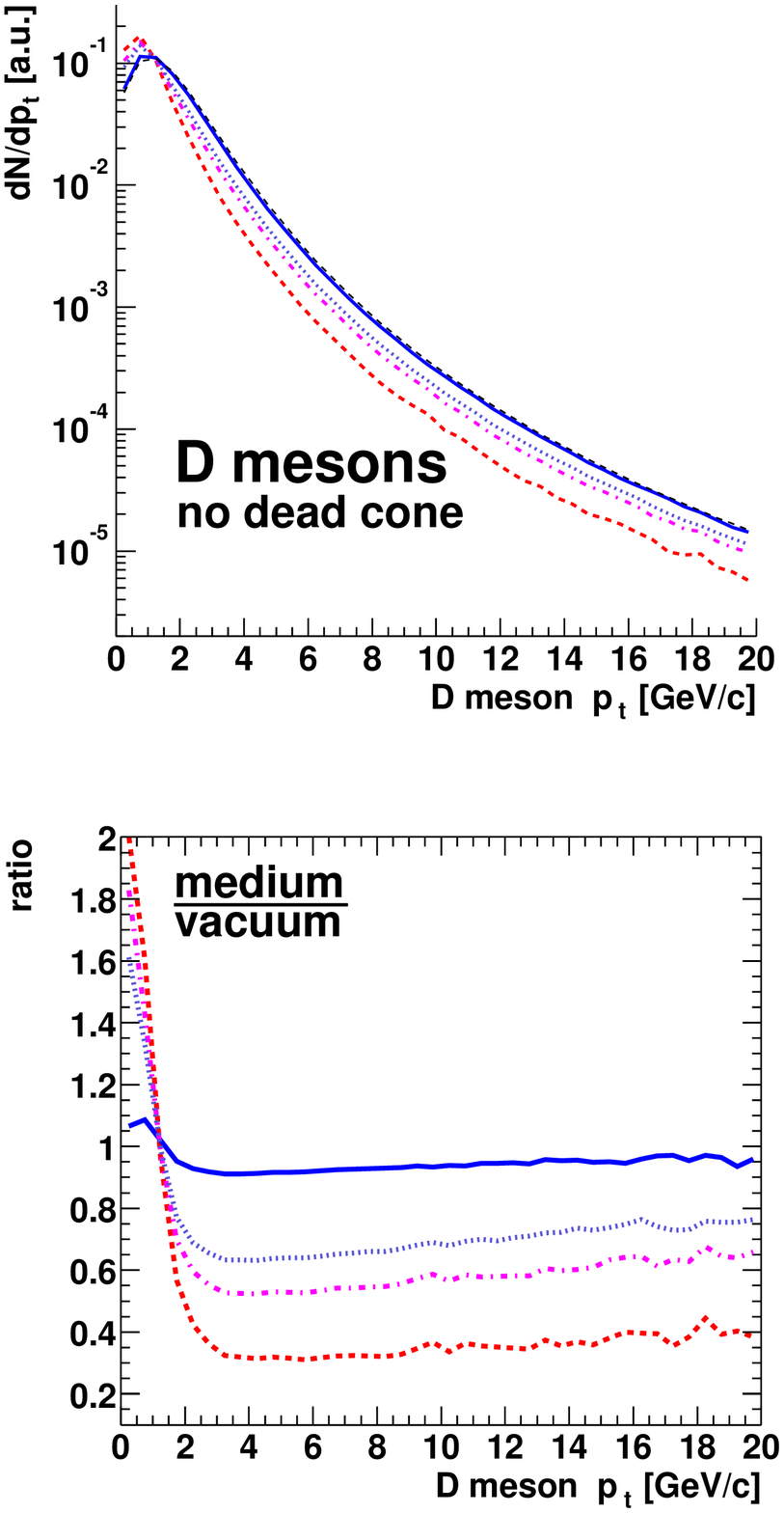}
    \includegraphics[width=.49\textwidth]{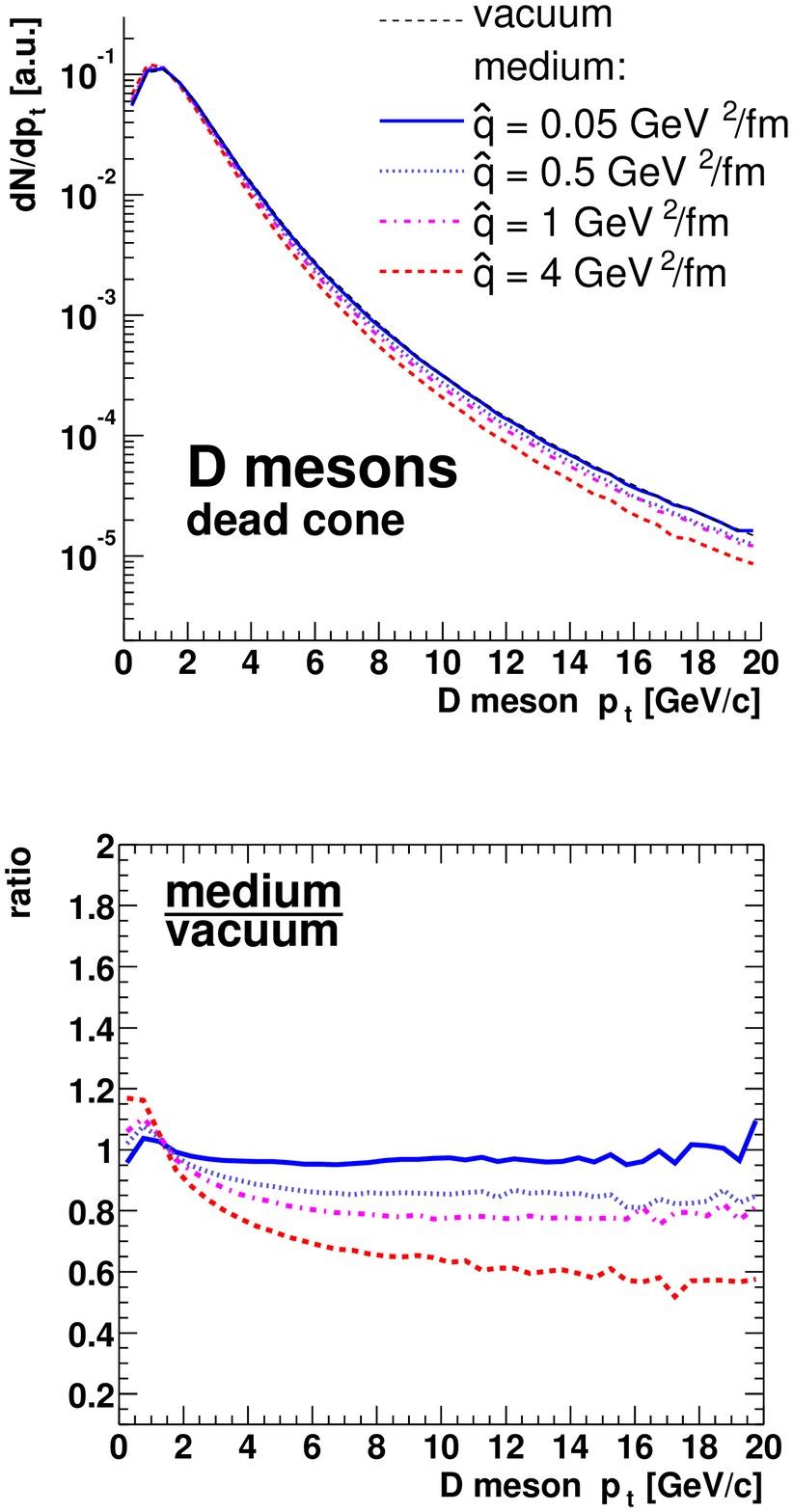}
    \caption{Transverse momentum distributions of D mesons in the vacuum 
             and in the medium (upper panels) and medium-to-vacuum ratios 
             (lower panels). The dead cone correction is not included 
             in the left-side panels and included in the right-side panels.} 
    \label{fig:ptDquench}
  \end{center}
\end{figure}

The dead cone correction, in the form a simple factor $F_{\rm H/L}$ 
that multiplies the radiated gluon energy distribution d$I/$d$\omega$,
can be directly implemented in the SW quenching weights as they are calculated 
in Ref.~\cite{carlosurs}. However, this way of proceeding would be much more
CPU-expensive than the simple folding we outlined above and also 
non-rigorous from a theoretical point of view~\cite{carlosurs3}. The results 
on the energy loss probability distribution obtained with the two methods, 
`folding $F_{\rm H/L}\otimes P(\Delta E)$' and `$F_{\rm H/L}$ in the SW
weights', were compared for
$\hat{q}=1~\gev^2/\fm$, $L=5~\fm$, $m_{\rm c}=1.2~\gev$ and 
$\pt=10,~20,~30~\gev/c$. The comparison for $\pt=10~\gev/c$, 
reported in Fig.~\ref{fig:deadconecorr} (right), is quite satisfactory.
A similar agreement is found for $\pt=20$ and $30~\gev/c$. 

\begin{figure}[!t]
  \begin{center}
    \includegraphics[width=.7\textwidth]{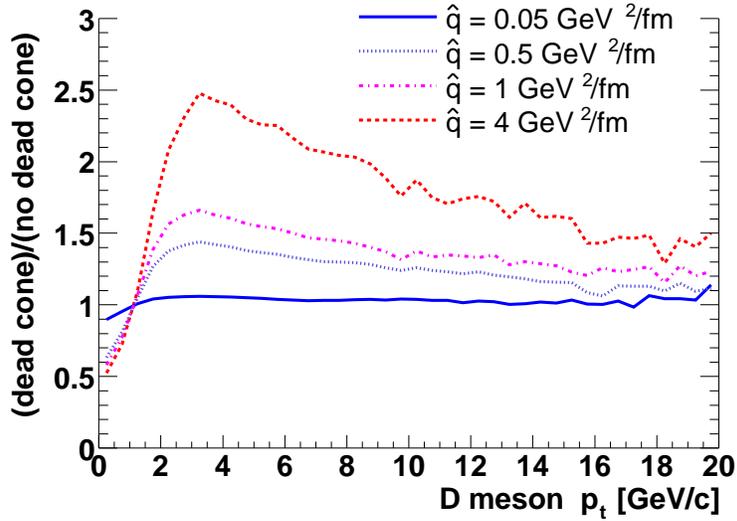}
    \caption{Ratio of the D mesons $\pt$ distributions with and without 
             dead cone effect.} 
    \label{fig:Ddeadcone}
  \end{center}
\end{figure}

Figure~\ref{fig:avdE-LightCharm} (left) reports the average energy 
loss as a function of the transport coefficient for light quarks and for 
charm quarks with $\pt= 1$-$2,~10,~20~\gev/c$, as obtained with the 
described dead cone correction ($\pt$-dependent
$P(\Delta E)\otimes F_{\rm H/L}$ folding). With $\hat{q}=4~\gev^2/\fm$, 
for c quarks of $1$-$2,~10,~20~\gev/c$ $\av{\Delta E}$ is about 2\%, 10\% 
and 20\%, respectively, of the average loss for massless quarks. 

Remarkably, for charm quarks we find that, for given $\hat{q}$,
the average energy loss is approximately proportional to the quark 
energy, $\av{\Delta E}\propto E$, 
(see right panel of Fig.~\ref{fig:avdE-LightCharm}), while the BDMPS average 
energy loss for massless partons is independent of the parton energy. 
On the basis of this 
observation {\sl we expect that not only the magnitude but also 
the $\pt$-dependence of the nuclear modification factor of {\rm D} mesons 
will be significantly affected by the dead cone. }

The transverse momentum distributions of charm mesons for $\hat{q}=0$ 
(vacuum), 0.05 (cold nuclear matter), 0.5, 1 and $4~\gev^2/\fm$ are shown 
in Fig.~\ref{fig:ptDquench} without (left) and with (right) dead cone 
correction. For $\hat{q}=4~\gev^2/\fm$ the ratio medium/vacuum, also 
reported in the figure, is of order 0.3 and rather flat with $\pt$ if the 
dead cone is not accounted for and it is of order 0.7-0.5, decreasing 
with $\pt$, with the dead cone correction. For $\pt<1~\gev/c$,
the ratio is larger than 1 due to the thermalized c quark component,
which accumulates at low momenta.
 
In order to better quantify the effect of the dead cone we plot the ratio 
of the $\pt$ distributions with and without the correction included in the 
quenching weights (Fig.~\ref{fig:Ddeadcone}). The ratio is clearly sensitive
to the density of the medium, e.g $\simeq 2$-$2.5$ for $\hat{q}=4~\gev^{2}/\fm$
and $\simeq 1.5$ for $\hat{q}=1~\gev^2/\fm$, and it decreases with $\pt$
as the mass of the charm quark becomes negligible. This 
(dead cone)/(no dead cone) ratio should be reflected by the the 
D/$hadrons$ observable, which we will calculate in the last section of the 
chapter.

\mysection{Results (I): nuclear modification factor $\RAA$ for ${\rm D}$ 
           mesons}
\label{CHAP8:RAA}

\begin{figure}[!t]
  \begin{center}
    \includegraphics[width=.9\textwidth]{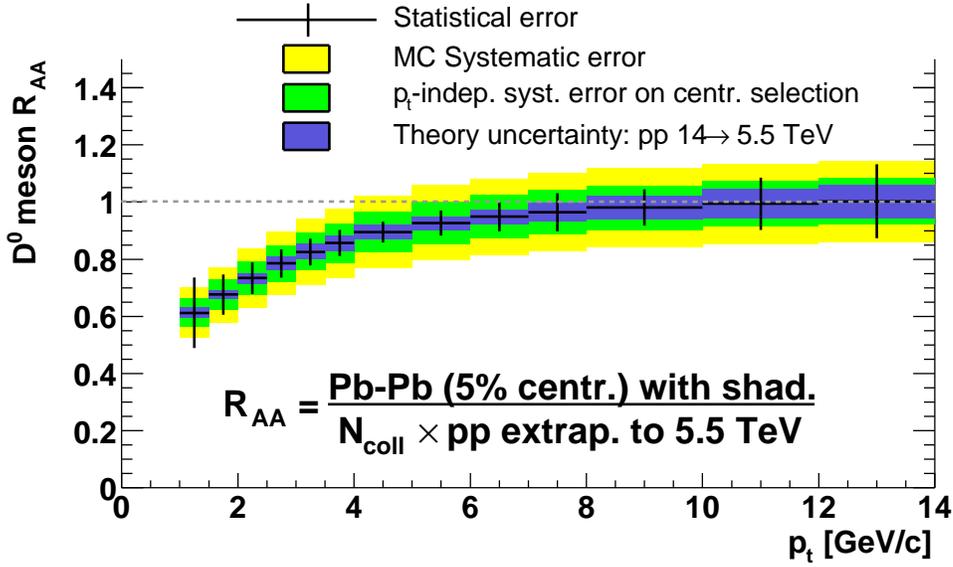}
    \caption{Nuclear modification factor for $\Dz$ mesons. Nuclear shadowing 
             and intrinsic $k_{\rm t}$ broadening are included. Energy loss
             is not included.} 
    \label{fig:RAAbare}
  \end{center}
\end{figure}

The nuclear modification factor for $\Dz$ mesons
\begin{equation}
\label{eq:raa2}
  R_{\rm AA}(\pt)=
    \frac{{\rm d}\sigma^{\rm AA}/{\rm d}\pt/{\rm binary~NN~collision}}
       {{\rm d}\sigma^{\rm pp}/{\rm d}\pt}
\end{equation}
as obtained including only nuclear shadowing and intrinsic parton transverse
momentum broadening and no parton energy loss, is shown in 
Fig.~\ref{fig:RAAbare}. 
The estimated error contributions reported in the figure are:
\begin{itemize}
\item {\sl statistical error}, obtained by adding in quadrature the 
      statistical errors estimated for \PbPb~and for pp collisions
      (see Fig.~\ref{fig:allerrors}); it amounts to 20\% at $\pt=1~\gev/c$,
      4\% at $4~\gev/c$ and 13\% at $14~\gev/c$;
\item {\sl systematic error for MC corrections}, 15\%, from quadratic sum of 
      \PbPb~and pp contributions (see Fig.~\ref{fig:allerrors}); we remind
      that this error has been conservatively set to 10\%, but may well be
      lower;
\item {\sl systematic error on centrality selection and nuclear parameters},
      $8\%\oplus 5\%=9.5\%$, discussed in Section~\ref{CHAP7:systxsec};
      the \NN~inelastic cross section uncertainty, which can be 
      reasonably assumed to be the same for \PbPb~and pp, cancels out in the 
      ratio;
\item {\sl systematic error on the extrapolation of the {\rm pp} results 
      from $14~\tev$ to $5.5~\tev$}, 3-6\% depending on $\pt$, as shown in 
      Fig.~\ref{fig:energyextrapolation}.
\end{itemize}
The systematic uncertainties on the $\DtoKpi$ branching ratio and on the 
correction for feed-down from ${\rm B} \to \Dz+X$ decays (i.e. uncertainty 
on $\bbbar$ cross section) can be assumed to be the same for the 
two colliding systems\footnote{The $\bbbar$ cross section in \PbPb~collisions 
has the additional uncertainty on the effect of nuclear shadowing. However,
given the large mass of the b quark, this uncertainty is relevant only for 
very low-$\pt$ production.} 
and neglected in the ratio.

\begin{figure}[!t]
  \begin{center}
    \includegraphics[width=.32\textwidth]{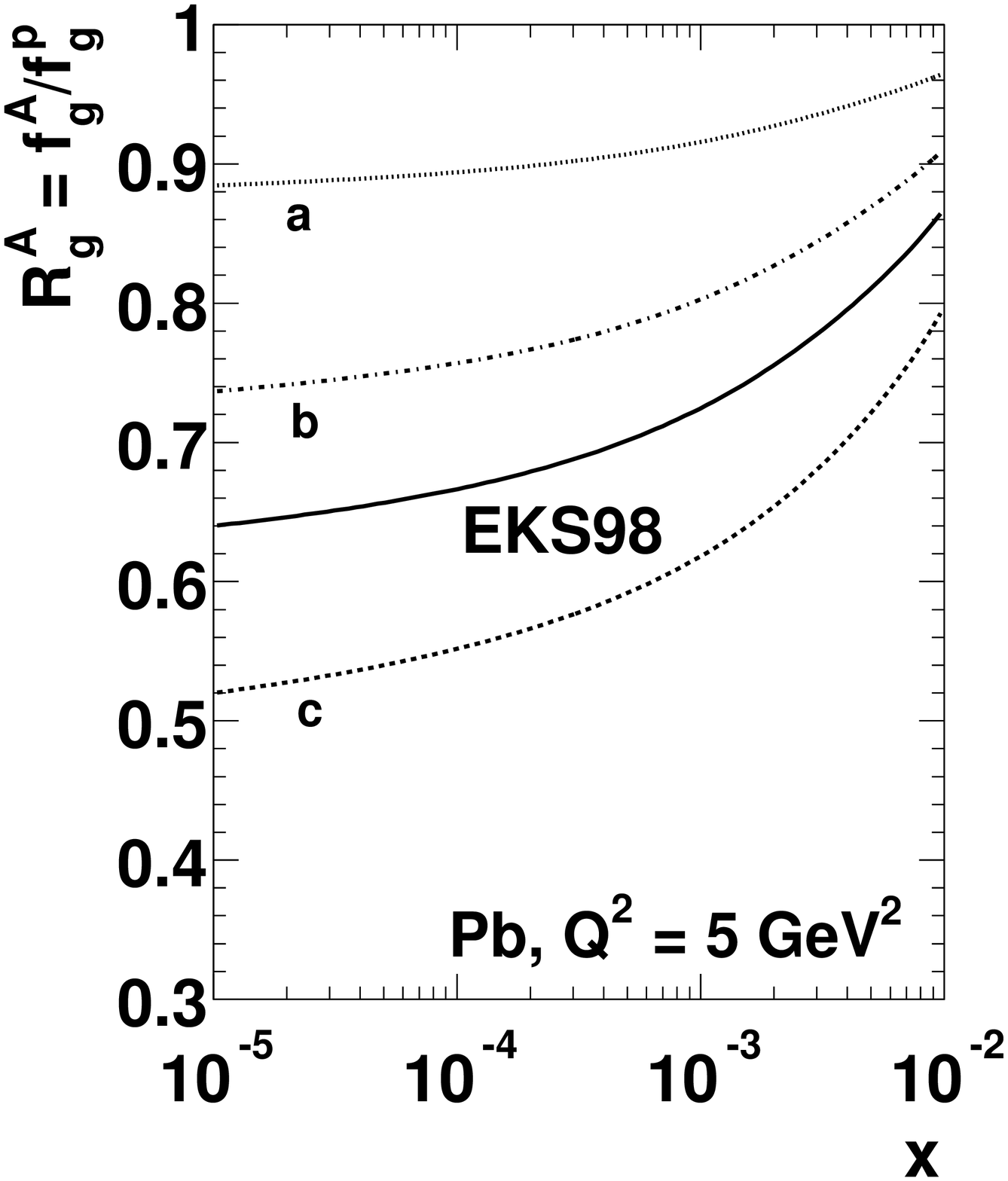}
    \includegraphics[width=.66\textwidth]{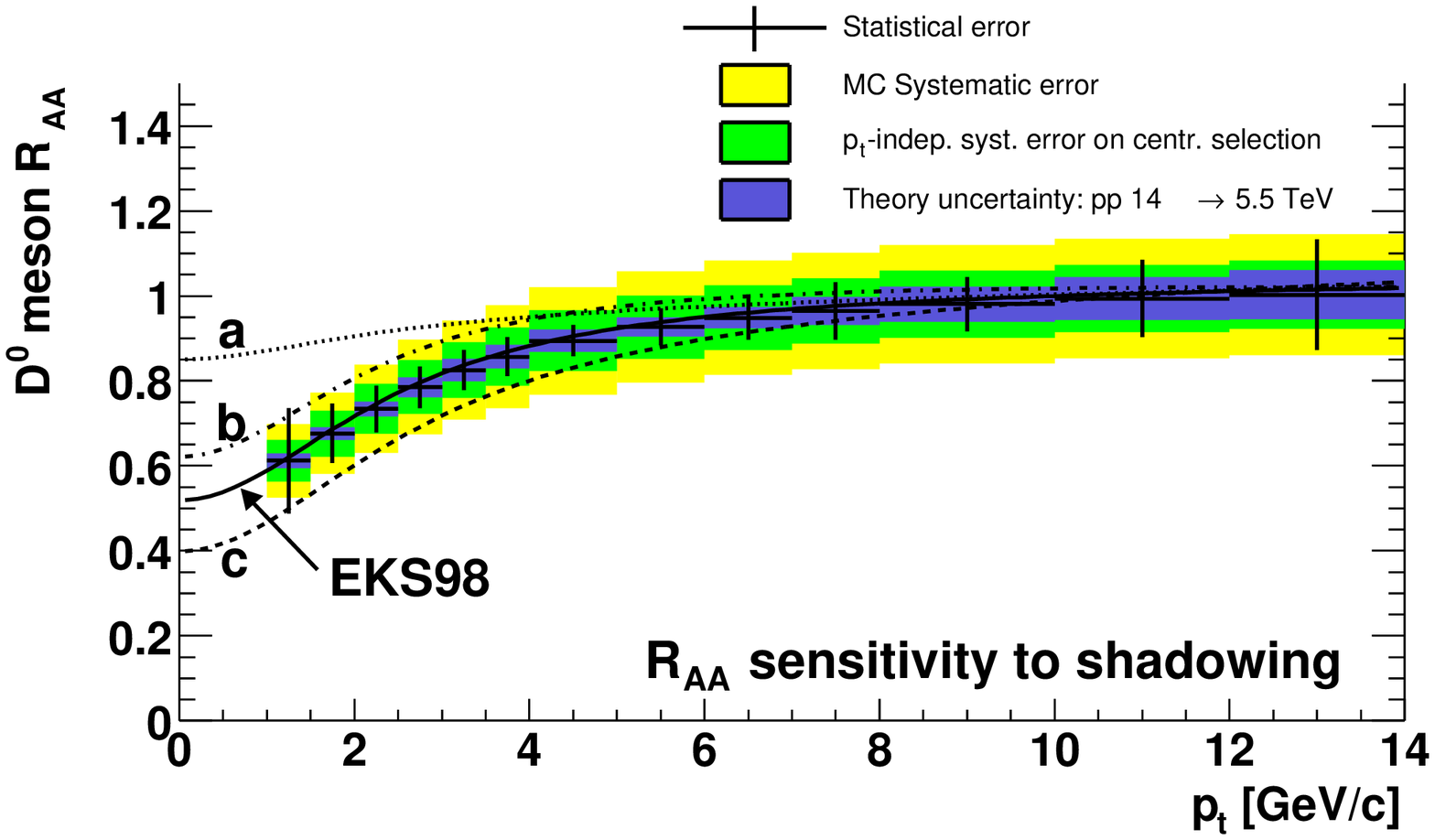}
    \caption{Nuclear modification of the gluon PDF (left) ---the valence and
             sea quark PDFs were also changed accordingly--- and its effect 
             on $\RAA$ of $\Dz$ mesons (right).} 
    \label{fig:RAAshad}
  \end{center}
\end{figure}

The effect of shadowing, introduced via the EKS98 parameterization~\cite{EKS},
is visible as a suppression of $\RAA$ at low transverse momenta, 
corresponding to small Bjorken $x$. As estimated in 
Section~\ref{CHAP2:charminhic}, the effect is negligible for 
$\pt>6$-$7~\gev/c$. Since there is significant uncertainty on the 
magnitude of shadowing in the low-$x$ region (see 
Fig.~\ref{fig:shadowingmodels}), we have studied the effect of such
uncertainty on $\RAA$ by varying the nuclear modification of 
parton distribution functions (PDFs). In Fig.~\ref{fig:RAAshad} we show 
different modifications of the gluon PDF for Pb nuclei and 
$Q^2=5~\gev^2\simeq (2\,m_{\rm c})^2$ ---the valence and sea quark PDFs
were changed accordingly--- and the resulting $\RAA$. Also in the 
case of shadowing 50\% stronger than in EKS98 (curve labelled ``c''),
we find $\RAA>0.93$ for $\pt>7~\gev/c$. We can, thus, confirm that {\sl for 
$\pt>7~\gev/c$ only (possible) parton energy loss is expected to affect 
the nuclear modification factor of {\rm D} mesons.}

\begin{figure}[!t]
  \begin{center}
    \includegraphics[width=\textwidth]{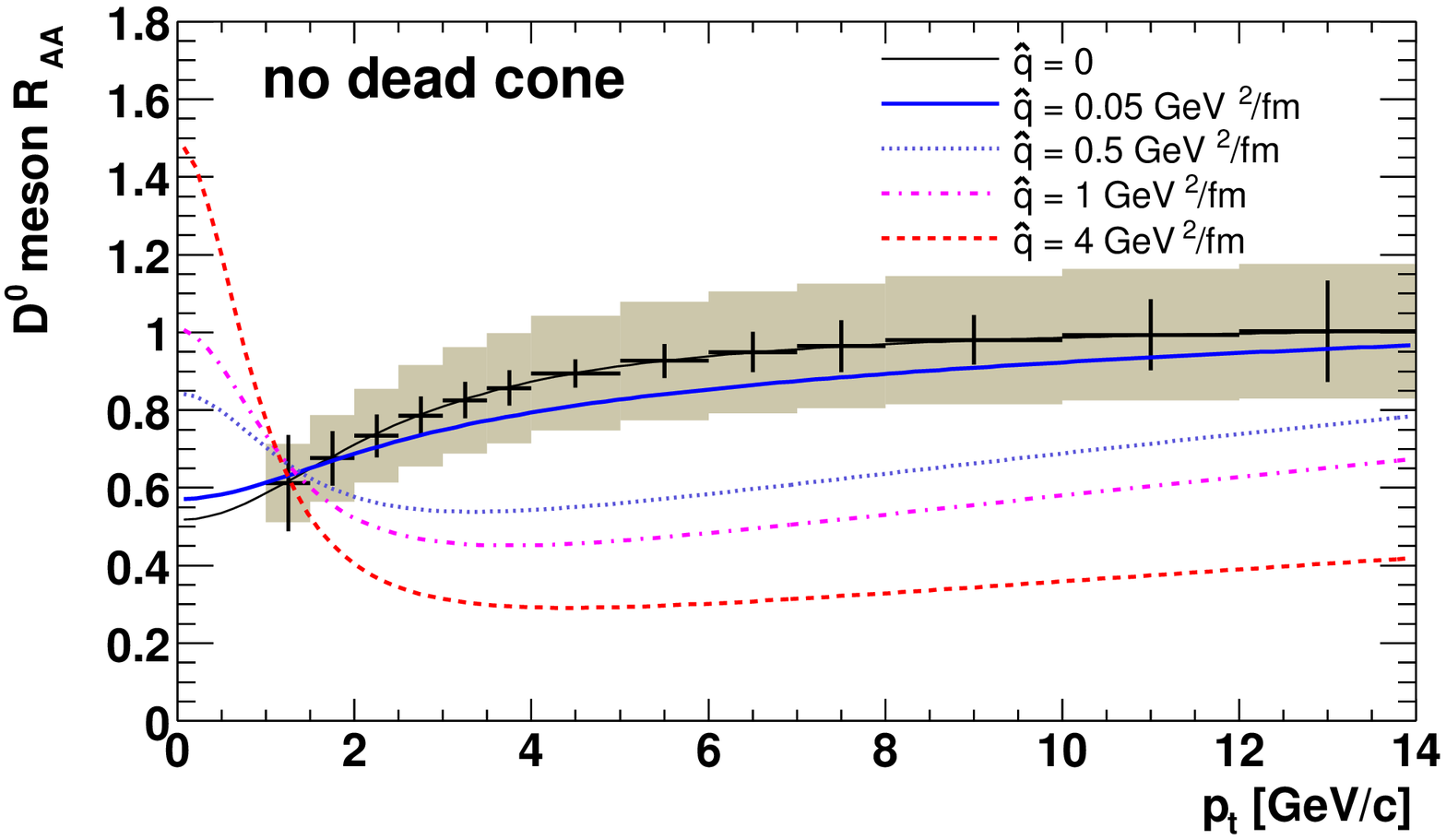}
    \vglue0.5cm
    \includegraphics[width=\textwidth]{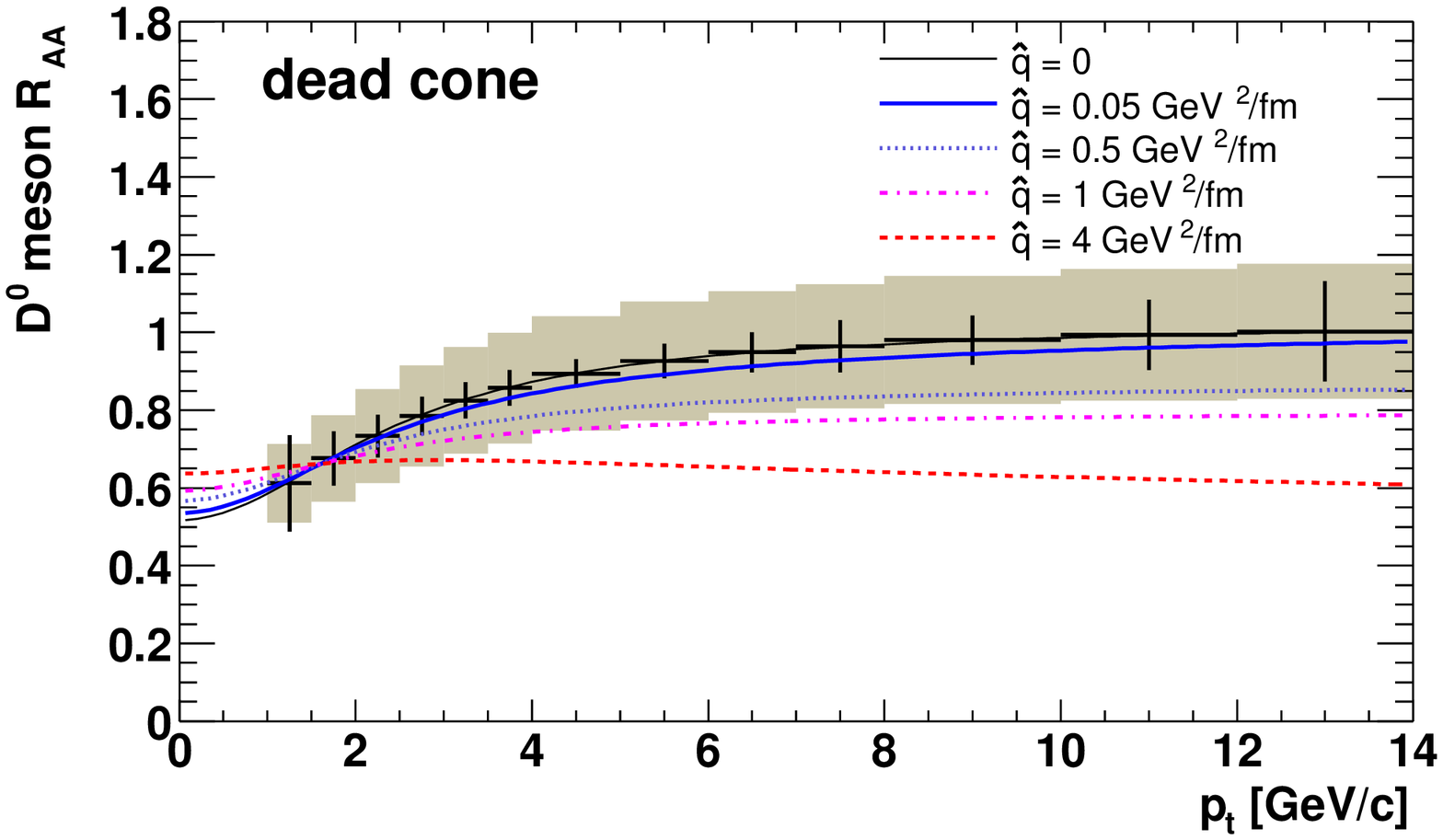}
    \caption{Nuclear modification factor for $\Dz$ mesons with shadowing, 
             intrinsic $k_{\rm t}$ broadening and parton energy loss.
             Upper panel: without dead cone correction; lower panel: with 
             dead cone correction. Errors corresponding to the curve 
             for $\hat{q}=0$ are shown: bars = statistical, 
             shaded area = systematic.} 
    \label{fig:RAAquench1}
  \end{center}
\end{figure}

Figure~\ref{fig:RAAquench1} presents the nuclear modification factor in the 
vacuum and in the medium for different values of the transport coefficient.
The result obtained without (with) dead cone is shown in the upper (lower) 
panel. All systematic uncertainties were added in quadrature.
 
\begin{figure}[!t]
  \begin{center}
    \includegraphics[width=\textwidth]{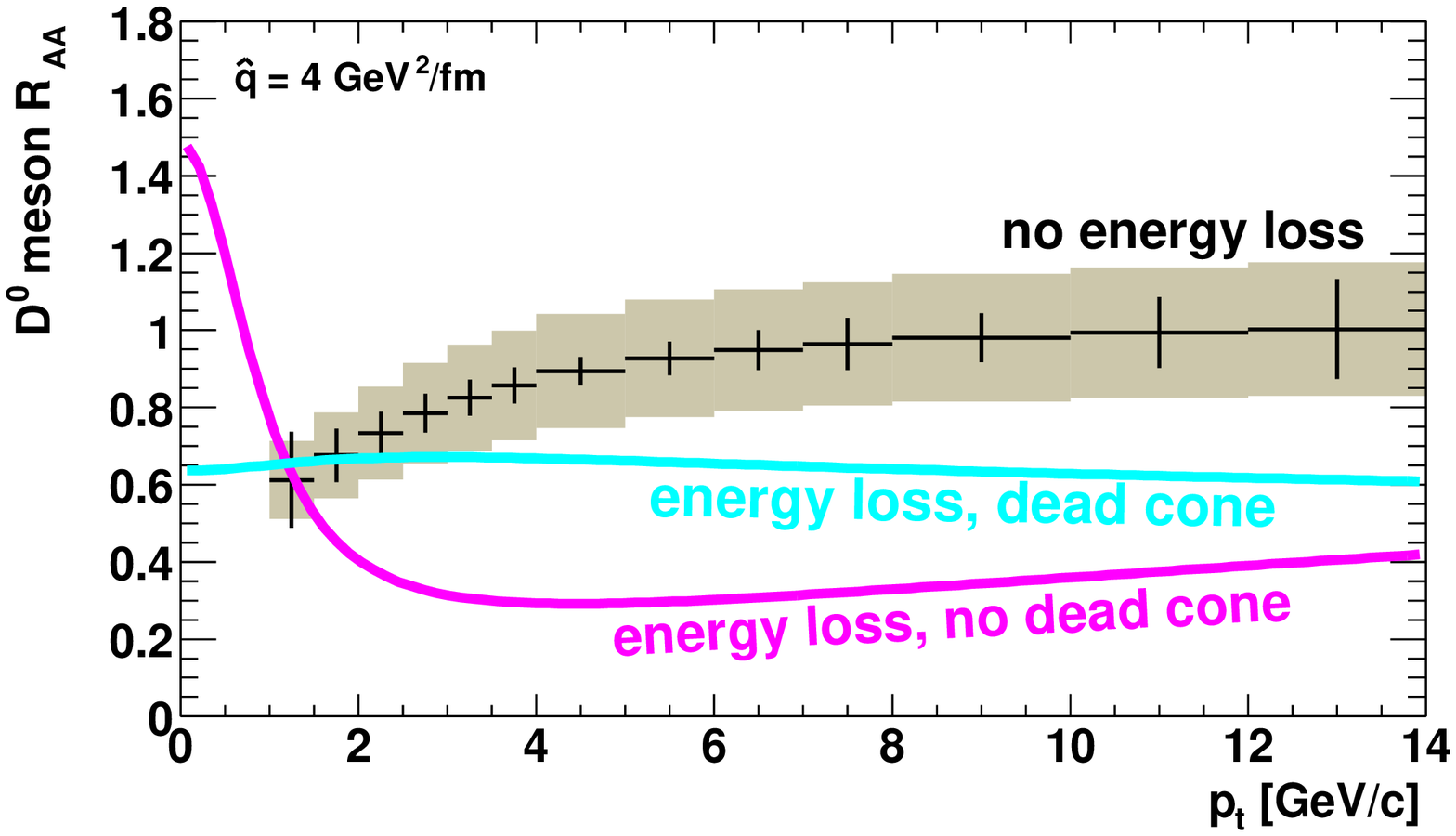}
    \caption{Nuclear modification factor for $\Dz$ mesons with shadowing, 
             intrinsic $k_{\rm t}$ broadening and parton energy loss for 
             $\hat{q}=4~\gev^2/\fm$.
             Errors: bars = statistical, 
             shaded area = systematic.} 
    \label{fig:RAAquench2}
  \end{center}
\end{figure}

For $\hat{q}=4~\gev^2/\fm$ and no dead cone, we find $\RAA$ 
reduced, with respect to 1, by a factor about 3 and slightly increasing 
with $\pt$, from 0.3 at $6~\gev/c$ to 0.4 at $14~\gev/c$. Even for 
a transport coefficient lower by a factor 4, $\hat{q}=1~\gev^2/\fm$,
$\RAA$ is significantly reduced (0.5-0.6). When the dead cone effect is
taken into account, the $\RAA$ reduction due to quenching is found to
be lower by about a factor 1.5-2.5, depending on $\hat{q}$ and $\pt$,
as already seen in Fig.~\ref{fig:Ddeadcone}. For our reference 
transport coefficient, $4~\gev^2/\fm$, $\RAA$ with dead cone is 
equal to 0.6 and essentially flat as a function of $\pt$. The expectations
without and with dead cone are compared directly in Fig.~\ref{fig:RAAquench2}, 
for $\hat{q}=4~\gev^2/\fm$.

We point out that the different $\pt$ dependence appears to be the only 
distinctive feature between a scenario with moderate quenching and 
negligible dead cone effect (e.g. $\hat{q}=1~\gev^2/\fm$ in the upper panel 
of Fig.~\ref{fig:RAAquench1}) and a scenario with large quenching but also 
strong dead cone effect (e.g. $\hat{q}=4~\gev^2/\fm$ in the lower panel).
However, the estimated systematic uncertainty of about 
18\% may prevent from giving a clear statement on the $\pt$ dependence of 
$\RAA$. As remarked in Chapter~\ref{CHAP2}, the comparison of the quenching
of charm-quark-originated mesons and massless-parton-originated hadrons
will be the best suited tool to disentangle the relative importance 
of energy loss and dead cone effects. 

{\small
Before moving to the results on the D/$hadrons$ ratio, we shortly comment on 
the errors presented in Figs.~\ref{fig:RAAquench1} and~\ref{fig:RAAquench2}.
The relative systematic error ($\simeq 18\%$) is larger than the statistical 
error up to $\pt=14~\gev/c$ for $\hat{q}=0$. However, as $\hat{q}$ is 
increased the statistics $S$ of $\Dz$ mesons at large transverse momenta 
in \PbPb~collisions decreases. Consequently, the relative statistical error 
increases roughly as $1/\sqrt{S}$. For example, for $\hat{q}=4~\gev^2/\fm$ and
no dead cone effect the factor 3 reduction in $\RAA$ is accompanied 
by an increase of a factor $\approx\sqrt{3}$ in the statistical error, 
which becomes larger than the systematic one at high $\pt$. Due to the 
lower quenching, the effect is less important if the dead cone is included.
}

\mysection{Results (II): ${\rm D}/hadrons$ ratio}
\label{CHAP8:RDh}

For the comparison of energy loss of heavy quarks and of massless partons, 
we consider the ratio ($\Dz$ mesons)/(charged hadrons), hereafter indicated 
as D/$h$, rather than the ratio ($\Dz$ mesons)/(pions), the reason 
being that the former can be measured with smaller systematic uncertainty. 
Experimentally, the $\Dz$-to-pions ratio can be defined up to 
$\pt\simeq 15~\gev/c$ only using neutral pions, which decay as 
$\pi^0\to\gamma\gamma$ and can be identified by means of 
invariant mass analysis of pairs of photons reconstructed in an 
electro-magnetic calorimeter (e.g. PHOS)\footnote{Charged pions can be 
separated, via $\dEdx$ or time-of-flight, from heavier hadrons (mainly kaons 
and protons) only at relatively low transverse momenta ($\pt<2$-$3~\gev$)
and are, therefore, not usable.}. However, this would not allow to 
cancel out the `MC systematic uncertainties' of the $\Dz$ and $\pi^0$ $\pt$
distributions, because they are measured using different detectors that have 
different systematics. On the other hand, the use of unidentified charged 
hadrons reconstructed by tracking in TPC and ITS allows to partially cancel 
out the uncertainties introduced by acceptance and efficiency corrections 
of these detectors. Part of the error will still remain as each $\Dz$ is 
reconstructed as a pair of opposite-charge tracks, while a charged hadron 
is obviously only one track. 

\begin{figure}[!t]
  \begin{center}
    \includegraphics[width=\textwidth]{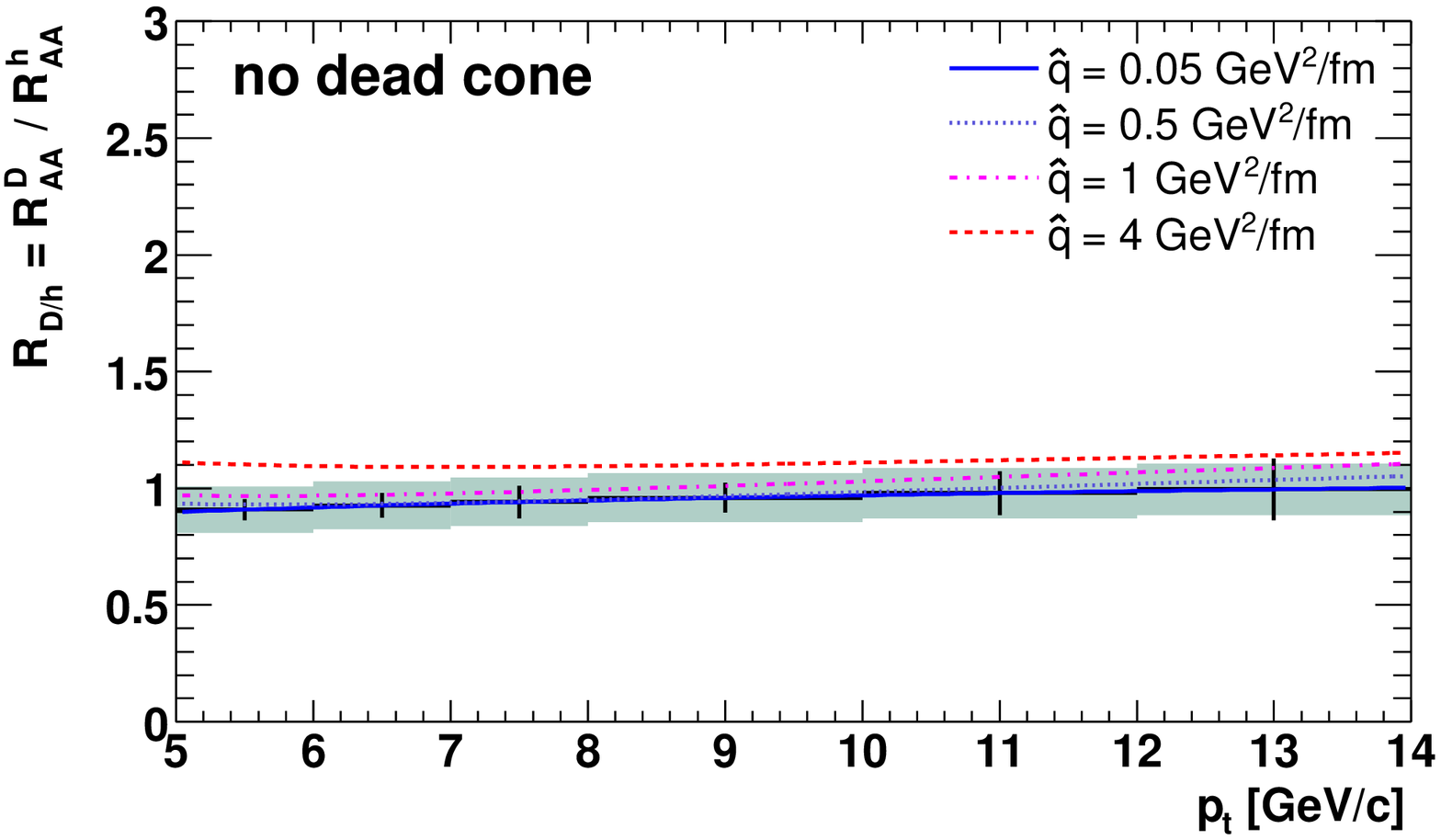}
    \vglue0.5cm
    \includegraphics[width=\textwidth]{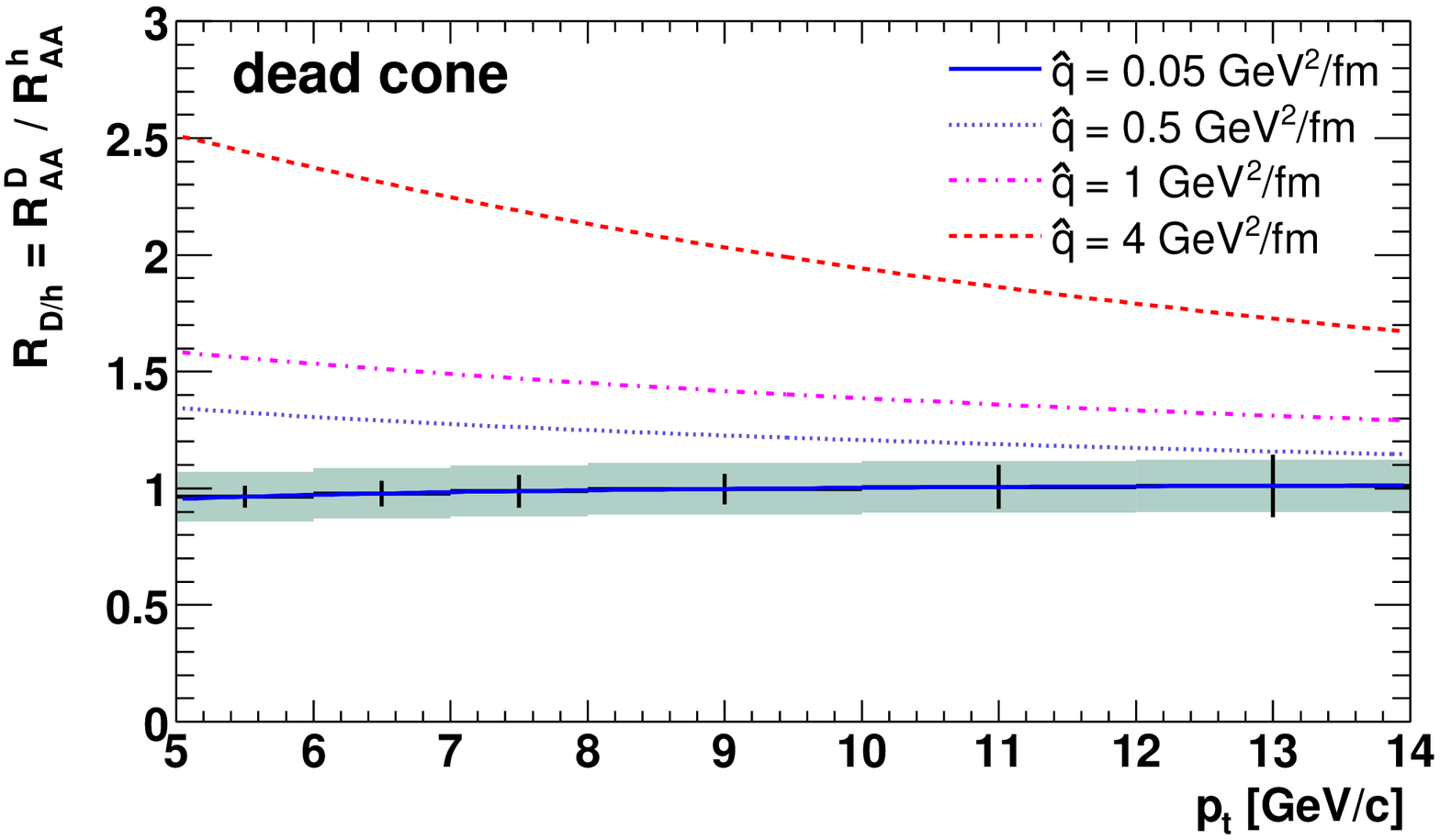}
    \caption{Ratio of the nuclear modification factors for $\Dz$ mesons 
             and for charged hadrons.
             Upper panel: without dead cone correction; lower panel: with 
             dead cone correction. Errors corresponding to the curve 
             for $\hat{q}=0.05~\gev^2/\fm$ are shown: bars = statistical, 
             shaded area = systematic.} 
    \label{fig:RDh}
  \end{center}
\end{figure}

The D/$h$ ratio, defined as
\begin{equation}
     R_{{\rm D}/h}(\pt)=\frac{R_{\rm AA}^{\rm D^0}(\pt)}{R_{\rm AA}^h(\pt)}=
    \frac{{\rm d}\sigma^{\rm AA}(\Dz)/{\rm d}\pt}
       {{\rm d}\sigma^{\rm pp}(\Dz)/{\rm d}\pt} \times
    \frac{{\rm d}\sigma^{\rm pp}(h)/{\rm d}\pt}
       {{\rm d}\sigma^{\rm AA}(h)/{\rm d}\pt},
\end{equation}
is presented in Fig.~\ref{fig:RDh} for the range $5<\pt<14~\gev/c$. 
We used $\RAA^h$ for hadrons as reported in the left panel of 
Fig.~\ref{fig:RAA-hadrons} and $\RAA^{\rm D^0}$ for $\Dz$ mesons, 
without and with dead cone, as reported in Fig.~\ref{fig:RAAquench1}.

The estimated errors on $\RDh$ are:
\begin{itemize}
\item {\sl statistical error}; the same as for $\RAA^{\rm D^0}$, since the 
      statistical error on the $\pt$ distributions of charged hadrons 
      will be negligible for $\pt<14~\gev/c$;
\item {\sl systematic error for MC corrections}; this error partially cancels 
      out in the two ratios 
      $({\rm d}\sigma^{\rm AA}(\Dz)/{\rm d}\pt)/({\rm d}\sigma^{\rm AA}(h)/{\rm d}\pt)$ and 
      $({\rm d}\sigma^{\rm pp}(\Dz)/{\rm d}\pt)/({\rm d}\sigma^{\rm pp}(h)/{\rm d}\pt)$; we consider a residual error of 10\%, due to fact that $\Dz$ is two 
      tracks while $h$ is one track.
\item {\sl systematic error on the extrapolation of {\rm pp} results from 
      $14~\tev$ to $5.5~\tev$}; both the $\Dz$ and the $h$ $\pt$ distributions
      measured in pp collisions will have to be extrapolated to $5.5~\tev$ by 
      means of pQCD; we assume a $\pt$-independent error of 5\% on 
      $\RDh$, slightly lower than what used for $\RAA^{\rm D^0}$, considering
      that the pQCD uncertainties for $\Dz$ and for $h$ will be partially 
      correlated.
\end{itemize}
The errors on the centrality selection in \PbPb~cancel out because the
$\Dz$ and $h$ analyses will be performed on the same sample of events.
The total systematic error is $10\%\oplus 5\%\simeq 11\%$ and it is shown by 
the shaded bands in Fig.~\ref{fig:RDh}.

We find that, if the dead cone correction for c quarks is not included, 
$\RDh$ is essentially 1 in the considered $\pt$ range, independently of the 
value of the transport coefficient, i.e. of the magnitude of the energy 
loss effect.
When the dead cone is taken into account, $\RDh$ is enhanced of a factor
strongly dependent on the transport coefficient of the medium:
e.g. 2-2.5 for $\hat{q}=4~\gev^2/\fm$ and 1.5 for $\hat{q}=1~\gev^2/\fm$.
The enhancement is decreasing with $\pt$, as expected.

{\sl The $\RDh$ ratio is, therefore, found to be enhanced, with respect to 1,
only by the dead cone effect and, consequently, it appears as a very 
clean tool to investigate and quantify this effect.} We point out that the 
obtained $\RDh$ with dead cone is extremely similar to the ratio 
of the $\Dz$ transverse momentum distributions with and without dead cone, 
shown in Fig.~\ref{fig:Ddeadcone}.  

In Section~\ref{CHAP2:deadcone} we discussed how the D/$h$ ratio 
should, in principle, be enhanced also in absence of dead cone effect,
as a consequence of the larger energy loss of gluons with respect to quarks.
Such enhancement is essentially not observed in the obtained $\RDh$ 
because it is `compensated' by the harder fragmentation of charm quarks with 
respect to light quarks and, particularly, gluons. With $z$ the typical 
momentum fraction taken by the hadron in the fragmentation,
$\pt^{\rm hadron}=z\,\pt^{\rm parton}$, and $\Delta E$ the average
energy loss for the parton, $(\pt^{\rm parton})'=\pt^{\rm parton}-\Delta E$,
we have
\begin{equation}
  (\pt^{\rm hadron})'=\pt^{\rm hadron} - z\,\Delta E,
\end{equation}
meaning that the energy loss observed in the nuclear modification factor is, 
indeed, $z\,\Delta E$. We have, thus, to compare 
$z_{\rm c\to D}\,\Delta E_{\rm c}$
to $z_{\rm gluon \to hadron}\,\Delta E_{\rm gluon}$. With 
$z_{\rm gluon \to hadron}\approx 0.4$ (from Fig.~\ref{fig:PYTHIApartonsFF}, 
left), $z_{\rm c\to D}\approx 0.8$ for $\pt^{\rm D}>5~\gev/c$ 
(it is $\approx 0.75$ for $\pt^{\rm D}>0$, see Section~\ref{CHAP3:hadr})
and $\Delta E_{\rm c}=\Delta E_{\rm gluon}/2.25$ (without dead cone),
we obtain 
\begin{equation}
  z_{\rm c\to D}\,\Delta E_{\rm c}\approx 0.9\, z_{\rm gluon \to hadron}\,\Delta E_{\rm gluon}.
\end{equation} 
This simple estimate confirms that the quenching for D mesons is 
almost the same as for (non-charm) charged hadrons, if the dead cone 
effect is not considered. 

The errors reported in Fig.~\ref{fig:RDh} show that ALICE is expected to have
good capabilities for the study of $\RDh$: in the range $5<\pt<10~\gev/c$
the enhancement due to the dead cone is an effect of more than $3~\sigma$
for $\hat{q}>1~\gev^2/\fm$. The comparison of the values for the 
transport coefficient extracted from the nuclear modification factor of 
charged hadrons (or neutral pions) and, {\sl independently}, from the 
D/$hadrons$ ratio shall provide an important test for the coherence of 
our understanding of the energy loss of hard probes propagating in the 
dense QCD medium formed in \PbPb~collisions at the LHC.

In conclusion, we remark that, presently, the procedure we used and equivalent 
ones~\cite{lokhtinHVQ} are the only way to 
keep into account (though with an approximation) the predicted dead cone 
effect in a simulation of energy loss. This topic now attracts more and more
attention from the theoretical side and many of the present uncertainties 
should be reduced in the near future.
These developments will be deployed to improve our studies. 

\clearpage
\pagestyle{plain}

\pagestyle{plain}
\chapter*{\centering Conclusions}
\addcontentsline{toc}{chapter}{Conclusions}

\pagestyle{plain}

This work was aimed at studying the performance of the ALICE detector 
for measuring charm production in heavy ion collisions at the LHC
and for investigating the properties of the deconfined quark--gluon
medium formed in central \AA~reactions by a comparison of its `opacities' 
to massive (charm) quarks and massless partons.

We carried out a detailed simulation study for the detection of $\Dz$
mesons in the $\rm K\pi$ decay channel in central \PbPb~at 
$\sqrtsNN=5.5~\tev$ and pp collisions at $\sqrt{s}=14~\tev$. 
The detection strategy is based on the invariant mass 
analysis of fully reconstructed topologies originated in displaced
secondary decay vertices. This strategy exploits the excellent capabilities 
of the ALICE detector in track reconstruction, vertexing and particle 
identification.  
The signal selection was studied as a function of the transverse momentum 
$\pt$. 

The $\pt$-differential $\Dz$ production cross section can be 
measured in the range $1$-$14~\gev/c$ for \PbPb~and $0$-$14~\gev/c$ 
for pp collisions with statistical uncertainty better than $10$-$15\%$ 
and systematic uncertainty better than $15$-$20\%$.


The effect of parton energy loss in a high-opacity quark--gluon medium 
was simulated using a state-of-the-art calculation. Particular attention 
was devoted to the description of the collision geometry and, 
thus, of the path length traveled by fast partons in the medium. We found that
the use of an equivalent average length is not a good solution and can 
significantly bias the final results. We implemented a simulation chain 
that allows to include an algorithm accounting for the 
suppression of small-angle gluon radiation off massive quarks 
(dead cone effect). This enabled us to obtain and compare the quenched $\pt$ 
distributions of c quarks and D mesons, with and without dead cone effect.
With the medium density estimated on the basis of the
hadron quenching measured at RHIC,
D meson production in the range $3<\pt<15~\gev/c$ in central 
\PbPb~collisions is expected to be suppressed by a factor 3, with 
respect to binary scaling from pp collisions, if the dead cone is not 
included and by a factor 1.5 if it is included.

{\sl The dead cone effect can be studied by means of the 
D/$charged~hadrons$ ratio. This ratio is found to be enhanced, with respect 
to 1, only by the dead cone
effect and the enhancement, which is sensitive to the density of the medium,
may reach up to a factor 2-2.5 for the conditions expected at the LHC.}

Remarkably, we find that D/$charged~hadrons$ is essentially 1, 
independently of $\pt$,
if the dead cone is not included. The expected enhancement due to the 
larger energy loss for gluons, that generate most of the hadrons, 
than for (c) quarks, that generate D mesons, is not observed.
This is explained by the softer fragmentation of gluons with respect 
to charm quarks. 

We identified a range in transverse momentum, $7<\pt<14~\gev/c$, where
energy loss studies can be experimentally addressed using reconstructed 
$\DtoKpi$ decays. In fact, we showed that the other nuclear effects 
(shadowing and intrinsic parton $\pt$ broadening) should be limited to 
the region $\pt<7~\gev/c$. 

The estimated experimental uncertainties allow to conclude that ALICE has 
a good potential for studying, for gluons, light quarks and heavy quarks, 
the medium-induced energy loss effect, which is one of the main predictions of
QCD in hot and dense matter.

In order to fully exploit such excellent physics reach, it is now important
to proceed with the preparation of the tools for data analysis and, in 
parallel, to push further the physics feasibility studies, 
in close collaboration with the theoretical community.
The most promising and stimulating items emerged during this work are
in the direction of parton energy loss.

\clearpage
\pagestyle{plain}

\setcounter{chapter}{0}
\pagestyle{myheadings}
\myappendix
\mychapter{Kinematics of the $\DtoKpi$ decay}
\label{App:kine}

\pagestyle{myheadings}

{\small

In this appendix we consider the kinematics of the two-body decay 
$\DtoKpi$; in particular, 
we show how the experimental resolution on the invariant mass is proportional 
to the momentum resolution and we calculate the mean value of the impact 
parameter of the decay products. 
~\\

For two particles with four-momenta $(E_1,\vec{p}_1)$ and $(E_2,\vec{p}_2)$,
the invariant mass is defined as the modulus of the total four-momentum:
\begin{equation}
M=\sqrt{s_{12}}=\sqrt{(E_1+E_2)^2-(\vec{p}_1+\vec{p}_2)^2}.
\end{equation}
In the case of the $\DtoKpi$ decay, the invariant mass of the 
$\K\pi$ system is equal to the $\Dz$ mass. 

In the relativistic case, $E\simeq p$, we can write:
\begin{equation}
M^2\simeq(p_{\rm K}+p_\pi)^2-(\vec{p}_{\rm K}+\vec{p}_\pi)^2=
2\,p_{\rm K}\,p_\pi\,(1-\cos\psi_{\rm K\pi}), 
\end{equation} 
where $\psi_{\rm K\pi}$ is the angle between the two particles. Since the 
angles are usually measured with much better precision than the momenta
(in particular in a moderate magnetic field, as 
is the case in ALICE), 
the main contribution to the invariant mass resolution comes from 
the momentum resolution. By differentiation of the previous expression, 
we obtain:
\begin{equation}
\frac{\Delta M}{M}=\frac{1}{\sqrt{2}}\,\frac{\Delta p}{p}.
\end{equation} 
This relation holds with good approximation for $\Dz$ meson decays at the LHC, 
since the two decay products can be considered relativistic 
(Section~\ref{CHAP3:hadr}). 

In the rest frame of the $\Dz$ meson, the two decay products are emitted 
back-to-back with the same total momentum $p^\star$ and with energies 
$E_{\rm K}^\star=\sqrt{m_{\rm K}^2+(p^\star)^2}$ and $E_\pi^\star=\sqrt{m_\pi^2+(p^\star)^2}$. Since $\vec{p}^\star_{\rm K}=-\vec{p}^\star_\pi$, 
the expression of the invariant mass gives:
\begin{equation}
M_{\rm D^0}=E_{\rm K}^\star+E_\pi^\star.
\end{equation}
Using this relation and the masses of the involved particles~\cite{pdg}, 
the values $p^\star=0.86~\gev/c$, $E^\star_{\rm K}=0.99~\gev/c$ and 
$E^\star_\pi=0.87~\gev/c$ are obtained.
We introduce the decay angle $\theta^\star$, defined as the angle, 
in the $\Dz$ rest frame, between the direction of the momenta of the decay 
products and the $\Dz$ flight direction (see sketch in 
Fig.~\ref{fig:D0kine}, left). 
The momenta of the two particles can be decomposed
in a component perpendicular to the $\Dz$ flight direction, 
$q_{\rm t}^\star=p^\star\sin\theta^\star$, and one parallel to it, 
$q_{\rm l}^\star=\pm p^\star\cos\theta^\star$. 

\begin{figure}[!t]
  \begin{center}
    \includegraphics[width=.49\textwidth]{figures/chap6/DecayAngle.eps}
    \includegraphics[width=.49\textwidth]{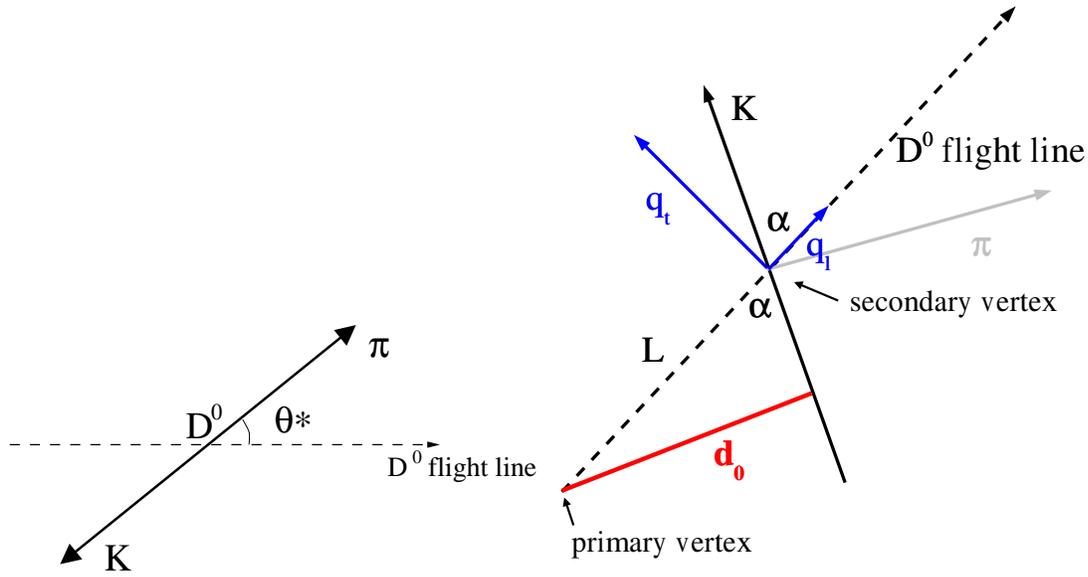}
    \caption{Left: definition of the decay angle $\theta^\star$ in the $\Dz$
    reference frame. Right: decay topology with the impact parameter $d_0$.} 
    \label{fig:D0kine}
  \end{center}
\end{figure}  

In the global reference frame, where the $\Dz$ has a velocity $\beta c$,
the momentum of the generic decay product, decomposed with respect to the 
$\Dz$ flight direction, is:
\begin{equation}
p=\sqrt{q_{\rm t}^2+q_{\rm l}^2},
\end{equation}
where:
\begin{equation}
q_{\rm t} = q_{\rm t}^\star~~~{\rm and}~~~q_{\rm l}=\gamma (q_{\rm l}^\star
+\beta E^\star),
\end{equation}
having applied the Lorentz boost to the component along the $\Dz$ 
flight direction. We remark that, since $E^\star_{\rm K}>E^\star_\pi$, 
the kaon has, on average, a momentum larger than that of the pion.

Let us now calculate the value of the impact parameter projection on 
the bending plane, that we indicate here as $d_0$.
With reference to the 
sketch in Fig.~\ref{fig:D0kine} (right), we have:
\begin{equation}
d_0=L\,\sin\alpha,
\end{equation}
where $L=ct\beta\gamma$ is the decay length of a $\Dz$ with proper decay time 
$t$, and $\alpha$ is the angle of the considered decay particle with 
respect to the $\Dz$ flight-direction. The effect of the magnetic field, $B$,
can be neglected because, for $B=0.4~$T and $\pt^{\rm K,\pi}=1~\gev/c$, 
the sagitta of the arc with length $\ell=100~\mum$ (the typical value 
of $L$) is $s=0.3 B\ell^2/8\pt\simeq 10^{-4}~\mum$. 

The `detectable' decays are those with large impact parameter Therefore,
for this estimation, we consider the situation 
in which the decay plane coincides with the 
bending plane and the decay angle $\theta^\star$ is $\pi/2$. This is the
configuration with higher probability to be detected.
 
With this assumption we have:
\begin{equation}
\sin\alpha=\frac{q_{\rm t}}{\sqrt{q_{\rm t}^2+q_{\rm l}^2}}=1\Bigg/{\sqrt{1+\left(\frac{q_{\rm l}}{q_{\rm t}}\right)^2}}.
\end{equation}
Then, using $p^\star\simeq E^\star$ and $\theta^\star=\pi/2$, we have:
\begin{equation}
\frac{q_{\rm l}}{q_{\rm t}}=\frac{\gamma(p^\star\cos\theta^\star+\beta E^\star)}{p^\star\sin\theta^\star}\simeq \frac{\beta\,\gamma\, p^\star}{p^\star}=\beta\,\gamma.
\end{equation}
We now write the impact parameter as
\begin{equation}
d_0=ct\,\beta\,\gamma\Bigg/{\sqrt{1+(\beta\,\gamma)^2}}=ct\Bigg/{\sqrt{1+1/(\beta\,\gamma)^2}}=ct\Bigg/{\sqrt{1+(M_{\rm D^0}/p_{\rm D^0})^2}}.
\end{equation}
Following the exponential distributions of the proper decay time $t$, the 
impact parameter has an exponential distribution with mean value:
\begin{equation}
\av{d_0}=c\tau\Bigg/{\sqrt{1+(M_{\rm D^0}/p_{\rm D^0})^2}}=124~\mum\Bigg/{\sqrt{1+(M_{\rm D^0}/p_{\rm D^0})^2}}.
\end{equation}
In Fig.~\ref{fig:avd0vsp} we plot the mean value of $d_0$ as a function of 
the momentum of the $\Dz$. The value (marked by the arrows) 
corresponding to the average momentum of $\Dz$ mesons with $|y|<1$ is 
$\approx 105~\mum$. For momenta of $\sim 1~\gev/c$, the average impact 
parameter is very small; therefore, in these cases, only mesons that decay 
with proper decay time $t$ significantly larger than the mean lifetime $\tau$ 
can be identified by means of a displaced vertex selection.

\clearpage
\begin{figure}[!t]
  \begin{center}
    \includegraphics[width=.8\textwidth]{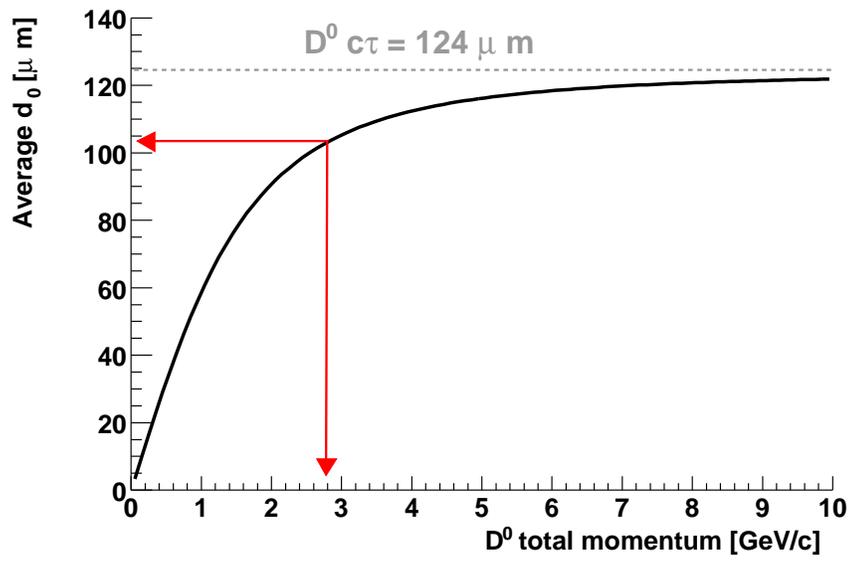}
    \caption{Average impact parameter in the transverse plane for the decay 
             products of a $\Dz$ meson, as a function of its momentum. 
             The arrows mark the value corresponding to the average momentum
             of $\Dz$ mesons with $|y|<1$ at LHC energies.} 
    \label{fig:avd0vsp}
  \end{center}
\end{figure}  

~\\
}

\clearpage
\pagestyle{plain}

\myappendix
\mychapter{PYTHIA parameters used for heavy quark generation at LHC energies}
\label{App:pythiahvq}

\pagestyle{myheadings}

{\small

In Table~\ref{tab:pythiaparams} we report the complete list of parameters 
used in the PYTHIA event generator~\cite{pythia} in order to reproduce the 
inclusive $\pt$ distribution for the heavy quarks given by the HVQMNR program 
based on NLO calculations by M.~Mangano, P.~Nason and 
G.~Ridolfi~\cite{MNRcode}. 
A detailed description of the parameters
can be found in Ref.~\cite{pythia}.

As specified in Section~\ref{CHAP3:generators}, the main parameter 
we tuned is the lower $\pt^{\rm hard}$ limit: the optimal value was 
found to be $2.1~\gev/c$ for charm production and $2.75~\gev/c$ for 
beauty production, both for \mbox{Pb--Pb} collisions at 
$\sqrtsNN=5.5~\tev$ and for pp collisions at $\sqrt{s}=14~\tev$.
Therefore, one can reasonably assume that the same values can be used 
also for \mbox{p--Pb} collisions at $\sqrtsNN=8.8~\tev$.    

The different values for the partonic intrinsic transverse momentum 
$k_{\rm t}$ in pp, \mbox{p--Pb} 
and \mbox{Pb--Pb} collisions were taken from Ref.~\cite{vogtnew}.

\begin{table}[!t]
\caption{PYTHIA parameters used for the generation of charm and
  beauty quarks in pp collisions at 14~TeV, \mbox{p--Pb} collisions at
  8.8~TeV and \mbox{Pb--Pb} collisions at 5.5~TeV. All non-specified parameters
  are left to PYTHIA 6.150 defaults.}
\label{tab:pythiaparams}
\begin{center}
\small
\begin{tabular}{c|c|c|c}
\hline
\hline
Description & Parameter & Charm & Beauty \\
\hline
\hline
Process types & MSEL & 1 & 1 \\
\hline
Quark mass $[\gev]$ & PMAS(4/5,1) & 1.2 & 4.75 \\
\hline
Minimum $\pt^{\rm hard}~[\gev/c]$ & CKIN(3) & 2.1 & 2.75 \\
\hline
CTEQ4L     & MSTP(51) & 4032 & 4032 \\
Proton PDF & MSTP(52) & 2 & 2 \\
\hline
Switch off & MSTP(81) & 0 & 0 \\
multiple & PARP(81) & 0 & 0 \\
interactions & PARP(82) & 0 & 0 \\
\hline
Initial/Final parton & MSTP(61) & 1 & 1 \\
shower on & MSTP(71) & 1 & 1 \\
\hline
2$^{\mathrm{nd}}$ order $\alpha_s$ & MSTP(2) & 2 & 2 \\
\hline
QCD scales & MSTP(32) & 2 & 2 \\
for hard scattering & PARP(34) & 1 & 1 \\
and parton shower & PARP(67) & 1 & 1 \\
& PARP(71) & 4 & 1 \\
\hline
Intrinsic $k_{\rm t}$ &  &  &  \\
from gaussian distr. with mean 0 & MSTP(91)      &  1    & 1 \\
   $\sigma$ $[\gev/c]$    & PARP(91) & 1.00 (pp) & 1.00 (pp) \\
                &          & 1.16 (p--Pb)   &  1.60 (p--Pb)   \\
                &          & 1.30 (Pb--Pb)   &  2.04 (Pb--Pb)   \\
upper cut-off (at 5 $\sigma$) $[\gev/c]$ & PARP(93) & 5.00 (pp)  & 5.00 (pp) \\
                &          & 5.81 (p--Pb)    & 8.02 (p--Pb)   \\
                &          & 6.52 (Pb--Pb)   &  10.17 (Pb--Pb)   \\
\hline
\hline
\end{tabular}
\end{center}
\end{table}

}

\clearpage
\pagestyle{plain}

\myappendix
\mychapter{Parameterization of the TPC tracking response:
           validation tests}
\label{App:tpc}

\pagestyle{myheadings}

This appendix is an extract from:\\
A.~Dainese and N.~Carrer, ALICE Internal Note ALICE-INT-2003-011 (2003).

After defining the track parameters used in the TPC, 
we report the results of the tests that were performed in order to validate
the parameterization tool for the use in physics studies. 
Refer to the complete 
note~\cite{notetpcparam} for details on the implementation of the method.

{\small

\mysection{Description of the track parameters used in the TPC}

We remind that the ALICE TPC is a cylinder shell with inner and outer 
radii of $\approx 85~\cm$ and $\approx 250~\cm$, respectively, and length of 
$\approx 500~\cm$. The read-out planes are azimuthally segmented in 18
sectors, covering an angle of $20^\circ$ each. At the inner radius 
every sector is $30~\cm$ 
($\simeq 85~\cm\times\tan 20^\circ$) wide. The non-active region between 
two adjacent sectors is approximately $2.7~\cm$ wide.

The ALICE global reference frame has the $z$ axis parallel to the beam line, 
and the $x$ and $y$ axes in the transverse plane, horizontal and vertical 
respectively. During track reconstruction,
the helix track model is defined in the local reference frame obtained 
from the global one with a rotation around the $z$ axis:
\begin{equation}
\begin{array}{l}
X = x\cdot \cos\alpha+y\cdot\sin\alpha \\
Y = -x\cdot \sin\alpha+y\cdot\cos\alpha \\
Z = z
\end{array}
\end{equation}
being $\alpha$ the azimuthal angle (w.r.t. the $x$ axis) of the TPC sector in 
which the 
track lies. Thus, $X$ is the radial coordinate in the sector. The 
track 
parameters are given at a reference plane for $X=X_{\rm ref}$.

The five track parameters are:
\begin{itemize}
\item $Y$: local $y$ coordinate of the track at the reference plane 
      $X=X_{\rm ref}$;
\item $Z$: local $z$ coordinate of the track at the reference plane 
      $X=X_{\rm ref}$; 
\item $k$: track curvature. $k=1/R$, where $R$ is the radius of the circle 
      obtained 
      projecting the track on the bending plane;  
\item $\gamma$: $k\cdot X_0$, being $X_0$ the
      local $x$ coordinate of the centre of the circle on the bending plane;
\item $\tan\lambda$: tangent of the track angle with the bending plane 
     ($\tan\lambda = p_z/\pt$).
\end{itemize}

\mysection{Validation tests and results}

In this section we present a comparison of the performances of the track 
reconstruction using:
\begin{itemize}
\item standard Kalman filter reconstruction in the TPC and in the ITS;
\item parameterization of the tracking in the TPC and standard Kalman 
filter in the ITS.
\end{itemize}
We compare tracking efficiency and resolution on the track parameters.
At the end of the section we report the gain in CPU time and 
disk space obtained performing the simulation with the parameterization 
tool we have implemented.

\subsection{Tracking efficiency in the TPC and in the ITS}

We start with a comparison of the total number of tracks found in the TPC:
when using the parameterization this number is lower by $15$-$16\%$ than the 
number obtained with standard reconstruction. The loss has two sources:
\begin{enumerate}
\item With the standard reconstruction we get about $2$-$3\%$ of \textsl{fake} 
tracks in the TPC ---a track is defined as \textsl{fake} if it has more 
than $10\%$ of incorrectly assigned clusters, typically the decay product 
of a particle that decays inside the TPC can originate a \textsl{fake} track 
(e.g. a charged pion from $\rm K_S^0\to \pi^+\pi^-$). Clearly, the 
parameterization will not give these tracks.
\item Tracks that enter the TPC in a non-active region (between two 
adjacent sectors) are not given by the parameterization, as they do not 
have a {\sl hit} in the inner pad row. Nevertheless, a part of them can then 
leave signal on a sufficient number of pad rows and be reconstructed by the 
Kalman filter. For this reason, using the parameterization we have a 
loss that amounts to about $12\%$. In Fig.~\ref{loss} we plot the 
entrance position, in the local 
coordinates $(Y,\,Z)$, of the tracks that are reconstructed by the Kalman 
filter and not given by the parameterization. They are concentrated at the 
edge of the TPC sectors, for $|Y|\simeq 15\ \cm$.
Recently, the simulation software was modified in order to allow to retrieve 
these tracks with the parameterization, but such change did not became 
available on time to be used for the studies presented in this thesis. 
\end{enumerate}

The lower `tracking efficiency' in the TPC 
introduced by the parameterization does not
limit the usefulness of the tool as, in principle, a correction could be 
applied. However, we chose not to apply such a 
correction and use the effect as a safety factor for the results of our 
simulation studies.

\begin{figure}[!t]
  \begin{center}
    \leavevmode
    \includegraphics[clip,angle=-90,width=.6\textwidth]{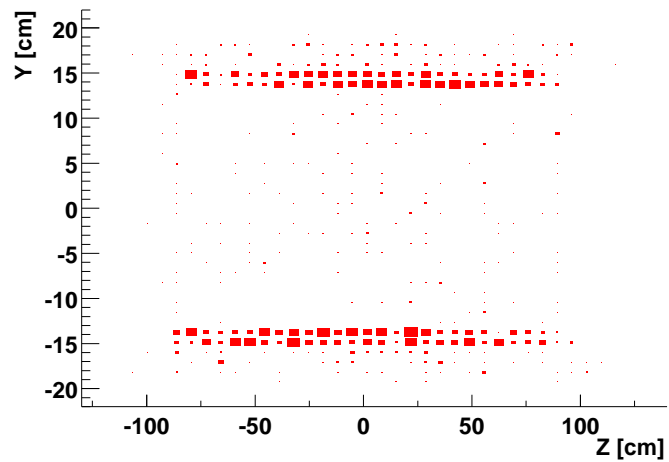}
    \caption{Entering position the TPC sector for the tracks found by the 
             Kalman filter and not given by the parameterization: they are 
             concentrated at the edges of the sector.} 
    \label{loss}
  \end{center}
\end{figure}
\begin{figure}[!t]
  \begin{center}
    \leavevmode
    \includegraphics[clip,width=.65\textwidth]{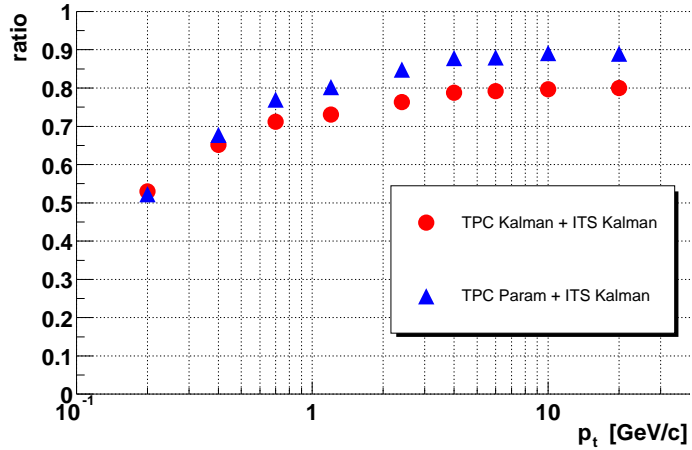}
    \caption{Ratio of the number of TPC tracks prolonged in the ITS to the 
             total number of tracks found in the TPC 
             (circles: Kalman filter in 
             the TPC, triangles: parameterization in the TPC).} 
    \label{ITStoTPC}
  \end{center}
\end{figure}

The next step is the comparison of the tracking efficiency in the ITS, 
defined as the ratio 
of the number of TPC tracks prolonged in the ITS to the total number of tracks 
found in the TPC. In this way we can understand if the performance of the 
Kalman filter in the ITS is altered by the fact of using the parameterization 
for the TPC. Track reconstruction in the ITS was performed with the standard 
requirement of having one assigned cluster in each of the 6 layers.
Only pions coming from the primary vertex were used for the comparison. The 
result is presented in Fig.~\ref{ITStoTPC}: the overall agreement is quite 
satisfactory; the slightly higher efficiency found when using the 
parameterization in the TPC is probably due to the fact that with the 
parameterization we do not have `low quality' tracks (e.g. \textsl{fake} 
tracks or tracks that do not have {\sl hits} in the inner pad rows of the 
TPC). 

\subsection{Resolution on the track parameters in TPC--ITS}

In Fig.~\ref{resolutions} we report the comparison of the resolutions on 
the main track parameters, $\pt,\ p_z,\ \tan\lambda,\ d_0(r\phi)$, after the 
reconstruction in the ITS. For all of these parameters we find that, 
using the parameterization of the tracking 
in the TPC, we reproduce the resolutions given by the 
standard reconstruction. Therefore, we conclude that the tool is suitable 
to be used for the simulation and analysis of physics cases.    

\begin{figure}[!h]
  \begin{center}
    \leavevmode
    \includegraphics[clip,width=.49\textwidth]{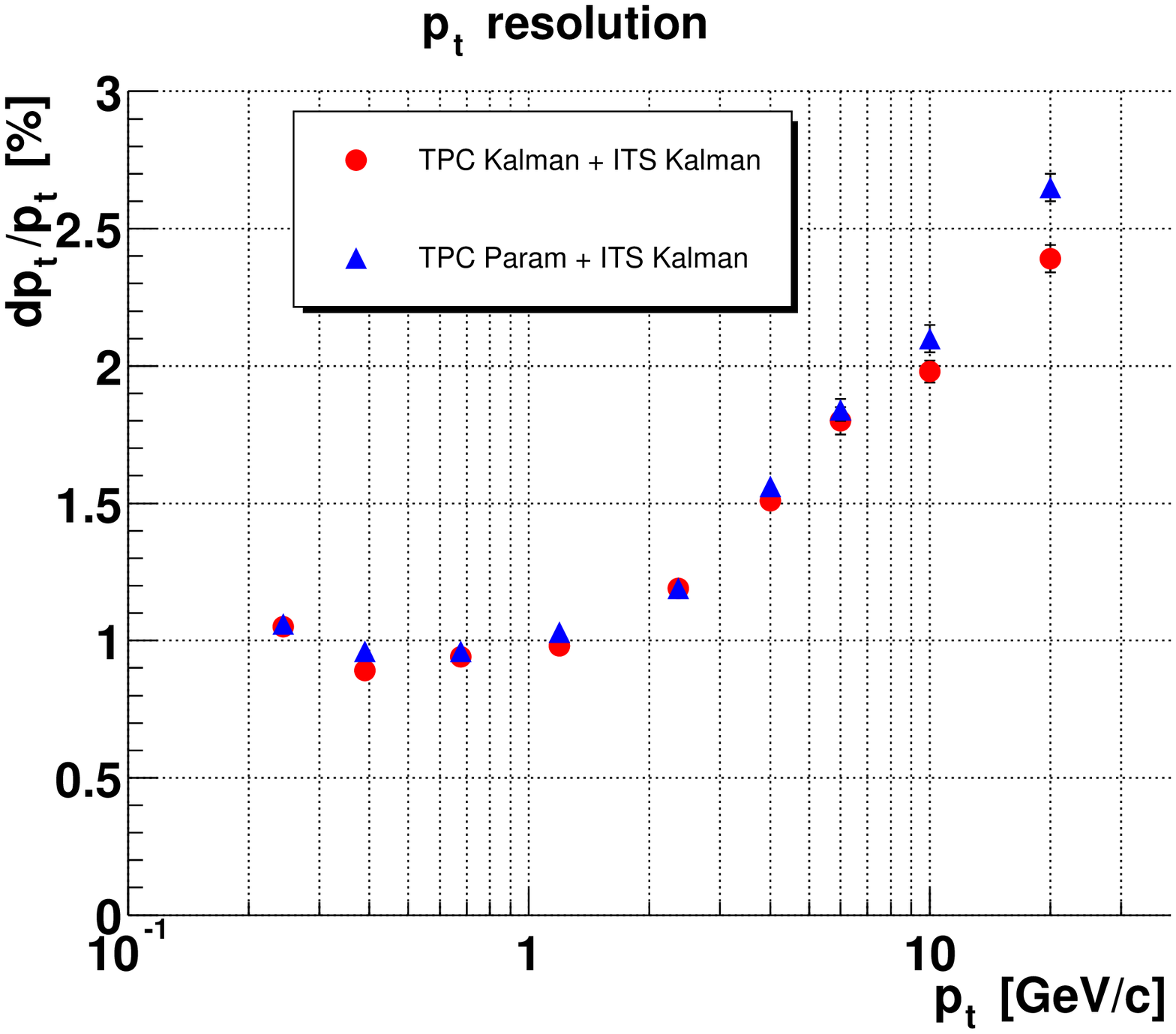}
    \includegraphics[clip,width=.49\textwidth]{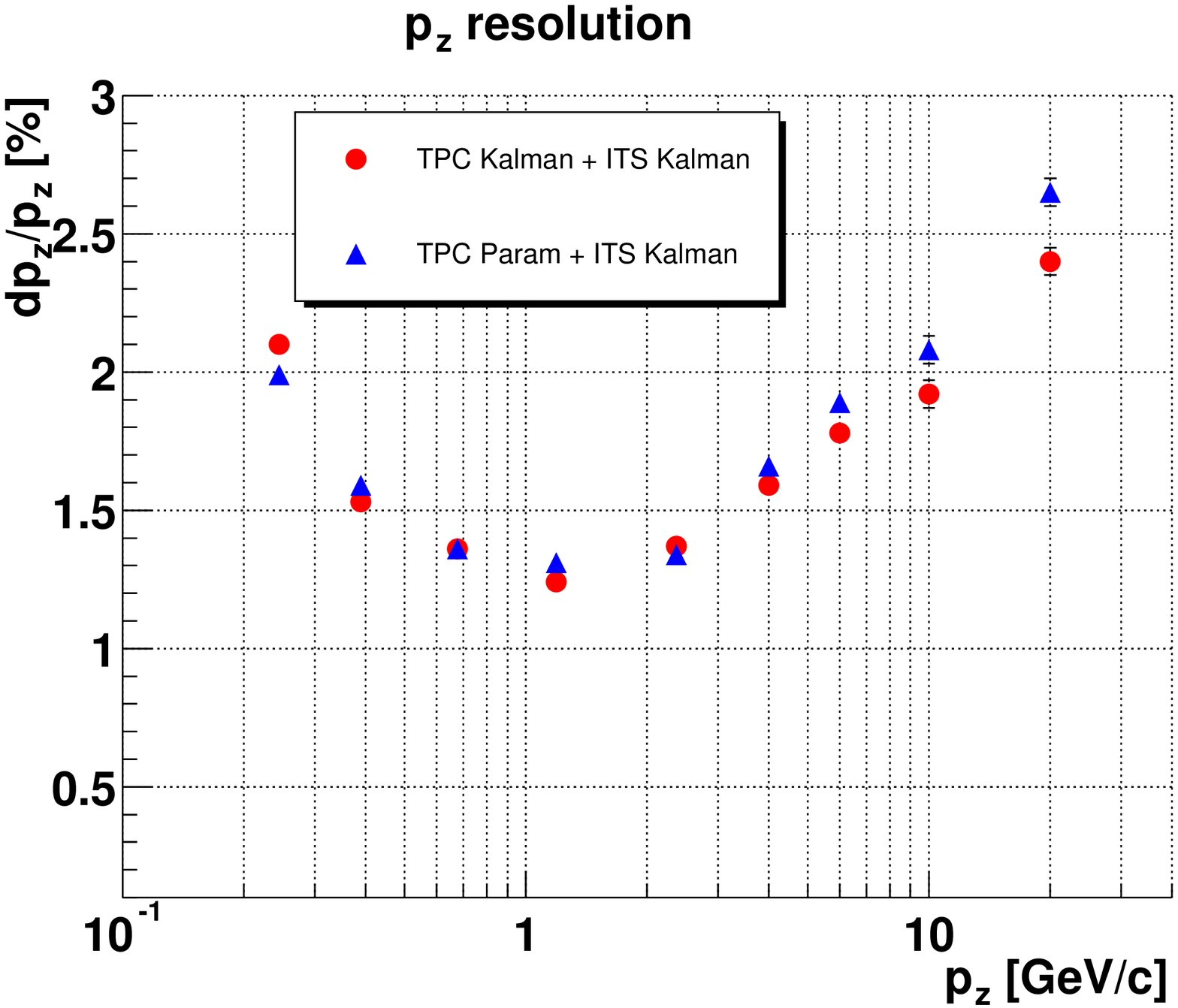}
    \includegraphics[clip,width=.49\textwidth]{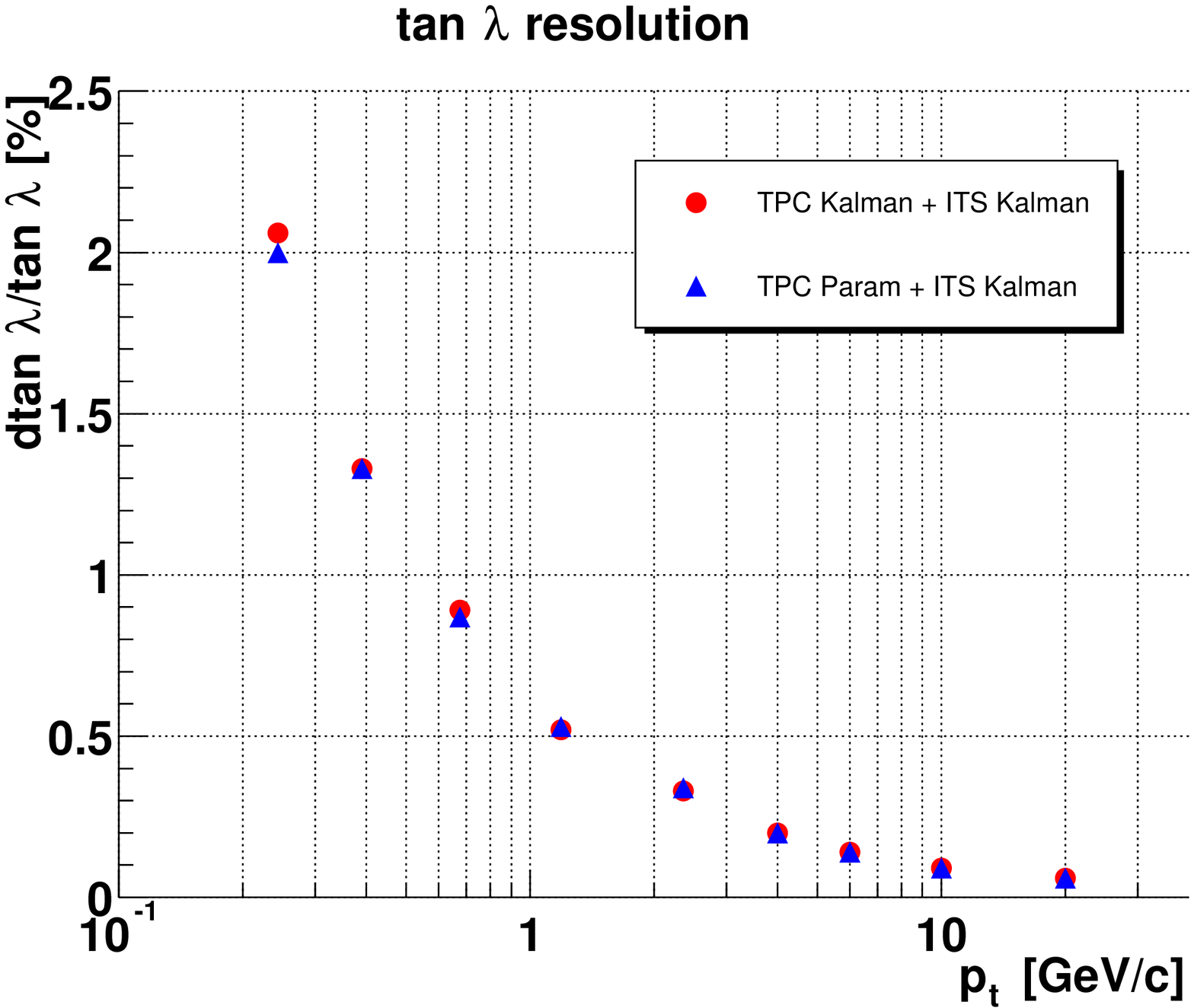}
    \includegraphics[clip,width=.49\textwidth]{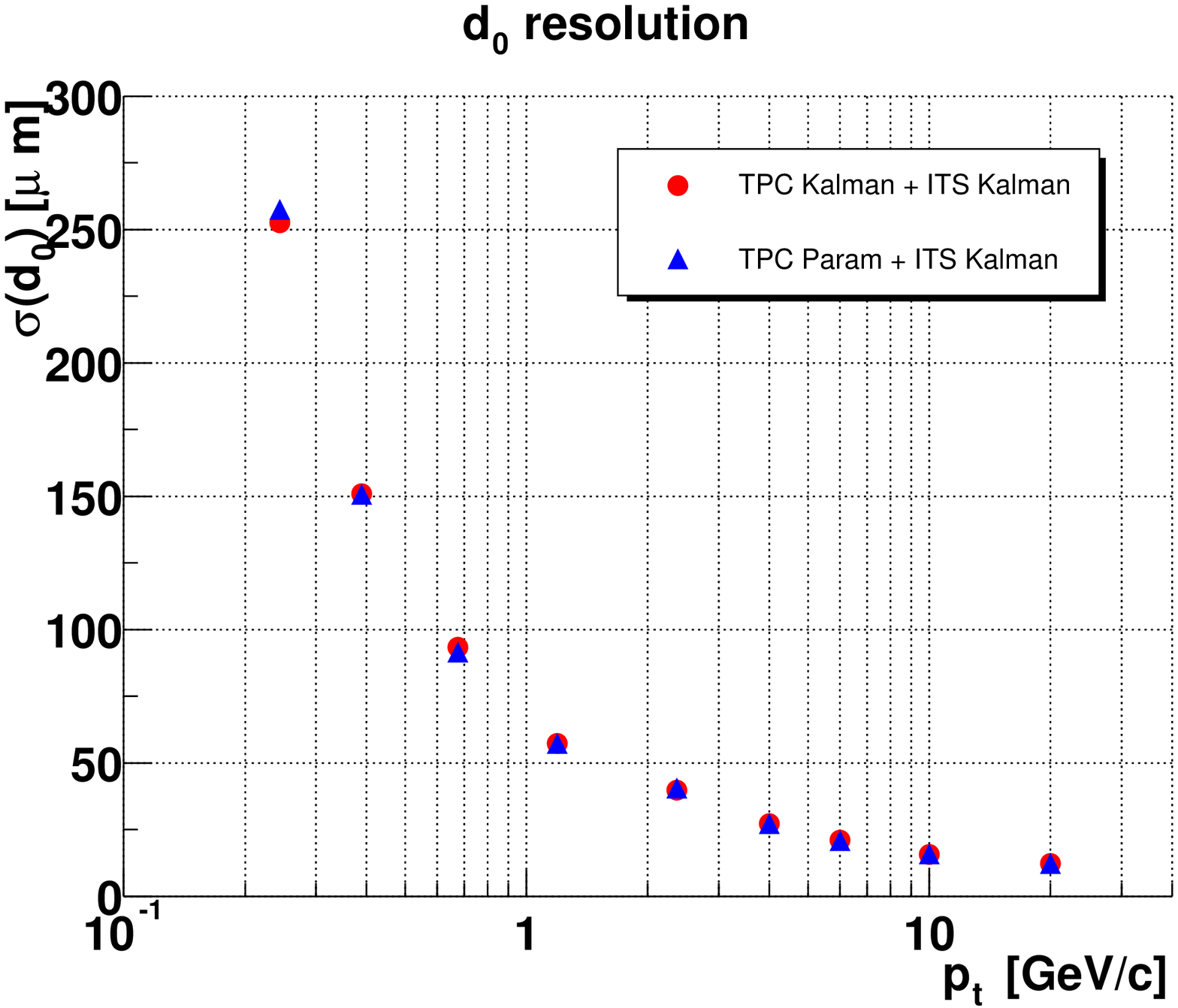}
    \caption{Comparison of the resolution on the main track parameters 
             (circles: Kalman 
             filter in the TPC, triangles: parameterization in the TPC).} 
    \label{resolutions}
  \end{center}
\end{figure}

\subsection{Effect on the required computing resources}

The CPU time and disk space required for the simulation of an event 
with 6000 charged particles per unit of rapidity are reported in 
Table~\ref{tabperf} (only particles in the rapidity interval $|y|<2$ were 
transported through the detector).
The reduction factor achieved with the introduction of the TPC tracking 
parameterization is 35.

The tool was integrated in the AliRoot~\cite{alisoft} 
framework (class \texttt{AliTPCtrackerParam}).

\begin{table}[!h]
\begin{center}
  \caption{CPU time (on a 800 MHz PC) and disk space for 1 
           central \PbPb~collision ($\dNdy=6000$), with particle transport 
           restricted to the rapidity interval $|y|<2$.}
  \begin{tabular}{llcc}
    \hline
    \hline
         &  & Full simulation and & Full simulation in $r<90~\cm$, \\
         &  & standard reconstruction &  parameterization in TPC +\\
         &  & in ITS--TPC     & standard reconstruction \\
         &  &             & in ITS \\
    \hline
    CPU time & [hours] & 9 & 0.25\\
    Disk space & [Mbytes] & 1200 & 35\\
    \hline
    \hline	
  \end{tabular}
  \label{tabperf}
\end{center}
\end{table}

}

\clearpage
\pagestyle{plain}

\pagestyle{plain}
\myappendix
\mychapter{Further studies \mbox{on the impact parameter resolution}}
\label{App:d0}

\pagestyle{myheadings}

{\small 

In this appendix we report some studies aimed at understanding the 
$\pt$-dependence and the high-$\pt$ values of the impact parameter
resolution. We also consider the dependence of the resolution on the 
thickness of the Silicon Pixel Detectors.
~\\

We start with the $\pt$-dependence of the resolutions on $d_0(r\phi)$
and $d_0(z)$, shown in Fig.~\ref{fig:d0resMain}.

For the $r\phi$ projection this dependence reflects what is expected 
from a contribution due 
to the spatial precision of the detectors added to a contribution due to the 
multiple scattering:
\begin{equation}
  \sigma(d_0) = \sigma_{\rm det.~res.} \oplus \sigma_{\rm scattering}.
\end{equation}
In this decomposition, the first term can be reasonably assumed to 
be independent of the particle momentum and direction, while the 
second depends on the momentum and on the amount of material
crossed by the particle (hence, also on its polar direction).

If we consider the simplified case of particles moving in the transverse plane
($p_z\simeq 0$), the material thickness can be taken as 
constant and the multiple scattering term should be proportional 
to $1/\pt$ (from $\theta^{RMS}_{\rm scattering} \propto 1/p$~\cite{pdg} 
and $p\simeq \pt$). From a fit of the $\pt$-dependence for pions in 
$|\eta|<0.1$ we have (see Fig.~\ref{fig:d0fit}):
\begin{equation}
  \sigma(d_0(r\phi))~[\mum]= 17 \oplus \frac{57}{\pt~[\gev/c]}.
\end{equation}

\begin{figure}[!t]
  \begin{center}
    \leavevmode
    \includegraphics[width=.65\textwidth]{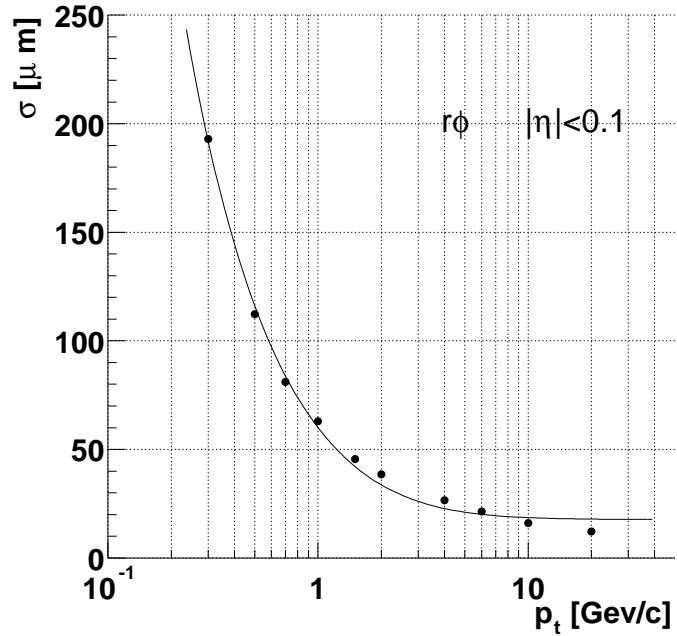}
    \caption{Impact parameter resolution in $r\phi$ for primary charged 
             pions in the pseudorapidity range $|\eta|<0.1$, fitted 
             to the expression $a\oplus  b/\pt$.} 
    \label{fig:d0fit}
  \end{center}
\end{figure}

For the $z$ projection the $\pt$-dependence is different 
and the two contributions cannot be clearly separated. In this case, 
another important effect has to be considered: the measurement of the 
$z$ impact parameter is strongly correlated to the measurement of the
track polar direction (i.e. of the track {\sl dip angle} 
$\lambda=\arctan(p_z/\pt)$). As it can be seen 
from Table~\ref{tab:its}, in the ITS, the resolution along the $z$ direction
is very good in the two central layers (SDD), while it is 
poorer in the innermost (SPD) and outermost (SSD) layers. 
This implies that the ITS alone does not provide precise 
information on the track {\sl dip angle}. This angle is essentially measured 
in the TPC, with a resolution that depends on the scatterings 
the particle has undergone in the ITS. The combination of the error
on the polar direction and the errors on the position of the 
clusters in the ITS determines the resolution on the $z$ projection 
of the impact parameter.
~\\

Since the multiple scattering contribution to the impact parameter resolution 
is negligible for high-$\pt$ tracks, the values of the resolutions 
at $\pt=20~\gev/c$, $\simeq 12~\mum$ for the $r\phi$ projection and 
$\simeq 40~\mum$ for the $z$ projection, can be estimated from the 
radial positions and spatial resolutions of the 6 ITS layers.
This kind of cross-checks are important in order to understand the results 
given by the detailed detector simulation and by the track reconstruction 
algorithm.

Because the number of points is quite large (6), 
we choose a Monte Carlo approach,
rather than an analytical one. High-$\pt$ tracks can be approximated to 
straight lines in the ITS; in fact, for a track of $\pt=20~\gev/c$ in 
a magnetic field of 0.4~T, the radius of curvature is $R=167~\m$ and 
the sagitta corresponding to a cord of length $\ell=0.45~\m$ 
(outer radius of the ITS) is $\ell^2/8R\simeq 150~\mum$. 
We, therefore, use the following procedure for the estimation of the 
impact parameter resolution:
\begin{itemize}
  \item generate a track as a straight line from the origin $(0,0,0)$ 
        (primary vertex) with a random {\sl dip angle} $\lambda<45^\circ$;
  \item assign the 6 hits in the ITS layers as the intersection points 
        with 6 cylinders having the radii of the ITS layers (from 
        Table~\ref{tab:its});
  \item assign the 6 `clusters' by a gaussian smearing of the positions 
        of the hits in $r\phi$ and in $z$ according to the spatial resolutions
        of the different detector types (Table~\ref{tab:its});
  \item fit a straight line to the 6 `clusters';
  \item determine the two impact parameter projections as the distances 
        of the fitted line to the origin in the transverse plane 
        and in the $z$ direction;
  \item iterate the previous steps for $10^4$ tracks and estimate the 
        resolutions as the dispersion of the obtained distributions 
        of impact parameters.
\end{itemize}  

Figure~\ref{fig:toyITS1} presents the distributions we obtained: the 
gaussian fits give $\sigma(d_0(r\phi))=10~\mum$ and 
$\sigma(d_0(z))=70~\mum$. While for the $r\phi$ projection the 
value is very close to that given by the full simulation, for the 
$z$ projection we get an estimated resolution larger by a factor 
2. This difference can be explained by observing that, as already 
pointed out, the resolution on 
$d_0(z)$ is strongly dependent on the resolution on the track 
{\sl dip angle} $\lambda$, which is not so good if only the ITS points 
are considered. The information from the TPC is 
essential to determine the polar direction of the track. For high-$\pt$ 
tracks ($>20~\gev/c$), the overall resolution (TPC--ITS) on $\tan\lambda$ is 
of about $0.05\%$. If we repeat the same procedure 
using also this information in the fit, we
obtain an estimate of $34~\mum$ for the resolution on the $z$ projection 
of $d_0$ (Fig.~\ref{fig:toyITS2}). Now both estimated resolutions
are compatible with the results of the full simulation.   
~\\

\begin{figure}[!t]
  \begin{center}
    \leavevmode
    \includegraphics[width=\textwidth]{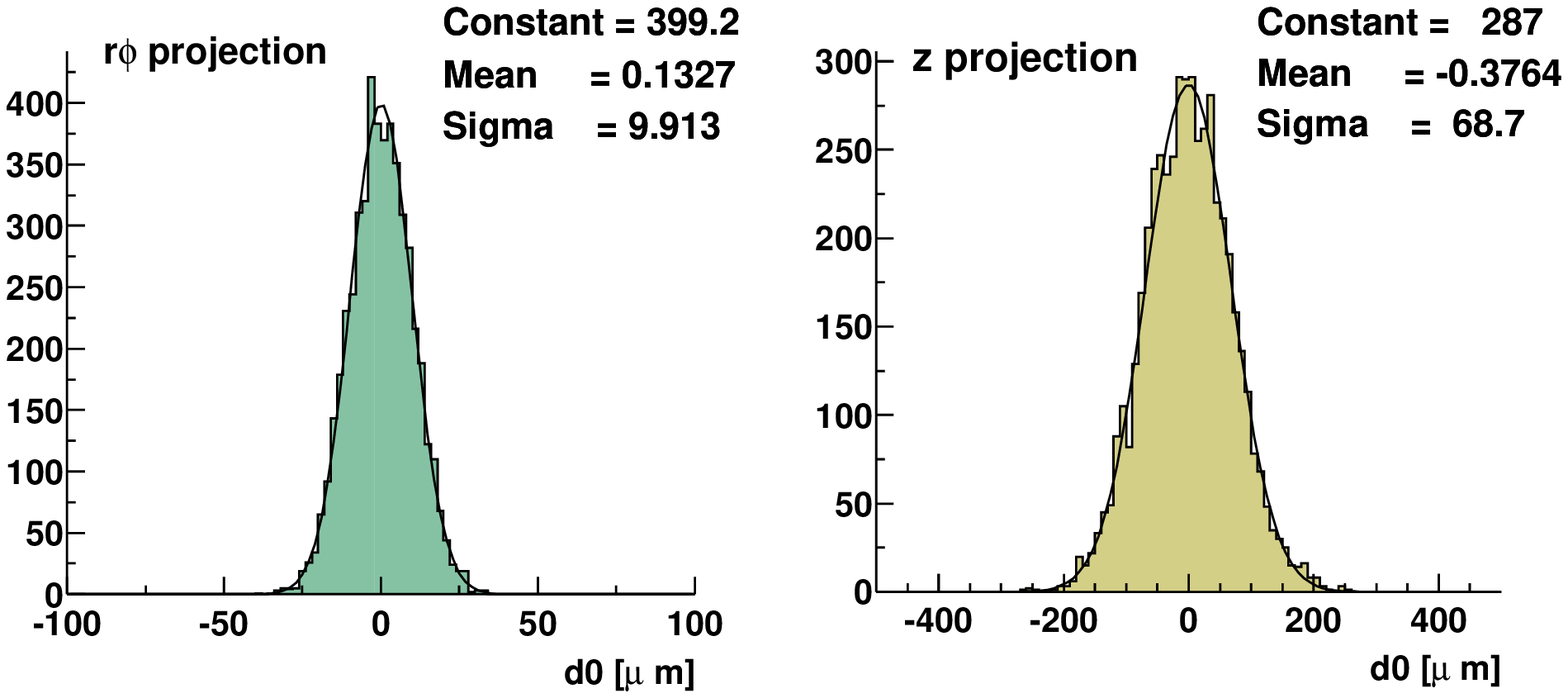}
    \caption{Impact parameters for high-$\pt$ tracks estimated with a 
             Monte Carlo approach based on a simple modeling of the 
             ITS.}   
    \label{fig:toyITS1}
    \vglue1cm
    \includegraphics[width=\textwidth]{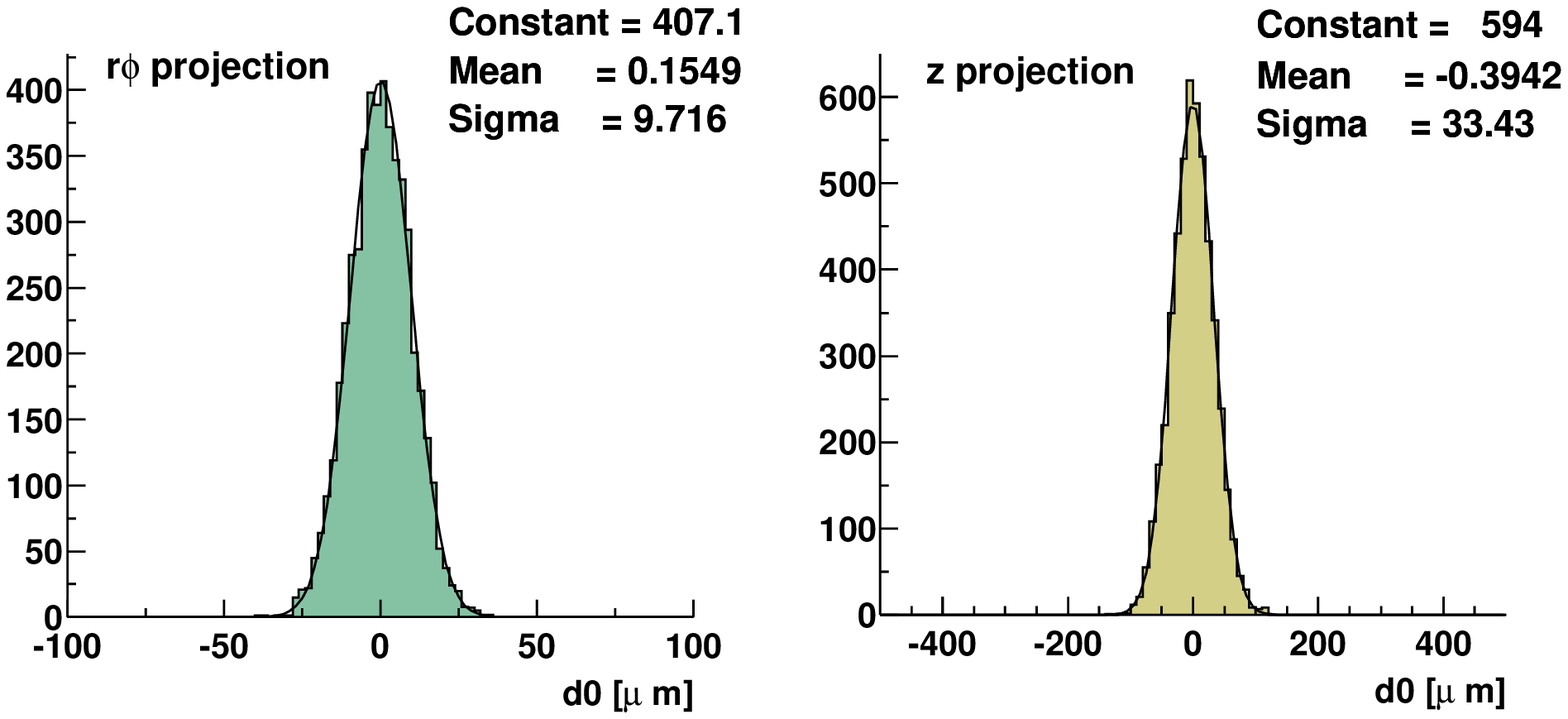}
    \caption{Same as Fig.~\ref{fig:toyITS1}. In this case the information 
             on track polar direction given by the TPC is included.}   
    \label{fig:toyITS2}
  \end{center}
\end{figure}

As a last point, we study the dependence of the impact parameter resolution 
on the amount of material in the two pixel layers. In fact, since these 
detectors are the closest to the interaction point, it is very important 
to know how much their thickness affects the $d_0$ resolution.
 
All the results presented in this thesis were 
obtained using in the simulation a total thickness of $400~\mum$ 
of silicon per pixel layer ($200~\mum$ for the detector plus $200~\mum$ 
for the readout electronics), which is the current design value. 
However, only prototypes with a thickness of $600~\mum$ have been produced 
and successfully tested up to now and it is not yet clear if the design 
parameters can be reached.
 
In Fig.~\ref{fig:spdthick} we show the effect of thicker pixel layers 
($600~\mum$) on the resolutions for pions in \PbPb: the resolution 
on the $r\phi$ projection is worse by $\simeq 10\%$ at $\pt=1~\gev/c$. 

\clearpage
\begin{figure}[!t]
  \begin{center}
    \leavevmode
    \includegraphics[width=.65\textwidth]{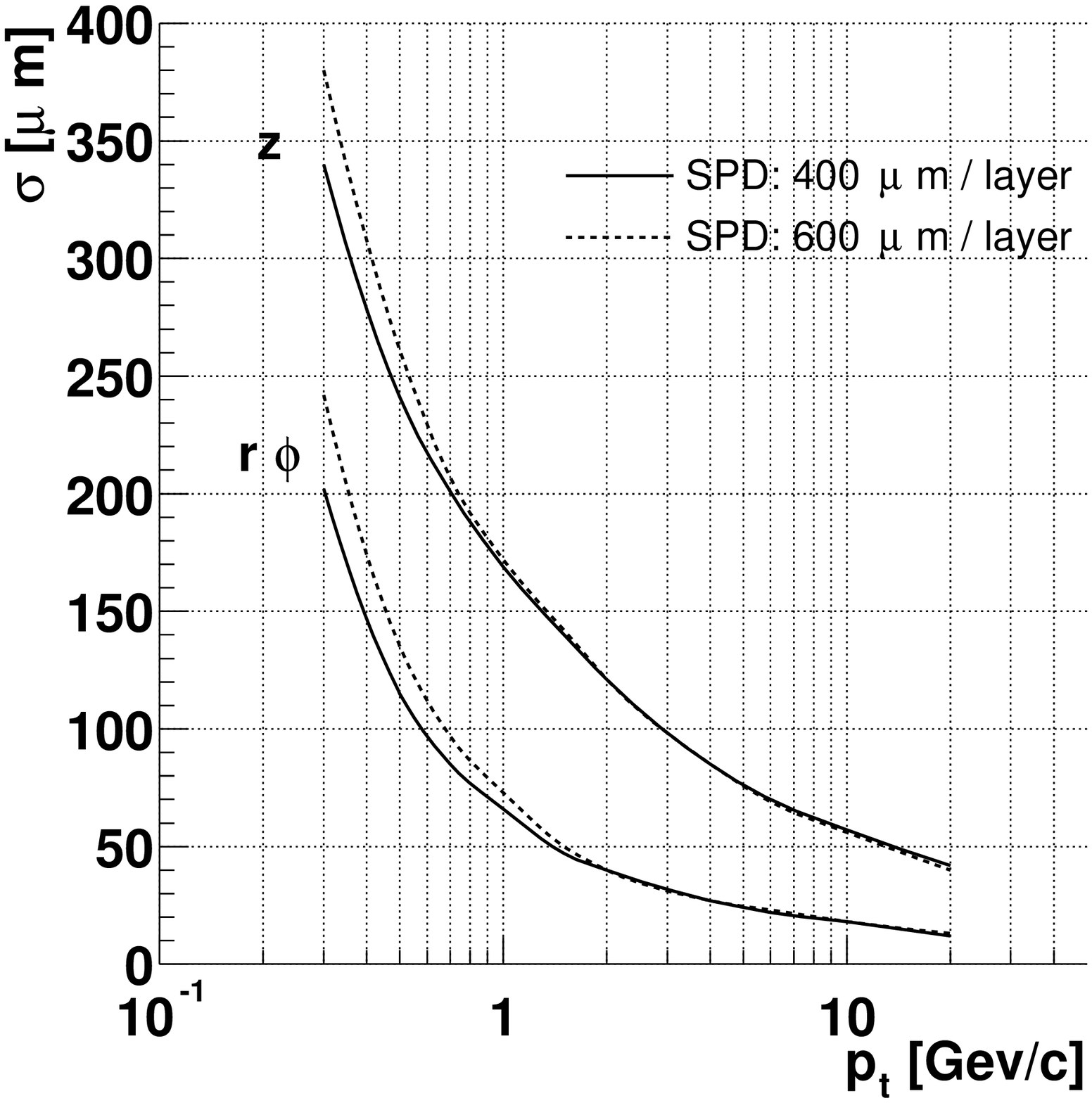}
    \caption{Impact parameter resolutions for pions in \PbPb~collisions for
             two values of the thickness of the Silicon Pixel Detectors: 
             $400~\mum$/layer (current design value) and $600~\mum$/layer.}   
    \label{fig:spdthick}
  \end{center}
\end{figure}
~\\

} 

\clearpage
\pagestyle{plain}

\pagestyle{empty}

\chapter*{\centering \LARGE{\sl Acknowledgements}}
\addcontentsline{toc}{chapter}{Acknowledgements}

\thispagestyle{empty}

I would like to thank my supervisor, Maurizio Morando, for his constant
support and, in particular, for encouraging me to attend schools
and conferences and for giving me the possibility to spend long 
and fruitful periods at CERN during the past three years. There, I was
guided by Nicola Carrer and Karel 
$\check{\mathrm{S}}$afa$\check{\mathrm{r}}$\'\i k. 
I am extremely grateful to them 
for their teachings and their support. 

I thank Federico Antinori, Emanuele Quercigh and Rosario Turrisi, 
for their many useful suggestions that improved my work and this write-up, 
and all the ALICE group in Padova.

During my stays at CERN I enjoyed many stimulating discussions with
Andreas Morsch, Yura Belikov, Jurgen Schukraft, Jean-Pierre Revol, Guy Paic, 
Yiota Foka, Andres Sandoval, Ramona Vogt, Marco Monteno,
Massimo Masera, Fabrizio Pierella, Philippe Crochet, Peter Hristov,
Carlos Salgado, Urs Wiedemann and Nestor Armesto. I wish to thank them all 
and I look forward to further collaborate with them.
~\\

The love and hard work of my parents allowed me to pursue my studies and 
researches: thank You.

Many friends accompanied me during these years: Andrea, Andrea, Andrea,
Paola, Ilaria, Riccardo, Dario, Patrick, Bernardo, Peter, Toni, Anna, 
Carlos, Nicola, Rosario, Marco, Alessandro, 
Lorenzo, Carla, Daniele, Massimo, Francesca.
  
I want also to thank Caf\'e Vieux Carouge for its delicious fondues
and Gino Spagnolo, my landlord, for the wonderful armchair, sitting on which 
most of this thesis was written. 
~\\

All my love goes to a very special person whose charm and sweetness
are irreplaceable to me: Barbara.

\end{document}